\pdfoutput=1
\newcommand*{\ATLASLATEXPATH}{}
\documentclass[PAPER, UKenglish, texlive=2016, LANGEDIT=true, LANGSHOW=false,cernpreprint]{\ATLASLATEXPATH atlasdoc}
\LEcontact{Steve Lloyd s.l.lloyd@qmul.ac.uk}
 
\usepackage[subfigure=true]{\ATLASLATEXPATH atlaspackage}
\usepackage{\ATLASLATEXPATH atlasbiblatex}
 
\usepackage{\ATLASLATEXPATH atlasphysics}
 
\graphicspath{{logos/}{figures/}}
 
\usepackage{ANA-TOPQ-2018-15-PAPER-defs}
\usepackage{lscape}
\usepackage{pdflscape}
\usepackage{multirow}
\usepackage{floatrow}
\usepackage{siunitx}
\usepackage{multirow}
 
\AtlasTitle{Measurements of  top-quark pair differential and double-differential cross-sections in the $\ell$+jets channel with $pp$ collisions at $\sqrt{s}=13$ TeV using the ATLAS detector}

\AtlasAbstract{
Single- and double-differential cross-section measurements are presented for the production of top-quark pairs, in the lepton + jets channel at particle and parton level. Two topologies, resolved and boosted, are considered and the results are presented as a function of several kinematic variables characterising the top and $t\bar{t}$ system and jet multiplicities. The study was performed using data from $pp$ collisions at centre-of-mass energy of 13 TeV collected  in 2015 and 2016 by the ATLAS detector at the CERN Large Hadron Collider (LHC), corresponding to an integrated luminosity of $36~\mathrm{fb}^{-1}$. 
Due to the large $t\bar{t}$ cross-section at the LHC, such measurements allow a detailed study of the properties of  top-quark production and decay, enabling precision tests of several Monte Carlo generators and fixed-order Standard Model predictions. 
Overall, there is good agreement between the theoretical predictions and the data.
}
 
\author{The ATLAS Collaboration}
 
\AtlasRefCode{ANA-TOPQ-2018-15}
 
\PreprintIdNumber{CERN-EP-2019-149}

\AtlasJournal{\EPJC}
\AtlasJournalRef{\EPJC 79 (2019) 1028}
\AtlasDOI{10.1140/epjc/s10052-019-7525-6}

\AtlasCoverSupportingNote{Measurements of  top-quark pair differential cross-sections in the $\ell+$ jets channel in $pp$ collisions at $\sqrt{s}=13$ TeV using the ATLAS detector}{https://cds.cern.ch/record/2304451}
 
\AtlasCoverCommentsDeadline{17 July 2019}
 
\AtlasCoverAnalysisTeam{Petr Baron, Lorenzo Bellagamba, Veronique Boisvert, Riccardo Di Sipio, Federica Fabbri, Michele Faucci Giannelli, Jiri Kvita, Francesco La Ruffa, Matteo Negrini, Serena Palazzo, Marino Romano, Matteo Scornajenghi, Francesco Span\`o, Enrico Tassi}

\AtlasCoverEdBoardMember{Martine Bosman (chair), Katharina Bierwagen, Marija Vranjes Milosavljevic}

\AtlasCoverEgroupEditors{atlas-topq-2018-15-editors@cern.ch}
 
\AtlasCoverEgroupEdBoard{atlas-topq-2018-15-editorial-board@cern.ch}
 
\hypersetup{pdftitle={ATLAS document},pdfauthor={The ATLAS Collaboration}}
 
\begin{document}
 
\maketitle
 
\tableofcontents

 
\section{Introduction}
\label{sec:intro}
 
The detailed studies of the characteristics of top-quark pair ($\ttbar{}$) production as a function of different kinematic variables that can now be performed at the Large Hadron Collider (LHC) provide a~unique opportunity to test the Standard Model (SM) at the~\TeV{} scale.
Furthermore, extensions to the SM may modify the \ttb{} differential cross-sections in ways that an inclusive cross-section measurement~\cite{Frederix:2009} is not sensitive to. In particular, such effects may distort the top-quark momentum distribution, especially at higher momentum~\cite{Atwood:1994vm,Englert:2012by}.
Therefore, a precise measurement of the \ttb{} differential cross-sections has the potential to enhance the sensitivity to possible effects beyond the SM, as well as to challenge theoretical predictions that now reach next-to-next-to-leading-order (NNLO) accuracy in perturbative quantum chromodynamics (pQCD)~\cite{Czakon:2015owf,Czakon:2016dgf,Catani:2019hip}.
Moreover, the differential distributions are sensitive to the differences between Monte Carlo (MC) generators and their settings, representing a valuable  input to the tuning of the MC parameters. This aspect is relevant for all the searches and measurements that are limited by the accuracy of the modelling of $\ttbar{}$ production.

The ATLAS~\cite{TOPQ-2011-07,TOPQ-2012-08,TOPQ-2013-07,TOPQ-2014-15,TOPQ-2015-06,TOPQ-2015-07,TOPQ-2016-04,TOPQ-2016-01,TOPQ-2017-01}
and  CMS~\cite{CMS-TOP-11-013,CMS-TOP-12-028,CMS-TOP-14-012,CMS-TOP-16-008,CMS-TOP-14-013,CMS-TOP-17-014,CMS-TOP-18-004} Collaborations have published measurements of $\ttbar{}$ differential cross-sections at centre-of-mass energies ($\sqrt{s}$) of 7~\TeV{}, 8~\TeV{} and 13~\TeV{} in $pp$ collisions using final states containing leptons,  both in the full phase-space using parton-level variables and in fiducial phase-space regions using observables constructed from final-state particles (particle-level).
These results have been largely used to improve the modelling of MC generators~\cite{ATL-PHYS-PUB-2016-016,ATL-PHYS-PUB-2016-020,ATL-PHYS-PUB-2017-007,ATL-PHYS-PUB-2018-009,CMS-GEN-17-001-003} and to reduce the uncertainties in the gluon parton distribution function (PDF)~\cite{Czakon:2016olj}.
 
The results presented in this paper probe the top-quark kinematic properties at $\sqrt{s}=13$~\TeV{}
and complement recent measurements involving leptonic final states by ATLAS~\cite{TOPQ-2016-04,TOPQ-2016-01,TOPQ-2017-01} and CMS~\cite{CMS-TOP-16-008,CMS-TOP-17-014} by measuring single- and double-differential cross-sections in the selected fiducial phase-spaces and extrapolating the results to the full phase-space at the parton level. 
 
In the SM, the top quark decays almost exclusively into a~\Wboson boson and a~$b$-quark. The signature of a~\ttbar{} decay is therefore determined by the \Wboson boson decay modes. This analysis makes use of the \ljets{} \ttbar{} decay mode, also called the semileptonic channel, where one \Wboson boson decays into an electron or a~muon and a~neutrino, and the other \Wboson boson decays into a quark--antiquark pair, with the two decay modes referred to as the $e$+jets and $\mu$+jets channels, respectively. Events in which the \Wboson boson decays into an electron or muon through a $\tau$-lepton decay may also meet the selection criteria. Since the reconstruction of the top quark depends on its decay products, in the following the two top quarks are referred to as `hadronically (or leptonically) decaying top quarks' (or alternatively `hadronic/leptonic top' ), depending on the \Wboson boson decay mode.
 
Two complementary topologies of the \ttbar{} final state in the \ljets{} channel are exploited, referred to as `resolved' and `boosted', where the decay products of the hadronically decaying top quark are either angularly well separated or collimated into a~single large-radius jet reconstructed in the calorimeter, respectively.
As the jet selection efficiency of the resolved analysis decreases with increasing top-quark transverse momentum, the boosted selection allows events with higher-momentum hadronically decaying top quarks to be efficiently selected.

The differential cross-sections for \ttbar{} production are measured as a function of a large number of variables  (described in Section~\ref{sec:observables}) including, for the first time in this channel in ATLAS, double-differential distributions. Moreover, the amount of data and the reduced detector uncertainties compared to previous publications also allows, for the first time, double differential measurements in the boosted topology to be made.
 
The analysis investigates a list of variables that characterise various aspects of the \ttbar{} system production. In particular, the variables selected are sensitive to the kinematics of the top and anti-top quarks and of the \ttbar{} system or are sensitive to initial- and final-state radiation effects. Furthermore, the variables are sensitive to the differences among PDFs and possible beyond the SM effects. Both normalised and absolute differential cross-sections are measured, with more emphasis given to the discussion of the normalised results.

Differential cross-sections are measured as a function of different variables in the fiducial and full phase-spaces, since they serve different purposes: the particle-level cross-sections in the fiducial phase-space are particularly suited to MC tuning while the parton-level cross-sections, extrapolated to the full phase-space, are the observables to be used for stringent tests of higher-order pQCD predictions and for the determination of the proton PDFs and the top-quark pole mass in pQCD analyses.

\FloatBarrier
 
\section{ATLAS detector}\label{sec:detector}
 
\newcommand{\AtlasCoordFootnote}{ATLAS uses a right-handed coordinate system with its origin at the nominal
interaction point in the centre of the detector. The positive $x$-axis is defined by the direction from the interaction point to the centre of the LHC ring, with the positive $y$-axis pointing upwards, while the beam direction defines the $z$-axis. Cylindrical coordinates $(r,\phi)$ are used in the transverse plane, $\phi$ being the azimuthal angle around the $z$-axis. The pseudorapidity $\eta$ is defined in terms of the polar angle $\theta$ by $\eta=-\ln\tan(\theta/2)$. Rapidity is defined as $y$ = $0.5 \ln[(E + p_z)/(E-p_z)]$ where $E$ denotes the energy and $p_z$ is the component of the momentum along the beam direction. The angular distance $\DeltaR$ is defined as $\sqrt{(\Delta y)^2 + (\Delta \phi)^2}$.}
 
ATLAS is a multipurpose detector~\cite{PERF-2007-01} that provides nearly full solid angle\footnote{\AtlasCoordFootnote}
coverage around the interaction point.
Charged-particle trajectories with pseudorapidity $|\eta| <2.5$ are reconstructed in the inner detector, which comprises a silicon pixel detector, a silicon microstrip detector and a~transition radiation tracker (TRT). The innermost pixel layer, the insertable B-layer~\cite{ATLAS-TDR-19,PIX-2018-001}, was added before the start of 13~\TeV{} LHC operation at an average~radius of 33~mm around a~new, thinner beam pipe. The inner detector is embedded in a superconducting solenoid generating a~2~T axial magnetic field, allowing precise measurements of charged-particle momenta. The calorimeter system covers the pseudorapidity range \(|\eta| < 4.9\).
Within the region \(|\eta|< 3.2\), electromagnetic calorimetry is provided by barrel and endcap high-granularity lead/liquid-argon (LAr) calorimeters, with an additional thin LAr presampler covering \(|\eta| < 1.8\), to correct for energy loss in material upstream of the calorimeters.
Hadronic calorimetry is provided by the steel/scintillating-tile calorimeter, segmented into three barrel structures within \(|\eta| < 1.7\), and two copper/LAr hadronic endcap calorimeters. The solid angle coverage is completed with forward copper/LAr and tungsten/LAr calorimeter modules optimised for electromagnetic and hadronic measurements respectively. The calorimeters are surrounded by a~muon spectrometer within a~magnetic field provided by air-core toroid magnets with a~bending integral of about 2.5~Tm in the barrel and up to 6~Tm in the endcaps.
Three stations of precision drift tubes and cathode-strip chambers provide an accurate measurement of the muon track curvature in the region  $|\eta| < 2.7$.
Resistive-plate and thin-gap chambers provide muon triggering capability up to $|\eta| = 2.4$.
 
Data were selected from inclusive $pp$ interactions using a~two-level trigger system~\cite{TRIG-2016-01}. A~hardware-based trigger uses custom-made hardware and coarser-granularity detector data to initially reduce the trigger rate to approximately $100\,$kHz from the original 40~MHz LHC bunch crossing rate.
A software-based high-level trigger, which has access to full detector granularity, is applied to further reduce the event rate to $1\,$kHz.
 
\FloatBarrier
\section{Data and simulation} \label{sec:samples}
The differential cross-sections are measured using data collected during the 2015 and 2016 LHC $pp$ stable collisions at $\rts =13$~\TeV{} with 25~ns bunch spacing and an average number of $pp$ interactions per bunch crossing  $\left\langle\mu\right\rangle$ of around 23.
The selected data sample, satisfying beam, detector and data-taking quality criteria, correspond to an integrated luminosity of \lumitot.

The data were collected using single-muon or single-electron triggers. For each lepton type, multiple trigger conditions were combined to maintain good efficiency in the full momentum range, while controlling the trigger rate. Different transverse momentum (\pt{}) thresholds were applied in the 2015 and 2016 data taking. In the data sample collected in 2015, the \pt{} thresholds for the electrons were 24~\GeV{}, 60~\GeV{} and 120~\GeV{}, while for muons the thresholds were 20~\GeV{} and 50~\GeV{}; in the data sample collected in 2016,  the \pt{} thresholds for the electrons were  26~\GeV{}, 60~\GeV{} and 140~\GeV{}, while for muons the thresholds were 26~\GeV{} and 50~\GeV{}.
Different $\pt$ thresholds were employed since tighter isolation or identification requirements were applied to the triggers with lowest \pt{} thresholds.

The signal and background processes are modelled with various MC event generators described below and summarised in Table~\ref{tab:MC}. Multiple overlaid $pp$ collisions were simulated with the soft QCD processes of \PYTHIAV{8.186}~\cite{Sjostrand:2007gs} using parameter values from the A2 set of tuned parameters (tune)~\cite{ATL-PHYS-PUB-2012-003} and the MSTW2008LO~\cite{Martin:2009iq} set of PDFs  to account for the effects of additional collisions from the same and nearby bunch crossings (pile-up). Simulation samples are reweighted so that their pile-up profile matches the one observed in data. The simulated samples are always reweighted to have the same integrated luminosity of the data.
 
The \EvtGen{} v1.2.0 program~\cite{EvtGen} was used to simulate the decay of bottom and charm hadrons for all event generators except for  \SHERPA\cite{sherpa2.2}. The detector response was simulated~\cite{SOFT-2010-01} in \GEANT4~\cite{geant4}. A `fast simulation'~\cite{ATL-PHYS-PUB-2010-013} (denoted by AFII in the plots throughout the paper), utilising parameterised showers in the calorimeter~\cite{ATL-PHYS-PUB-2010-013}, but with full simulation of the inner detector and muon spectrometer, was used in the samples generated to estimate \ttb{} modelling uncertainties. The data and MC events were reconstructed with the same software algorithms.
 
\subsection{Signal simulation samples} \label{sec:signalMC}
In this section the MC generators used for the simulation of \ttb{} event samples are described for the nominal sample, the alternative samples used to estimate systematic uncertainties and the other samples used in the comparisons of the measured differential cross-sections~\cite{ATL-PHYS-PUB-2016-004}. The top-quark mass ($m_t$) and width were set to 172.5 \GeV{} and 1.32 \GeV~\cite{PhysRevLett.110.042001}, respectively, in all MC event generators.
 
For the generation of \ttb{} events, the \POWHEGBOXV{v2}~\cite{powheg-box,Nason:2004rx,Frixione:2007vw,Alioli:2010xd} generator with the NNPDF30NLO PDF sets~\cite{NNPDF3.0} in the matrix element (ME) calculations was used. Events where both top quarks decayed hadronically  were not included.
The parton shower, fragmentation, and the underlying events were simulated using \PYTHIAV{8.210}~\cite{Sjostrand:2007gs} with the NNPDF23LO PDF~\cite{NNPDF} sets and the A14 tune ~\cite{ATL-PHYS-PUB-2014-021}. The \hdamp{} parameter, which controls the \pt{} of the first gluon or quark emission beyond the Born configuration in \POWHEGBOXV{v2}, was set to $1.5\,m_t$~\cite{ATL-PHYS-PUB-2016-020}. The main effect of this parameter is to regulate the high-\pt{} emission against which the \ttbar{} system recoils. Signal \ttb{} events generated with those settings are referred to as the nominal signal sample. In all the following figures and tables the predictions based on this MC sample are referred to as `\textsc{Pwg+Py8}'.

The uncertainties affecting the description of the hard gluon radiation are evaluated using two samples with different factorisation and renormalisation scales relative to the nominal sample, as well as a different \hdamp{} parameter value~\cite{ATL-PHYS-PUB-2018-009}. For one sample, the factorisation and renormalisation scales were reduced by a factor of 0.5,  the \hdamp{} parameter was increased to $3 m_t$ and the Var3cUp eigentune from the A14 tune was used. In all the following figures and tables the predictions based on this MC sample are referred to as `\textsc{Pwg+Py8} Rad.~Up'. For the second sample, the factorisation and renormalisation scales were increased by a factor of 2.0 while the \hdamp{} parameter was unchanged and the Var3cDown eigentune from the A14 tune was used. In all the following figures and tables the predictions based on this MC sample are referred to as `\textsc {Pwg+Py8} Rad.~Down'.
 
The effect of the simulation of the parton shower and hadronisation is studied using the \POWHEGBOXV{v2} generator with the NNPDF30NLO PDF interfaced to \HERWIGV{7.0.1}~\cite{Bellm:Herwig,Bahr:Herwig1} for the showering, using the MMHT2014lo68cl PDF set~\cite{MMHT2014} and the H7-UE-MMHT tune~\cite{Seymour:2013qka}. In all the following figures and tables the predictions based on this MC sample are referred to as `\textsc{Pwg+H7}'.
 
The impact of the generator choice, including matrix element calculation, matching procedure, parton-shower and hadronisation model, is evaluated using events generated with \SHERPAV{2.2.1}~\cite{sherpa2.2}, which models the zero and one additional-parton process at next-to-leading-order (NLO) accuracy
and up to four additional partons at leading-order (LO) accuracy  using the \textsc{MEPS@NLO}\ prescription~\cite{Hoeche:2012yf}, with the NNPDF3.0NNLO PDF set~\cite{NNPDF3.0}. The calculation uses its own parton-shower tune and hadronisation model.  In all the following figures and tables the predictions based on this MC sample are referred to as `\Sherpa{}'.

All the $t\bar{t}$ samples described are normalised to the NNLO+NNLL in pQCD by the means of a $k$-factor. The cross-section used to evaluate the $k$-factor is $\sigma_{t\bar{t}} = 832^{+20}_{-29}(\mathrm{scale})~\pm 35~(\mathrm{PDF}, \alphas)$~pb, as calculated with the Top++2.0 program to NNLO in pQCD, including soft-gluon resummation to next-to-next-to-leading-log order (NNLL)~\cite{Czakon:2011xx,Beneke:2011mq,Cacciari:2011hy,Baernreuther:2012ws,Czakon:2012zr,Czakon:2012pz,Czakon:2013goa}, and assuming $m_t$ = 172.5~\GeV{}. The first uncertainty comes from the independent variation of the factorisation and renormalisation scales, $\mu_{\mathrm{F}}$ and $\mu_{\mathrm{R}}$, while the second one is associated with variations in the PDF and $\alphas$, following the PDF4LHC prescription with the MSTW2008 68\% CL NNLO, CT10 NNLO and NNPDF2.3 5f FFN PDF sets, described in Refs.~\cite{Martin:2009bu,CT10,PDF4LHC,NNPDF}.
 
\subsection{Background simulation samples} \label{sec:BackgroundMC}
Several processes can produce the same final state as the $\ttbar{}$ \ljets{} channel. The events produced by these backgrounds need to be estimated and subtracted from the data to determine the top-quark pair cross-sections. They are all estimated by using MC simulation with the exception of the background events containing a fake or non-prompt lepton, for which data-driven techniques  are employed. The processes considered are $W$+jets, $Z$+jets production, diboson final states and single top-quark production, in the $t$-channel, in association with a $W$ boson and in the $s$-channel. The contributions from top and $\ttbar{}$ produced in association with weak bosons and \ttbar\ttbar{} are also considered. The overall contribution of these processes is denoted by $t+X$.
 
For the generation of single top quarks in the $tW$ channel and $s$-channel the \POWHEGBOXV{v1}~\cite{Re:2010bp,Alioli:2009je} generator with the CT10 PDF~\cite{CT10} sets in the ME calculations was used.
Electroweak $t$-channel single-top-quark events were generated using the \POWHEGBOXV{v1} generator. This generator uses the four-flavour scheme for the NLO ME calculations~\cite{Frederix:2012dh} together with the fixed four-flavour PDF set CT10f4. For these processes the parton shower, fragmentation, and the underlying event were simulated using \PYTHIAV{6.428}~\cite{Sjostrand:2006za} with the CTEQ6L1 PDF~\cite{Nadolsky:2008zw} sets and the corresponding Perugia 2012 tune (P2012)~\cite{Skands:2010ak}. The single-top-quark cross-sections for the $tW$ channel were normalised using its NLO+NNLL prediction, while the $t$- and $s$-channels were normalised using their NLO predictions~\cite{Kidonakis:2011wy,Kidonakis:2010ux,Kidonakis:2010tc,Kidonakis:2013zqa,Aliev:2010zk,Kant:2014oha}.
 
The modelling uncertainties related to the additional radiation in the generation of single top quarks in the $tW$- and $t$-channels are assessed using two alternative samples for each channel, generated with different factorisation and renormalisation scales and different P2012 tunes relative to the nominal samples. In the first two samples, the factorisation and renormalisation scales were reduced by a factor of 0.5 and the radHi tune was used. For the second two samples, the factorisation and renormalisation scales were increased by a factor of 2.0 and the radLo tune was used.
An additional sample is used to assess the uncertainty due to the method used in the subtraction of the overlap of $tW$ production of single top quarks and production of $\ttbar$ pairs from the $tW$ sample~\cite{Frixione:2008yi}. In the nominal sample the diagram removal method (DR) is used, while the alternative sample is generated using the diagram subtraction (DS) one. All the other settings are identical in the two samples.

Events containing $W$ or $Z$ bosons associated with jets were simulated using the \SHERPAV{2.2.1}~\cite{sherpa2.2} generator. Matrix elements were calculated for up to two partons at NLO and four partons at LO using the Comix~\cite{Gleisberg:2008fv} and OpenLoops~\cite{Cascioli:2011va} ME generators and merged with the \Sherpa parton shower~\cite{Schumann:2007mg} using the ME+PS@NLO prescription~\cite{Hoeche:2012yf}. The  NNPDF3.0NNLO PDF set was used in conjunction with dedicated parton-shower tuning. The $W/Z+$jets events were normalised to the NNLO cross-sections~\cite{ATL-PHYS-PUB-2017-006,Anastasiou:2003ds}.

Diboson processes with one of the bosons decaying hadronically and the other leptonically were simulated using the \SHERPAV{2.2.1} generator. They were calculated for up to one ($ZZ$) or zero ($WW$, $WZ$) additional partons at NLO and up to three additional partons at LO using the Comix and OpenLoops ME generators and merged with the \SHERPA parton shower using the ME+PS@NLO prescription. The CT10 PDF set was used in conjunction with dedicated parton-shower tuning. The samples were normalised to the NLO cross-sections evaluated by the generator.
 
The $t\bar{t}W$ and $t\bar{t}Z$ samples were simulated using \mgamcatnlo{} and the NNPDF23NNLO PDF set~\cite{NNPDF} for the ME\@. In addition to the $t\bar{t}W$ and $t\bar{t}Z$ samples, the predictions for $tZ$, \ttb{}\ttb{}, \ttb{}$WW$ and $tWZ$ are included in the $t+X$ background.
These processes have never been observed at the LHC, except  for strong evidence for $tZ$~\cite{TOPQ-2016-14,CMS-TOP-16-020}, and have a cross-section significantly smaller than for $t\bar{t}W$ and $t\bar{t}Z$ production, providing a subdominant contribution to the $t+X$ background.
The simulation of the $tZ$, $\ttbar WW$ and $\ttb\ttb$ samples was performed using \madgraph while the simulation of the $tWZ$ sample was obtained with \mgamcatnlo.
For all the samples in the $t+X$ background, \PYTHIAV{8.186}~\cite{Sjostrand:2007gs} and the PDF set NNPDF23LO with the A14 tune were used for the showering and hadronisation.
 
\begin{table*}[t]
\scriptsize
\centering
\begin{tabular}{|l|l|c|c|c|c|} \hline
Physics process                  & Generator             & PDF set for  & Parton shower   & Tune         & Cross-section   \\
&                       & hard process &                 &              & normalisation   \\ \hline
$\ttbar$ signal                 & \PowhegBox v2         & NNPDF3.0NLO  & \Pythia 8.186   & A14          & \NNLO+\NNLL     \\
$\ttbar$ PS syst.                & \PowhegBox v2         & NNPDF3.0NLO  & \herwig{}7.0.1  & H7-UE-MMHT   & \NNLO+\NNLL     \\
$\ttbar$ generator syst.                & \Sherpa 2.2.1         & NNPDF3.0NNLO  &  \Sherpa  & \Sherpa      & \NNLO+\NNLL           \\
$\ttbar$ rad. syst.              & \PowhegBox v2         & NNPDF3.0NLO  & \Pythia 8.186   & Var3cDown/Var3cUp   & \NNLO+\NNLL     \\
Single top:  $t$-channel        & \PowhegBox v1         & CT10f4       & \Pythia 6.428   & Perugia2012  & \NLO            \\
Single top:  $t$-channel syst.         & \PowhegBox v1         & CT10f4       & \Pythia 6.428   & Perugia2012 radHi/radLo  & \NLO            \\
Single top:  $s$-channel         & \PowhegBox v1         & CT10         & \Pythia 6.428   & Perugia2012  & \NLO            \\
Single top: $tW$ channel       & \PowhegBox v1         & CT10         & \Pythia 6.428   & Perugia2012  & \NLO+\NNLL      \\
Single top: $tW$ channel syst.        & \PowhegBox v1         & CT10         & \Pythia 6.428   & Perugia2012 radHi/radLo & \NLO+\NNLL      \\
Single top: $tW$ channel DS        & \PowhegBox v1         & CT10         & \Pythia 6.428   & Perugia2012  & \NLO+\NNLL      \\
$t+X$                            &  \textsc{MadGraph5} & NNPDF2.3LO   & \Pythia 8.186   & A14          & \NLO            \\
$W(\to \ell \nu) $+ jets         & \Sherpa 2.2.1         & NNPDF3.0NNLO         & \Sherpa         & \Sherpa      & \NNLO           \\
$Z(\to \ell {\bar \ell}) $+ jets & \Sherpa 2.2.1         & NNPDF3.0NNLO         & \Sherpa         & \Sherpa      & \NNLO           \\
$WW, WZ, ZZ$                     & \Sherpa 2.1.1         & NNPDF3.0NNLO         & \Sherpa         & \Sherpa      & \NLO            \\ \hline
\end{tabular}
\caption{Summary of MC samples used for the nominal measurement and to assess the systematic uncertainties,
showing the event generator for the hard-scattering process, the order in pQCD of the cross-section used for normalisation, PDF choice, as well as the parton-shower generator and the corresponding tune used in the analysis.}
\label{tab:MC}
\end{table*}
 
\FloatBarrier
 
\section{Object reconstruction and event selection} \label{sec:reco}
 
The following sections describe the detector- and particle-level objects used to characterise the final-state event topology and to define the fiducial phase-space regions for the measurements.

\subsection{Detector-level object reconstruction}\label{sec:objects:reco}
 
Primary vertices are formed from reconstructed tracks that are spatially compatible with the interaction region. The hard-scatter primary vertex is chosen to be the one with at least two associated tracks and the highest $\sum \pt^2$, where the sum extends over all tracks with $\pt > 0.4\,\GeV$ matched to the vertex.
 
Electron candidates are reconstructed by matching tracks in the inner detector to energy deposits in the EM calorimeter. They must satisfy a `tight' likelihood-based identification criterion based on shower shapes in the EM calorimeter, track quality and detection of transition radiation produced in the TRT detector~\cite{PERF-2017-01}. The reconstructed EM clusters are required to have a~transverse energy $\ET> 27\,\GeV$ and a pseudorapidity $|\eta| < 2.47$, excluding the transition region between the barrel and endcap calorimeters ($1.37 < |\eta| < 1.52$). The longitudinal impact parameter $z_0$ of the associated track is required to satisfy $|\Delta z_0\, {\sin} \theta|<0.5$~mm, where $\theta$ is the polar angle of the track, and the transverse impact parameter significance $|d_0|/\sigma(d_0)<5$, where $d_0$ is the transverse impact parameter and $\sigma(d_0)$ is its uncertainty. The impact parameters $d_0$ and $z_0$ are calculated relative to the beam spot and the beam line, respectively.
Isolation requirements based on calorimeter and tracking quantities are used to reduce the background from jets misidentified as prompt leptons (fake leptons) or due to semileptonic decays of heavy-flavour hadrons (non-prompt real leptons)~\cite{TOPQ-2015-09}. The isolation criteria are \pT{}- and $\eta$-dependent, and ensure an efficiency of 90\% for electrons with \pT{} of 25~\GeV{} and 99\% efficiency for electrons with $\pt{}$ of 60~\GeV{}. The identification, isolation and trigger efficiencies are measured using electrons from $Z$ boson decays~\cite{PERF-2017-01}.
 
Muon candidates are identified by matching tracks in the muon spectrometer to tracks in the inner detector~\cite{PERF-2015-10}. The track $\pt$ is determined through a~global fit to the hits, which takes into account the energy loss in the calorimeters. Muons are required to have $\pt > 27\,\GeV$ and $|\eta|<2.5$.
To reduce the background from muons originating from heavy-flavour decays inside jets, muons are required to be isolated using track-quality and isolation criteria similar to those applied to electrons.
 
Jets are reconstructed using the anti-$k_{t}$ algorithm~\cite{akt1} with radius parameter $R = 0.4$ as implemented in the  \textsc{FastJet} package~\cite{Fastjet}. Jet reconstruction in the calorimeter starts from topological clustering of individual calorimeter cell~\cite{PERF-2014-07} signals. They are calibrated to be consistent with electromagnetic cluster shapes using corrections determined in simulation and inferred from test-beam data. Jet four-momenta are then corrected for pile-up effects using the jet-area method~\cite{Cacciari:2008gn}. To reduce the number of jets originating from pile-up, an additional selection criterion based on a~jet-vertex tagging (JVT) technique is applied. The jet-vertex tagging is a~likelihood discriminant that combines information from several track-based variables~\cite{PERF-2014-03} and the criterion is only applied to jets with $\pT < 60 \,\GeV$ and $|\eta|< 2.4$.
The jets' energy and direction are calibrated using an energy- and $\eta$-dependent simulation-based calibration scheme with \textit{in situ} corrections based on data~\cite{PERF-2016-04}, and are accepted if they have $\pT > 25\,\GeV$ and $|\eta| < 2.5$.
 
To identify  jets containing $b$-hadrons, a multivariate discriminant (MV2c10)~\cite{PERF-2012-04,ATL-PHYS-PUB-2016-012} is used, combining information about the secondary vertices,  impact parameters and the reconstruction of the full $b$-hadron decay chain~\cite{Piacquadio_2008}.  Jets are considered as $b$-tagged if the value of the multivariate  analysis (MVA) discriminant is larger than a certain threshold. The thresholds are chosen to provide a 70\% $b$-jet tagging efficiency in an inclusive \ttbar{} sample, corresponding to rejection factors for charm quark and light-flavour parton initiated jets of 12 and 381, respectively.

Large-$R$ jets are reconstructed using the reclustering approach~\cite{Nachman:2014kla}: the anti-$k_t$ algorithm, with radius parameter $R = 1$, is applied directly to the calibrated small-$R$ ($R = 0.4$) jets, defined in the previous paragraph.
Applying this technique, the small-$R$ jet calibrations and uncertainties can be directly propagated in the dense environment of the reclustered jet, without additional corrections or systematic uncertainties~\cite{ATLAS-CONF-2017-062}. 
The reclustered jets rely mainly on the technique and cuts applied to remove the pile-up contribution in the calibration of the small-$R$ jets. However, a trimming technique~\cite{trim} is applied to the reclustered jet to remove soft small-$R$ jets that could originate entirely from pile-up. The trimming procedure removes all the small-$R$ jets with fraction of $\pT$ smaller than 5\% of the reclustered jet $\pT$~\cite{SUSY-2015-10,SUSY-2015-02}. Only reclustered jets with $\pt > 350\,\GeV$  and $|\eta| < 2.0$ are considered in the analysis. The reclustered jets are considered $b$-tagged if at least one of the constituent small-$R$ jets is $b$-tagged. To top-tag the reclustered jets the jet mass is required to be $120 <m_{\mathrm{jet}}< 220\,\GeV{}$. This selection has an efficiency of ~60\%, evaluated by only considering reclustered jets with a top quark satisfying $\Delta R\left(\text{reclustered jet},t^\text{had}\right) < 0.75$, where $t^\text{had}$ is the generated top quark that decays hadronically.
 
For objects satisfying more than one selection criteria, a procedure called `overlap removal' is applied to assign a unique hypothesis to each object. If a muon shares a track with an electron, it is likely to have undergone bremsstrahlung and hence the electron is not selected. To prevent double-counting of electron energy deposits as jets, the jet closest to a reconstructed electron is discarded if $\Delta R(\mathrm{jet},e) < 0.2$.
Subsequently, to reduce the impact of non-prompt electrons, if $\Delta R\left(\mathrm{jet},e\right) < 0.4$, then that electron is removed.
In case a jet is within $\Delta R\left(\mathrm{jet},\mu\right) = 0.4$ of a muon, if the jet has fewer than three tracks the jet is removed whereas if the jet has at least three tracks the muon is removed.

The missing transverse momentum $\met$ is defined as the magnitude of the $\vec{p}_{\mathrm{T}}^{\mathrm{miss}}$ vector  computed from the negative sum of the transverse momenta of the reconstructed calibrated physics objects (electrons, photons, hadronically decaying $\tau$-leptons, small-$R$ jets and muons) together with an additional soft term constructed with all tracks that are associated with the primary vertex but not with these objects~\cite{PERF-2016-07,ATLAS-CONF-2018-023}.
 
\subsection{Particle-level object definition}\label{sec:objects:particle}
Particle-level objects are defined in simulated events using only stable particles, i.e.\ particles with a mean lifetime $\tau > 30$~ps. The fiducial phase-spaces used for the measurements in the resolved and boosted topologies are defined using a series of requirements applied to particle-level objects analogous to those used in the selection of the detector-level objects, described above.
 
Stable electrons and muons are required to not originate from a generated hadron in the MC event, either directly or through a $\tau$-lepton decay. This ensures that the lepton is from an electroweak decay without requiring a direct match to a $W$ boson. Events where the $W$ boson decays into a leptonically decaying $\tau$-lepton are accepted.
The four-momenta of the bare leptons are then modified by adding the four-momenta of all photons, not originating from hadron decay, within a cone of size $\DeltaR=0.1$, to take into account final-state photon radiation.
Such `dressed leptons' are then required to have $\pT > 27\,\GeV$ and $|\eta| < 2.5$.

Neutrinos from hadron decays either directly or via a $\tau$-lepton decay are rejected. The particle-level missing transverse momentum is calculated from the four-vector sum of the selected neutrinos.
 
Particle-level jets are reconstructed using the same anti-$k_{t}$ algorithm used at the detector level. The jet-reconstruction procedure takes as input all stable particles, except for charged leptons and neutrinos not from hadron decay as described above, inside a radius $R = 0.4$. Particle-level jets are required to have $\pT > 25\,\GeV$ and $|\eta| < 2.5$.
A jet is identified as a $b$-jet if a hadron containing a $b$-quark is matched to the jet through a ghost-matching technique described in Ref.~\cite{Cacciari:2008gn}; the hadron must have $\pT > 5\,\GeV$.
 
The reclustered jets are reconstructed at particle level using the anti-$k_{t}$ algorithm with $R=1$ starting from the particle-level jets with $R=0.4$. The same trimming used at detector level is also applied at particle level: subjets of the reclustered jets with $\pT<$ 5$\%$ of the jet $\pT$ are removed from the jet.
The reclustered jets are considered $b$-tagged if at least one of the constituent small-$R$ jets is $b$-tagged. As in the case of detector-level jets, only reclustered jets with $\pt > 350\,\GeV$  and $|\eta| < 2.0$ are considered and the jet is tagged as coming from a boosted top quark if $120<m_{\mathrm{jet}}<220$~\GeV.

Particle-level objects are subject to different overlap removal criteria than reconstructed objects. After dressing and jet reclustering, muons and electrons with separation $\Delta R < 0.4$ from a jet are excluded. Since the electron--muon overlap removal at detector level is dependent on the detector-level reconstruction of these objects, it is not applied at particle level.

\subsection{Parton-level objects and full phase-space definition}
\label{sec:objects:parton}
Parton-level objects are defined for simulated events. Only top quarks decaying directly into a $W$ boson and a $b$-quark in the simulation are considered. The full phase-space for the measurements presented in this
paper is defined by the set of \ttbar{} pairs in which one top quark decays leptonically (including $\tau$-leptons)
and the other decays hadronically. In the boosted topology, to avoid a complete dependence on the MC predictions due to the extrapolation into regions not covered by the detector-level selection, the parton-level measurement is limited to the region where the top quark is produced with $\pt > 350\,\GeV$.
This region represents less than 2\% of the entire phase-space. The measurement in the resolved topology covers the entire phase-space.
 
Events in which both top quarks decay leptonically are removed from the parton-level signal simulation.

\subsection{Particle- and detector-level event selection}\label{sec:selection}
The event selection comprises a~set of requirements based on the general event quality and on the reconstructed objects, defined above, that characterise the final-state event topology. The analysis applies two exclusive event selections: one corresponding to a~resolved topology and another targeting a boosted topology, where all the decay products of the hadronic top quark are collimated in a single reclustered jet. The same selection cuts are applied to the reconstructed- and particle-level objects.
 
For both selections, events are required to have a reconstructed primary vertex with two or more associated tracks and contain exactly one reconstructed lepton candidate with $\pt > 27\,\GeV{}$  geometrically matched to a corresponding object at trigger level. The requirements on the primary vertex and trigger matching are applied only at detector level.
 
For the resolved event selection, each event is also required to contain at least four small-$R$ jets with $\pt > 25\,\GeV$ and $|\eta| < 2.5$ of which at least two must be tagged as $b$-jets. As discussed in Section~\ref{sec:reconstruction:resolved}, the strategy employed to reconstruct the detector-level kinematics of the  \ttb{} system in the resolved topology, when performing the measurement at parton level, is a kinematic likelihood fit. When this method is applied, a further selection requirement on the likelihood of the best permutation is introduced, i.e.\ it must satisfy $\log L > -52$. The selection criteria for the resolved topology are summarised in Table~\ref{tab:seln:resolved}.
 
For the boosted event selection, at least one reclustered top-tagged jet with $\pt > 350\,\GeV$  and at least one small-$R$ jet close to the lepton and far from the reclustered jet, i.e.\ with $\Delta R\left(\mathrm{jet}_{R=0.4}, \ell\right) < 2.0$ and $\Delta R\left(\text{reclustered jet}, \mathrm{jet}_{R=0.4}\right) > 1.5$, are required. All the small-$R$ jets fulfilling these requirements are considered associated with the lepton.
The reclustered jet must be well separated from the lepton, with $\Delta \phi \left(\text{reclustered jet}, \ell\right)> 1.0$. In the boosted selection, only one $b$-tagged jet is required in the final state, to reduce the loss of signal due to the decrease in $b$-tagging efficiency in the high $\pt$ region. This jet must fulfil additional requirements: it is either among the components of the reclustered jet, or it is one of the small-$R$ jets associated with the lepton.
To suppress the multijet background in the boosted topology, where only one $b$-tagged jet is required, the missing transverse momentum is required to be $\met>20$~\GeV{} and the sum of \met{} and the transverse mass of the $W$ boson 
is required to be $\met + m_{\textrm T}^W>60$~\GeV{}, with $m_{\textrm T}^{W} = \sqrt{2 p_{\textrm T}^{\ell} E_{\textrm T}^{\textrm{miss}} \left(1 - \cos \Delta \phi\left(\ell, \vec{p}_{\mathrm{T}}^{\mathrm{miss}}\right)\right)}$. The selection criteria for the boosted topology are summarised in Table~\ref{tab:seln:boosted}.

\begin{table*} [!htbp]
\centering
\noindent\makebox[\textwidth]{
\resizebox{\textwidth}{!}{
\begin{tabular}{|l|l|l|l|}
\hline
Selection & \multicolumn{2}{c|}{Detector level} & Particle level \\
\hline
& $\eplus$ & $\muplus$ & \\
\hline\hline
Leptons & \parbox[c][8em][c]{5.1cm}{One electron, no muons\\$|d_0|/\sigma(d_0)< 5\,$ \\ $|\Delta z_0\,\sin\theta | < 0.5\,$mm
\\ Track and calorimeter isolation
\\ $|\eta|<$1.37 or $1.52<|\eta|<2.47$
\\ $E_{\mathrm{T}} > 27\,\GeV$} & \parbox[c][8em][c]{5.2cm}{One muon, no electrons\\  $|d_0|/\sigma(d_0)<$ 3 \\ $|\Delta z_0\,\sin\theta | < 0.5\,$mm
\\Track and calorimeter isolation
\\ $|\eta|<2.5$ \\ $p_{\mathrm{T}} > 27\,\GeV$ }
& \parbox[c][7em][c]{2.1cm}{ One lepton\\ ($e/\mu$)\\ $|\eta|<2.5$ \\
$p_{\mathrm{T}} > 27\,\GeV$} \\
\hline
Anti-$k_t$ $R=0.4$ jets &
\multicolumn{2}{c|}{\parbox[c][8em][c]{6.4cm}{\centering $N^\mathrm{jets}\geq 4$\\  $|\eta|<2.5$ \\ $p_{\mathrm{T}} >  25\,\GeV$
\\ JVT cut (if $p_{\mathrm{T}}< 60\,\GeV$ and $|\eta| < 2.4$) \\ $b$-tagging: $\geq 2$ jets with MV2c10 at 70\%} }
& \parbox[c][4em][c]{3.5cm}{$N^\mathrm{jets}\geq 4$\\$|\eta|<2.5$\\ $p_{\mathrm{T}} >  25\,\GeV$ \\ $b$-tagging: \\ Ghost-matched \\ $b$-hadron } \\
\hline
Overlap removal &

\multicolumn{2}{c|}{\parbox[c][10em][c]{10cm}{\centering If an electron shares a track with a muon: electron removed\\
If $\DeltaR(e,\mathrm{jet}_{R=0.4})<0.2$: jet removed\\
then\\
If $\DeltaR(e,\mathrm{jet}_{R=0.4})<0.4$: $e$ removed\\
If $\DeltaR(\mu,\mathrm{jet}_{R=0.4})<0.4$ and $n_{\mathrm{tracks}}^{\mathrm{jet}}\geq3$: $\mu$ removed \\
If $\DeltaR(\mu,\mathrm{jet}_{R=0.4})<0.4$ and $n_{\mathrm{tracks}}^{\mathrm{jet}}<3$: jet is removed}}   & \parbox[c][10em][c]{3.8cm}{If $\DeltaR(e,\mathrm{jet}_{R=0.4})<0.4$: $e$ removed\\
If $\DeltaR(\mu,\mathrm{jet}_{R=0.4})<0.4$: $\mu$ removed} \\

\hline
Top reconstruction quality &

\multicolumn{2}{c|}{\parbox[c][7em][c]{7.1cm}{\centering Remove events passing boosted selection.\\
Parton level measurement: $\log L> -52$ for the best permutation from the kinematic fit}} & \parbox[c][4em][c]{3.8cm}{} \\

\hline
\end{tabular}
}
}\caption{Summary of the requirements for detector-level and MC-generated particle-level events in the resolved topology.}
\label{tab:seln:resolved}
\end{table*}
 
\begin{table} [!htbp]
\centering
\noindent\makebox[\textwidth]{
\resizebox{\textwidth}{!}{
\begin{tabular}{|l|l|l|l|}
\hline
Selection & \multicolumn{2}{c|}{Detector level} & Particle level \\
\hline
& $\eplus$ & $\muplus$ & \\
\hline\hline
Leptons & \parbox[c][8em][c]{5cm}{One electron, no muons\\$|d_0|/\sigma(d_0)< 5\,$ \\ $|\Delta z_0\,\sin\theta | < 0.5\,$mm
\\ Track and calorimeter isolation
\\ $|\eta|<$1.37 or $1.52<|\eta|<2.47$
\\ $E_{\mathrm{T}} > 27\,\GeV$} & \parbox[c][8em][c]{5cm}{  One muon, no electrons\\$|d_0|/\sigma(d_0)<$ 3 \\ $|\Delta z_0\,\sin\theta | < 0.5\,$mm
\\Track and calorimeter isolation
\\ $|\eta|<2.5$ \\ $p_{\mathrm{T}} > 27\,\GeV$ }
& \parbox[c][7em][c]{2.1cm}{One lepton\\ ($e/\mu$)\\ $|\eta|<2.5$ \\
$p_{\mathrm{T}} > 27\,\GeV$} \\
\hline
Reclustered $R\! =\! 1.0$ jet & \multicolumn{3}{c|}{ $\pt> 350\,\GeV$, $|\eta|<2.0$ } \\
\hline
Anti-\kt{} $R=0.4$ jets &
\multicolumn{2}{c|}{\parbox[c][8em][c]{8cm}{\centering $\geq 1$ jet\\ $p_{\mathrm{T}} >  25\,\GeV$
\\  $|\eta|<2.5$ \\ JVT cut (if $p_{\mathrm{T}}< 60\,\GeV$ and $|\eta| < 2.4$) \\ $b$-tagging: $\geq 1$ jets with MV2c10 at 70\%} }
& \parbox[c][8em][c]{3.5cm}{$\geq 1$ jet \\ $|\eta|<2.5$,\\ $\pt >  25\,\GeV$ \\ $b$-tagging:\\ Ghost-matched\\  $b$-hadron } \\

\hline
 
Overlap removal &

\multicolumn{2}{c|}{\parbox[c][10em][c]{10cm}{\centering If an electron shares a track with a muon: electron removed\\
If $\DeltaR(e,\mathrm{jet}_{R=0.4})<0.2$: jet removed\\
then\\
If $\DeltaR(e,\mathrm{jet}_{R=0.4})<0.4$: $e$ removed\\
If $\DeltaR(\mu,\mathrm{jet}_{R=0.4})<0.4$ and $n_{\mathrm{tracks}}^{\mathrm{jet}}\geq3$: $\mu$ removed \\
If $\DeltaR(\mu,\mathrm{jet}_{R=0.4})<0.4$ and $n_{\mathrm{tracks}}^{\mathrm{jet}}<3$: jet is removed}}   & \parbox[c][10em][c]{3.5cm}{If $\DeltaR(e,\mathrm{jet}_{R=0.4})<$0.4: $e$ removed\\
If $\DeltaR(\mu,\mathrm{jet}_{R=0.4})<$0.4: $\mu$ removed} \\
 
\hline
$\met$, $\mtw$ & \multicolumn{3}{c|}{ $\met > 20\,\GeV$, $\met+\mtw> 60\,\GeV$}   \\
\hline
Hadronic top &
\multicolumn{3}{c|}{ \parbox[c][4em][c]{14.5cm}{\centering Top-tagging on the leading reclustered jet: $120\,\gev < m_{\mathrm{jet}} < 220\,\gev$, \\ $|\Delta \phi(\ell,\mathrm{jet}_{R=1.0})| > 1.0$}} \\
\hline
Leptonic top & \multicolumn{3}{c|}{ \parbox[c][4em][c]{14.5cm}{ \centering At least one anti-\kt{} $R = 0.4$ jet with $\Delta R(\ell,\mathrm{jet}_{R=0.4}) < 2.0$,\\ $\Delta R(\mathrm{jet}_{R=1.0},\mathrm{jet}_{R=0.4}) > 1.5$ }}\\
\hline
$b$-jets & \multicolumn{3}{c|}{ \parbox[c][7.5em][c]{14.5cm}{
At least one of:
\begin{enumerate}
\item one of the anti-\kt{} $R = 0.4$ jet with $\Delta R(\ell,\mathrm{jet}_{R=0.4}) < 2.0$ and $\Delta R(\mathrm{jet}_{R=1.0},\mathrm{jet}_{R=0.4}) > 1.5$ is $b$-tagged;
\item one of the anti-\kt{} $R = 0.4$ jet, component of the top-tagged  reclustered jet, is $b$-tagged.
\end{enumerate}}
}   \\
\hline
\end{tabular}
}
}
\caption{Summary of the requirements for detector-level and MC-generated particle-level events, for the boosted event selection. The description of the particle-level selection is in Section~\ref{sec:objects:particle}.}
\label{tab:seln:boosted}
\end{table}
 
Finally, to make the resolved and boosted topologies statistically independent, an additional requirement is defined at detector level: all events passing both the resolved and the boosted selection are removed from the resolved topology. The net effect of this requirement is a reduction in the overall event yield of the order of 2\% in the resolved topology.
\FloatBarrier

\section{Background determination}
\label{sec:backgrouds}
After the event selection, various backgrounds, mostly involving real leptons, contribute to the event yields. Data-driven techniques are used to estimate backgrounds that derive from events containing jets mimicking the signature of charged leptons or leptons from hadron decay, for which precise enough simulations are not available.
 
The single-top-quark background, comprising $t$-channel, $s$-channel and $tW$ production modes, is the largest background contribution in the resolved topology, amounting to 4.3\% and 4.2\% of the total event yield and 39\% and 30\% of the total background estimate in the resolved and boosted topologies, respectively. Shapes of all distributions of this background are modelled using MC simulation, and the event yields are normalised using calculations of its cross-section, as described in Section~\ref{sec:samples}.
 
The $W$+jets background represents the largest background in the boosted topology, amounting to approximately 3\% and 7\% of the total event yield, corresponding to approximately 25\% and 44\% of the total background estimate in the resolved and boosted topologies, respectively.
The estimation of this background is performed using MC simulations as described in Section~\ref{sec:samples}.
 
Multijet production processes, including production of hadronically decaying \ttbar{} pairs, have a~large cross-section and mimic the \ljets{} signature due to fake leptons or non-prompt real leptons. The multijet background is estimated directly from data using a~matrix method~\cite{TOPQ-2011-02}. The estimate is based on the introduction of a `loose' lepton definition, obtained by removing the isolation requirement and loosening the likelihood-based identification criteria in the electron case, compared to the `tight'  lepton definition given in Section~\ref{sec:objects:reco}. The number of fake and non-prompt leptons contained in the signal region is evaluated by inverting the matrix that relates the number of `loose' and `tight' leptons to the number of real and fake leptons. This matrix is built using the efficiencies for fake leptons and real leptons to pass the `tight' selection. The fake-lepton efficiency is measured using data in control regions dominated by  multijet background with the real-lepton contribution subtracted using MC simulation. The real-lepton efficiency is extracted by applying a~tag-and-probe technique using leptons from $Z$ boson decays.
The multijet background contributes approximately 3\% and 2\% to the total event yield, corresponding to approximately 24\% and 15\% of the total background estimate in the resolved and boosted topologies, respectively.

The background contributions from $Z$+jets, diboson and $t+X$ events are obtained from MC generators, and the event yields are normalised as described in Section~\ref{sec:samples}. The total contribution from these processes is approximately 1.4\% and 2.1\%, corresponding to approximately 12\% and 15\% of the total background estimate in the resolved and boosted topologies, respectively.

Dilepton top-quark pair events (including decays into $\tau$-leptons) can satisfy the event selection and are considered in the analysis as signal at both the detector and particle levels. They contribute to the \ttbar{} yield with a fraction of approximately 13\% (8\% after applying the cut on the likelihood of the kinematic fit described in Section~\ref{sec:reconstruction:resolved}) in the resolved topology and 6\% in the boosted topology.
In the full phase-space analysis at parton level, events where both top quarks decay leptonically are considered as background and a correction factor is applied to the detector-level spectra to account for this background.
 
In the fiducial phase-space analysis at particle level, all the \ttbar{} semileptonic events that could pass the fiducial selection described in Section~\ref{sec:selection} are considered as signal. For this reason, the  leptonic top-quark decays into $\tau$-leptons are considered as signal only if the $\tau$-lepton decays leptonically. Cases where both top quarks decay into a~$\tau$-lepton, which in turn decays into a quark--antiquark pair, are accounted for in the multijet background.
The full phase-space analysis at parton level includes all semileptonic decays of the \ttbar{} system, consequently the $\tau$-leptons from the leptonically decaying $W$ bosons are considered as signal, regardless of the $\tau$-lepton decay mode.


\begin{table}[hbpt]
\renewcommand*{\arraystretch}{1.1}
\centering
\begin{tabular}{l
rC{0.5mm}r
rC{0.5mm}r}
\hline\hline
Process & \multicolumn{6}{c}{\ \ \ Yield\ \ \ } \\
\hline
& \multicolumn{3}{c}{Resolved} & \multicolumn{3}{c}{Boosted} \\
\hline
$t\bar{t}$ & \num{1120000}&$\pm$&\num{90000} & \num{44700}&$\pm$&\num{1900} \\
Single top & \num{54000}&&$^{+\num{10000}}_{\num{-11000}}$& \num{2000}&$\pm$&\num{900} \\
Multijet &  \num{34000}&$\pm$&\num{16000} &  \num{1000}&$\pm$&\num{400} \\
$W$+jets & \num{34000}&$\pm$&\num{20000} &  \num{3200}&$\pm$&\num{1500} \\
$Z$+jets &   \num{12000}&$\pm$&\num{6000}  & \num{380}&$\pm$&\num{210} \\
$t+X$ &  \num{3800}&$\pm$&\num{500} &  \num{440}&$\pm$&\num{60} \\
Diboson & \num{1680}&&$^{+\num{220}}_{\num{-190}}$ & \num{194}&&$^{+\num{19}}_{\num{-21}}$ \\
\hline
Total prediction & \num{1260000}&$\pm$&\num{100000} & \num{52000}&$\pm$&\num{2900} \\[2.3pt]
\hline
Data & \multicolumn{3}{l}{\num{1252692}} & \multicolumn{3}{l}{\num{47600}}  \\ \hline
Data/Prediction & 0.99&$\pm$& 0.08& 0.92&$\pm$&0.05 \\   \hline\hline
\end{tabular}
\caption{Event yields after the resolved and boosted selections. Events that satisfy both the resolved and boosted selections are removed from the resolved selection. The cut on the kinematic fit likelihood has not been applied. The signal model, denoted \ttbar{} in the table, was generated using \Powheg{}+\PythiaEight{}, normalised to NNLO calculations.  The uncertainties include the combined statistical and systematic uncertainties, excluding the systematic uncertainties related to the modelling of the \ttbar{} system, as described in Section~\ref{sec:uncertainties}.
}
\label{tab:yields}
\end{table}

 
As the individual \ejets{} and \mujets{} channels have very similar corrections (as described in~Section~\ref{sec:xsec_calculation}) and give consistent results at detector level, they are combined by summing the distributions.
The event yields, in the resolved and boosted regimes, are shown in Table~\ref{tab:yields} for data, simulated signal, and backgrounds. The selection leads to a sample with an expected background of $11$\%  and $15$\% for the resolved and boosted topologies, respectively. The overall difference between data and prediction is $1$\% and $8$\% in the resolved and boosted topologies, respectively. In the resolved topology this is in good agreement within the experimental systematic uncertainties, while in the boosted topology the predicted event yield overestimates the data.
 
\begin{figure*}[p]
\centering
\subfigure[]{ \includegraphics[width=0.45\textwidth]{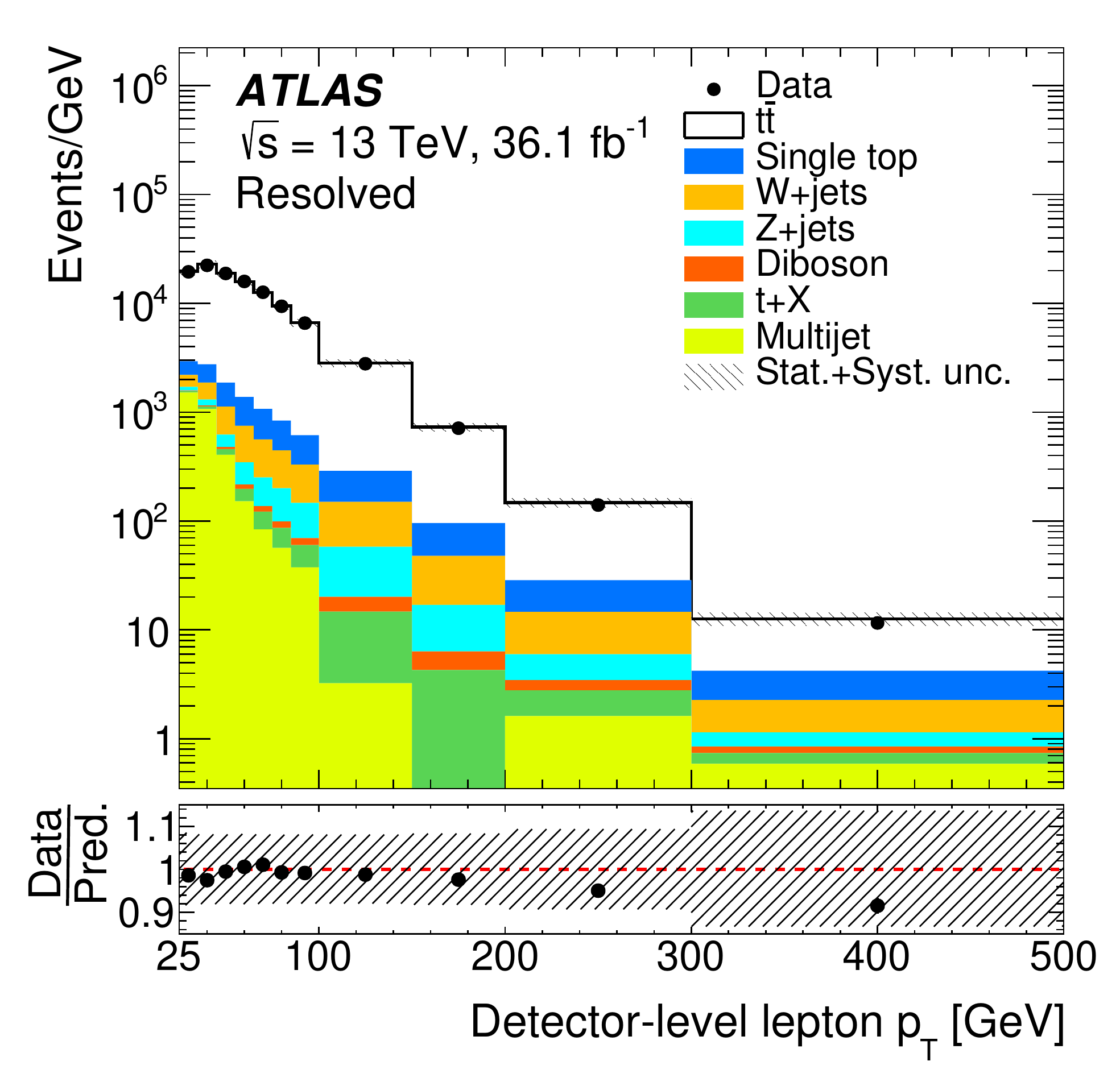}\label{fig:lep_pt_co}}
\subfigure[]{ \includegraphics[width=0.45\textwidth]{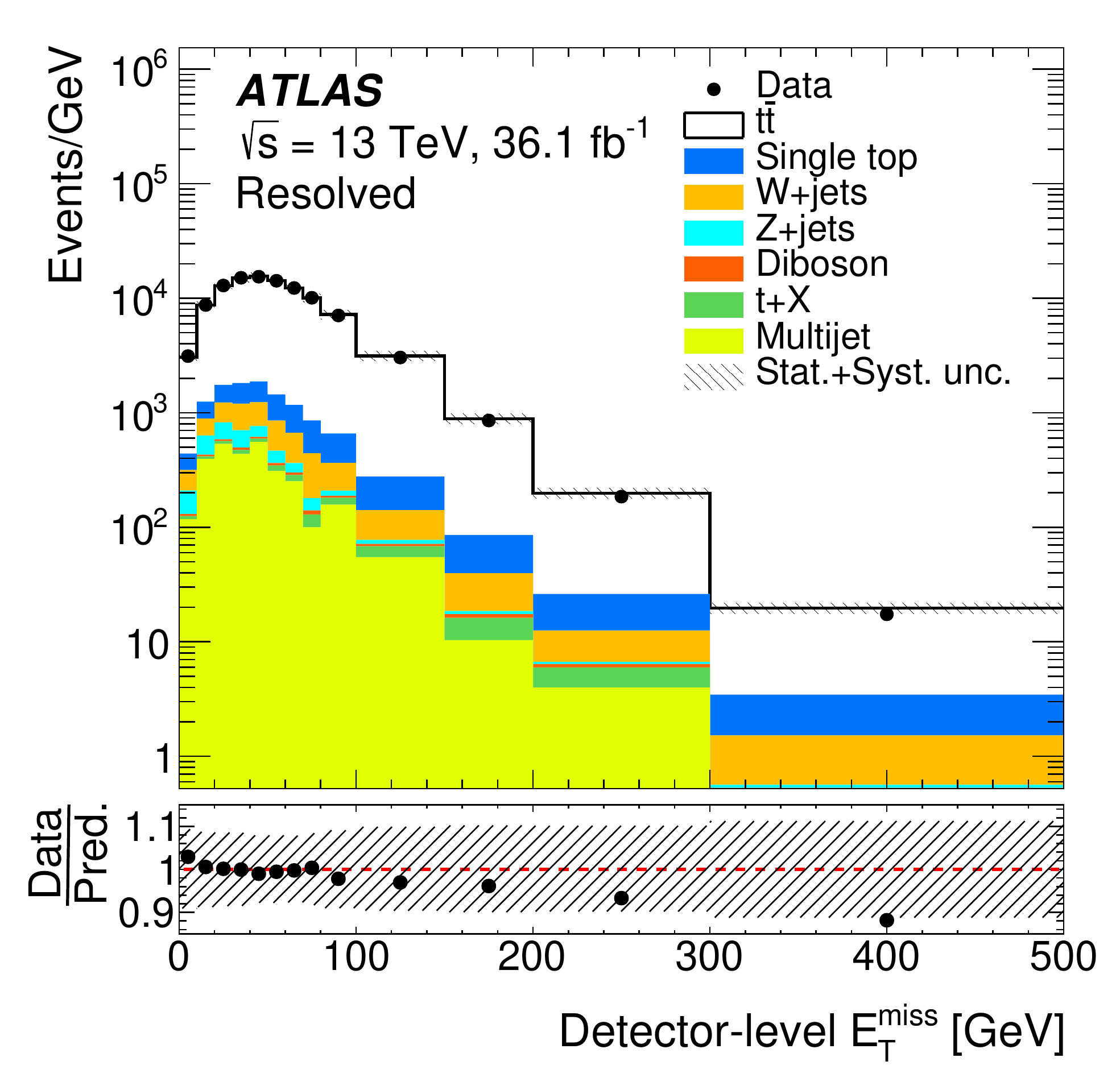}\label{fig:met_co}}
\subfigure[]{ \includegraphics[width=0.45\textwidth]{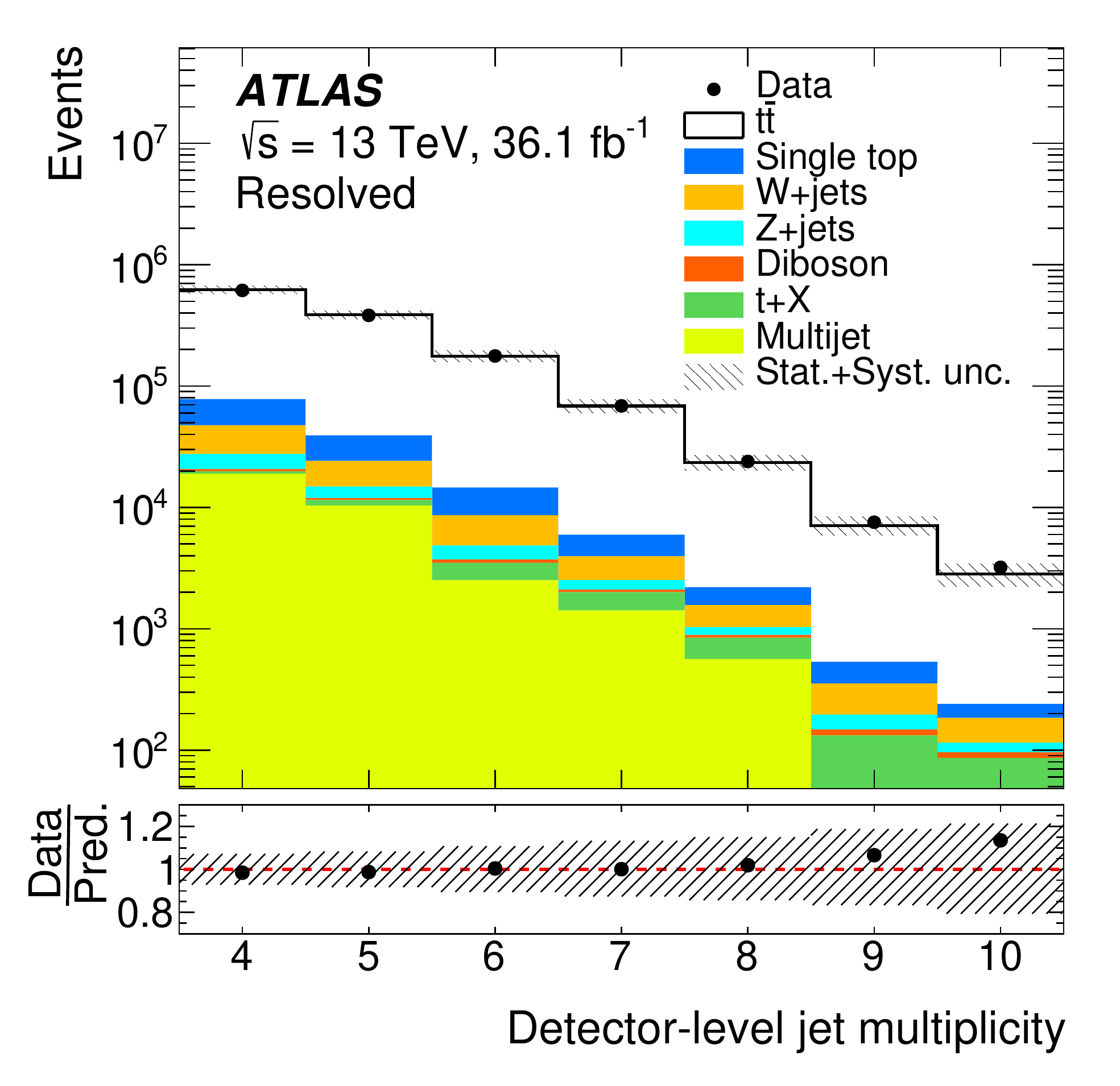}\label{fig:jet_n_co}}
\subfigure[]{ \includegraphics[width=0.45\textwidth]{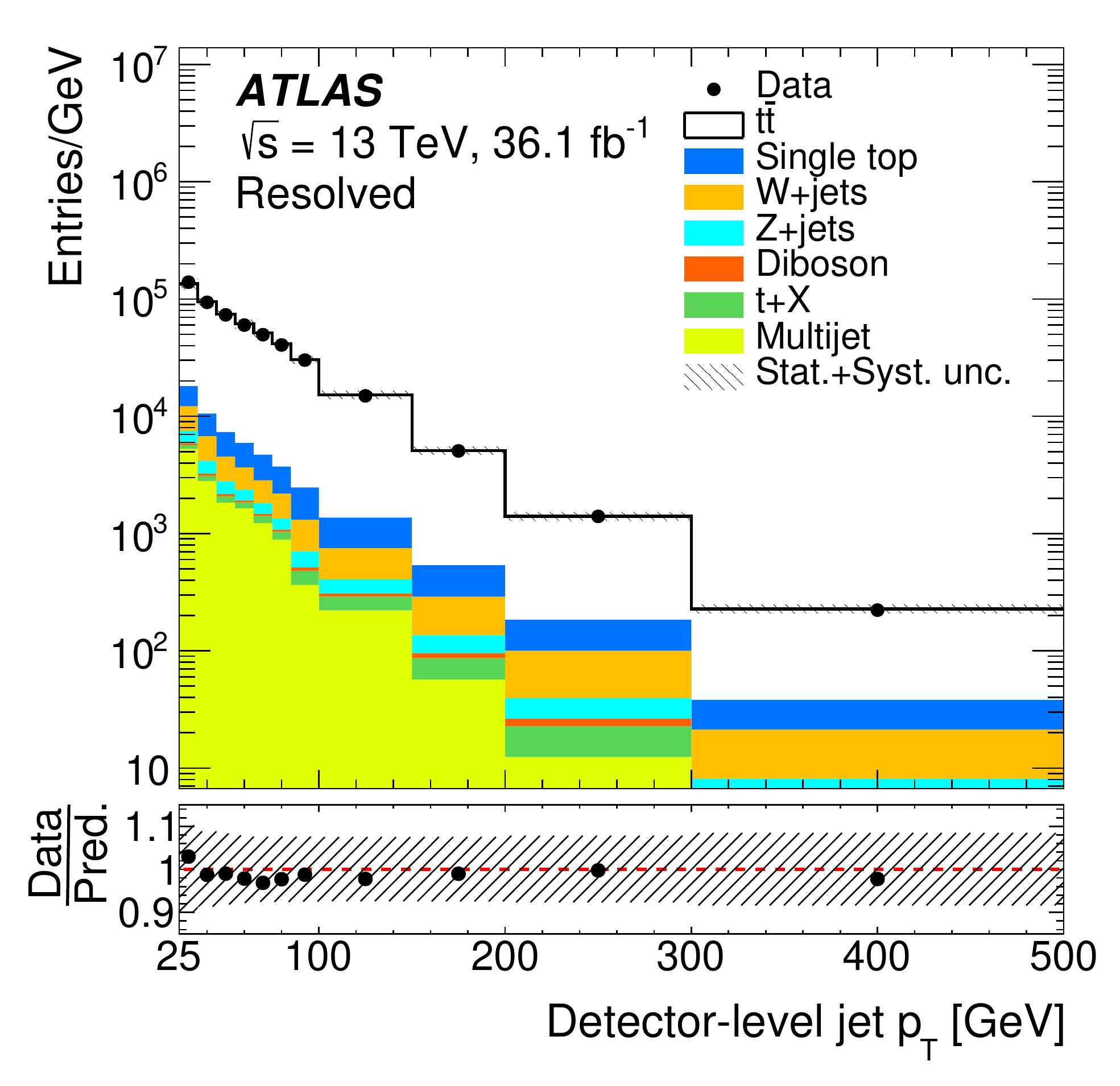}\label{fig:jet_pt_co}}
\caption{Kinematic distributions in the \ljets{} channel in the resolved topology at detector-level: \subref{fig:lep_pt_co}~lepton transverse momentum and \subref{fig:met_co}~missing transverse momentum \Etmiss{}, \subref{fig:jet_n_co}~jet multiplicity and \subref{fig:jet_pt_co}~transverse momenta of selected jets. Data distributions are compared with predictions using \Powheg+\PythiaEight{} as the \ttbar{} signal model. The hatched area represents the combined statistical and systematic uncertainties (described in Section~\ref{sec:uncertainties}) in the total prediction, excluding systematic uncertainties related to the modelling of the \ttbar{} events. Underflow and overflow events, if any, are included in the first and last bins. The lower panel shows the ratio of the data to the total prediction.}
\label{fig:controls_4j2b_detector}
\end{figure*}

\begin{figure*}[p]
\centering
\subfigure[]{ \includegraphics[width=0.45\textwidth]{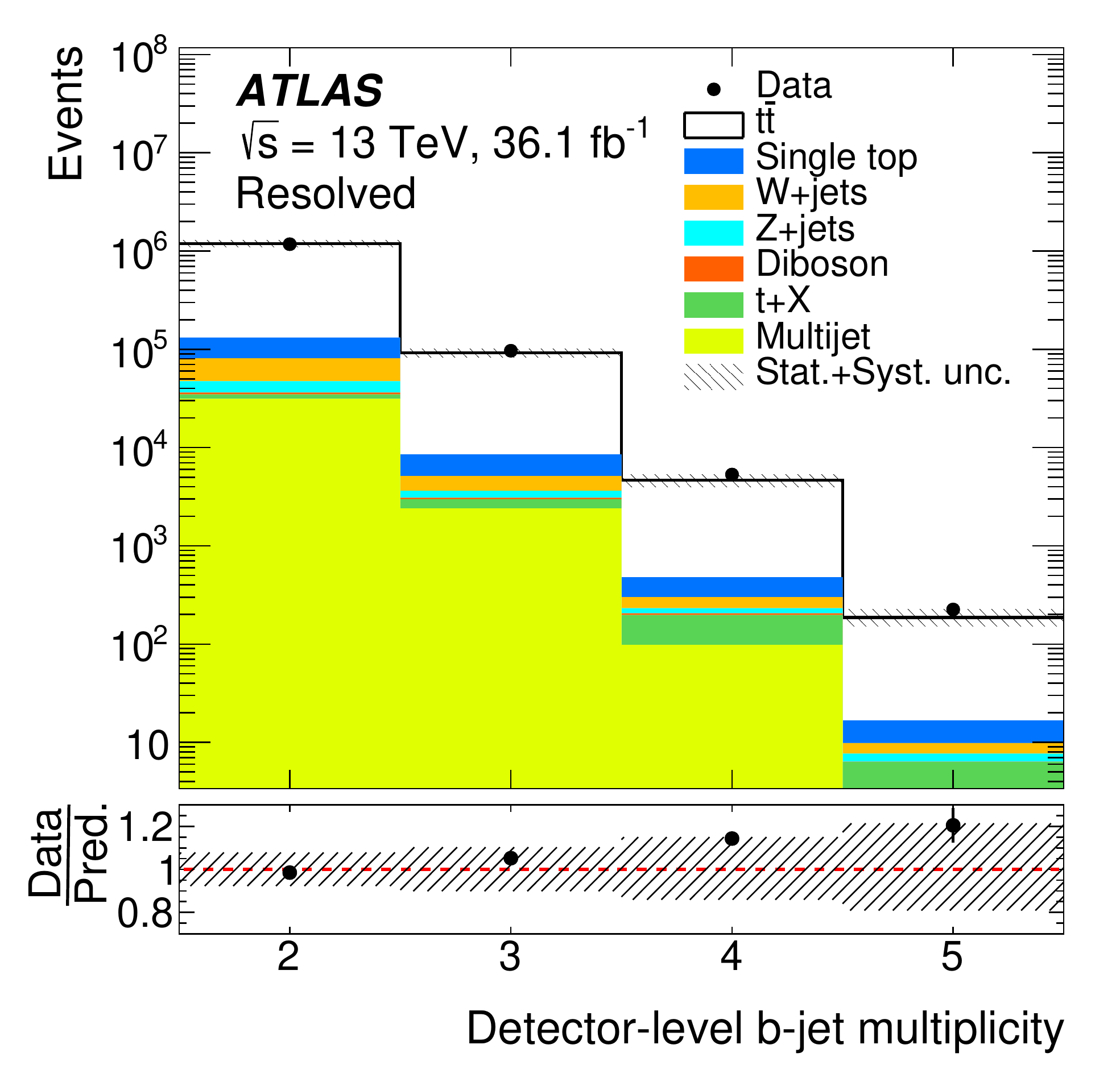}\label{fig:bjet_n_co}}
\subfigure[]{ \includegraphics[width=0.45\textwidth]{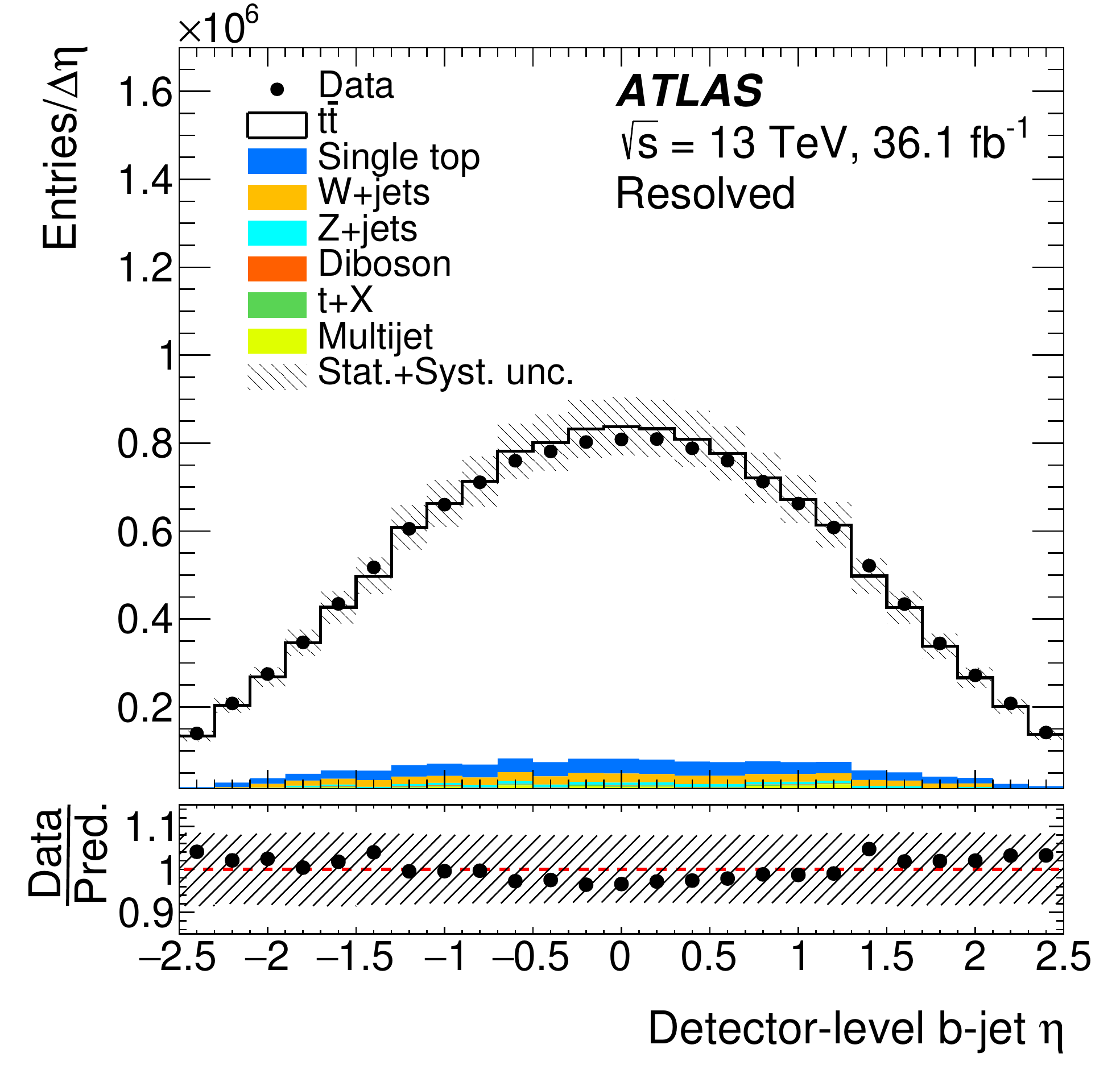}\label{fig:bjet_eta_co}}
\caption{Kinematic distributions in the \ljets{} channel in the resolved topology at detector-level: \subref{fig:bjet_n_co}~number of $b$-tagged jets and  \subref{fig:bjet_eta_co}~$b$-tagged jet pseudorapidity. Data distributions are compared with predictions using \Powheg+\PythiaEight{} as the \ttbar{} signal model. The hatched area represents the combined statistical and systematic uncertainties (described in Section~\ref{sec:uncertainties}) in the total prediction, excluding systematic uncertainties related to the modelling of the \ttbar{} events. Underflow and overflow events, if any, are included in the first and last bins. The lower panel shows the ratio of the data to the total prediction.}
\label{fig:controls_4j2b_detector_2}
\end{figure*}

\begin{figure}[p]
\centering
\subfigure[]{\includegraphics[width=0.45\textwidth]{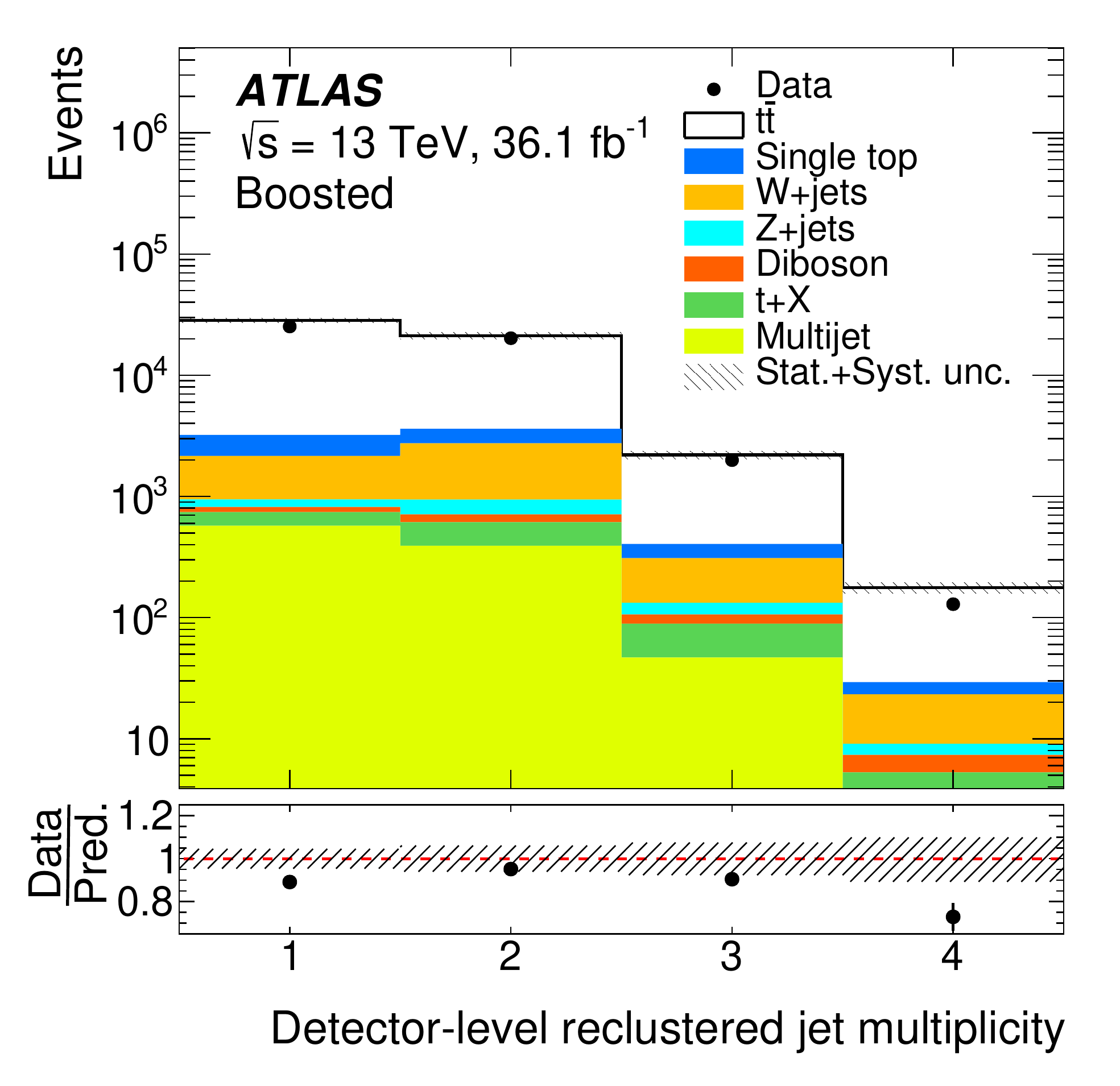} \label{fig:largejet_n}}
\subfigure[]{\includegraphics[width=0.45\textwidth]{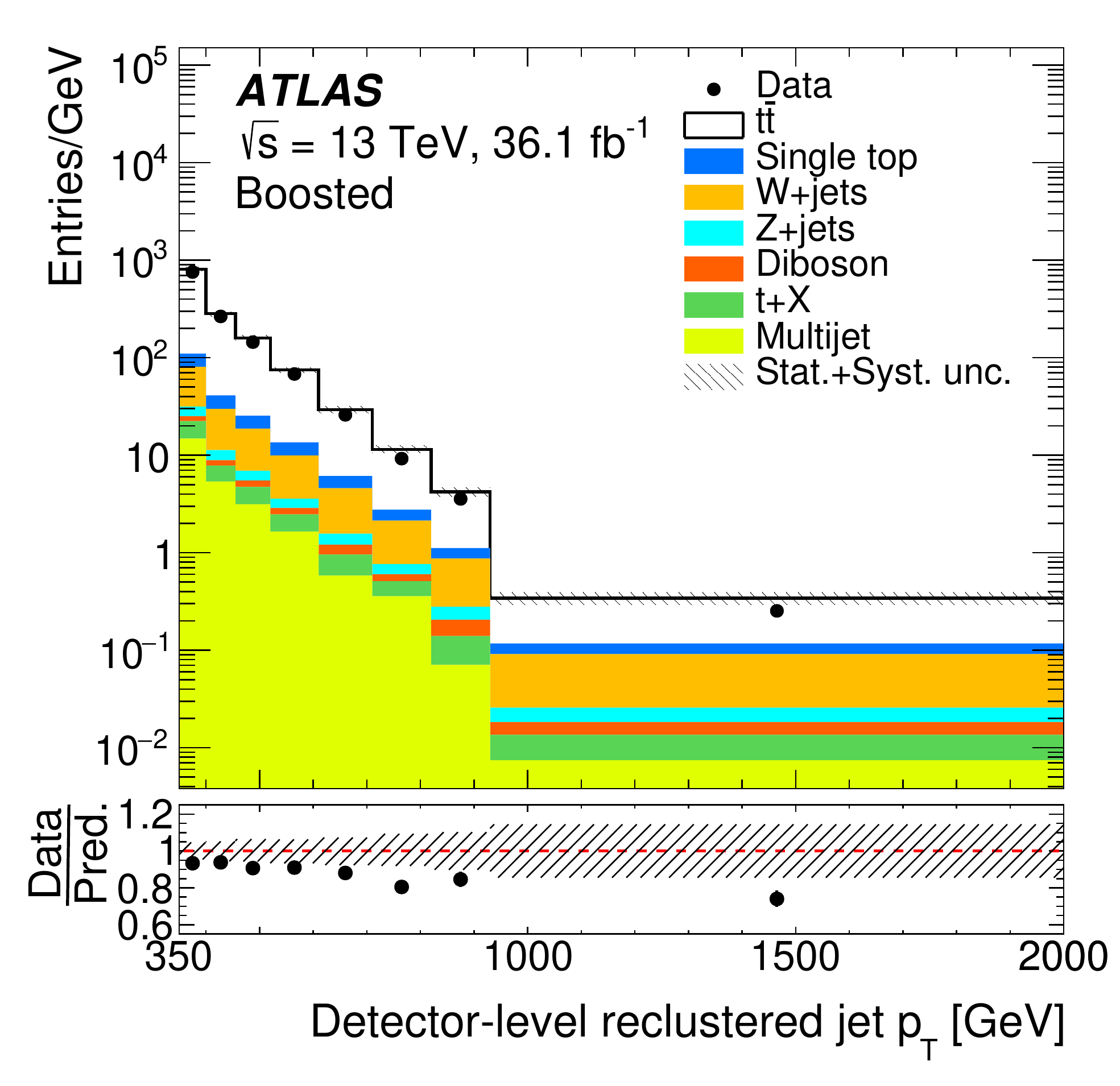}\label{fig:largejet_pt}}
\caption{Kinematic distributions in the \ljets{} channel in the boosted topology at detector-level: \subref{fig:largejet_n} number of reclustered jets and \subref{fig:largejet_pt} reclustered jet $\pt{}$. Data distributions are compared with predictions using \Powheg+\PythiaEight{} as the \ttbar{} signal model. The hatched area represents the combined statistical and systematic uncertainties (described in Section~\ref{sec:uncertainties}) in the total prediction, excluding systematic uncertainties related to the modelling of the \ttbar{} events. Underflow and overflow events, if any, are included in the first and last bins. The lower panel shows the ratio of the data to the total prediction.}
\label{fig:controls_1fj1b_detector}
\end{figure}

\begin{figure}[p]
\centering
\subfigure[]{\includegraphics[width=0.45\textwidth]{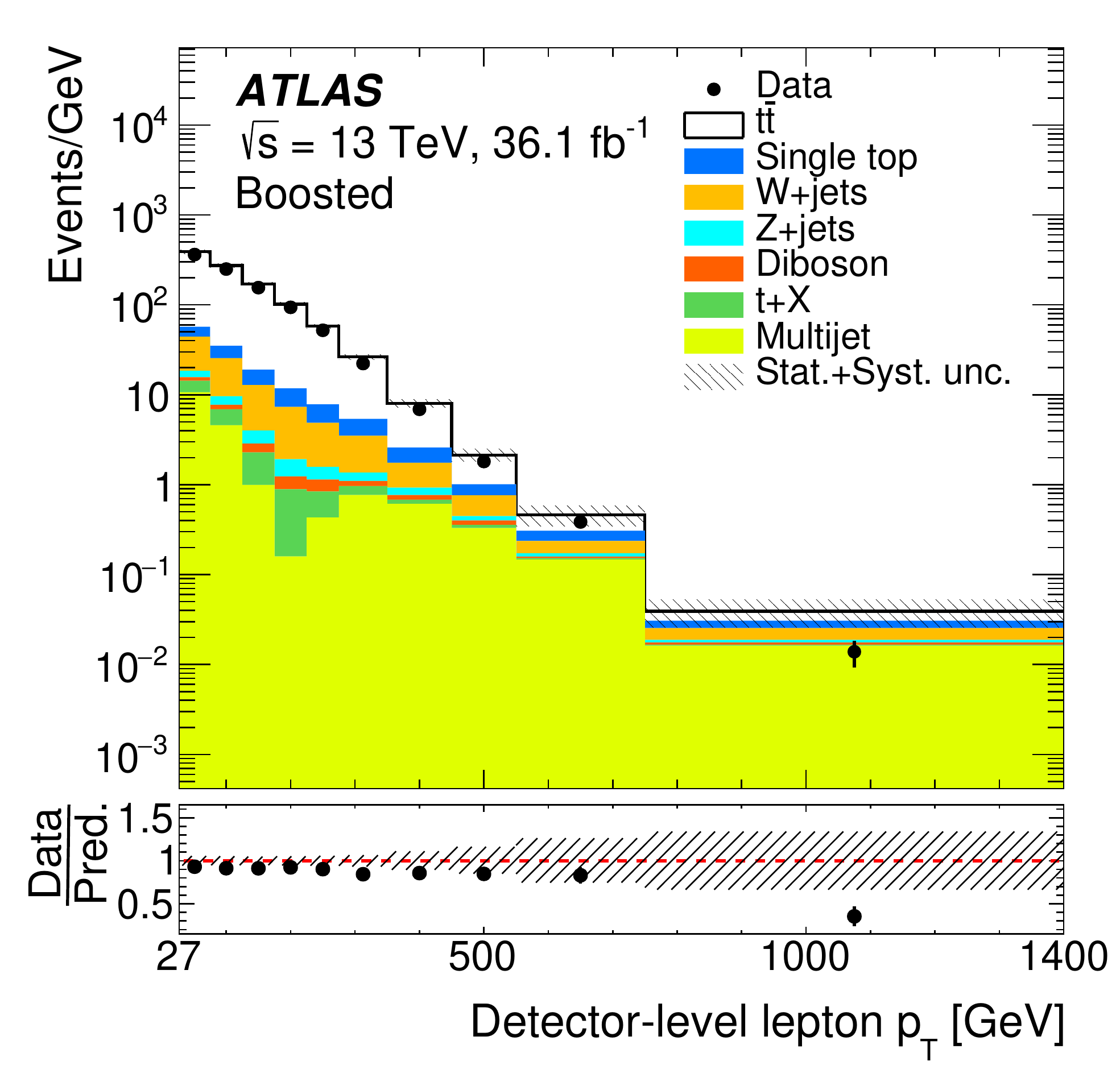} \label{fig:largejet_lep_pt}}
\subfigure[]{\includegraphics[width=0.45\textwidth]{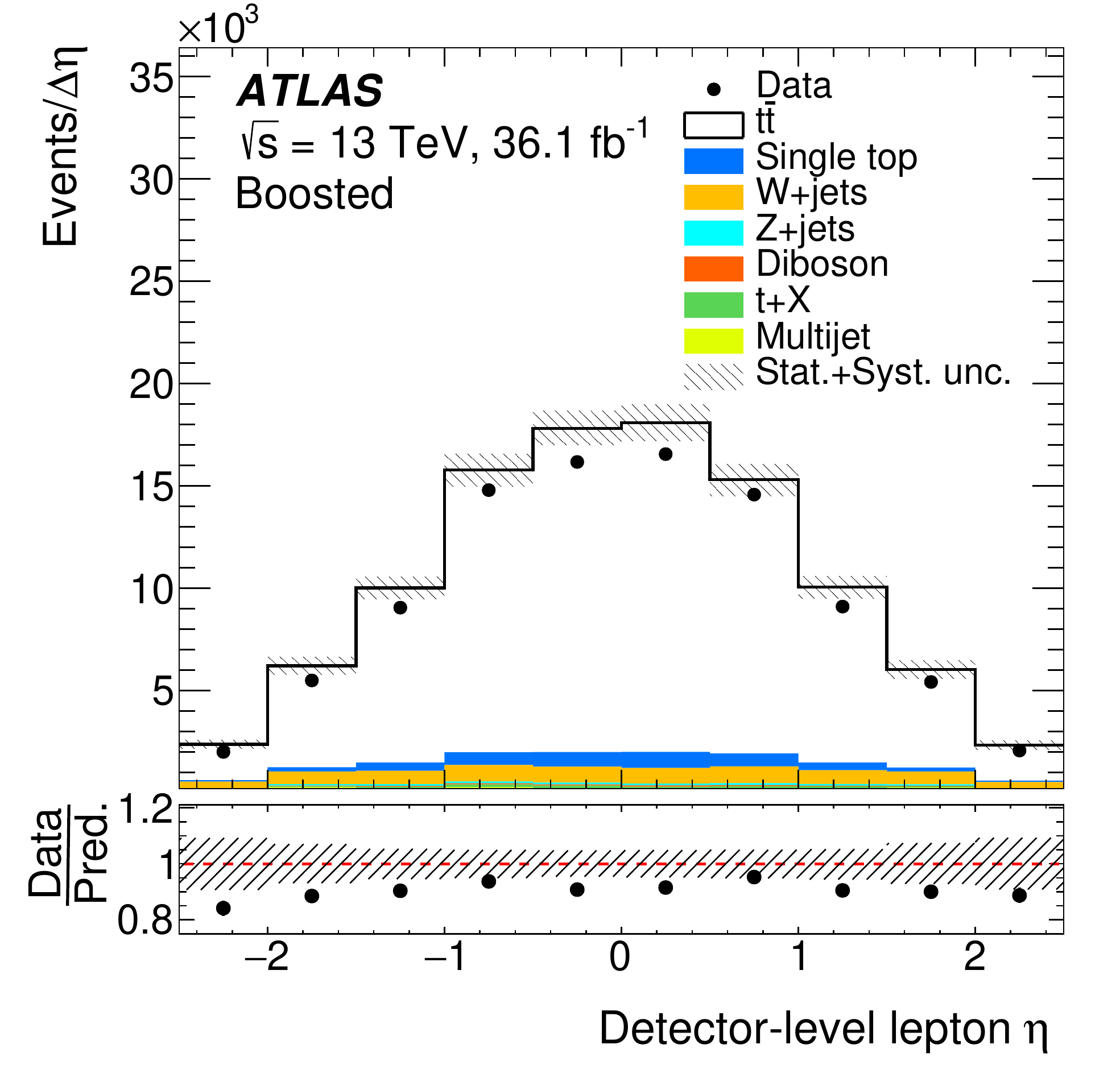} \label{fig:largejet_lep_eta}}
\subfigure[]{\includegraphics[width=0.45\textwidth]{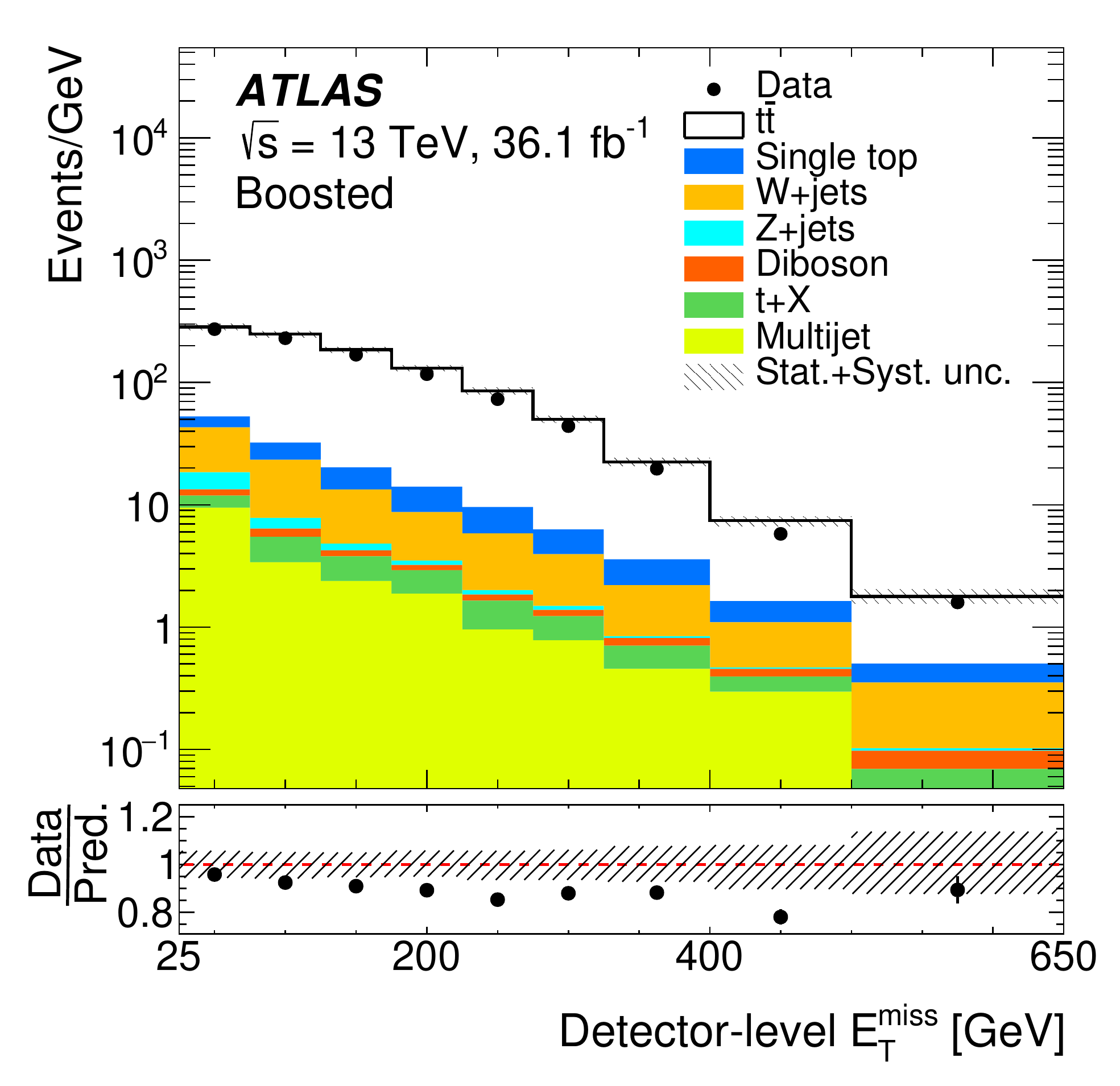} \label{fig:largejet_met}}
\subfigure[]{\includegraphics[width=0.45\textwidth]{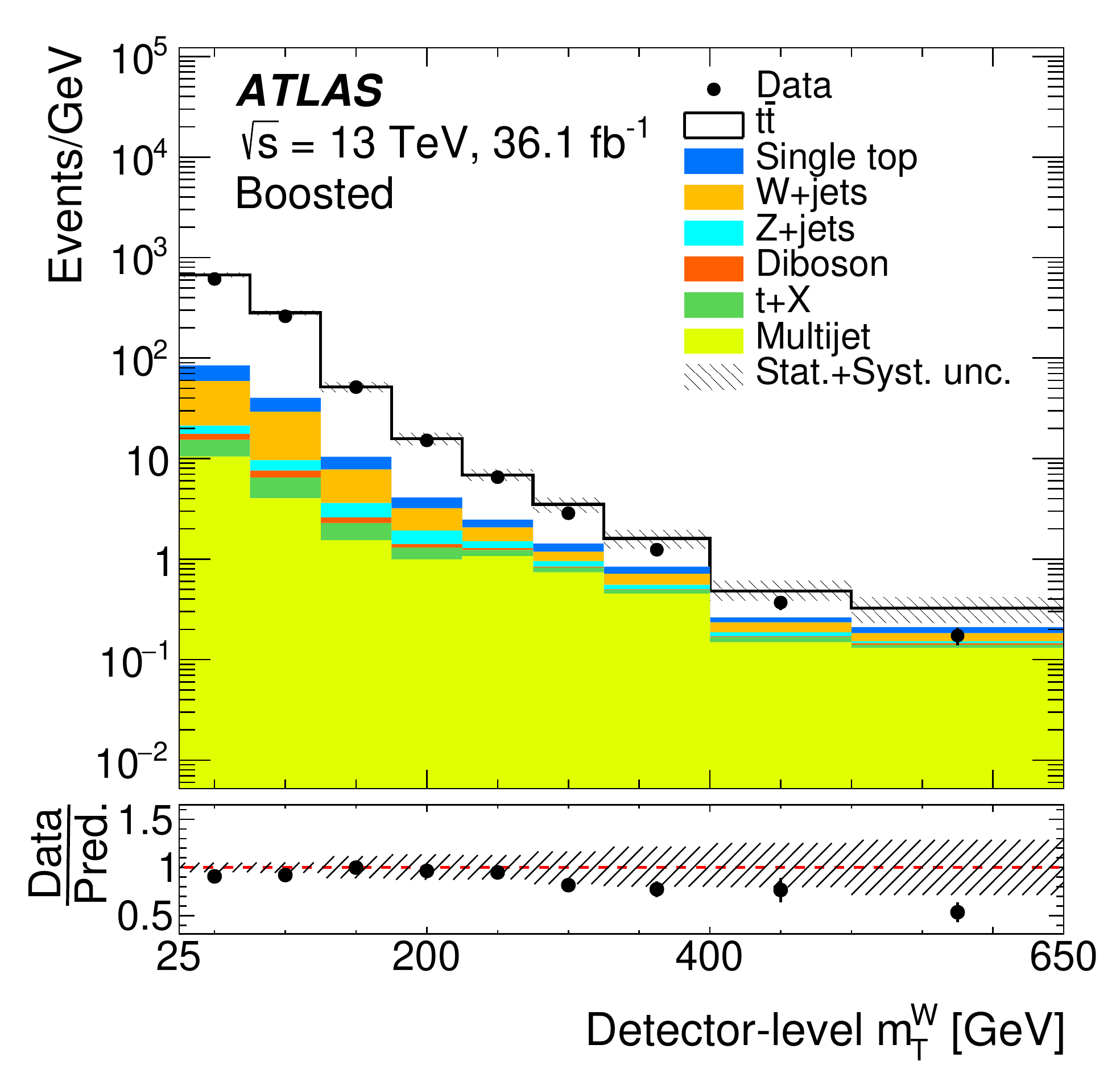} \label{fig:largejet_mtw}}
\caption{Kinematic distributions in the \ljets{} channel in the boosted topology at detector-level: \subref{fig:largejet_lep_pt} lepton $\pt{}$ and \subref{fig:largejet_lep_eta} pseudorapidity, \subref{fig:largejet_met} missing transverse momentum \Etmiss{} and \subref{fig:largejet_mtw} transverse mass of the $W$ boson. Data distributions are compared with predictions using \Powheg+\PythiaEight{} as the \ttbar{} signal model. The hatched area represents the combined statistical and systematic uncertainties (described in Section~\ref{sec:uncertainties}) in the total prediction, excluding systematic uncertainties related to the modelling of the \ttbar{} events. Underflow and overflow events, if any, are included in the first and last bins. The lower panel shows the ratio of the data to the total prediction.}
\label{fig:controls_1fj1b_detector_2}
\end{figure}

Figures~\ref{fig:controls_4j2b_detector}--\ref{fig:controls_1fj1b_detector_2} show,\footnote{Throughout this paper, all data as well as theory points are plotted at the bin centre of the $x$-axis. Moreover, the bin contents of all the histograms are divided by the corresponding bin width.} for different distributions, the comparison between data and predictions. The reconstructed distributions, in the resolved topology, of the \pt{} of the lepton, \etmiss, jet multiplicty and \pt{} are presented in Figure~\ref{fig:controls_4j2b_detector} and the \bjet{} multiplicity and $\eta$ in Figure~\ref{fig:controls_4j2b_detector_2}. The reconstructed distributions, in the boosted topology, of the reclustered jet multiplicity and jet \pt{} are shown in Figure~\ref{fig:controls_1fj1b_detector} and the \pt{} and $\eta$ of the lepton, \etmiss{} and \mtw{} in Figure~\ref{fig:controls_1fj1b_detector_2}.
In the resolved topology, good agreement between the prediction and the data is observed in all the distributions shown, while in the boosted topology the agreement lies at the edge of the uncertainty band. This is due to the overestimate of the predicted rate of events of about ~10\%, varying with the top quark \pT{}, reflected in all the distributions.


\FloatBarrier
 
\section{Kinematic reconstruction of the \ttbar{} system}
\label{sec:reconstruction}
Since the \ttbar{} production differential cross-sections are measured as a function of observables involving
the top quark and the \ttbar{} system, an event reconstruction is performed in each topology.
 
\subsection{Resolved topology}
\label{sec:reconstruction:resolved}
For the resolved topology, two reconstruction methods are employed: the pseudo-top algorithm~\cite{TOPQ-2013-07} is used to reconstruct the objects to be used in the particle-level measurement; a kinematic likelihood fitter (KLFitter)~\cite{KLFit:2013} is used to fully reconstruct the \ttb{} kinematics in the parton-level measurement. This approach performs better than the pseudo-top method in terms of resolution and bias for the reconstruction of the parton-level kinematics.

The pseudo-top algorithm reconstructs the four-momenta of the top quarks and their complete decay chain from final-state objects, namely the charged lepton (electron or muon), missing transverse momentum, and four jets, two of which are $b$-tagged. In events with more than two $b$-tagged jets, only the two with the highest transverse momentum values are considered as $b$-jets from the decay of the top quarks. The same algorithm is used to reconstruct the kinematic properties of top quarks as detector- and  particle-level objects. The pseudo-top  algorithm starts with the reconstruction of the neutrino four-momentum. While the $x$ and $y$ components of the neutrino momentum are set to the corresponding components of the missing transverse momentum, the $z$ component is calculated by imposing the $W$ boson mass constraint on the invariant mass of the charged-lepton--neutrino system. If the resulting quadratic equation has two real solutions, the one with the smaller  value of $|p_z|$ is chosen. If the discriminant is negative, only the real part is considered.
The leptonically decaying $W$ boson is reconstructed from the charged lepton and the neutrino. The leptonic top quark is reconstructed from the leptonic $W$ and the \btagged jet closest in $\Delta R$ to the charged lepton. The hadronic $W$ boson is reconstructed from the two non-$b$-tagged jets whose invariant mass is closest to the mass of the $W$ boson. This choice yields the best performance of the algorithm in terms of the
correspondence between the detector and particle levels.
Finally, the hadronic top quark is reconstructed from the hadronic $W$ boson and the other \bjet.
The advantage of using this method at particle level is that any dependence on the parton-level top quark is removed from the reconstruction and it is possible to have perfect consistency among the techniques used to reconstruct the top quarks at particle level and detector level.
 
\begin{figure*}[t]
\centering
\includegraphics[width=0.45\textwidth]{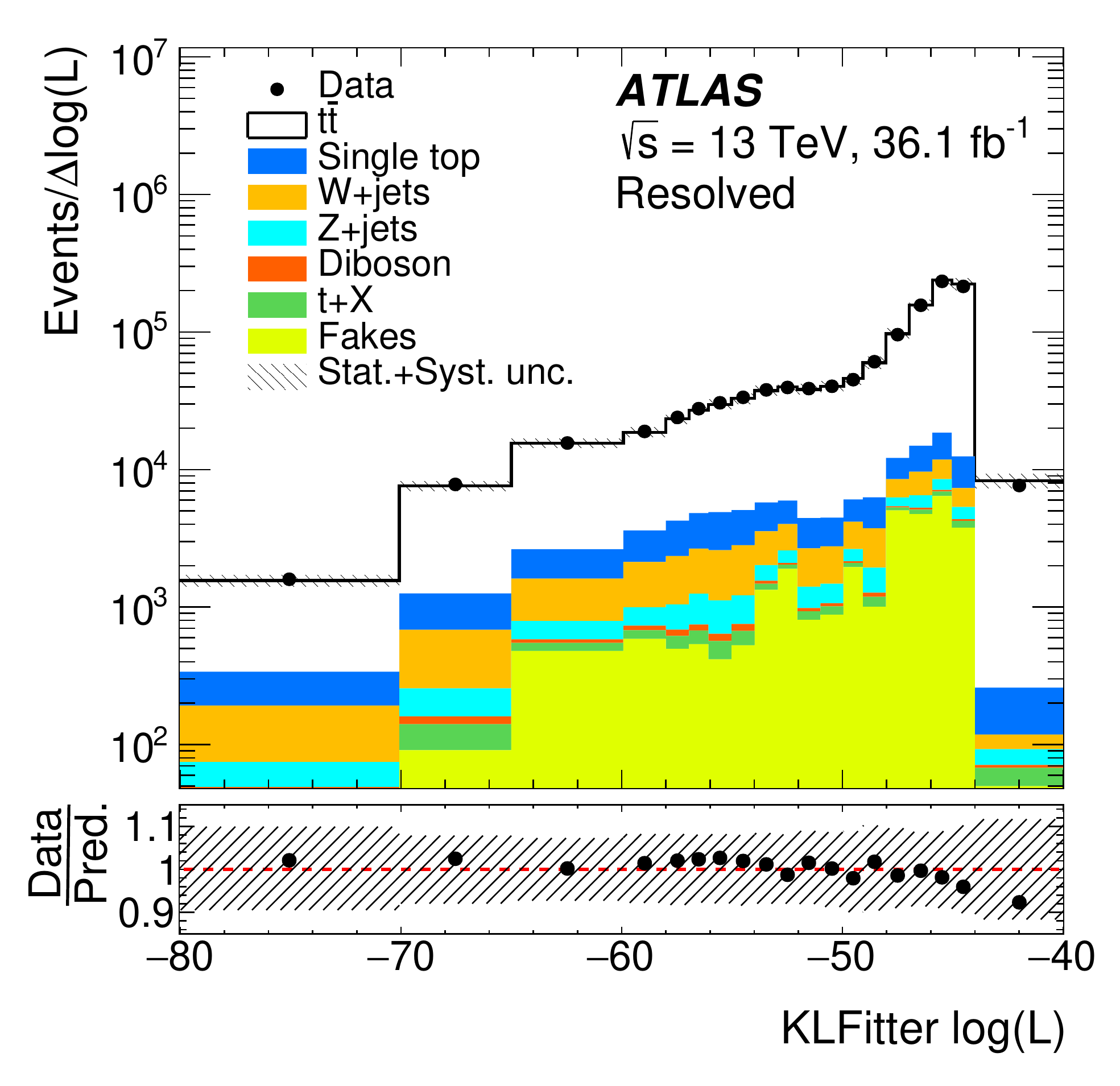}
 
\caption{Distribution in the \ljets{}~channel of the logarithm of the likelihood obtained from the kinematic fit in the resolved topology. Data distributions are compared with predictions using \Powheg+\PythiaEight{} as the \ttbar{} signal model. The hatched area represents the combined statistical and systematic uncertainties in the total prediction, excluding systematic uncertainties related to the modelling of the \ttbar{} events. Underflow and overflow events are included in the first and last bins. The lower panel shows the ratio of the data to the total prediction. Only events with $\log L > -52$ are considered in the parton-level measurement in resolved topology.}
\label{fig:logL}
\end{figure*}

The kinematic likelihood fit algorithm used for the parton-level measurements relates the measured kinematics of the reconstructed objects (lepton, jets and \met) to the leading-order representation of the \ttbar{} system decay. Compared to the pseudo-top algorithm, this procedure leads to better resolution (with an improvement of the order of 10\% for the \pt{} of \ttb{} system) in the reconstruction of the kinematics of the parton-level top quark. The kinematic likelihood fit has not been employed for the particle-level measurement because its likelihood, described in the following, is designed to improve the jet-to-quark associations and so is dependent on parton-level information. The likelihood is constructed as the product of Breit--Wigner distributions and  transfer functions that associate the energies of parton-level objects with those at the detector level.
Breit--Wigner distributions associate the missing transverse momentum, lepton, and jets with $W$ bosons and top quarks, and make use of their known widths and masses, with the top-quark mass fixed to 172.5~\GeV. The transfer functions represent the experimental resolutions in terms of the probability that the given true energy  for each of the \ttbar{} decay products produces the observed  energy at the detector level. The missing transverse momentum is used as a starting value for the neutrino transverse momentum, with its longitudinal component ($\pz^\nu$) as a free parameter in the  kinematic likelihood fit. Its starting value is computed from the $W$ mass constraint. If there are no real solutions for $\pz^\nu$ then zero is used as a starting value. Otherwise, if there are two real solutions, the one giving the larger likelihood is used. The five highest-\pt{} jets (or four if  there are only four jets in the event) are used as input to the likelihood fit. The input jets are defined by giving priority to the $b$-tagged jets and then adding the hardest remaining light-flavour jets. If there are more than four jets
in the event satisfying $\pt{} >$ 25\,\GeV{} and $|\eta|<2.5$, all subsets of four jets
from the five-jets collection are considered. The likelihood is maximised as a function of the energies of the $b$-quarks, the quarks from the hadronic $W$ boson decay, the charged lepton, and the components of the neutrino three-momentum. The maximisation is performed for each possible matching of jets to partons and the combination with the highest likelihood is retained. The event likelihood must satisfy $\log L > -52$. This requirement provides good separation between well and poorly reconstructed events and improves the purity of the sample. Distributions of $\log L$ in the resolved topology for data and simulation are shown in Figure~\ref{fig:logL} in the \ljets{} channel. The efficiency of the likelihood requirement in data is found to be well modelled by the simulation.
 
\subsection{Boosted topology}
\label{sec:reconstruction:boosted}
 
In the boosted topology, the same detector-level reconstruction procedure is applied for both the particle- and parton-level measurements. The leading reclustered jet that passes the selection described in Section~\ref{sec:reco} is  considered the hadronic top quark. 
Once the hadronic top-quark candidate is identified, the leptonic top quark is reconstructed using the leading $b$-tagged jet that fulfils the following requirements:
\begin{itemize}
\item $\deltaR\left(\ell, \bjet\right) < 2.0$;
\item $\deltaR\left(\mathrm{jet}_{R=1.0},\bjet\right)> 1.5$.
\end{itemize}
If there are no $b$-tagged jets that fulfil these requirements then the leading $\pt$ jet is used.
The procedure for the reconstruction of the leptonically decaying $W$ boson starting from the lepton and the missing transverse momentum  is analogous to the pseudo-top reconstruction described in Section~\ref{sec:reconstruction:resolved}.
 
 
\section{Observables}
\label{sec:observables}
A set of measurements of the \ttbar{} production cross-sections is presented as a function of kinematic observables.
In the following, the indices \emph{had} and \emph{lep} refer to the hadronically and leptonically decaying
top quarks, respectively. The indices 1 and 2 refer respectively to the leading and subleading top quark, where \emph{leading} refers to the top quark with the largest transverse momentum.
 
First, a set of baseline observables is presented: transverse momentum (\ptt) and absolute value of the rapidity ($|\yt|$) of the top quarks, and the transverse momentum (\pttt), absolute value of the rapidity ($|\ytt|$) and invariant mass (\mtt) of the \ttbar{} system and the transverse momentum of the leading ($\pt^{t,1}$) and subleading ($\pt^{t,2}$) top quarks. For parton-level measurements, the \pt{} and rapidity of the top quark are measured from the \pt{} and rapidity of the reconstructed hadronic top quarks. The differential cross-sections as a function of all these observables, with the exception of the \pt{} of the leading and subleading top quarks, were previously measured in the fiducial phase-space in the resolved topology by the ATLAS Collaboration using 13~\TeV\ data~\cite{TOPQ-2016-01}, while in the boosted topology only \ptth{} and $|\yth|$ were measured. The differential cross-sections as a function of the \pt{} of the leading and subleading top quarks were previously measured, at particle- and parton-level, only in the boosted topology in the fully hadronic channel~\cite{TOPQ-2016-09}.
 
\begin{figure*}[bp]
\centering
\subfigure[]{ \includegraphics[width=0.45\textwidth]{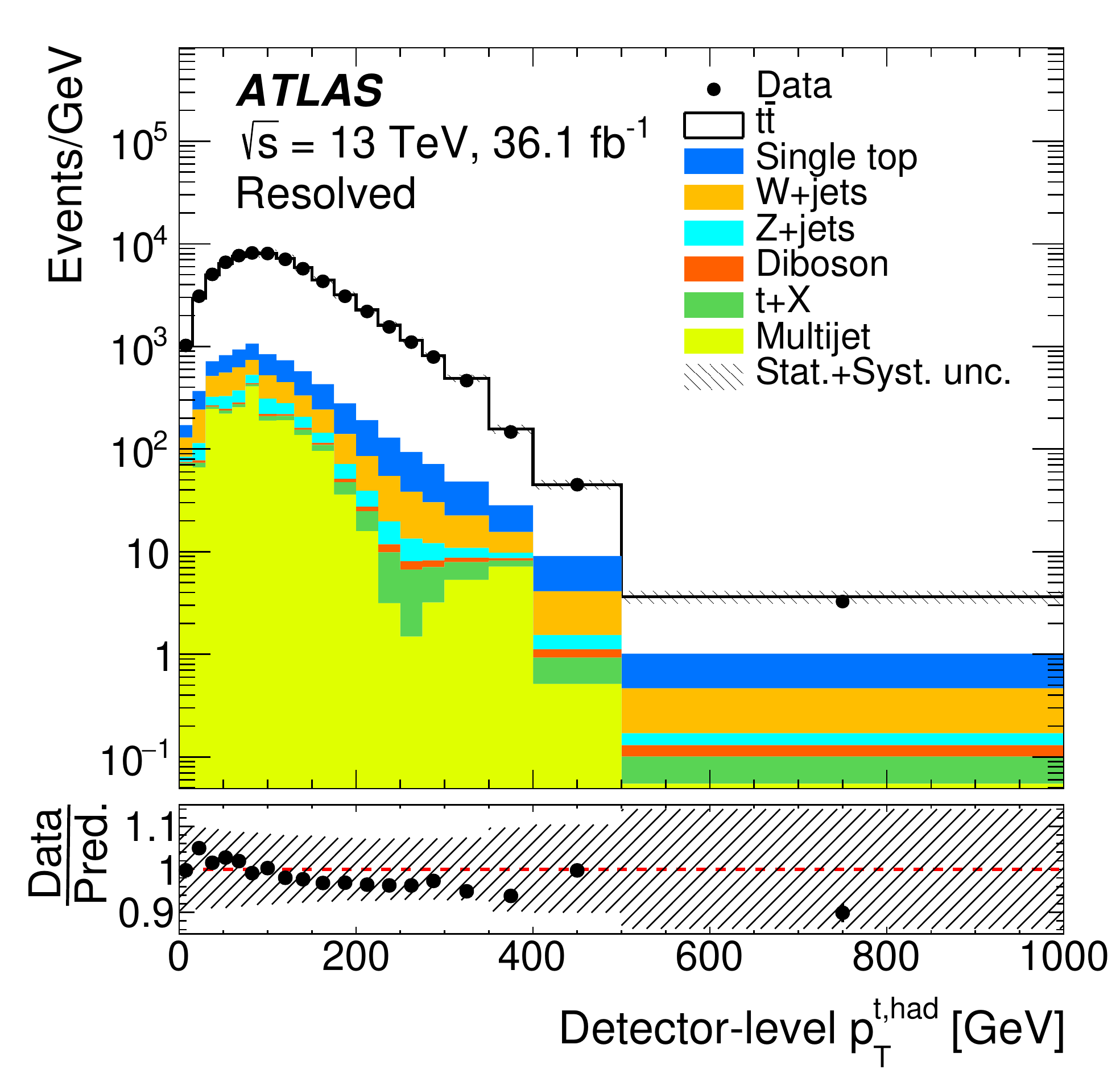}\label{fig:PseudoTop_Reco_top_had_pt}}
\subfigure[]{ \includegraphics[width=0.45\textwidth]{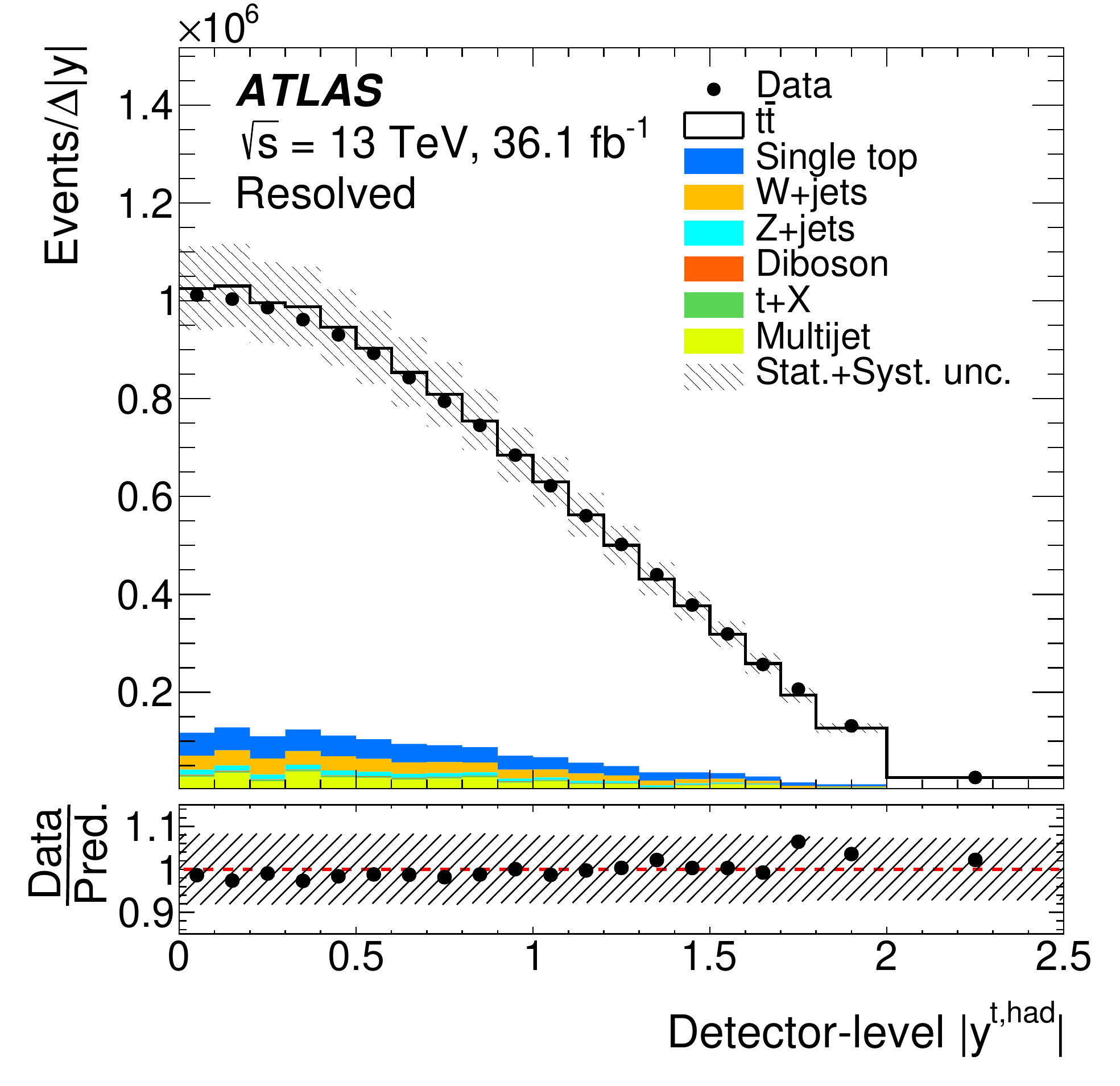}\label{fig:PseudoTop_Reco_top_had_abs_y}}\\
\subfigure[]{ \includegraphics[width=0.45\textwidth]{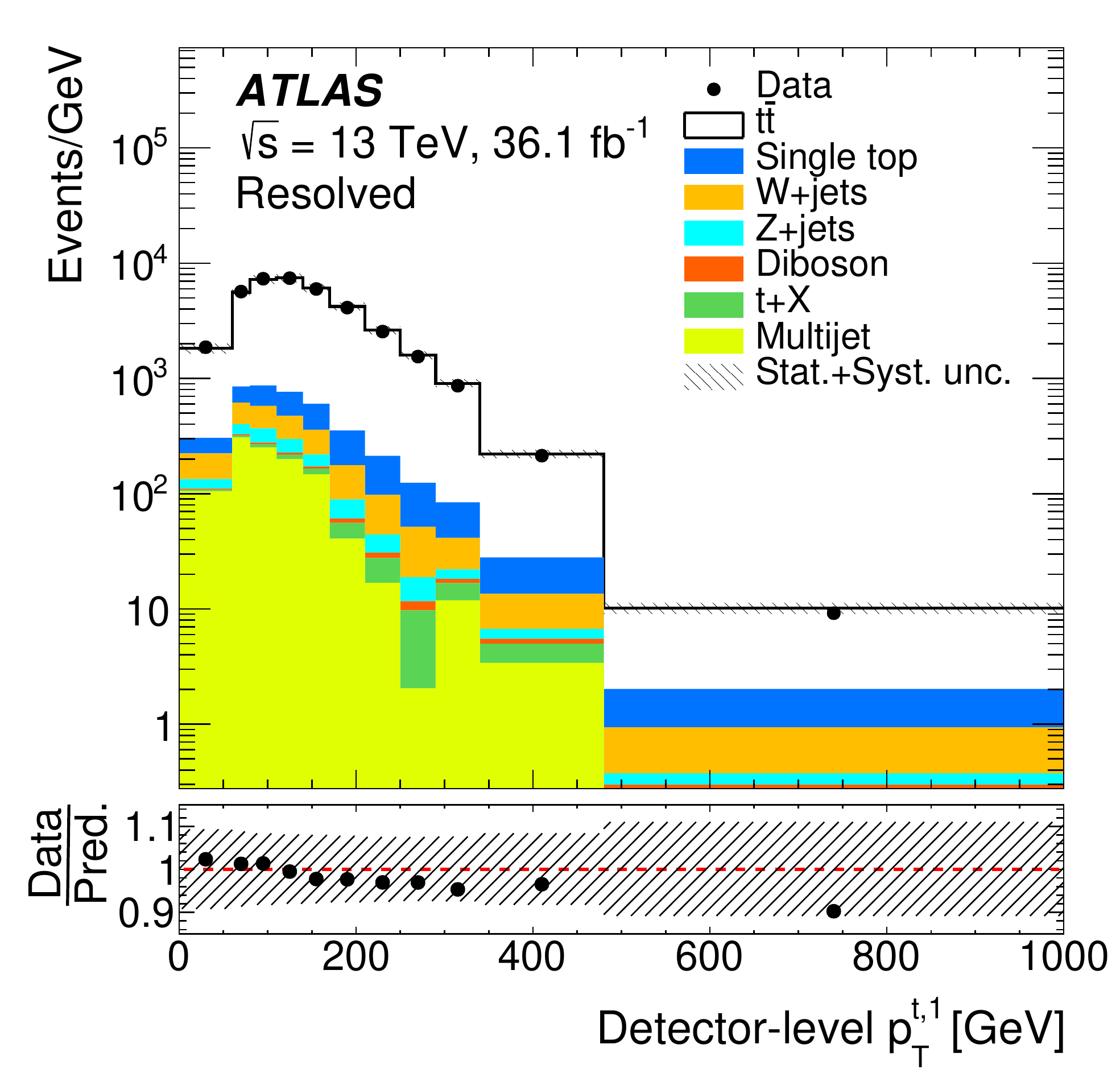}\label{fig:PseudoTop_Reco_leading_top_pt}}
\subfigure[]{ \includegraphics[width=0.45\textwidth]{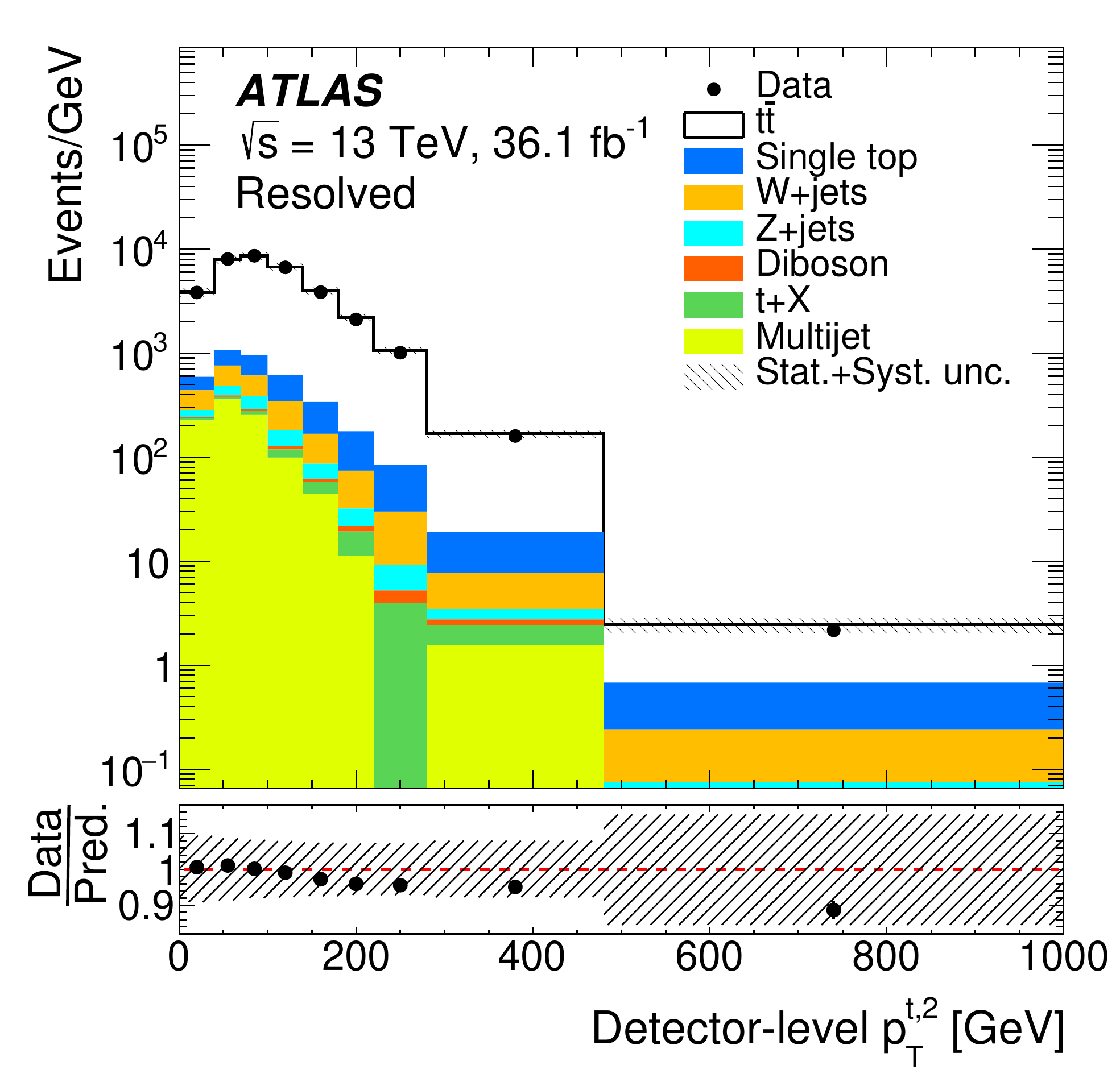}\label{fig:PseudoTop_Reco_subleading_top_pt}}
\caption{
Distributions of observables in the \ljets{} channel reconstructed with the pseudo-top algorithm  in the resolved topology at detector-level: \protect\subref{fig:PseudoTop_Reco_top_had_pt} transverse momentum   and \protect\subref{fig:PseudoTop_Reco_top_had_abs_y}  absolute value of the rapidity of the hadronic top quark, \protect\subref{fig:PseudoTop_Reco_leading_top_pt}~transverse momentum of the leading top quark and \protect\subref{fig:PseudoTop_Reco_subleading_top_pt}~transverse momentum of the subleading top quark. Data distributions are compared with
predictions, using \Powheg+\PythiaEight as the \ttbar{} signal model. The hatched area represents the combined statistical and
systematic uncertainties (described in Section~\ref{sec:uncertainties}) in the total prediction, excluding systematic uncertainties related to the modelling of the \ttbar{} events. Underflow and overflow events, if any, are included in the first and last bins. The lower panel shows the ratio of the data to the total prediction.
}
\label{fig:controls_resolved_detector:top}
\end{figure*}

\begin{figure*}[t]
\centering
\subfigure[]{ \includegraphics[width=0.45\textwidth]{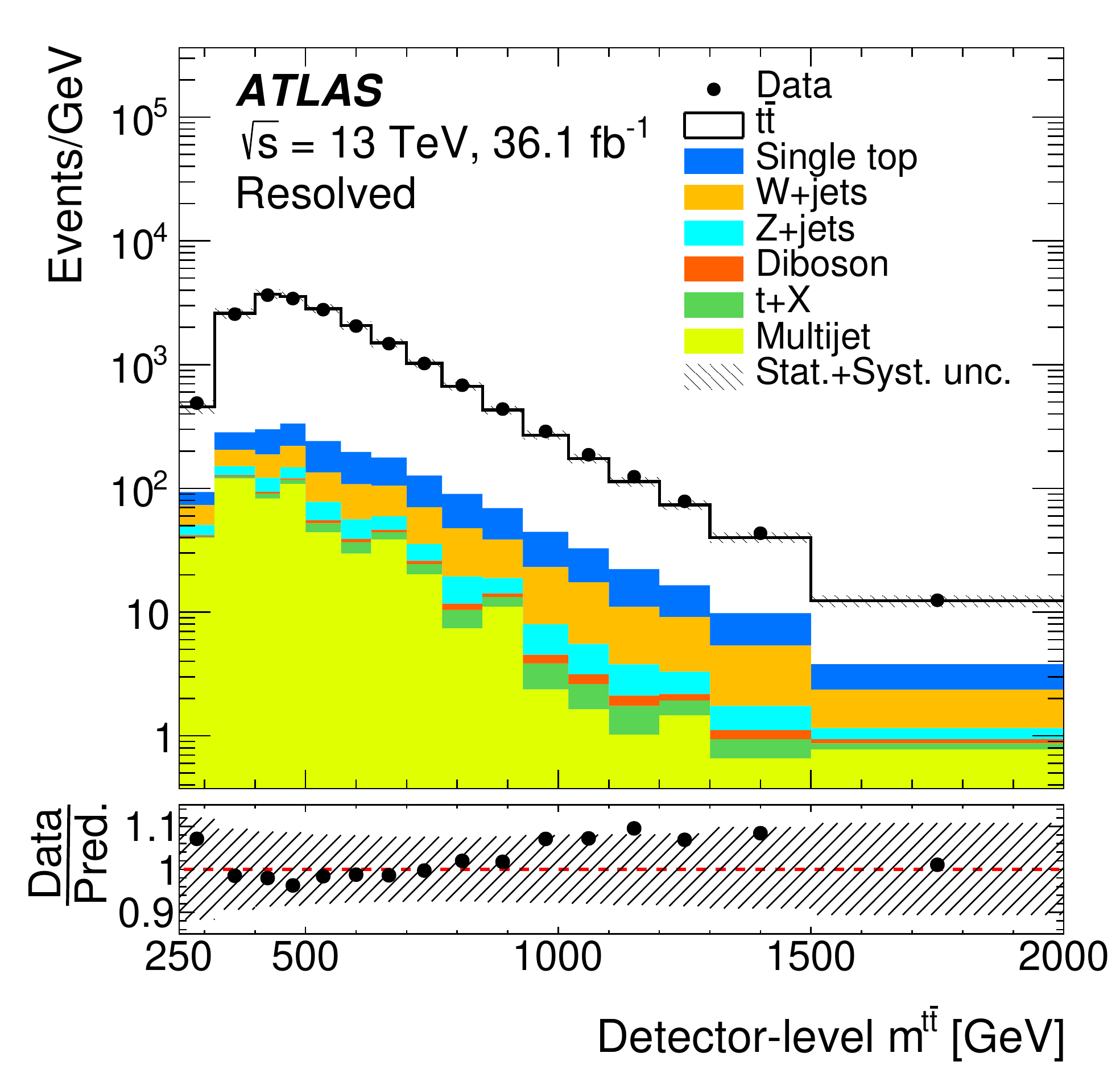}\label{fig:PseudoTop_Reco_ttbar_m}}
\subfigure[]{ \includegraphics[width=0.45\textwidth]{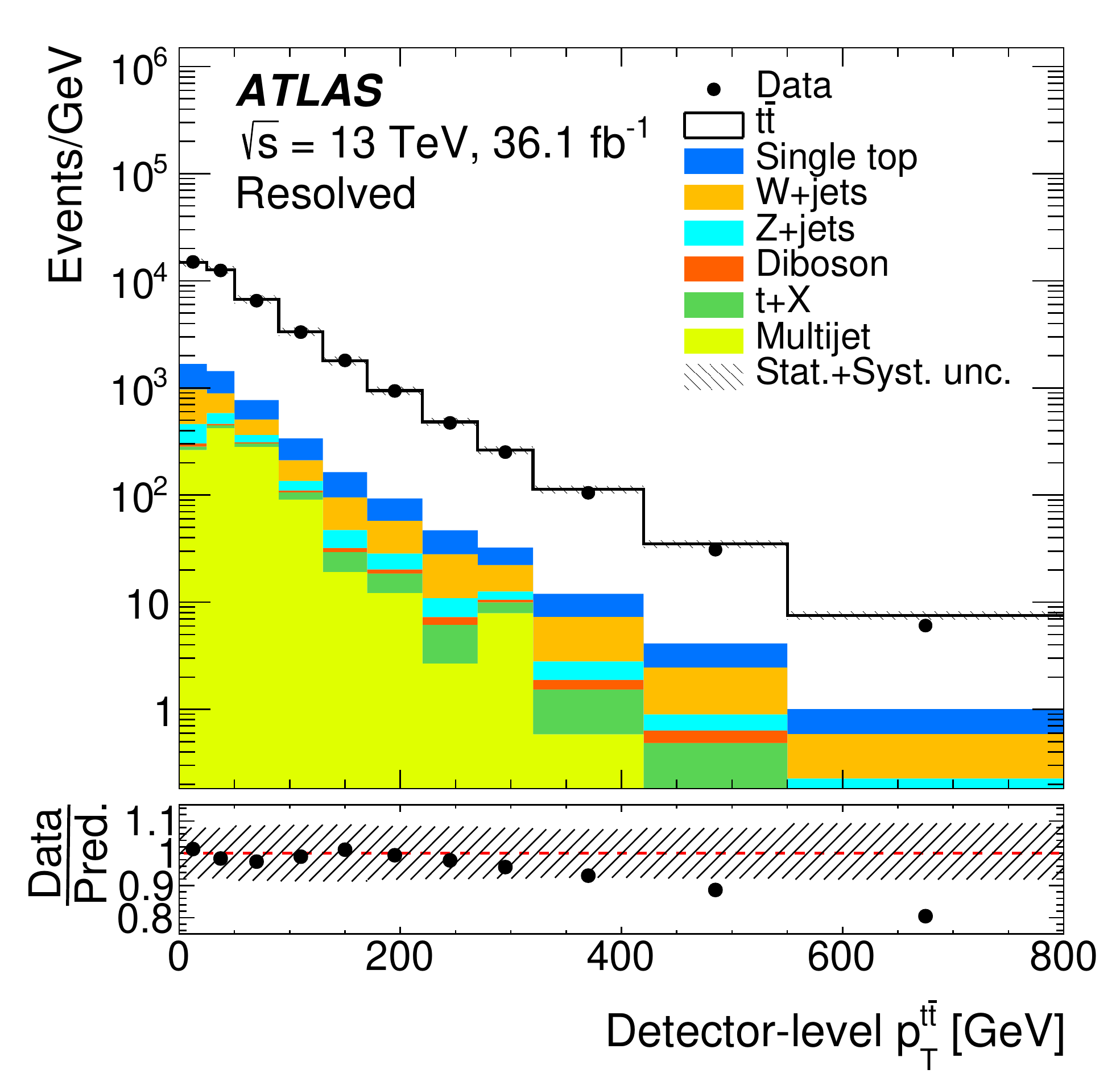}\label{fig:PseudoTop_Reco_ttbar_pt}}
\subfigure[]{ \includegraphics[width=0.45\textwidth]{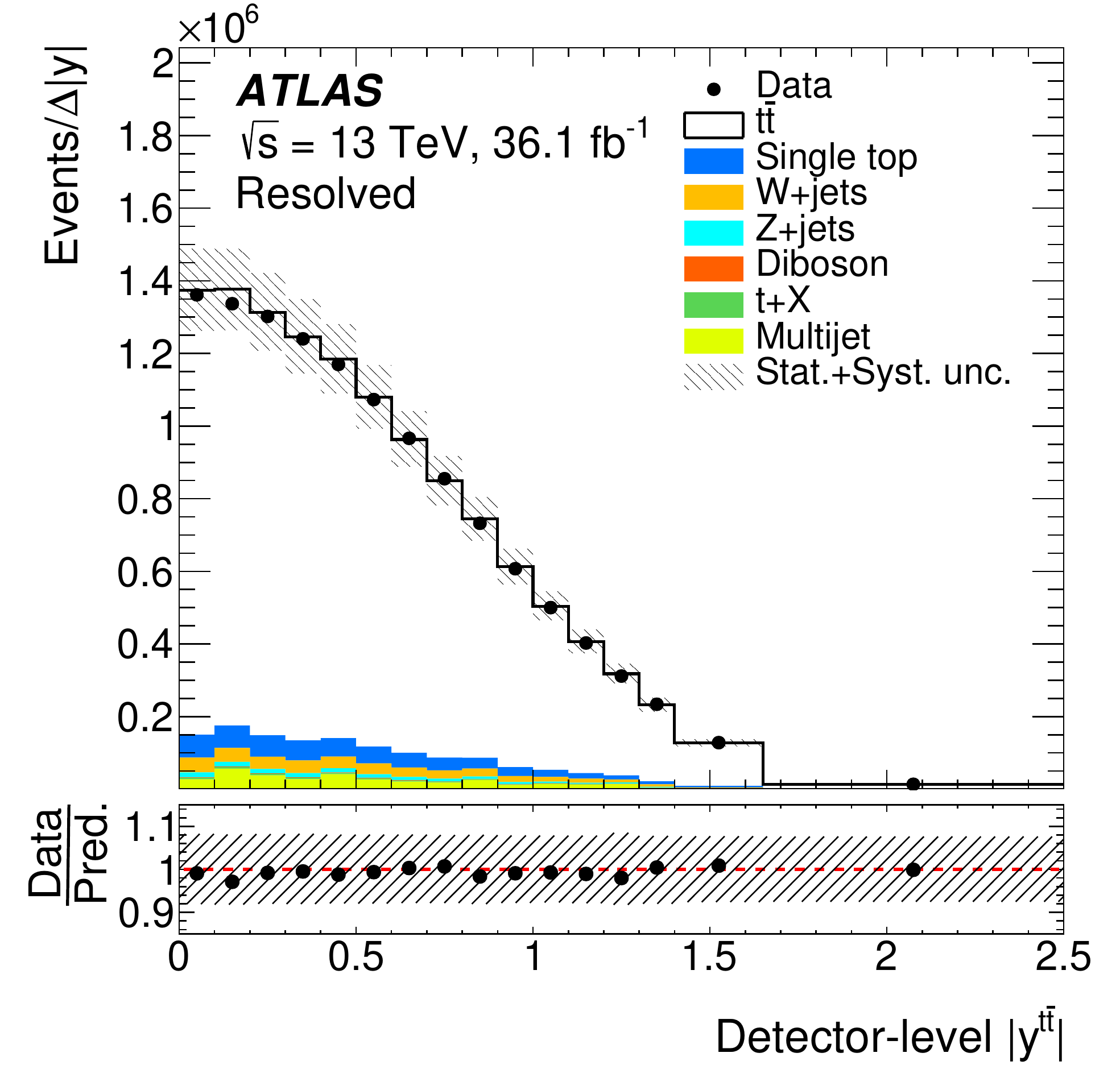}\label{fig:PseudoTop_Reco_ttbar_abs_y}}
\caption{
Distributions of observables in the \ljets{} channel reconstructed with the pseudo-top algorithm  in the resolved topology at detector-level:  \protect\subref{fig:PseudoTop_Reco_ttbar_m}  invariant mass,  \protect\subref{fig:PseudoTop_Reco_ttbar_pt}   transverse momentum
and \protect\subref{fig:PseudoTop_Reco_ttbar_abs_y} absolute value of the rapidity of the \ttbar{} system. Data distributions are compared with
predictions, using \Powheg+\PythiaEight as the \ttbar{} signal model. The hatched area represents the combined statistical and
systematic uncertainties (described in Section~\ref{sec:uncertainties}) in the total prediction, excluding systematic uncertainties related to the modelling of the \ttbar{} events. Underflow and overflow events, if any, are included in the first and last bins. The lower panel shows the ratio of the data to the total prediction.
}
\label{fig:controls_resolved_detector:ttbar}
\end{figure*}

\begin{figure*}[t]
\centering
\subfigure[]{ \includegraphics[width=0.45\textwidth]{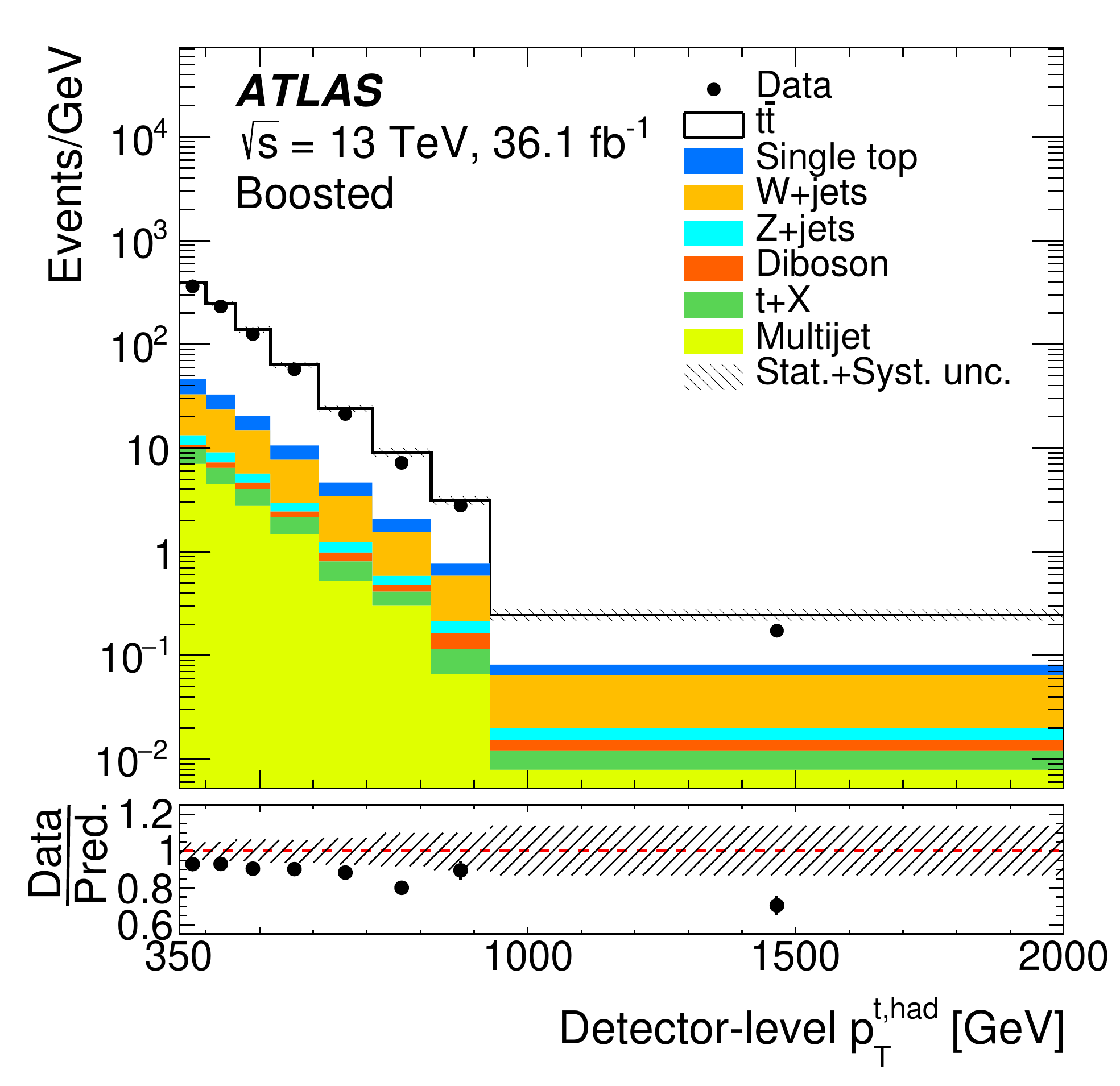}\label{fig:hadTop_boosted_rc_pt}}
\subfigure[]{ \includegraphics[width=0.45\textwidth]{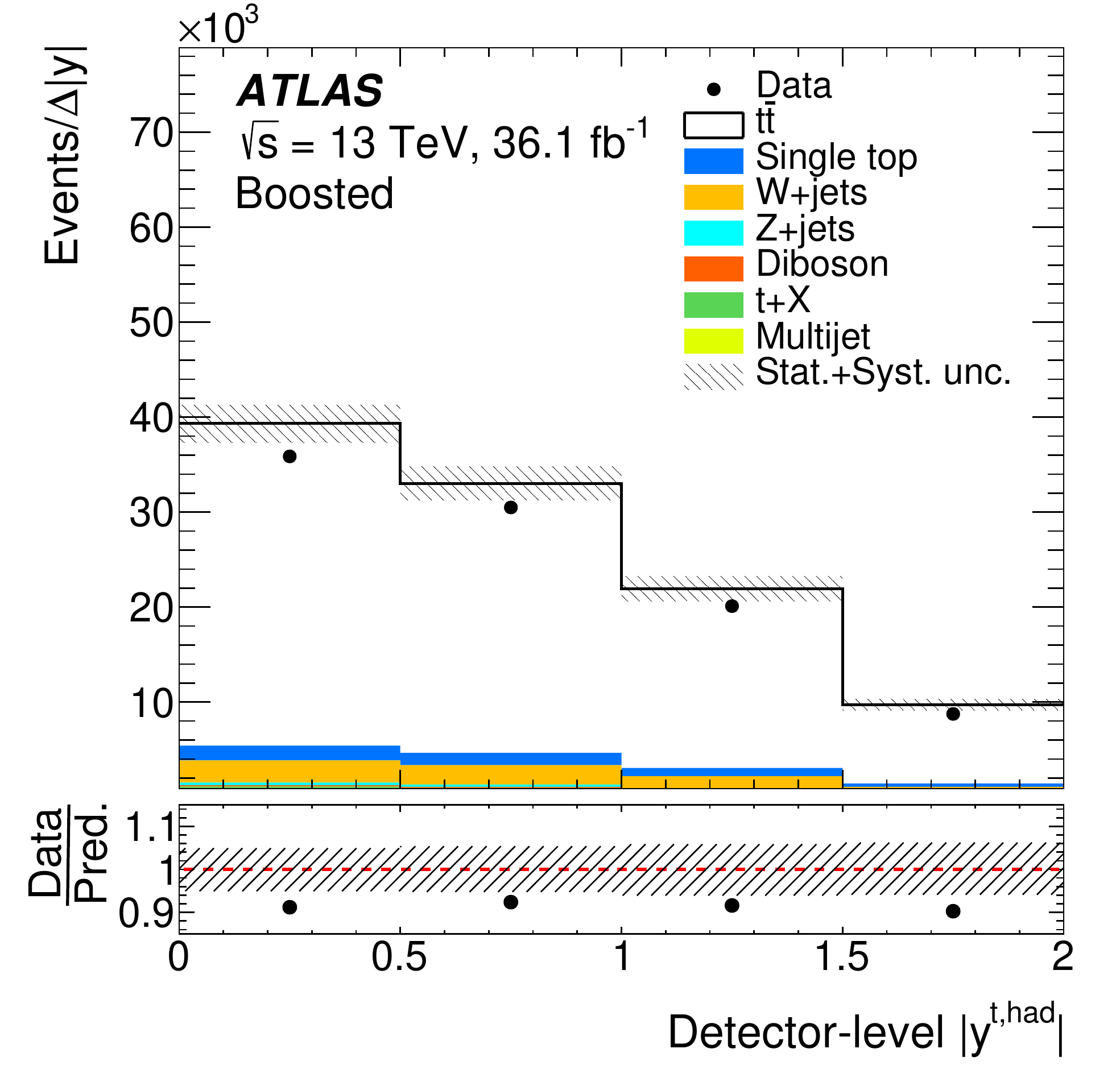}\label{fig:hadTop_boosted_rc_y}}
\subfigure[]{ \includegraphics[width=0.45\textwidth]{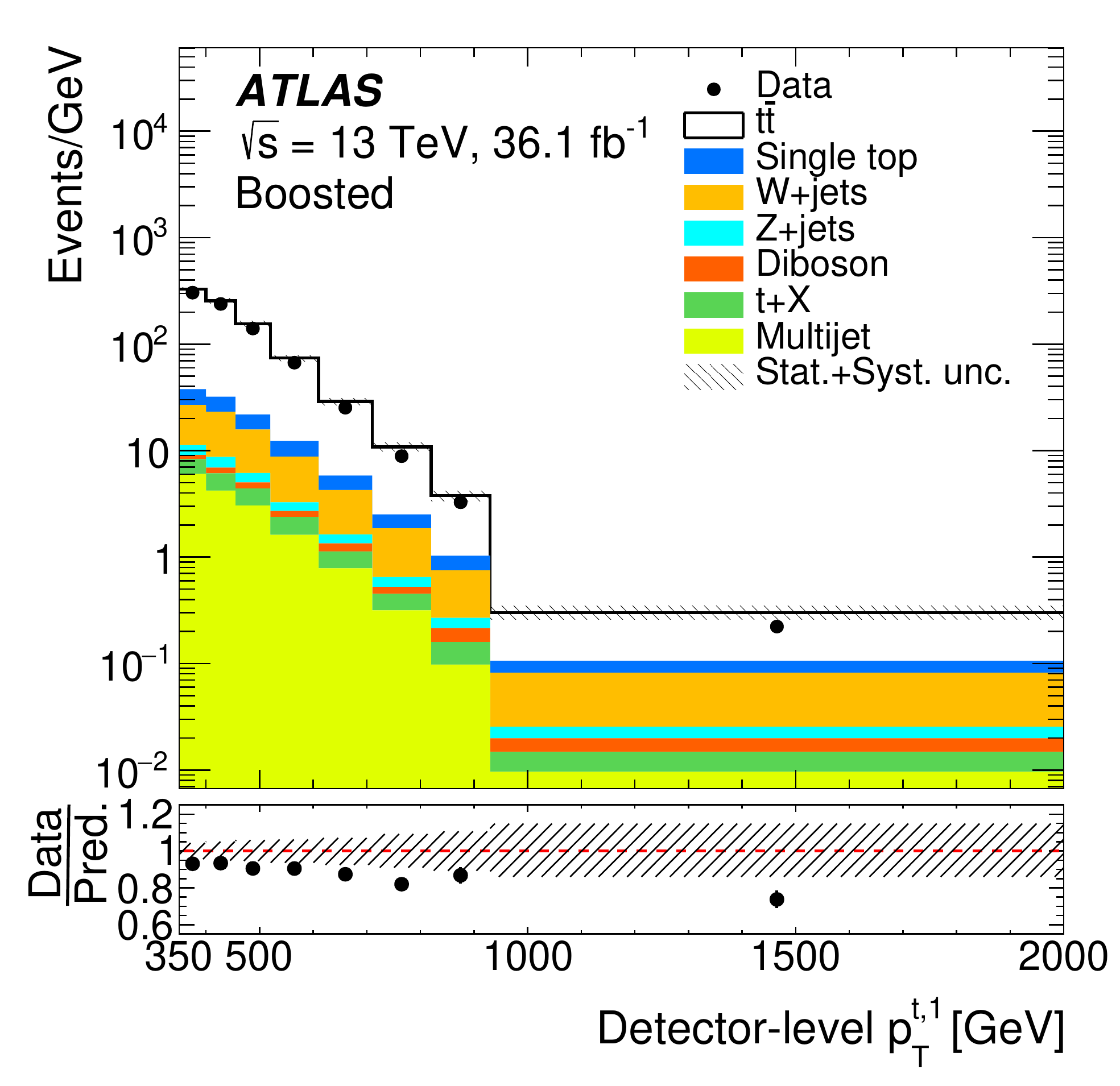}\label{fig:hadTop_boosted_rc_pt_leading}}
\subfigure[]{ \includegraphics[width=0.45\textwidth]{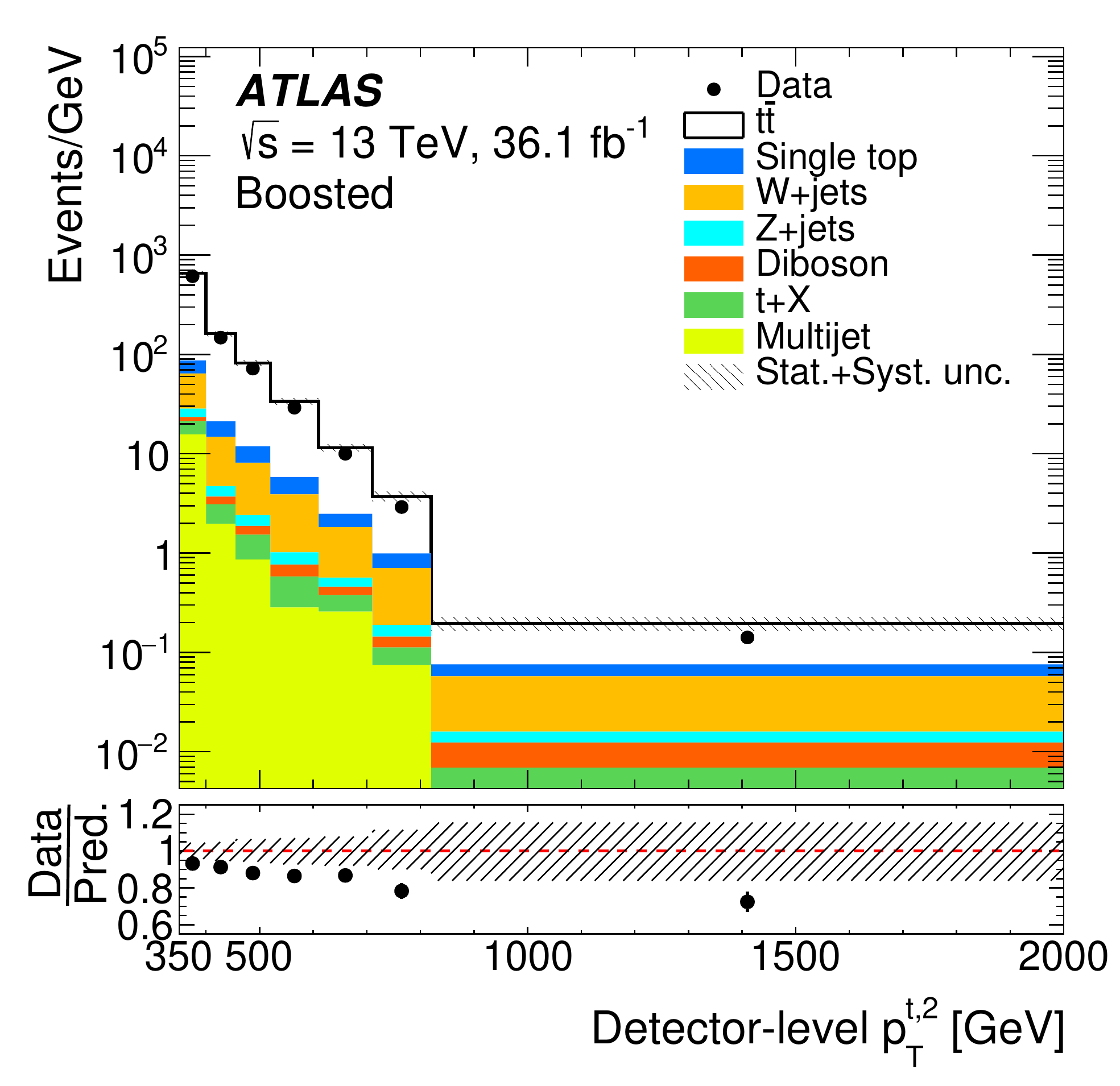}\label{fig:hadTop_boosted_rc_pt_subleading}}
\caption{Distributions of observables in the \ljets{} channel  in the boosted topology at detector-level: \protect\subref{fig:hadTop_boosted_rc_pt}   transverse momentum and \protect\subref{fig:hadTop_boosted_rc_y} absolute value of the rapidity of the  hadronic top quark, \protect\subref{fig:hadTop_boosted_rc_pt_leading}~transverse momentum of the leading top quark and \protect\subref{fig:hadTop_boosted_rc_pt_subleading}~transverse momentum of the subleading top quark. Data distributions are compared with
predictions, using \Powheg+\PythiaEight as the \ttbar{} signal model. The hatched area represents the combined statistical and
systematic uncertainties (described in Section~\ref{sec:uncertainties}) in the total prediction, excluding systematic uncertainties related to the modelling of the \ttbar{} events. Underflow and overflow events, if any, are included in the first and last bins. The lower panel shows the ratio of the data to the total prediction.
}
\label{fig:controls_boosted_detector:top}
\end{figure*}

\begin{figure*}[t]
\centering
\subfigure[]{ \includegraphics[width=0.45\textwidth]{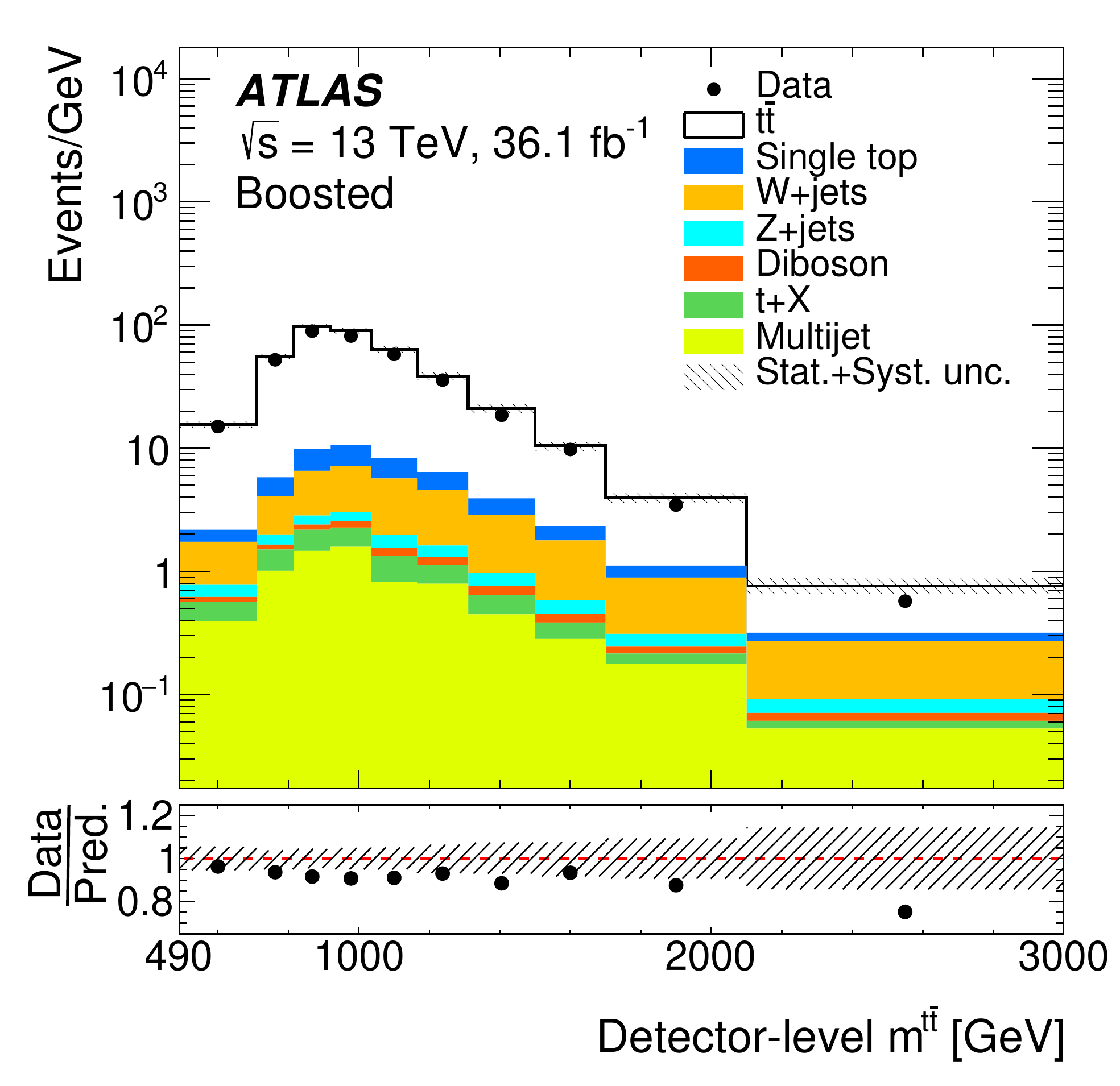}\label{fig:ttbar_boosted_rc_m}}
\subfigure[]{ \includegraphics[width=0.45\textwidth]{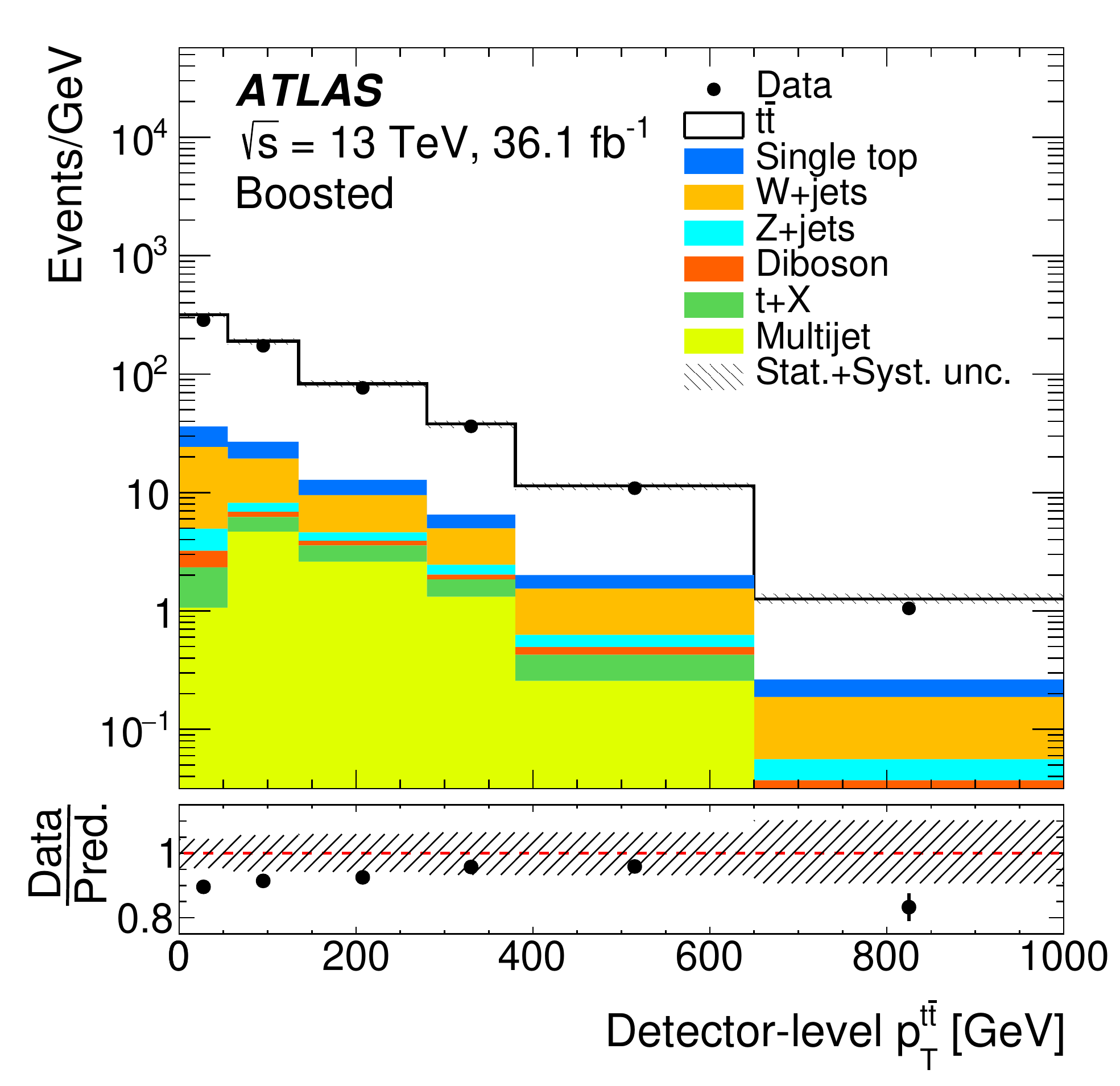}\label{fig:ttbar_boosted_rc_pt}}
\subfigure[]{ \includegraphics[width=0.45\textwidth]{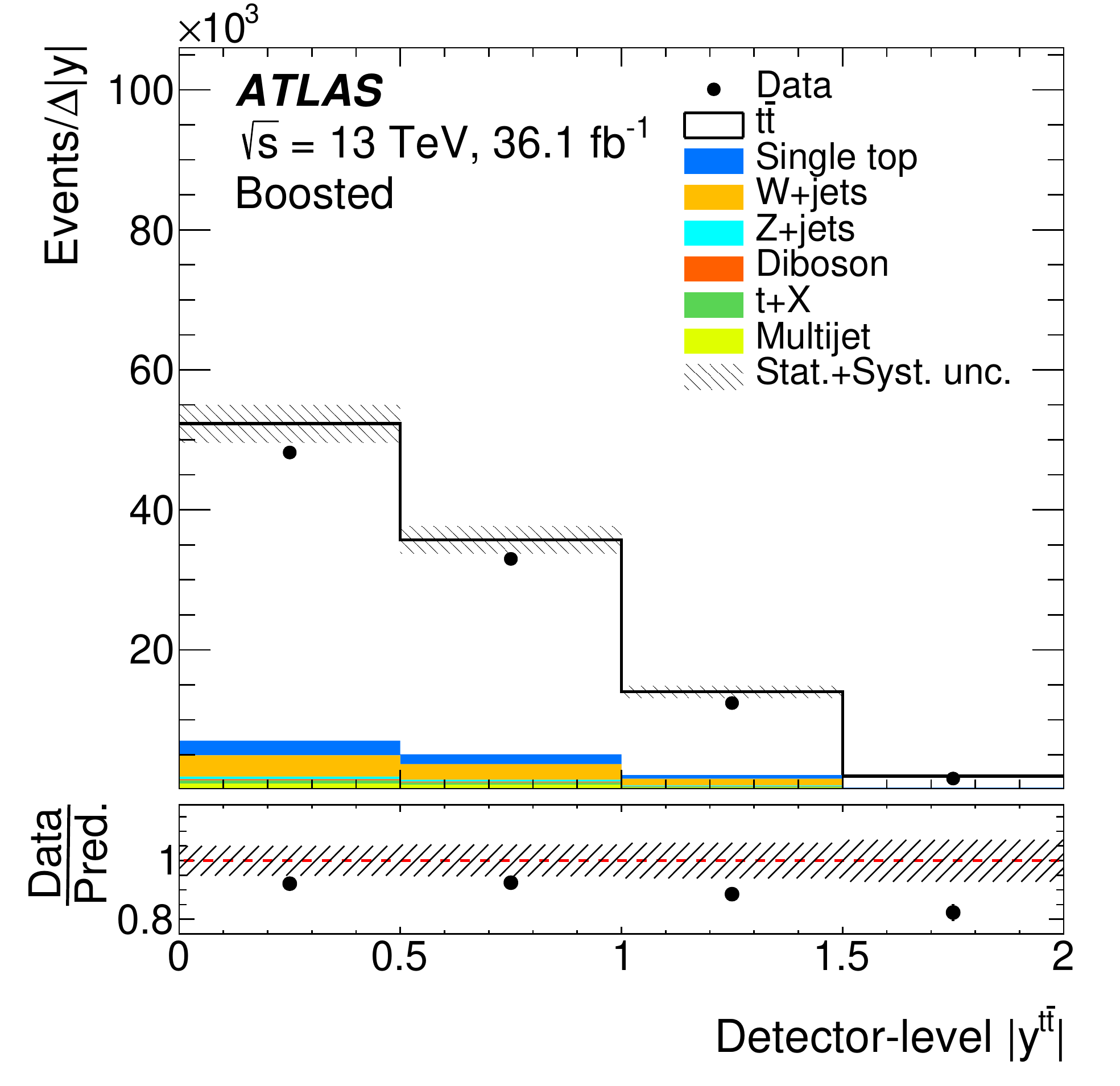}\label{fig:ttbar_boosted_rc_y}}
\caption{Kinematic distributions in the \ljets{} channel in the boosted topology at detector-level: \protect\subref{fig:ttbar_boosted_rc_m}   invariant mass, \protect\subref{fig:ttbar_boosted_rc_pt}    transverse momentum
and~\protect\subref{fig:ttbar_boosted_rc_y} absolute value of the rapidity of the \ttbar{} system. Data distributions are compared with
predictions, using \Powheg+\PythiaEight as the \ttbar{} signal model. The hatched area represents the combined statistical and
systematic uncertainties (described in Section~\ref{sec:uncertainties}) in the total prediction, excluding systematic uncertainties related to the modelling of the \ttbar{} events. Underflow and overflow events, if any, are included in the first and last bins. The lower panel shows the ratio of the data to the total prediction.
}
\label{fig:controls_boosted_detector:ttbar}
\end{figure*}

The detector-level distributions of the kinematic variables of the top quark and \ttb{} system in the resolved topology are presented in Figures~\ref{fig:controls_resolved_detector:top} and \ref{fig:controls_resolved_detector:ttbar}, respectively. The detector-level distributions of the same observables, reconstructed in the boosted topology, are shown in Figures~\ref{fig:controls_boosted_detector:top} and~\ref{fig:controls_boosted_detector:ttbar}.

Furthermore, angular variables sensitive to the momentum imbalance in the transverse plane (\pout{}), i.e.\ to the emission
of radiation associated with the production of the top-quark pair, are used to investigate the central
production region~\cite{fermilab:dijet}. The angle between the two top quarks is sensitive to non-resonant
contributions from hypothetical new particles exchanged in the $t$-channel~\cite{EXOT-2014-15}. The rapidities of the two
top quarks in the \ttbar{} centre-of-mass frame are $y^*=\frac{1}{2}\left(\yth-\ytl\right)$ and $-y^*$. The longitudinal motion of the \ttbar{} system in the laboratory frame is described by  the rapidity boost $\yboost = \frac{1}{2}\left(\yth+\ytl\right)$.
The production polar angle is closely related to the variable $\chitt$, defined as $\chitt=\mathrm{e}^{2\left|y^*\right|}$, which is included in the measurement since many signals due to processes not included in the SM are predicted to peak at low values of this distribution~\cite{EXOT-2014-15}. Finally, observables depending on the transverse momentum of the decay products of the top quark are sensitive to higher-order corrections~\cite{WWbb:NNLO,ttbar:offshell:nlo}.
In summary, the following additional observables are measured:
\begin{itemize}
\item The absolute value of the azimuthal angle between the two top quarks (\deltaPhitt).
\item The out-of-plane momentum, i.e.\ the projection of the top-quark three-momentum
onto the direction perpendicular to the plane defined by the other top quark and the beam axis
($z$) in the laboratory frame~\cite{fermilab:dijet}:
\begin{equation*}
\Poutthad=\vec{p}^{\,t\mathrm{,had}} \cdot \frac {\vec{p}^{\,t\mathrm{,lep}}\times\vec{e}_z} {\left|\vec{p}^{\,t\mathrm{,lep}}\times\vec{e}_z\right|} \mathrm{,}
\end{equation*}
\begin{equation*}
\Pouttlep=\vec{p}^{\,t\mathrm{,lep}} \cdot \frac {\vec{p}^{\,t\mathrm{,had}}\times\vec{e}_z} {\left|\vec{p}^{\,t\mathrm{,had}}\times\vec{e}_z\right|}
\end{equation*}
In particular, $|\Poutthad|$, introduced in Ref.~\cite{TOPQ-2015-06}, is used in the resolved topology, while in the boosted topology, where an asymmetry between $p^{t,\mathrm{had}}$ and $p^{t,\mathrm{lep}}$ exists by construction, the variable $|\Pouttlep|$ is measured. This reduces the correlation between $p_{\mathrm{out}}$ and $p^{t,\mathrm{had}}$, biased toward high values by construction, while keeping the sensitivity to the momentum imbalance.
\item The longitudinal boost of the \ttbar{} system in the laboratory frame (\yboost)~\cite{EXOT-2014-15}.
\item $\chitt=\mathrm{e}^{2\left|y^*\right|}$~\cite{EXOT-2014-15}, closely related to the production polar angle.
\item The scalar sum of the transverse momenta of the hadronic and leptonic top quarks (\Htt $= {p}^{\,t\mathrm{,had}}_{\mathrm{T}} + {p}^{\,t\mathrm{,lep}}_{\mathrm{T}}$ )~\cite{WWbb:NNLO,ttbar:offshell:nlo}.
 
\end{itemize}
 
\begin{figure*}[t]
\centering
\subfigure[]{ \includegraphics[width=0.45\textwidth]{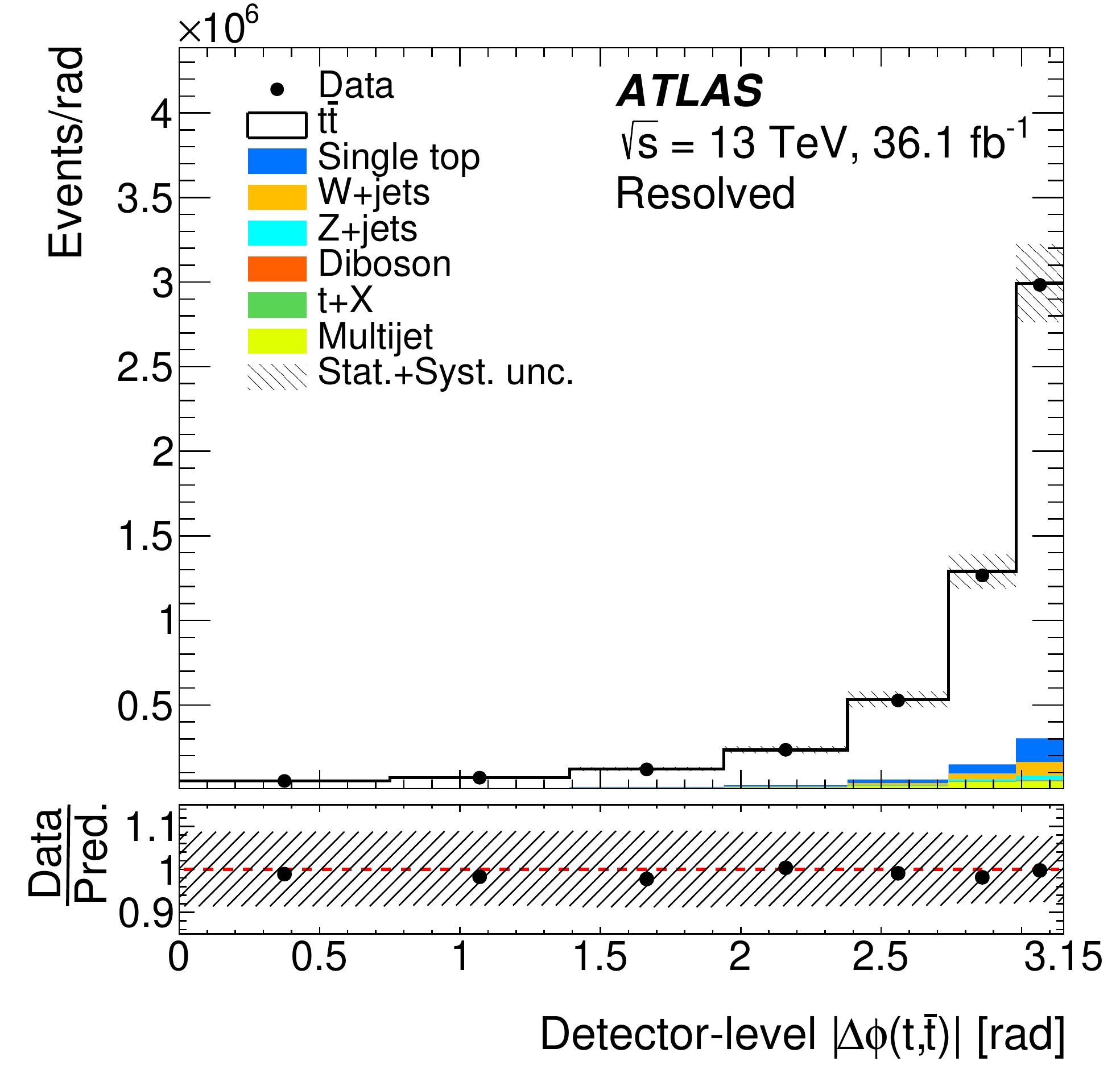}\label{fig:PseudoTop_Reco_deltaPhi_tt}}
\subfigure[]{ \includegraphics[width=0.45\textwidth]{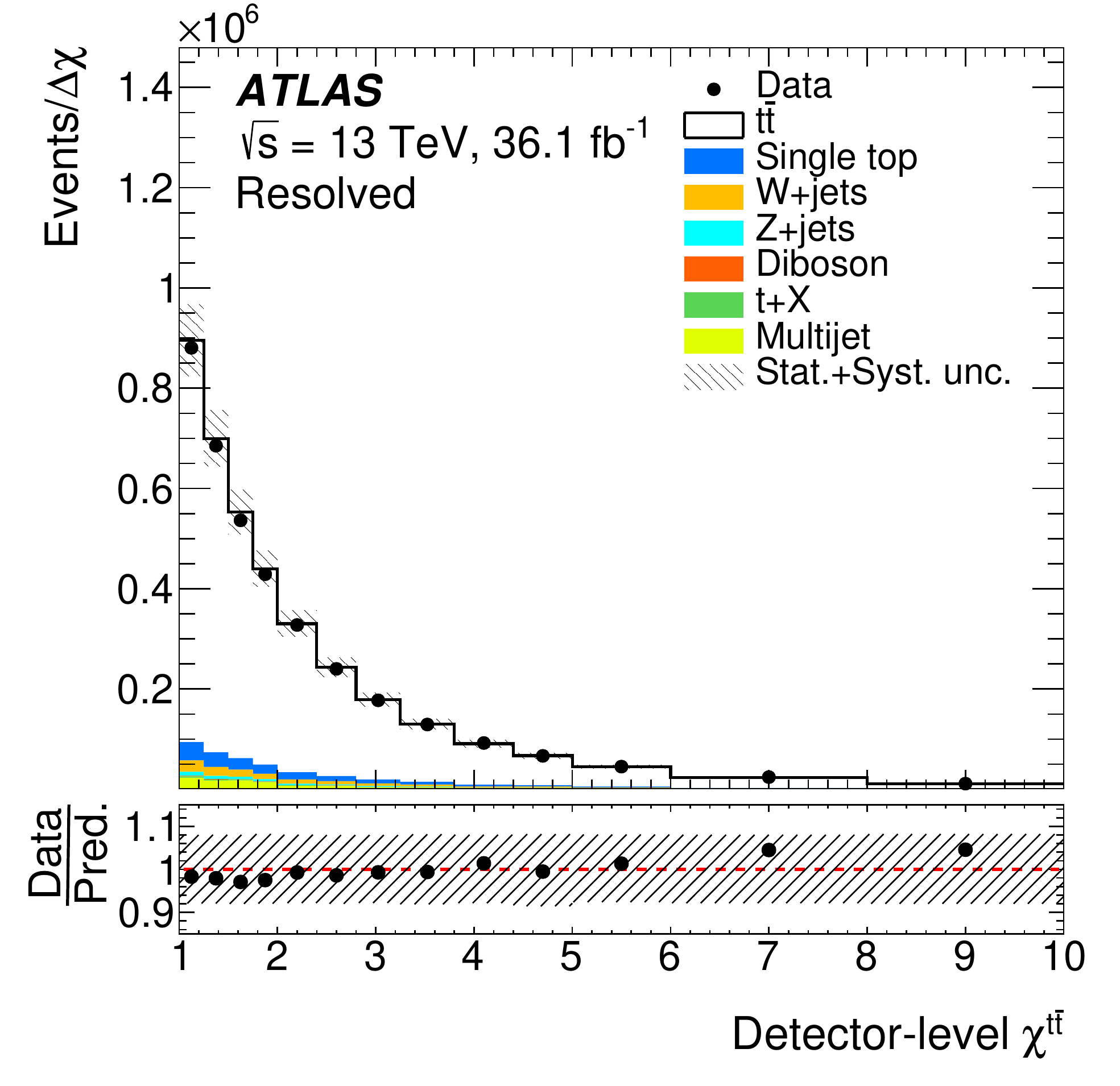}\label{fig:PseudoTop_Reco_chi_tt}}
\subfigure[]{ \includegraphics[width=0.45\textwidth]{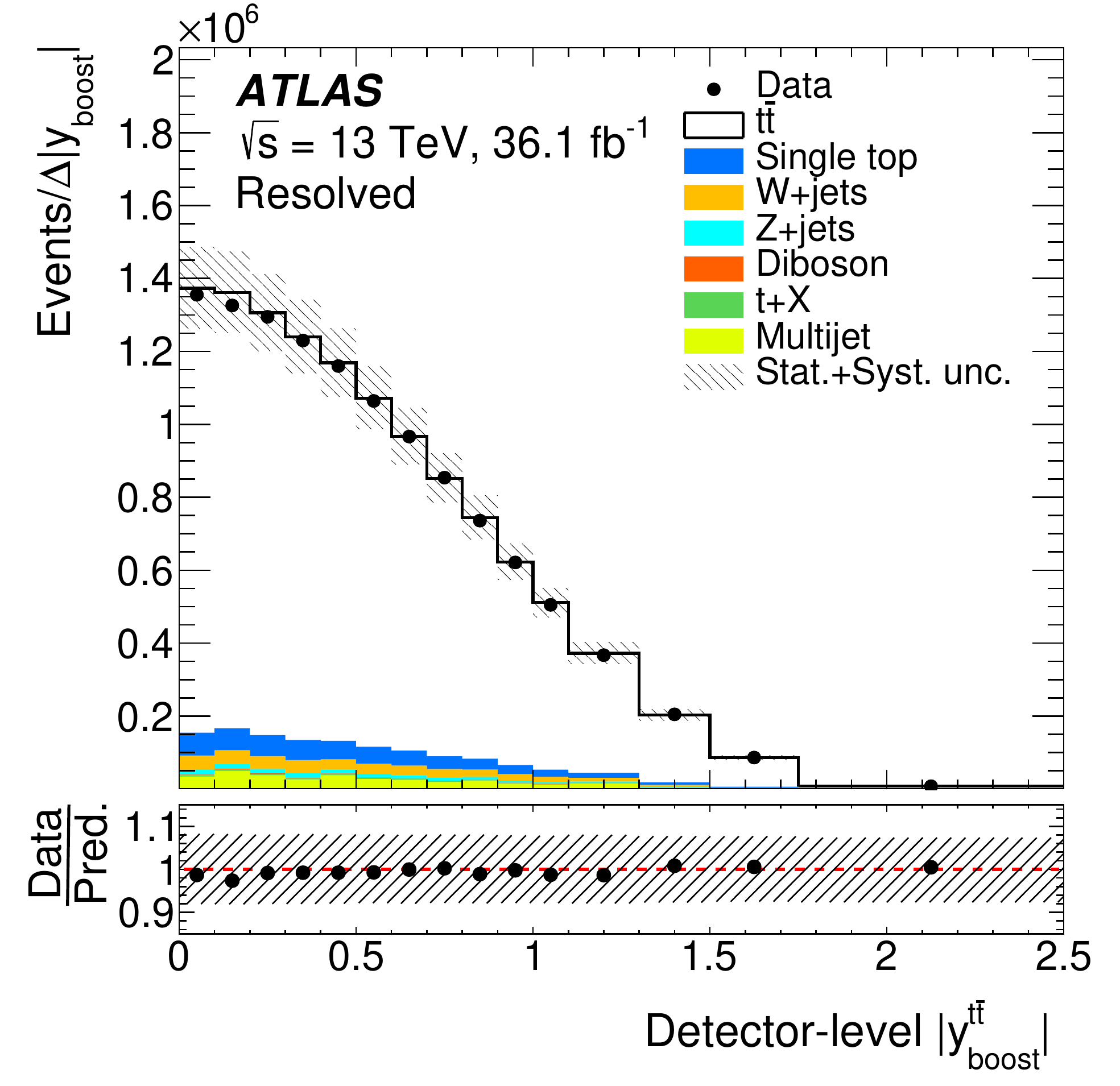}\label{fig:PseudoTop_Reco_abs_yboost}}
\caption{ Distributions of observables in the \ljets{} channel reconstructed with the pseudo-top algorithm  in the resolved topology at detector-level: \protect\subref{fig:PseudoTop_Reco_deltaPhi_tt}  azimuthal angle between the two top quarks \deltaPhitt, \protect\subref{fig:PseudoTop_Reco_chi_tt} production
angle \chitt and   \protect\subref{fig:PseudoTop_Reco_abs_yboost} absolute value of the longitudinal boost \yboost{}. Data distributions are compared with
predictions, using \Powheg+\PythiaEight as the \ttbar{} signal model. The hatched area represents the combined statistical and
systematic uncertainties (described in Section~\ref{sec:uncertainties}) in the total prediction, excluding systematic uncertainties related to the modelling of the \ttbar{} events. Underflow and overflow events, if any, are included in the first and last bins. The lower panel shows the ratio of the data to the total prediction.
}
\label{fig:controls_resolved_detector:additional:1}
\end{figure*}

\begin{figure*}[t]
\centering
 
\subfigure[]{ \includegraphics[width=0.45\textwidth]{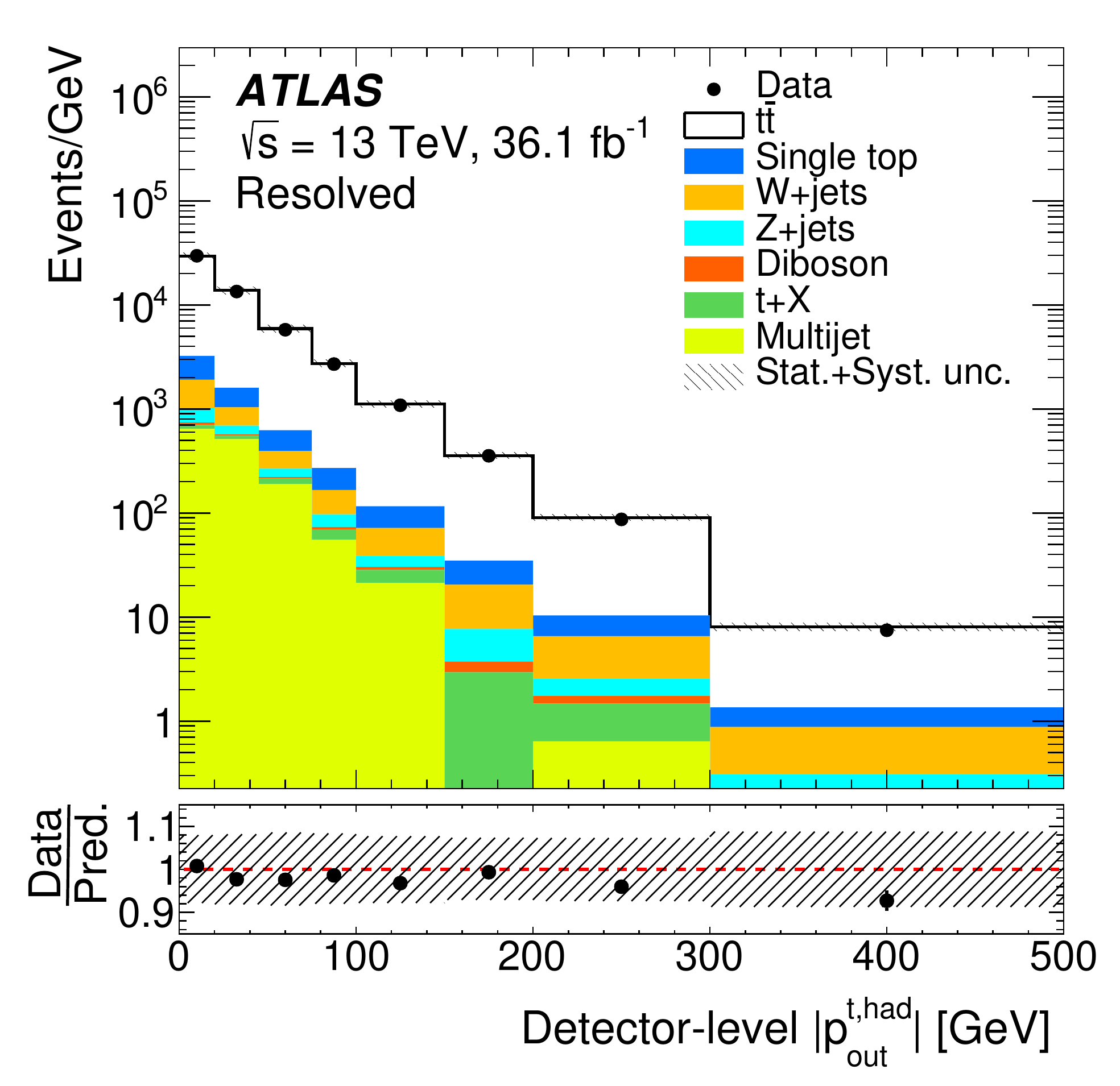}\label{fig:PseudoTop_Reco_absPout}}
\subfigure[]{ \includegraphics[width=0.45\textwidth]{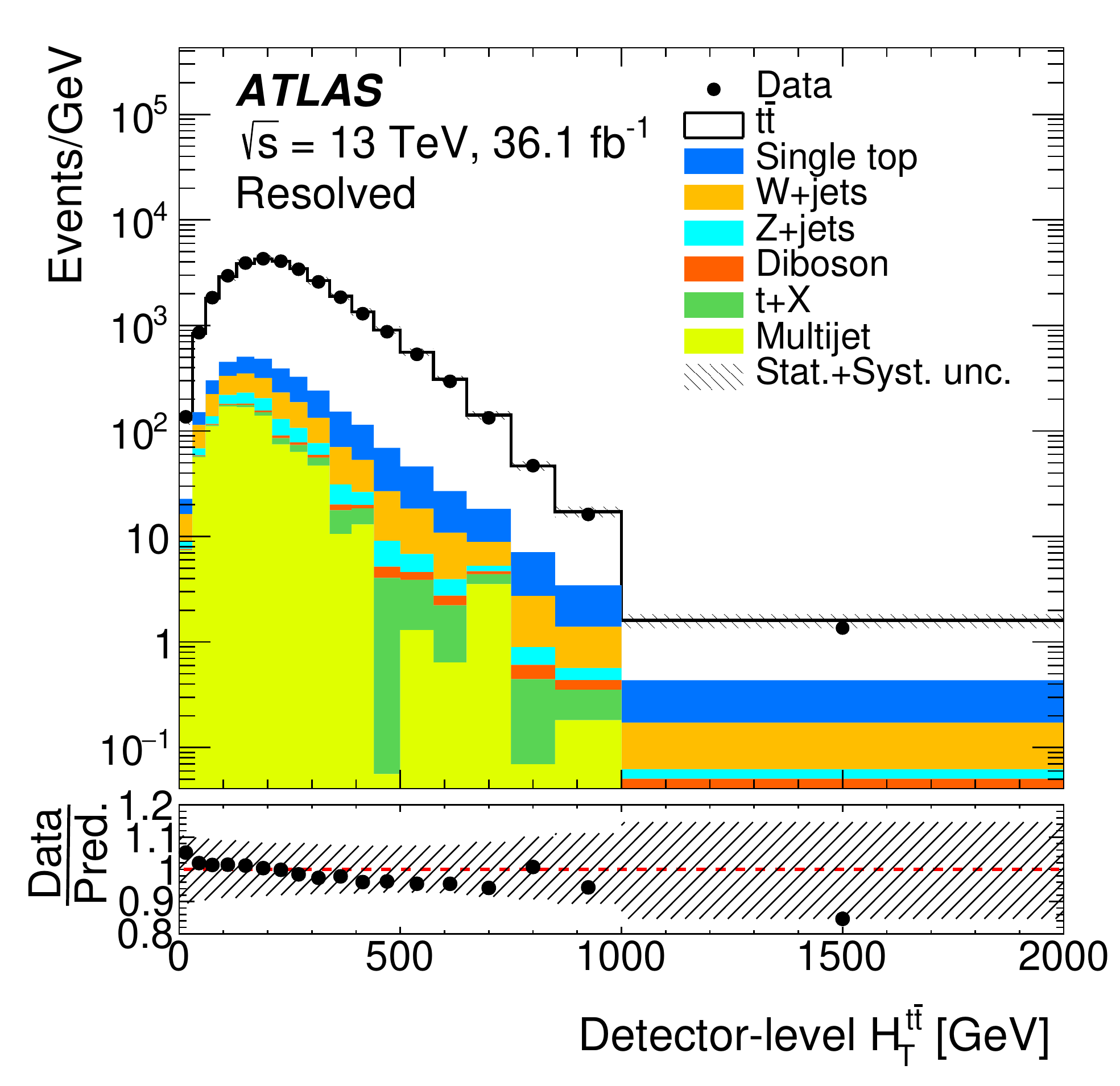}\label{fig:PseudoTop_Reco_HT_tt}}
\caption{ Kinematic distributions in the \ljets{} channel in the resolved topology reconstructed with the pseudo-top algorithm at detector-level: \protect\subref{fig:PseudoTop_Reco_absPout}  absolute value of the out-of-plane momentum $|\Poutthad|$
and  \protect\subref{fig:PseudoTop_Reco_HT_tt} scalar sum of the transverse momenta of the hadronic and leptonic top quarks  \Htt. Data distributions are compared with
predictions, using \Powheg+\PythiaEight as the \ttbar{} signal model. The hatched area represents the combined statistical and
systematic uncertainties (described in Section~\ref{sec:uncertainties}) in the total prediction, excluding systematic uncertainties related to the modelling of the \ttbar{} events. Underflow and overflow events, if any, are included in the first and last bins. The lower panel shows the ratio of the data to the total prediction.
}
\label{fig:controls_resolved_detector:additional:2}
\end{figure*}

\begin{figure*}[t]
\centering
\subfigure[]{ \includegraphics[width=0.45\textwidth]{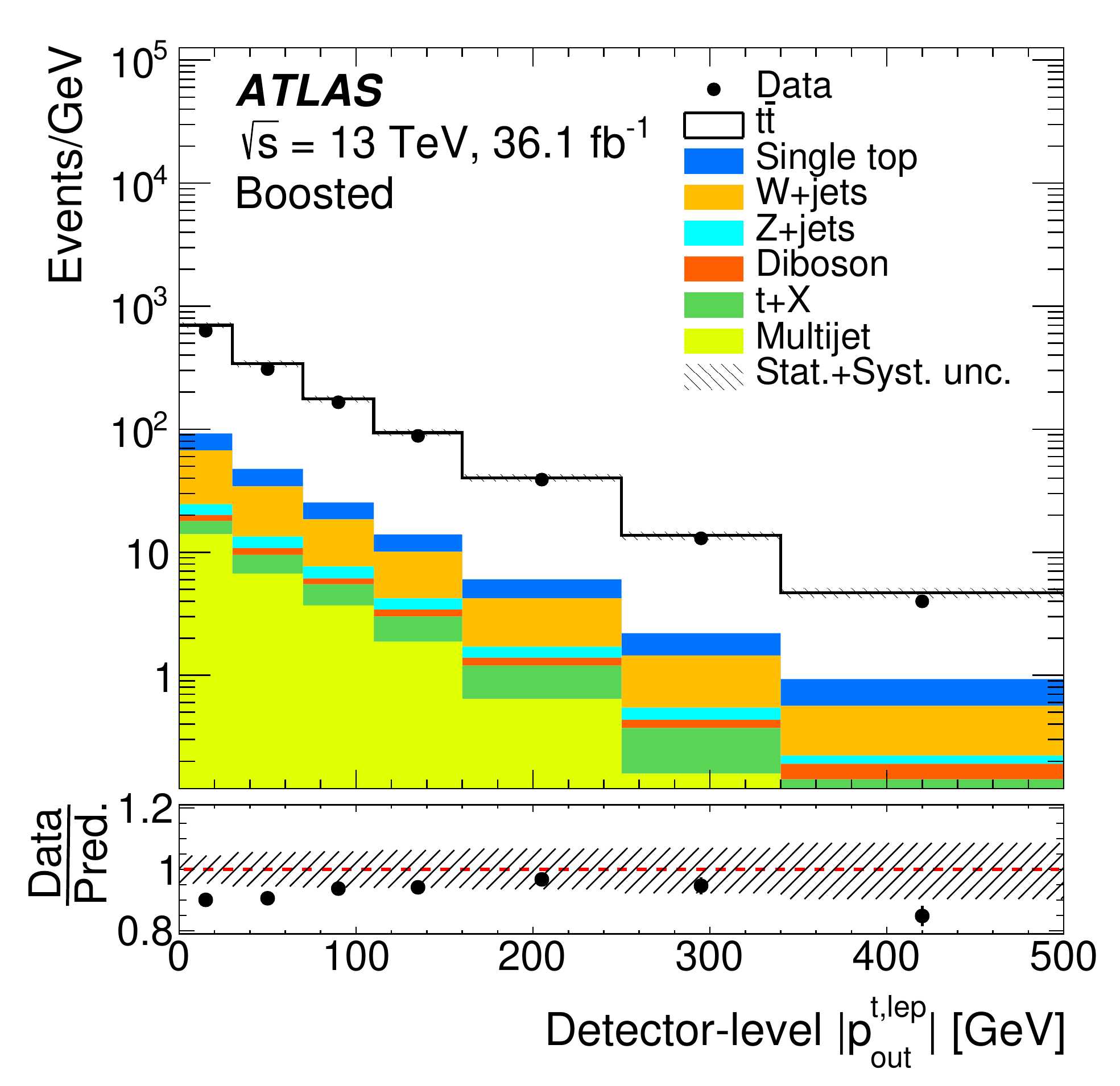}\label{fig:hadTop_boosted_rc_Pout}}
\subfigure[]{ \includegraphics[width=0.45\textwidth]{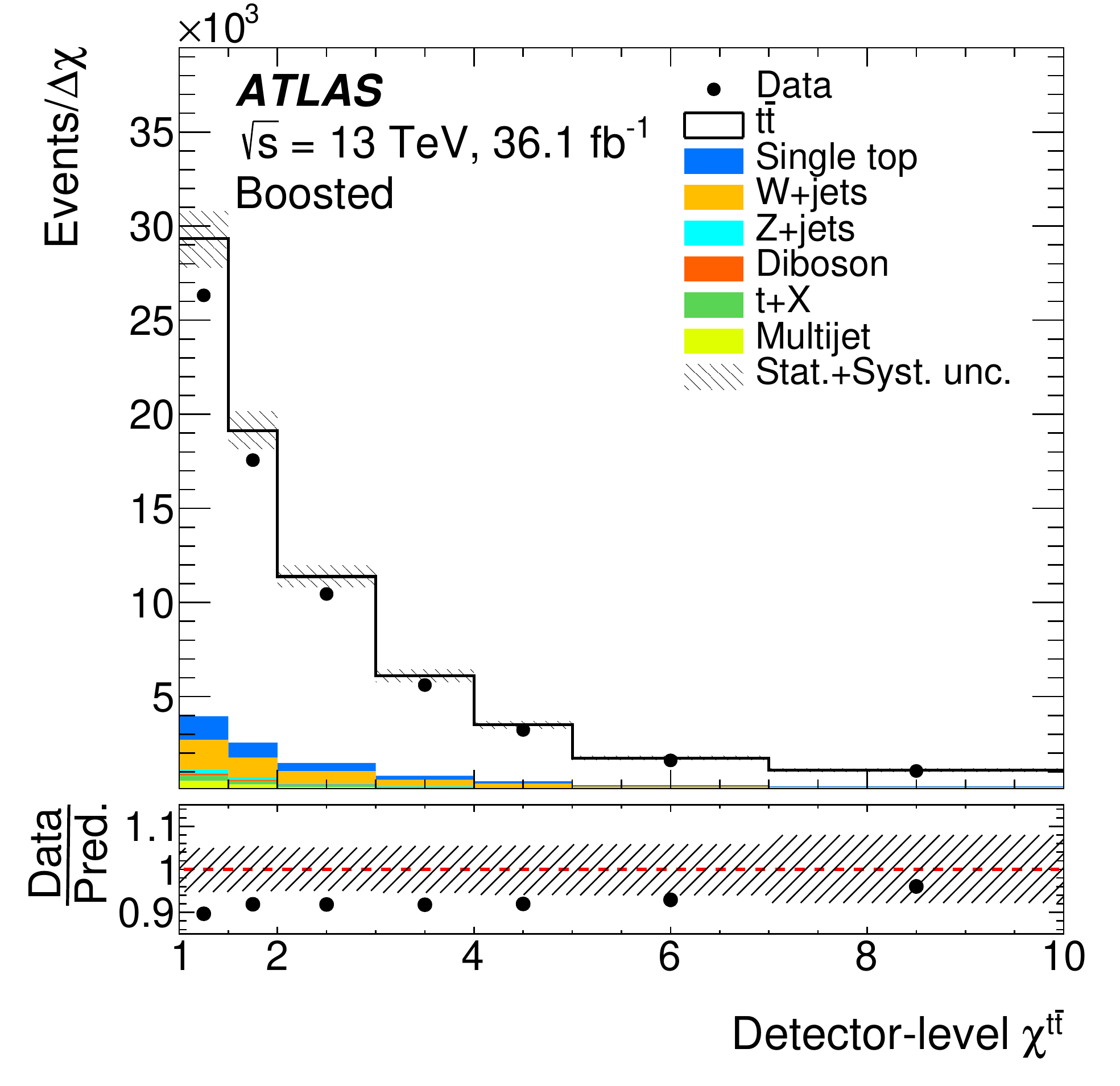}\label{fig:hadTop_boosted_rc_chi_tt}}
\subfigure[]{ \includegraphics[width=0.45\textwidth]{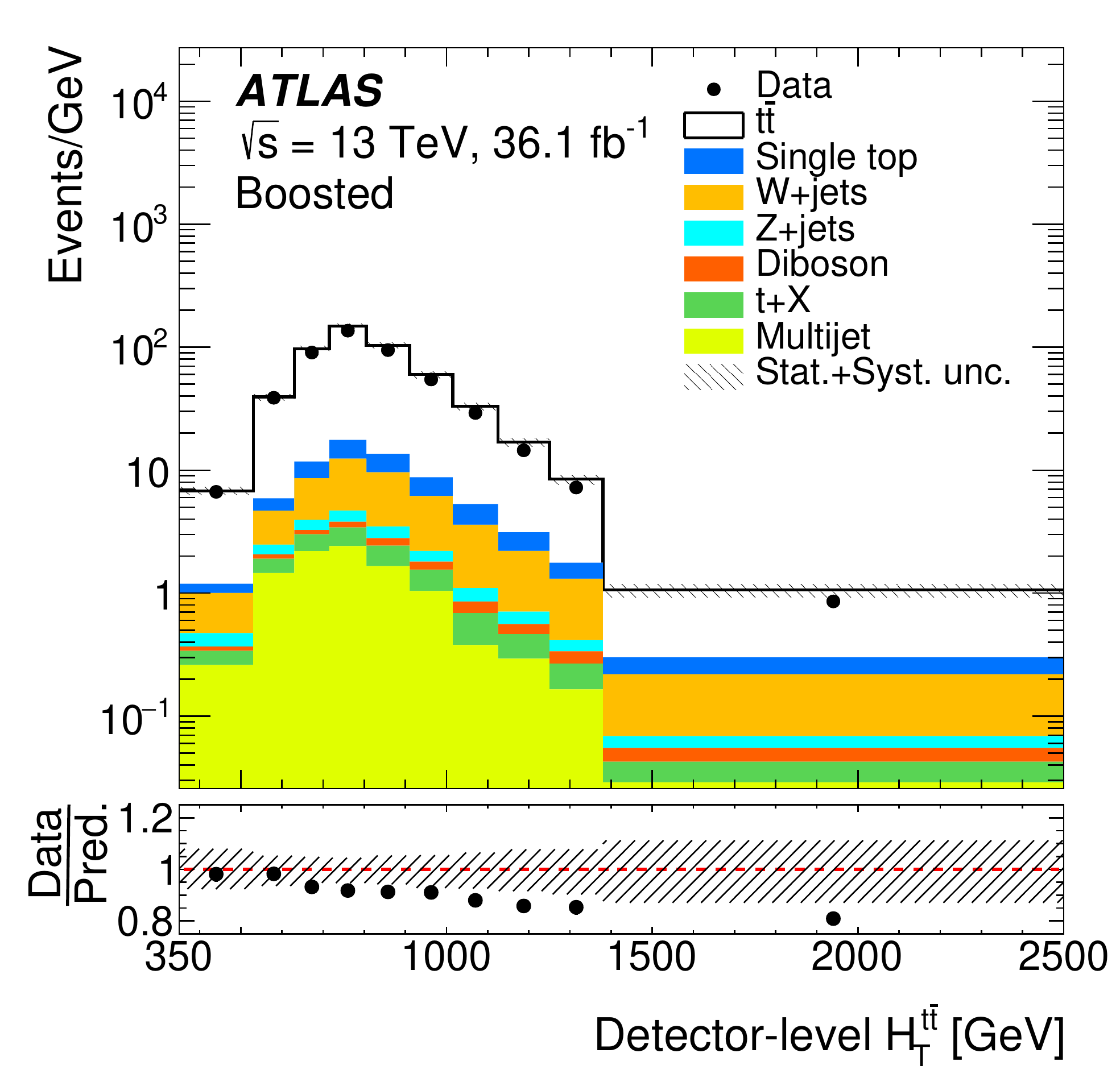}\label{fig:ttbar_boosted_rc_HT}}
\caption{Distributions of observables in the \ljets{} channel  in the boosted topology at detector-level: \protect\subref{fig:hadTop_boosted_rc_Pout}  absolute value of the out-of-plane momentum $|\Pouttlep|$, \protect\subref{fig:hadTop_boosted_rc_chi_tt} production angle \chitt{} and  \protect\subref{fig:ttbar_boosted_rc_HT} scalar sum of the transverse momenta of the hadronic and leptonic top quarks \Htt. Data distributions are compared with predictions, using \Powheg+\PythiaEight as the \ttbar{} signal model. The hatched area represents the combined statistical and systematic uncertainties (described in Section~\ref{sec:uncertainties}) in the total prediction, excluding systematic uncertainties related to the modelling of the \ttbar{} events. Underflow and overflow events, if any, are included in the first and last bins. The lower panel shows the ratio of the data to the total prediction.
}
\label{fig:controls_boosted_detector:additional}
\end{figure*}

These observables were previously measured in the resolved topology by the
ATLAS Collaboration using 8~\TeV\ data~\cite{TOPQ-2015-06} and, using 13~\TeV\ data, as a function of the jet multiplicity~\cite{TOPQ-2017-01}. Figures~\ref{fig:controls_resolved_detector:additional:1} and~\ref{fig:controls_resolved_detector:additional:2} show the distributions of these additional variables at detector-level in the resolved topology, while the distributions of $|\Pouttlep|$, \chitt{} and \Htt{} in the boosted topology are shown in Figure~\ref{fig:controls_boosted_detector:additional}.
 
Finally, differential cross-sections have been measured at particle level as a function of the number of jets not employed in \ttbar{} reconstruction  in the resolved and boosted topology ($N^\mathrm{extrajets}$).  In addition, in the boosted topology, the cross-section as a function of the number of small-$R$ jets clustered inside a top candidate ($N^\mathrm{subjets}$) is measured.
 
In the resolved topology, as shown in Figures~\ref{fig:controls_resolved_detector:top}, \ref{fig:controls_resolved_detector:ttbar}, \ref{fig:controls_resolved_detector:additional:1} and~\ref{fig:controls_resolved_detector:additional:2}, good agreement between the prediction and the data is observed. Trends of deviations at the boundaries of the uncertainty bands are seen for high values of \mtt{} and \pttt{}. In the boosted topology, the predicted rate of events is overestimated at the level of 8.5\%, leading to a corresponding offset in most distributions, as shown in Figures~\ref{fig:controls_boosted_detector:top}, \ref{fig:controls_boosted_detector:ttbar} and~\ref{fig:controls_boosted_detector:additional}.

A trend is observed in the \Htt{} distribution, where the predictions tend to overestimate the data at high values. This is more pronounced in the boosted topology, where the agreement lies outside the error band towards high values of \Htt{}.
A summary of the observables measured in the particle and parton phase-spaces is given in Tables~\ref{tab:variables:resolved:particle}--\ref{tab:variables:resolved:parton} for the resolved topology and in Tables~\ref{tab:variables:boosted:particle}--\ref{tab:variables:boosted:parton} for the boosted topology.

\begin{table}[ht]
\centering
\small
\begin{tabular}{|l|l|}
\hline
1D observables & 2D combinations\\
\hline
\mtt{} & in bins of: $|\ytt|$ and $N^{\mathrm{extrajets}}$\\
\pttt{} & in bins of: \mtt{}, $|\ytt|$ and $N^{\mathrm{extrajets}}$\\
$|\ytt|$ & in bins of: $N^{\mathrm{extrajets}}$\\
\ptth{} & in bins of: \mtt{}, \pttt{}, $|\yth|$ and $N^{\mathrm{extrajets}}$\\
$|\yth|$ & in bins of: $N^{\mathrm{extrajets}}$\\
$\pt^{t, 1}$  &\\
$\pt^{t, 2}$  & \\
\chitt &  in bins of: $N^{\mathrm{extrajets}}$\\
$|\yboost|$  & \\
\deltaPhitt{} & in bins of: $N^{\mathrm{extrajets}}$\\
\Htt{} & in bins of: $N^{\mathrm{extrajets}}$\\
$|\Poutthad|$    & in bins of: $N^{\mathrm{extrajets}}$\\
$N^{\mathrm{extrajets}}$ &\\
\hline
\end{tabular}
\caption{ The single- and double-differential spectra, measured in the resolved topology at particle level.}
\label{tab:variables:resolved:particle}
\end{table}

\begin{table}[ht]
\centering
\small
\begin{tabular}{|l|l|}
\hline
1D observables & 2D combinations\\
\hline
\mtt{} & in bins of: $|\ytt|$ \\
\pttt{} & in bins of: \mtt{} and $|\ytt|$\\
$|\ytt|$ & \\
\ptth{} & in bins of: \mtt{}, \pttt{} and $|\yth|$ \\
$|\yth|$ & \\
\chitt &  \\
$|\yboost|$  & \\
\Htt{} & \\
\hline
\end{tabular}
\caption{ The single-differential and double-differential spectra, measured in the resolved topology at parton level.}
\label{tab:variables:resolved:parton}
\end{table}
 
\begin{table}[ht]
\centering
\small
\begin{tabular}{|l|l|}
\hline
1D observables & 2D combinations\\
\hline
\mtt{} & in bins of: $\Htt{}$, $|\ytt|$, \pttt{} and $N^{\mathrm{extrajets}}$\\
\pttt{} & in bins of: $N^{\mathrm{extrajets}}$\\
$|\ytt|$ & \\
\ptth{} & in bins of: \mtt{}, \pttt{}, $|\yth|$, $|\ytt|$ and $N^{\mathrm{extrajets}}$\\
$|\yth|$ &\\
$\pt^{t, 1}$  &\\
$\pt^{t, 2}$  & \\
\chitt & \\
\Htt{} & \\
$|p_\mathrm{out}^{t,\mathrm{lep}}|$    &\\
$N^{\mathrm{extrajets}}$ &\\
$N^{\mathrm{subjets}}$ &\\
\hline
\end{tabular}
\caption{ The single- and double-differential spectra, measured in the boosted topology at particle level.}
\label{tab:variables:boosted:particle}
\end{table}

\begin{table}[ht]
\centering
\small
\begin{tabular}{|l|l|}
\hline
1D observables & 2D combinations\\
\hline
\mtt{} & in bins of: $\ptth{}$ \\
\ptth{}  & \\
\hline
\end{tabular}
\caption{ The single-differential and double-differential spectra, measured in the boosted topology at parton level.}
\label{tab:variables:boosted:parton}
\end{table}
\FloatBarrier
 
\section{Cross-section extraction}
\label{sec:xsec_calculation}
The underlying differential cross-section distributions are obtained from the detector-level events using an
unfolding technique that corrects for detector effects. The iterative Bayesian method~\cite{unfold:bayes} as implemented in RooUnfold~\cite{Adye:2011gm} is used.
 
Once the detector-level distributions are unfolded, the single- and double-differential cross-sections are extracted using the following equations:

\begin{equation*}
\frac{\textrm{d}\sigma}{\textrm{d}X_i} \equiv \frac{1}{\mathcal{L} \cdot \Delta X_i} \cdot  N^{\mathrm{unf}}_{i}
\end{equation*}
 
\begin{equation*}
\frac{\textrm{d}^{2}\sigma}{\textrm{d}X_i \textrm{d}Y_j} \equiv \frac{1}{\mathcal{L} \cdot \Delta X_i \Delta Y_j} \cdot  N^{\mathrm{unf}}_{ij}
\end{equation*}

where the index $i\;(j)$ iterates over bins of $X\;(Y)$ at generator level, $\Delta X_i\;(\Delta Y_j)$ is the bin width, $\mathcal{L}$ is the integrated luminosity and $N^{\mathrm{unf}}$ represents the unfolded distribution, obtained as described in the following sections. Overflow and underflow events are never considered when evaluating $N^{\mathrm{unf}}$, with the exception of the distributions as a function of jet multiplicities.

The unfolding procedure described in the following is applied to both the single- and double-differential distributions, the only difference being the creation of concatenated distributions in the double-differential case. In particular, $N^{\mathrm{unf}}$ is derived by introducing a new vector of size $m =\sum_{i=1}^{n_X}n_{Y,i}$, where $n_{X}$ is the number of bins of the variable $X$ and $n_{Y,i}$ is the number of bins of the variable $Y$ in the $i$-th bin of the variable $X$. The vector is constructed by concatenating all the bins of the original two-dimensional distribution.

The total cross-section is obtained by integrating the unfolded differential cross-section over the kinematic bins, and its value is used to compute the normalised differential cross-section $1/\sigma \cdot\textrm{d}\sigma / \textrm{d}X^i$.

\subsection{Particle level in the fiducial phase-space}
\label{sec:fiducial:space}
The unfolding procedure aimed to evaluate the particle-level distributions starts from the detector-level event distribution ($N_{\mathrm{detector}})$, from which the expected number of background events ($N_{\mathrm{bkg}}$)
is subtracted. Next, the bin-wise acceptance correction $f_{\mathrm{acc}}$, defined as
\begin{equation*}
f_{\mathrm{acc}}=\frac{N_{\mathrm{particle}\,\wedge\mathrm{detector}}}{N_{\mathrm{detector}}}\mathrm{,}\label{eq:f_acc:particle}
\end{equation*}
with $N_{\mathrm{particle}\,\wedge\mathrm{detector}}$ being the number of detector-level events that satisfy the particle-level selection, corrects for events that are generated outside the fiducial phase-space but satisfy the detector-level selection.
 
In the resolved topology, to separate resolution and combinatorial effects, distributions evaluated using a MC simulation are corrected to the level where detector- and particle-level objects forming the pseudo-top quarks are angularly well matched. The matching is performed using geometrical criteria based on the distance \DeltaR. Each particle-level $e$ ($\mu$) is required to be matched
to the detector-level $e$ ($\mu$) within $\DeltaR = 0.02$. Particle-level jets forming the particle-level hadronic top are required to be matched to the jets from the detector-level hadronic top within $\DeltaR  = 0.4$. The same procedure is applied to the particle- and detector-level \bjet{} from the leptonically decaying top quark. If a detector-level jet is not matched to a particle-level jet, it is assumed to be either from pile-up or from matching inefficiency and is ignored. If two jets are reconstructed
with a $\DeltaR  < 0.4$ from a single particle-level jet, the detector-level jet with smaller \DeltaR{} is matched to the particle-level jet and the other detector-level jet is unmatched. The matching correction $f_{\mathrm{match}}$, which accounts for the corresponding efficiency, is defined as:
\begin{equation*}
f_{\mathrm{match}}=\frac{N_{\mathrm{particle}\,\wedge\mathrm{detector}\,\wedge\mathrm{match}}}{N_{\mathrm{particle}\,\wedge\mathrm{detector}}}\mathrm{,}\label{eq:match:particle}
\end{equation*}
where $N_{\mathrm{particle}\,\wedge\mathrm{detector}\,\wedge\mathrm{match}}$ is the number of detector-level events that satisfy the particle-level selection and satisfy the matching requirement.
 
\begin{figure*}[t]
\centering
\subfigure[]{ \includegraphics[width=0.45\textwidth]{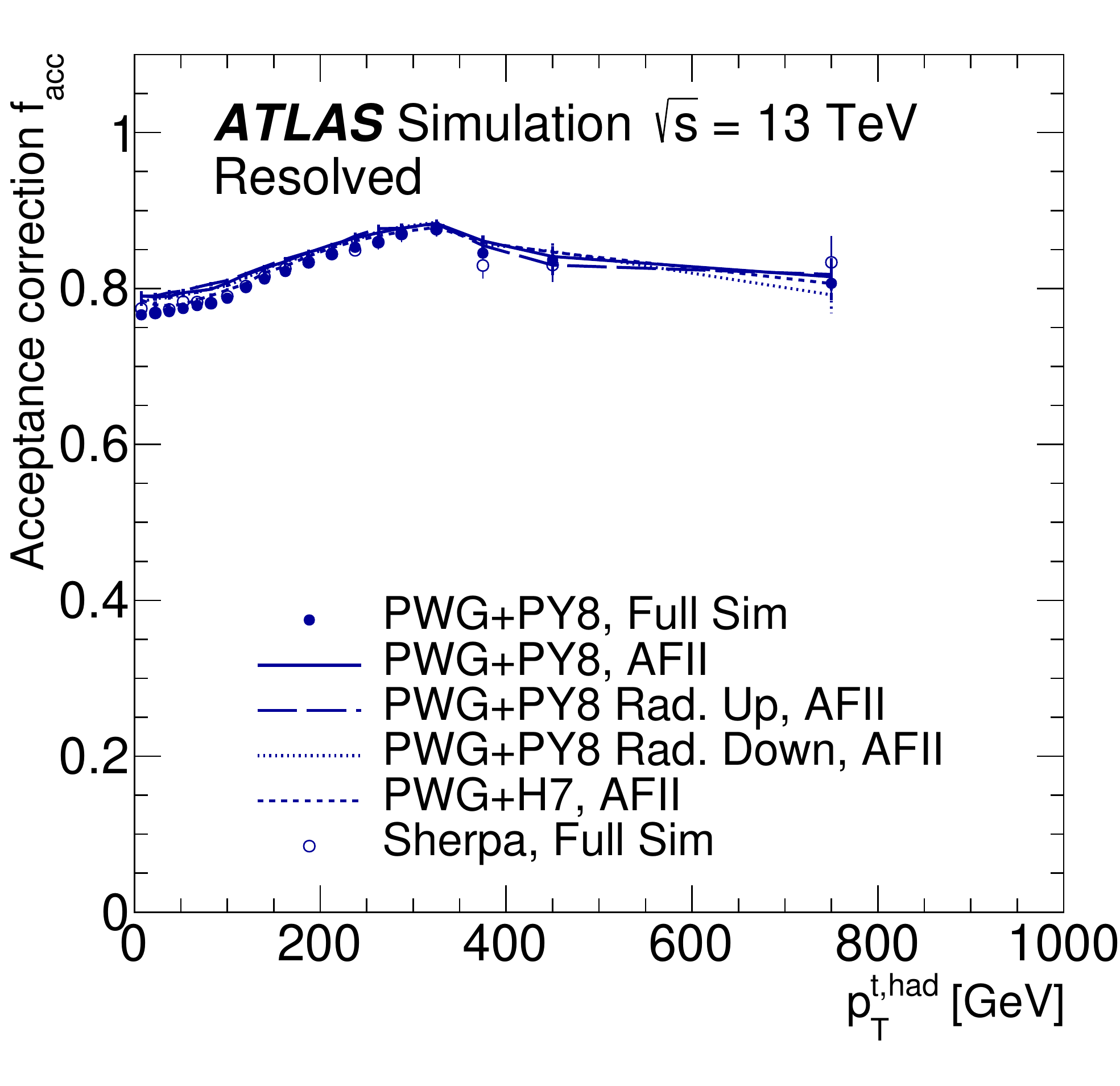}\label{fig:acc_top_had_pt_ljet}}
\subfigure[]{ \includegraphics[width=0.45\textwidth]{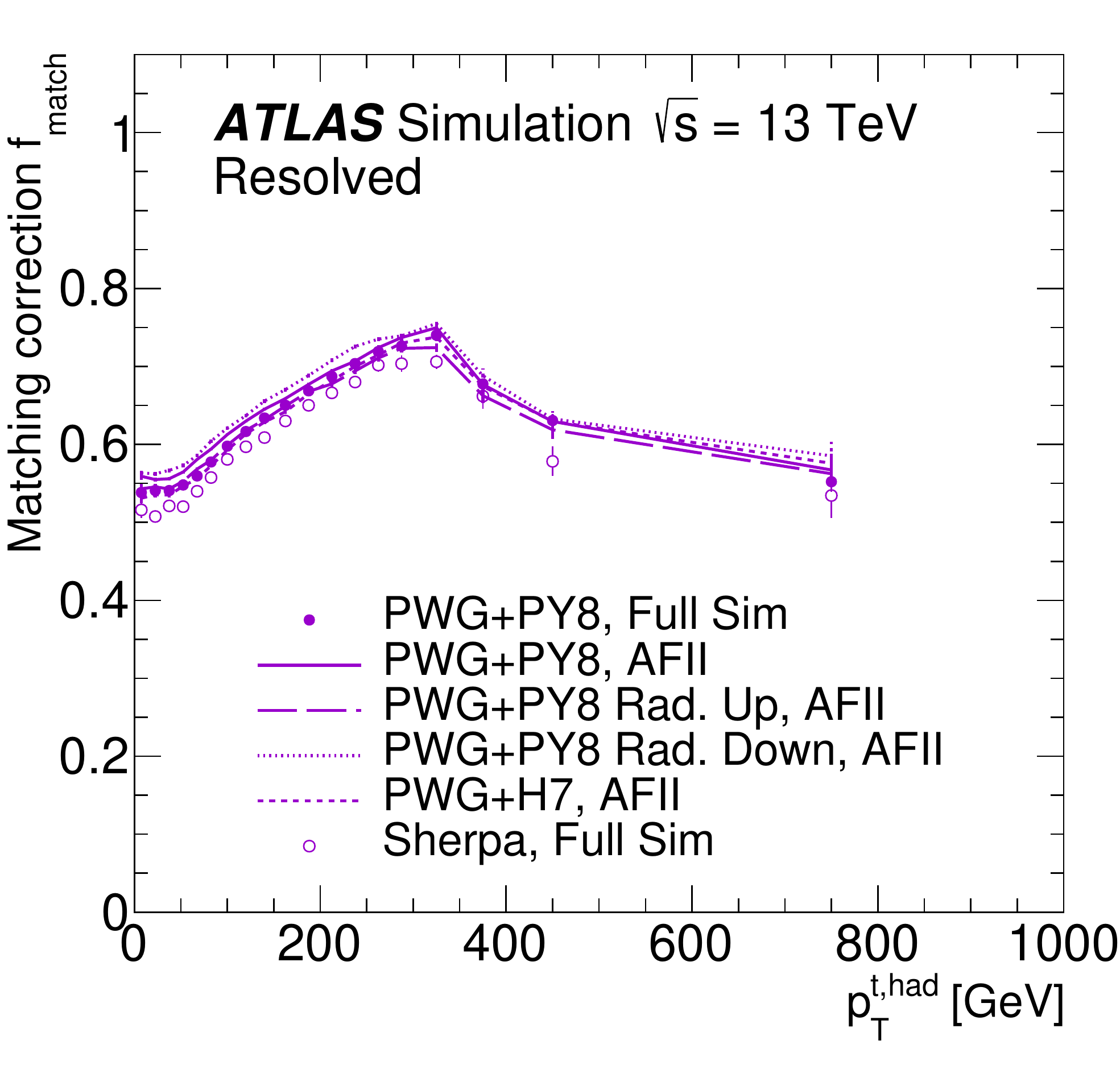}\label{fig:match_top_had_pt_ljet}}
\subfigure[]{ \includegraphics[width=0.45\textwidth]{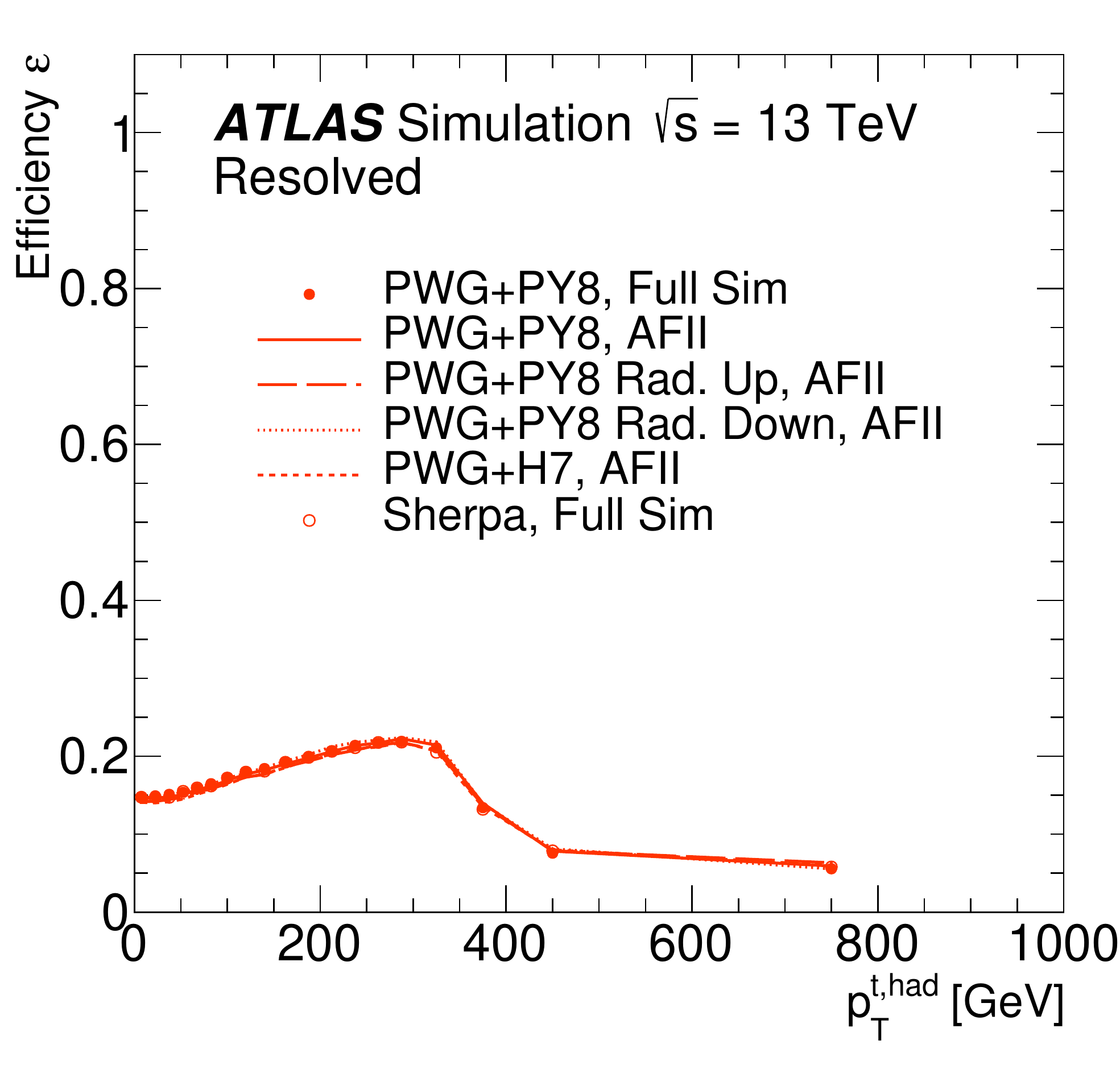}\label{fig:eff_top_had_pt_ljet}}
\subfigure[]{ \includegraphics[width=0.45\textwidth]{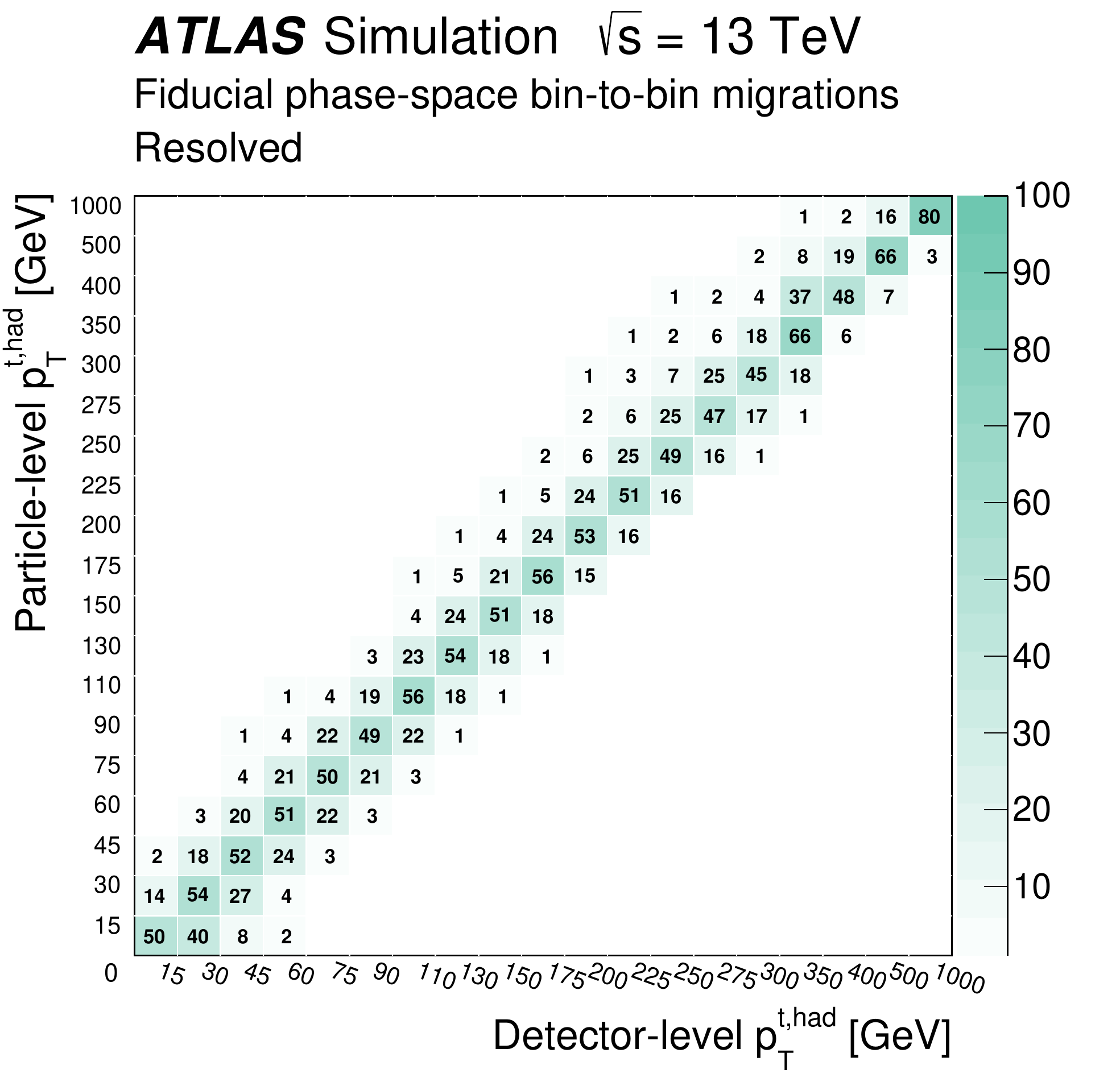}\label{fig:PseudoTop_Particle_top_had_pt_vs_PseudoTop_Reco_top_had_pt}}
\caption{  The \protect\subref{fig:acc_top_had_pt_ljet} acceptance $f_\textrm{acc}$, \protect\subref{fig:match_top_had_pt_ljet}  matching $f_\textrm{match}$ and  \protect\subref{fig:eff_top_had_pt_ljet} efficiency $\varepsilon$ corrections (evaluated with the Monte Carlo samples used to assess the signal modelling uncertainties, as described in Section~\ref{sec:uncertainties:signal}), and  \protect\subref{fig:PseudoTop_Particle_top_had_pt_vs_PseudoTop_Reco_top_had_pt} the  migration matrix (evaluated with the nominal \Powheg+\PythiaEight simulation sample)  for the hadronic top-quark transverse momentum in the resolved topology at particle level.
}
\label{fig:corrections_resolved_detector:particle}
\end{figure*}

The unfolding step uses a migration matrix ($\mathcal{M}$) derived from simulated \ttbar{} events that maps the binned
generated particle-level events to the binned matched detector-level events. The probability for particle-level events
to remain in the same bin is therefore represented by the diagonal elements, and the off-diagonal
elements describe the fraction of particle-level events that migrate into other bins. Therefore, the elements
of each row add up to unity as shown, for example, in Figure~\ref{fig:PseudoTop_Particle_top_had_pt_vs_PseudoTop_Reco_top_had_pt}. The binning is chosen such that the fraction of
events in the diagonal bins is always greater than 50\%. The unfolding is performed using four iterations
to balance the dependence on the prediction used to derive the corrections\footnote{At every iteration the result of the previous iteration is taken as prior. This allows information derived from the data to be introduced into the prior and hence reduce the dependence on the prediction.} and the statistical uncertainty. The effect of varying the number of iterations
by one is negligible. Finally, the efficiency correction $1/\varepsilon$ corrects for events that satisfy the particle-level selection but are not reconstructed at the detector level. The efficiency is defined as the ratio
\begin{equation*}
\varepsilon = \frac{N_{\mathrm{particle}\,\wedge\,\mathrm{detector}}}{N_{\mathrm{particle}}}\mathrm{,}\label{eq:eff:particle}
\end{equation*}
where $N_{\mathrm{particle}}$ is the total number of particle-level events. In the resolved topology, to account for the matching requirement, the numerator  is replaced with $N_{\mathrm{particle}\,\wedge\,\mathrm{detector}\,\wedge\,\mathrm{match}}$. The inclusion of the matching requirement, in conjunction with the requirement on 2 $b$-tagged jets, identified with 70\% efficiency, reflects in an overall efficiency below 25\% in the resolved topology. This is lower than in the boosted topology, where the efficiency ranges between 35\% and 50\% thanks to the request of only one $b$-tagged jet and the absence of the matching correction.
 
\begin{figure*}[t]
\centering
\subfigure[]{ \includegraphics[width=0.450\textwidth]{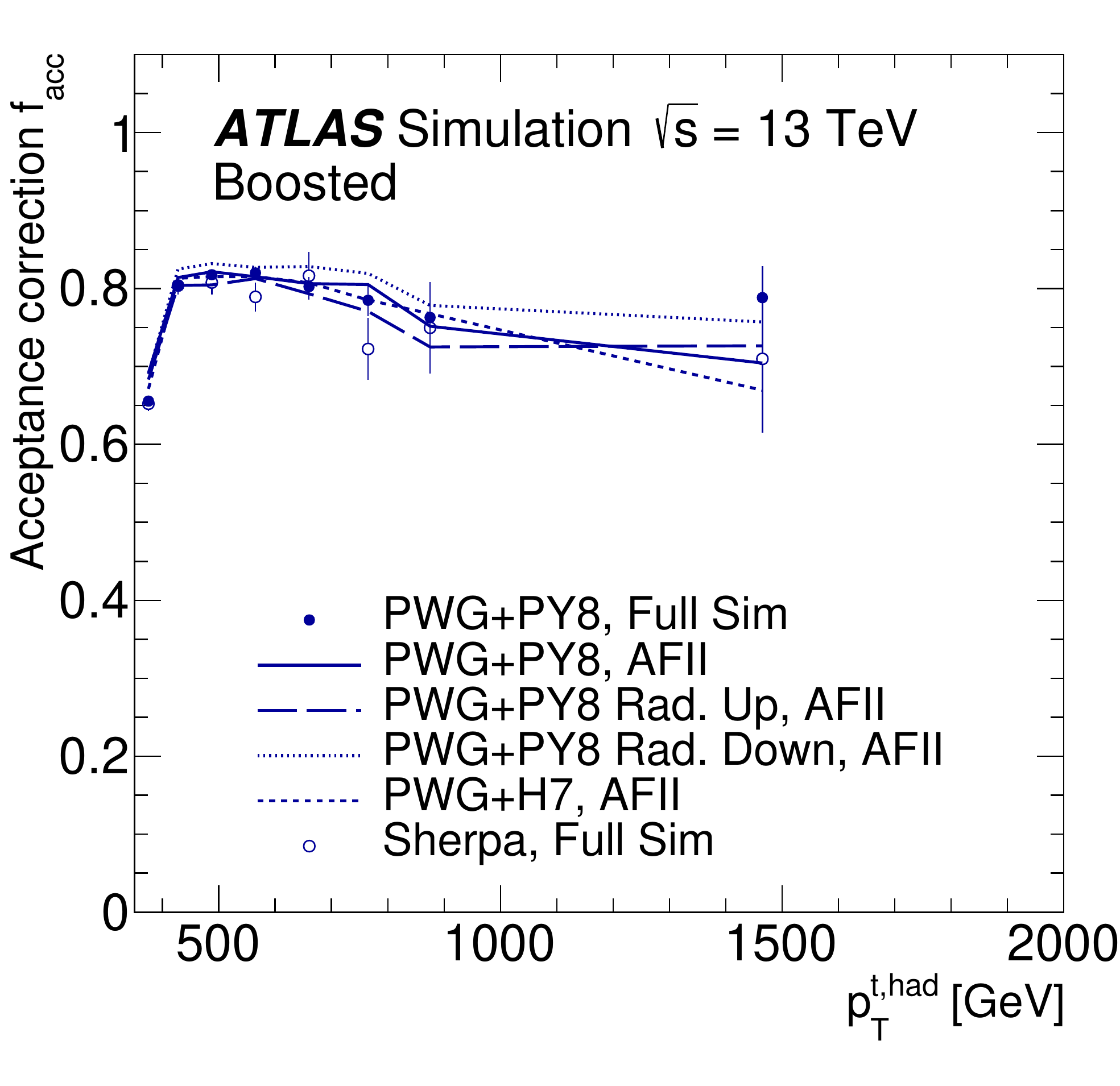}\label{fig:particle:boosted:acc_hadTop_pt_ljet}}
\subfigure[]{ \includegraphics[width=0.450\textwidth]{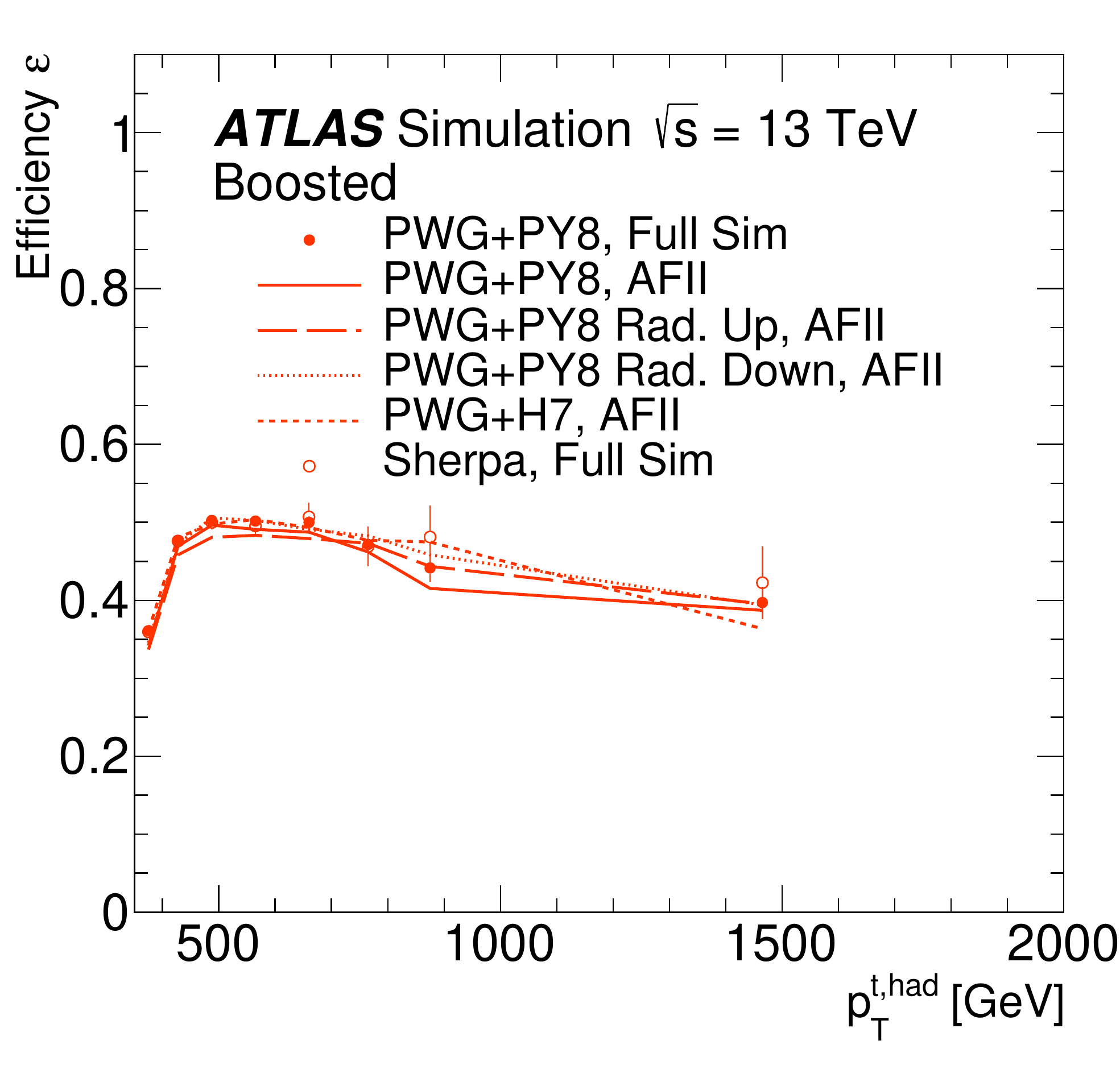}\label{fig:particle:boosted:eff_hadTop_pt_ljet}}
\subfigure[]{ \includegraphics[width=0.450\textwidth]{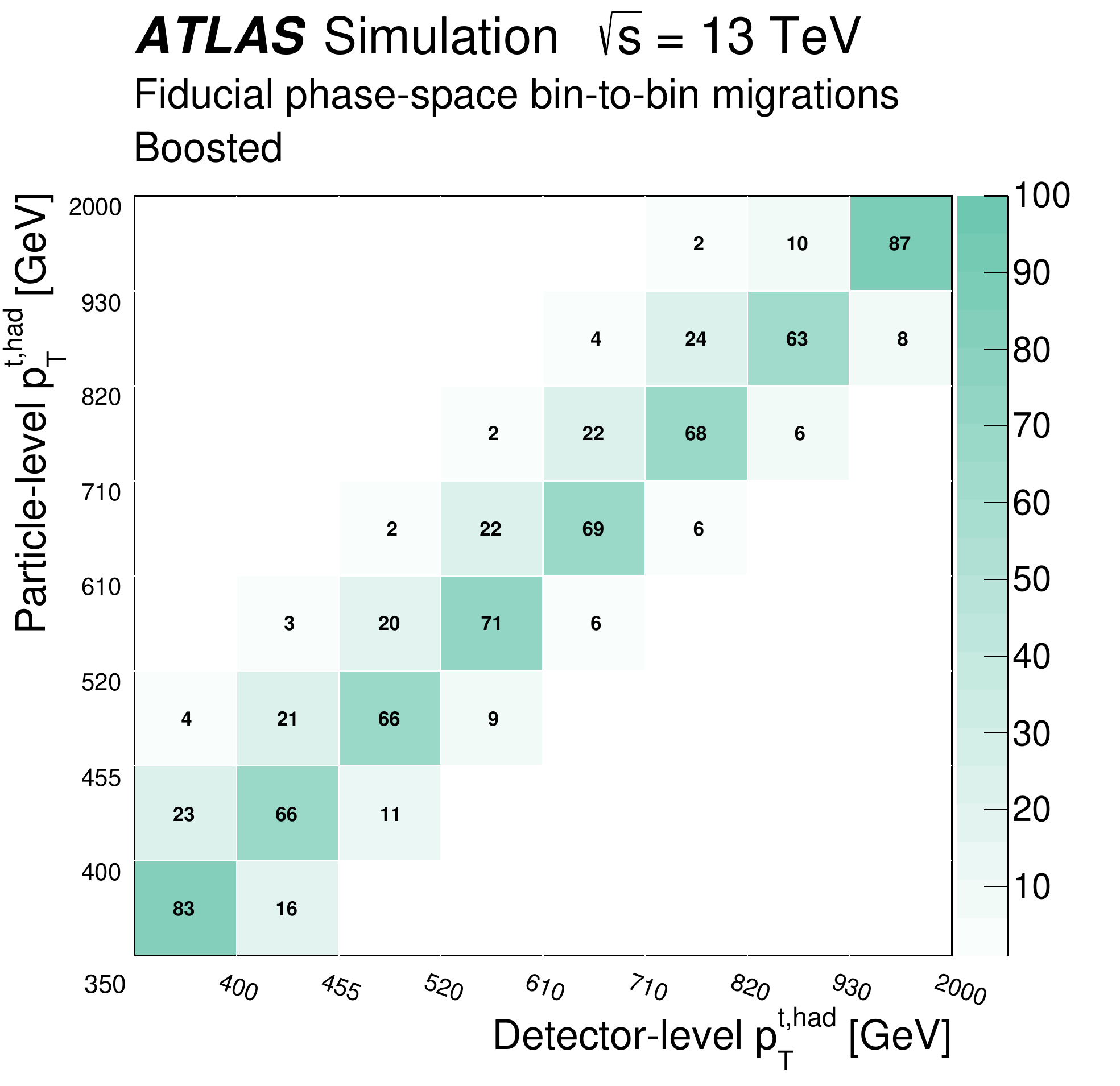}\label{fig:hadTop_boosted_rc_pt_vs_hadTop_boosted_rc_pt}}
\caption{  The \protect\subref{fig:particle:boosted:acc_hadTop_pt_ljet} acceptance $f_\textrm{acc}$ and  \protect\subref{fig:particle:boosted:eff_hadTop_pt_ljet} efficiency $\varepsilon$ corrections (evaluated with the Monte Carlo samples used to assess the signal modelling uncertainties, as described in Section~\ref{sec:uncertainties:signal}), and  \protect\subref{fig:hadTop_boosted_rc_pt_vs_hadTop_boosted_rc_pt}  the migration matrix (evaluated with the nominal \Powheg+\PythiaEight simulation sample)  for the hadronic top-quark transverse momentum in the boosted topology at particle level.
}
\label{fig:corrections_boosted_detector:particle}
\end{figure*}

All corrections ($f_{\mathrm{acc}}$, $f_{\mathrm{match}}$ and $\varepsilon$) and the migration matrices are evaluated  with simulated events for all the distributions to be measured. As an example, Figures~\ref{fig:corrections_resolved_detector:particle} and~\ref{fig:corrections_boosted_detector:particle} show the corrections and migration matrices for the case of the \pt of the hadronically decaying top quark, in the resolved and boosted topologies, respectively. This variable is particularly representative since the kinematics of the decay products of the top quark change substantially in the observed range. In the resolved topology, the decrease in the efficiency at high values is primarily due to the increasingly large fraction of non-isolated leptons and to the partially or totally overlapping jets in events with high top-quark \pt. An additional contribution is caused by the event veto removing   the events passing the boosted selection from the resolved topology, as described in Section~\ref{sec:selection}. This loss of efficiency is recovered by the measurement performed in the boosted topology.

The unfolded distribution for an observable $X$ at particle level is given by:
 
\begin{equation*}
N^{\mathrm{unf}}_i \equiv \frac{1}{\varepsilon^i} \cdot \sum_j \mathcal{M}_{ij}^{-1} \cdot f_\textrm{match}^j \cdot  f_\textrm{acc}^j \cdot \left(N_\textrm{detector}^j - N_\textrm{bkg}^j\right)\hbox{,}
\end{equation*}

where the index $j$ iterates over bins of $X$ at detector level, while the $i$ index labels bins at particle level. The Bayesian unfolding is symbolised by $\mathcal{M}_{ij}^{-1}$. No matching correction is applied in the boosted case ($f_\textrm{match}$ =1).

\FloatBarrier
 
\subsection{Parton level in the full phase-space}
\label{sec:full:space}
The measurements are extrapolated to the full phase-space of the parton-level \ttbar{} system using a procedure similar to
the one described in Section~\ref{sec:fiducial:space}. At detector level, the only difference is in the definition of the reconstructed objects for the measurement in the resolved topology, where the event reconstruction uses the kinematic fit method instead of the pseudo-top method.
 
To define \ljets{} final states at the parton level, the contribution of \ttbar{} pairs decaying dileptonically (in all combinations of electrons, muons and $\tau$-leptons) is removed by applying a bin-wise correction factor $f_{\mathrm{dilep}}$ (dilepton correction) defined as
\begin{equation*}
f_{\mathrm{dilep}}=\frac{N_{\textrm{detector}\,\wedge\,\ell+\textrm{jets}}}{N_\textrm{detector}}\mathrm{,}
\end{equation*}
 
which represents the fraction of the detector-level \ttbar{} single-lepton events ($N_{\textrm{detector}\,\wedge\,\ell+\textrm{jets}}$) in the total detector-level \ttb{} sample ($N_{\textrm{detector}}$), where the lepton can be either an electron, muon or $\tau$-lepton. The cross-section measurements correspond to the top quarks before decay (parton level) and after QCD radiation. Observables related to top quarks are extrapolated to the full phase-space starting from top quarks decaying hadronically at the detector level.
 
The acceptance correction $f_{\mathrm{acc}}$ corrects for detector-level events that are generated at parton level outside the  range of the given variable, and is defined by a formula similar to the particle-level acceptance described in Section~\ref{sec:fiducial:space}. The migration matrix ($\mathcal{M}$) is derived from simulated \ttbar{} events
decaying in the single-lepton channel and the efficiency correction $1/\varepsilon$ corrects for events that did not
satisfy the detector-level selection where
\begin{equation*}
\varepsilon=\frac{N_{\textrm{detector}\,\wedge\,\ell+\textrm{jets}}}{N_{\ell+\textrm{jets}}}\mathrm{,}
\end{equation*}
$N_{\textrm{detector}\,\wedge\,\ell+\textrm{jets}}$ is the number of parton-level events in the \ljets{} channel passing the detector-level selection and $N_{\ell+\textrm{jets}}$ is the total number of  events at parton level, as defined in Section~\ref{sec:objects:parton}.

\begin{figure*}[t]
\centering
\subfigure[]{ \includegraphics[width=0.45\textwidth]{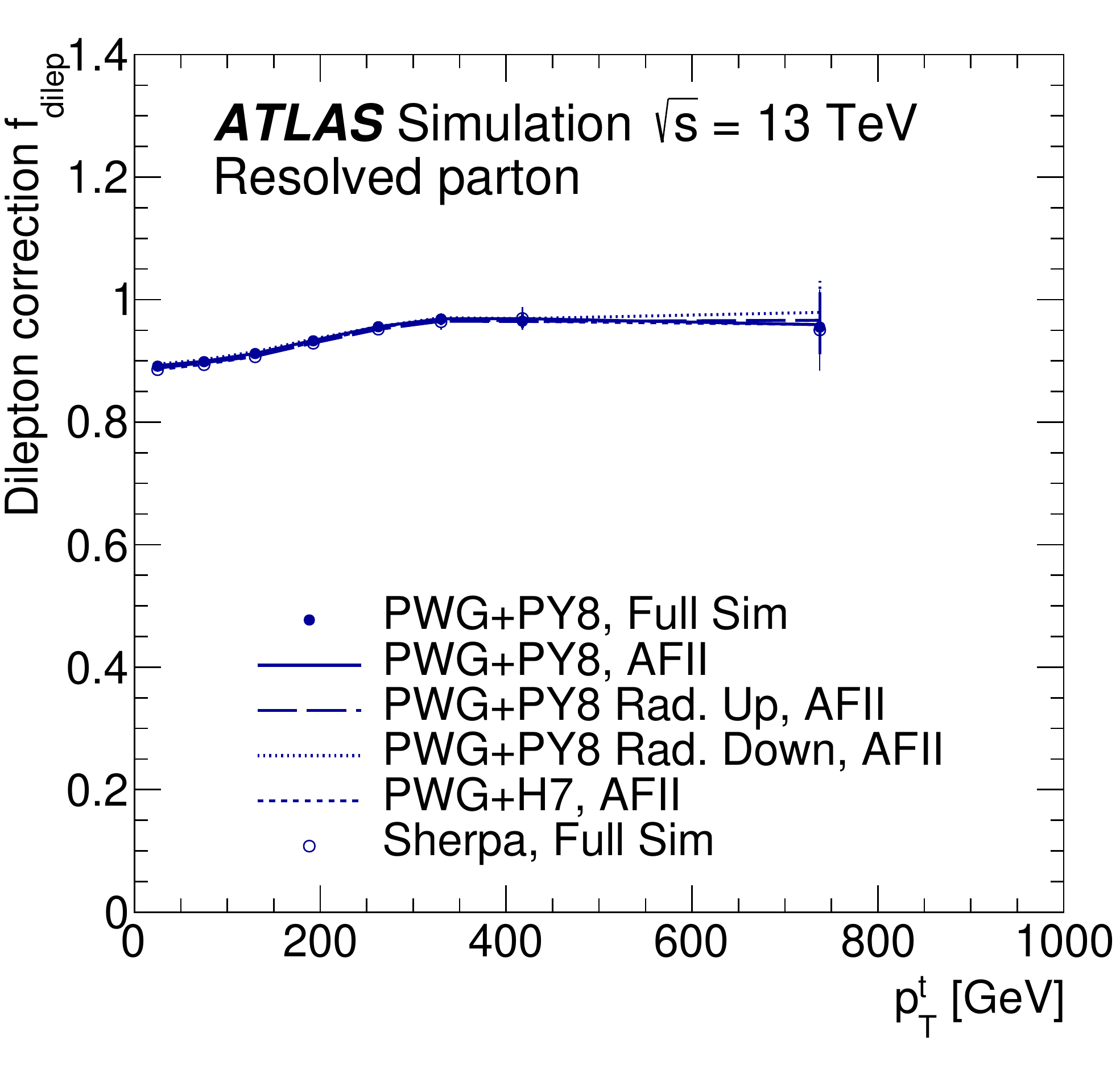}\label{fig:acc_topHad_pt_ljet_lhoodcut}}
\subfigure[]{ \includegraphics[width=0.45\textwidth]{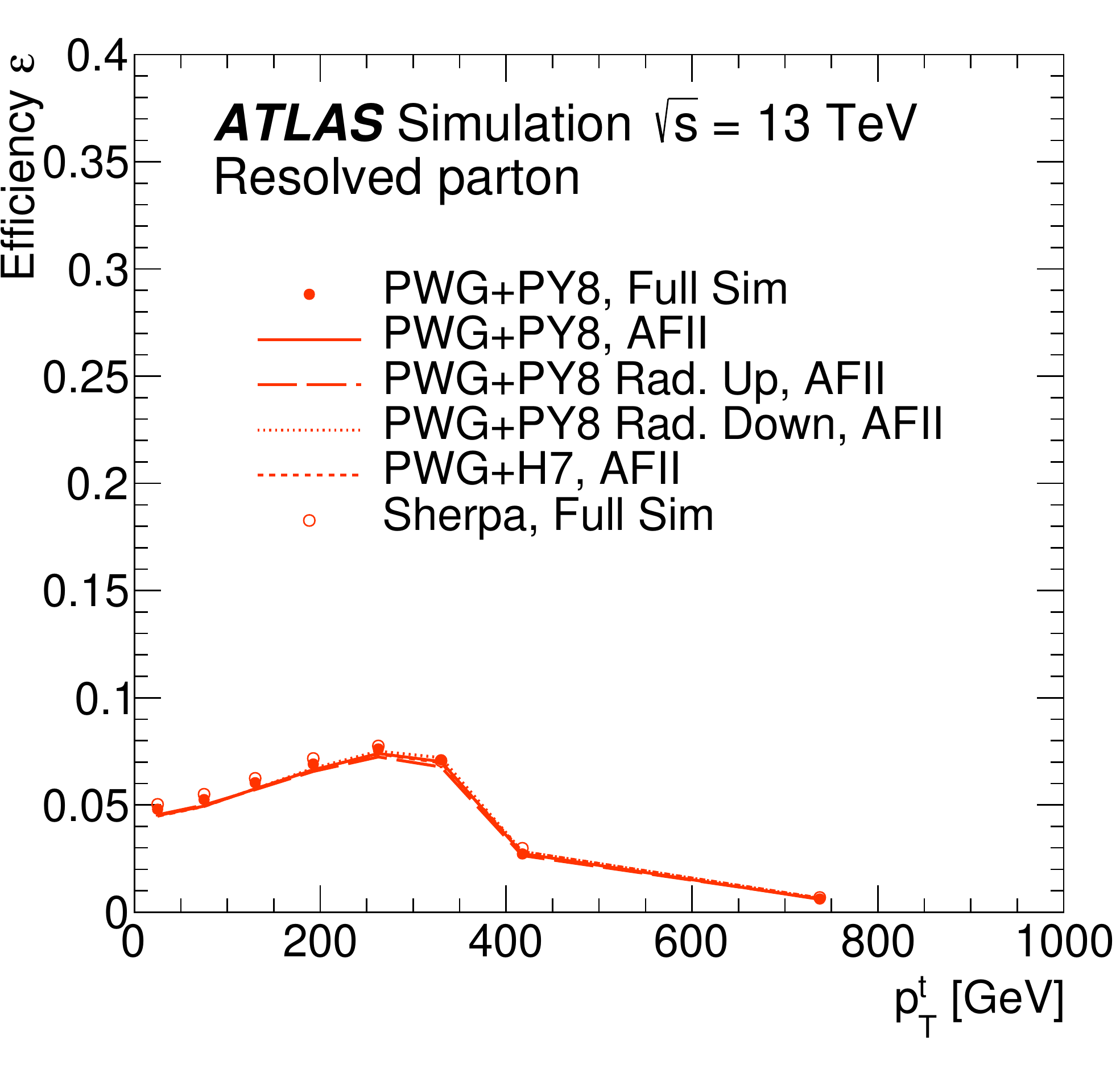}\label{fig:eff_topHad_pt_ljet_lhoodcut}}
\subfigure[]{ \includegraphics[width=0.45\textwidth]{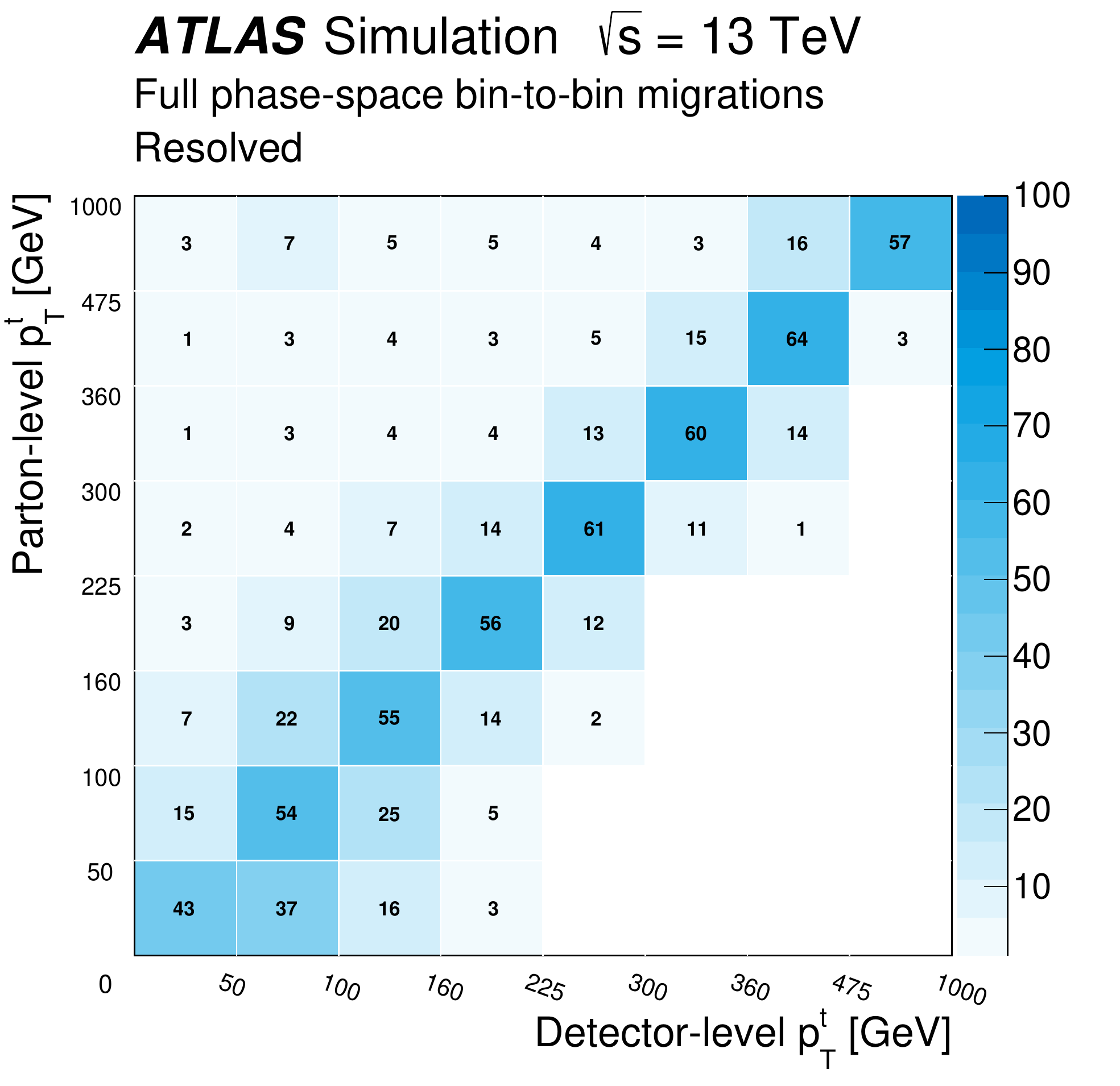}\label{fig:MC_thad_afterFSR_pt_vs_klfitter_bestPerm_topHad_pt}}
\caption{  The \protect\subref{fig:acc_topHad_pt_ljet_lhoodcut} dilepton $f_{\mathrm{dilep}}$ and \protect\subref{fig:eff_topHad_pt_ljet_lhoodcut} efficiency $\varepsilon$ corrections (evaluated with the Monte Carlo samples used to assess the signal modelling uncertainties, as described in Section~\ref{sec:uncertainties:signal}), and   \protect\subref{fig:MC_thad_afterFSR_pt_vs_klfitter_bestPerm_topHad_pt} the migration matrix (evaluated with the nominal \Powheg+\PythiaEight simulation sample)  for the hadronic top-quark transverse momentum in the resolved topology at parton level, for events selected with the kinematic likelihood cut.
}
\label{fig:corrections_resolved_detector:parton}
\end{figure*}

\begin{figure*}[t]
\centering
\subfigure[]{ \includegraphics[width=0.450\textwidth]{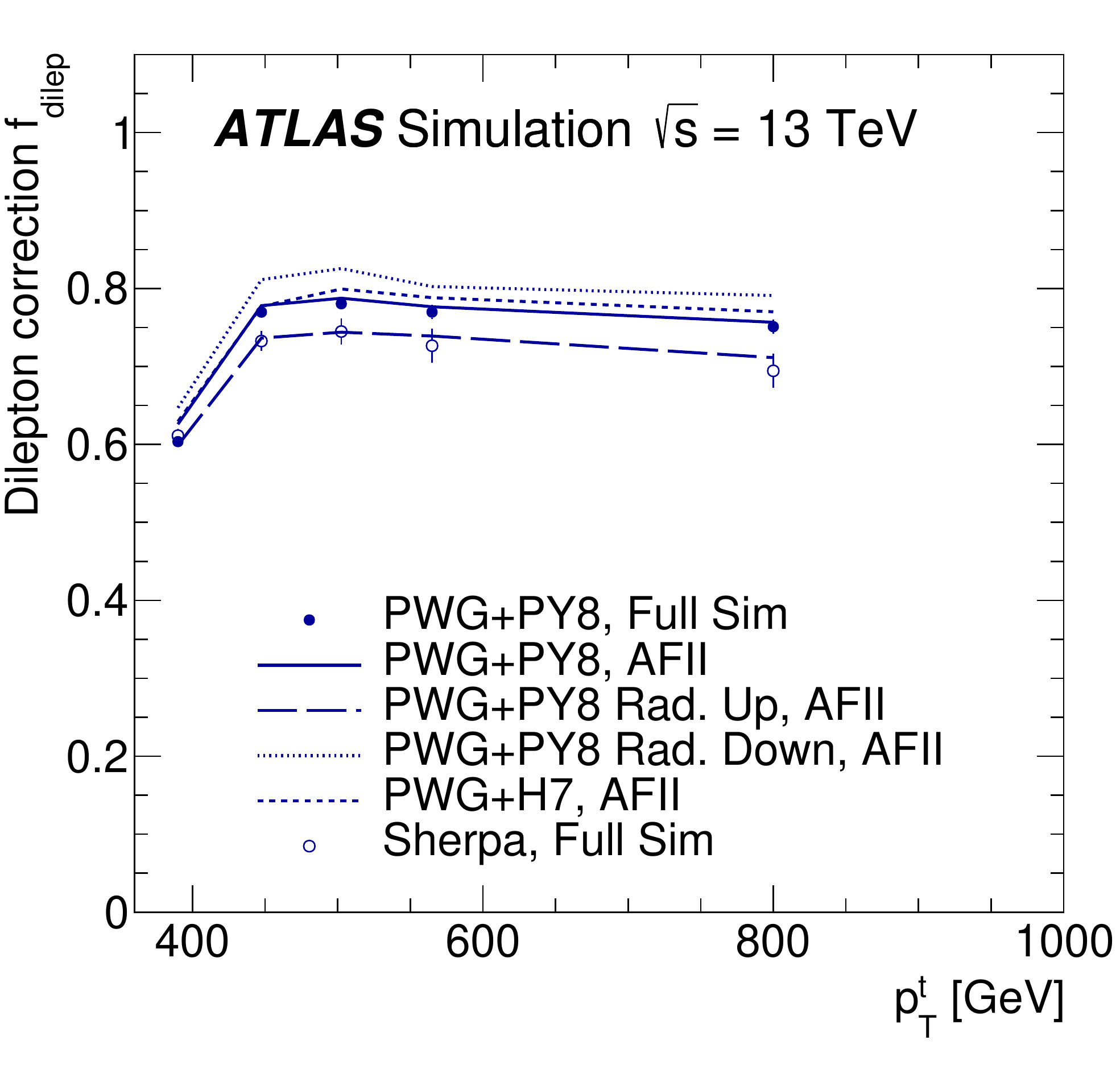}\label{fig:parton:boosted:acc_hadTop_pt_ljet}}
\subfigure[]{ \includegraphics[width=0.450\textwidth]{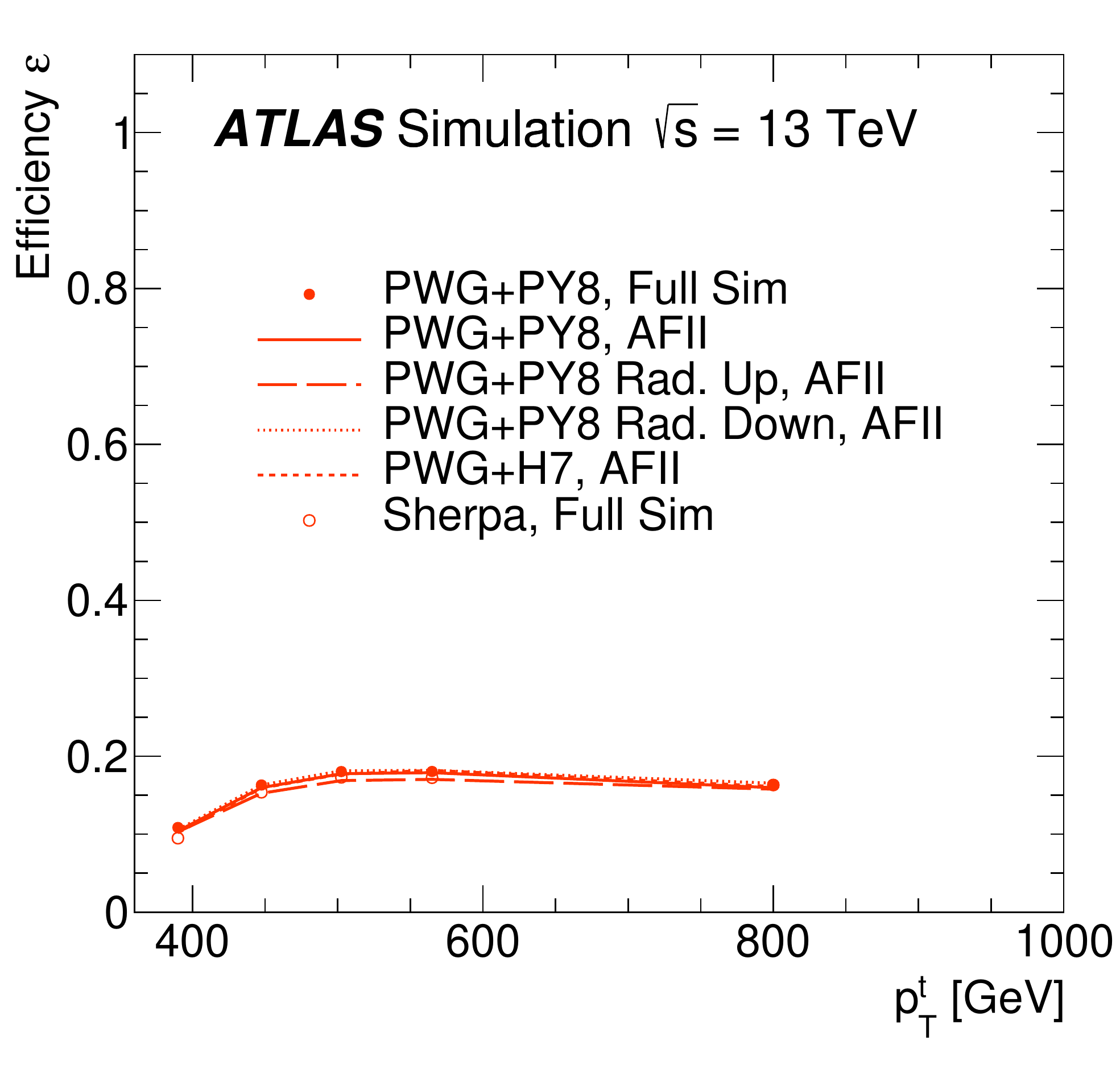}\label{fig:parton:boosted:eff_hadTop_pt_ljet}}
\subfigure[]{ \includegraphics[width=0.450\textwidth]{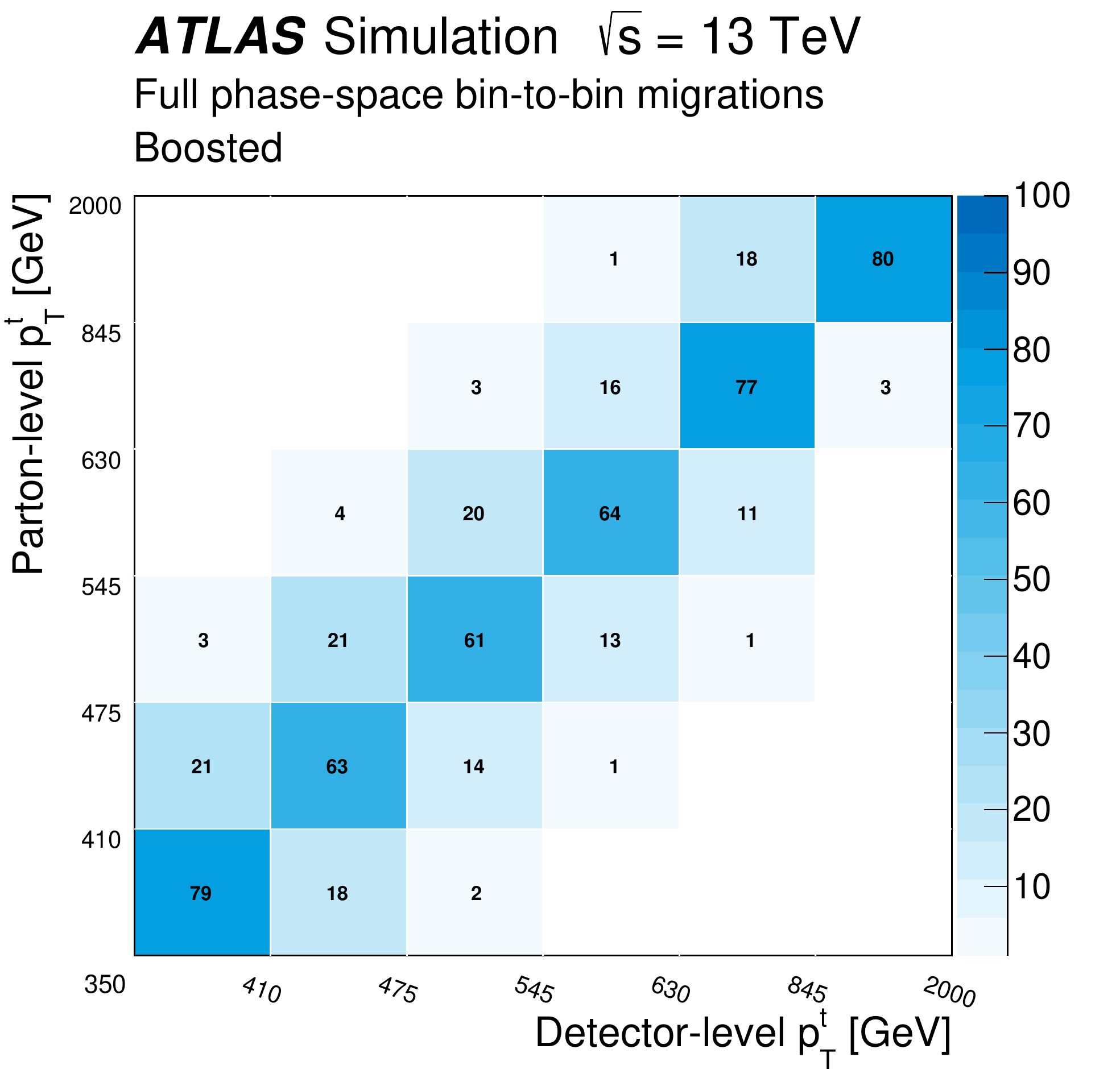}\label{fig:migration_MC_hadTop_boosted_rc_pt_vs_hadTop_boosted_rc_parton_pt_migra_parton_det_ljet}}
\caption{  The \protect\subref{fig:parton:boosted:acc_hadTop_pt_ljet} dilepton $f_{\mathrm{dilep}}$ and \protect\subref{fig:parton:boosted:eff_hadTop_pt_ljet} efficiency $\varepsilon$ corrections (evaluated with the Monte Carlo samples used to assess the signal modelling uncertainties, as described in Section~\ref{sec:uncertainties:signal}), and   \protect\subref{fig:migration_MC_hadTop_boosted_rc_pt_vs_hadTop_boosted_rc_parton_pt_migra_parton_det_ljet} the  migration matrix (evaluated with the nominal \Powheg+\PythiaEight simulation sample)  for the hadronic top-quark transverse momentum in the boosted topology at parton level.
}
\label{fig:corrections_boosted_detector:parton}
\end{figure*}

All corrections and the migration matrices for the parton-level measurement are evaluated with simulated events. As an example, Figures~\ref{fig:corrections_resolved_detector:parton} and~\ref{fig:corrections_boosted_detector:parton} show the corrections and migration matrices for the case of the \pt of the top quark, in the resolved and boosted topologies, respectively.
 
The unfolding procedure is summarised by the expression
\begin{equation*}
N^{\mathrm{unf}}_i \equiv \frac{1}{\mathcal{B}} \cdot \frac{1}{\varepsilon^i} \cdot \sum_j \mathcal{M}_{ij}^{-1} \cdot f_\textrm{dilep}^j \cdot  f_\textrm{acc}^j \cdot \left(N_\textrm{detector}^j - N_\textrm{bkg}^j\right)\hbox{,}
\end{equation*}
where the index $j$ iterates over bins of the observable at the detector level, while the $i$ index labels the bins at
the parton level,  $\mathcal{B} = 0.438$ is the \ljets{} branching ratio~\cite{PDG} and $\mathcal{M}_{ij}^{-1}$ represents the Bayesian unfolding.

\FloatBarrier
 
\subsection{Unfolding validation}
\label{sec:unfolding:validation}
The statistical stability of the unfolding procedure has been tested with closure tests. With these tests it is checked that the unfolding procedure is able to correctly recover a statistically independent sample generated with the same modelling used in the production of the unfolding corrections. These tests, performed on all the measured differential cross-sections, confirm that good statistical stability is achieved for all the spectra.
 
\begin{figure*}[t]
\centering
\subfigure[]{ \includegraphics[width=0.450\textwidth]{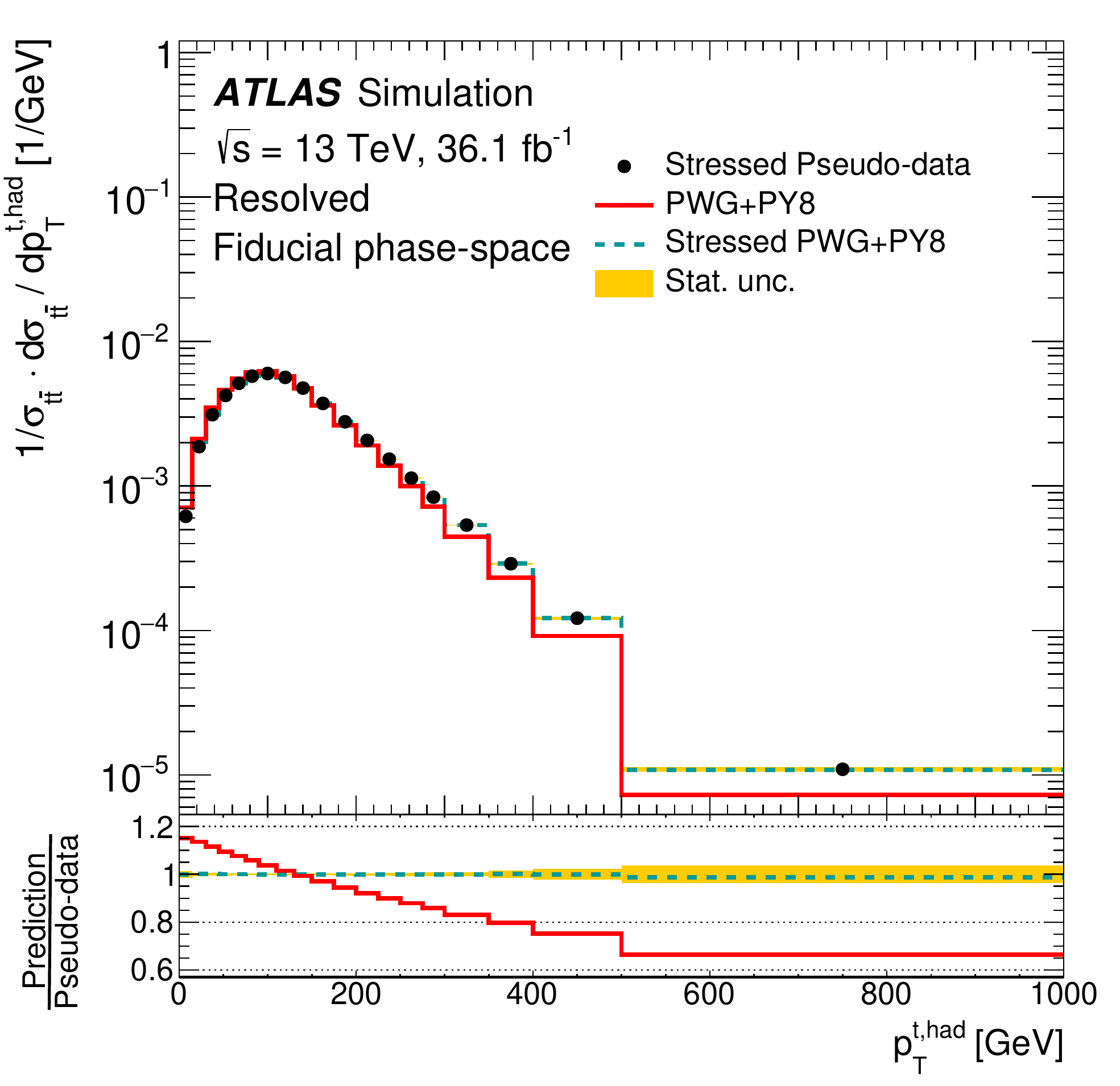}\label{fig:stress:resolved}}
\subfigure[]{ \includegraphics[width=0.450\textwidth]{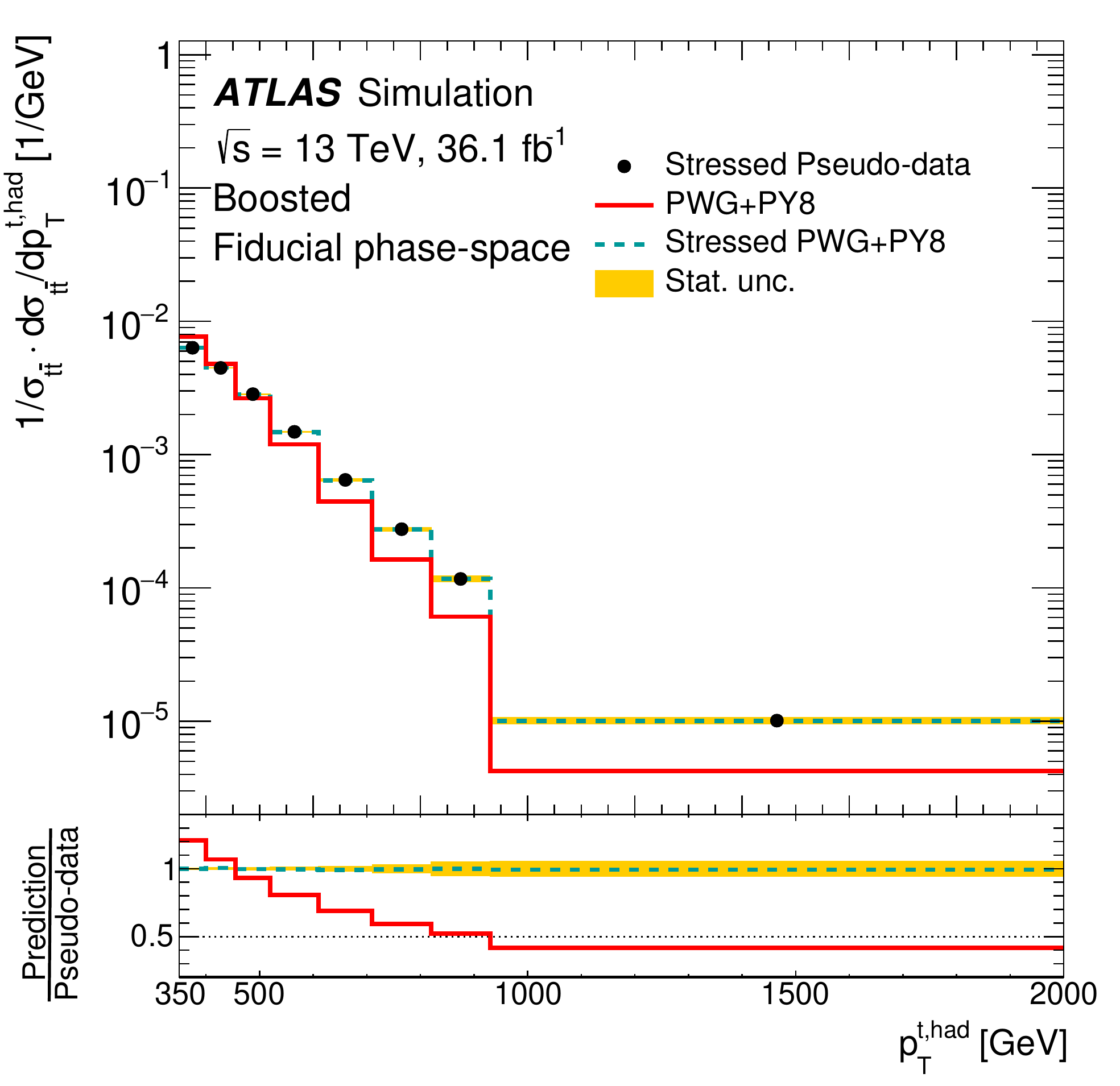}\label{fig:stress:boosted}}
\caption{ Stress tests for the particle-level normalised differential cross-sections  as a function of the \pt{} of the hadronically decaying top in \protect\subref{fig:stress:resolved}~the resolved and \protect\subref{fig:stress:boosted}~the boosted topologies. The pseudo-data are obtained by reweighting the detector-level distributions obtained with \Powheg+\PythiaEight{} generator using a linear function of the particle-level \ptth{} and unfolded using the nominal corrections. The pseudo-data are compared to the nominal prediction and the prediction obtained by reweighting the particle-level distribution. The bands represent the uncertainty due to the Monte Carlo statistics. Pseudo-data points are placed at the centre of each bin. The lower panel shows the ratios of the predictions to pseudo-data.
}
\label{fig:stress}
\end{figure*}

To ensure that the results are not biased by the MC generator used for the unfolding procedure, a study
is performed in which the particle-level and parton-level spectra in the \Powheg+\PythiaEight{} simulation are altered by changing the shape of the distributions using continuous functions of the particle-level and parton-level $\pt^{t}$ and of the actual data/MC ratio observed  at detector level.
These tests are performed on all the measured distributions using the final binning and employing the entire MC statistics available, and are referred to as stress tests.  An additional stress test is performed on the distributions depending on $\mtt{}$, where the spectra are modified  to simulate the presence of a new resonance. Examples of stress tests performed by changing the distribution of the \pt{} of the hadronic top employing a linear function of the particle-level \ptth{} are presented, for both the resolved and boosted topologies, in Figure~\ref{fig:stress}.  The studies confirm that these altered shapes are preserved within statistical uncertainties by the unfolding procedure based on
the nominal corrections.

\FloatBarrier

\section{Systematic uncertainties} \label{sec:uncertainties}
 
This section describes the estimation of systematic uncertainties related to object reconstruction and calibration, MC generator modelling and background estimation. As a result of the studies described in Section~\ref{sec:unfolding:validation} no systematic uncertainty has been associated to the unfolding procedure.
 
To evaluate the impact of each uncertainty after the unfolding, the reconstructed signal and background distributions in simulation are varied and unfolded using corrections from the nominal \Powheg+\PythiaEight{} signal sample. The unfolded distribution is compared with the corresponding particle- and parton-level spectrum and the relative difference is assigned as the uncertainty in the measured distribution.
All detector- and background-related systematic uncertainties  are evaluated using the same generator, while alternative generators and generator set-ups are employed to assess modelling systematic uncertainties. In these cases, the corrections, derived from the nominal generator, are used to unfold the detector-level spectra of the alternative generator and the comparison between the unfolded distribution and the alternative particle- or parton-level spectrum is used to assess the corresponding  uncertainty.

The covariance matrices of the statistical and systematic uncertainties are obtained for each observable by evaluating the covariance between the kinematic bins using pseudo-experiments, as explained in Section~\ref{sec:results}.

\subsection{Object reconstruction and calibration}\label{sec:uncertainties:detector}
The small-$R$ jet energy scale (JES) uncertainty is derived using a~combination of simulations,
test-beam data and \textit{in situ} measurements~\cite{PERF-2012-01, PERF-2016-04}. Additional contributions
from jet flavour composition, $\eta$-intercalibration, punch-through, single-particle response, calorimeter response to different jet flavours and pile-up are taken into account,  resulting in 29 independent subcomponents of systematic uncertainty, including the uncertainties in the jet energy resolution obtained with an \textit{in situ} measurement of the jet response in dijet events~\cite{ATL-PHYS-PUB-2015-015}. This uncertainty is found to be in the range of 5\%--10\%, depending on the variable, increasing to $20$\% in regions with high jet multiplicity.

The efficiency to tag jets containing $b$-hadrons is corrected in simulated events by applying $b$-tagging scale factors, extracted from a~$\ttbar$ dilepton sample, to account for the residual difference between data and simulation. Scale factors are also applied for jets originating from light quarks that are misidentified as $b$-jets.
The associated flavour-tagging systematic uncertainties, split into eigenvector components, are computed by varying the scale factors  within their uncertainties~\cite{PERF-2012-04,ATLAS-CONF-2018-001}.  The uncertainties due to the $b$-tagging efficiencies are constant for most of the measured distributions, amounting to $10\%$ and $2\%$ for the absolute differential cross-sections in the resolved and boosted topologies, respectively, and become negligible in most of the normalised differential cross-sections.
 
The lepton reconstruction efficiency in simulated events is corrected by scale factors
derived from measurements of these efficiencies in data using a control region enriched in~$Z \to \ee$ and $Z \to \mumu$
events.
The lepton trigger and reconstruction efficiency scale factors, energy scale and energy resolution are varied
within their uncertainties~\cite{PERF-2017-01,PERF-2017-03,PERF-2015-10} 
derived using the same sample.
 
The uncertainty associated with $\Etmiss$ is calculated by propagating the
energy scale and resolution systematic uncertainties to all jets and leptons in the
$\Etmiss$ calculation. Additional $\Etmiss$ uncertainties arising from energy deposits not associated with any reconstructed objects are also included~\cite{PERF-2016-07,ATLAS-CONF-2018-023}.
 
The systematic uncertainties due to the lepton and \Etmiss{} reconstruction are generally subdominant (around 2\%--3\%) in both the resolved and boosted topologies.

\subsection{Signal modelling}\label{sec:uncertainties:signal}
Uncertainties in the signal modelling
affect the kinematic properties of simulated $\ttbar$ events as well as detector- and particle-level efficiencies.
 
To assess the uncertainty related to the choice of MC generator for the $\ttbar$ signal process, events simulated with \SHERPAV{2.2.1} are unfolded using the migration matrix and correction factors derived from the nominal \Powheg+\PythiaEight{} sample. \SHERPAV{2.2.1} includes its own parton-shower and hadronisation model, which are consequently included in the variation and considered in the systematic uncertainty. This variation is indicated as `generator' uncertainty.  The symmetrised full difference between the unfolded distribution and the generated particle- and parton-level distribution of the \SHERPA sample is assigned as the relative uncertainty in the distributions. This uncertainty is found to be in the range of 5\%--10\%, depending on the variable, increasing to 20\% at very low \mtt{} at particle level, and at high \pt{} at parton level, in both the boosted and resolved topologies.
 
To assess the impact of different parton-shower and hadronisation models, unfolded results using events simulated with \Powheg+\PythiaEight{}  are compared with events simulated with \Powheg+\herwig{}7, 
with the same procedure as described above to evaluate the uncertainty related to the \ttbar generator. This variation is indicated as `hadronisation' uncertainty. The resulting systematic uncertainties, taken as the symmetrised full difference, are found to be  typically at the level of 2\%--5\% in the resolved and boosted topologies, increasing to 20\% at high top and \ttbar{} transverse momentum.
 
To evaluate the uncertainty related to the modelling of additional radiations (Rad.), two $\ttbar$ MC samples with modified \hdamp{}, scales and showering tune are used. The MC samples used for the evaluation of this uncertainty were generated using the \PowhegBox{} generator interfaced to the \Pythia{} shower model, where the parameters are varied as described in Section~\ref{sec:samples}.
This uncertainty is found to be in the range of 5\%--10\% for the absolute spectra in both the resolved and the boosted topology, increasing to 20\% at  high \pt{} at parton level.
 
The estimation of the uncertainty due to different parton-shower models and additional radiation modelling is performed using samples obtained with the `fast' simulation, introduced in Section~\ref{sec:samples}.
In most of the distributions the fast simulation gives the same result as the full simulation, and consequently the corrections obtained with the two samples are consistent as shown in Figure~\ref{fig:parton:boosted:acc_hadTop_pt_ljet}, comparing the two versions of \Powheg{}+\PythiaEight.
However, in some distributions a difference between fast and full simulation is observed, as shown in Figure~\ref{fig:acc_top_had_pt_ljet} in the low $\pt$ range. To completely disentangle this effect from the modelling uncertainties estimate, the AFII version of \Powheg{}+\PythiaEight is used to calculate the unfolding corrections when the alternative samples, used to evaluate the systematic uncertainty, are produced with the fast simulation.

The impact of the uncertainty related to the PDF is assessed using the nominal signal  sample generated with \PowhegBox{} interfaced to \PythiaEight.  Acceptance, matching, efficiency and dilepton corrections and migration matrices for the unfolding procedure are obtained by reweighting the  $\ttbar$ sample using the 30 eigenvectors of the PDF4LHC15 PDF set~\cite{PDF4LHC15}. Using these corrections, the detector-level \Powheg+\PythiaEight{} distribution, obtained with the central eigenvector of the PDF4LHC15 set, is unfolded and the relative deviation from the expected  particle- or parton-level spectrum obtained with the same PDF set is computed. The total uncertainty is then obtained by adding these relative differences in quadrature. This procedure, obtained applying the recommendation given in Ref~\cite{PDF4LHC15} to unfolded measurements, differs from the approach used for the other modelling uncertainties, where nominal corrections are used to unfold detector-level distributions obtained with alternative generators. In addition, a further source of uncertainty derived from the choice of the PDF set is considered. This is estimated in a~similar way to the other component but comparing the central distribution of PDF4LHC15 and NNPDF3.0NLO sets. The two components are added in quadrature.
The total PDF-induced uncertainty is found to be less than $1$\% in most of the bins of the measured cross-sections.
 
\subsection{Background modelling}\label{sec:uncertainties:background}
 
Systematic uncertainties affecting the backgrounds are evaluated by varying the background distribution, while keeping the signal unchanged, in the input to the unfolding procedure.
The shift between the resulting unfolded distribution and the nominal one is used to estimate the size of the uncertainty.
 
For the single-top-quark background, three kind of uncertainties are considered:
\begin{enumerate}
\item Total normalisation uncertainty: the cross-section of the single-top-quark process is varied within its uncertainty for the $t$-channel (5\%)~\cite{Kidonakis:2011wy}, $s$-channel (3.6\%)~\cite{Kidonakis:2010tc} and $tW$ production (5.3\%)~\cite{Kidonakis:2010ux}.
\item Additional radiation uncertainty: single-top-quark ($tW$ production and $t$-channel) MC samples with modified scales and showering tunes are used in a similar way to those for estimating the equivalent systematic uncertainty for the signal sample. The samples are described in Section~\ref{sec:samples}.
\item Diagram subtraction versus diagram removal (DR/DS) uncertainty: the uncertainty due to the overlap of $tW$ production of single top quarks and production of $\ttbar$ pairs is evaluated by comparing the single-top-quark samples obtained using the diagram removal and diagram subtraction schemes~\cite{Frixione:2008yi}, using the samples described in Section~\ref{sec:samples}.
\end{enumerate}
In the final measurement, the sum of these components, dominated by the DR/DS uncertainty, gives a small contribution in the low \pt{} region, while it reaches 9\% and 12\% in the high \pt{} region of the resolved and boosted topologies, respectively.

For the $W$+jets process, two different uncertainty components are constructed from two \alphas{} variations of $\pm 0.002$ around the
nominal value of 0.118 and from an envelope formed from 7-point scale variations of the renormalisation
and factorisation scales, following the prescriptions described in Ref.~\cite{ATL-PHYS-PUB-2017-006}. The uncertainty due to the PDF variations is found to be subdominant and consequently not included.
An additional uncertainty in the fraction of the heavy-flavour components is considered. This uncertainty is evaluated by applying a 50\% shift to the cross-section of the samples in which the $W$ boson is produced in association with at least one $b$-quark and also rescaling the other samples to keep the total $W$+jets cross-section constant.
This uncertainty is considered sufficient to cover a possible mismodelling of the heavy-flavour composition since no disagreements among predictions and data are observed.
The $W$+jets uncertainty on the final result ranges from 2\% to 4\% in the resolved topology, depending on the variable and phase-space, and between 2\% and 12\% in the boosted topology.

The uncertainty due to the background from non-prompt and fake leptons is evaluated by
changing the parameterisation of the real- and fake-lepton efficiencies used in the matrix method calculation.
In addition, an extra 50\% uncertainty is assigned to this background to account for the remaining mismodelling observed in various control regions.
The combination of all these components also affects the shape of this background and the overall impact of these systematic uncertainties on the measurement is at the 2\% level in both topologies, increasing to  almost 4\% in the low \pt{} region in the resolved topology.
 
In the case of the $Z$+jets processes, a  global uncertainty, binned in jet multiplicity and based on $\alphas$, PDF and scale variations calculated in Ref.~\cite{ATL-PHYS-PUB-2017-006}, is applied to the MC prediction of the $Z$+jets background components.

For diboson  backgrounds, a 40\% uncertainty is applied, including the uncertainty in the cross-section
and a contribution due to the presence of at least two additional jets. For the \ttV{} background, an overall uncertainty of 14\% is applied, covering the uncertainties related to the scale, \alphas{} and PDF for the \ttb{} + $W$ and $Z$ components.
 
The overall impact of these additional background uncertainties on the final result is less than $1$\%, and the largest contribution comes from the $Z$+jets background.
 
\subsection{Statistical uncertainty of the Monte Carlo samples}
To account for the finite number of simulated events, test distributions based on total predictions are varied in each bin according to their statistical uncertainty, excluding the data-driven fake-lepton background.
The effect on the measured differential cross-sections is at most 1\% in the resolved and boosted topologies,  peaking at 6\% in the highest top-quark $\pt$ bins in the boosted topology.
 
\subsection{Integrated luminosity}
The uncertainty in the combined 2015--2016 integrated luminosity is 2.1\%~\cite{ATLAS-CONF-2019-021}, obtained using the LUCID-2 detector~\cite{LUCID2} for the primary luminosity measurements.
This uncertainty is not dominant for the absolute differential cross-section results and it mostly cancels out for the normalised differential cross-section results.
 
\subsection{Systematic uncertainties summary}
 
\begin{figure*}[t]
\centering
\subfigure[]{  \includegraphics[width=0.45\textwidth]{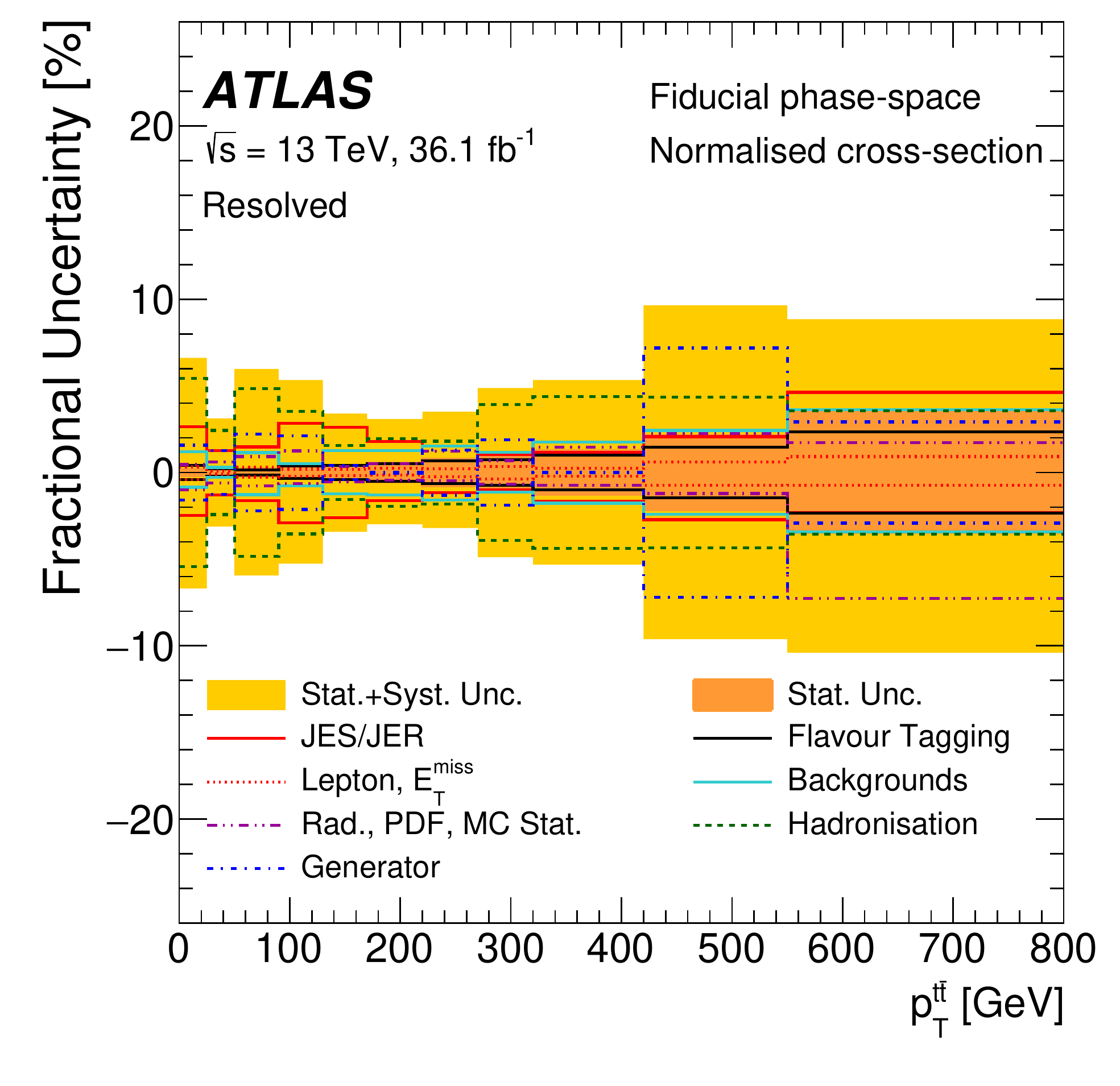}\label{fig:unc_particle:resolved:top_had_pt:rel}}
\subfigure[]{  \includegraphics[width=0.45\textwidth]{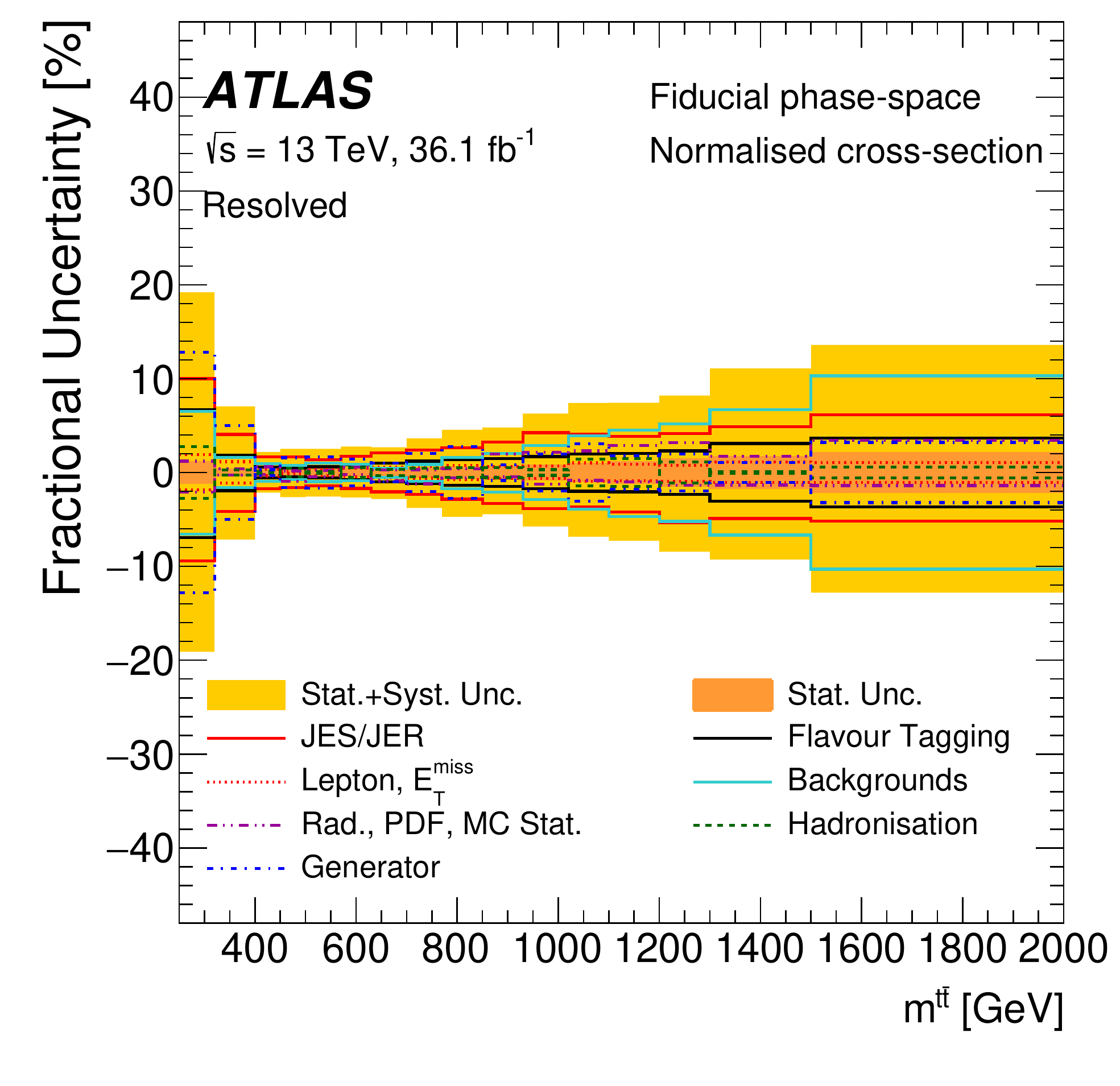}\label{fig:unc_particle:resolved:ttbar_m:rel}}
\subfigure[]{  \includegraphics[width=0.90\textwidth]{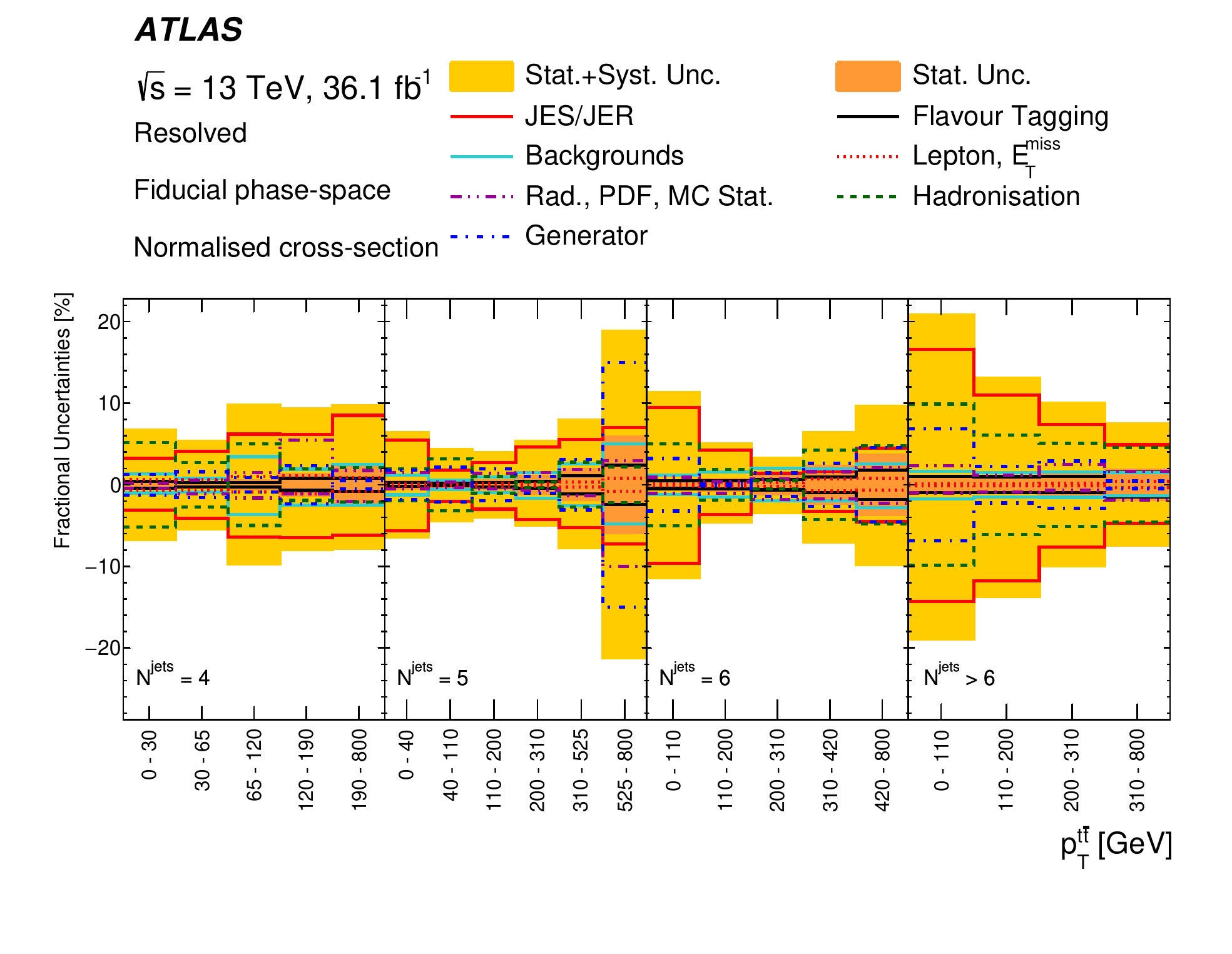}\label{fig:unc_particle:resolved:ttbar_pt:jet_n:rel}}
 
\caption{Uncertainties in the particle-level normalised differential cross-sections as a~function of \subref{fig:unc_particle:resolved:top_had_pt:rel}~the transverse momentum, \subref{fig:unc_particle:resolved:ttbar_m:rel} the mass of the \ttbar{} system, and \subref{fig:unc_particle:resolved:ttbar_pt:jet_n:rel} the transverse momentum of the \ttbar{} system as a function of the jet multiplicity in the resolved topology. The bands represent the statistical and total uncertainty in the data. }
\label{fig:unc_results:particle:resolved:rel}
\end{figure*}

\begin{figure*}[t]
\centering
\subfigure[]{  \includegraphics[width=0.45\textwidth]{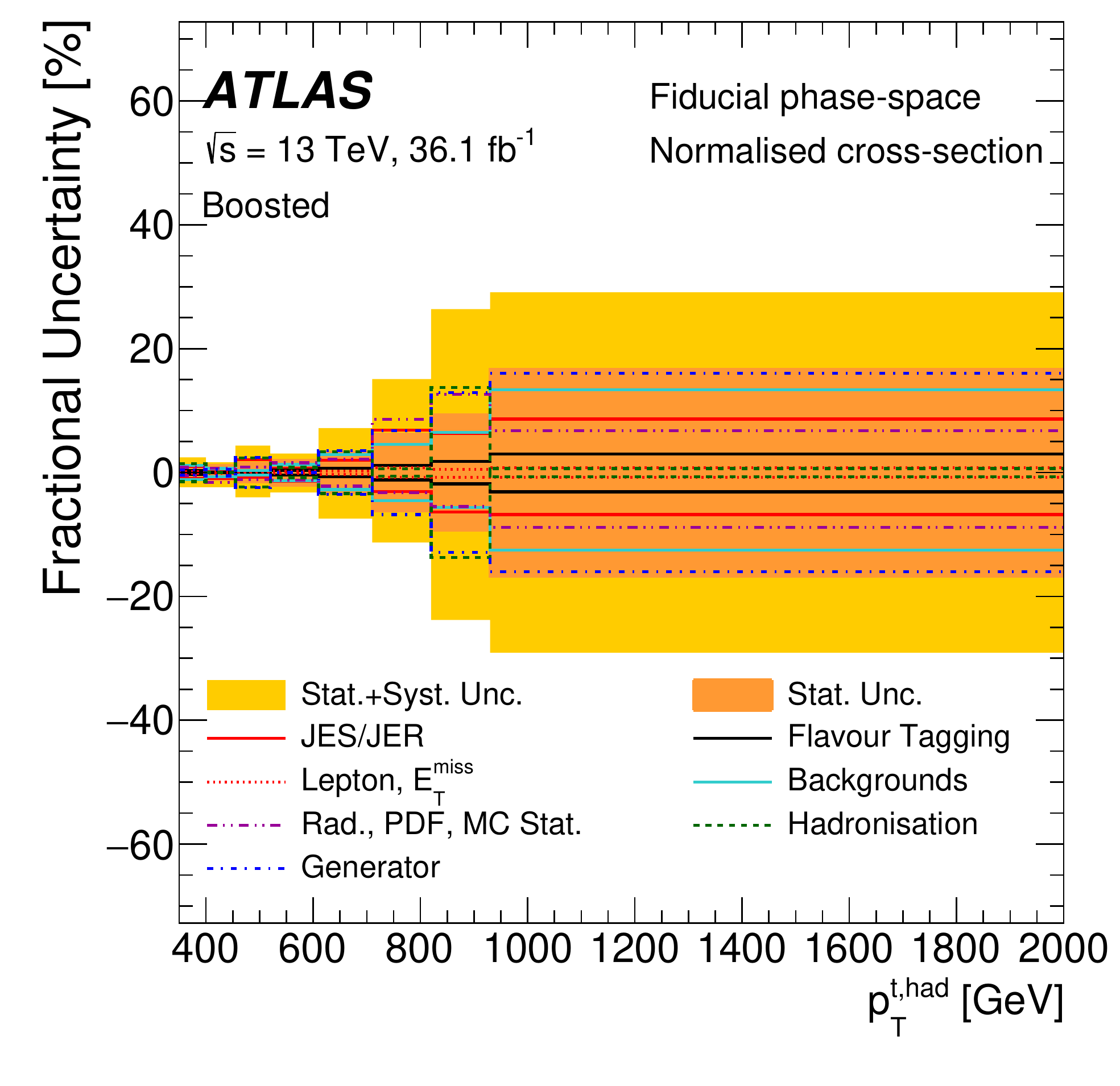}\label{fig:unc_particle:boosted:top_had_pt:rel}}
\subfigure[]{  \includegraphics[width=0.45\textwidth]{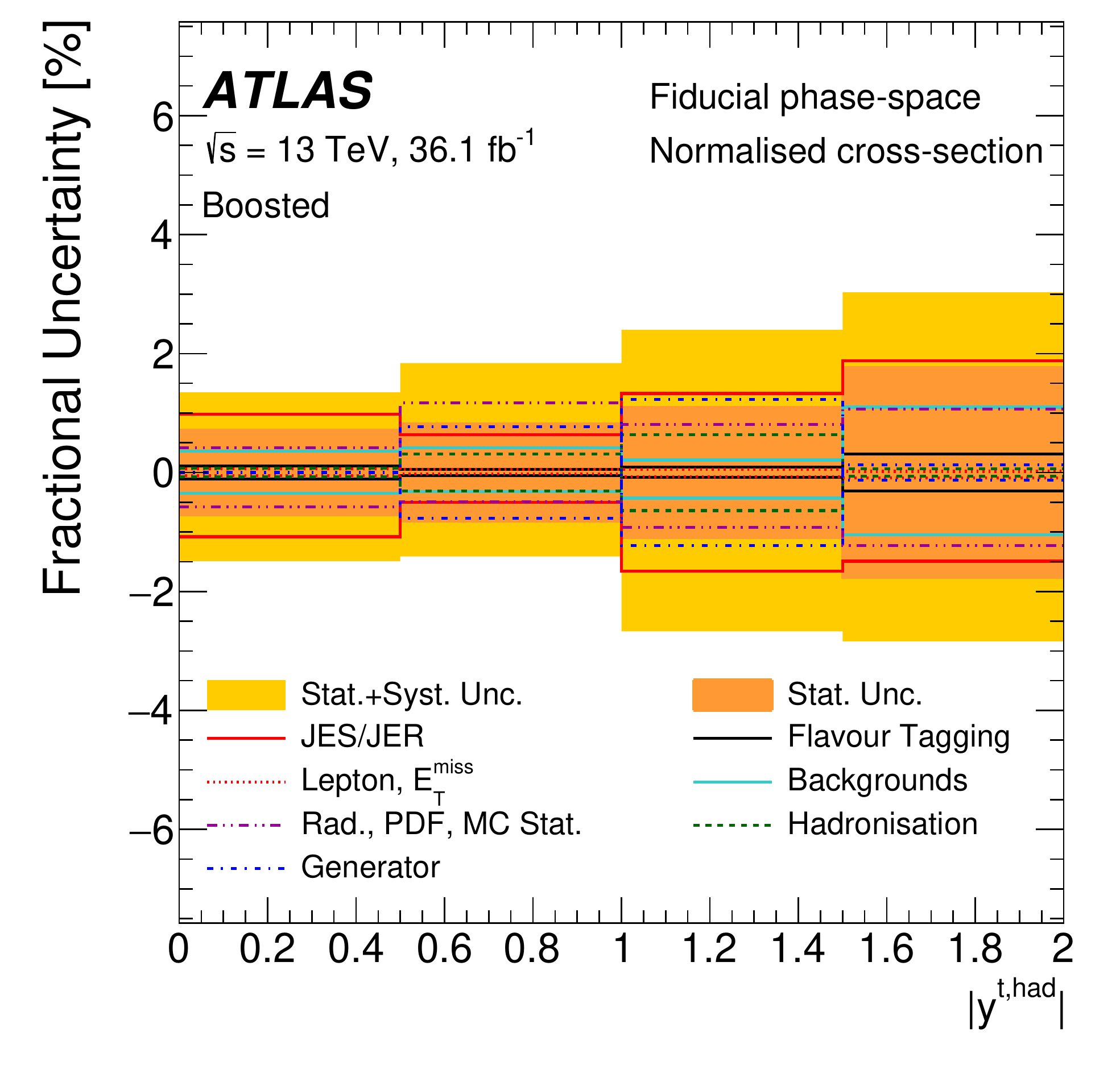}\label{fig:unc_particle:boosted:top_had_y:rel}}
\subfigure[]{  \includegraphics[width=0.90\textwidth]{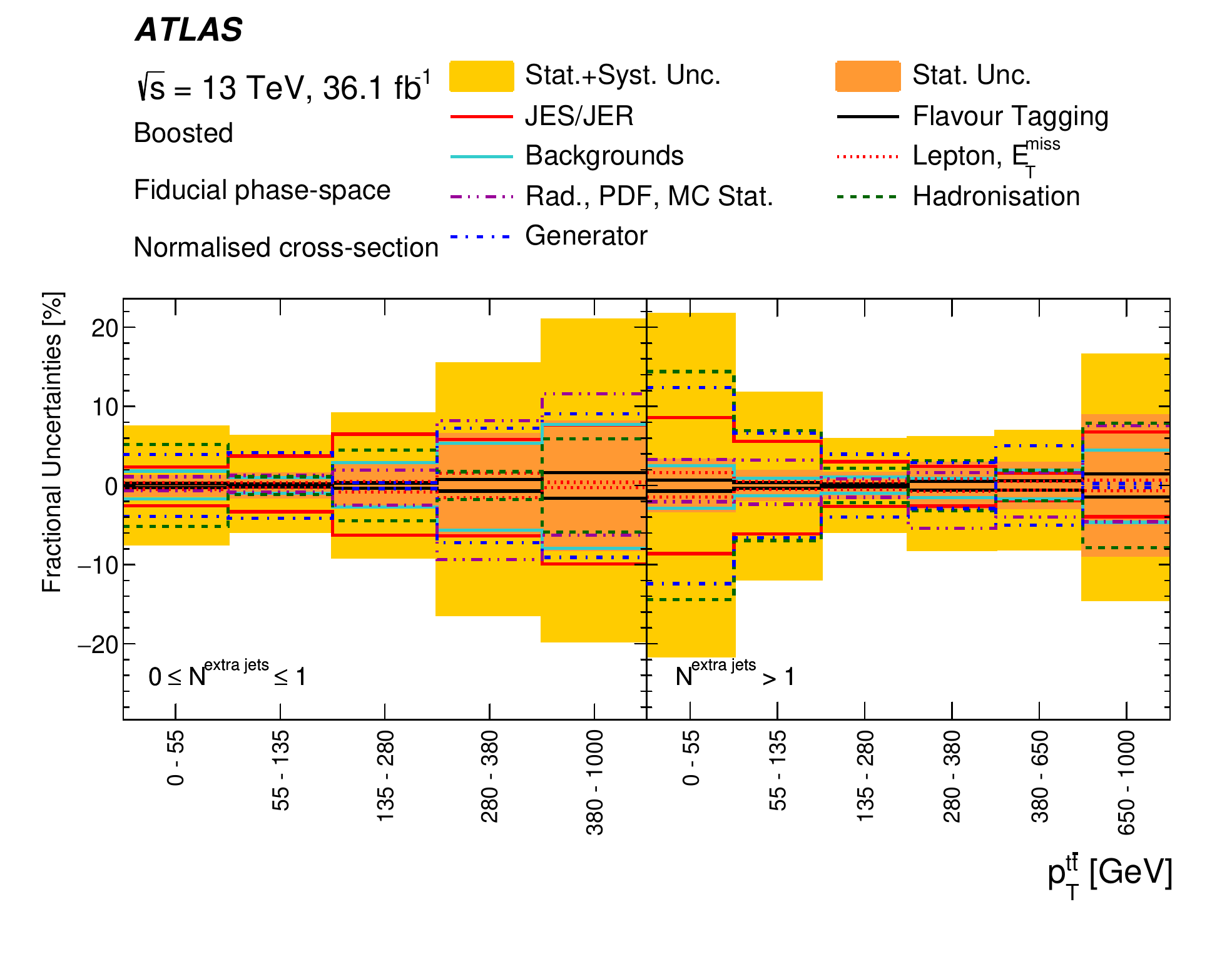}\label{fig:unc_particle:boosted:ttbar_pt:jet_n:rel}}
 
\caption{Uncertainties in the particle-level normalised differential cross-sections as a~function of \subref{fig:unc_particle:boosted:top_had_pt:rel}~the transverse momentum, \subref{fig:unc_particle:boosted:top_had_y:rel}~the rapidity of the hadronically decaying top quark  and~\subref{fig:unc_particle:boosted:ttbar_pt:jet_n:rel} the \pt{}  of the \ttbar{} system as a function of the number of additional jets in the boosted topology. The bands represent the statistical and total uncertainty in the data.}
\label{fig:unc_results:particle:boosted:rel}
\end{figure*}

\begin{figure*}[t]
\centering
\subfigure[]{  \includegraphics[width=0.45\textwidth]{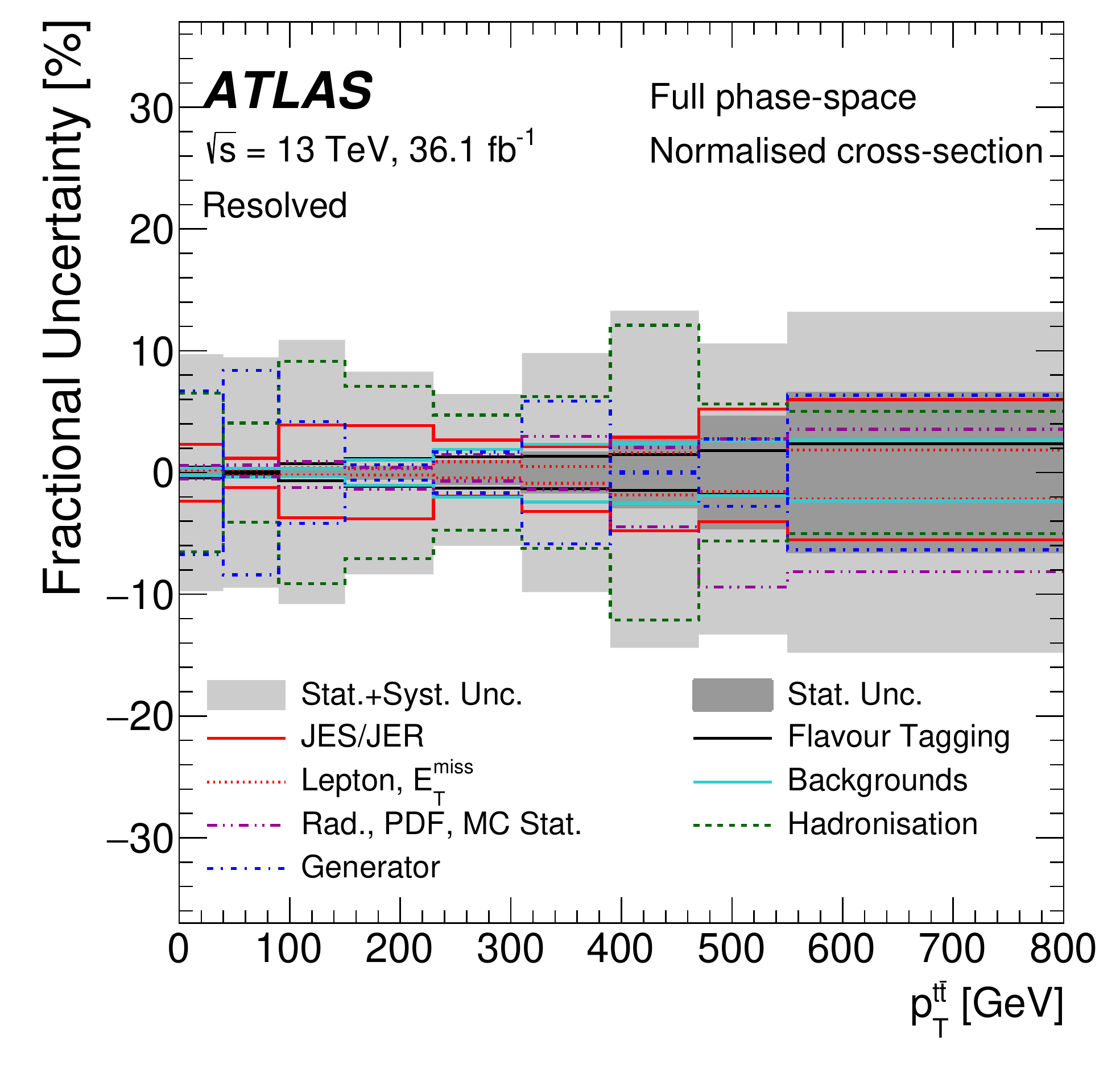}\label{fig:unc_parton:resolved:ttbar_pt:rel}}
\subfigure[]{  \includegraphics[width=0.45\textwidth]{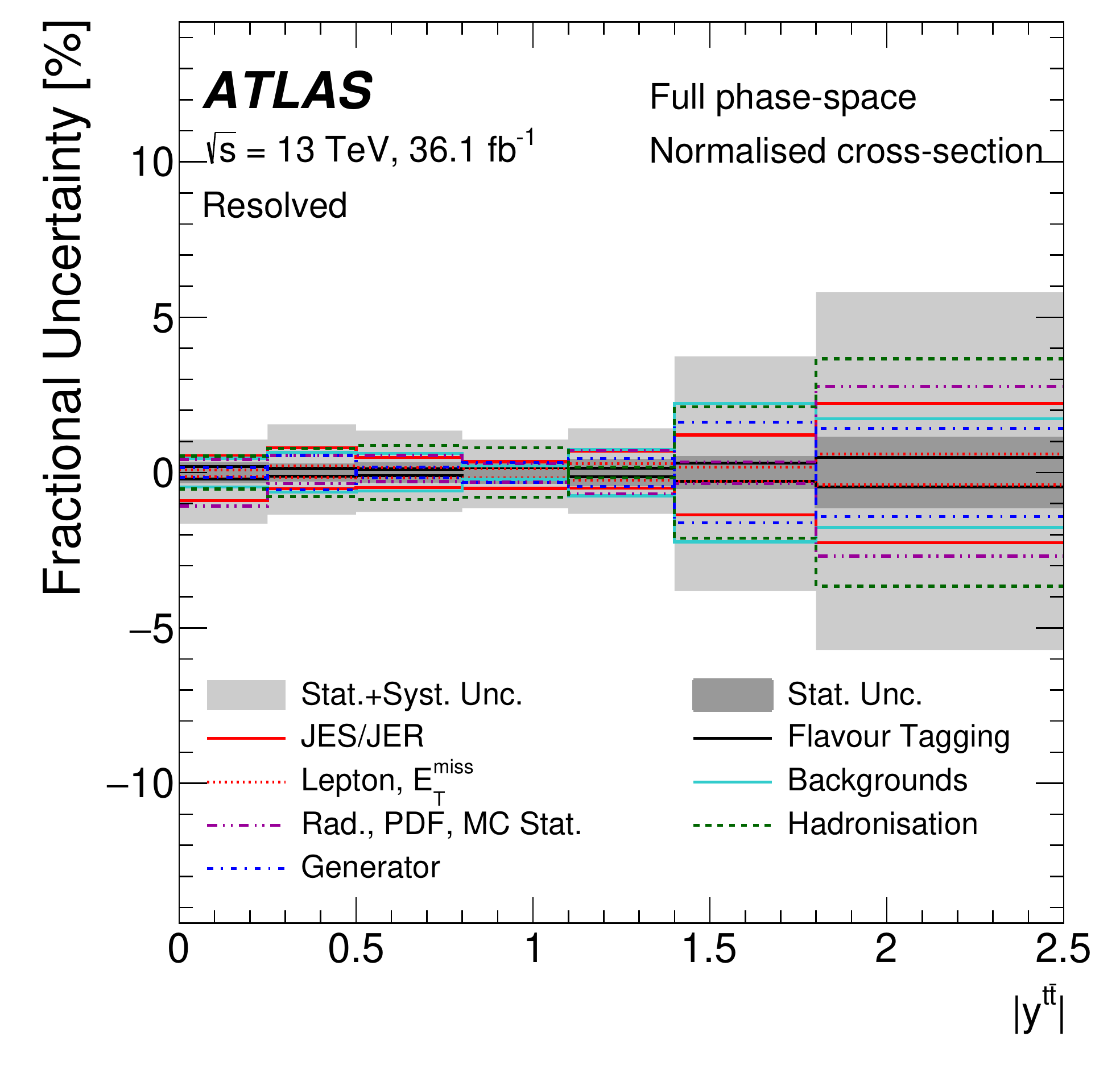}\label{fig:unc_parton:resolved:ttbar_abs_y:rel}}
\subfigure[]{  \includegraphics[width=0.90\textwidth]{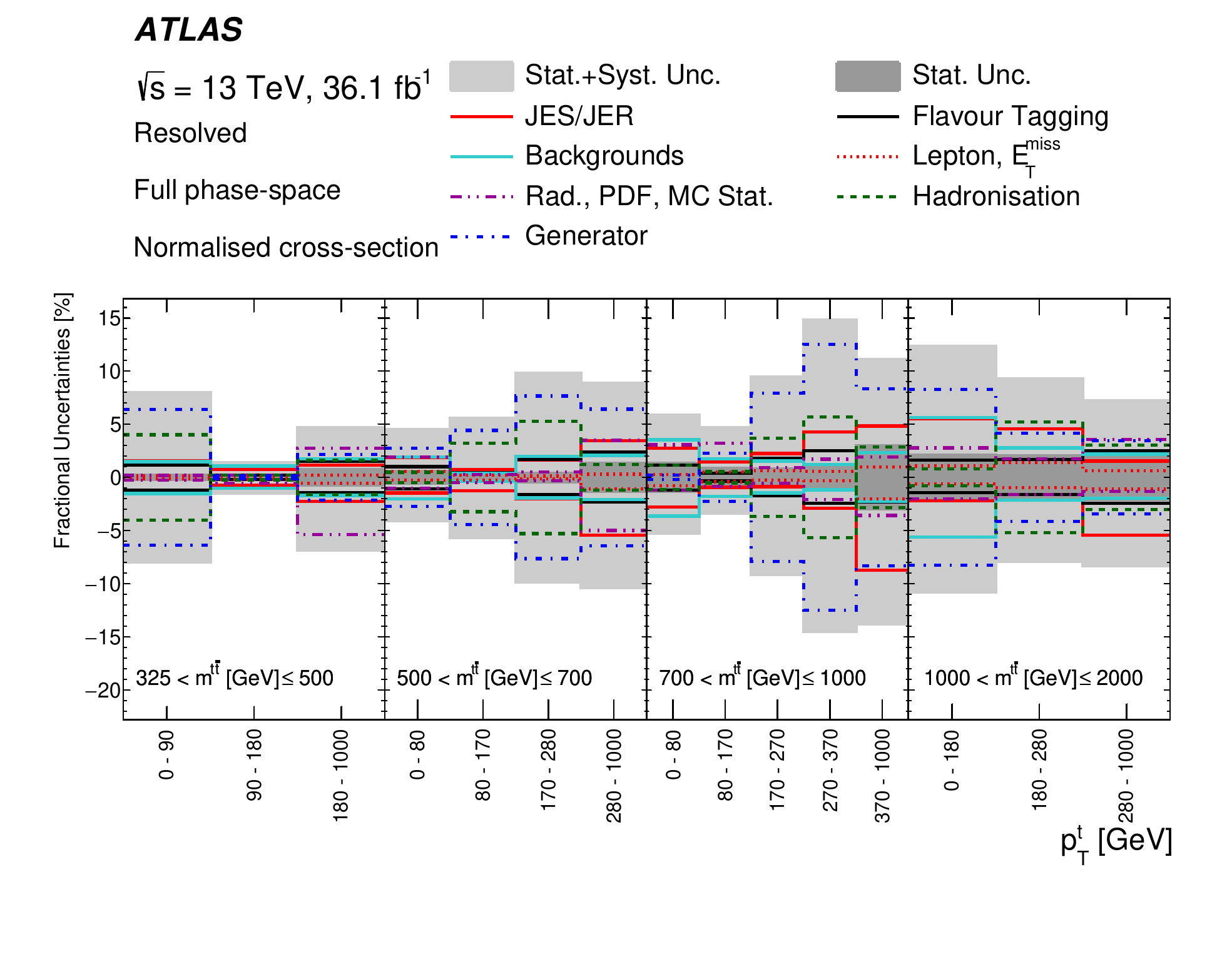}\label{fig:unc_parton:resolved:top_pt:ttbar_m:rel}}
 
\caption{Uncertainties in the parton-level normalised differential cross-sections as a~function of \subref{fig:unc_parton:resolved:ttbar_pt:rel}~the \ttbar{} system  transverse momentum \subref{fig:unc_parton:resolved:ttbar_abs_y:rel} the absolute value of the rapidity and \subref{fig:unc_parton:resolved:top_pt:ttbar_m:rel} the transverse momentum of the top quark as a function of the mass of the \ttbar{} system in the resolved topology. The bands represent the statistical and total uncertainty in the data. }
\label{fig:unc_results:parton:resolved:rel}
\end{figure*}

\begin{figure*}[t]
\centering
\subfigure[]{  \includegraphics[width=0.45\textwidth]{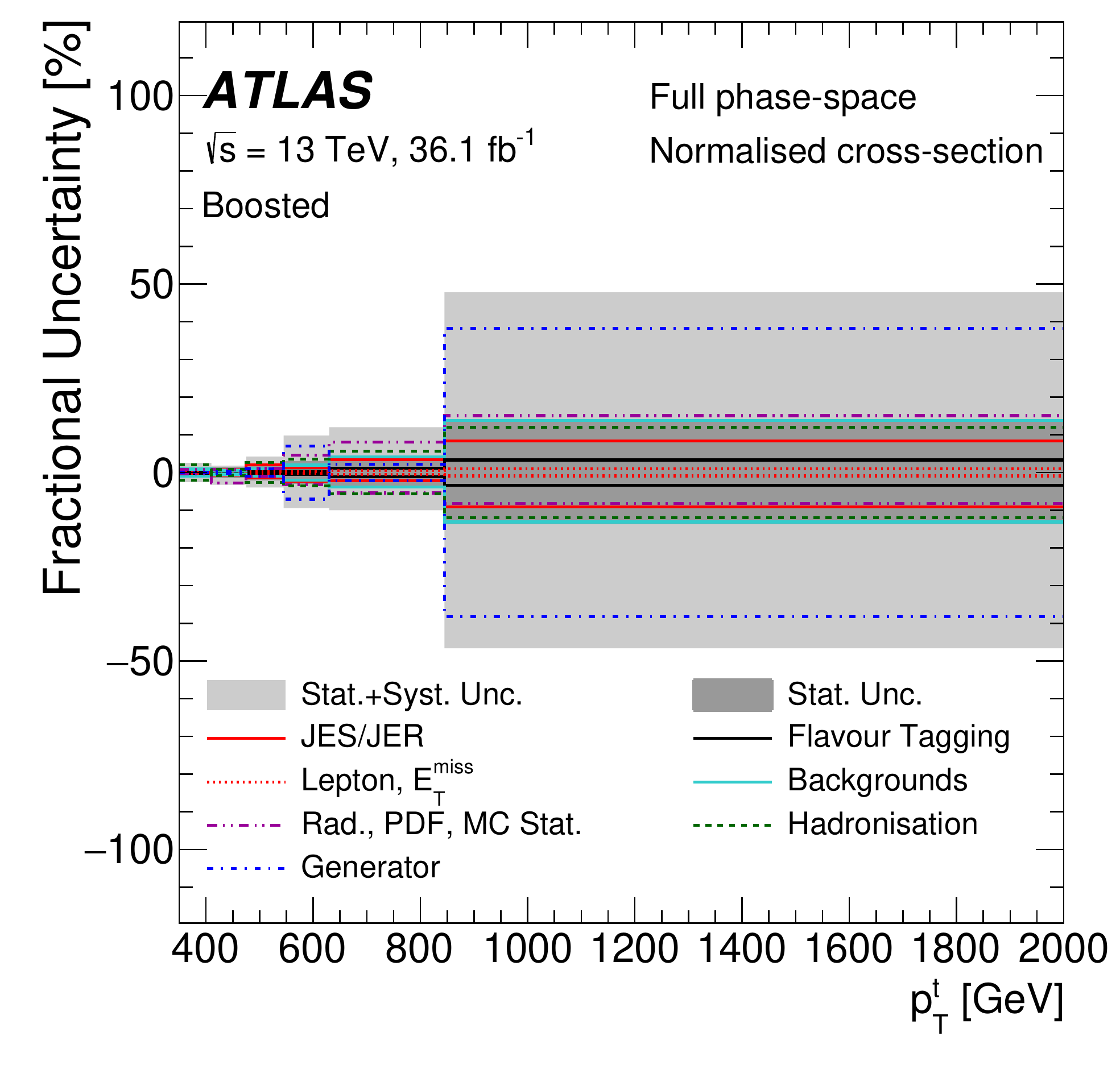}\label{fig:unc_parton:boosted:top_had_pt:rel}}
\subfigure[]{  \includegraphics[width=0.45\textwidth]{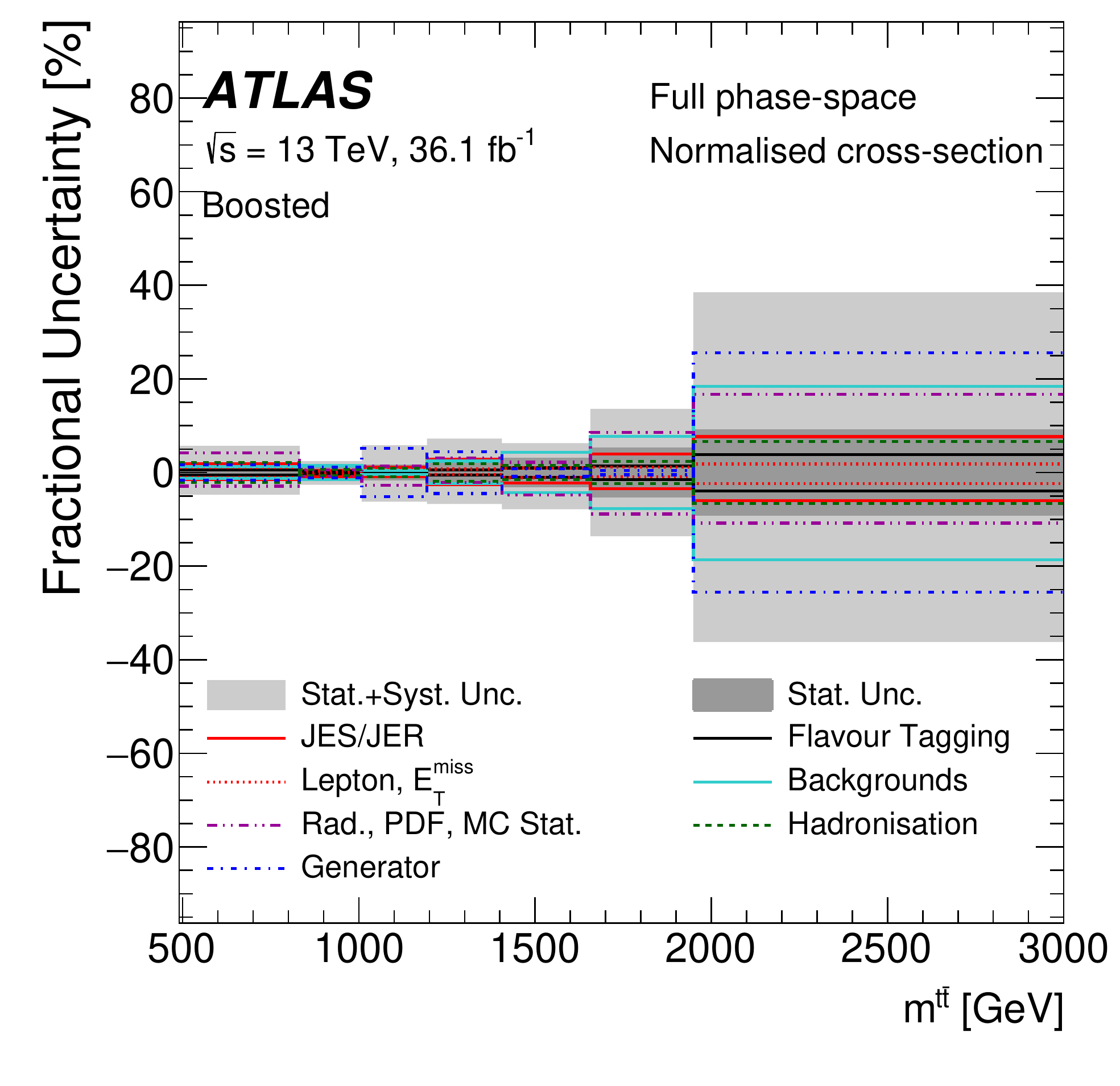}\label{fig:unc_parton:boosted:ttbar_m:rel}}
\subfigure[]{  \includegraphics[width=0.90\textwidth]{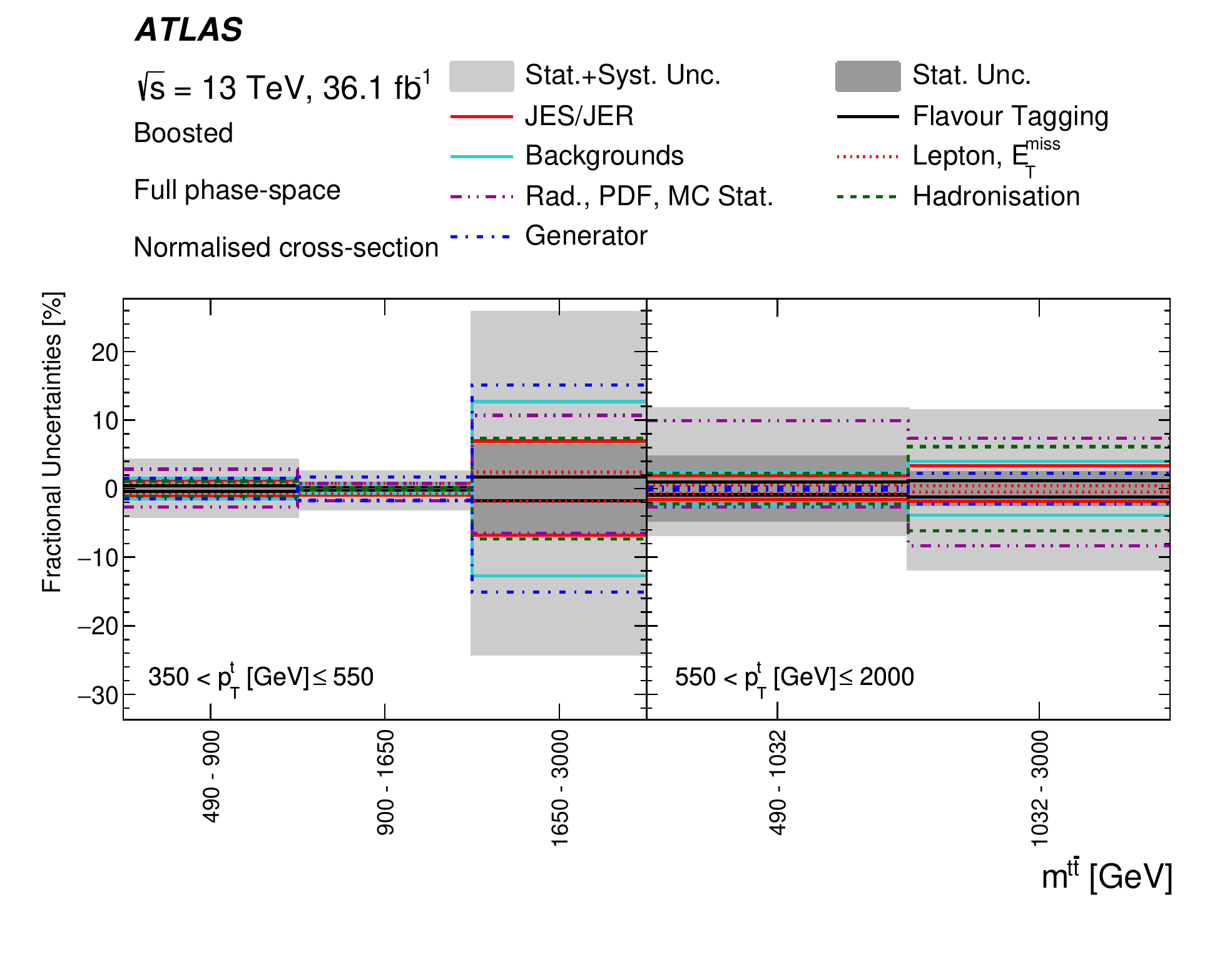}\label{fig:unc_parton:boosted:ttbar_m:top_pt:rel}}
 
\caption{Uncertainties in the parton-level normalised  differential cross-sections as a~function of \subref{fig:unc_parton:boosted:top_had_pt:rel}~the transverse momentum of the top quark, \subref{fig:unc_parton:boosted:ttbar_m:rel}~the mass of the \ttbar system and \subref{fig:unc_parton:boosted:ttbar_m:top_pt:rel} the mass  of the \ttbar{} system as a function of the \pt   of the top quark in the boosted topology. The bands represent the statistical and total uncertainty in the data.}
\label{fig:unc_results:parton:boosted:rel}
\end{figure*}

Figures~\ref{fig:unc_results:particle:resolved:rel}--\ref{fig:unc_results:parton:boosted:rel} present the uncertainties in the particle- and parton-level normalised differential cross-sections as a~function of some of the different observables in the resolved and boosted topologies, respectively.

The dominant systematic uncertainties in many measured normalised differential cross-sections in the resolved topology are those related to the jet energy scale and resolution, especially for differential cross-sections sensitive to the jet multiplicity. While negligible in the normalised spectra, the uncertainties related to the flavour tagging become dominant when measuring inclusive and absolute differential cross-sections. Other significant uncertainties, dominant in the boosted topology, include those from the signal modelling with, depending on the observable, either the generator, hadronisation or the additional radiation component being the most dominant.
 
For most distributions in the resolved topology and in large parts of the phase-space, the measurements have a precision of the order of 10\%--15\%, while for the boosted topology the precision obtained varies from 7\% to about 30\% at particle level, increasing to 40\% at parton level.
 
\FloatBarrier

\clearpage
\section{Results} \label{sec:results}
 
In this section, comparisons between the measured single- and double-differential cross-sections and several
SM predictions are presented for the observables discussed in Section~\ref{sec:observables}. The results are
presented for both the resolved and boosted topologies, at  particle level in the fiducial phase-spaces and
at  parton level in the full phase-space.

For the comparisons at the particle level, the predictions are obtained using different MC generators.
The \PowhegBox generator, denoted `PWG' in the figures, is used with two different parton-shower
and hadronisation models, as implemented in  \PythiaEight{} and \herwig{}7, as well as two extra settings for the
radiation modelling. In addition the \SHERPAV{2.2.1} generator is also  compared with the data.
All the MC samples are detailed in Section~\ref{sec:signalMC}.
 
The measured differential cross-sections at the parton level are compared with \NNLO pQCD theoretical predictions~\cite{Czakon:2015owf,Czakon:2016dgf}. An additional comparison is performed, for a subset of the  differential parton-level cross-sections, with existing fixed-order predictions at \NNLO{} pQCD accuracy and including electroweak (EW) corrections~\cite{Czakon:2017wor}.
 
To quantify the level of agreement between the measured cross-sections and the different theoretical
predictions, $\chi^2$ values are calculated, using the total covariance matrices evaluated for the measured cross-sections, according to the following relation
\begin{equation*}
\chi^2 = V_{N_{\textrm b}}^{\textrm T} \cdot {\textrm{Cov}}_{N_{\textrm b}}^{-1} \cdot V_{N_{\textrm b}} \,,
\end{equation*}
where $N_{\textrm b}$ is the number of bins of the spectrum under consideration, $V_{N_{\textrm b}}$ is the
vector of differences between the measured and predicted cross-sections and ${\textrm{Cov}}_{N_{\textrm b}}$
represents the covariance matrix. This includes both the statistical and systematic uncertainties and is evaluated by performing 10~000 pseudo-experiments, following the procedure described in Ref.~\cite{TOPQ-2016-01}. No uncertainties in the theoretical predictions are included in the $\chi^2$ calculation.
The $p$-values are then evaluated from the $\chi^2$ and the number of degrees of freedom (NDF).

For normalised cross-sections, $V_{N_{\rm b}}$ must be replaced with $V_{N_{\rm b}-1}$, which is the vector of differences between data and prediction obtained by discarding one of the $N_{\rm b}$ elements and, consequently,  ${\rm Cov}_{N_{\rm b}-1}$ is the $(N_{\rm b}-1) \times (N_{\rm b}-1)$ sub-matrix derived from the full covariance matrix discarding the corresponding row and column. The sub-matrix obtained in this way is invertible and allows the $\chi^2$ to be computed. The $\chi^2$ value does not depend on the choice of the element discarded for the vector $V_{N_{\rm b}-1}$ and the corresponding sub-matrix ${\rm Cov}_{N_{\rm b}-1}$.
 
The determination of statistical correlations within each spectrum and among different spectra are evaluated using the Bootstrap Method~\cite{bootstrap}. The method is based on the extraction of 1000 Bootstrap samples (pseudo-experiments) obtained by reweighting the measured data sample on an event-by-event basis with a Poisson distribution.

To allow comparisons to be made between the shapes of the measured cross-sections and the predictions, all the results included in this section are presented as normalised cross-sections: the measurement of the normalised cross-sections significantly reduces the contribution of uncertainties common to all bins of the distributions, highlighting shape differences relative to the absolute case. Examples to illustrate this features are presented in Section~\ref{sec:results:fiducial}, while the results of $\chi^2$ and $p$-value calculations are always reported for both the normalised and absolute cross-sections.
 
\subsection{Results at particle level in the fiducial phase-spaces}
\label{sec:results:fiducial}
\subsubsection{Resolved topology}
\label{sec:results:fiducial:resolved}
 
\begin{figure*}[t]
\centering
\subfigure[]{\includegraphics[width=0.45\textwidth]{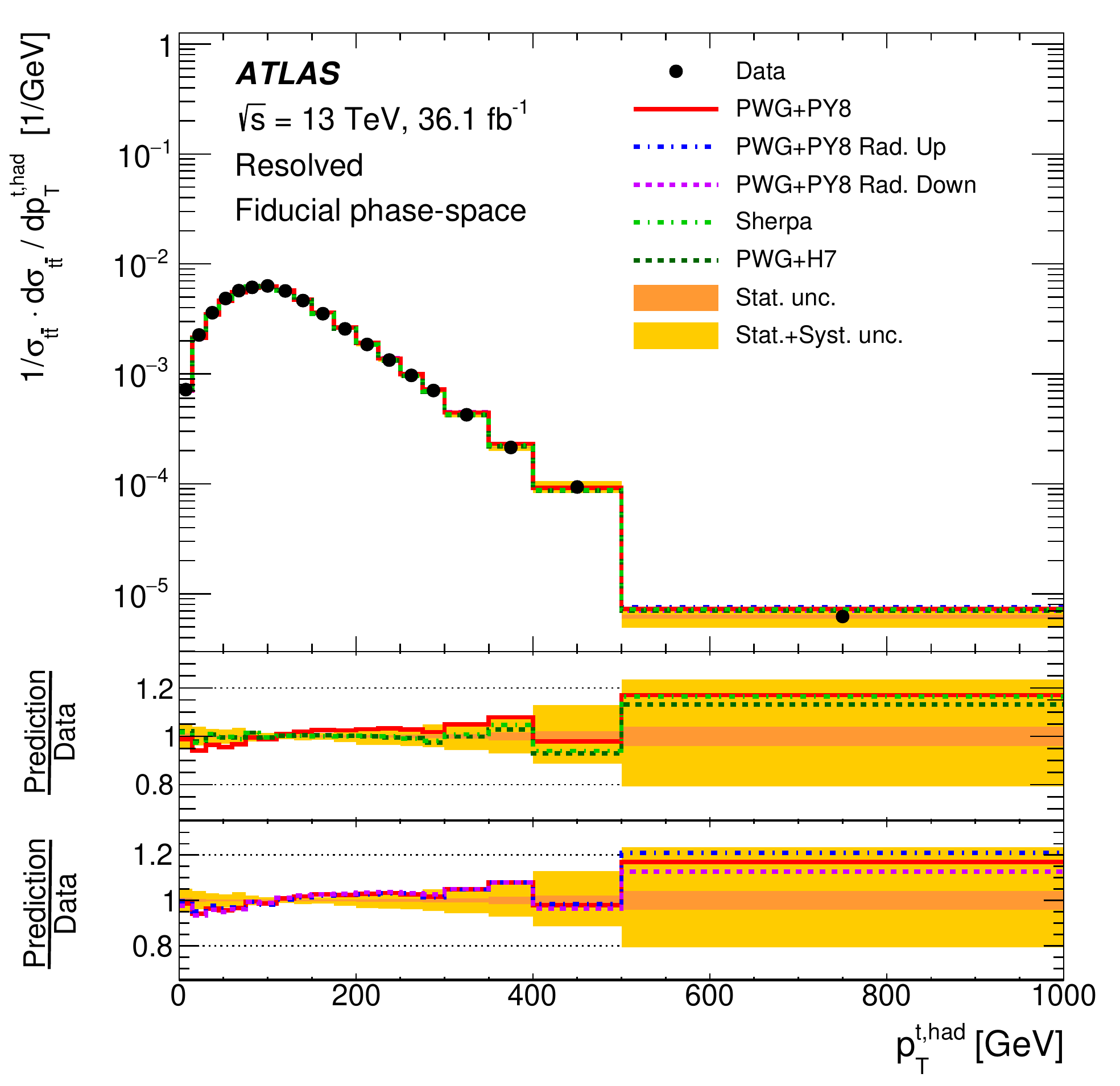}
\label{fig:results_particle:resolved:topHad_pt:rel}}
\subfigure[]{\includegraphics[width=0.45\textwidth]{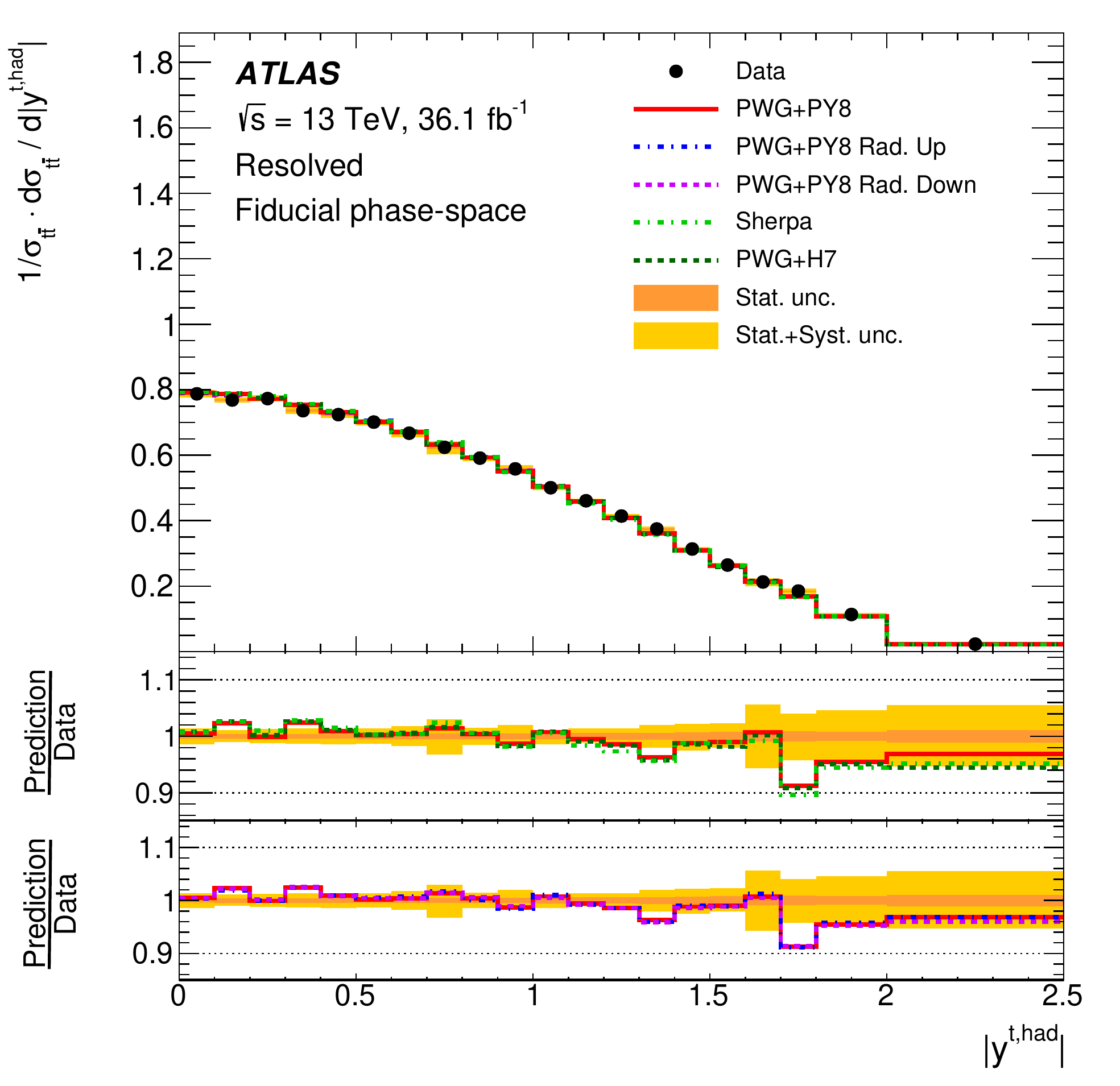}
\label{fig:results_particle:resolved:topHad_abs_y:rel}}
\caption{\small{Particle-level normalised differential cross-sections as a function of \subref{fig:results_particle:resolved:topHad_pt:rel}
the transverse momentum and \subref{fig:results_particle:resolved:topHad_abs_y:rel} the absolute value of the rapidity of the hadronically decaying
top quark in the resolved topology, compared with different Monte Carlo predictions. The bands represent the statistical and total uncertainty in the data. Data points are placed at the centre of each bin. The lower panel shows the ratios of the simulations to data.}}
\label{fig:results:rel:particle:resolved:1D:top_had}
\end{figure*}

\begin{figure*}[t]
\centering
\subfigure[]{\includegraphics[width=0.45\textwidth]{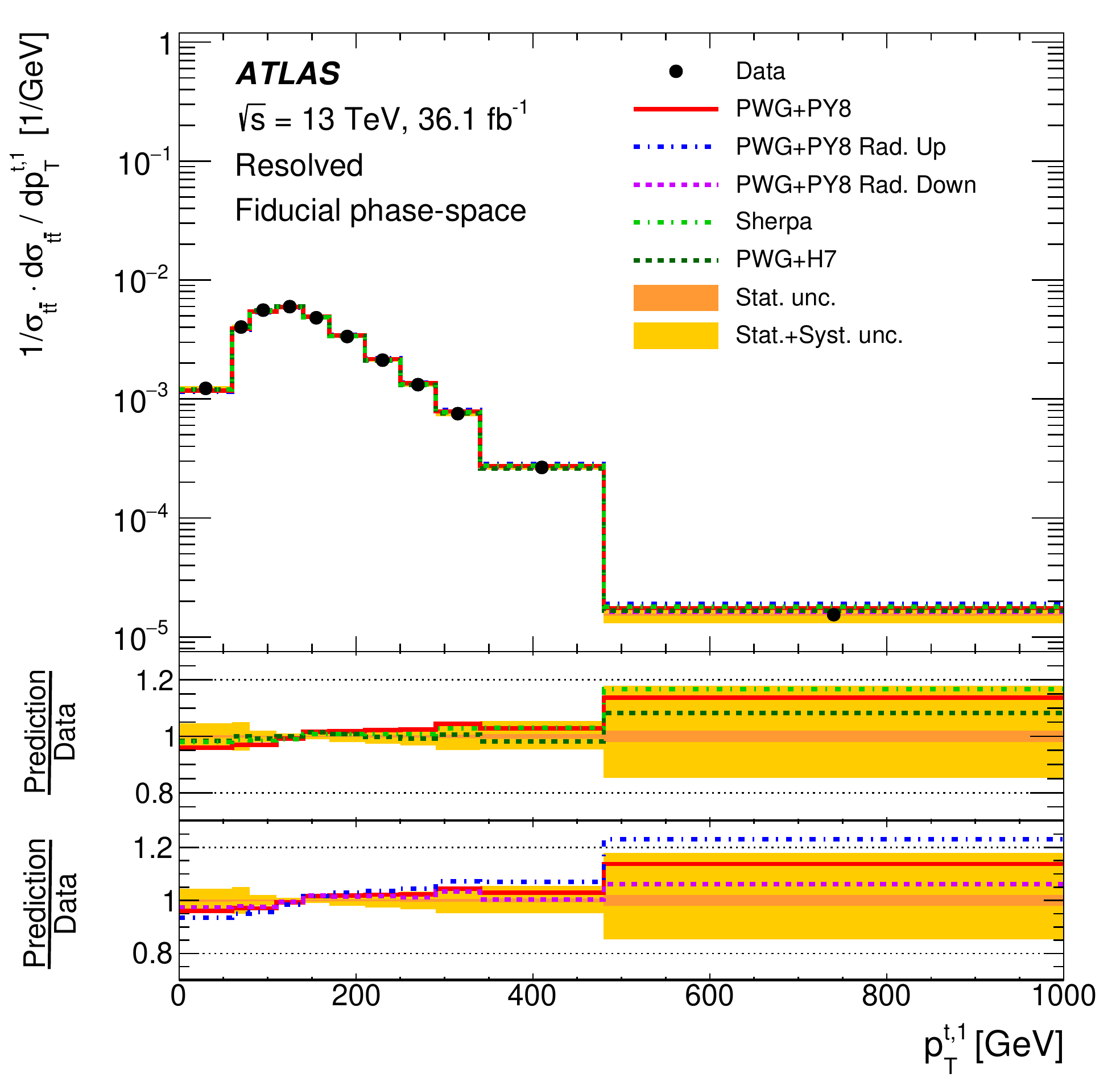}
\label{fig:results_particle:resolved:topHad_pt:isleading:rel}}
\subfigure[]{\includegraphics[width=0.45\textwidth]{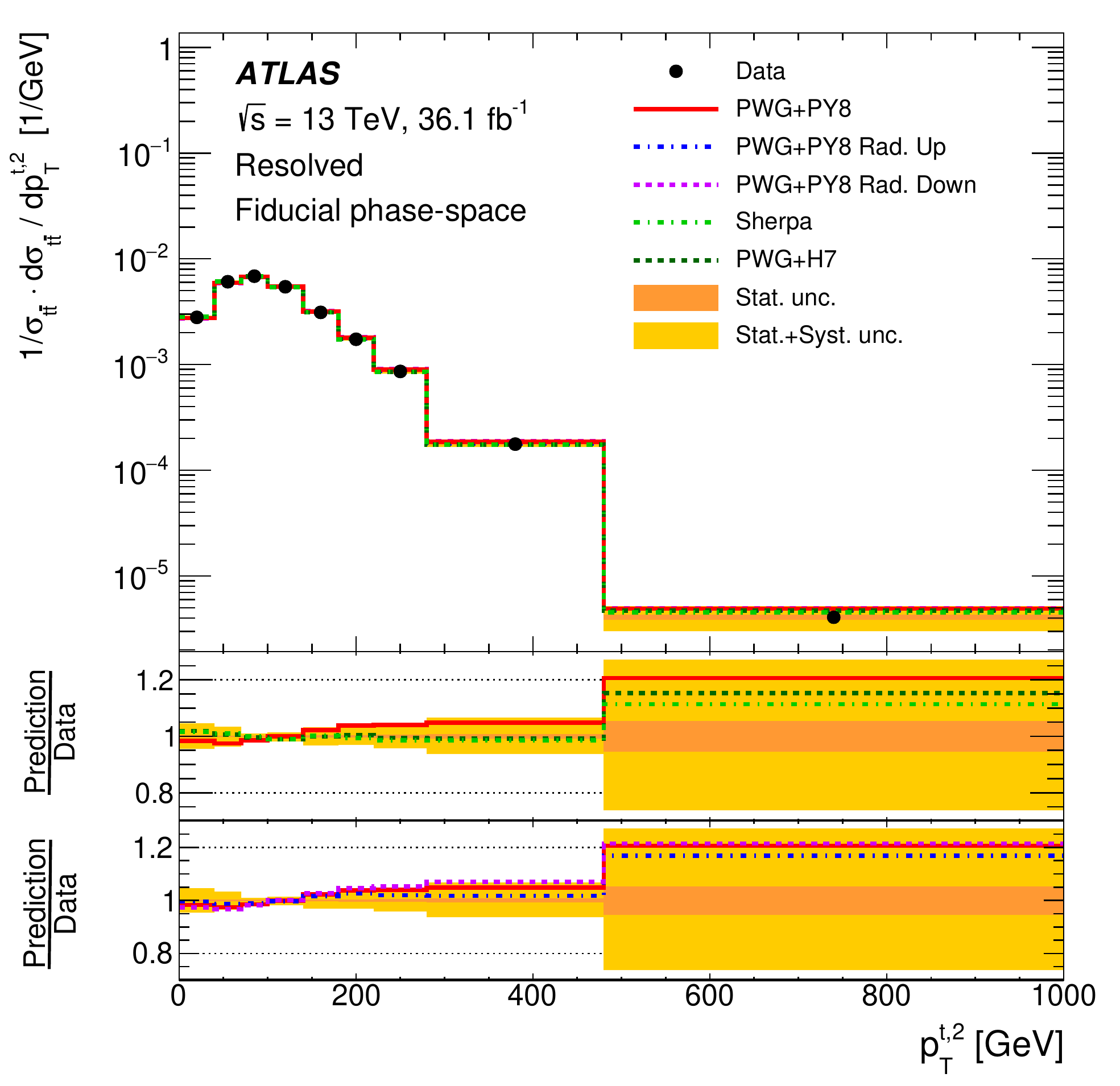}
\label{fig:results_particle:resolved:topHad_pt:issubleading:rel}}
\caption{\small{Particle-level normalised differential cross-sections as a function of the transverse momentum of~\subref{fig:results_particle:resolved:topHad_pt:isleading:rel} the leading and~\subref{fig:results_particle:resolved:topHad_pt:issubleading:rel}  the subleading top quark in the resolved topology,
compared with different Monte Carlo predictions. The bands represent the statistical and total uncertainty in the data.  Data points are placed at the centre of each bin. The lower panel shows the ratios of the simulations to data.}}
\label{fig:results:rel:particle:resolved:1D:top_leading}
\end{figure*}

\begin{figure*}[t]
\centering
\subfigure[]{\includegraphics[width=0.45\textwidth]{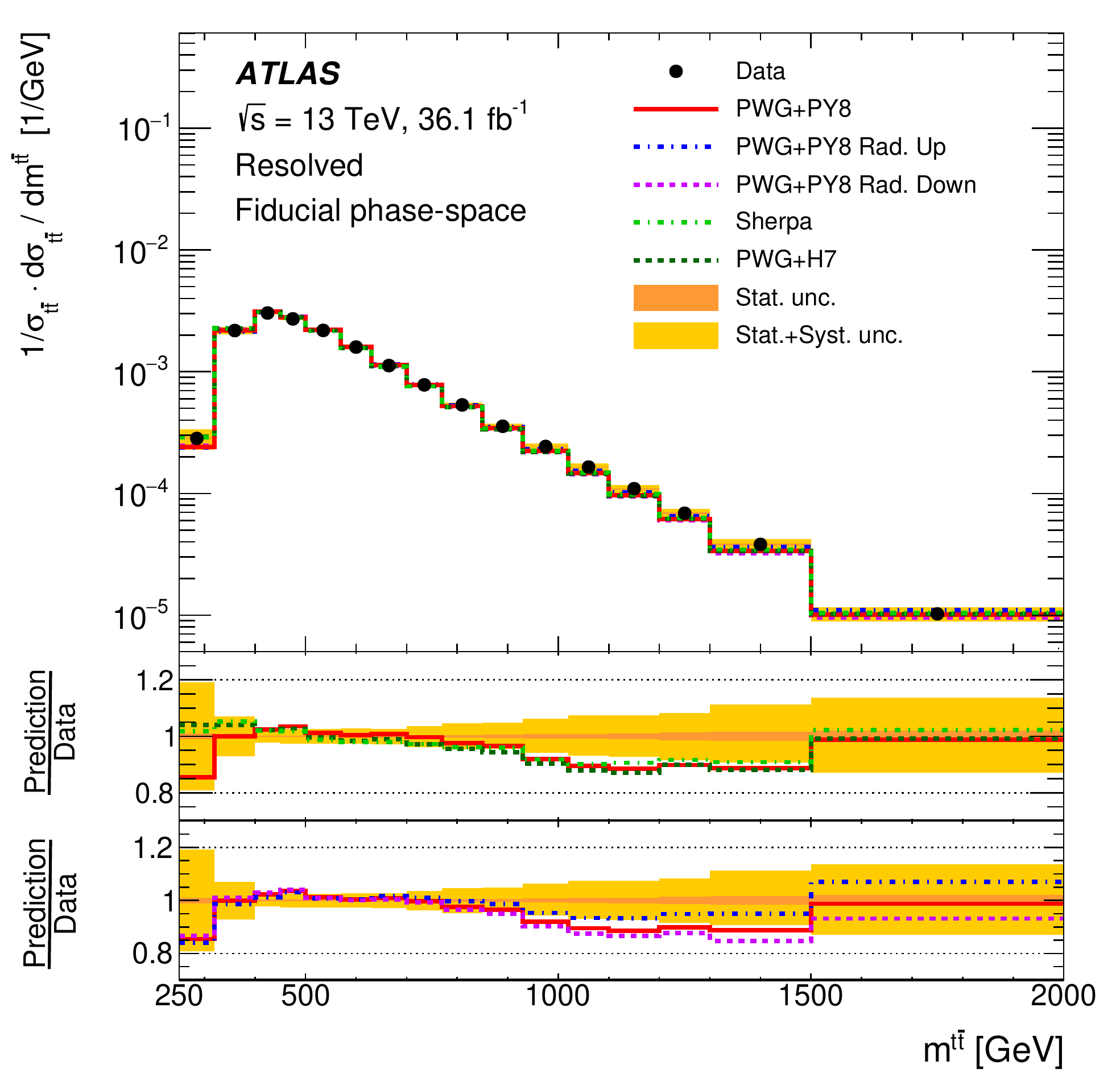}
\label{fig:results_particle:resolved:ttbar_m:rel}}
\subfigure[]{\includegraphics[width=0.45\textwidth]{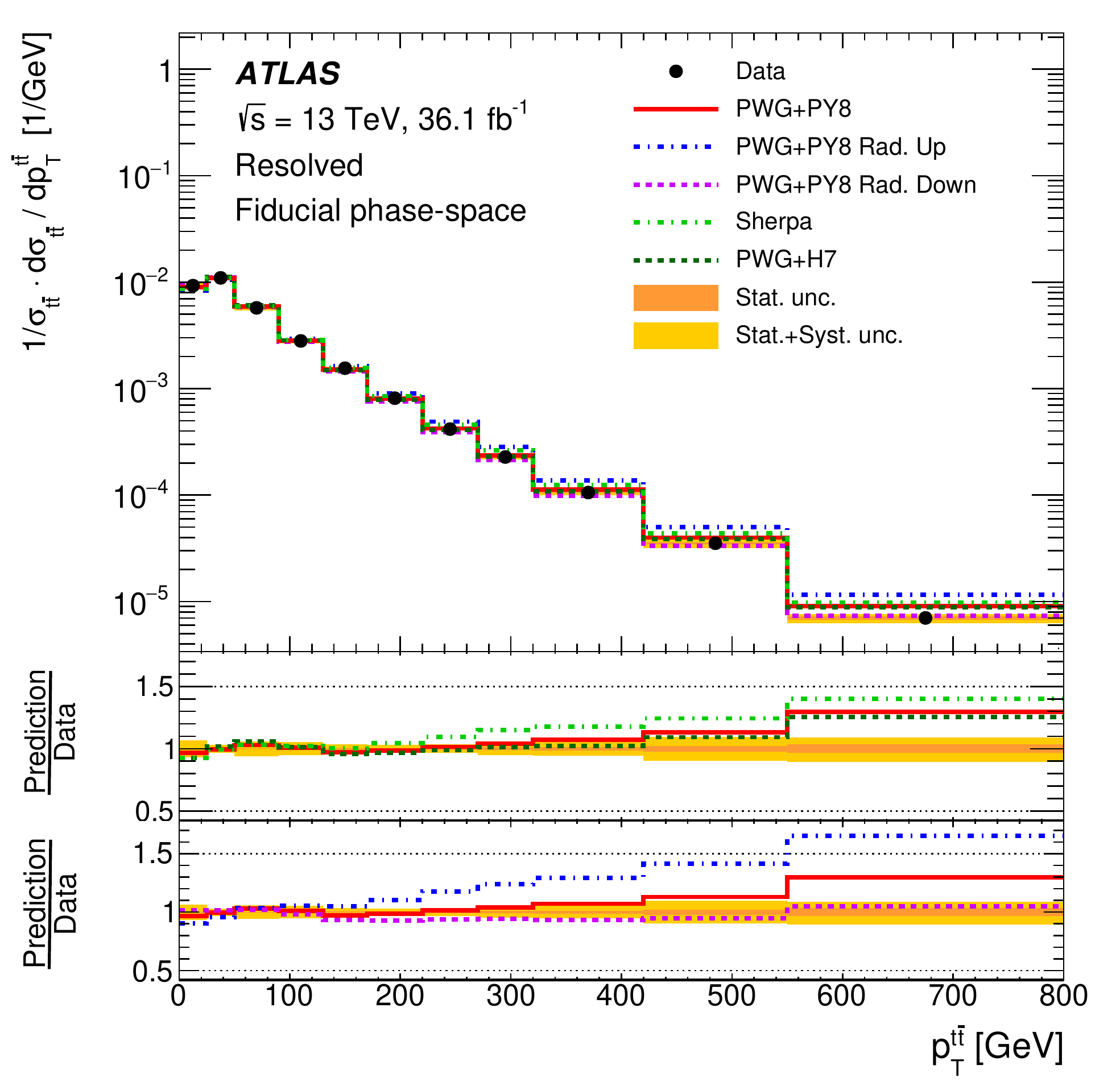}
\label{fig:results_particle:resolved:ttbar_pt:rel}}\\
\caption{\small{Particle-level normalised differential cross-sections as a function of~\subref{fig:results_particle:resolved:ttbar_m:rel} the mass and \subref{fig:results_particle:resolved:ttbar_pt:rel} the transverse momentum of the \ttbar{} system in the resolved topology, compared with different Monte Carlo
predictions. The bands represent the statistical and total uncertainty in the data.
Data points are placed at the centre of each bin. The lower panel shows the ratios of the simulations to data.}}
\label{fig:results:rel:particle:resolved:1D:ttbar}
\end{figure*}

\begin{figure*}[t]
\centering
\subfigure[]{\includegraphics[width=0.45\textwidth]{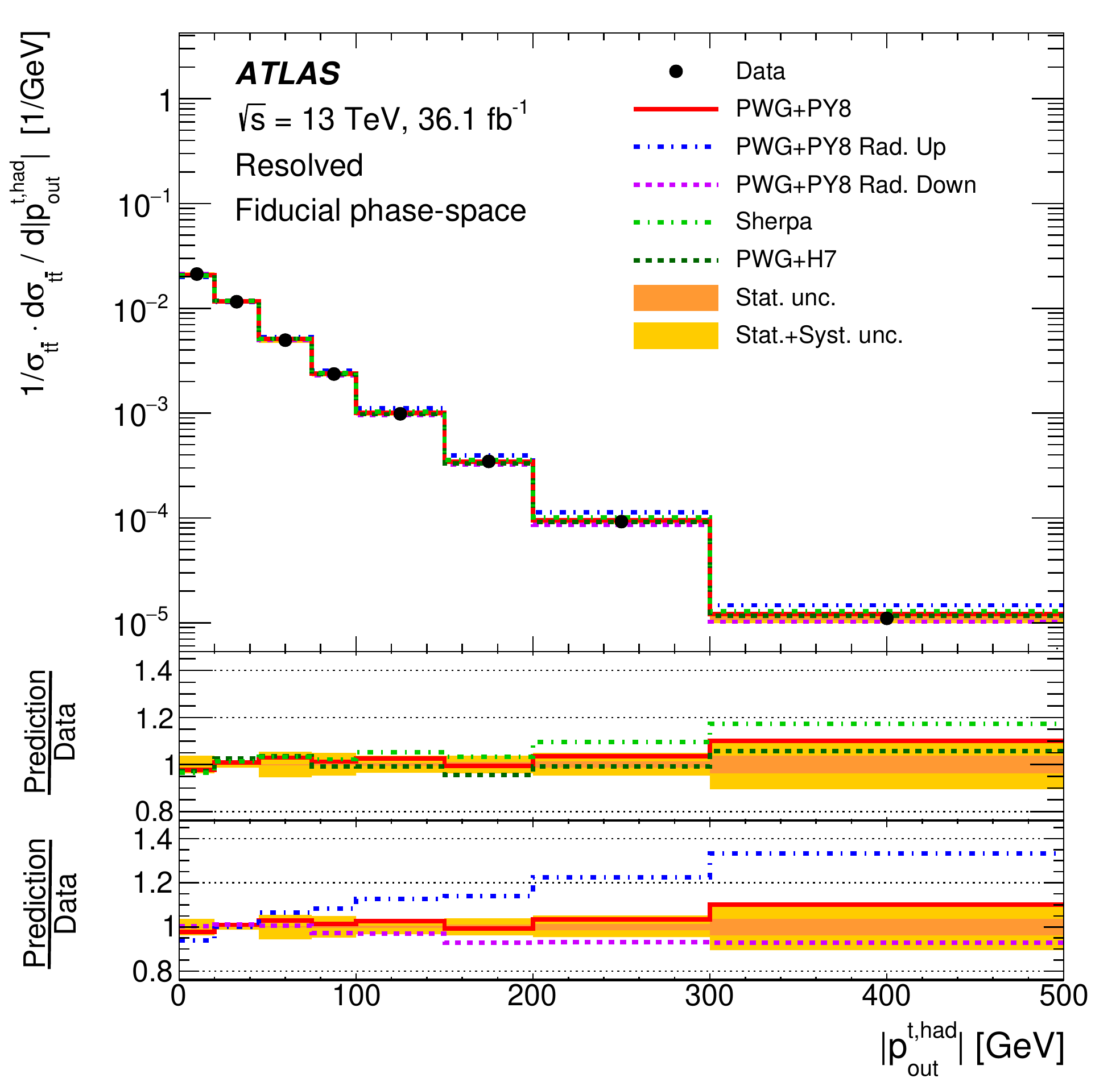}
\label{fig:results_particle:resolved:absPout:rel}}
\subfigure[]{\includegraphics[width=0.45\textwidth]{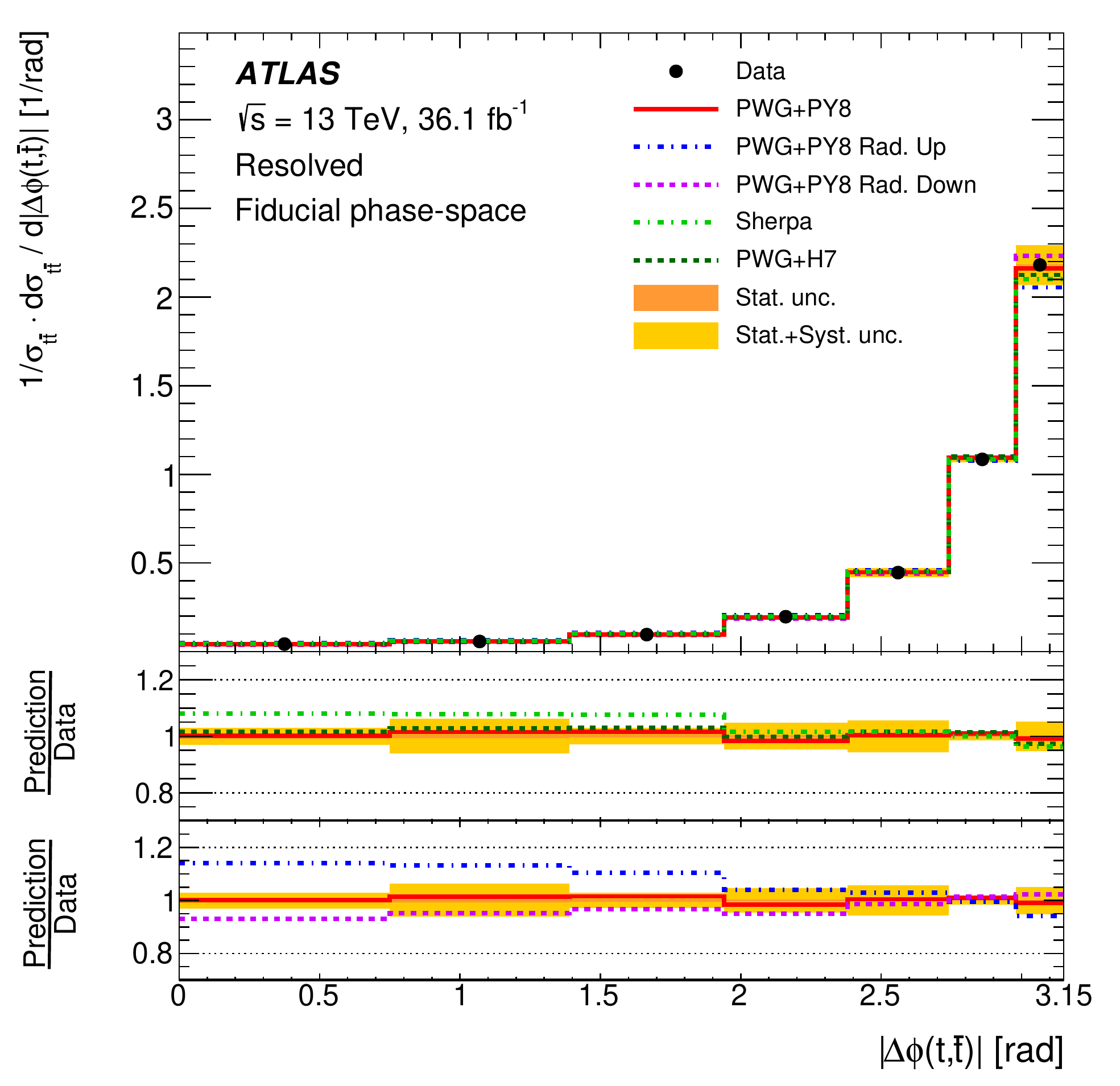}
\label{fig:results_particle:resolved:deltaPhi_tt:rel}}\\
\subfigure[]{\includegraphics[width=0.45\textwidth]{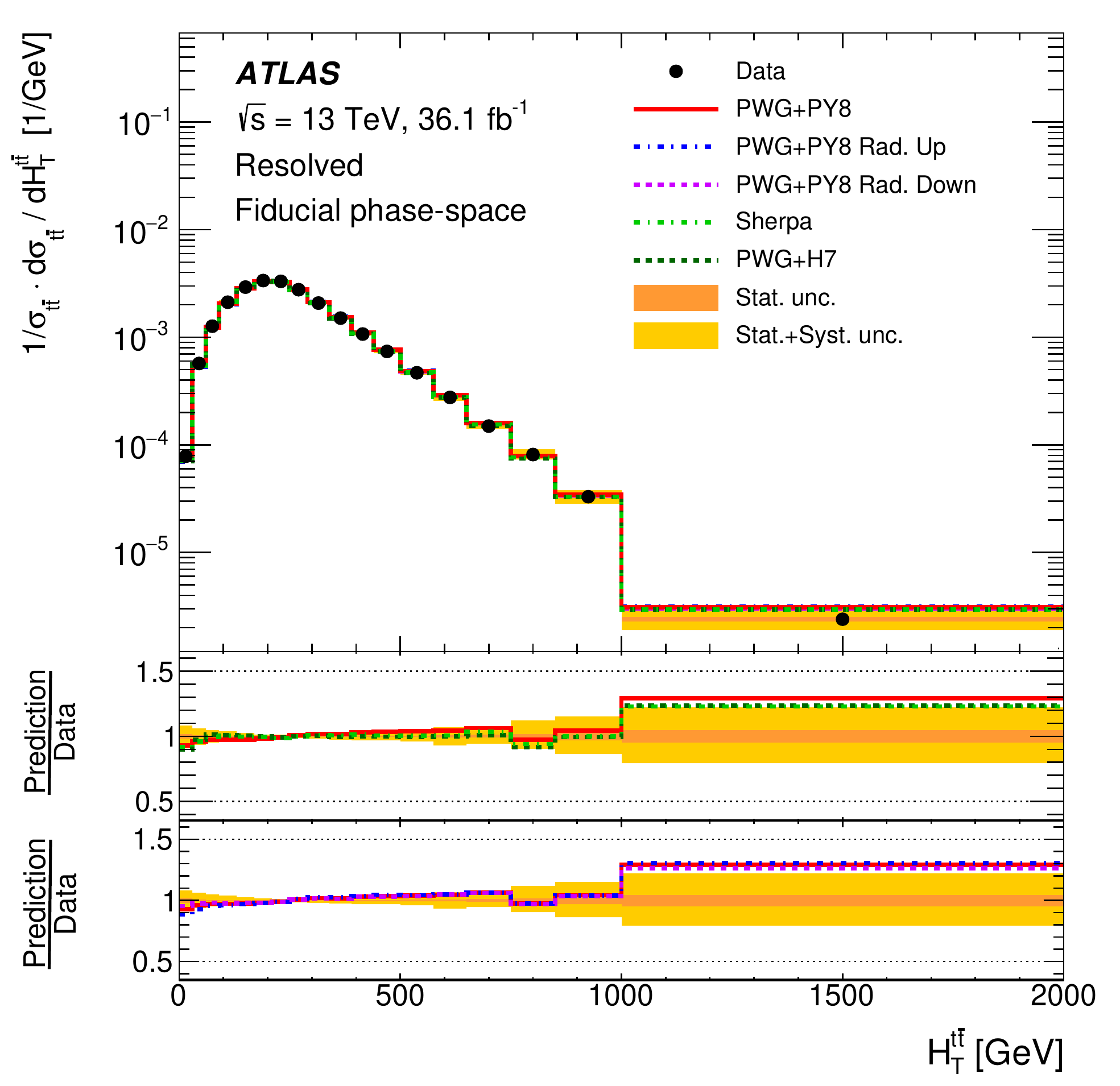}
\label{fig:results_particle:resolved:HT_tt:rel}}
\subfigure[]{\includegraphics[width=0.45\textwidth]{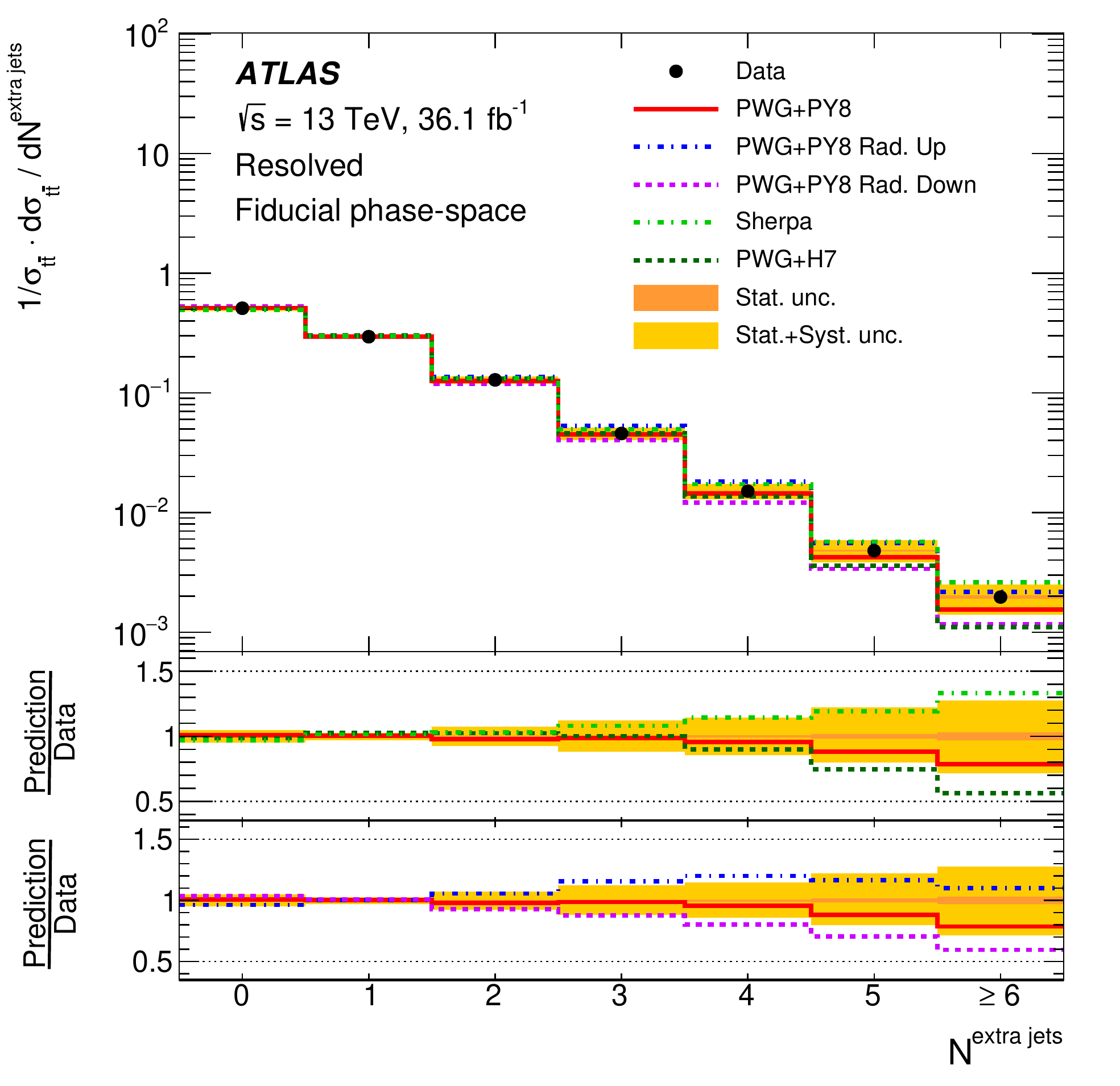}
\label{fig:results_particle:resolved:jet_n:rel}}
\caption{\small{Particle-level normalised differential cross-sections as a function of~\subref{fig:results_particle:resolved:absPout:rel} \absPoutthad{}, \subref{fig:results_particle:resolved:deltaPhi_tt:rel} \dPhittbar{}, \subref{fig:results_particle:resolved:HT_tt:rel} \HTtt{} and \subref{fig:results_particle:resolved:jet_n:rel}~additional jet multiplicity in the resolved topology, compared with different Monte Carlo predictions. The bands represent the statistical and total uncertainty in the data.  Data points are placed
at the centre of each bin. The lower panel shows the ratios of the simulations to data.}}
\label{fig:results:rel:particle:resolved:1D:additional_variables}
\end{figure*}

The normalised single-differential cross-sections are measured as a function of the transverse momentum and absolute value of the rapidity of the hadronically decaying top quark, as well as of the mass and transverse momentum of the \ttb{} system and of the additional variables \absPoutthad{}, \dPhittbar{}, \Htt{} and jet multiplicity.  Moreover, the differential cross-section as a function of the \pt{} of the top quark is measured separately for the leading and subleading top quark. The results are shown in Figures~\ref{fig:results:rel:particle:resolved:1D:top_had}--\ref{fig:results:rel:particle:resolved:1D:additional_variables}. The quantitative comparisons among the particle-level results and predictions, obtained with a $\chi^2$ test statistic, are shown in Tables~\ref{tab:chisquare:relative:1D:allpred:resolved:particle} and~\ref{tab:chisquare:absolute:1D:allpred:resolved:particle}, for normalised and absolute  single-differential cross-sections, respectively.
 
\begin{figure*}[t]
\centering
\subfigure[]{\includegraphics[width=0.38\textwidth]{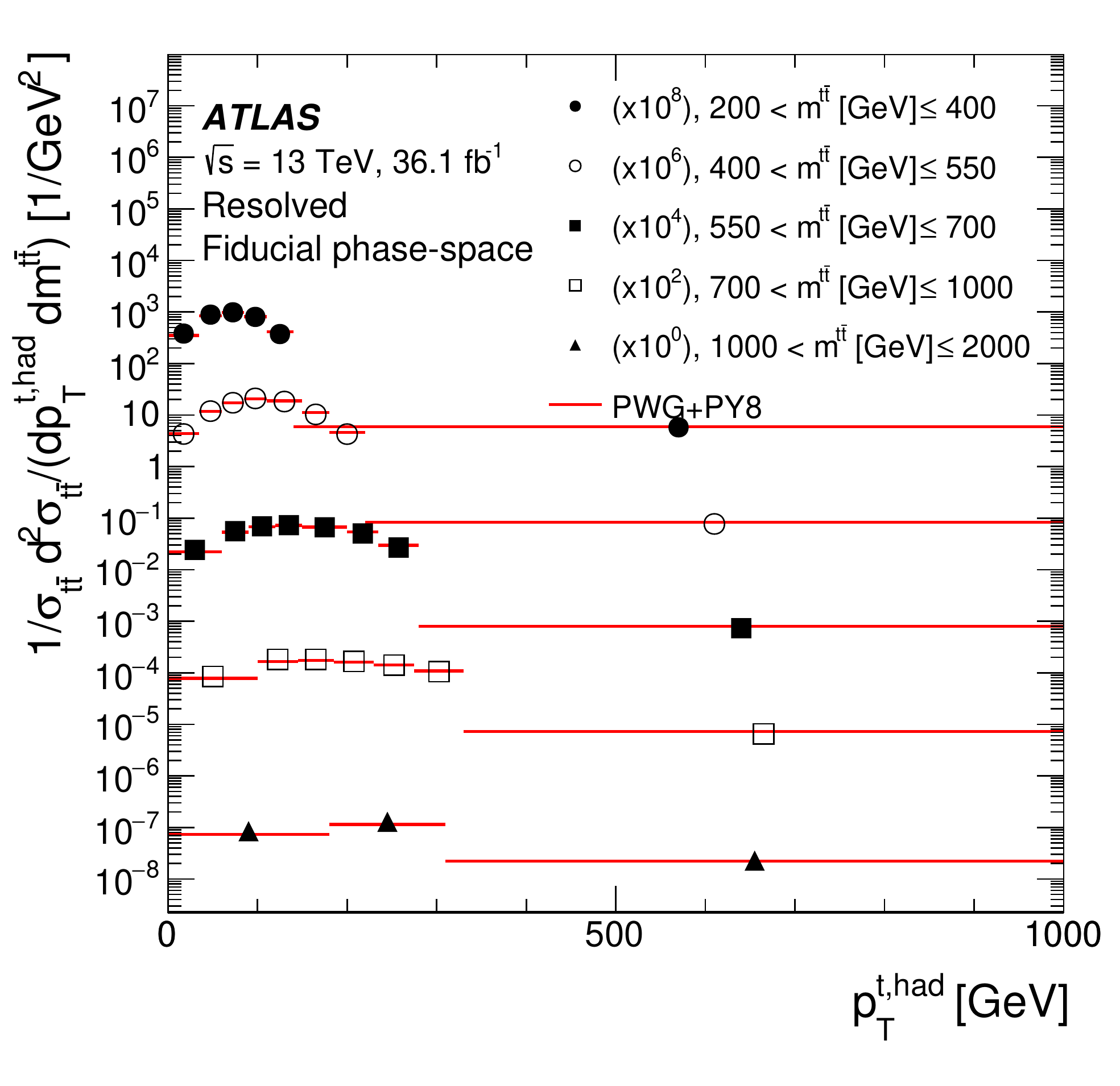}\label{fig:results:particle:resolved:top_had_pt:ttbar_m:rel}}
\subfigure[]{\includegraphics[width=0.58\textwidth]{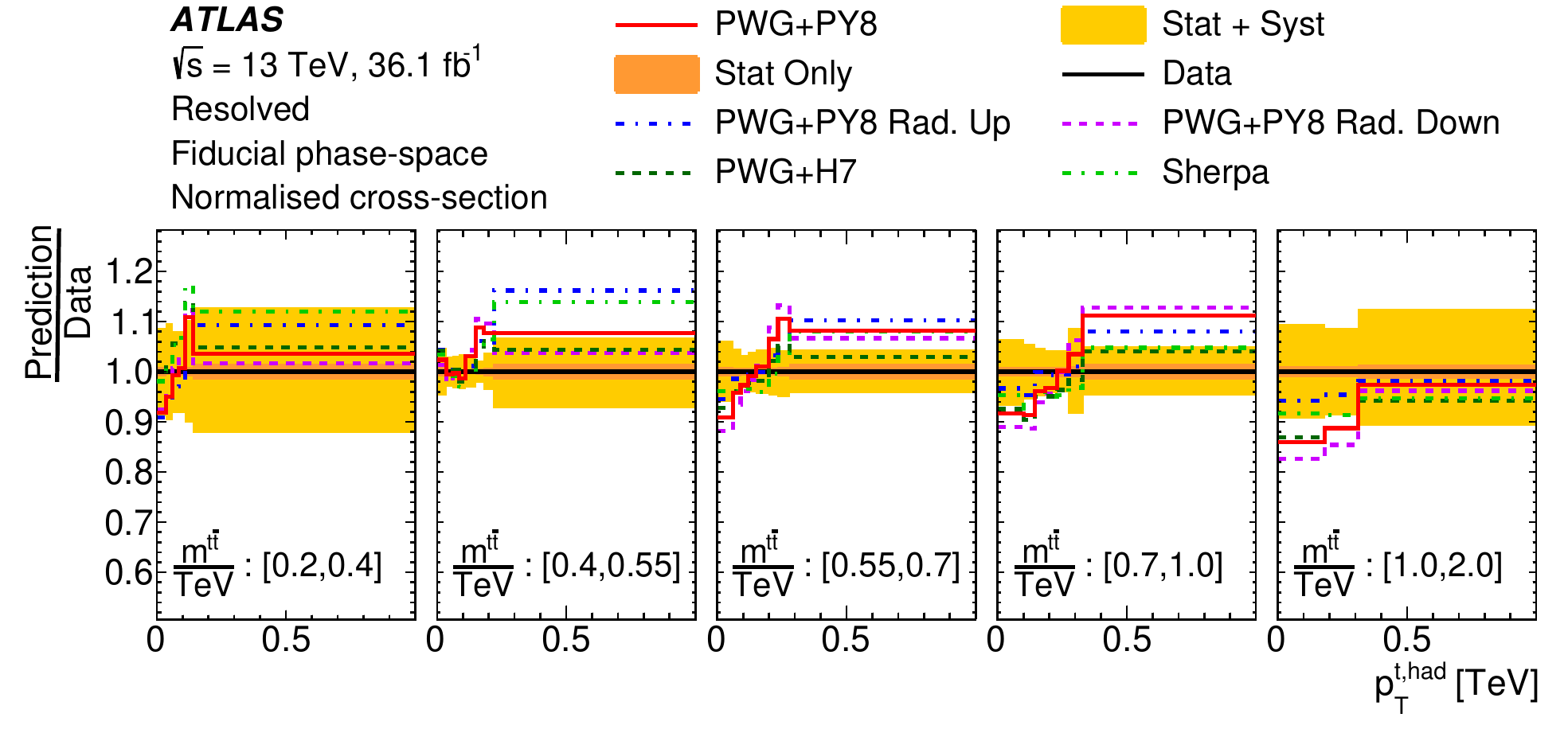}\label{fig:results:particle:resolved:top_had_pt:ttbar_m:rel:ratio}}
\caption{\small{\subref{fig:results:particle:resolved:top_had_pt:ttbar_m:rel} Particle-level normalised differential cross-section as a function of \ptth{} in bins of \mtt{} in the resolved topology compared with the prediction obtained with the \Powheg+\PythiaEight{} MC generator.  Data points are placed at the centre of each bin. \subref{fig:results:particle:resolved:top_had_pt:ttbar_m:rel:ratio} The ratio of the measured cross-section to  different Monte Carlo predictions.  The bands represent the statistical and total uncertainty in the data.}}
\label{fig:results:rel:particle:resolved:2D:ttbar_m:top_had_pt}
\end{figure*}

\begin{figure*}[t]
\centering
\subfigure[]{\includegraphics[width=0.38\textwidth]{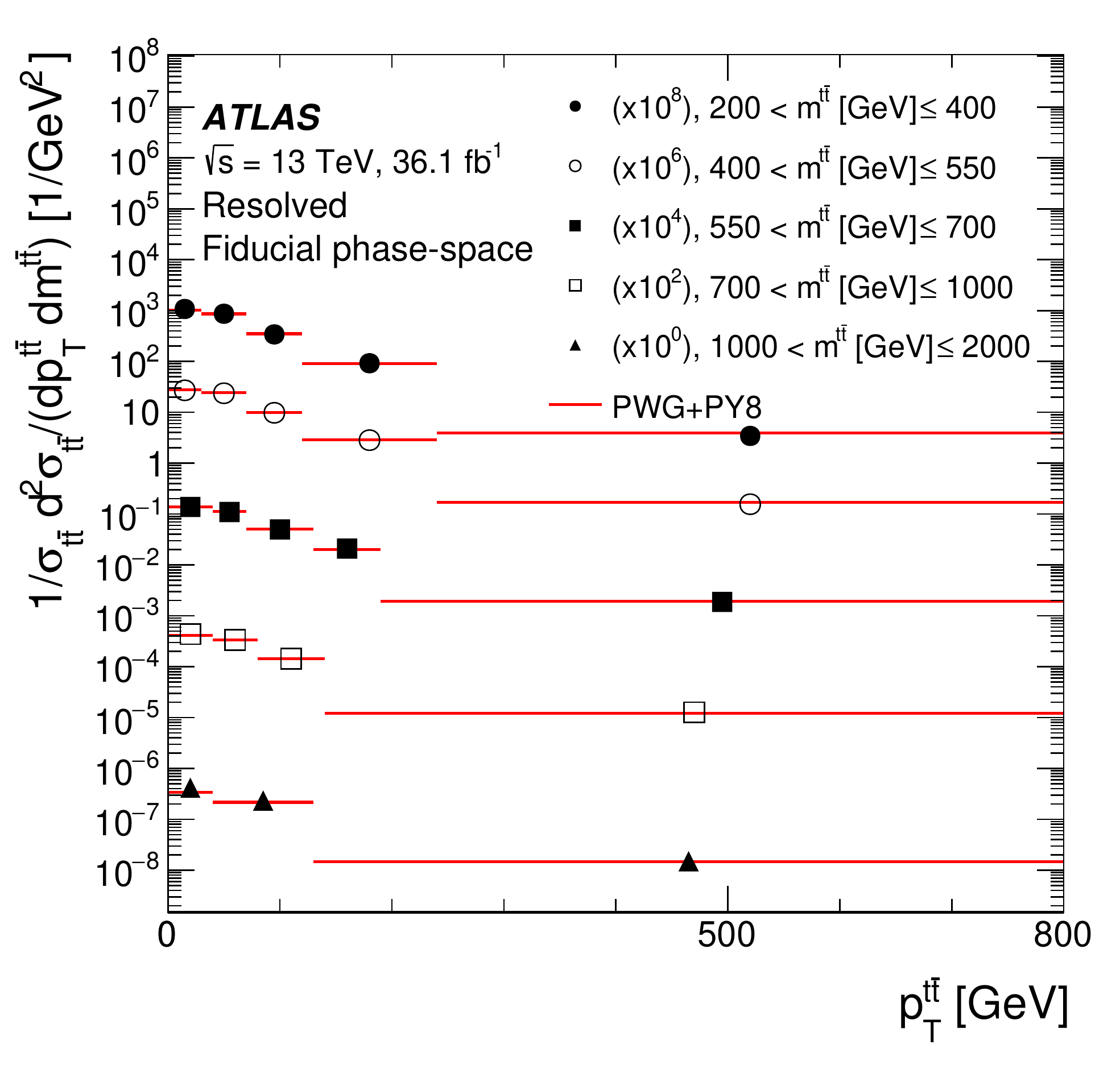}\label{fig:results:particle:resolved:ttbar_pt:ttbar_m:rel}}
\subfigure[]{\includegraphics[width=0.58\textwidth]{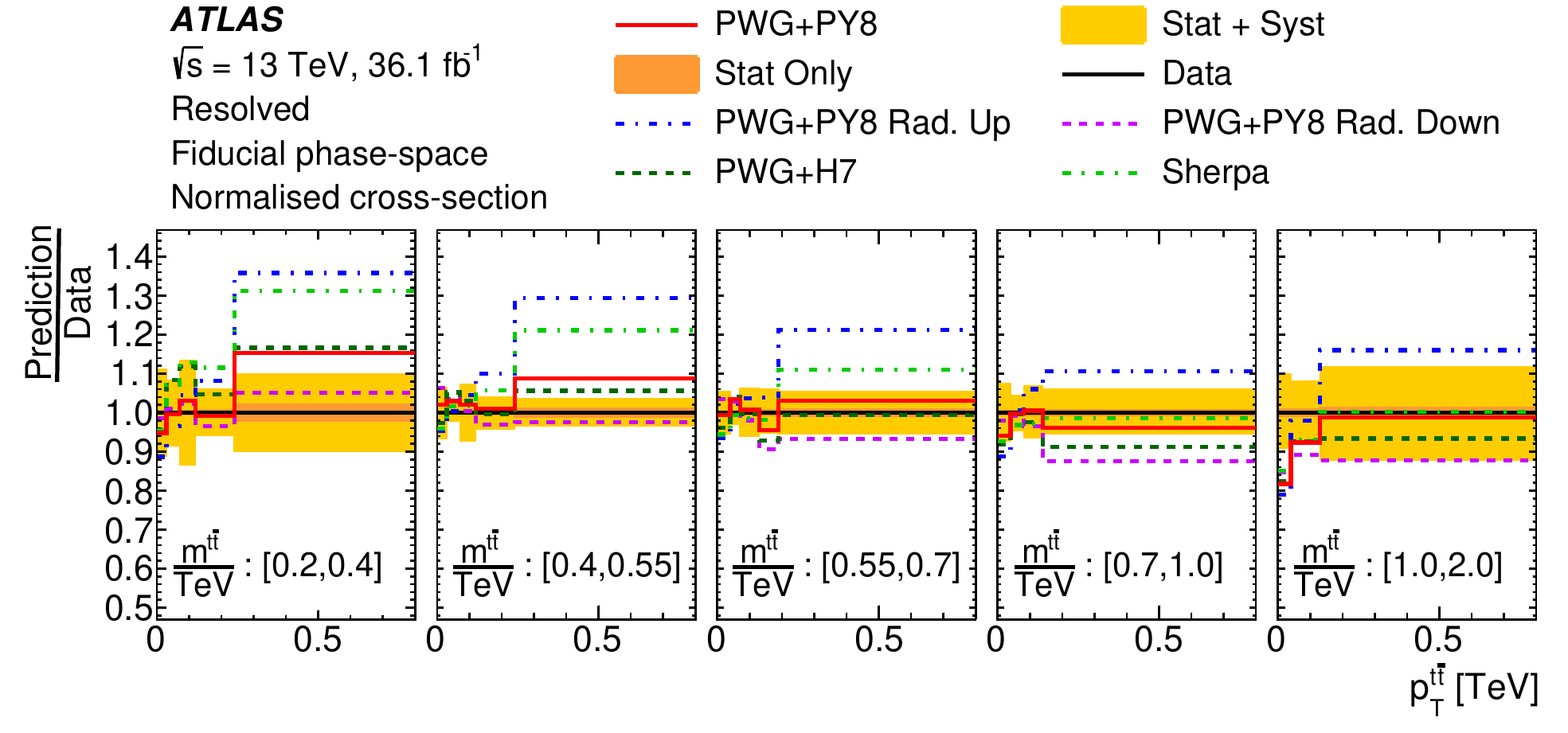}\label{fig:results:particle:resolved:ttbar_pt:ttbar_m:rel:ratio}}
\caption{\small{\subref{fig:results:particle:resolved:ttbar_pt:ttbar_m:rel} Particle-level normalised differential cross-section as a function of \pttt{} in bins of \mtt{} in the resolved topology compared with the prediction obtained with the \Powheg+\PythiaEight{} MC generator.  Data points are placed at the centre of each bin. \subref{fig:results:particle:resolved:ttbar_pt:ttbar_m:rel:ratio} The ratio of the measured cross-section to different Monte Carlo predictions.  The bands represent the statistical and total uncertainty in the data.}}
\label{fig:results:rel:particle:resolved:2D:ttbar_m:ttbar_pt}
\end{figure*}

\begin{figure*}[t]
\centering
\subfigure[]{\includegraphics[width=0.38\textwidth]{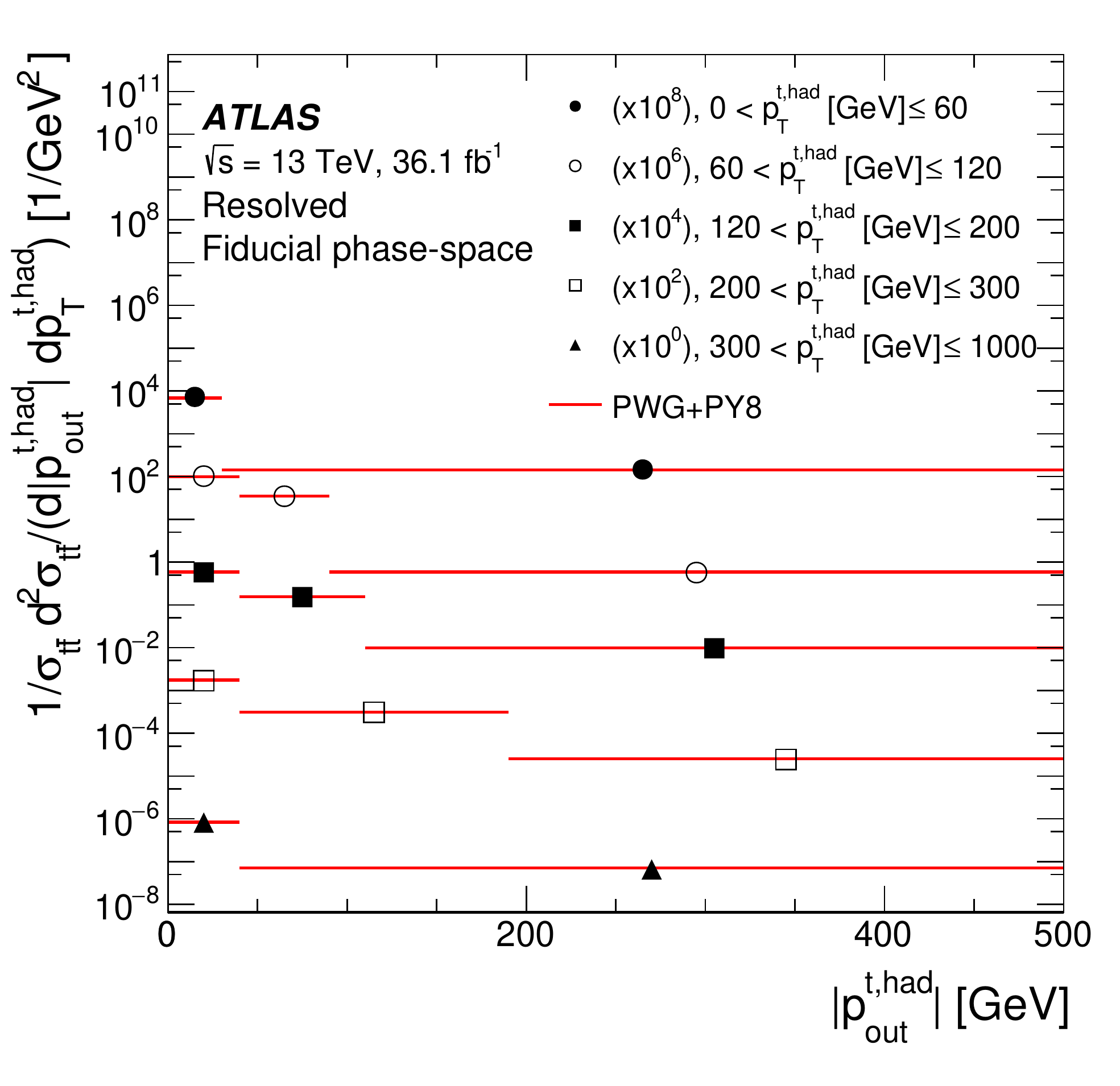}\label{fig:results:particle:resolved:absPout:top_had_pt:rel}}
\subfigure[]{\includegraphics[width=0.58\textwidth]{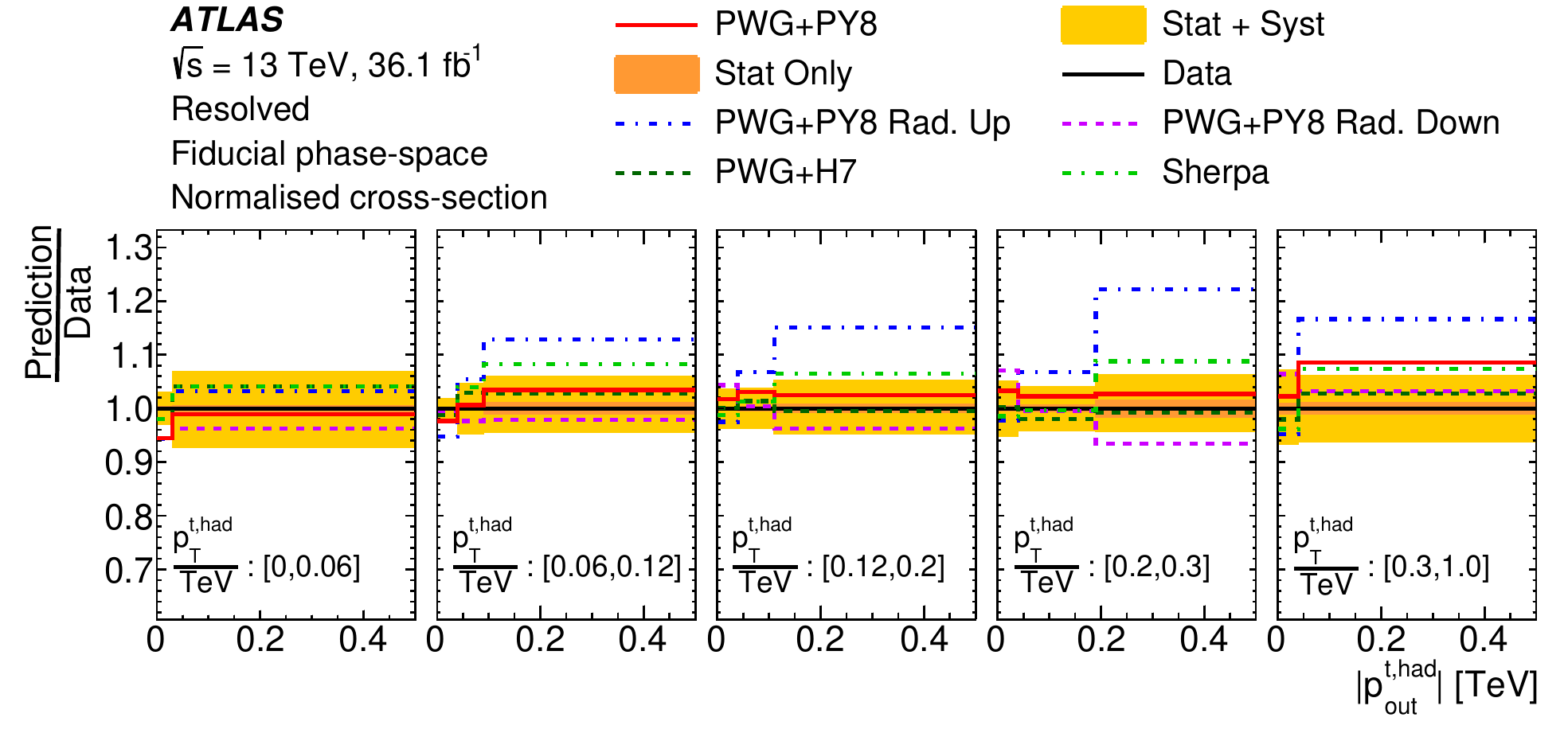}\label{fig:results:particle:resolved:absPout:top_had_pt:rel:ratio}}
\caption{\small{\subref{fig:results:particle:resolved:absPout:top_had_pt:rel} Particle-level normalised differential cross-section as a function of \absPoutthad{} in bins of \ptth{} in the resolved topology compared with the prediction obtained with the \Powheg+\PythiaEight{} MC generator.  Data points are placed at the centre of each bin. \subref{fig:results:particle:resolved:absPout:top_had_pt:rel:ratio} The ratio of the measured cross-section to  different Monte Carlo predictions.  The bands represent the statistical and total uncertainty in the data.}}
\label{fig:results:rel:particle:resolved:2D:top_had_pt:absPout}
\end{figure*}

\begin{figure*}[t]
\centering
\subfigure[]{\includegraphics[width=0.38\textwidth]{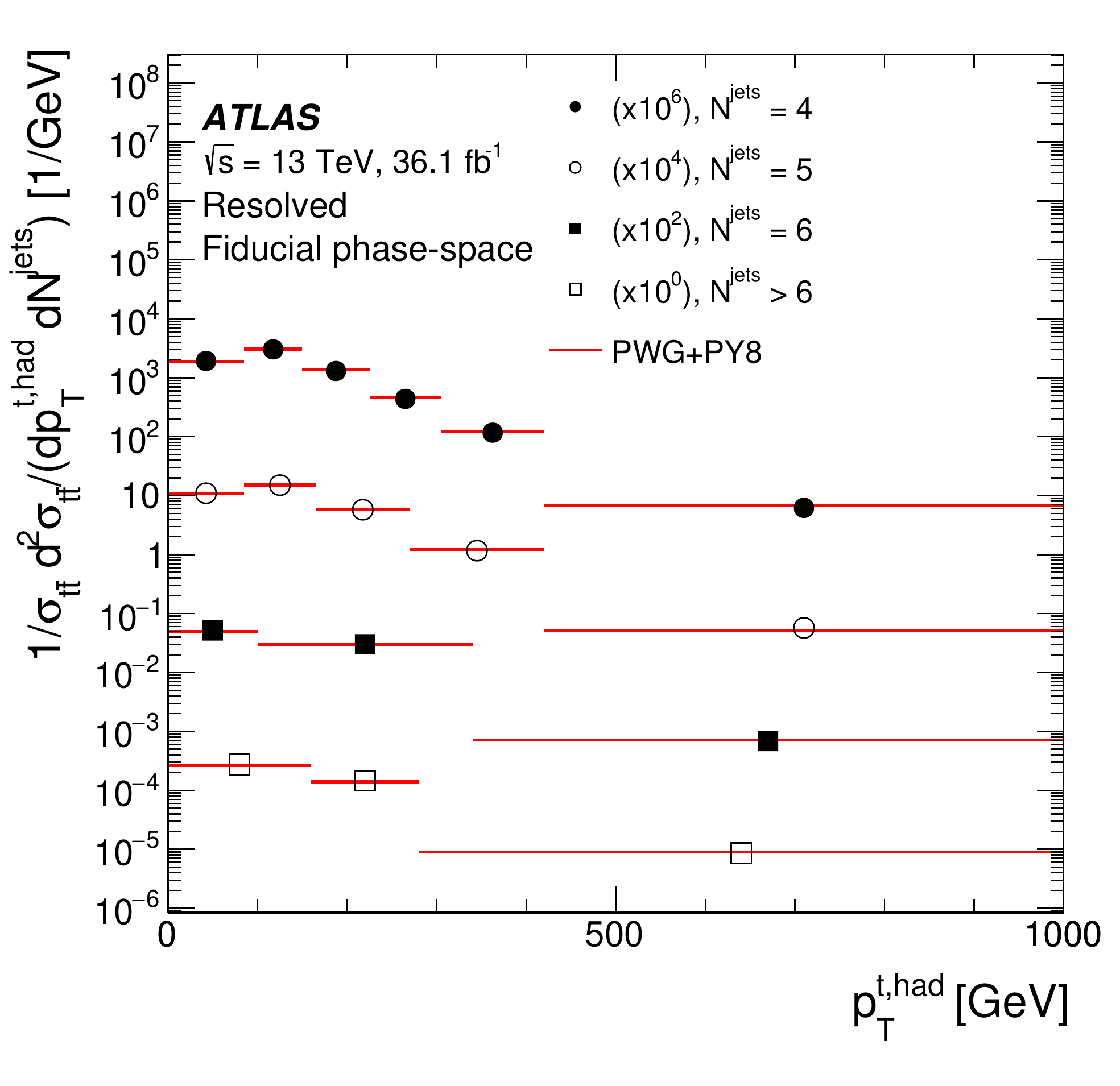}\label{fig:results:particle:resolved:top_had_pt:jet_n:rel}}
\subfigure[]{\includegraphics[width=0.58\textwidth]{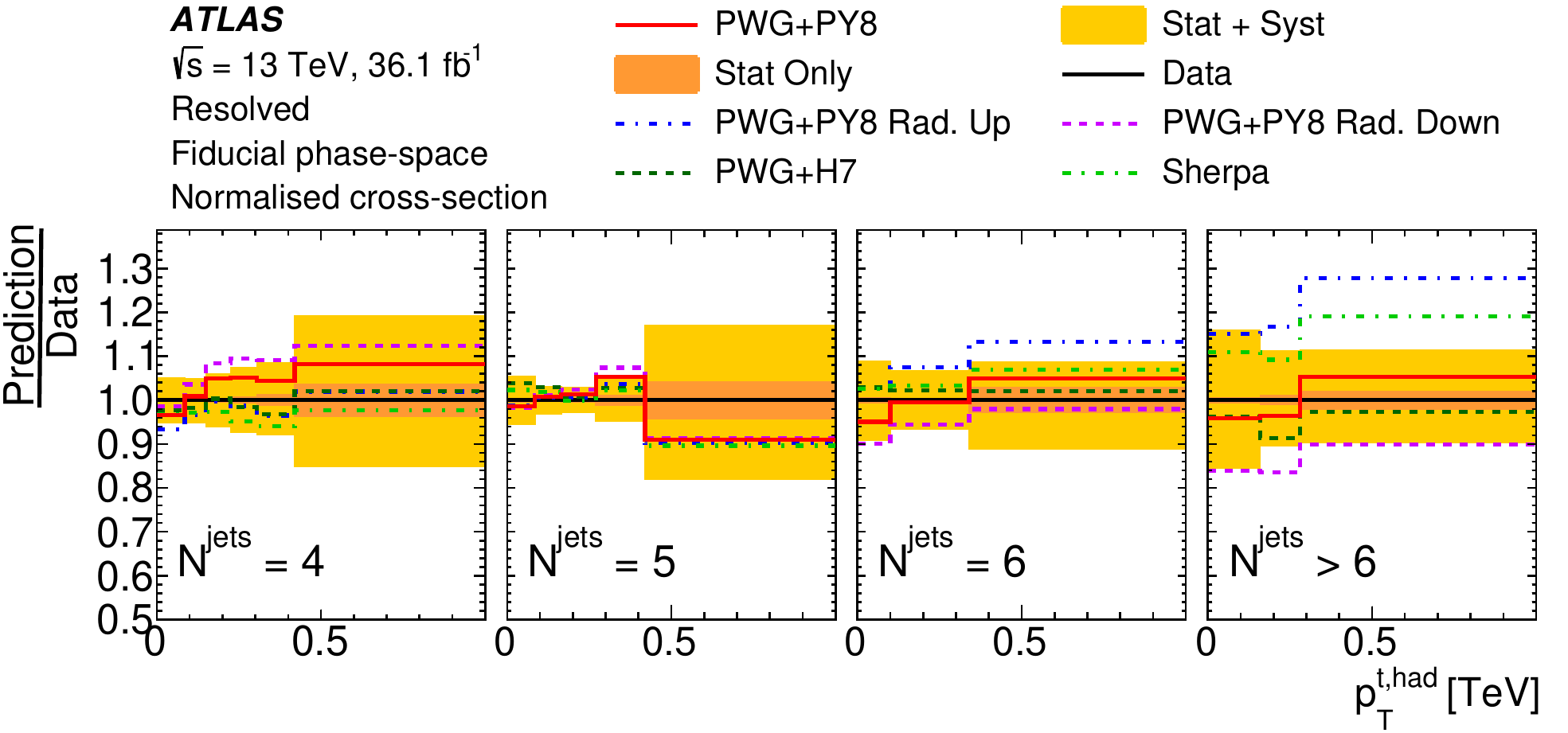}\label{fig:results:particle:resolved:top_had_pt:jet_n:rel:ratio}}
\caption{\small{\subref{fig:results:particle:resolved:top_had_pt:jet_n:rel} Particle-level normalised differential cross-section as a function of \ptth{} in bins of the jet multiplicity in the resolved topology compared with the prediction obtained with the \Powheg+\PythiaEight{} MC generator.  Data points are placed at the centre of each bin. \subref{fig:results:particle:resolved:top_had_pt:jet_n:rel:ratio} The ratio of the measured cross-section to  different Monte Carlo predictions.  The bands represent the statistical and total uncertainty in the data.}}
\label{fig:results:rel:particle:resolved:2D:jet_n:top_had_pt}
\end{figure*}

\begin{figure*}[t]
\centering
\subfigure[]{\includegraphics[width=0.38\textwidth]{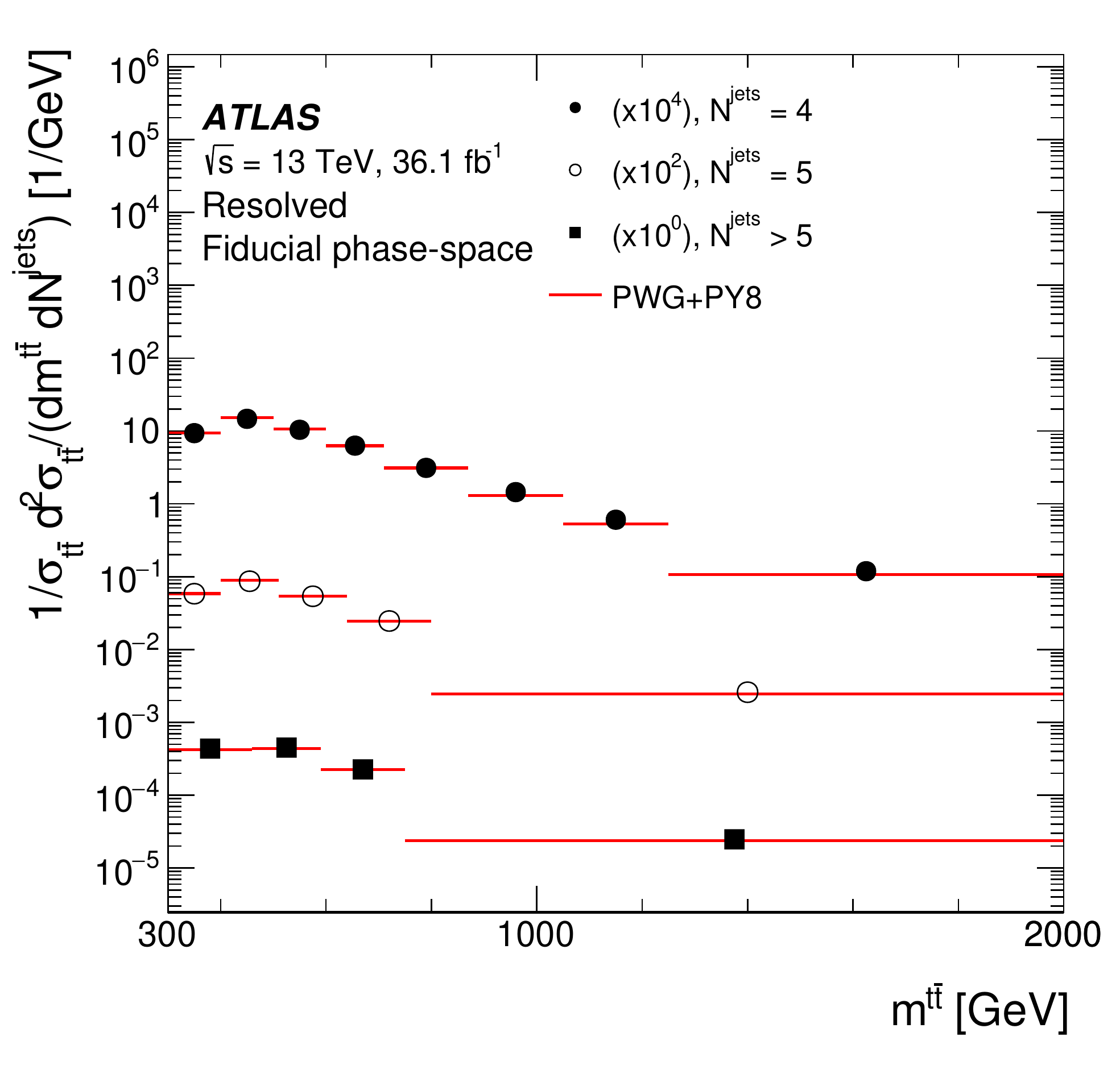}\label{fig:results:particle:resolved:ttbar_m:jet_n:rel}}
\subfigure[]{\includegraphics[width=0.58\textwidth]{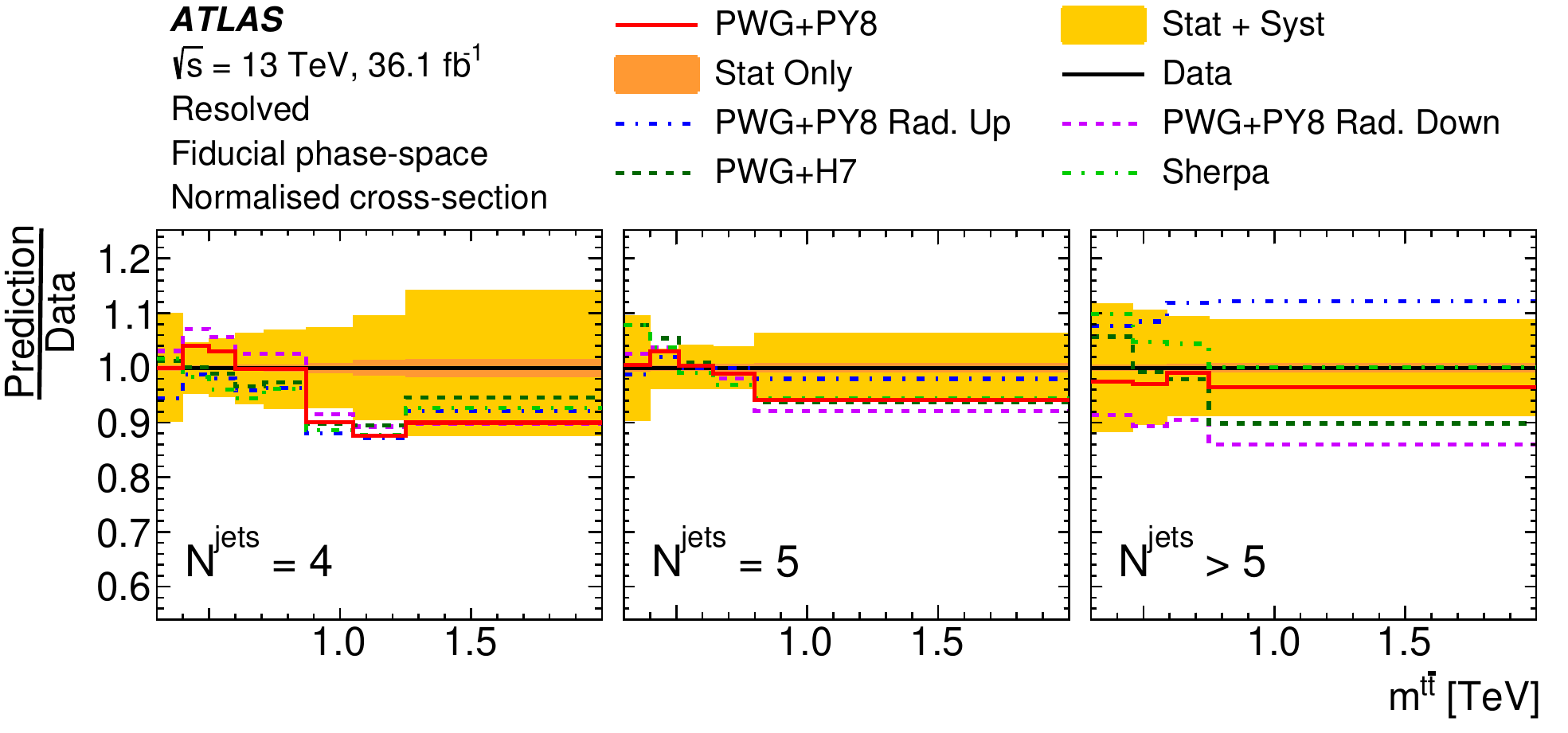}\label{fig:results:particle:resolved:ttbar_m:jet_n:rel:ratio}}
\caption{\small{\subref{fig:results:particle:resolved:ttbar_m:jet_n:rel} Particle-level normalised differential cross-section as a function of \mtt{} in bins of the jet multiplicity in the resolved topology compared with the prediction obtained with the \Powheg+\PythiaEight{} MC generator.  Data points are placed at the centrer of each bin. \subref{fig:results:particle:resolved:ttbar_m:jet_n:rel:ratio} The ratio of the measured cross-section to  different Monte Carlo predictions.  The bands represent the statistical and total uncertainty in the data.}}
\label{fig:results:rel:particle:resolved:2D:jet_n:ttbar_m}
\end{figure*}

\begin{figure*}[t]
\centering
\subfigure[]{\includegraphics[width=0.38\textwidth]{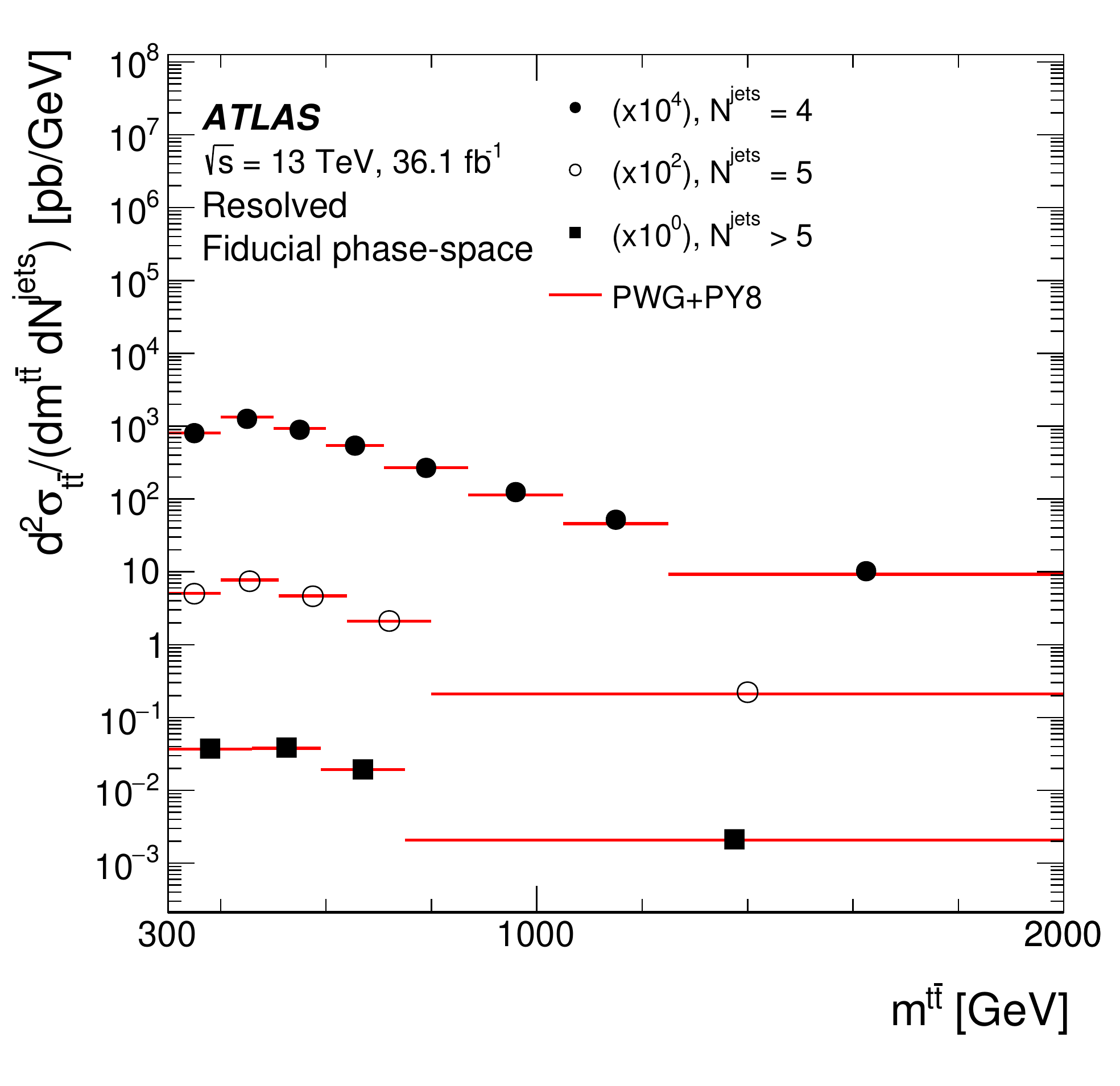}\label{fig:results:particle:resolved:ttbar_m:jet_n:abs}}
\subfigure[]{\includegraphics[width=0.58\textwidth]{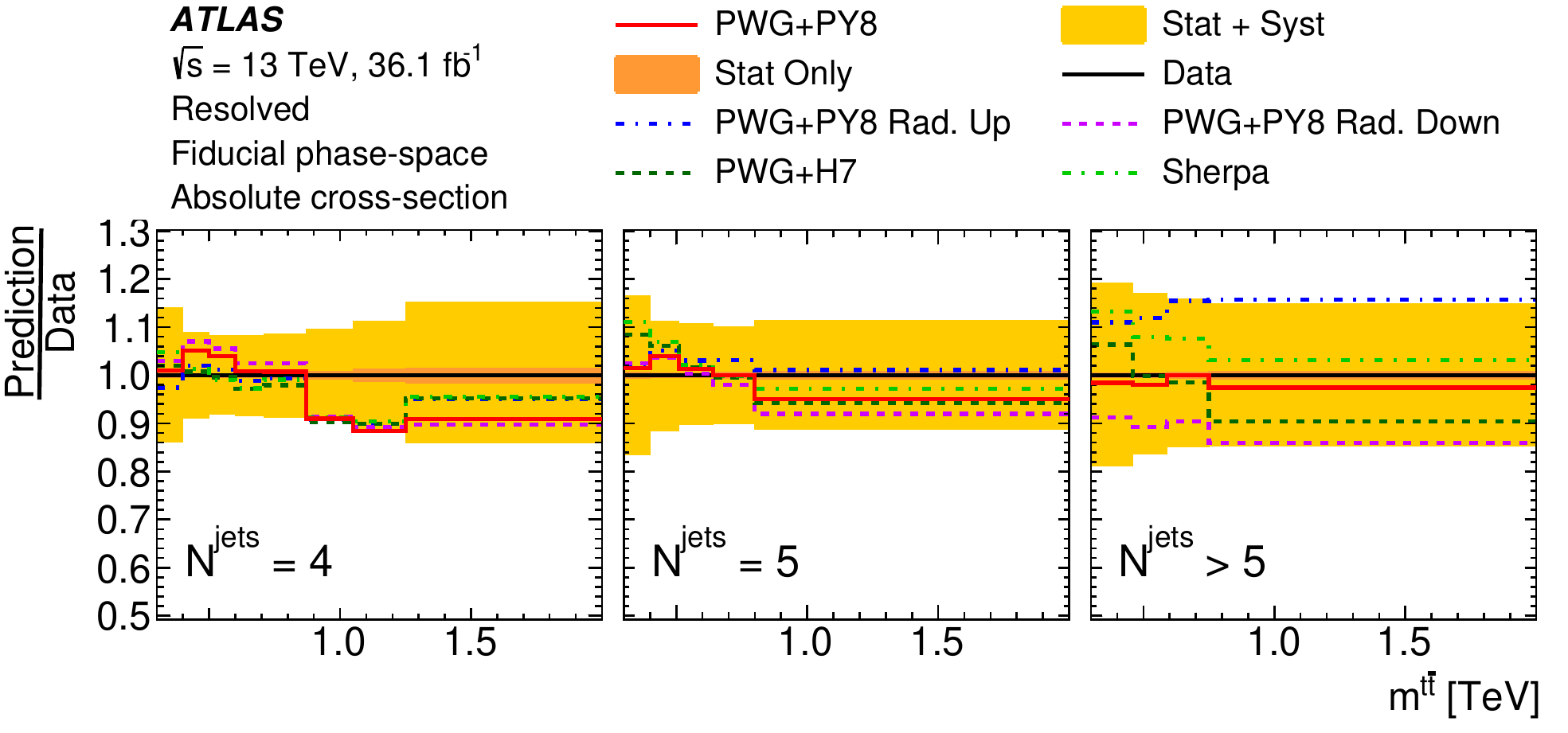}\label{fig:results:particle:resolved:ttbar_m:jet_n:abs:ratio}}
\caption{\small{\subref{fig:results:particle:resolved:ttbar_m:jet_n:abs} Particle-level absolute differential cross-section as a function of \mtt{} in bins of the jet multiplicity in the resolved topology compared with the prediction obtained with the \Powheg+\PythiaEight{} MC generator.  Data points are placed at the centre of each bin. \subref{fig:results:particle:resolved:ttbar_m:jet_n:rel:ratio} The ratio of the measured cross-section to  different Monte Carlo predictions.  The bands represent the statistical and total uncertainty in the data.}}
\label{fig:results:abs:particle:resolved:2D:jet_n:ttbar_m}
\end{figure*}

\begin{figure*}[t]
\centering
\subfigure[]{\includegraphics[width=0.38\textwidth]{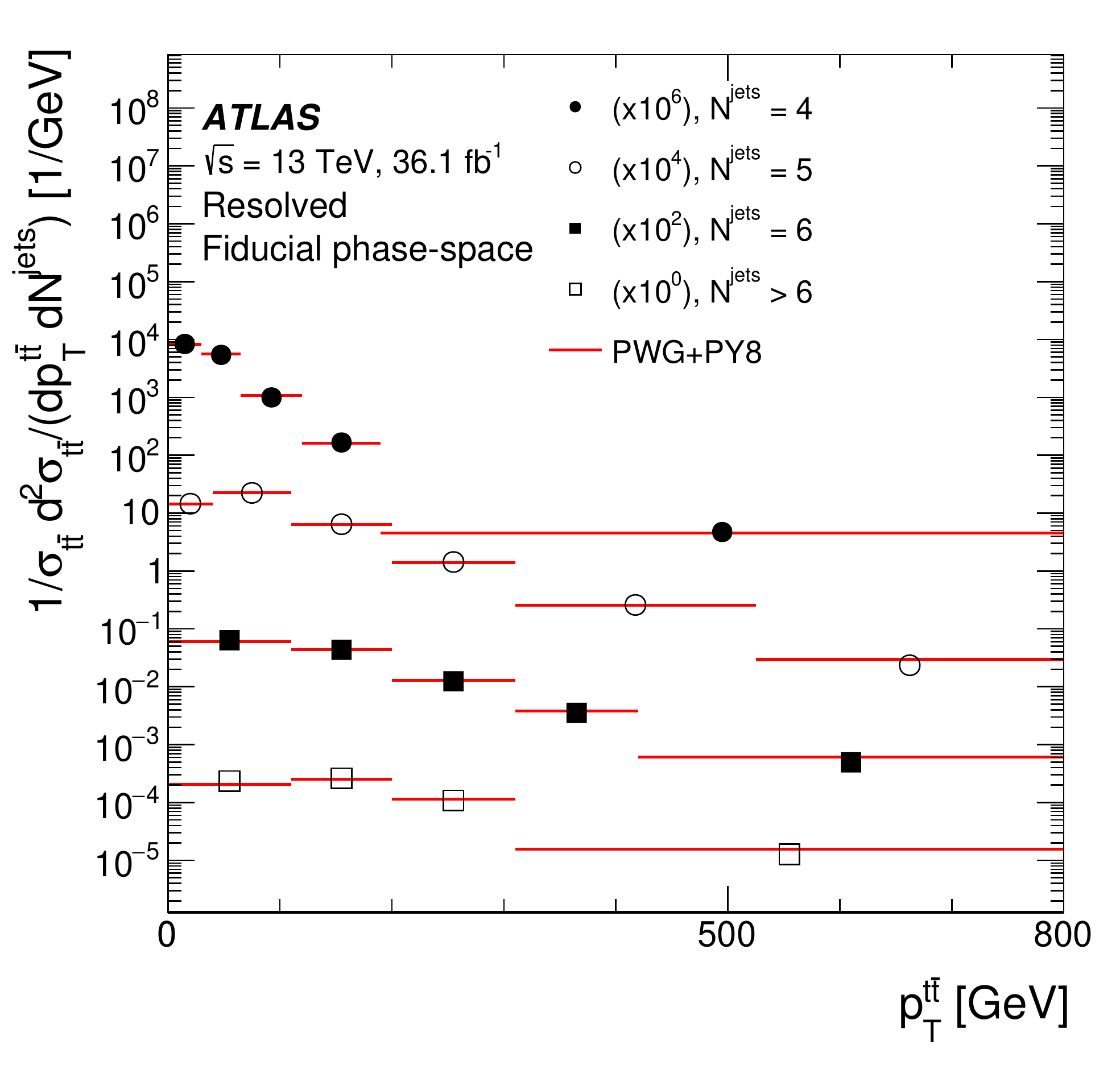}\label{fig:results:particle:resolved:ttbar_pt:jet_n:rel}}
\subfigure[]{\includegraphics[width=0.58\textwidth]{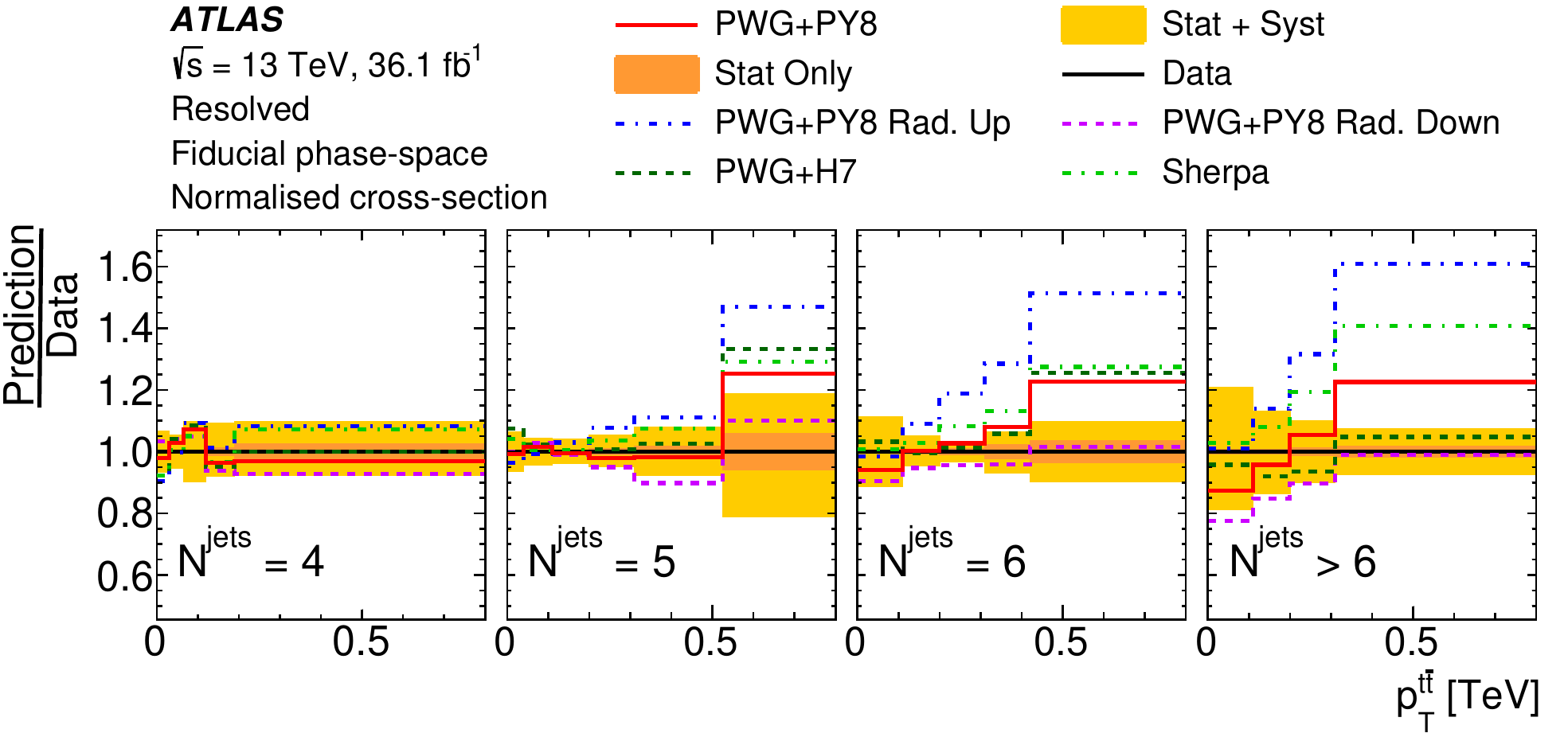}\label{fig:results:particle:resolved:ttbar_pt:jet_n:rel:ratio}}
\caption{\small{\subref{fig:results:particle:resolved:ttbar_pt:jet_n:rel} Particle-level normalised differential cross-section as a function of \pttt{} in bins of the jet multiplicity in the resolved topology compared with the prediction obtained with the \Powheg+\PythiaEight{} MC generator.  Data points are placed at the centre of each bin. \subref{fig:results:particle:resolved:ttbar_pt:jet_n:rel:ratio} The ratio of the measured cross-section to  different Monte Carlo predictions.  The bands represent the statistical and total uncertainty in the data.}}
\label{fig:results:rel:particle:resolved:2D:jet_n:ttbar_pt}
\end{figure*}

\begin{figure*}[t]
\centering
\subfigure[]{\includegraphics[width=0.38\textwidth]{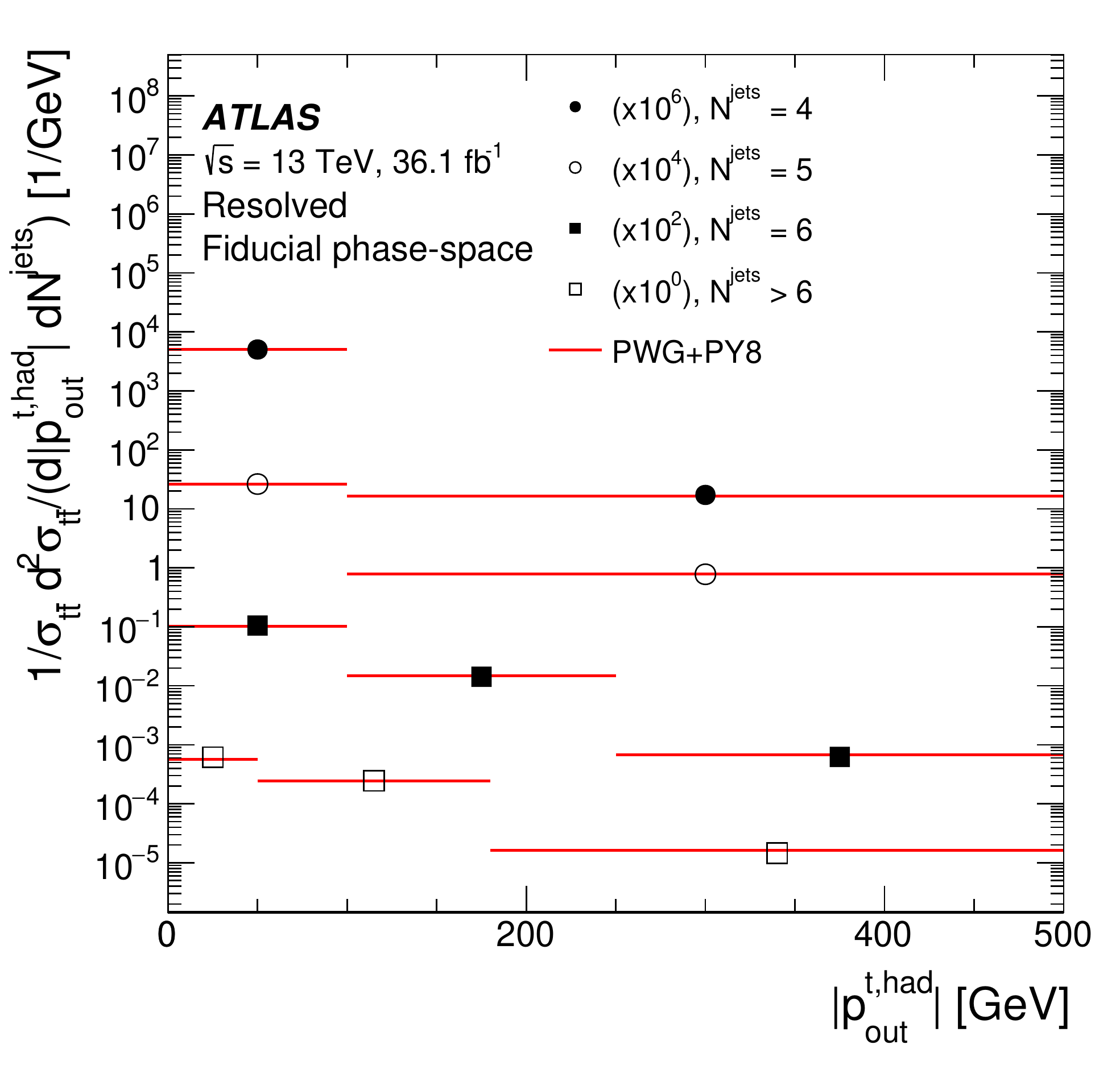}\label{fig:results:particle:resolved:absPout:jet_n:rel}}
\subfigure[]{\includegraphics[width=0.58\textwidth]{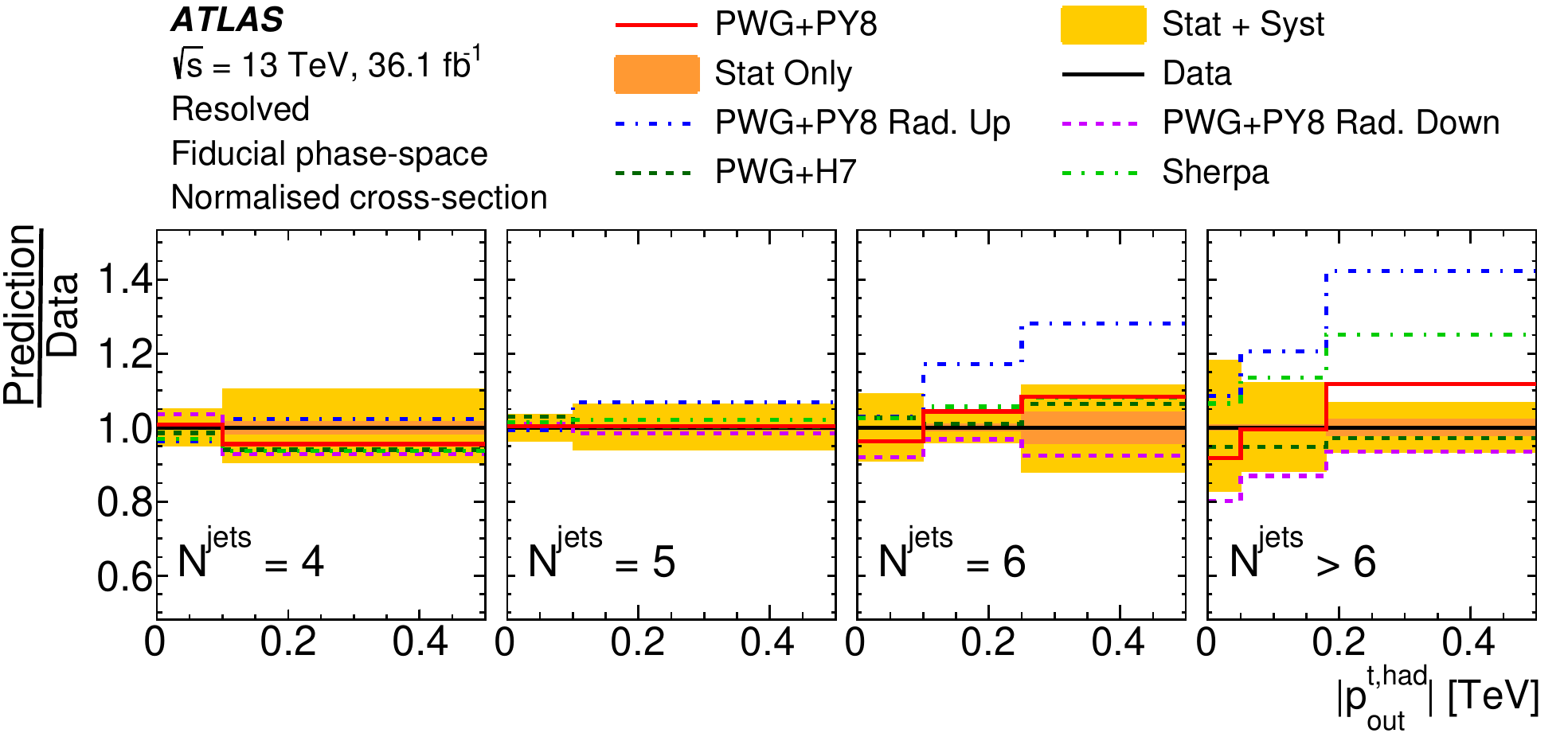}\label{fig:results:particle:resolved:absPout:jet_n:rel:ratio}}
\caption{\small{\subref{fig:results:particle:resolved:absPout:jet_n:rel} Particle-level normalised differential cross-section as a function of \absPoutthad{} in bins of the jet multiplicity in the resolved topology compared with the prediction obtained with the \Powheg+\PythiaEight{} MC generator.  Data points are placed at the centre of each bin. \subref{fig:results:particle:resolved:absPout:jet_n:rel:ratio} The ratio of the measured cross-section to  different Monte Carlo predictions.  The bands represent the statistical and total uncertainty in the data.}}
\label{fig:results:rel:particle:resolved:2D:jet_n:absPout}
\end{figure*}

\begin{figure*}[t]
\centering
\subfigure[]{\includegraphics[width=0.38\textwidth]{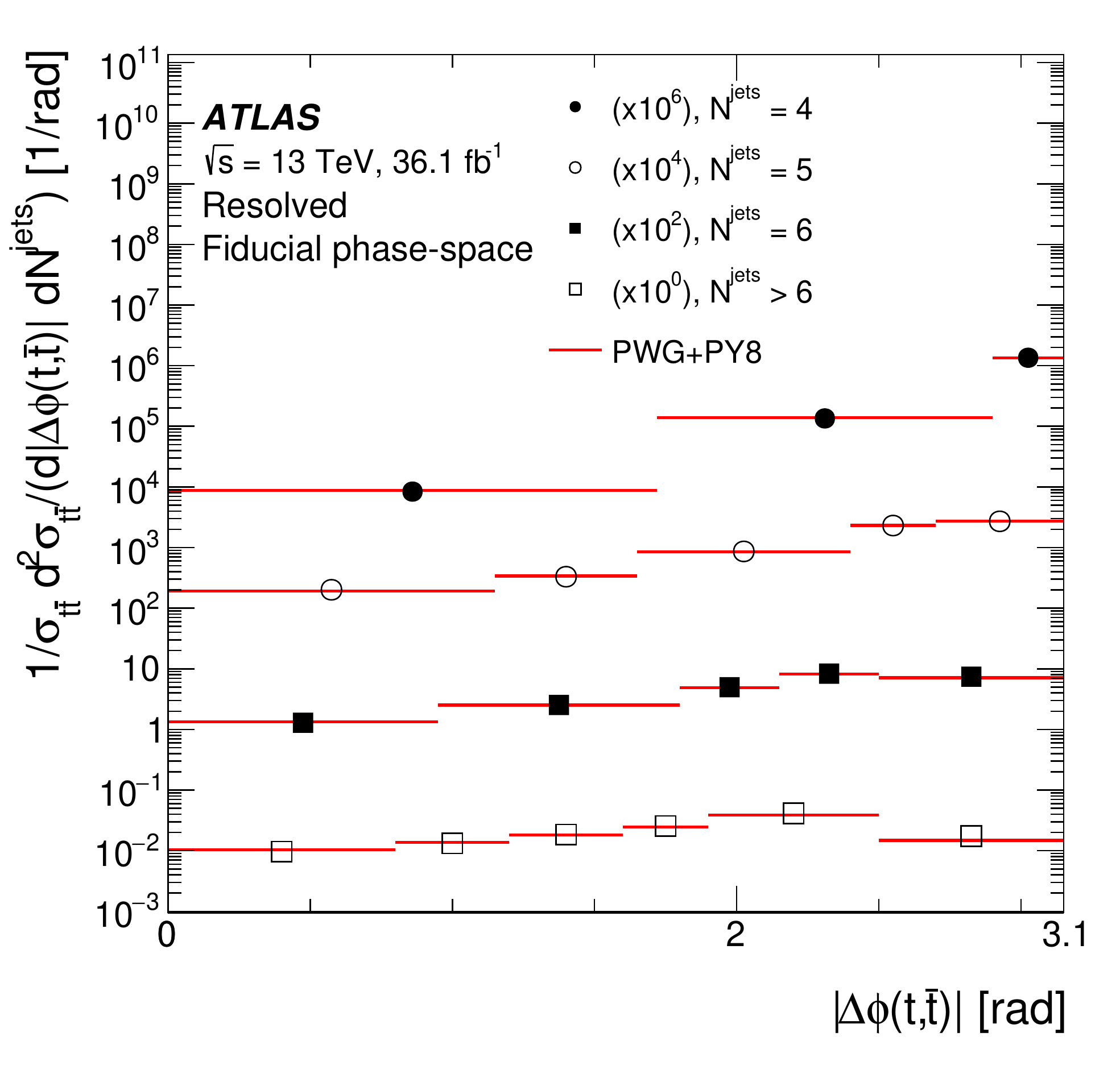}\label{fig:results:particle:resolved:deltaPhi_tt:jet_n:rel}}
\subfigure[]{\includegraphics[width=0.58\textwidth]{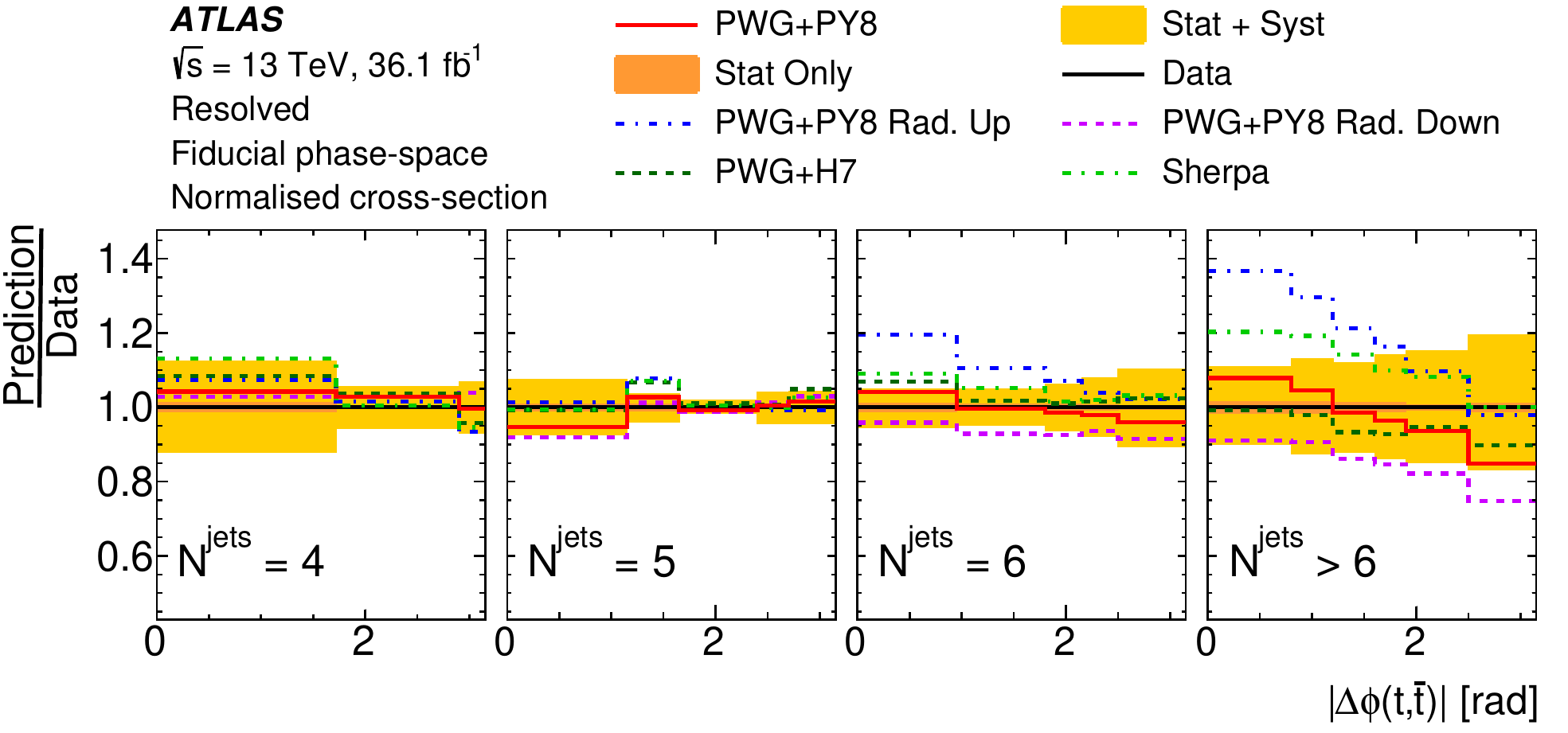}\label{fig:results:particle:resolved:deltaPhi_tt:jet_n:rel:ratio}}
\caption{\small{\subref{fig:results:particle:resolved:deltaPhi_tt:jet_n:rel} Particle-level normalised differential cross-section as a function of \deltaPhittbar{} in bins of the jet multiplicity in the resolved topology compared with the prediction obtained with the \Powheg+\PythiaEight{} MC generator.  Data points are placed at the centre of each bin. \subref{fig:results:particle:resolved:deltaPhi_tt:jet_n:rel:ratio} The ratio of the measured cross-section to  different Monte Carlo predictions.  The bands represent the statistical and total uncertainty in the data.}}
\label{fig:results:rel:particle:resolved:2D:jet_n:deltaPhi_tt}
\end{figure*}

\begin{figure*}[t]
\centering
\subfigure[]{\includegraphics[width=0.38\textwidth]{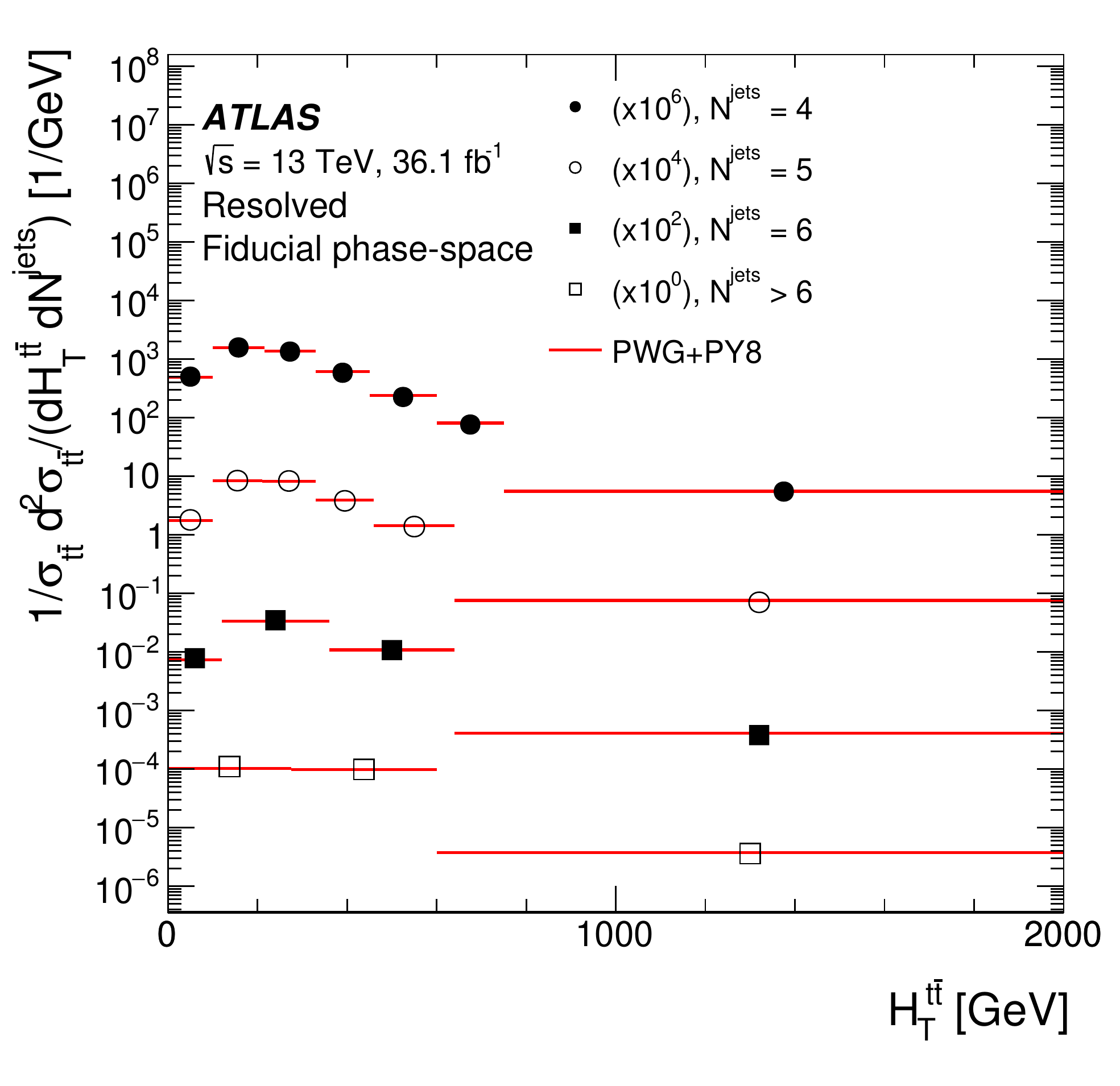}\label{fig:results:particle:resolved:HT_tt:jet_n:rel}}
\subfigure[]{\includegraphics[width=0.58\textwidth]{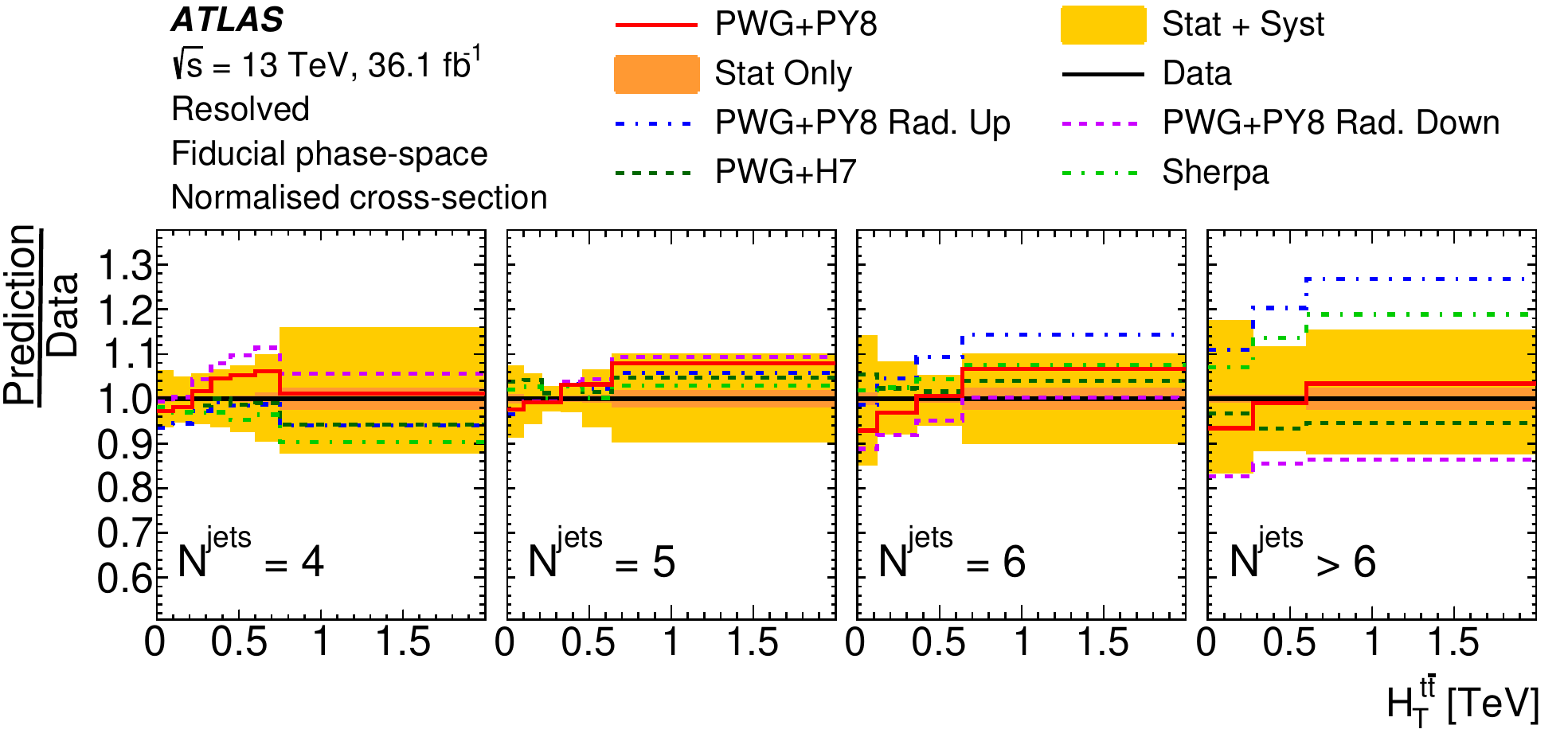}\label{fig:results:particle:resolved:HT_tt:jet_n:rel:ratio}}
\caption{\small{\subref{fig:results:particle:resolved:HT_tt:jet_n:rel} Particle-level normalised differential cross-section as a function of \Httbar{} in bins of the jet multiplicity in the resolved topology compared with the prediction obtained with the \Powheg+\PythiaEight{} MC generator.  Data points are placed at the centre of each bin. \subref{fig:results:particle:resolved:HT_tt:jet_n:rel:ratio} The ratio of the measured cross-section to  different Monte Carlo predictions.  The bands represent the statistical and total uncertainty in the data.}}
\label{fig:results:rel:particle:resolved:2D:jet_n:HT_tt}
\end{figure*}

The normalised double-differential cross-sections, presented in Figures~\ref{fig:results:rel:particle:resolved:2D:ttbar_m:top_had_pt}--\ref{fig:results:rel:particle:resolved:2D:jet_n:HT_tt}, are measured as a function of the \pt{} of the hadronically decaying top quark and
of the \ttb{} system in bins of the mass the \ttbar{} system, as a function of \absPoutthad{} in bins of the \pt{} of the hadronically decaying top quark  and finally as a function of \ptth{}, \mtt{}, \pttt{}, \absPoutthad{}, \dPhittbar{} and \Htt{} in bins of jet multiplicity.
The quantitative comparisons among the particle-level results and predictions, obtained with a $\chi^2$ test statistic, are shown in Tables~\ref{tab:chisquare:relative:2D:allpred:resolved:particle} and~\ref{tab:chisquare:absolute:2D:allpred:resolved:particle}, for normalised and absolute  double-differential cross-sections, respectively.   An example of an absolute differential cross-section, as a function of \mtt{} in bins of jet multiplicity, is given in Figure~\ref{fig:results:abs:particle:resolved:2D:jet_n:ttbar_m}. In this case, the total uncertainty is larger than the uncertainty in the corresponding normalised differential cross-section, as shown Figure~\ref{fig:results:rel:particle:resolved:2D:jet_n:ttbar_m}.

\begin{figure*}[t]
\centering
 
\includegraphics[width=0.58\textwidth]{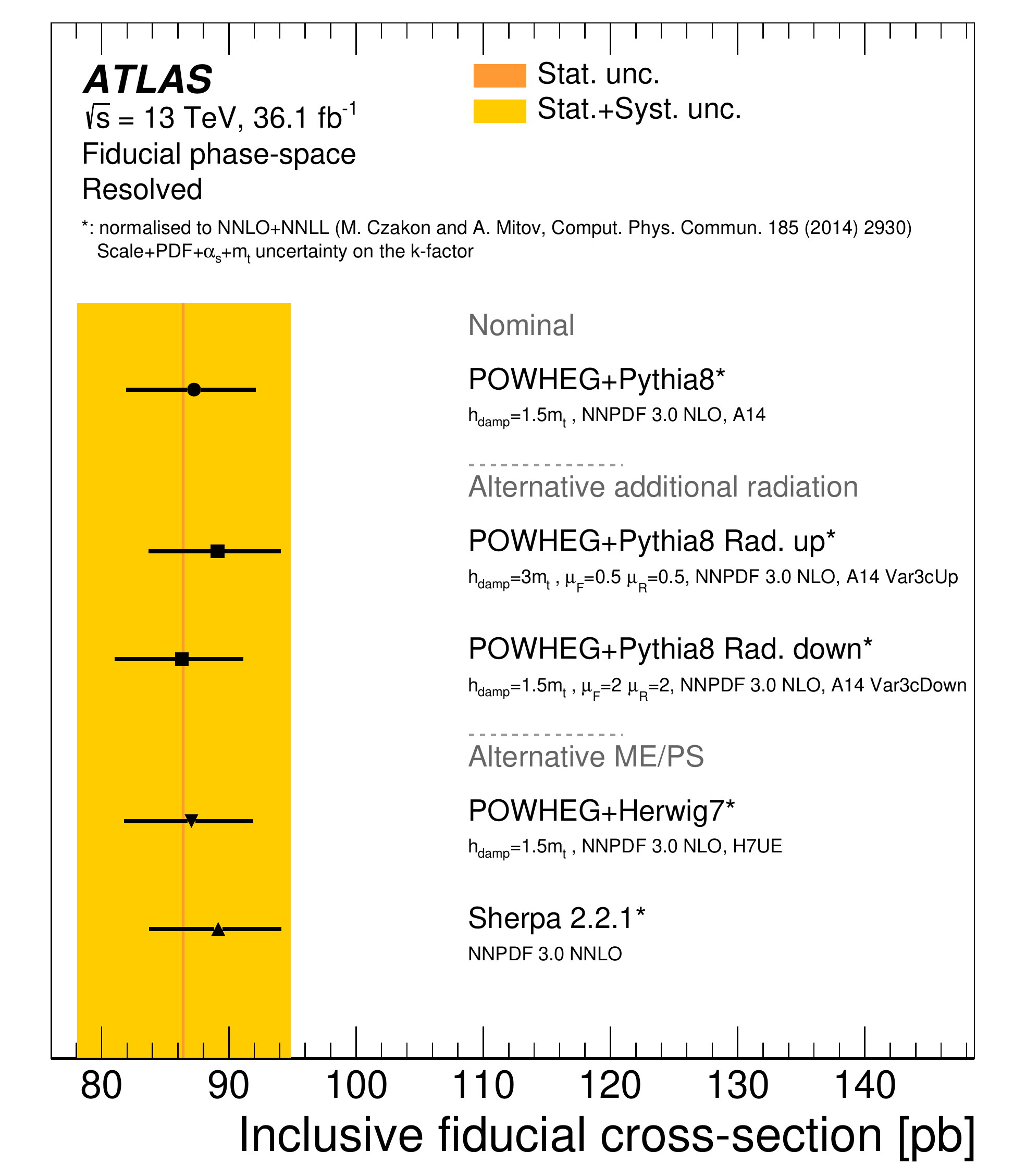}
\caption{Comparison of the measured inclusive fiducial cross-section in the resolved topology with the predictions from several MC generators. The bands represent the statistical and total uncertainty in the data. The uncertainty on the cross-section predicted by each NLO MC generator only includes the uncertainty (due to PDFs, $m_{t}$ and $\alpha_{s}$)  affecting the $k$-factor used in the normalisation.}
\label{fig:results_particle:resolved:totalXs}
 
\end{figure*}

Additionally,  the total cross-section is measured in the fiducial phase-space of the resolved topology and is compared with the MC predictions previously described, as shown in Figure~\ref{fig:results_particle:resolved:totalXs}. The total cross-section predicted by each NLO MC generator is normalised to the NNLO+NNLL prediction as quoted in Ref.~\cite{Czakon:2011xx} and the corresponding uncertainty only includes the uncertainty affecting the $k$-factor used  in the normalisation. The differences between the quoted fiducial cross-sections hence result from different acceptance predictions from each model.

All the measured differential cross-sections are compared with the MC predictions. Overall, these MC predictions give a good description of the measured single-differential cross-sections. Poorer agreement is observed in specific regions of the probed
phase-space. In Figures \ref{fig:results_particle:resolved:ttbar_pt:rel} and \ref{fig:results_particle:resolved:absPout:rel}, showing the differential cross-sections as a function of \pttt{} and  $|\Poutthad|$, the predictions overestimate the data in the high \pttt{} region, with the exception of \Powheg+\PythiaEight{} prediction with the Var3cDown tuning, and several generators overestimate the high $|\Poutthad|$ region.
A similar trend is observed in the double-differential cross-sections as a function of the \pt{} of the \ttbar{} system in bins of jet
multiplicity (Figure~\ref{fig:results:rel:particle:resolved:2D:jet_n:ttbar_pt}), in particular for bins of higher jet multiplicities.
The Var3cUp tuning of \Powheg+\PythiaEight{}, in combination with the increase of the \hdamp{} value to $3m_t$, is the prediction that shows the largest disagreement with the data. Overall, the NLO+PS generator that gives the better description of several double-differential distributions is \Powheg+\PythiaEight{}.
 
The  measured single- and double-differential cross-sections are often able to discriminate between the different features exhibited by the MC predictions and this sensitivity is hence relevant for the tuning of the MC generators and will contribute to improving the description of the $t\bar{t}$ final state and to reducing the systematic uncertainties related to top-quark modelling.
A relevant example is the fiducial single-differential cross-section as a function of \mtt{} and \ptth{} that is well described by all the NLO MC predictions, as shown in Figures~\ref{fig:results_particle:resolved:topHad_pt:rel} and~\ref{fig:results_particle:resolved:ttbar_m:rel}  and Table~\ref{tab:chisquare:relative:1D:allpred:resolved:particle}, while the double-differential cross-section where these two variables are combined shows strong disagreement with several predictions, as shown in Figure~\ref{fig:results:rel:particle:resolved:2D:ttbar_m:top_had_pt}.
The comparison of the NLO MC predictions with the measured double-differential cross-sections reveals, overall, poorer agreement than in the single-differential case. In particular, it is observed that no generator is able to describe any double-differential observable that includes \pttt{} as a probed variable.

\begin{table}[t]
\footnotesize
\centering\noindent\makebox[\textwidth]{
\renewcommand*{\arraystretch}{1.2}\begin{tabular}{|c | r @{/} l r  | r @{/} l r  | r @{/} l r  | r @{/} l r  | r @{/} l r |}
\hline
Observable
& \multicolumn{3}{c|}{\textsc{Pwg+Py8}}& \multicolumn{3}{c|}{\textsc{Pwg+Py8} Rad.~Up}& \multicolumn{3}{c|}{\textsc{Pwg+Py8} Rad.~Down}& \multicolumn{3}{c|}{\textsc{Pwg+H7}}& \multicolumn{3}{c|}{\textsc{Sherpa} 2.2.1}\\
& \multicolumn{2}{c}{$\chi^{2}$/NDF} &  ~$p$-value& \multicolumn{2}{c}{$\chi^{2}$/NDF} &  ~$p$-value& \multicolumn{2}{c}{$\chi^{2}$/NDF} &  ~$p$-value& \multicolumn{2}{c}{$\chi^{2}$/NDF} &  ~$p$-value& \multicolumn{2}{c}{$\chi^{2}$/NDF} &  ~$p$-value\\
\hline
\hline
$H_{\mathrm{T}}^{t\bar{t}}$ &{\ } 9.5 & 17 & 0.92 & {\ } 12.3 & 17 & 0.78 & {\ } 12.1 & 17 & 0.80 & {\ } 7.6 & 17 & 0.97 & {\ } 7.7 & 17 & 0.97\\
$|p_{\mathrm{out}}^{t,\mathrm{had}}|$ &{\ } 6.3 & 7 & 0.51 & {\ } 71.3 & 7 & $<$0.01 & {\ } 6.3 & 7 & 0.51 & {\ } 12.9 & 7 & 0.07 & {\ } 24.6 & 7 & $<$0.01\\
$|y_{\mathrm{boost}}^{t\bar{t}}|$ &{\ } 5.9 & 14 & 0.97 & {\ } 7.4 & 14 & 0.92 & {\ } 5.1 & 14 & 0.98 & {\ } 8.4 & 14 & 0.87 & {\ } 7.8 & 14 & 0.90\\
$     \chi^{t\bar{t}}$ &{\ } 18.1 & 12 & 0.11 & {\ } 10.5 & 12 & 0.57 & {\ } 36.0 & 12 & $<$0.01 & {\ } 14.6 & 12 & 0.26 & {\ } 22.7 & 12 & 0.03\\
$|\Delta\phi(t,\bar{t})|$ &{\ } 3.3 & 6 & 0.77 & {\ } 45.8 & 6 & $<$0.01 & {\ } 8.0 & 6 & 0.24 & {\ } 5.7 & 6 & 0.46 & {\ } 21.6 & 6 & $<$0.01\\
$p_{\mathrm{T}}^{t,1}$ &{\ } 6.0 & 10 & 0.81 & {\ } 10.0 & 10 & 0.44 & {\ } 6.8 & 10 & 0.74 & {\ } 3.1 & 10 & 0.98 & {\ } 3.0 & 10 & 0.98\\
$p_{\mathrm{T}}^{t,2}$ &{\ } 4.2 & 8 & 0.84 & {\ } 3.4 & 8 & 0.91 & {\ } 5.3 & 8 & 0.73 & {\ } 1.9 & 8 & 0.98 & {\ } 0.9 & 8 & 1.00\\
$|y^{t,\mathrm{had}}|$ &{\ } 9.1 & 19 & 0.97 & {\ } 9.6 & 19 & 0.96 & {\ } 9.0 & 19 & 0.97 & {\ } 10.4 & 19 & 0.94 & {\ } 14.6 & 19 & 0.74\\
$p_{\mathrm{T}}^{t,\mathrm{had}}$ &{\ } 11.7 & 18 & 0.86 & {\ } 11.1 & 18 & 0.89 & {\ } 14.3 & 18 & 0.71 & {\ } 6.4 & 18 & 0.99 & {\ } 6.8 & 18 & 0.99\\
$      |y^{t\bar{t}}|$ &{\ } 8.2 & 15 & 0.91 & {\ } 11.1 & 15 & 0.75 & {\ } 7.4 & 15 & 0.95 & {\ } 9.1 & 15 & 0.87 & {\ } 10.6 & 15 & 0.78\\
$        m^{t\bar{t}}$ &{\ } 16.0 & 15 & 0.38 & {\ } 14.8 & 15 & 0.46 & {\ } 19.8 & 15 & 0.18 & {\ } 14.7 & 15 & 0.48 & {\ } 15.3 & 15 & 0.43\\
$p_{\mathrm{T}}^{t\bar{t}}$ &{\ } 19.6 & 10 & 0.03 & {\ } 165.0 & 10 & $<$0.01 & {\ } 17.5 & 10 & 0.07 & {\ } 28.6 & 10 & $<$0.01 & {\ } 71.2 & 10 & $<$0.01\\
$N^{\mathrm{extra jets}}$ &{\ } 5.8 & 6 & 0.44 & {\ } 14.4 & 6 & 0.03 & {\ } 29.2 & 6 & $<$0.01 & {\ } 94.0 & 6 & $<$0.01 & {\ } 8.8 & 6 & 0.19\\
\hline
\end{tabular}}
\caption{ Comparison of the measured particle-level normalised single-differential cross-sections in the resolved topology with the predictions from several MC generators. For each prediction a $\chi^2$ and a $p$-value are calculated using the covariance matrix of the measured spectrum. The NDF is equal to the number of bins in the distribution minus one. 
}
\label{tab:chisquare:relative:1D:allpred:resolved:particle}
\end{table}
 
\begin{table}[t]
\footnotesize
\centering\noindent\makebox[\textwidth]{
\renewcommand*{\arraystretch}{1.2}\begin{tabular}{|c | r @{/} l r  | r @{/} l r  | r @{/} l r  | r @{/} l r  | r @{/} l r |}
\hline
Observable
& \multicolumn{3}{c|}{\textsc{Pwg+Py8}}& \multicolumn{3}{c|}{\textsc{Pwg+Py8} Rad.~Up}& \multicolumn{3}{c|}{\textsc{Pwg+Py8} Rad.~Down}& \multicolumn{3}{c|}{\textsc{Pwg+H7}}& \multicolumn{3}{c|}{\textsc{Sherpa} 2.2.1}\\
& \multicolumn{2}{c}{$\chi^{2}$/NDF} &  ~$p$-value& \multicolumn{2}{c}{$\chi^{2}$/NDF} &  ~$p$-value& \multicolumn{2}{c}{$\chi^{2}$/NDF} &  ~$p$-value& \multicolumn{2}{c}{$\chi^{2}$/NDF} &  ~$p$-value& \multicolumn{2}{c}{$\chi^{2}$/NDF} &  ~$p$-value\\
\hline
\hline
$H_{\mathrm{T}}^{t\bar{t}}$ &{\ } 11.1 & 18 & 0.89 & {\ } 17.7 & 18 & 0.48 & {\ } 10.5 & 18 & 0.91 & {\ } 11.4 & 18 & 0.88 & {\ } 11.9 & 18 & 0.85\\
$|p_{\mathrm{out}}^{t,\mathrm{had}}|$ &{\ } 9.2 & 8 & 0.32 & {\ } 97.3 & 8 & $<$0.01 & {\ } 8.3 & 8 & 0.41 & {\ } 11.2 & 8 & 0.19 & {\ } 27.8 & 8 & $<$0.01\\
$|y_{\mathrm{boost}}^{t\bar{t}}|$ &{\ } 7.0 & 15 & 0.96 & {\ } 8.7 & 15 & 0.89 & {\ } 6.1 & 15 & 0.98 & {\ } 9.8 & 15 & 0.83 & {\ } 10.2 & 15 & 0.81\\
$     \chi^{t\bar{t}}$ &{\ } 20.4 & 13 & 0.09 & {\ } 12.3 & 13 & 0.51 & {\ } 38.3 & 13 & $<$0.01 & {\ } 17.7 & 13 & 0.17 & {\ } 22.5 & 13 & 0.05\\
$|\Delta\phi(t,\bar{t})|$ &{\ } 3.0 & 7 & 0.89 & {\ } 57.7 & 7 & $<$0.01 & {\ } 12.3 & 7 & 0.09 & {\ } 4.7 & 7 & 0.70 & {\ } 22.1 & 7 & $<$0.01\\
$p_{\mathrm{T}}^{t,1}$ &{\ } 9.2 & 11 & 0.60 & {\ } 15.0 & 11 & 0.18 & {\ } 8.8 & 11 & 0.64 & {\ } 7.8 & 11 & 0.73 & {\ } 6.5 & 11 & 0.84\\
$p_{\mathrm{T}}^{t,2}$ &{\ } 5.3 & 9 & 0.80 & {\ } 5.2 & 9 & 0.81 & {\ } 6.0 & 9 & 0.74 & {\ } 2.5 & 9 & 0.98 & {\ } 2.1 & 9 & 0.99\\
$|y^{t,\mathrm{had}}|$ &{\ } 12.7 & 20 & 0.89 & {\ } 13.5 & 20 & 0.86 & {\ } 12.5 & 20 & 0.90 & {\ } 13.2 & 20 & 0.87 & {\ } 19.5 & 20 & 0.49\\
$p_{\mathrm{T}}^{t,\mathrm{had}}$ &{\ } 19.0 & 19 & 0.46 & {\ } 23.3 & 19 & 0.23 & {\ } 18.0 & 19 & 0.52 & {\ } 15.0 & 19 & 0.72 & {\ } 14.5 & 19 & 0.75\\
$      |y^{t\bar{t}}|$ &{\ } 9.2 & 16 & 0.90 & {\ } 11.5 & 16 & 0.78 & {\ } 8.3 & 16 & 0.94 & {\ } 9.8 & 16 & 0.88 & {\ } 13.5 & 16 & 0.64\\
$        m^{t\bar{t}}$ &{\ } 17.8 & 16 & 0.34 & {\ } 16.4 & 16 & 0.43 & {\ } 20.2 & 16 & 0.21 & {\ } 15.5 & 16 & 0.49 & {\ } 17.1 & 16 & 0.38\\
$p_{\mathrm{T}}^{t\bar{t}}$ &{\ } 23.1 & 11 & 0.02 & {\ } 196.0 & 11 & $<$0.01 & {\ } 16.9 & 11 & 0.11 & {\ } 33.4 & 11 & $<$0.01 & {\ } 88.0 & 11 & $<$0.01\\
$N^{\mathrm{extra jets}}$ &{\ } 9.5 & 7 & 0.22 & {\ } 7.7 & 7 & 0.36 & {\ } 28.3 & 7 & $<$0.01 & {\ } 104.0 & 7 & $<$0.01 & {\ } 11.5 & 7 & 0.12\\
\hline
\end{tabular}}
\caption{ Comparison of the measured particle-level absolute single-differential cross-sections in the resolved topology with the predictions from several MC generators. For each prediction a $\chi^2$ and a $p$-value are calculated using the covariance matrix of the measured spectrum. The NDF is equal to the number of bins in the distribution.}
\label{tab:chisquare:absolute:1D:allpred:resolved:particle}
\end{table}

\begin{table}[t]
\footnotesize
\centering\noindent\makebox[\textwidth]{
\renewcommand*{\arraystretch}{1.2}\begin{tabular}{|c | r @{/} l r  | r @{/} l r  | r @{/} l r  | r @{/} l r  | r @{/} l r |}
\hline
Observable
& \multicolumn{3}{c|}{\textsc{Pwg+Py8}}& \multicolumn{3}{c|}{\textsc{Pwg+Py8} Rad.~Up}& \multicolumn{3}{c|}{\textsc{Pwg+Py8} Rad.~Down}& \multicolumn{3}{c|}{\textsc{Pwg+H7}}& \multicolumn{3}{c|}{\textsc{Sherpa} 2.2.1}\\
& \multicolumn{2}{c}{$\chi^{2}$/NDF} &  ~$p$-value& \multicolumn{2}{c}{$\chi^{2}$/NDF} &  ~$p$-value& \multicolumn{2}{c}{$\chi^{2}$/NDF} &  ~$p$-value& \multicolumn{2}{c}{$\chi^{2}$/NDF} &  ~$p$-value& \multicolumn{2}{c}{$\chi^{2}$/NDF} &  ~$p$-value\\
\hline
\hline
$H_{\mathrm{T}}^{t\bar{t}}\textrm{ vs }N^{\mathrm{extra jets}}$ &{\ } 9.7 & 19 & 0.96 & {\ } 57.9 & 19 & $<$0.01 & {\ } 19.4 & 19 & 0.43 & {\ } 48.7 & 19 & $<$0.01 & {\ } 27.4 & 19 & 0.10\\
$|p_{\mathrm{out}}^{t,\mathrm{had}}|\textrm{ vs }N^{\mathrm{extra jets}}$ &{\ } 10.8 & 9 & 0.29 & {\ } 89.2 & 9 & $<$0.01 & {\ } 31.9 & 9 & $<$0.01 & {\ } 32.6 & 9 & $<$0.01 & {\ } 19.2 & 9 & 0.02\\
$\chi^{t\bar{t}} \textrm{ vs }N^{\mathrm{extra jets}}$ &{\ } 37.6 & 19 & $<$0.01 & {\ } 31.6 & 19 & 0.03 & {\ } 88.9 & 19 & $<$0.01 & {\ } 84.8 & 19 & $<$0.01 & {\ } 23.7 & 19 & 0.21\\
$|\Delta\phi(t,\bar{t})| \textrm{ vs }N^{\mathrm{extra jets}}$ &{\ } 21.8 & 18 & 0.24 & {\ } 125.0 & 18 & $<$0.01 & {\ } 31.0 & 18 & 0.03 & {\ } 44.4 & 18 & $<$0.01 & {\ } 36.7 & 18 & $<$0.01\\
$|y^{t,\mathrm{had}}|\textrm{ vs }N^{\mathrm{extra jets}}$ &{\ } 9.5 & 12 & 0.66 & {\ } 19.1 & 12 & 0.09 & {\ } 26.8 & 12 & $<$0.01 & {\ } 30.8 & 12 & $<$0.01 & {\ } 10.4 & 12 & 0.58\\
$|y^{t,\mathrm{had}}|\textrm{ vs }p_{\mathrm{T}}^{t,\mathrm{had}}$ &{\ } 14.9 & 12 & 0.25 & {\ } 11.9 & 12 & 0.45 & {\ } 18.1 & 12 & 0.11 & {\ } 8.4 & 12 & 0.75 & {\ } 9.4 & 12 & 0.67\\
$p_{\mathrm{T}}^{t,\mathrm{had}}\textrm{ vs }|p_{\mathrm{out}}^{t,\mathrm{had}}|$ &{\ } 10.5 & 12 & 0.57 & {\ } 74.5 & 12 & $<$0.01 & {\ } 25.3 & 12 & 0.01 & {\ } 13.4 & 12 & 0.34 & {\ } 22.4 & 12 & 0.03\\
$p_{\mathrm{T}}^{t,\mathrm{had}}\textrm{ vs }N^{\mathrm{extra jets}}$ &{\ } 14.2 & 16 & 0.58 & {\ } 45.7 & 16 & $<$0.01 & {\ } 37.3 & 16 & $<$0.01 & {\ } 67.5 & 16 & $<$0.01 & {\ } 13.9 & 16 & 0.60\\
$|y^{t\bar{t}}|\textrm{ vs }N^{\mathrm{extra jets}}$ &{\ } 8.2 & 12 & 0.77 & {\ } 14.6 & 12 & 0.26 & {\ } 25.4 & 12 & 0.01 & {\ } 55.5 & 12 & $<$0.01 & {\ } 13.9 & 12 & 0.30\\
$|y^{t\bar{t}}|\textrm{ vs }m^{t\bar{t}}$ &{\ } 18.0 & 14 & 0.21 & {\ } 12.0 & 14 & 0.60 & {\ } 23.1 & 14 & 0.06 & {\ } 13.2 & 14 & 0.51 & {\ } 14.8 & 14 & 0.40\\
$|y^{t\bar{t}}|\textrm{ vs }p_{\mathrm{T}}^{t\bar{t}}$ &{\ } 28.5 & 12 & $<$0.01 & {\ } 149.0 & 12 & $<$0.01 & {\ } 23.2 & 12 & 0.03 & {\ } 31.8 & 12 & $<$0.01 & {\ } 70.7 & 12 & $<$0.01\\
$m^{t\bar{t}}\textrm{ vs }N^{\mathrm{extra jets}}$ &{\ } 29.1 & 16 & 0.02 & {\ } 25.5 & 16 & 0.06 & {\ } 49.6 & 16 & $<$0.01 & {\ } 24.6 & 16 & 0.08 & {\ } 11.5 & 16 & 0.78\\
$m^{t\bar{t}}\textrm{ vs }p_{\mathrm{T}}^{t,\mathrm{had}}$ &{\ } 58.9 & 31 & $<$0.01 & {\ } 51.4 & 31 & 0.01 & {\ } 92.3 & 31 & $<$0.01 & {\ } 35.6 & 31 & 0.26 & {\ } 44.8 & 31 & 0.05\\
$m^{t\bar{t}}\textrm{ vs }p_{\mathrm{T}}^{t\bar{t}}$ &{\ } 43.6 & 21 & $<$0.01 & {\ } 260.0 & 21 & $<$0.01 & {\ } 47.0 & 21 & $<$0.01 & {\ } 44.7 & 21 & $<$0.01 & {\ } 149.0 & 21 & $<$0.01\\
$p_{\mathrm{T}}^{t\bar{t}}\textrm{ vs }N^{\mathrm{extra jets}}$ &{\ } 69.1 & 19 & $<$0.01 & {\ } 283.0 & 19 & $<$0.01 & {\ } 58.5 & 19 & $<$0.01 & {\ } 82.8 & 19 & $<$0.01 & {\ } 102.0 & 19 & $<$0.01\\
$p_{\mathrm{T}}^{t\bar{t}}\textrm{ vs }p^{t,\mathrm{had}}_\mathrm{T}$ &{\ } 39.2 & 19 & $<$0.01 & {\ } 282.0 & 19 & $<$0.01 & {\ } 51.5 & 19 & $<$0.01 & {\ } 55.8 & 19 & $<$0.01 & {\ } 137.0 & 19 & $<$0.01\\
\hline
\end{tabular}}
\caption{ Comparison of the measured particle-level normalised double-differential cross-sections in the resolved topology with the predictions from several MC generators. For each prediction a $\chi^2$ and a $p$-value are calculated using the covariance matrix of the measured spectrum. The NDF is equal to the number of bins in the distribution minus one.
}
\label{tab:chisquare:relative:2D:allpred:resolved:particle}
\end{table}
 
\begin{table}[t]
\footnotesize
\centering\noindent\makebox[\textwidth]{
\renewcommand*{\arraystretch}{1.2}\begin{tabular}{|c | r @{/} l r  | r @{/} l r  | r @{/} l r  | r @{/} l r  | r @{/} l r |}
\hline
Observable
& \multicolumn{3}{c|}{\textsc{Pwg+Py8}}& \multicolumn{3}{c|}{\textsc{Pwg+Py8} Rad.~Up}& \multicolumn{3}{c|}{\textsc{Pwg+Py8} Rad.~Down}& \multicolumn{3}{c|}{\textsc{Pwg+H7}}& \multicolumn{3}{c|}{\textsc{Sherpa} 2.2.1}\\
& \multicolumn{2}{c}{$\chi^{2}$/NDF} &  ~$p$-value& \multicolumn{2}{c}{$\chi^{2}$/NDF} &  ~$p$-value& \multicolumn{2}{c}{$\chi^{2}$/NDF} &  ~$p$-value& \multicolumn{2}{c}{$\chi^{2}$/NDF} &  ~$p$-value& \multicolumn{2}{c}{$\chi^{2}$/NDF} &  ~$p$-value\\
\hline
\hline
$H_{\mathrm{T}}^{t\bar{t}}\textrm{ vs }N^{\mathrm{extra jets}}$ &{\ } 13.8 & 20 & 0.84 & {\ } 72.9 & 20 & $<$0.01 & {\ } 31.3 & 20 & 0.05 & {\ } 56.6 & 20 & $<$0.01 & {\ } 40.5 & 20 & $<$0.01\\
$|p_{\mathrm{out}}^{t,\mathrm{had}}|\textrm{ vs }N^{\mathrm{extra jets}}$ &{\ } 16.3 & 10 & 0.09 & {\ } 165.0 & 10 & $<$0.01 & {\ } 15.7 & 10 & 0.11 & {\ } 35.6 & 10 & $<$0.01 & {\ } 50.9 & 10 & $<$0.01\\
$\chi^{t\bar{t}} \textrm{ vs }N^{\mathrm{extra jets}}$ &{\ } 44.4 & 20 & $<$0.01 & {\ } 60.3 & 20 & $<$0.01 & {\ } 88.3 & 20 & $<$0.01 & {\ } 62.2 & 20 & $<$0.01 & {\ } 24.6 & 20 & 0.21\\
$|\Delta\phi(t,\bar{t})| \textrm{ vs }N^{\mathrm{extra jets}}$ &{\ } 41.6 & 19 & $<$0.01 & {\ } 183.0 & 19 & $<$0.01 & {\ } 43.6 & 19 & $<$0.01 & {\ } 44.2 & 19 & $<$0.01 & {\ } 60.0 & 19 & $<$0.01\\
$|y^{t,\mathrm{had}}|\textrm{ vs }N^{\mathrm{extra jets}}$ &{\ } 11.3 & 13 & 0.59 & {\ } 50.3 & 13 & $<$0.01 & {\ } 23.1 & 13 & 0.04 & {\ } 28.7 & 13 & $<$0.01 & {\ } 14.8 & 13 & 0.32\\
$|y^{t,\mathrm{had}}|\textrm{ vs }p_{\mathrm{T}}^{t,\mathrm{had}}$ &{\ } 13.3 & 13 & 0.42 & {\ } 12.9 & 13 & 0.45 & {\ } 15.6 & 13 & 0.27 & {\ } 8.7 & 13 & 0.80 & {\ } 9.8 & 13 & 0.71\\
$p_{\mathrm{T}}^{t,\mathrm{had}}\textrm{ vs }|p_{\mathrm{out}}^{t,\mathrm{had}}|$ &{\ } 8.6 & 13 & 0.80 & {\ } 79.6 & 13 & $<$0.01 & {\ } 28.8 & 13 & $<$0.01 & {\ } 9.7 & 13 & 0.72 & {\ } 16.0 & 13 & 0.25\\
$p_{\mathrm{T}}^{t,\mathrm{had}}\textrm{ vs }N^{\mathrm{extra jets}}$ &{\ } 19.3 & 17 & 0.31 & {\ } 59.5 & 17 & $<$0.01 & {\ } 43.3 & 17 & $<$0.01 & {\ } 65.3 & 17 & $<$0.01 & {\ } 24.7 & 17 & 0.10\\
$|y^{t\bar{t}}|\textrm{ vs }N^{\mathrm{extra jets}}$ &{\ } 7.0 & 13 & 0.90 & {\ } 26.7 & 13 & 0.01 & {\ } 22.1 & 13 & 0.05 & {\ } 51.5 & 13 & $<$0.01 & {\ } 31.5 & 13 & $<$0.01\\
$|y^{t\bar{t}}|\textrm{ vs }m^{t\bar{t}}$ &{\ } 22.3 & 15 & 0.10 & {\ } 15.0 & 15 & 0.45 & {\ } 29.8 & 15 & 0.01 & {\ } 15.8 & 15 & 0.40 & {\ } 19.1 & 15 & 0.21\\
$|y^{t\bar{t}}|\textrm{ vs }p_{\mathrm{T}}^{t\bar{t}}$ &{\ } 32.7 & 13 & $<$0.01 & {\ } 143.0 & 13 & $<$0.01 & {\ } 21.2 & 13 & 0.07 & {\ } 36.8 & 13 & $<$0.01 & {\ } 81.4 & 13 & $<$0.01\\
$m^{t\bar{t}}\textrm{ vs }N^{\mathrm{extra jets}}$ &{\ } 28.0 & 17 & 0.04 & {\ } 29.0 & 17 & 0.03 & {\ } 49.2 & 17 & $<$0.01 & {\ } 36.3 & 17 & $<$0.01 & {\ } 14.0 & 17 & 0.67\\
$m^{t\bar{t}}\textrm{ vs }p_{\mathrm{T}}^{t,\mathrm{had}}$ &{\ } 56.2 & 32 & $<$0.01 & {\ } 59.9 & 32 & $<$0.01 & {\ } 79.9 & 32 & $<$0.01 & {\ } 31.9 & 32 & 0.47 & {\ } 48.5 & 32 & 0.03\\
$m^{t\bar{t}}\textrm{ vs }p_{\mathrm{T}}^{t\bar{t}}$ &{\ } 49.0 & 22 & $<$0.01 & {\ } 310.0 & 22 & $<$0.01 & {\ } 53.3 & 22 & $<$0.01 & {\ } 55.1 & 22 & $<$0.01 & {\ } 175.0 & 22 & $<$0.01\\
$p_{\mathrm{T}}^{t\bar{t}}\textrm{ vs }N^{\mathrm{extra jets}}$ &{\ } 93.2 & 20 & $<$0.01 & {\ } 412.0 & 20 & $<$0.01 & {\ } 51.9 & 20 & $<$0.01 & {\ } 91.8 & 20 & $<$0.01 & {\ } 163.0 & 20 & $<$0.01\\
$p_{\mathrm{T}}^{t\bar{t}}\textrm{ vs }p^{t,\mathrm{had}}_\mathrm{T}$ &{\ } 38.6 & 20 & $<$0.01 & {\ } 294.0 & 20 & $<$0.01 & {\ } 66.5 & 20 & $<$0.01 & {\ } 46.1 & 20 & $<$0.01 & {\ } 128.0 & 20 & $<$0.01\\
\hline
\end{tabular}}
\caption{ Comparison of the measured particle-level absolute double-differential cross-sections in the resolved topology with the predictions from several MC generators. For each prediction a $\chi^2$ and a $p$-value are calculated using the covariance matrix of the measured spectrum. The NDF is equal to the number of bins in the distribution.}
\label{tab:chisquare:absolute:2D:allpred:resolved:particle}
\end{table}
 
\FloatBarrier
\subsubsection{Boosted topology}
\label{sec:results:fiducial:boosted}
\begin{figure*}[t]
\centering
\subfigure[]{\includegraphics[width=0.45\textwidth]{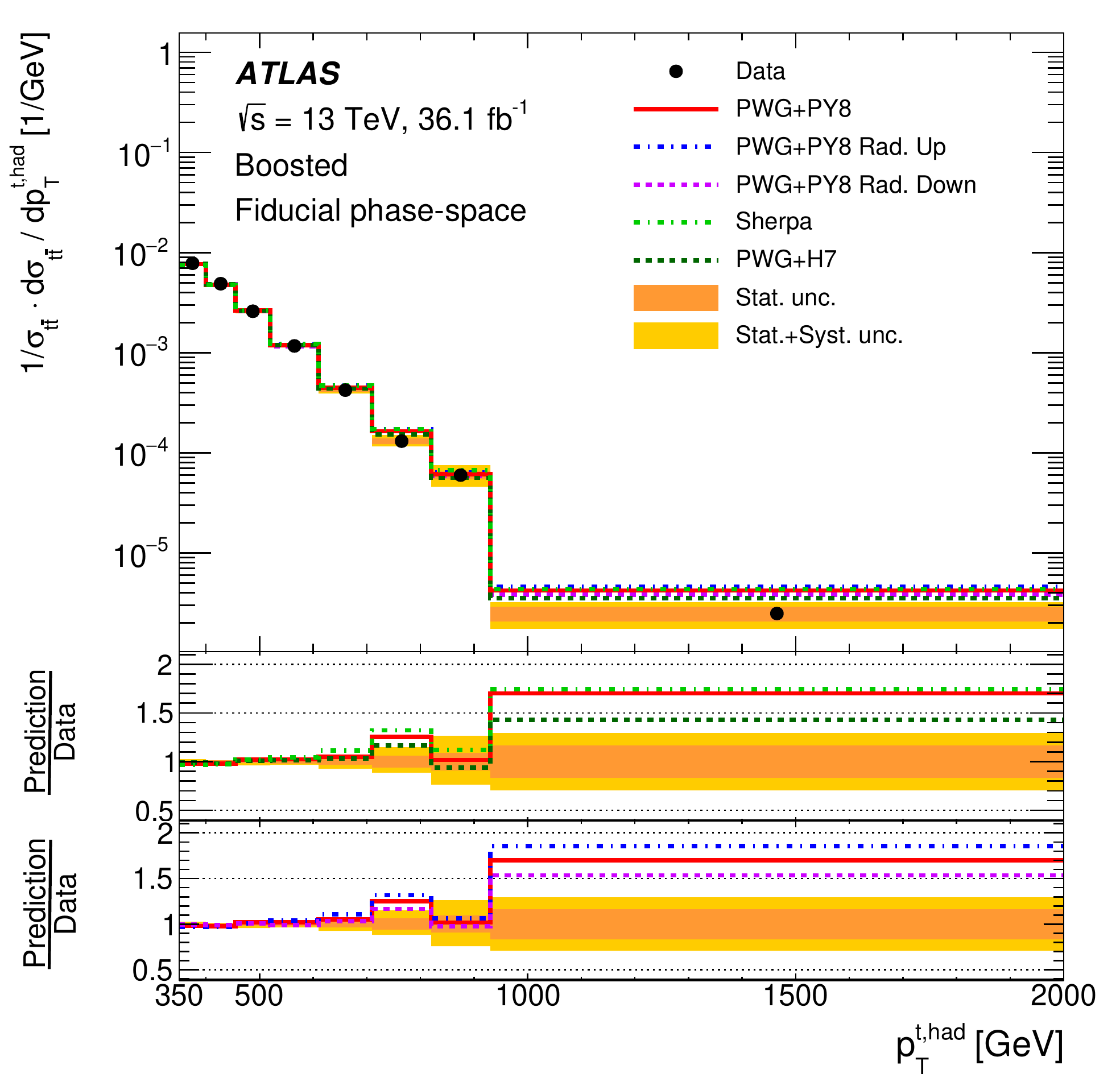}
\label{fig:results_particle:boosted:topHad_pt:rel}}
\subfigure[]{\includegraphics[width=0.45\textwidth]{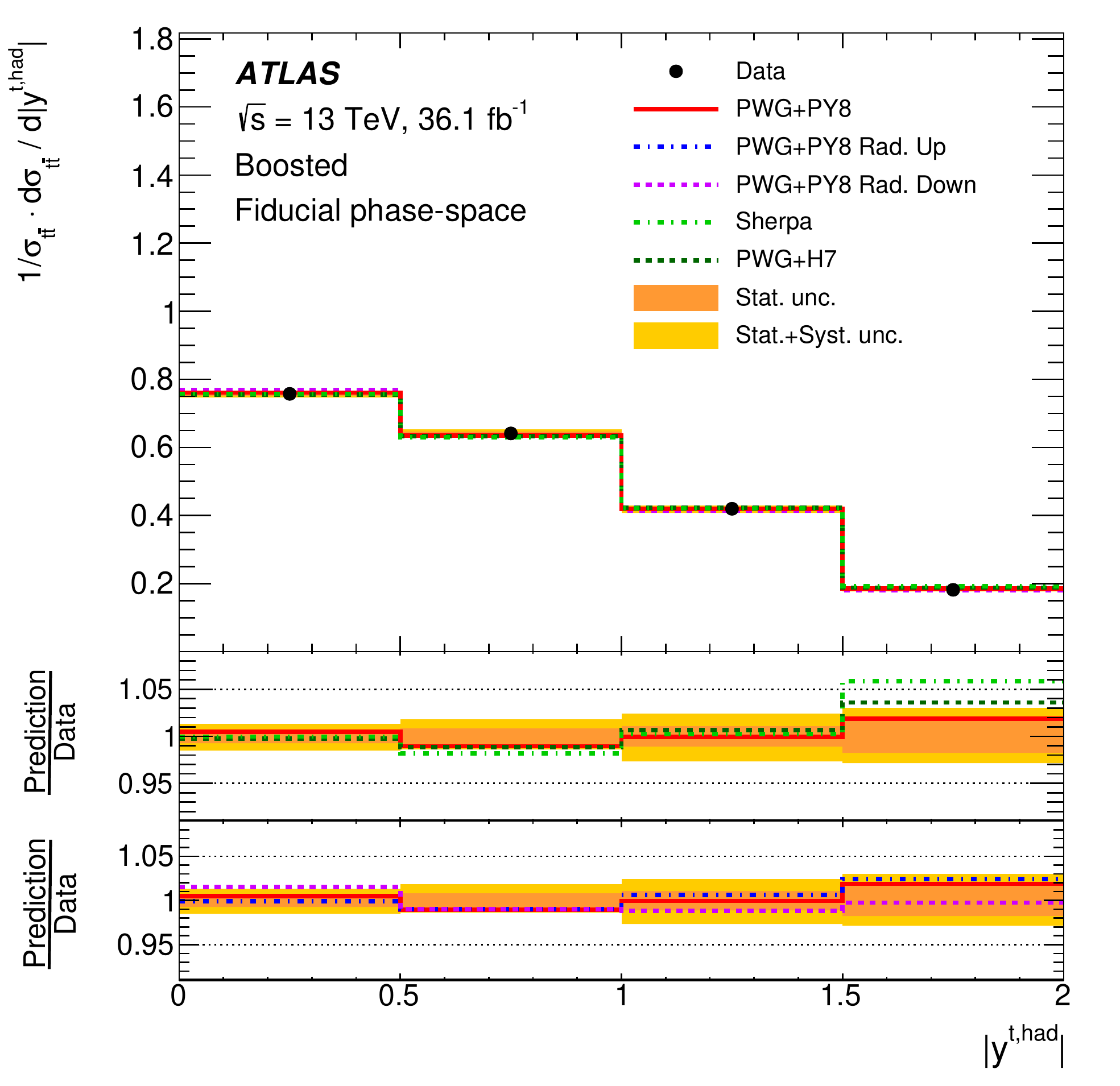}
\label{fig:results_particle:boosted:topHad_y:rel}}
\caption{\small{Particle-level normalised differential cross-sections as a function of \subref{fig:results_particle:boosted:topHad_pt:rel} the transverse momentum and \subref{fig:results_particle:boosted:topHad_y:rel} the absolute value of the rapidity of the hadronically decaying  top quark in the boosted topology, compared with different Monte Carlo predictions. The bands represent the statistical and total uncertainty in the data. Data points are placed at the centre of each bin. The lower panel shows the ratios of the simulations to data.}}
\label{fig:results:rel:particle:boosted:pt_y_thad}
\end{figure*}

\begin{figure*}[t]
\centering
\subfigure[]{\includegraphics[width=0.45\textwidth]{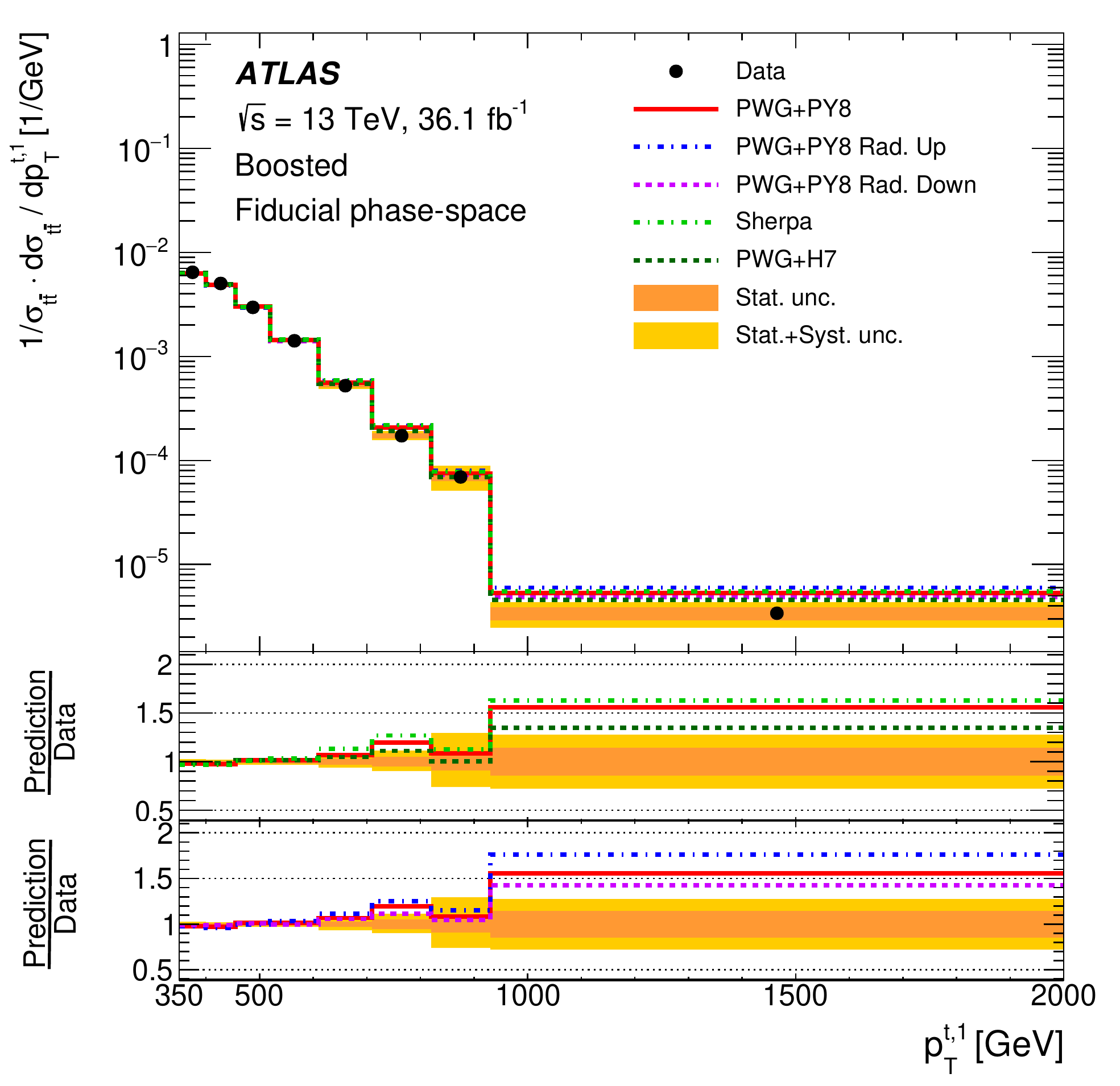}
\label{fig:results_particle:boosted:top_leading:rel}}
\subfigure[]{\includegraphics[width=0.45\textwidth]{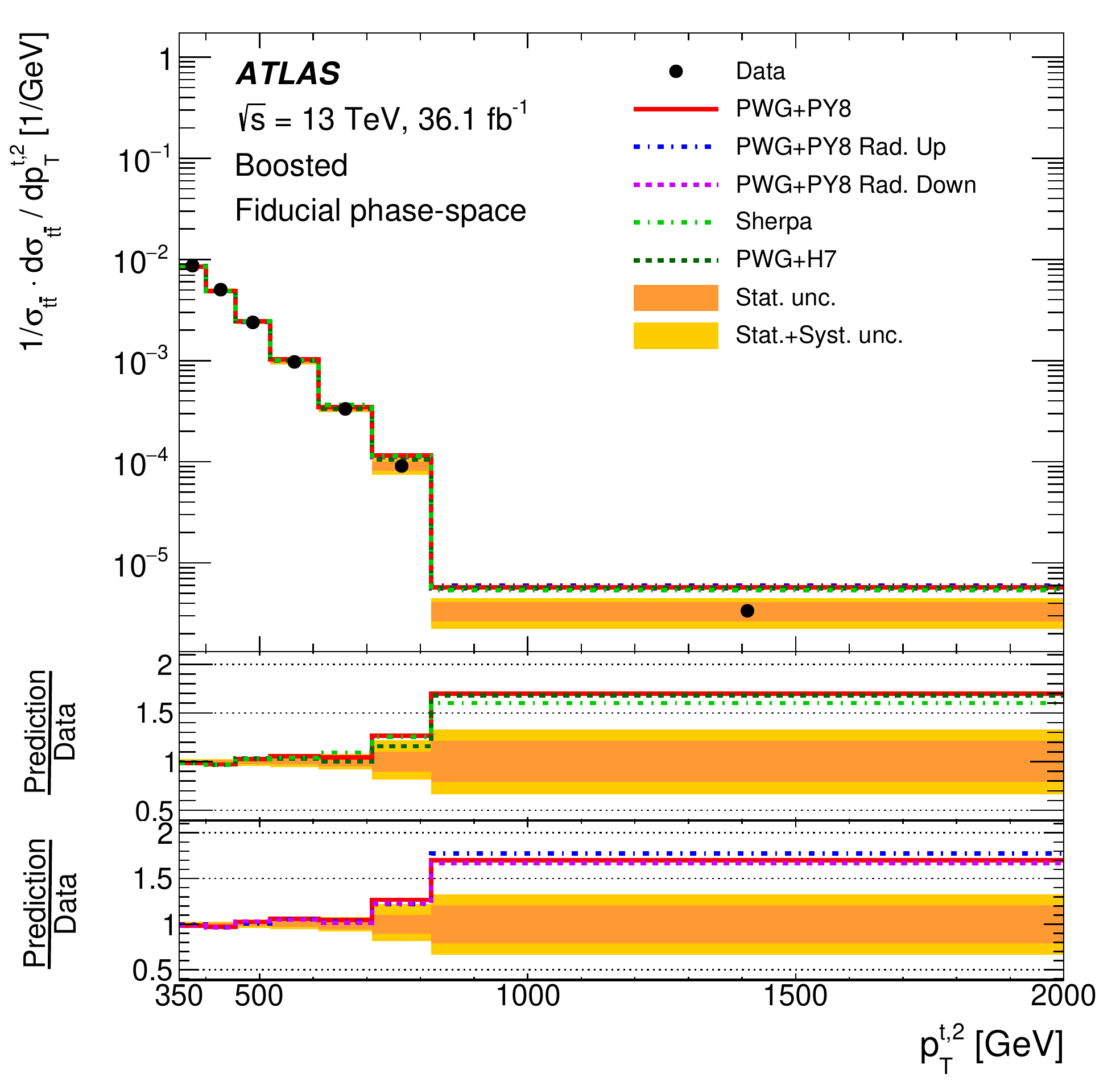}
\label{fig:results_particle:boosted:top_subleading:rel}}
\caption{\small{Particle-level normalised differential cross-sections as a function of the transverse momentum of \subref{fig:results_particle:boosted:top_leading:rel}~the leading and \subref{fig:results_particle:boosted:top_subleading:rel}~the subleading top quark in the boosted topology, compared with different Monte Carlo predictions. The bands represent the statistical and total uncertainty in the data. Data points are placed at the centre of each bin. The lower panel shows the ratios of the simulations to data.}}
\label{fig:results:rel:particle:boosted:leading}
\end{figure*}

\begin{figure*}[t]
\centering
\subfigure[]{\includegraphics[width=0.45\textwidth]{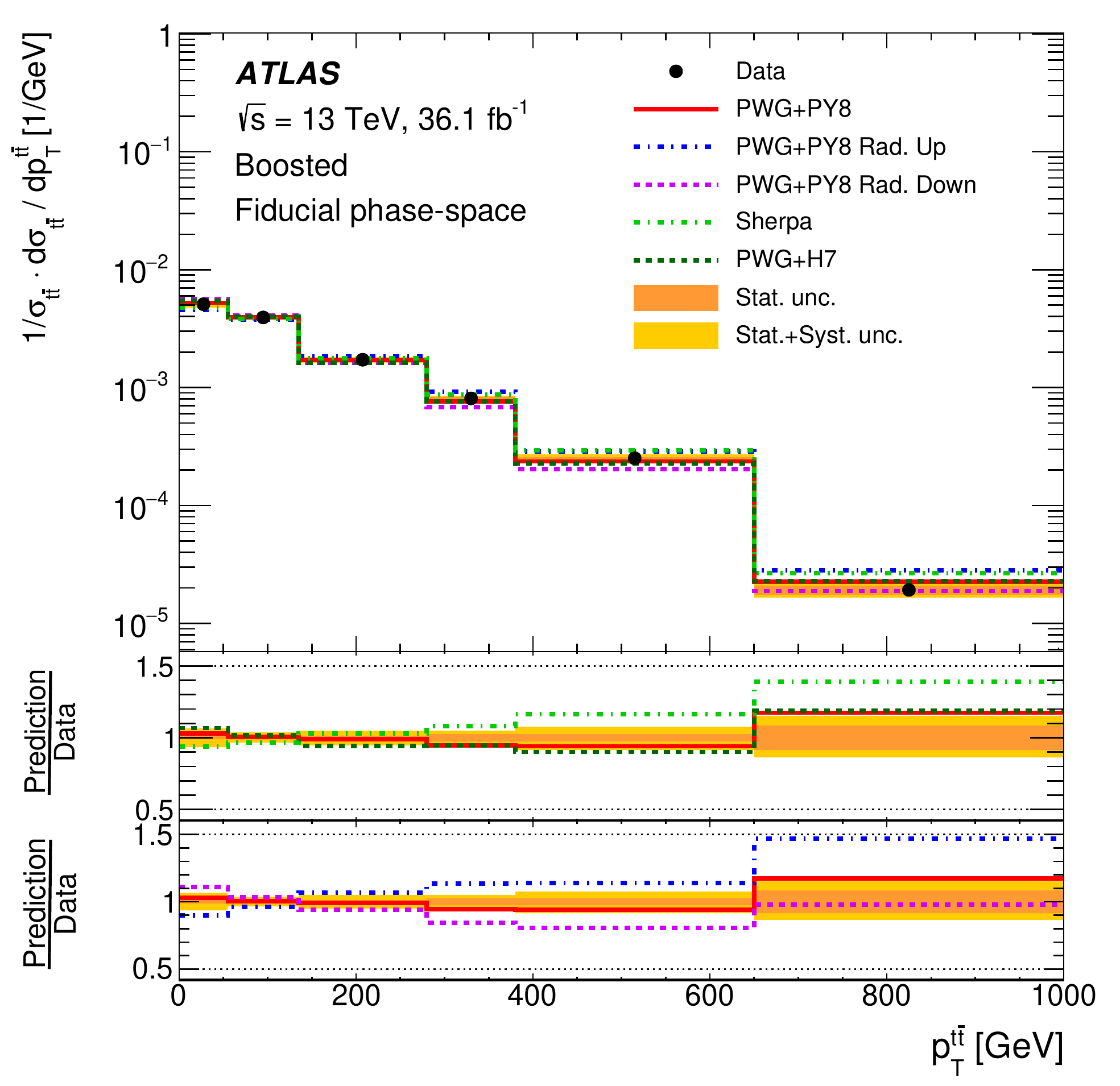}
\label{fig:results_particle:boosted:ttbar_pt:rel}}
\subfigure[]{\includegraphics[width=0.45\textwidth]{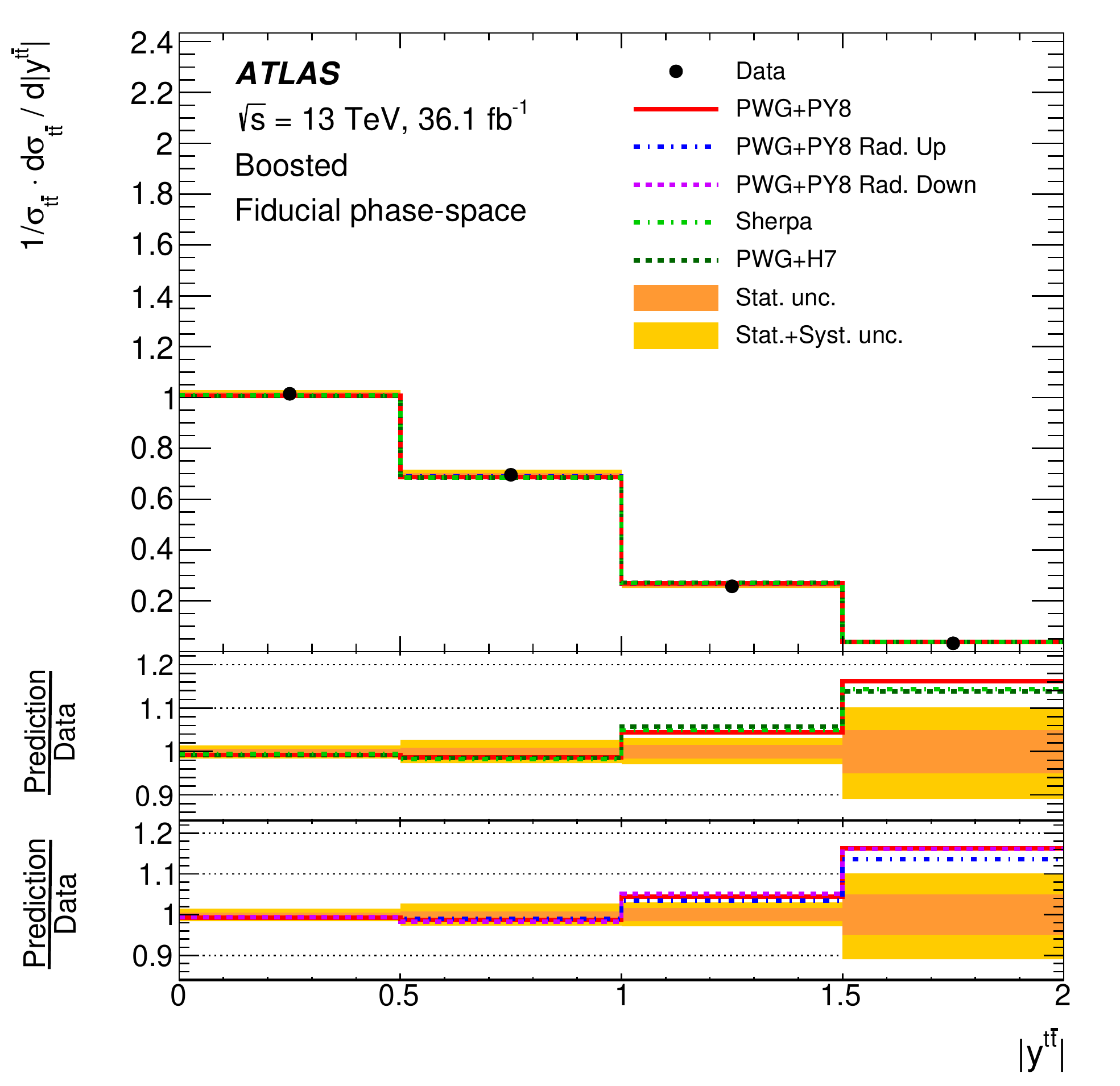}
\label{fig:results_particle:boosted:ttbar_eta:rel}}
\caption{\small{Particle-level normalised differential cross-sections as a function of \subref{fig:results_particle:boosted:ttbar_pt:rel} the transverse momentum and \subref{fig:results_particle:boosted:ttbar_eta:rel} the absolute value of the rapidity of the \ttb{} system in the boosted topology, compared with different Monte Carlo predictions. The bands represent the statistical and total uncertainty in the data. Data points are placed at the centre of each bin. The lower panel shows the ratios of the simulations to data.}}
\label{fig:results:rel:particle:boosted:ttbar}
\end{figure*}

\begin{figure*}[t]
\centering
\subfigure[]{\includegraphics[width=0.45\textwidth]{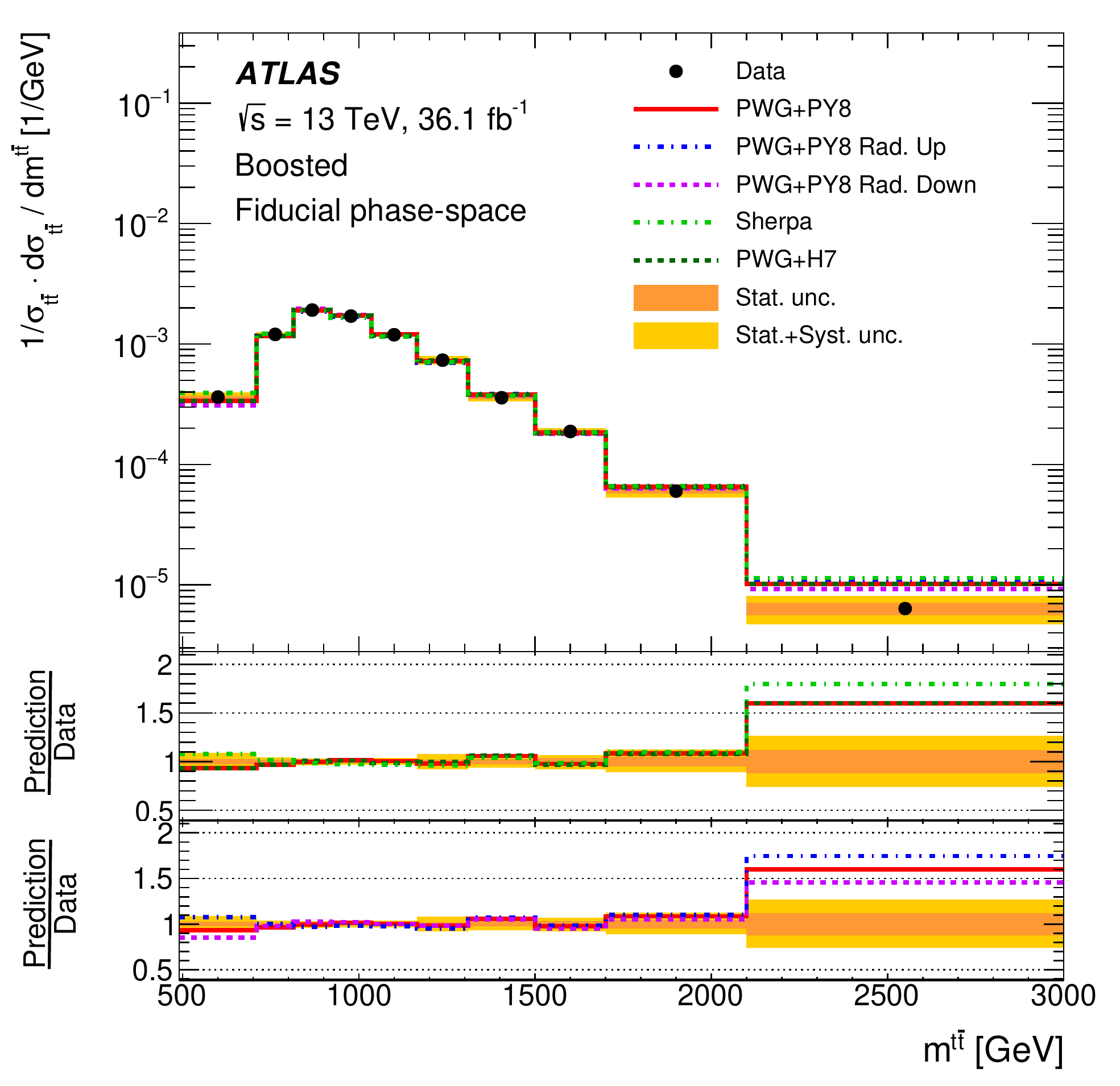}
\label{fig:results_particle:boosted:ttbar_m:rel}}
\subfigure[]{\includegraphics[width=0.45\textwidth]{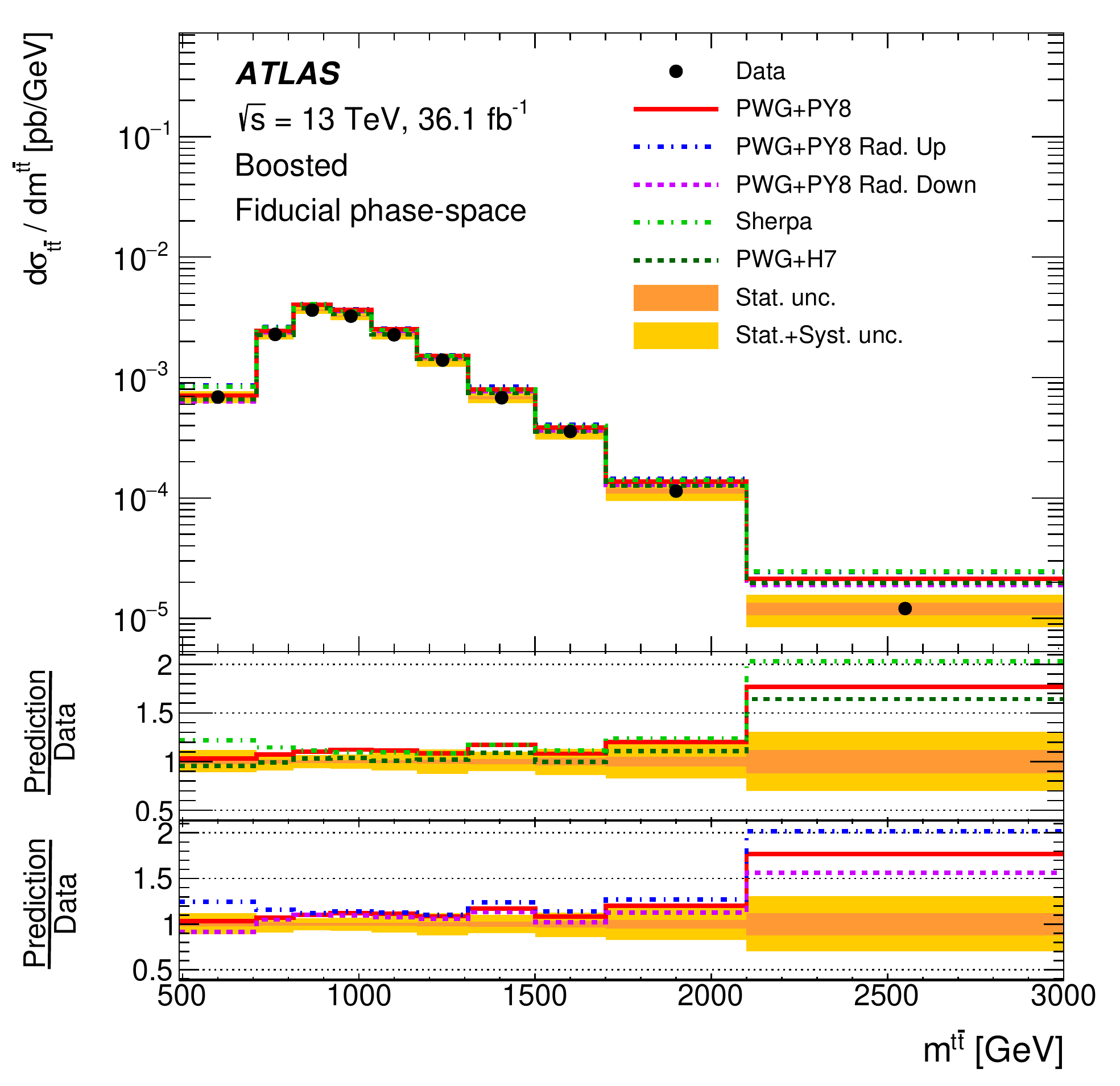}
\label{fig:results_particle:boosted:ttbar_m:abs}}
\caption{\small{Particle-level~\subref{fig:results_particle:boosted:ttbar_m:rel} normalised and~\subref{fig:results_particle:boosted:ttbar_m:abs} absolute  differential cross-sections as a function of $\mtt{}$ in the boosted topology, compared with different Monte Carlo predictions. The bands represent the statistical and total uncertainty in the data. Data points are placed at the centre of each bin. The lower panel shows the ratios of the simulations to data.}}
\label{fig:results:particle:boosted:ttbar_m}
\end{figure*}

\begin{figure*}[htbp]
\centering
\subfigure[]{\includegraphics[width=0.45\textwidth]{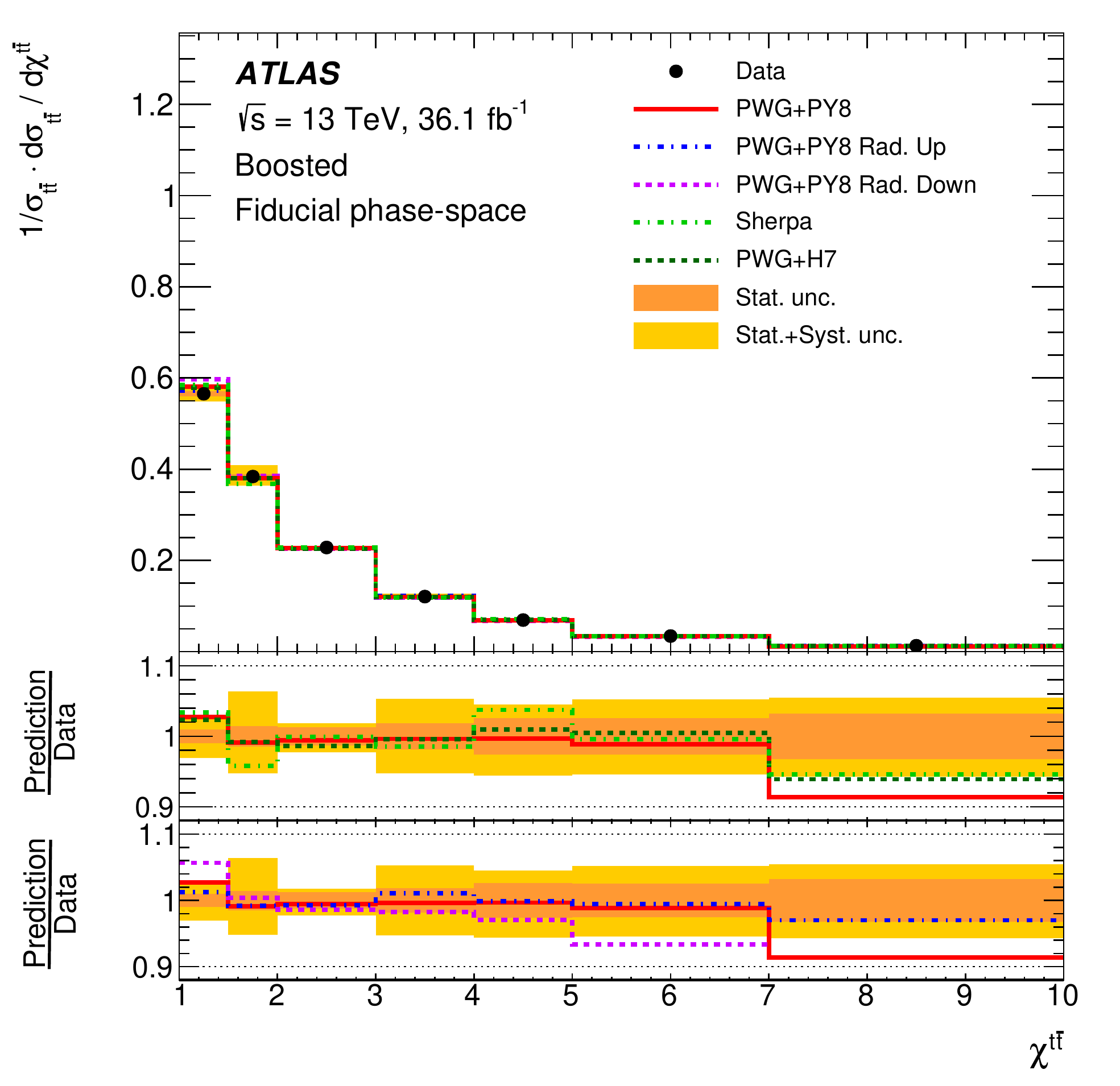}
\label{fig:results_particle:boosted:ttbar_chi:rel}}
\subfigure[]{\includegraphics[width=0.45\textwidth]{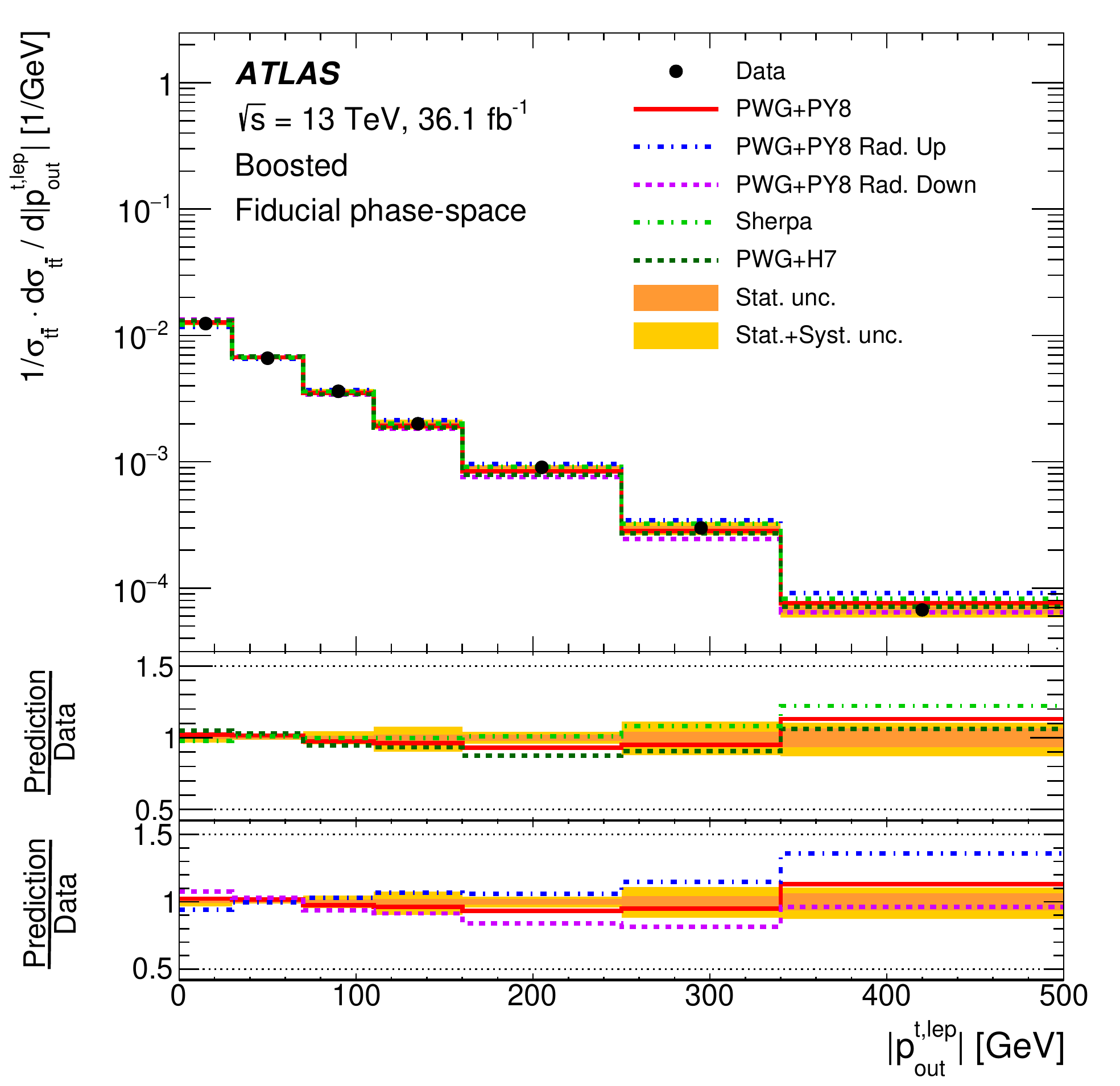}
\label{fig:results_particle:boosted:abs_pout:rel}}\\
\subfigure[]{\includegraphics[width=0.45\textwidth]{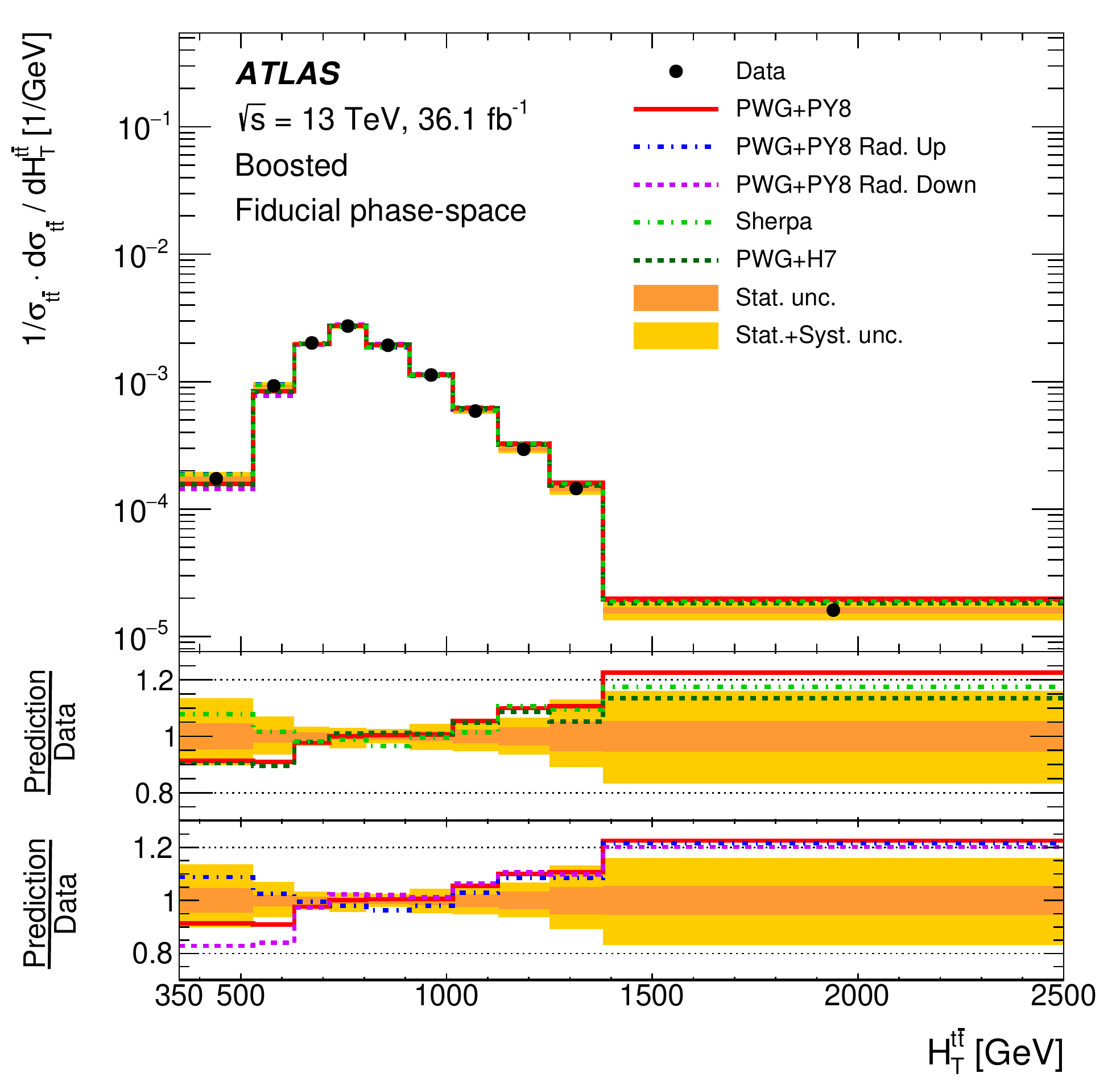}
\label{fig:results_particle:boosted:ht:rel}}
\caption{\small{Particle-level normalised differential cross-sections as a function of  \subref{fig:results_particle:boosted:ttbar_chi:rel}~\chitt, \subref{fig:results_particle:boosted:abs_pout:rel}~\absPouttlep{} and \subref{fig:results_particle:boosted:ht:rel}~\HTtt{} in the boosted topology, compared with different Monte Carlo predictions. The bands represent the statistical and total uncertainty in the data. Data points are placed at the centre of each bin. The lower panel shows the ratios of the simulations to data.}}
\label{fig:results:rel:particle:boosted:additionalvars}
\end{figure*}

\begin{figure*}[t]
\centering
\subfigure[]{\includegraphics[width=0.45\textwidth]{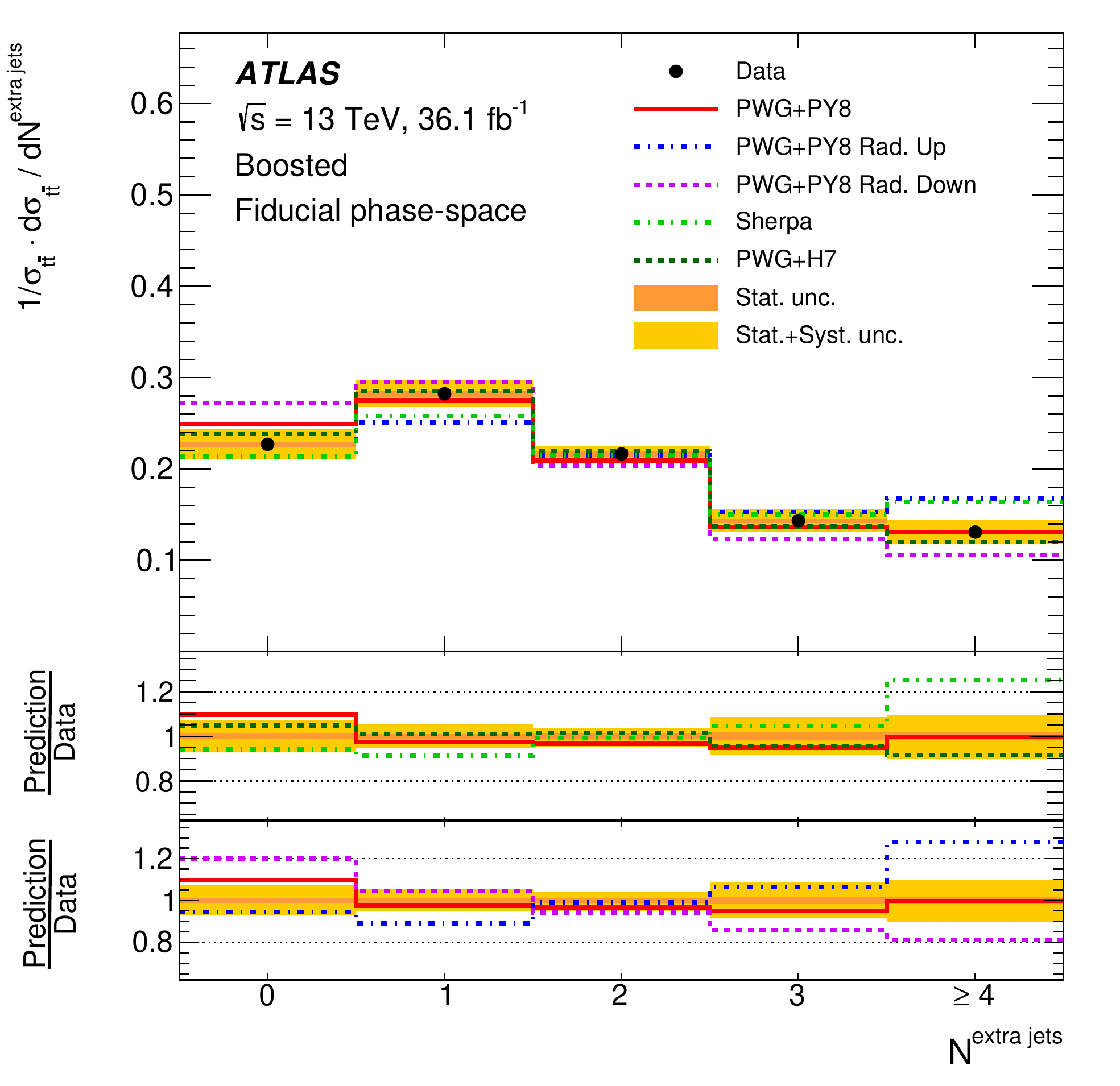}
\label{fig:results_particle:boosted:extraj:rel}}
\subfigure[]{\includegraphics[width=0.45\textwidth]{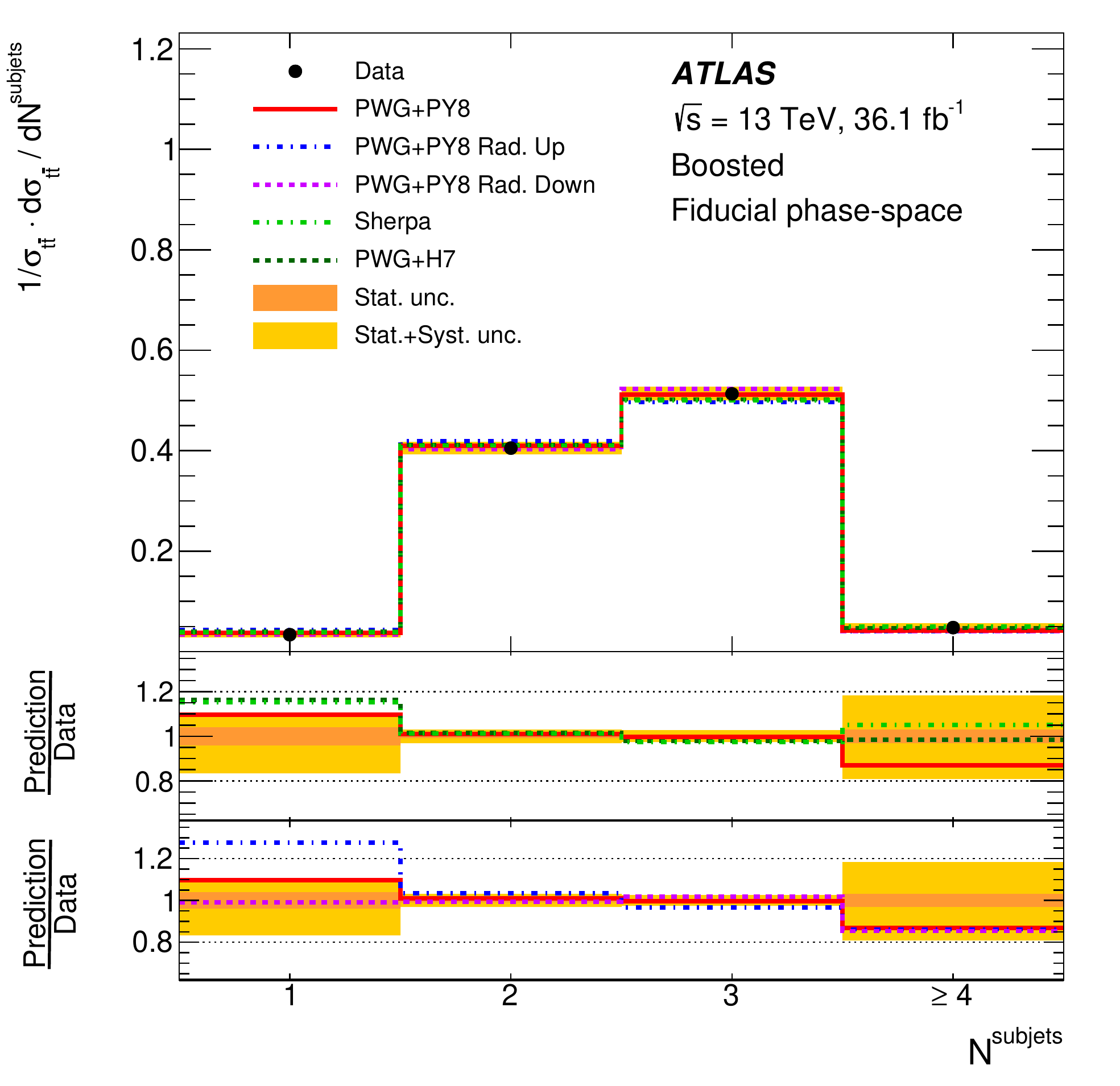}
\label{fig:results_particle:boosted:smallrj:rel}}
\caption{\small{
Particle-level normalised differential cross-sections as a function of \subref{fig:results_particle:boosted:ttbar_chi:rel}~the number of additional jets and \subref{fig:results_particle:boosted:smallrj:rel}~the number of small-$R$ jets composing the hadronically decaying top quark in the boosted topology, compared with different Monte Carlo predictions. The bands represent the statistical and total uncertainty in the data. Data points are placed at the centre of each bin. The lower panel shows the ratios of the simulations to data.}}
\label{fig:results:rel:particle:boosted:additionalvars_jets}
\end{figure*}

The single-differential cross-sections are measured as a function of the transverse momentum and absolute value of the rapidity  of the hadronically decaying top quark  as well as of the mass, transverse momentum and rapidity of the \ttb{} system and of the additional variables \absPouttlep{}, \Htt{}, \chitt{}, additional jet multiplicity and the number of small-$R$ jets reclustered inside the hadronic top. The differential cross-section as a function of the \pt{} of the top quark is also measured separately for the leading and subleading top quark. The results are shown in Figures~\ref{fig:results:rel:particle:boosted:pt_y_thad}--\ref{fig:results:rel:particle:boosted:additionalvars_jets}. The quantitative comparisons among the particle-level results and predictions, obtained with a $\chi^2$ test statistic, are shown in Tables~\ref{tab:chisquare:relative:allpred:1D:boosted:particle} and~\ref{tab:chisquare:absolute:allpred:1D:boosted:particle}, for normalised and absolute single-differential cross-sections, respectively.
In Figure~\ref{fig:results_particle:boosted:ttbar_m:abs} an example of an absolute differential cross-section in the boosted topology is given. The total uncertainty in the differential cross-section as a function of $\mtt{}$ is reduced relative to the corresponding normalised cross-section, Figure~\ref{fig:results_particle:boosted:ttbar_m:rel}. 
 
\begin{figure*}[t]
\centering
\subfigure[]{\includegraphics[width=0.38\textwidth]{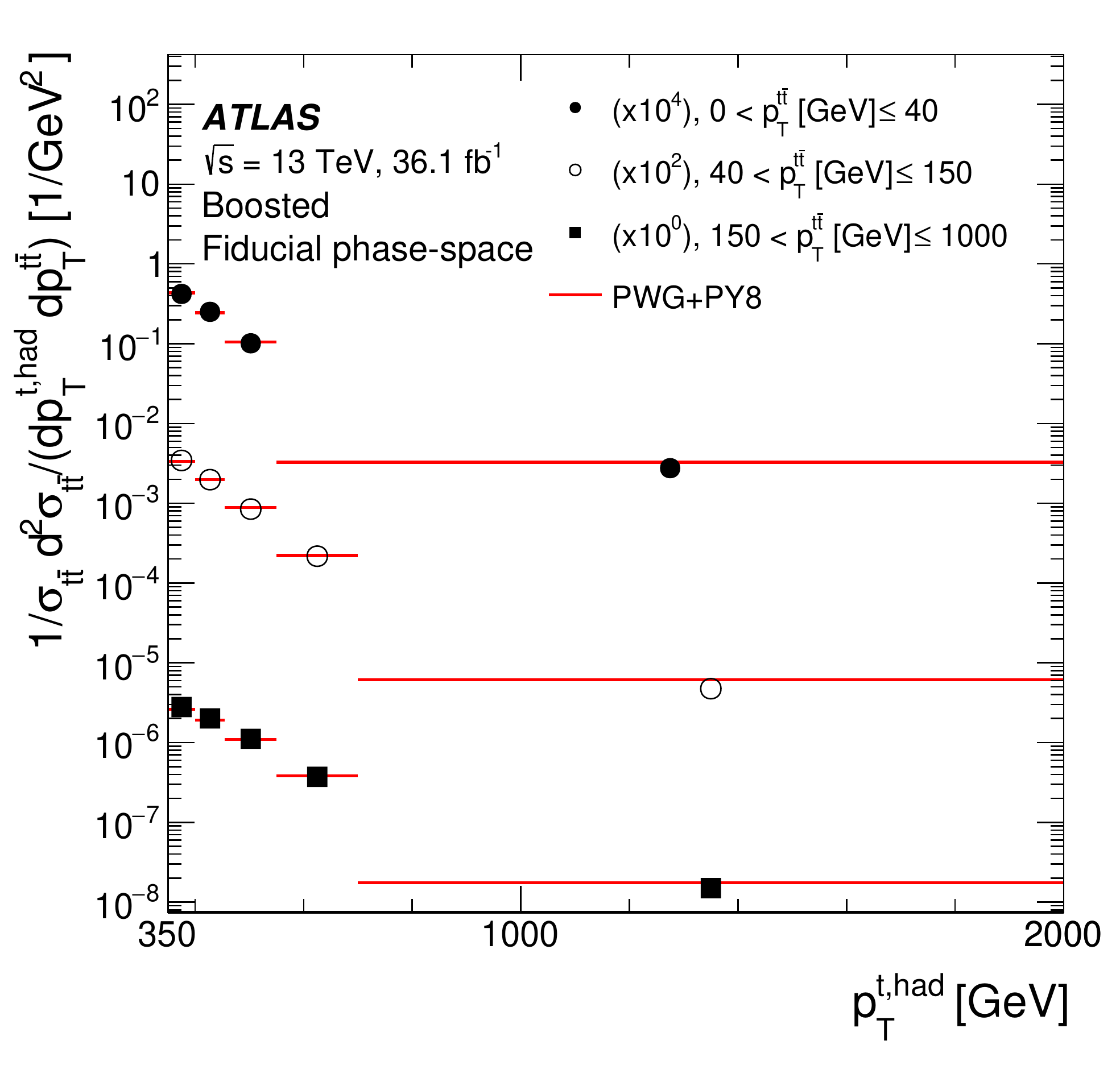}
\label{fig:results_particle:boosted:pthad:ptttbar:rel}}
\subfigure[]{\includegraphics[width=0.58\textwidth]{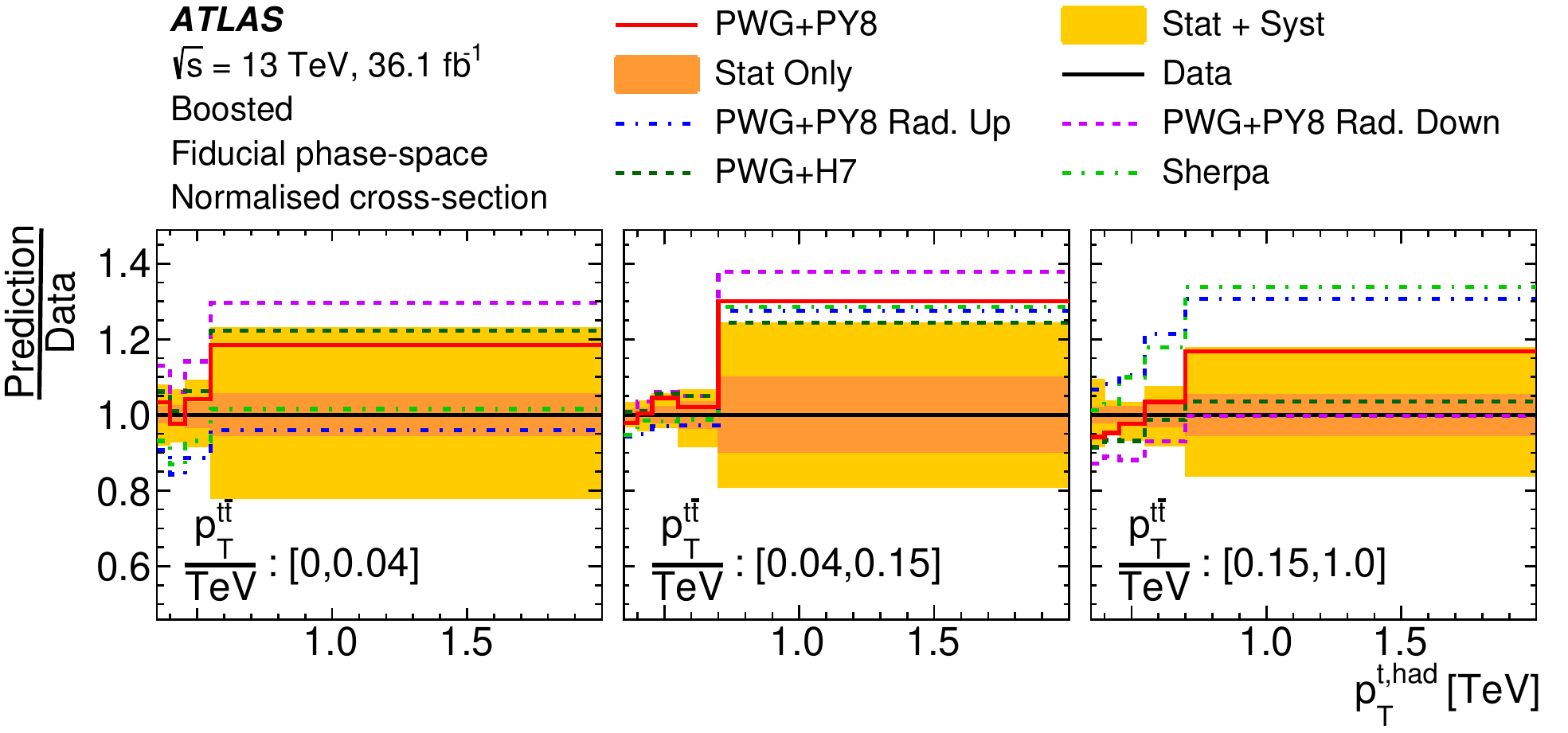}
\label{fig:results_particle:boosted:pthad:ptttbar:rel:ratio}}
\caption{\small{\subref{fig:results_particle:boosted:pthad:ptttbar:rel} Particle-level normalised differential cross-section as a function of \ptth{} in bins of \pttt{} in the boosted topology compared with the prediction obtained with the \Powheg+\PythiaEight{} MC generator.  Data points are placed at the centre of each bin. \subref{fig:results_particle:boosted:pthad:ptttbar:rel:ratio} The ratio of the measured cross-section to  different Monte Carlo predictions.  The bands represent the statistical and total uncertainty in the data.}}
\label{fig:results:rel:particle:boosted:rel:2D:comparisons:pthad_ptttbar}
\end{figure*}

\begin{figure*}[t]
\centering
\subfigure[]{\includegraphics[width=0.38\textwidth]{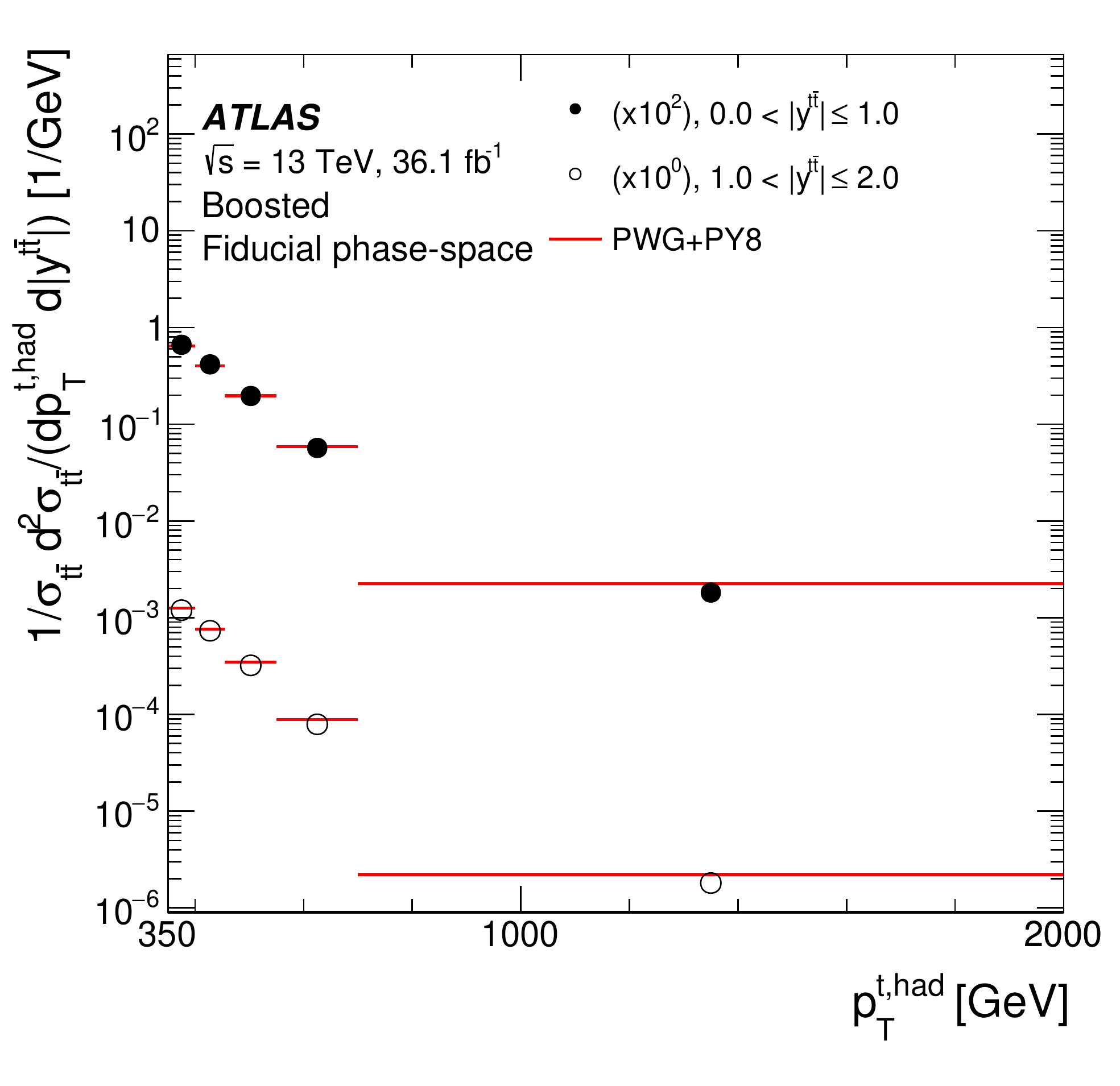}
\label{fig:results_particle:boosted:pthad:yttbar:rel}}
\subfigure[]{\includegraphics[width=0.58\textwidth]{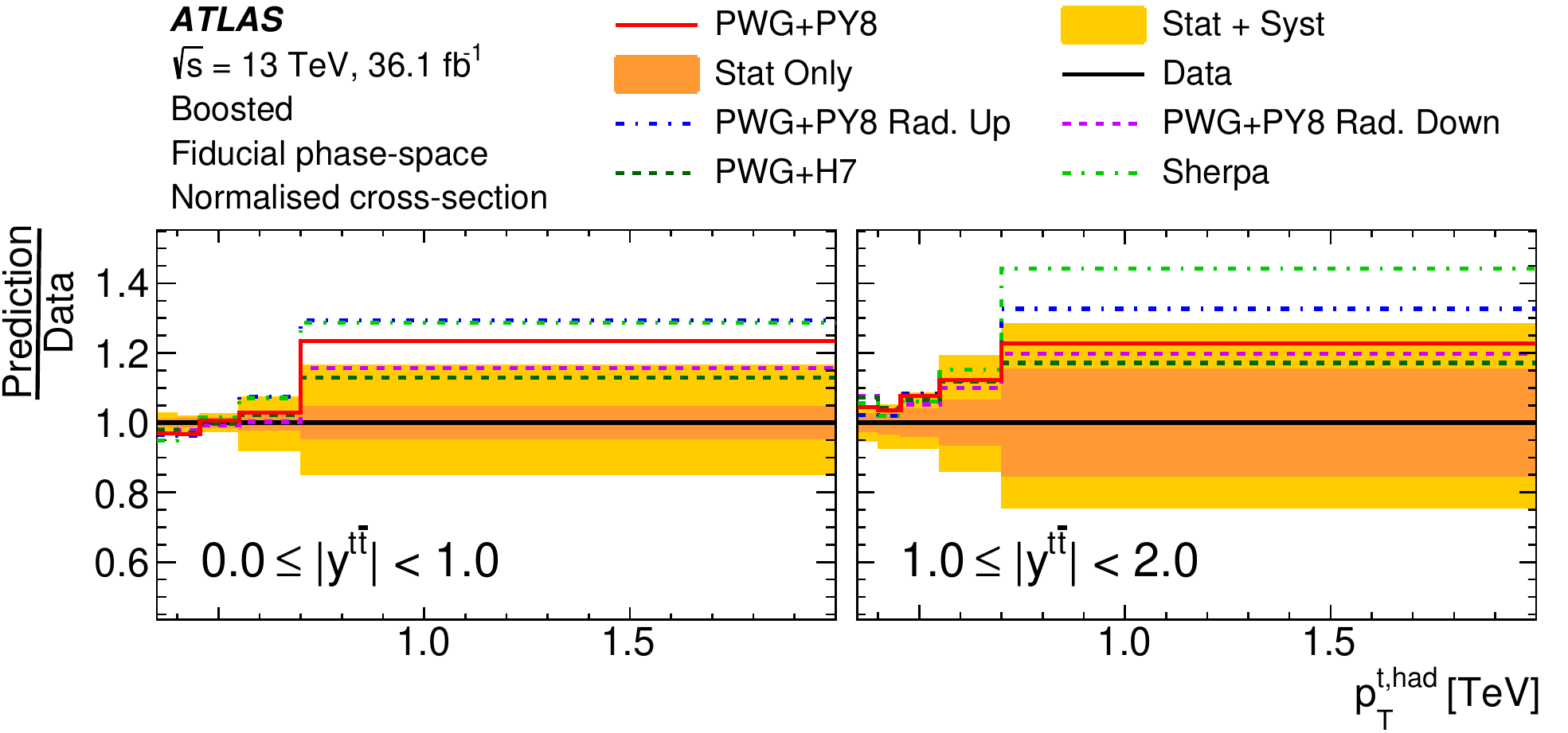}
\label{fig:results_particle:boosted:pthad:yttbar:rel:ratio}}
\caption{\small{\subref{fig:results_particle:boosted:pthad:yttbar:rel} Particle-level normalised differential cross-section as a function of \ptth{} in bins of the absolute value of the rapidity of the \ttb{} system in the boosted topology compared with the prediction obtained with the \Powheg+\PythiaEight{} MC generator.  Data points are placed at the centre of each bin. \subref{fig:results_particle:boosted:pthad:yttbar:rel:ratio} The ratio of the measured cross-section to  different Monte Carlo predictions.  The bands represent the statistical and total uncertainty in the data.}}
\label{fig:results:rel:particle:boosted:rel:2D:comparisons:pthad_yttbar}
\end{figure*}

\begin{figure*}[t]
\centering
\subfigure[]{\includegraphics[width=0.38\textwidth]{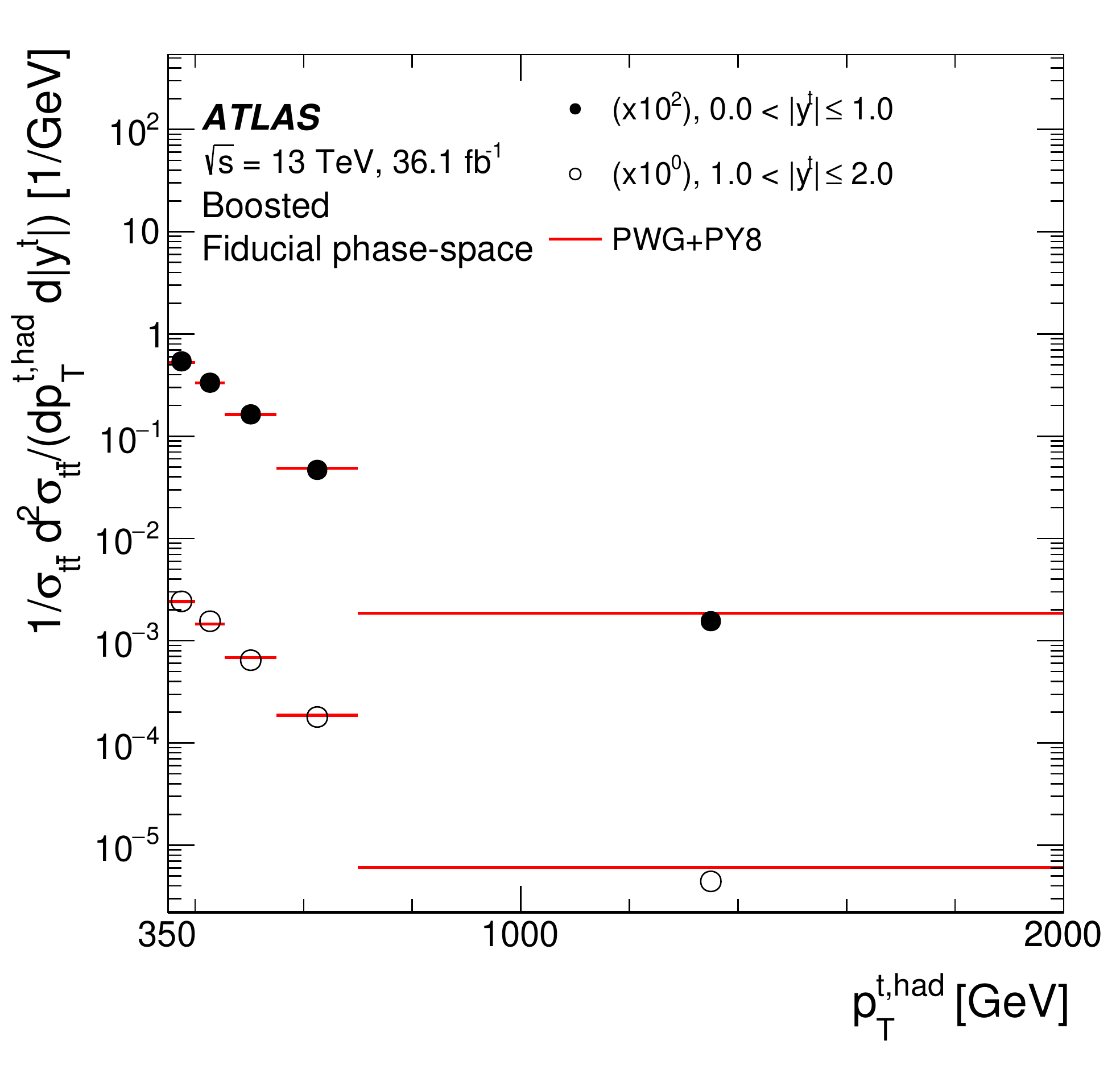}\label{fig:results_particle:boosted:pthad:yt:rel}}
\subfigure[]{\includegraphics[width=0.58\textwidth]{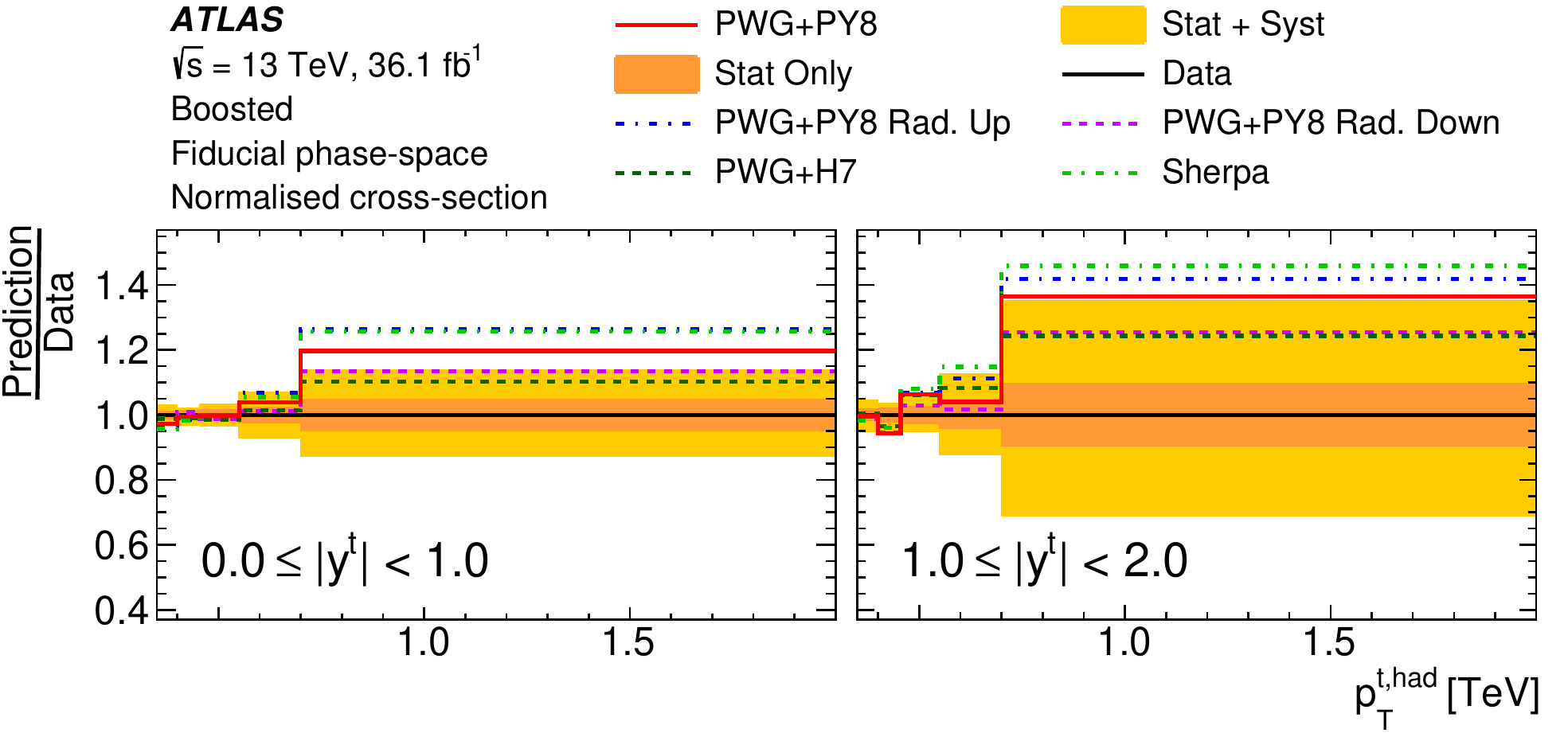}
\label{fig:results_particle:boosted:pthad:yt:rel:ratio}}
\caption{\small{\subref{fig:results_particle:boosted:pthad:yt:rel} Particle-level normalised differential cross-section as a function of \ptth{} in bins of the absolute value of the rapidity of the hadronically decaying top quark in the boosted topology compared with the prediction obtained with the \Powheg+\PythiaEight{} MC generator.  Data points are placed at the centre of each bin. \subref{fig:results_particle:boosted:pthad:yt:rel:ratio} The ratio of the measured cross-section to  different Monte Carlo predictions.  The bands represent the statistical and total uncertainty in the data.}}
\label{fig:results:rel:particle:boosted:rel:2D:comparisons:pthad_yt}
\end{figure*}

\begin{figure*}[t]
\centering
\subfigure[]{\includegraphics[width=0.38\textwidth]{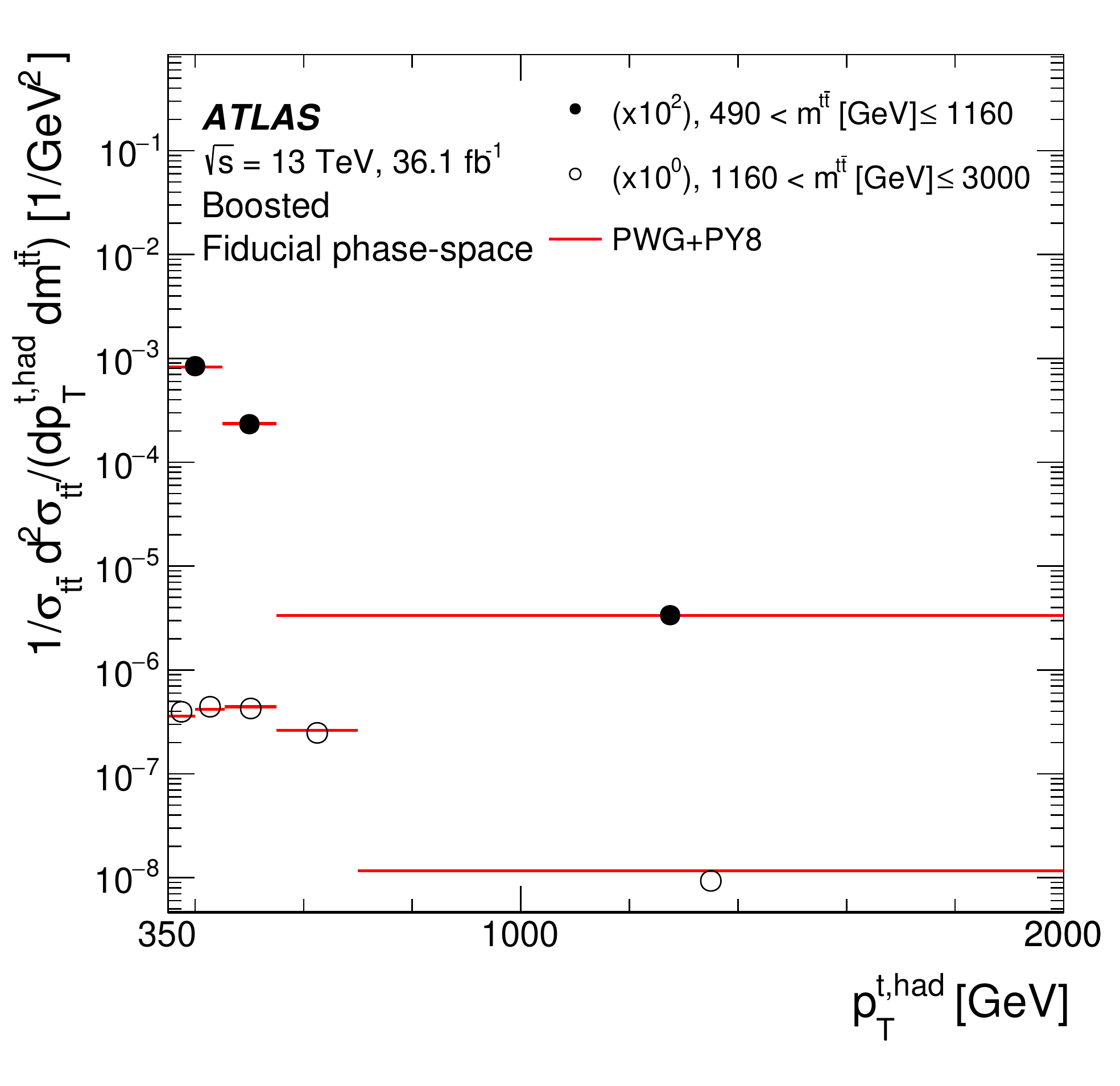}\label{fig:results_particle:boosted:pthad:mttbar:rel}}
\subfigure[]{\includegraphics[width=0.58\textwidth]{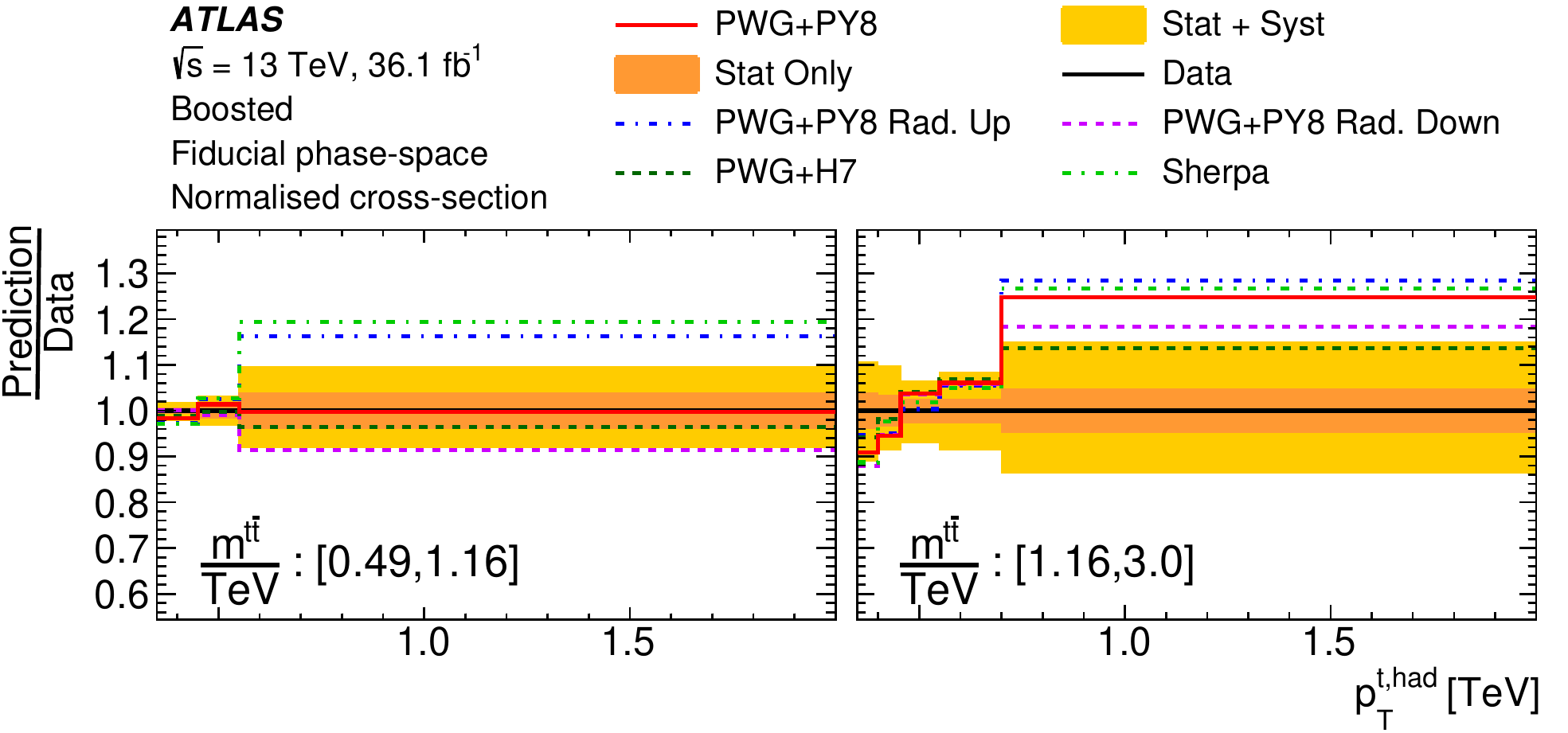}
\label{fig:results_particle:boosted:pthad:mttbar:rel:ratio}}
\caption{\small{\subref{fig:results_particle:boosted:pthad:mttbar:rel} Particle-level normalised differential cross-section as a function of \ptth{} in bins of the mass of the \ttb{} system in the boosted topology compared with the prediction obtained with the \Powheg+\PythiaEight{} MC generator.  Data points are placed at the centre of each bin. \subref{fig:results_particle:boosted:pthad:mttbar:rel:ratio} The ratio of the measured cross-section to  different Monte Carlo predictions.  The bands represent the statistical and total uncertainty in the data.}}
\label{fig:results:rel:particle:boosted:rel:2D:comparisons:pthad_mtttbar}
\end{figure*}

\begin{figure*}[t]
\centering
\subfigure[]{\includegraphics[width=0.38\textwidth]{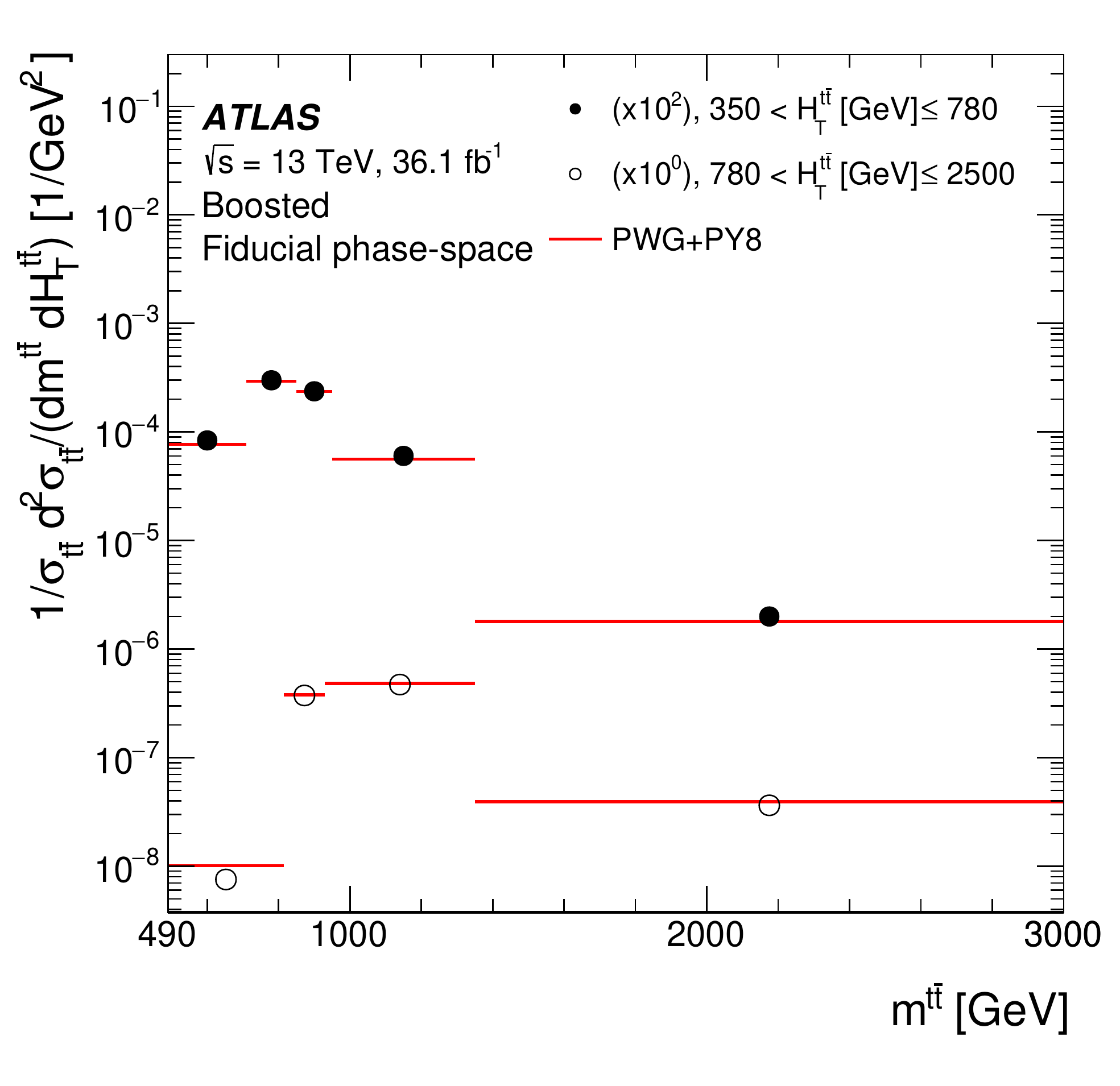}
\label{fig:results_particle:boosted:mttbar:ht:rel}}
\subfigure[]{\includegraphics[width=0.58\textwidth]{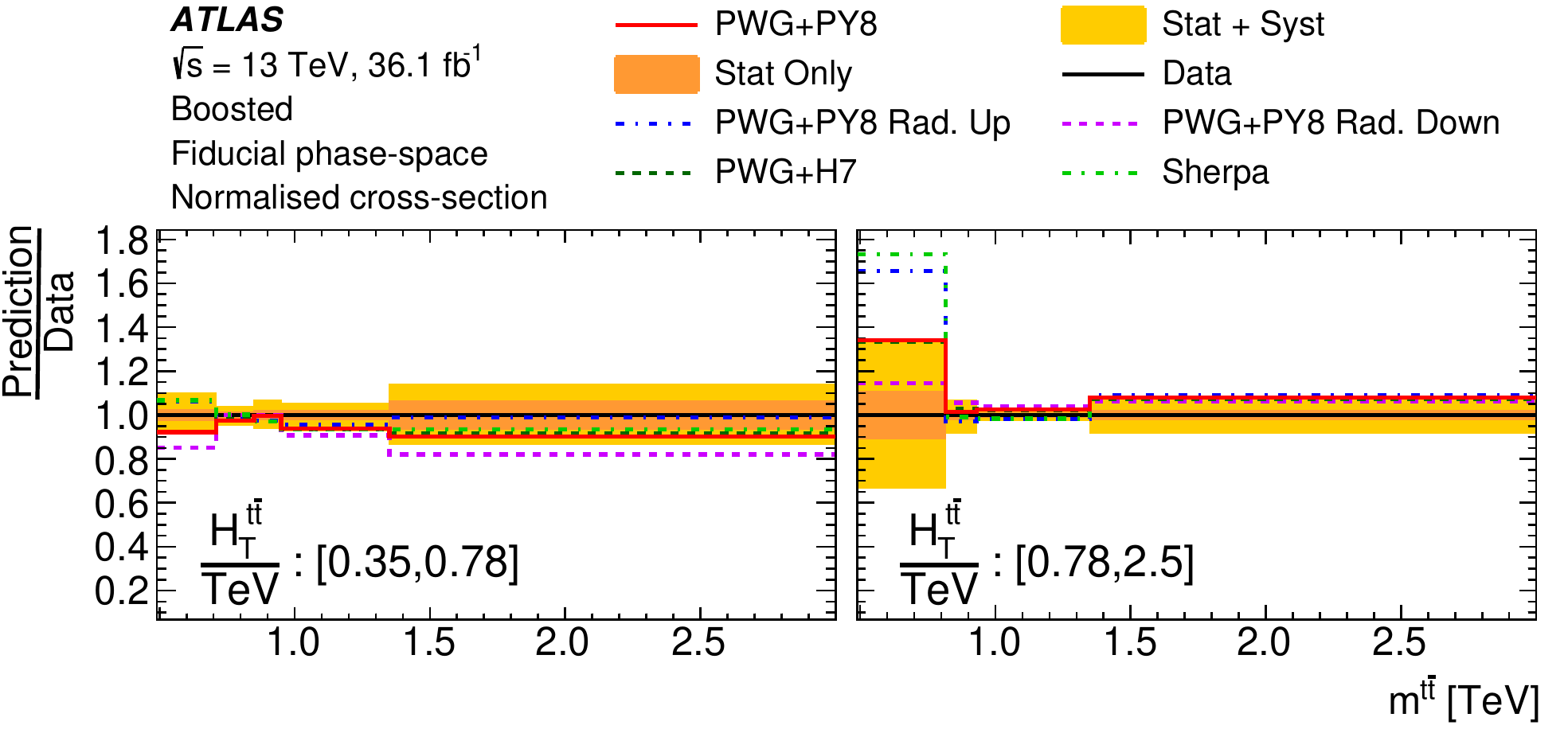}
\label{fig:results_particle:boosted:mttbar:ht:rel:ratio}}
\caption{\small{\subref{fig:results_particle:boosted:mttbar:ht:rel} Particle-level normalised differential cross-section as a function of the mass of the \ttb{} system in bins of \HTtt{} in the boosted topology compared with the prediction obtained with the \Powheg+\PythiaEight{} MC generator.  Data points are placed at the centre of each bin. \subref{fig:results_particle:boosted:mttbar:ht:rel:ratio} The ratio of the measured cross-section to  different Monte Carlo predictions.  The bands represent the statistical and total uncertainty in the data.}}
\label{fig:results:rel:particle:boosted:rel:2D:comparisons:mttbar_ht}
\end{figure*}

\begin{figure*}[t]
\centering
\subfigure[]{\includegraphics[width=0.38\textwidth]{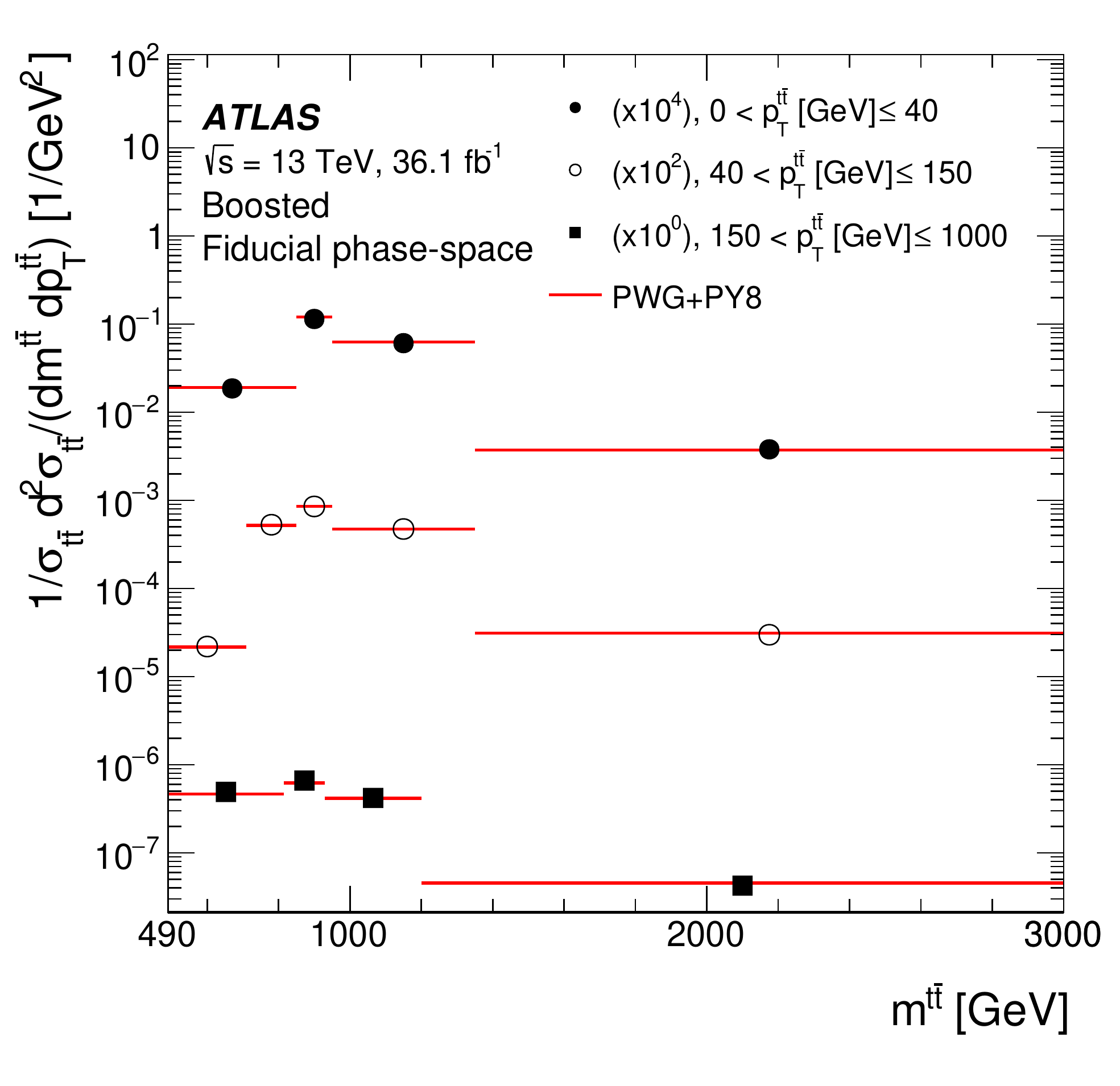}
\label{fig:results_particle:boosted:mttbar:ptttbar:rel}}
\subfigure[]{\includegraphics[width=0.58\textwidth]{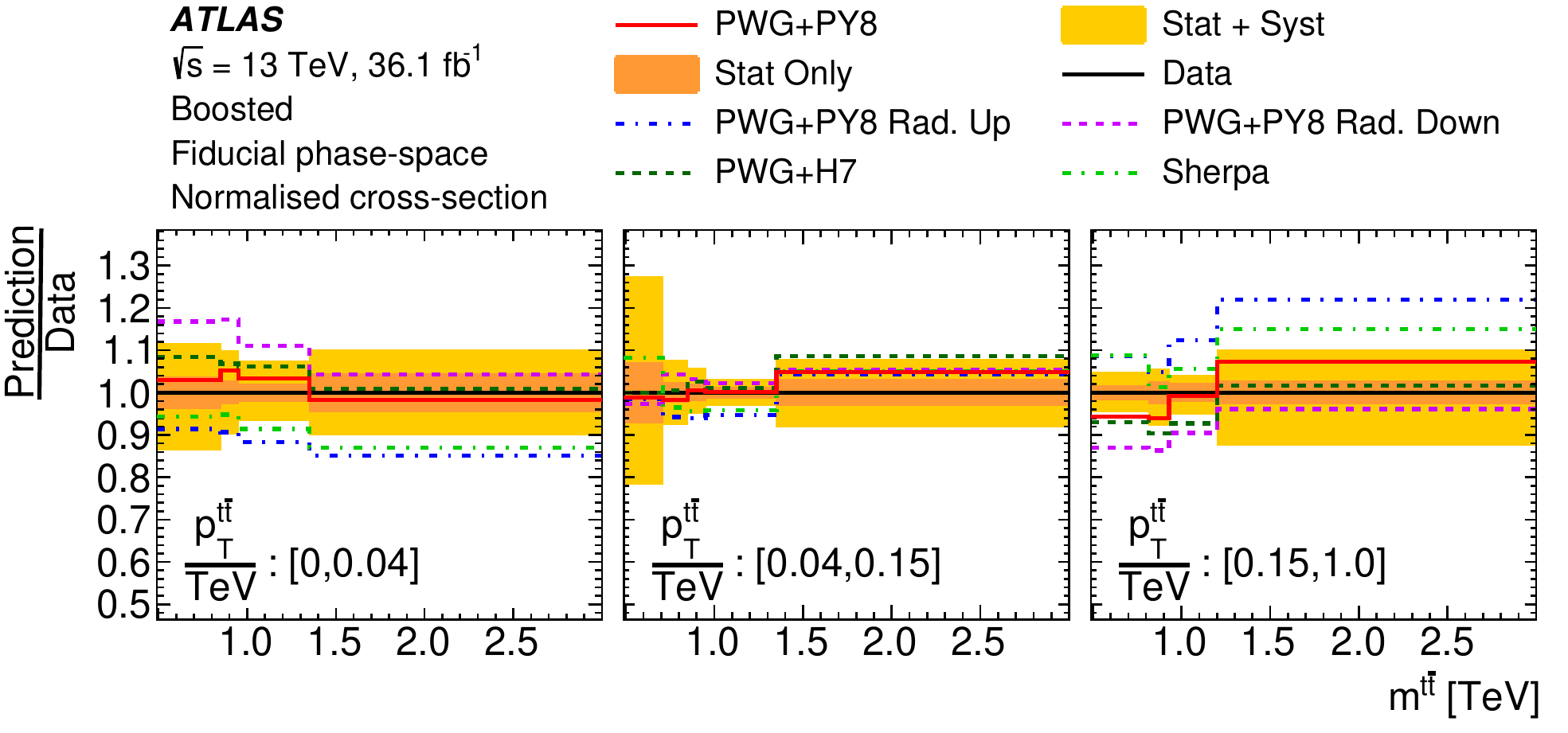}
\label{fig:results_particle:boosted:mttbar:ptttbar:rel:ratio}}
\caption{\small{\subref{fig:results_particle:boosted:mttbar:ptttbar:rel} Particle-level normalised differential cross-section as a function of the mass of the \ttb{} system in bins of \pttt{} in the boosted topology compared with the prediction obtained with the \Powheg+\PythiaEight{} MC generator.  Data points are placed at the centre of each bin. \subref{fig:results_particle:boosted:mttbar:ptttbar:rel:ratio} The ratio of the measured cross-section to  different Monte Carlo predictions.  The bands represent the statistical and total uncertainty in the data.}}
\label{fig:results:rel:particle:boosted:rel:2D:comparisons:mttbar_ptttbar}
\end{figure*}

\begin{figure*}[t]
\centering
\subfigure[]{\includegraphics[width=0.38\textwidth]{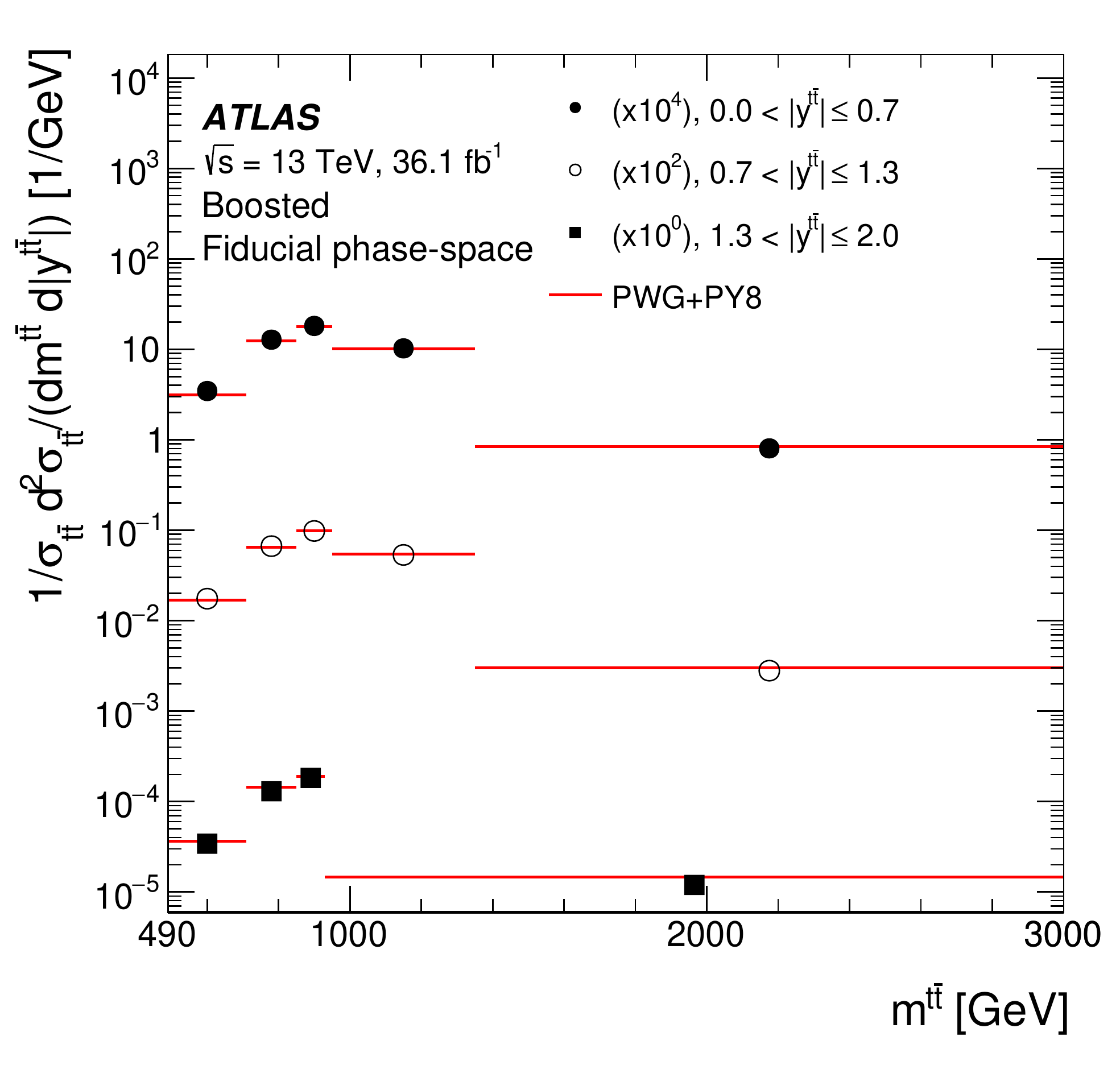}\label{fig:results_particle:boosted:mttbar:yttbar:rel}}
\subfigure[]{\includegraphics[width=0.58\textwidth]{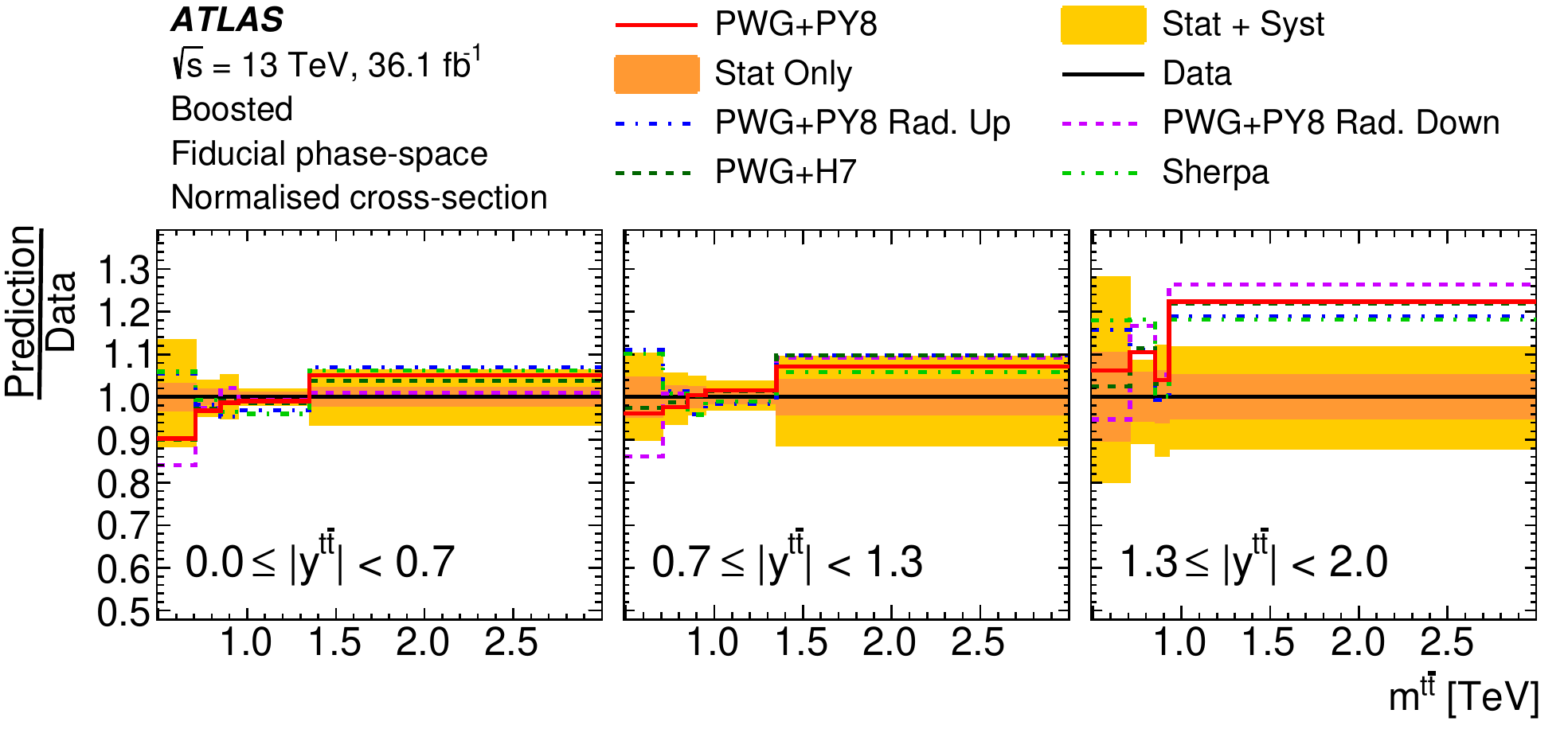}\label{fig:results_particle:boosted:mttbar:yttbar:rel:ratio}}
\caption{\small{\subref{fig:results_particle:boosted:mttbar:yttbar:rel} Particle-level normalised differential cross-section as a function of the mass of the \ttb{} system in bins of $|\ytt|$ in the boosted topology compared with the prediction obtained with the \Powheg+\PythiaEight{} MC generator.  Data points are placed at the centre of each bin. \subref{fig:results_particle:boosted:mttbar:yttbar:rel:ratio} The ratio of the measured cross-section to  different Monte Carlo predictions. The bands represent the statistical and total uncertainty in the data.}}
\label{fig:results:rel:particle:boosted:rel:2D:comparisons:mttbar_yttbar}
\end{figure*}

\begin{figure*}[t]
\centering
\subfigure[]{\includegraphics[width=0.38\textwidth]{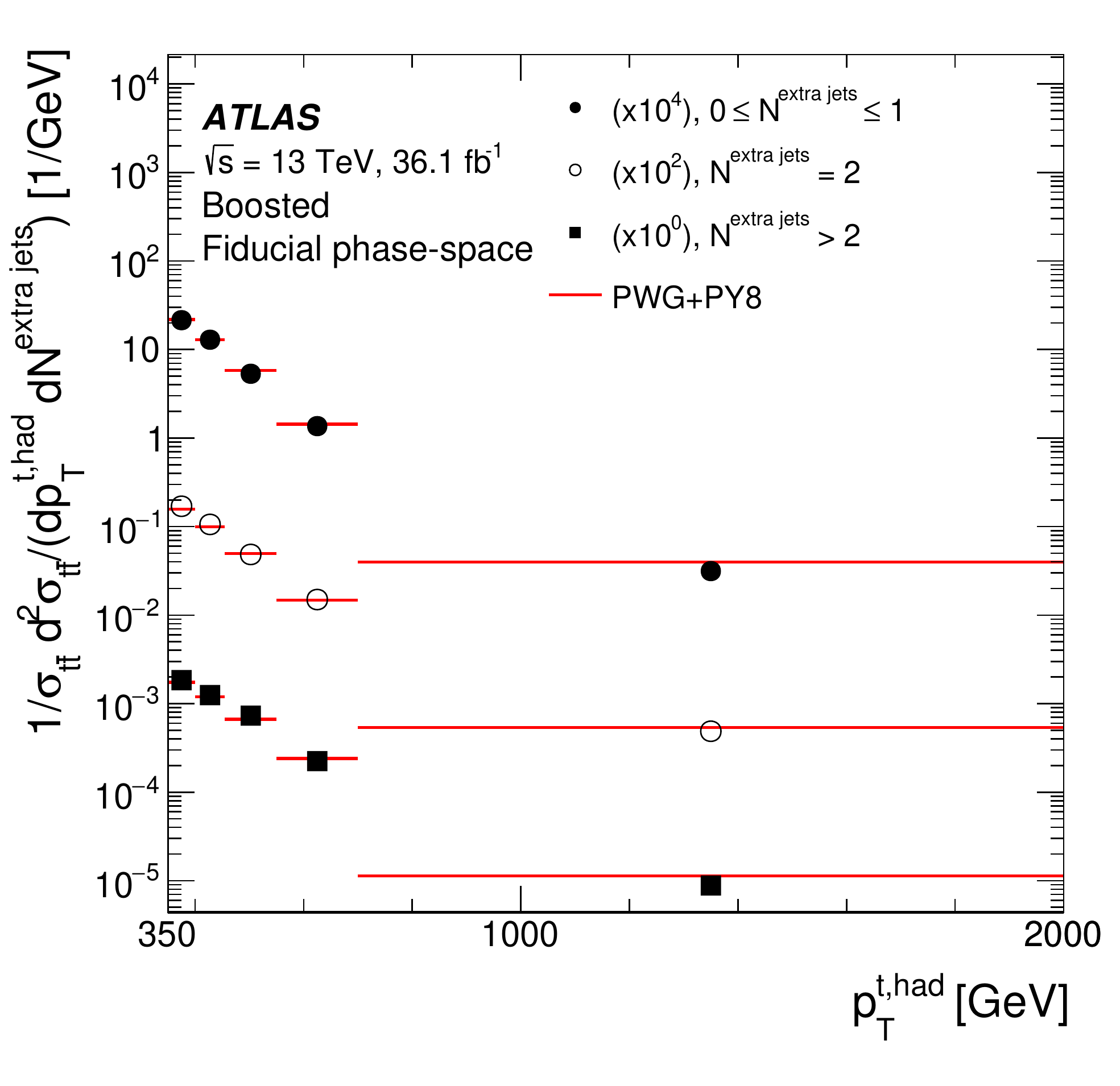}
\label{fig:results_particle:boosted:pthad:njet:rel}}
\subfigure[]{\includegraphics[width=0.58\textwidth]{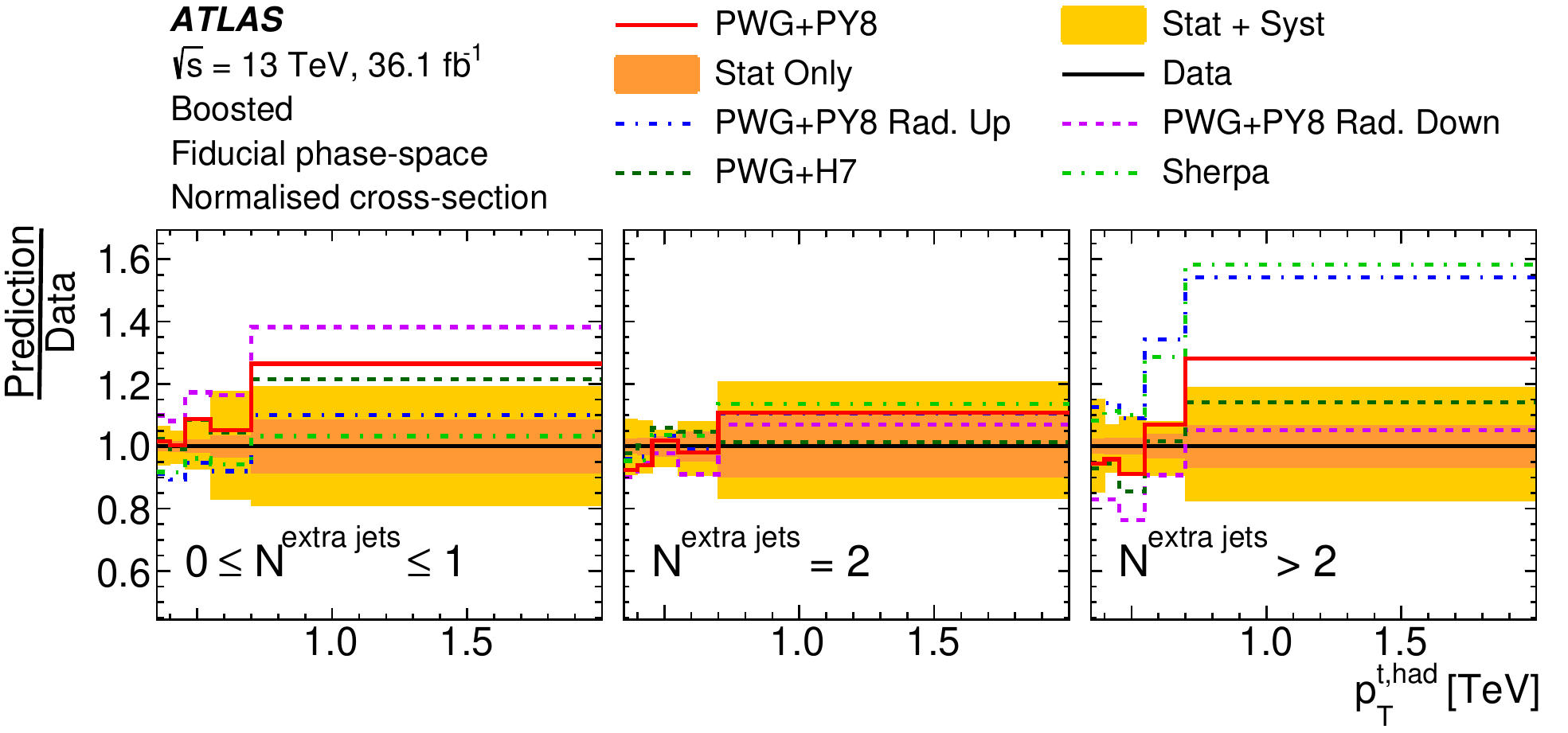}
\label{fig:results_particle:boosted:pthad:njet:rel:ratio}}
\caption{\small{\subref{fig:results_particle:boosted:pthad:njet:rel} Particle-level normalised differential cross-section as a function of the \pt{} of the hadronically decaying top quark in bins of the number of additional jets in the boosted topology compared with the prediction obtained with the \Powheg+\PythiaEight{} MC generator.  Data points are placed at the centre of each bin. \subref{fig:results_particle:boosted:pthad:njet:rel:ratio} The ratio of the measured cross-section to  different Monte Carlo predictions. The bands represent the statistical and total uncertainty in the data.}}
\label{fig:results:rel:particle:boosted:rel:2D:comparisons:pthad_njet}
\end{figure*}

\begin{figure*}[t]
\centering
\subfigure[]{\includegraphics[width=0.38\textwidth]{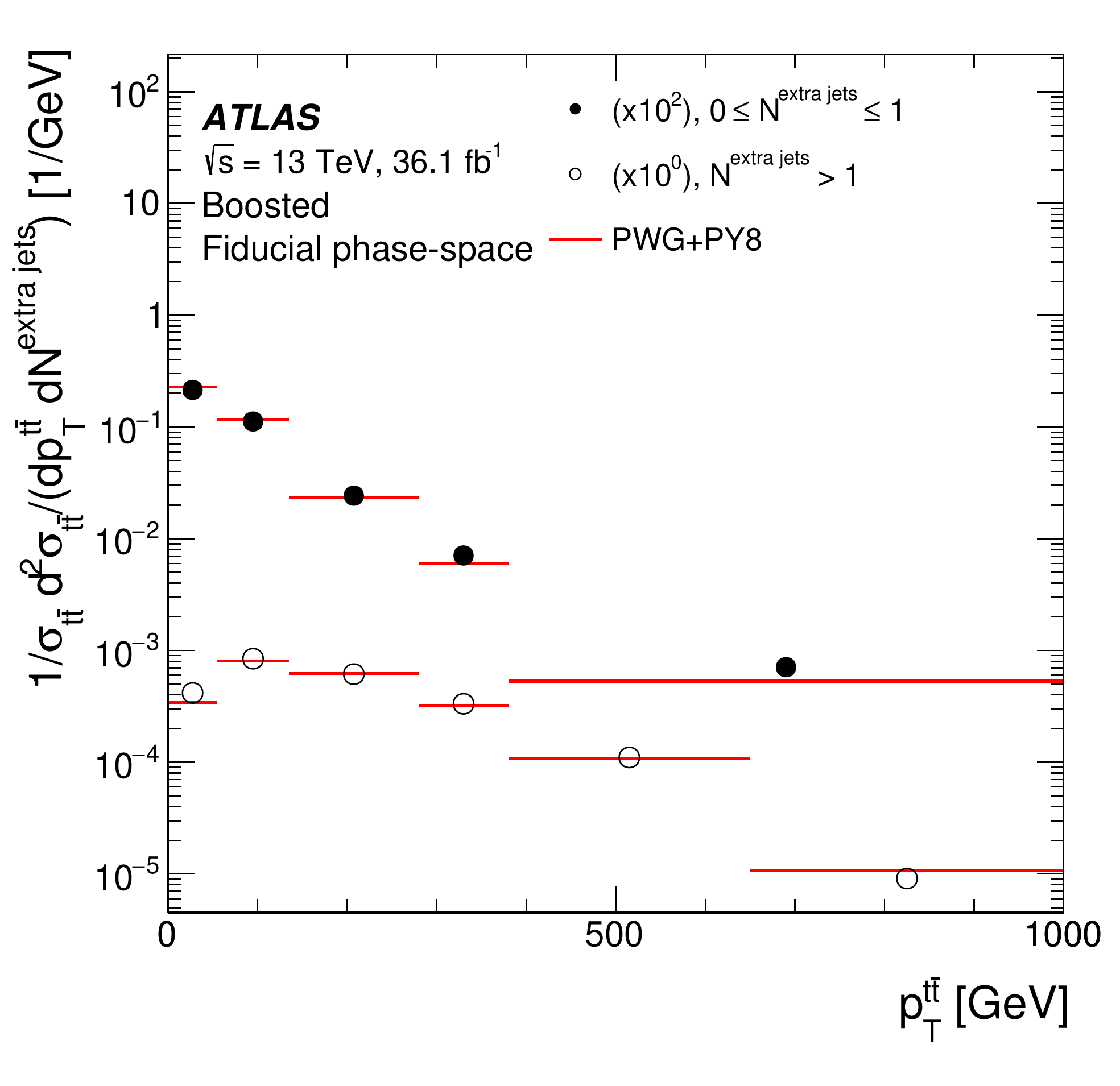}
\label{fig:results_particle:boosted:ptttbar:njet:rel}}
\subfigure[]{\includegraphics[width=0.58\textwidth]{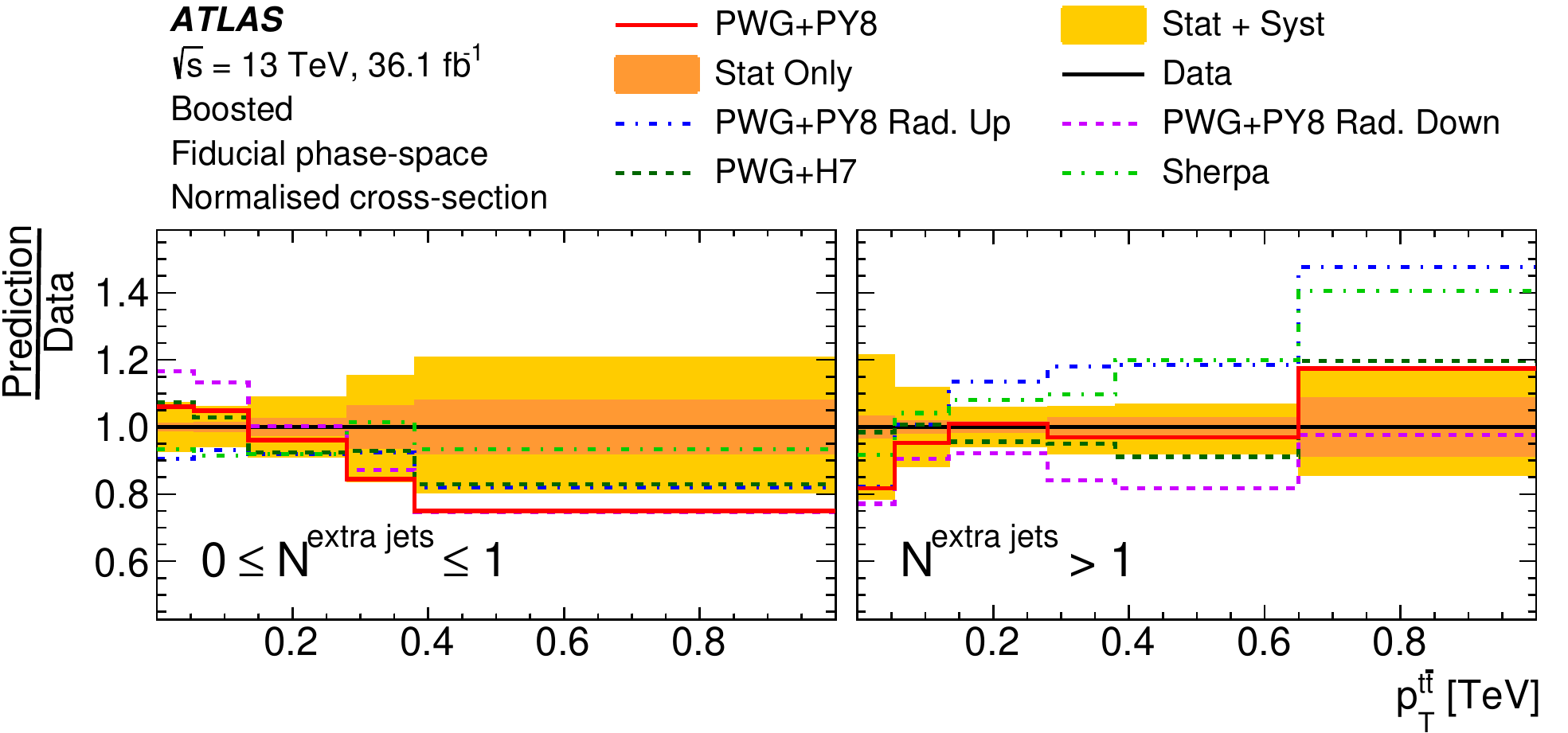}
\label{fig:results_particle:boosted:ptttbar:njet:rel:ratio}}
\caption{\small{\subref{fig:results_particle:boosted:ptttbar:njet:rel} Particle-level normalised differential cross-section as a function of the \pt{} of the \ttb{} system in bins of the number of additional jets in the boosted topology compared with the prediction obtained with the \Powheg+\PythiaEight{} MC generator.  Data points are placed at the centre of each bin. \subref{fig:results_particle:boosted:ptttbar:njet:rel:ratio} The ratio of the measured cross-section to  different Monte Carlo predictions. The bands represent the statistical and total uncertainty in the data.}}
\label{fig:results:rel:particle:boosted:rel:2D:comparisons:ptttbar_njet}
\end{figure*}

\begin{figure*}[t]
\centering
\subfigure[]{\includegraphics[width=0.38\textwidth]{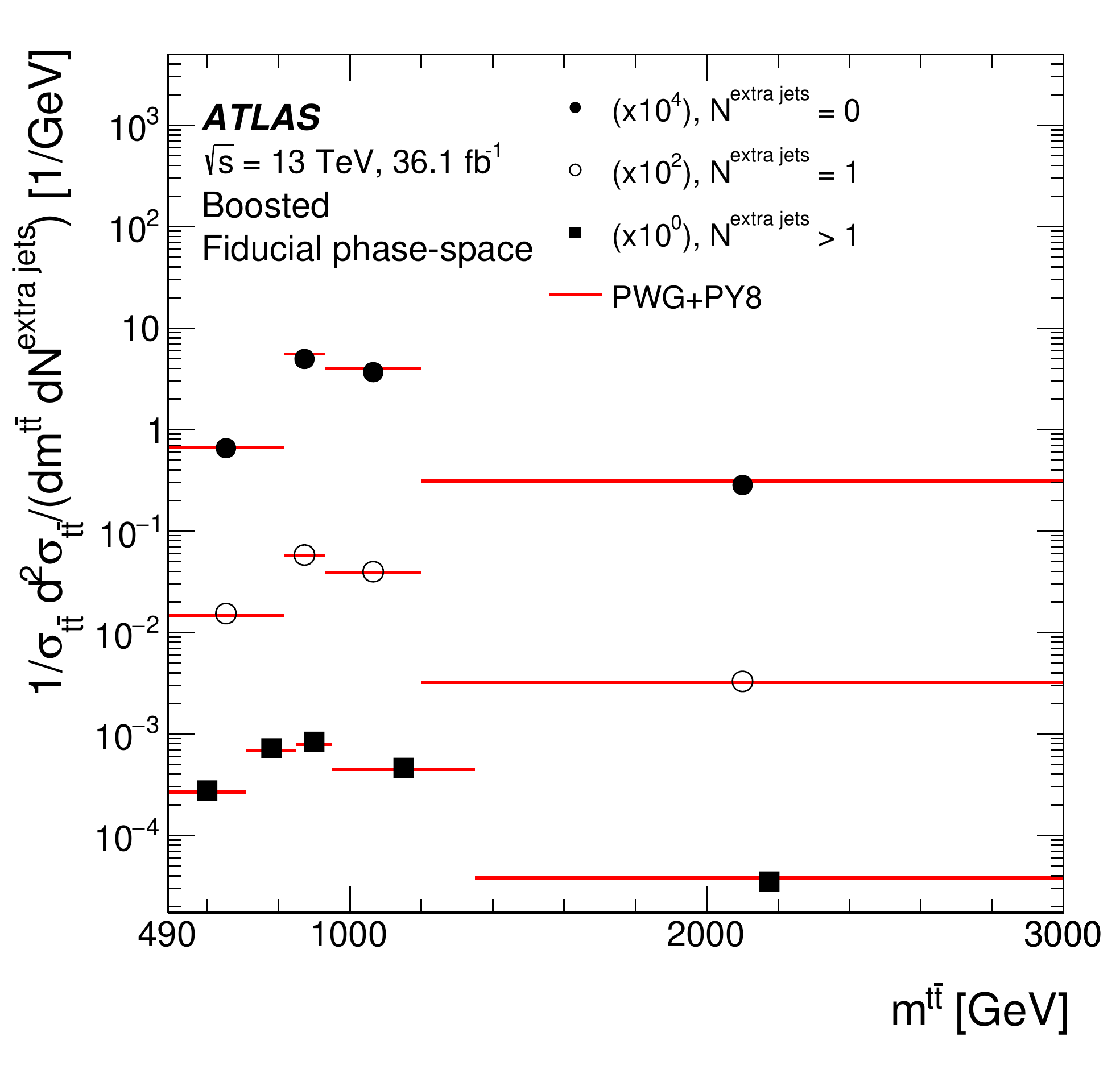}\label{fig:results_particle:boosted:mttbar:njet:rel}}
\subfigure[]{\includegraphics[width=0.58\textwidth]{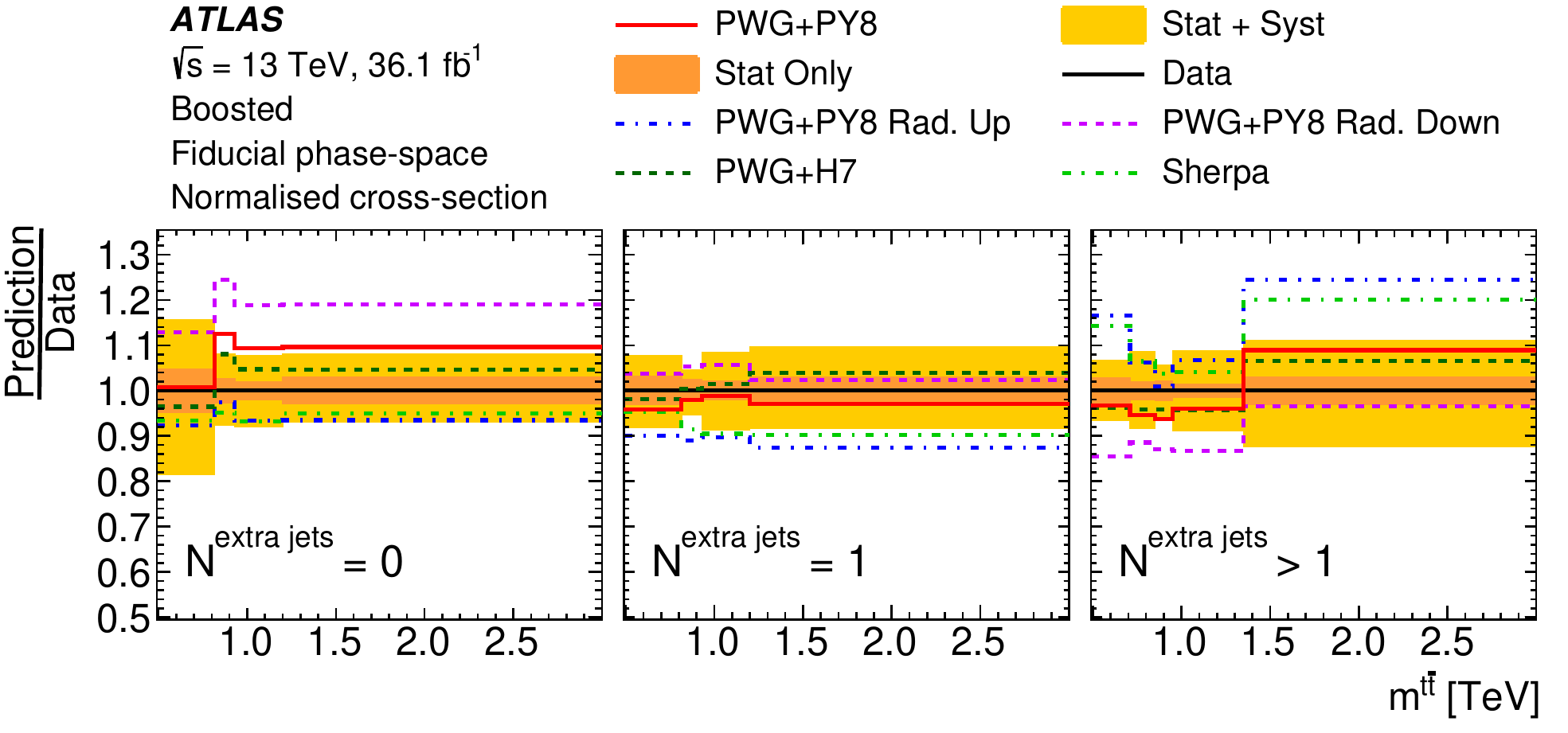}
\label{fig:results_particle:boosted:mttbar:njet:rel:ratio}}
\caption{\small{\subref{fig:results_particle:boosted:mttbar:njet:rel} Particle-level normalised differential cross-section as a function of the mass of the \ttb{} system in bins of the number of additional jets in the boosted topology compared with the prediction obtained with the \Powheg+\PythiaEight{} MC generator.  Data points are placed at the centre of each bin. \subref{fig:results_particle:boosted:mttbar:njet:rel:ratio} The ratio of the measured cross-section to  different Monte Carlo predictions. }}
\label{fig:results:rel:particle:boosted:rel:2D:comparisons:mttbar_njet}
\end{figure*}

The double-differential cross-sections, presented in Figures~\ref{fig:results:rel:particle:boosted:rel:2D:comparisons:pthad_ptttbar}--\ref{fig:results:rel:particle:boosted:rel:2D:comparisons:mttbar_njet}, are measured as a function of \ptth{} in bins of \pttt{},
\absyttbar{}, \absyt{} and \mtt{} as well as  a function of \mtt{} in bins of  \pttt{}, \absyttbar{} and \Htt{} and finally as a function of \ptth{}, \pttt{} and \mtt{} in bins of jet multiplicity. The quantitative comparisons among the particle-level results and predictions, obtained with a $\chi^2$ test statistic, are shown in Tables~\ref{tab:chisquare:relative:allpred:2D:boosted:particle} and~\ref{tab:chisquare:absolute:allpred:2D:boosted:particle}, for normalised and absolute double-differential cross-sections, respectively.

\begin{figure*}[t]
\centering
 
\includegraphics[width=0.58\textwidth]{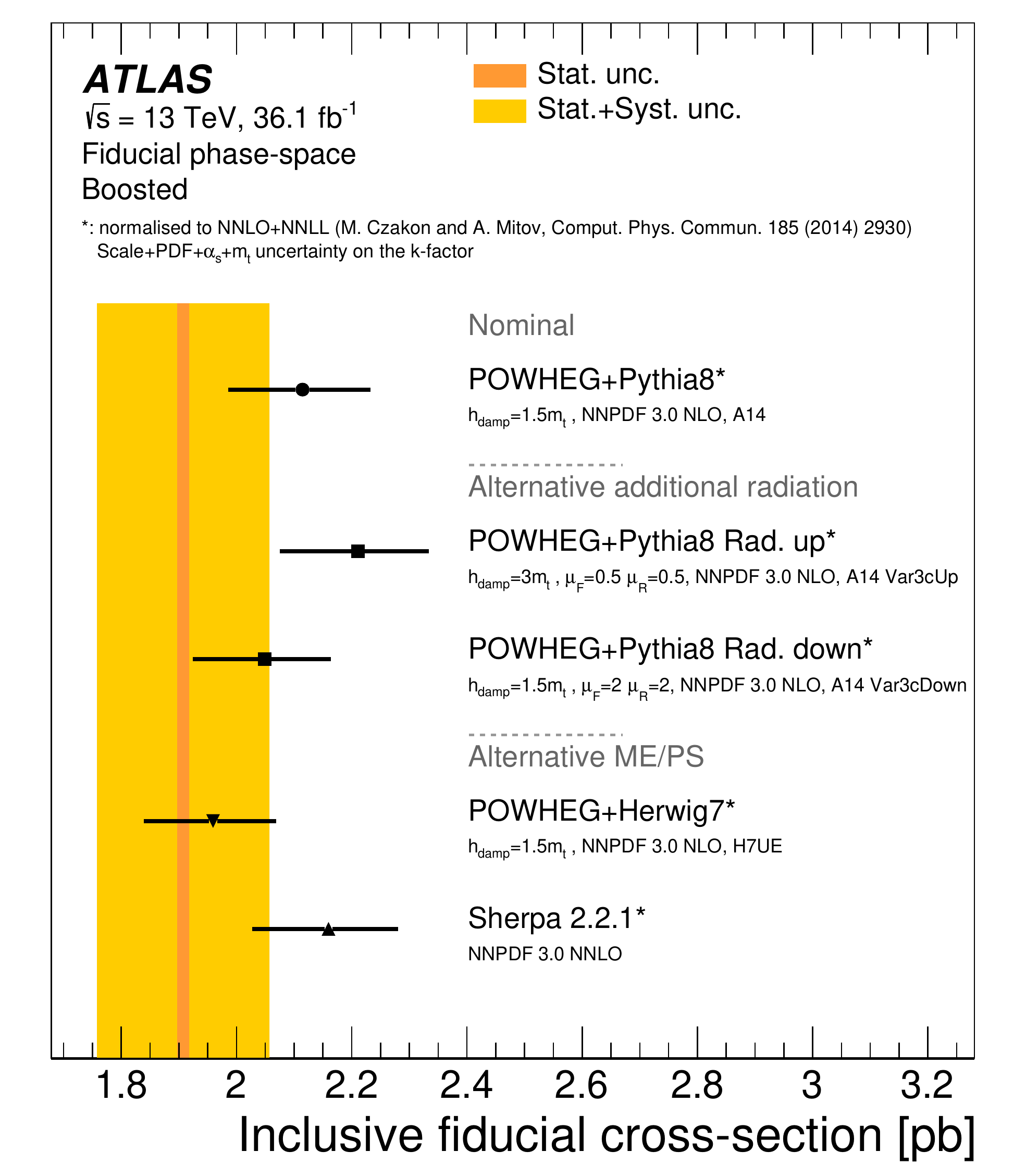}
\caption{Comparison of the measured inclusive fiducial cross-section in the boosted topology with the predictions from several MC generators. The bands represent the statistical and total uncertainty in the data. The uncertainty on the cross-section predicted by each NLO MC generator only includes the uncertainty (due to PDFs, $m_{t}$ and $\alpha_{s}$) affecting the $k$-factor used in the normalisation.}
\label{fig:results_particle:boosted:totalXs}
 
\end{figure*}

Additionally,  the total cross-section is measured in the fiducial phase-space of the boosted topology and is compared with the MC predictions previously described, as shown in Figure~\ref{fig:results_particle:boosted:totalXs}. The total cross-section predicted by each NLO MC generator is normalised to the NNLO+NNLL prediction as quoted in Ref.~\cite{Czakon:2011xx} and the corresponding uncertainty only includes the uncertainty affecting the $k$-factor used  in the normalisation. As in the case of the inclusive fiducial cross-section in the resolved topology,  the differences between the quoted fiducial cross-sections result from different acceptance predictions from each model. It is observed that several NLO+PS predictions, with the exception of \Powheg+\herwig{}7 and \Powheg+\PythiaEight{} Rad. down, overestimate the measurement of the inclusive cross-section.
 
The MC predictions are not always able to describe the measured single-differential cross-sections in the entire fiducial phase-space; mismodelling  is observed, in particular, for the differential cross-section as a function of the \pt{} of the hadronic top quark, shown in Figure~\ref{fig:results_particle:boosted:topHad_pt:rel}, for the differential cross-section as a function of \mtt{}, shown in Figure~\ref{fig:results_particle:boosted:ttbar_m:rel}, and for the observable \HTtt{}, shown in Figure~\ref{fig:results_particle:boosted:ht:rel},  where all the MC predictions tend to overestimate the data in the tails of the distributions. A similar trend is observed for the differential cross-sections as a function of the transverse momentum of the leading and subleading top quark (shown in Figure~\ref{fig:results:rel:particle:boosted:leading}). To a smaller extent, discrepancies are observed
at high values of \absyttbar{}, shown in  Figure~\ref{fig:results_particle:boosted:ttbar_eta:rel}, and in the tails of the \absPouttlep{} distribution, shown in Figure~\ref{fig:results_particle:boosted:abs_pout:rel}.
 
The tensions between the MC predictions and the data are observed also in the measured double-differential cross-sections, in particular for the cross-sections as a function  of \ptth{} in bins of \absyttbar{}, \absyt{} and \mtt{} (shown in Figures~\ref{fig:results:rel:particle:boosted:rel:2D:comparisons:pthad_yttbar}--\ref{fig:results:rel:particle:boosted:rel:2D:comparisons:pthad_mtttbar}) and as
a function of \mtt{} in bins of \absyttbar{} (shown in Figure~\ref{fig:results:rel:particle:boosted:rel:2D:comparisons:mttbar_yttbar}).
As in the case of the double-differential cross-sections in the resolved topology, the measurements allow discrimination between the different MC predictions.
Overall, for the double-differential cross-sections, the MC predictions obtained from \Powheg+\herwig{}7 provide the better
description of the data while those from \SHERPAV{2.2.1} and \Powheg+\PythiaEight{} with the Var3cDown tuning show a significant disagreement with  the data, as also observed in the resolved topology to a smaller extent.

\begin{figure}[t]
\centering
\includegraphics[width=0.75\textwidth]{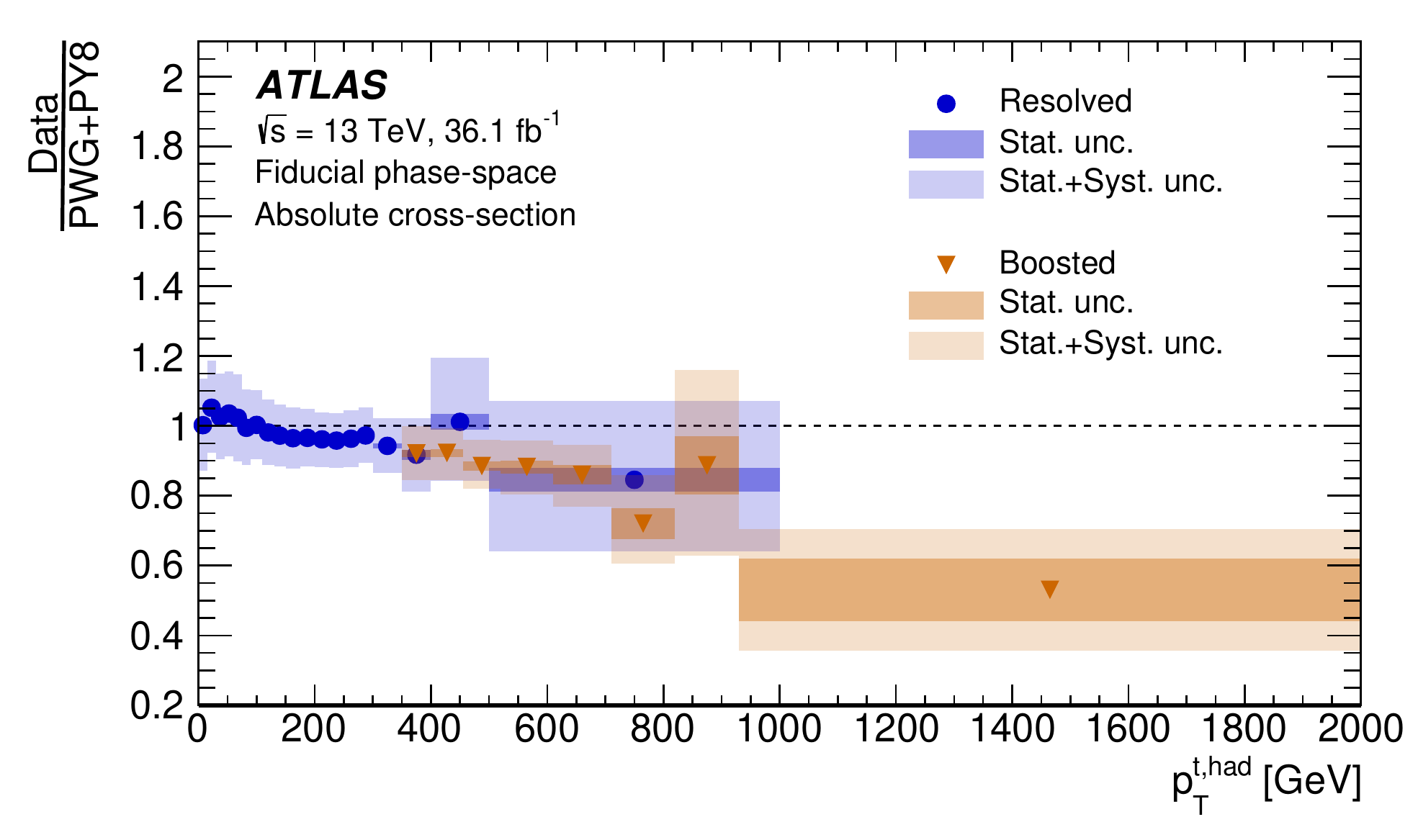}
\caption{\small{The ratios of the measured fiducial phase-space absolute differential cross-sections to the predictions obtained with the \Powheg+\PythiaEight{} MC generator in the resolved and boosted topologies as a function of the transverse momentum
of the hadronic top quark. The bands indicate the statistical and total uncertainties of the data in each bin. }}
\label{fig:resolved_boosted_ratio:particle}
\end{figure}

Since the definitions of the phase-space and the particle-level hadronic top quark differ between the resolved and boosted topologies, a direct comparison of the measured differential cross-sections is not possible. However, it can be seen in Figure~\ref{fig:resolved_boosted_ratio:particle} that the ratio of data to prediction is consistent between the measured absolute differential cross-sections in the overlap region of the two topologies.

\begin{table}[t]
\footnotesize
\centering\noindent\makebox[\textwidth]{
\renewcommand*{\arraystretch}{1.2}\begin{tabular}{|c | r @{/} l c  | r @{/} l c  | r @{/} l c  | r @{/} l c  | r @{/} l c |}
\hline
Observable
& \multicolumn{3}{c|}{\textsc{Pwg+Py8}}& \multicolumn{3}{c|}{\textsc{Pwg+Py8} Rad.~Up}& \multicolumn{3}{c|}{\textsc{Pwg+Py8} Rad.~Down}& \multicolumn{3}{c|}{\textsc{Pwg+H7}}& \multicolumn{3}{c|}{\textsc{Sherpa} 2.2.1}\\
& \multicolumn{2}{c}{$\chi^{2}$/NDF} &  ~$p$-value& \multicolumn{2}{c}{$\chi^{2}$/NDF} &  ~$p$-value& \multicolumn{2}{c}{$\chi^{2}$/NDF} &  ~$p$-value& \multicolumn{2}{c}{$\chi^{2}$/NDF} &  ~$p$-value& \multicolumn{2}{c}{$\chi^{2}$/NDF} &  ~$p$-value\\
\hline
\hline
$p_{\mathrm{T}}^{t,1}$ &{\ } 6.2 & 7 & 0.51 & {\ } 10.3 & 7 & 0.17 & {\ } 2.8 & 7 & 0.90 & {\ } 2.4 & 7 & 0.93 & {\ } 11.1 & 7 & 0.14\\
$p_{\mathrm{T}}^{t,2}$ &{\ } 4.0 & 6 & 0.68 & {\ } 3.9 & 6 & 0.69 & {\ } 4.1 & 6 & 0.66 & {\ } 3.2 & 6 & 0.78 & {\ } 4.4 & 6 & 0.62\\
$H_{\mathrm{T}}^{t\bar{t}}$ &{\ } 9.0 & 9 & 0.44 & {\ } 7.1 & 9 & 0.62 & {\ } 24.1 & 9 & $<$0.01 & {\ } 10.4 & 9 & 0.32 & {\ } 7.8 & 9 & 0.56\\
$\absPouttlep$ &{\ } 7.1 & 6 & 0.31 & {\ } 17.2 & 6 & $<$0.01 & {\ } 43.3 & 6 & $<$0.01 & {\ } 25.4 & 6 & $<$0.01 & {\ } 2.9 & 6 & 0.82\\
$     \chi^{t\bar{t}}$ &{\ } 3.5 & 6 & 0.74 & {\ } 1.0 & 6 & 0.98 & {\ } 18.4 & 6 & $<$0.01 & {\ } 3.2 & 6 & 0.79 & {\ } 8.9 & 6 & 0.18\\
$N^{\mathrm{extra jets}}$ &{\ } 5.5 & 4 & 0.24 & {\ } 15.7 & 4 & $<$0.01 & {\ } 17.0 & 4 & $<$0.01 & {\ } 2.5 & 4 & 0.64 & {\ } 8.6 & 4 & 0.07\\
$p_{\mathrm{T}}^{t,\mathrm{had}}$ &{\ } 6.2 & 7 & 0.52 & {\ } 11.0 & 7 & 0.14 & {\ } 3.2 & 7 & 0.86 & {\ } 3.5 & 7 & 0.83 & {\ } 10.6 & 7 & 0.16\\
$N^{\mathrm{subjets}}$ &{\ } 0.3 & 3 & 0.95 & {\ } 4.3 & 3 & 0.23 & {\ } 0.7 & 3 & 0.86 & {\ } 2.1 & 3 & 0.55 & {\ } 2.6 & 3 & 0.46\\
$|y^{t,\mathrm{had}}|$ &{\ } 0.6 & 3 & 0.90 & {\ } 0.5 & 3 & 0.93 & {\ } 1.5 & 3 & 0.68 & {\ } 0.6 & 3 & 0.90 & {\ } 1.2 & 3 & 0.75\\
$      |y^{t\bar{t}}|$ &{\ } 3.2 & 3 & 0.36 & {\ } 1.9 & 3 & 0.60 & {\ } 4.5 & 3 & 0.21 & {\ } 5.2 & 3 & 0.16 & {\ } 4.2 & 3 & 0.24\\
$        m^{t\bar{t}}$ &{\ } 7.5 & 9 & 0.59 & {\ } 11.8 & 9 & 0.23 & {\ } 16.2 & 9 & 0.06 & {\ } 8.1 & 9 & 0.52 & {\ } 8.3 & 9 & 0.50\\
$p_{\mathrm{T}}^{t\bar{t}}$ &{\ } 3.5 & 5 & 0.63 & {\ } 25.6 & 5 & $<$0.01 & {\ } 35.7 & 5 & $<$0.01 & {\ } 9.8 & 5 & 0.08 & {\ } 19.7 & 5 & $<$0.01\\
\hline
\end{tabular}}
\caption{ Comparison of the measured particle-level normalised single-differential cross-sections in the boosted topology with the predictions from several MC generators. For each prediction a $\chi^2$ and a $p$-value are calculated using the covariance matrix of the measured spectrum. The NDF is equal to the number of bins in the distribution minus one.  
}
\label{tab:chisquare:relative:allpred:1D:boosted:particle}
\end{table}
 
\begin{table}[t]
\footnotesize
\centering\noindent\makebox[\textwidth]{
\renewcommand*{\arraystretch}{1.2}\begin{tabular}{|c | r @{/} l r  | r @{/} l r  | r @{/} l r  | r @{/} l r  | r @{/} l r |}
\hline
Observable
& \multicolumn{3}{c|}{\textsc{Pwg+Py8}}& \multicolumn{3}{c|}{\textsc{Pwg+Py8} Rad.~Up}& \multicolumn{3}{c|}{\textsc{Pwg+Py8} Rad.~Down}& \multicolumn{3}{c|}{\textsc{Pwg+H7}}& \multicolumn{3}{c|}{\textsc{Sherpa} 2.2.1}\\
& \multicolumn{2}{c}{$\chi^{2}$/NDF} &  ~$p$-value& \multicolumn{2}{c}{$\chi^{2}$/NDF} &  ~$p$-value& \multicolumn{2}{c}{$\chi^{2}$/NDF} &  ~$p$-value& \multicolumn{2}{c}{$\chi^{2}$/NDF} &  ~$p$-value& \multicolumn{2}{c}{$\chi^{2}$/NDF} &  ~$p$-value\\
\hline
\hline
$p_{\mathrm{T}}^{t,1}$ &{\ } 7.8 & 8 & 0.46 & {\ } 14.1 & 8 & 0.08 & {\ } 3.9 & 8 & 0.86 & {\ } 2.8 & 8 & 0.95 & {\ } 12.9 & 8 & 0.11\\
$p_{\mathrm{T}}^{t,2}$ &{\ } 5.3 & 7 & 0.62 & {\ } 6.6 & 7 & 0.47 & {\ } 5.7 & 7 & 0.58 & {\ } 5.6 & 7 & 0.59 & {\ } 4.8 & 7 & 0.68\\
$H_{\mathrm{T}}^{t\bar{t}}$ &{\ } 10.9 & 10 & 0.37 & {\ } 10.5 & 10 & 0.40 & {\ } 15.5 & 10 & 0.12 & {\ } 7.0 & 10 & 0.72 & {\ } 11.4 & 10 & 0.33\\
$\absPouttlep$ &{\ } 24.2 & 7 & $<$0.01 & {\ } 21.7 & 7 & $<$0.01 & {\ } 72.0 & 7 & $<$0.01 & {\ } 31.9 & 7 & $<$0.01 & {\ } 9.9 & 7 & 0.19\\
$     \chi^{t\bar{t}}$ &{\ } 12.9 & 7 & 0.07 & {\ } 9.2 & 7 & 0.24 & {\ } 32.0 & 7 & $<$0.01 & {\ } 4.5 & 7 & 0.72 & {\ } 17.2 & 7 & 0.02\\
$N^{\mathrm{extra jets}}$ &{\ } 38.5 & 5 & $<$0.01 & {\ } 46.0 & 5 & $<$0.01 & {\ } 57.0 & 5 & $<$0.01 & {\ } 4.7 & 5 & 0.45 & {\ } 33.4 & 5 & $<$0.01\\
$p_{\mathrm{T}}^{t,\mathrm{had}}$ &{\ } 9.2 & 8 & 0.33 & {\ } 16.0 & 8 & 0.04 & {\ } 5.9 & 8 & 0.66 & {\ } 4.5 & 8 & 0.81 & {\ } 12.0 & 8 & 0.15\\
$N^{\mathrm{subjets}}$ &{\ } 7.6 & 4 & 0.11 & {\ } 11.2 & 4 & 0.02 & {\ } 8.1 & 4 & 0.09 & {\ } 1.3 & 4 & 0.87 & {\ } 3.6 & 4 & 0.46\\
$|y^{t,\mathrm{had}}|$ &{\ } 4.0 & 4 & 0.41 & {\ } 5.8 & 4 & 0.21 & {\ } 3.9 & 4 & 0.42 & {\ } 2.3 & 4 & 0.68 & {\ } 10.6 & 4 & 0.03\\
$      |y^{t\bar{t}}|$ &{\ } 8.8 & 4 & 0.07 & {\ } 10.3 & 4 & 0.04 & {\ } 8.1 & 4 & 0.09 & {\ } 6.7 & 4 & 0.15 & {\ } 10.5 & 4 & 0.03\\
$        m^{t\bar{t}}$ &{\ } 16.5 & 10 & 0.09 & {\ } 28.5 & 10 & $<$0.01 & {\ } 24.3 & 10 & $<$0.01 & {\ } 11.2 & 10 & 0.34 & {\ } 25.5 & 10 & $<$0.01\\
$p_{\mathrm{T}}^{t\bar{t}}$ &{\ } 21.0 & 6 & $<$0.01 & {\ } 59.3 & 6 & $<$0.01 & {\ } 107.0 & 6 & $<$0.01 & {\ } 27.8 & 6 & $<$0.01 & {\ } 38.4 & 6 & $<$0.01\\
\hline
\end{tabular}}
\caption{ Comparison of the measured particle-level absolute single-differential cross-sections in the boosted topology with the predictions from several MC generators. For each prediction a $\chi^2$ and a $p$-value are calculated using the covariance matrix of the measured spectrum. The NDF is equal to the number of bins in the distribution.}
\label{tab:chisquare:absolute:allpred:1D:boosted:particle}
\end{table}

\begin{table}[t]
\footnotesize
\centering\noindent\makebox[\textwidth]{
\renewcommand*{\arraystretch}{1.2}\begin{tabular}{|c | r @{/} l r  | r @{/} l r  | r @{/} l r  | r @{/} l r  | r @{/} l r |}
\hline
Observable
& \multicolumn{3}{c|}{\textsc{Pwg+Py8}}& \multicolumn{3}{c|}{\textsc{Pwg+Py8} Rad.~Up}& \multicolumn{3}{c|}{\textsc{Pwg+Py8} Rad.~Down}& \multicolumn{3}{c|}{\textsc{Pwg+H7}}& \multicolumn{3}{c|}{\textsc{Sherpa} 2.2.1}\\
& \multicolumn{2}{c}{$\chi^{2}$/NDF} &  ~$p$-value& \multicolumn{2}{c}{$\chi^{2}$/NDF} &  ~$p$-value& \multicolumn{2}{c}{$\chi^{2}$/NDF} &  ~$p$-value& \multicolumn{2}{c}{$\chi^{2}$/NDF} &  ~$p$-value& \multicolumn{2}{c}{$\chi^{2}$/NDF} &  ~$p$-value\\
\hline
\hline
$m^{t\bar{t}} \textrm{ vs }N^{\mathrm{extra jets}}$ &{\ } 14.3 & 12 & 0.28 & {\ } 30.4 & 12 & $<$0.01 & {\ } 28.7 & 12 & $<$0.01 & {\ } 5.4 & 12 & 0.94 & {\ } 19.1 & 12 & 0.09\\
$p_{\mathrm{T}}^{t\bar{t}} \textrm{ vs }N^{\mathrm{extra jets}}$ &{\ } 13.5 & 10 & 0.20 & {\ } 43.0 & 10 & $<$0.01 & {\ } 41.9 & 10 & $<$0.01 & {\ } 13.0 & 10 & 0.22 & {\ } 22.7 & 10 & 0.01\\
$m^{t\bar{t}} \textrm{ vs }H_{\mathrm{T}}^{t\bar{t}}$ &{\ } 7.3 & 8 & 0.51 & {\ } 16.5 & 8 & 0.04 & {\ } 15.7 & 8 & 0.05 & {\ } 7.1 & 8 & 0.53 & {\ } 20.8 & 8 & $<$0.01\\
$m^{t\bar{t}} \textrm{ vs }|y^{t\bar{t}}|$ &{\ } 4.8 & 13 & 0.98 & {\ } 11.5 & 13 & 0.57 & {\ } 15.9 & 13 & 0.26 & {\ } 5.8 & 13 & 0.95 & {\ } 16.4 & 13 & 0.23\\
$m^{t\bar{t}}\textrm{ vs }p_{\mathrm{T}}^{t\bar{t}}$ &{\ } 7.8 & 12 & 0.80 & {\ } 34.6 & 12 & $<$0.01 & {\ } 40.6 & 12 & $<$0.01 & {\ } 18.6 & 12 & 0.10 & {\ } 18.0 & 12 & 0.12\\
$p_{\mathrm{T}}^{t,\mathrm{had}}\textrm{ vs }|y^t|$ &{\ } 8.6 & 9 & 0.47 & {\ } 12.7 & 9 & 0.17 & {\ } 6.5 & 9 & 0.69 & {\ } 5.7 & 9 & 0.77 & {\ } 12.5 & 9 & 0.18\\
$p_{\mathrm{T}}^{t,\mathrm{had}} \textrm{ vs }|y^{t\bar{t}}|$ &{\ } 10.0 & 9 & 0.35 & {\ } 11.6 & 9 & 0.24 & {\ } 8.5 & 9 & 0.48 & {\ } 8.9 & 9 & 0.45 & {\ } 13.5 & 9 & 0.14\\
$p_{\mathrm{T}}^{t,\mathrm{had}} \textrm{ vs }N^{\mathrm{extra jets}}$ &{\ } 16.3 & 14 & 0.29 & {\ } 42.6 & 14 & $<$0.01 & {\ } 30.3 & 14 & $<$0.01 & {\ } 18.6 & 14 & 0.18 & {\ } 30.8 & 14 & $<$0.01\\
$p_{\mathrm{T}}^{t,\mathrm{had}} \textrm{ vs }m^{t\bar{t}}$ &{\ } 6.9 & 7 & 0.44 & {\ } 18.7 & 7 & $<$0.01 & {\ } 8.9 & 7 & 0.26 & {\ } 4.4 & 7 & 0.73 & {\ } 25.6 & 7 & $<$0.01\\
$p_{\mathrm{T}}^{t,\mathrm{had}} \textrm{ vs }p_{\mathrm{T}}^{t\bar{t}}$ &{\ } 16.1 & 13 & 0.24 & {\ } 50.4 & 13 & $<$0.01 & {\ } 63.2 & 13 & $<$0.01 & {\ } 26.0 & 13 & 0.02 & {\ } 33.9 & 13 & $<$0.01\\
\hline
\end{tabular}}
\caption{ Comparison of the measured particle-level normalised double-differential cross-sections in the boosted topology with the predictions from several MC generators. For each prediction a $\chi^2$ and a $p$-value are calculated using the covariance matrix of the measured spectrum. The NDF is equal to the number of bins in the distribution minus one.
}
\label{tab:chisquare:relative:allpred:2D:boosted:particle}
\end{table}

\begin{table}[t]
\footnotesize
\centering\noindent\makebox[\textwidth]{
\renewcommand*{\arraystretch}{1.2}\begin{tabular}{|c | r @{/} l r  | r @{/} l r  | r @{/} l r  | r @{/} l r  | r @{/} l r |}
\hline
Observable
& \multicolumn{3}{c|}{\textsc{Pwg+Py8}}& \multicolumn{3}{c|}{\textsc{Pwg+Py8} Rad.~Up}& \multicolumn{3}{c|}{\textsc{Pwg+Py8} Rad.~Down}& \multicolumn{3}{c|}{\textsc{Pwg+H7}}& \multicolumn{3}{c|}{\textsc{Sherpa} 2.2.1}\\
& \multicolumn{2}{c}{$\chi^{2}$/NDF} &  ~$p$-value& \multicolumn{2}{c}{$\chi^{2}$/NDF} &  ~$p$-value& \multicolumn{2}{c}{$\chi^{2}$/NDF} &  ~$p$-value& \multicolumn{2}{c}{$\chi^{2}$/NDF} &  ~$p$-value& \multicolumn{2}{c}{$\chi^{2}$/NDF} &  ~$p$-value\\
\hline
\hline
$m^{t\bar{t}} \textrm{ vs }N^{\mathrm{extra jets}}$ &{\ } 38.9 & 13 & $<$0.01 & {\ } 53.2 & 13 & $<$0.01 & {\ } 73.4 & 13 & $<$0.01 & {\ } 9.1 & 13 & 0.77 & {\ } 35.9 & 13 & $<$0.01\\
$p_{\mathrm{T}}^{t\bar{t}} \textrm{ vs }N^{\mathrm{extra jets}}$ &{\ } 41.6 & 11 & $<$0.01 & {\ } 86.5 & 11 & $<$0.01 & {\ } 102.0 & 11 & $<$0.01 & {\ } 25.4 & 11 & $<$0.01 & {\ } 45.9 & 11 & $<$0.01\\
$m^{t\bar{t}} \textrm{ vs }H_{\mathrm{T}}^{t\bar{t}}$ &{\ } 12.7 & 9 & 0.17 & {\ } 17.8 & 9 & 0.04 & {\ } 25.3 & 9 & $<$0.01 & {\ } 11.8 & 9 & 0.22 & {\ } 24.4 & 9 & $<$0.01\\
$m^{t\bar{t}} \textrm{ vs }|y^{t\bar{t}}|$ &{\ } 18.4 & 14 & 0.19 & {\ } 17.3 & 14 & 0.24 & {\ } 36.5 & 14 & $<$0.01 & {\ } 14.2 & 14 & 0.43 & {\ } 22.1 & 14 & 0.08\\
$m^{t\bar{t}}\textrm{ vs }p_{\mathrm{T}}^{t\bar{t}}$ &{\ } 15.5 & 13 & 0.28 & {\ } 70.1 & 13 & $<$0.01 & {\ } 86.4 & 13 & $<$0.01 & {\ } 27.8 & 13 & $<$0.01 & {\ } 28.8 & 13 & $<$0.01\\
$p_{\mathrm{T}}^{t,\mathrm{had}}\textrm{ vs }|y^t|$ &{\ } 11.2 & 10 & 0.34 & {\ } 15.9 & 10 & 0.10 & {\ } 7.3 & 10 & 0.70 & {\ } 6.7 & 10 & 0.75 & {\ } 15.3 & 10 & 0.12\\
$p_{\mathrm{T}}^{t,\mathrm{had}} \textrm{ vs }|y^{t\bar{t}}|$ &{\ } 9.7 & 10 & 0.47 & {\ } 10.6 & 10 & 0.39 & {\ } 8.1 & 10 & 0.62 & {\ } 8.5 & 10 & 0.58 & {\ } 13.4 & 10 & 0.20\\
$p_{\mathrm{T}}^{t,\mathrm{had}} \textrm{ vs }N^{\mathrm{extra jets}}$ &{\ } 35.7 & 15 & $<$0.01 & {\ } 74.2 & 15 & $<$0.01 & {\ } 61.1 & 15 & $<$0.01 & {\ } 22.5 & 15 & 0.09 & {\ } 59.6 & 15 & $<$0.01\\
$p_{\mathrm{T}}^{t,\mathrm{had}} \textrm{ vs }m^{t\bar{t}}$ &{\ } 14.8 & 8 & 0.06 & {\ } 29.8 & 8 & $<$0.01 & {\ } 16.4 & 8 & 0.04 & {\ } 4.4 & 8 & 0.82 & {\ } 32.6 & 8 & $<$0.01\\
$p_{\mathrm{T}}^{t,\mathrm{had}} \textrm{ vs }p_{\mathrm{T}}^{t\bar{t}}$ &{\ } 24.6 & 14 & 0.04 & {\ } 70.1 & 14 & $<$0.01 & {\ } 94.3 & 14 & $<$0.01 & {\ } 30.0 & 14 & $<$0.01 & {\ } 48.7 & 14 & $<$0.01\\
\hline
\end{tabular}}
\caption{ Comparison of the measured particle-level absolute double-differential cross-sections in the boosted topology with the predictions from several MC generators. For each prediction a $\chi^2$ and a $p$-value are calculated using the covariance matrix of the measured spectrum. The NDF is equal to the number of bins in the distribution.}
\label{tab:chisquare:absolute:allpred:2D:boosted:particle}
\end{table}

\FloatBarrier

\FloatBarrier
 
\subsection{Results at parton level in the full phase-space}
\subsubsection{Resolved topology}
\label{sec:results:full:resolved}
\begin{figure*}[t]
\centering
\subfigure[]{\includegraphics[width=0.45\textwidth]{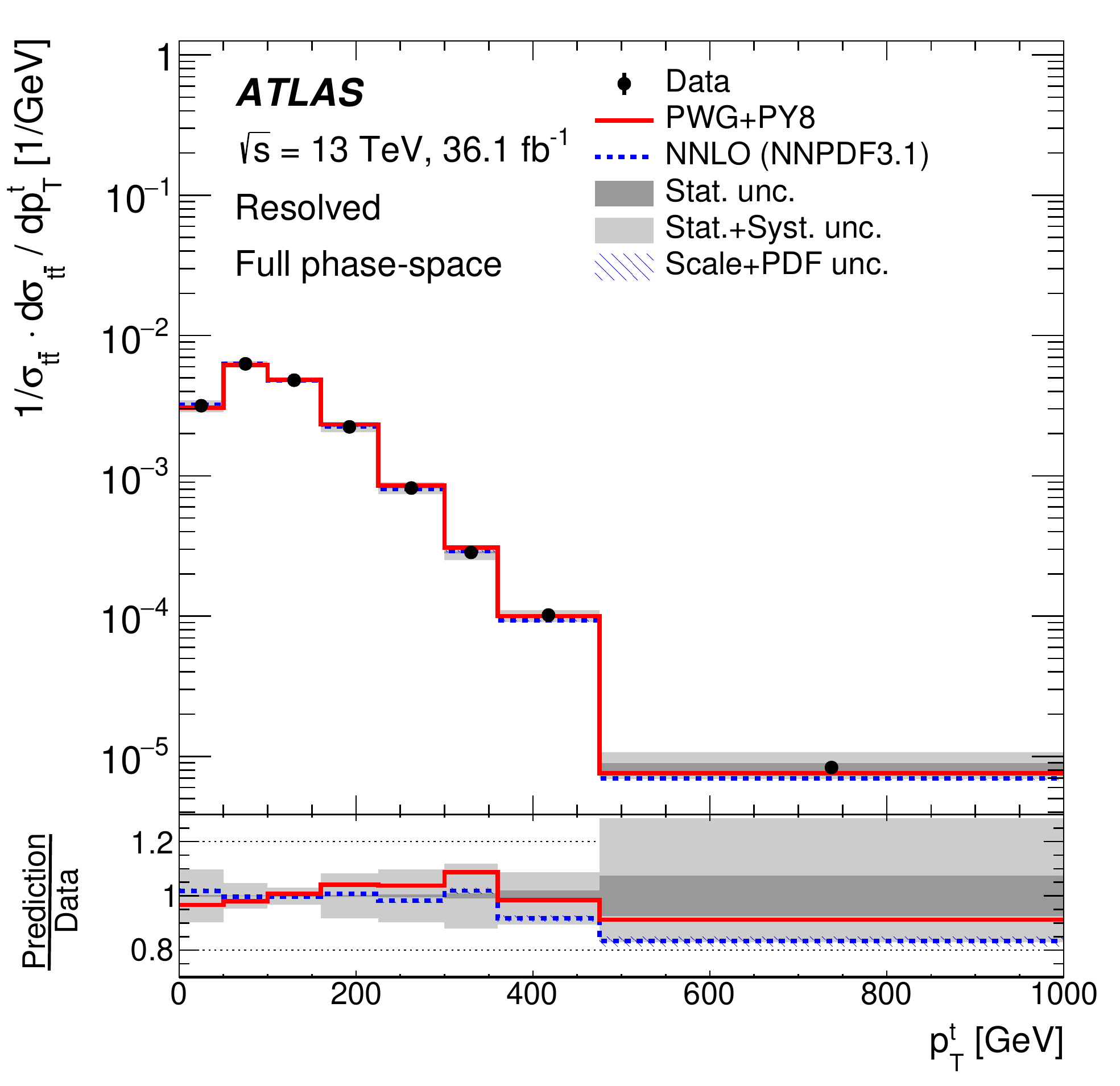}
\label{fig:results:NNLO:resolved:top_pt:rel}}
\subfigure[]{\includegraphics[width=0.45\textwidth]{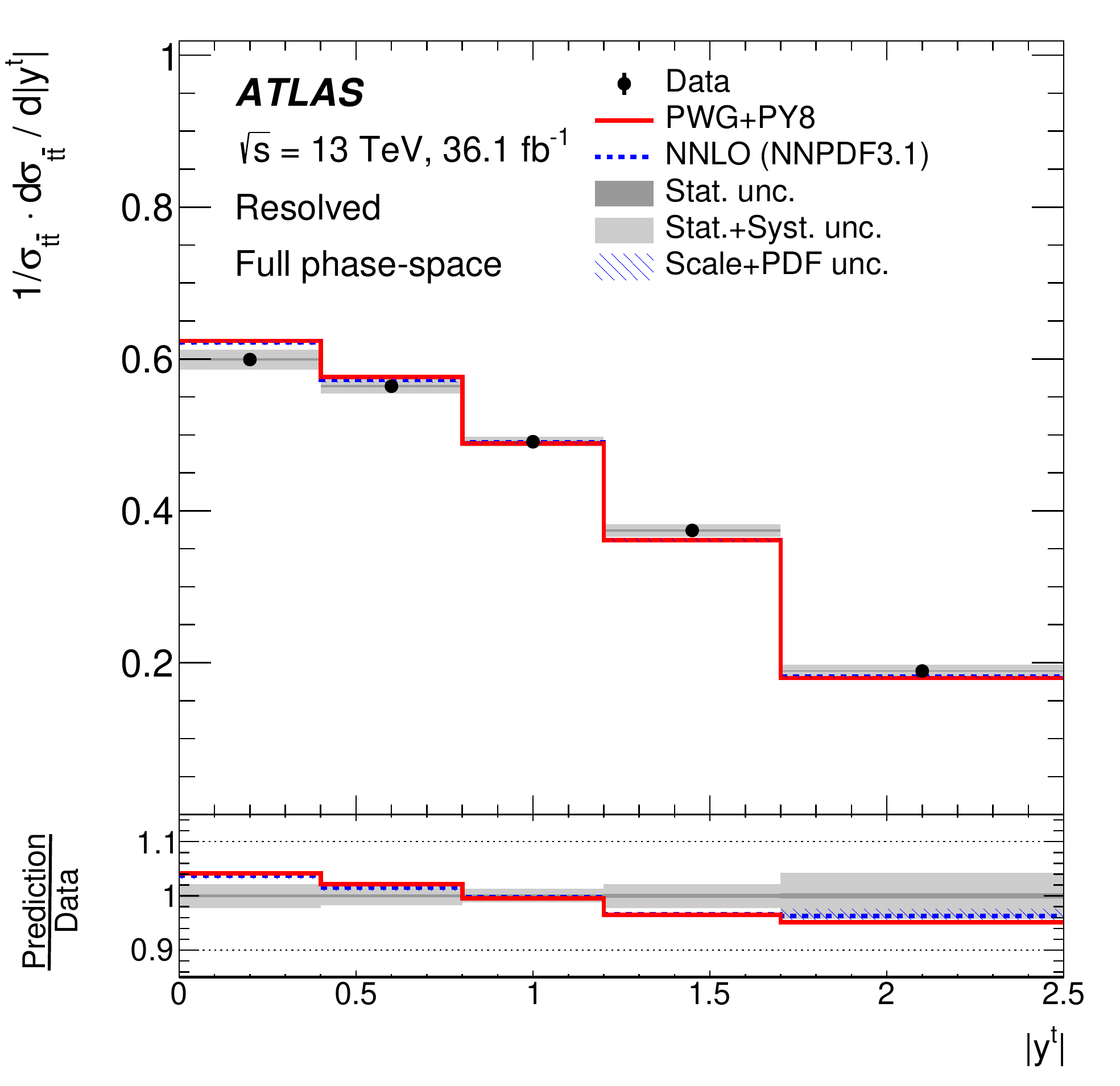}
\label{fig:results:NNLO:resolved:top_y:rel}}
\caption{Parton-level normalised differential cross-sections as a function of~\subref{fig:results:NNLO:resolved:top_pt:rel} the \pt{}
and~\subref{fig:results:NNLO:resolved:top_y:rel} normalised rapidity of the top in the resolved topology, compared with the NNLO predictions obtained
using the NNPDF3.1 NNLO PDF set and the  predictions obtained with the \Powheg+\PythiaEight{} MC generator. The hatched band represents the total uncertainty in the NNLO prediction. The solid bands represent the statistical and total uncertainty in the data. Data points are placed at the centre of each bin. The lower panel shows the ratios of the predictions to data.}
\label{fig:results:NNLO:rel:parton:resolved:1D:top}
\end{figure*}

\begin{figure*}[t]
\centering
\subfigure[]{\includegraphics[width=0.45\textwidth]{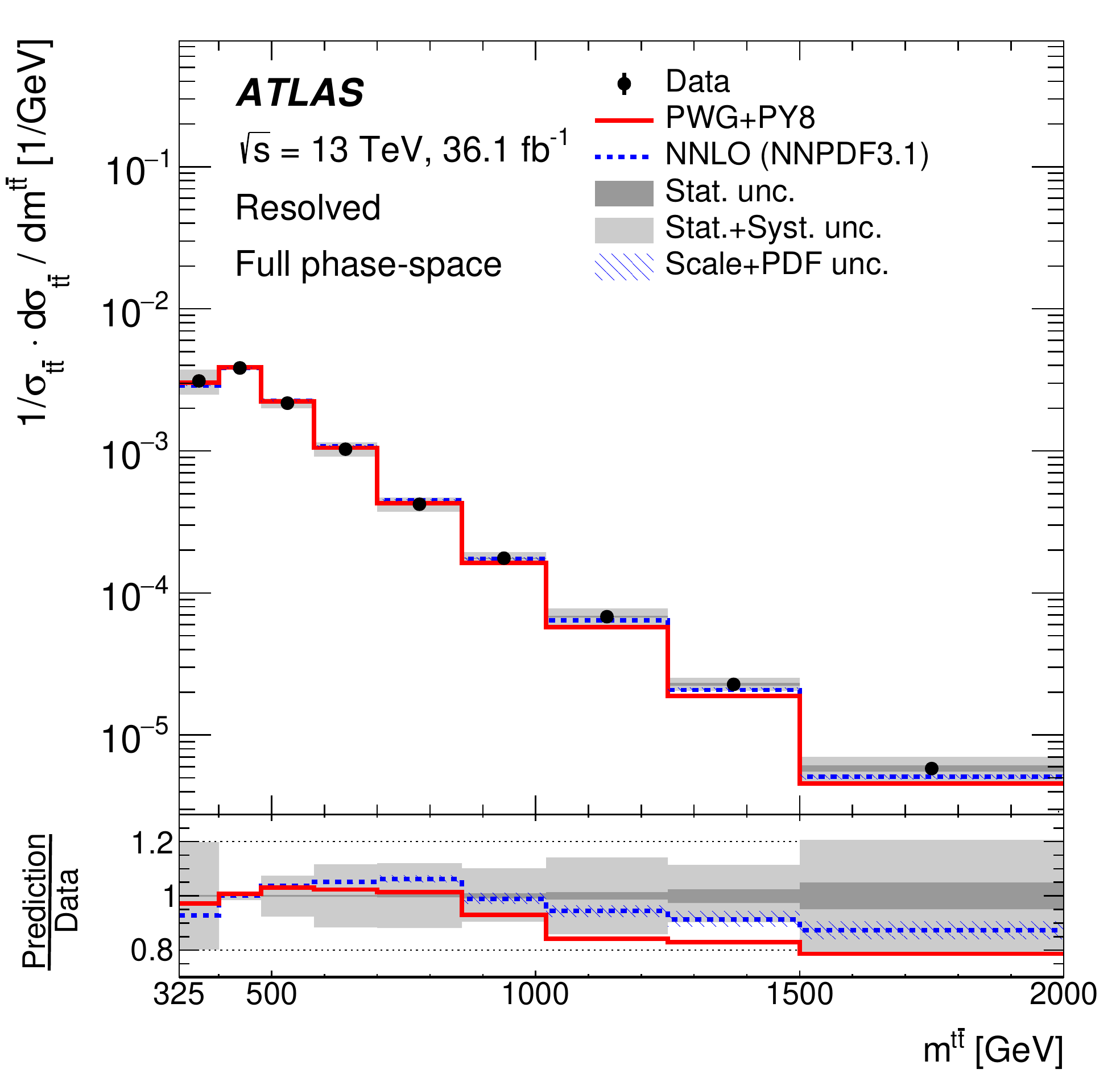}
\label{fig:results:NNLO:resolved:ttbar_m:rel}}
\subfigure[]{\includegraphics[width=0.45\textwidth]{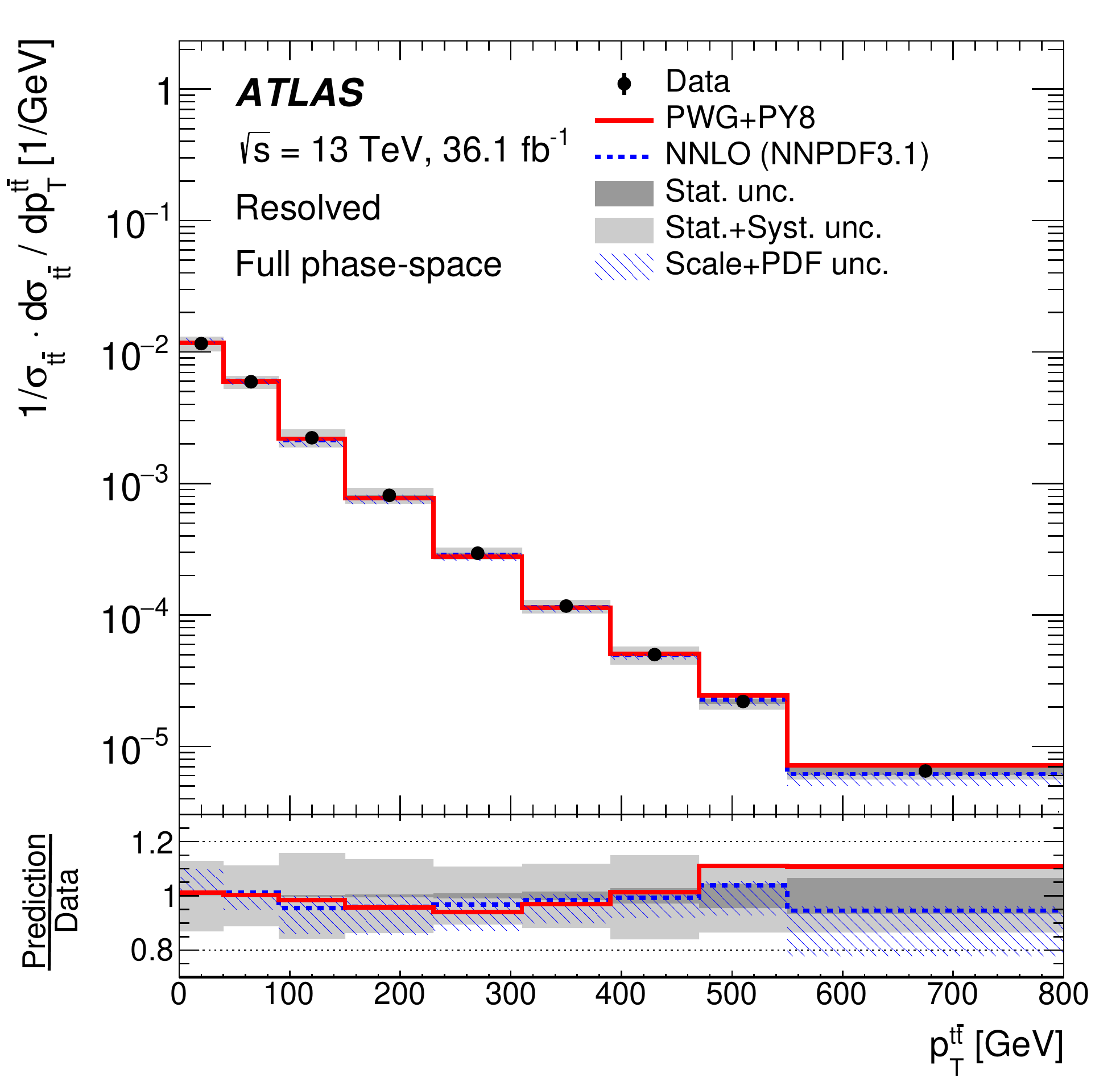}
\label{fig:results:NNLO:resolved:ttbar_pt:rel}}\\
\subfigure[]{\includegraphics[width=0.45\textwidth]{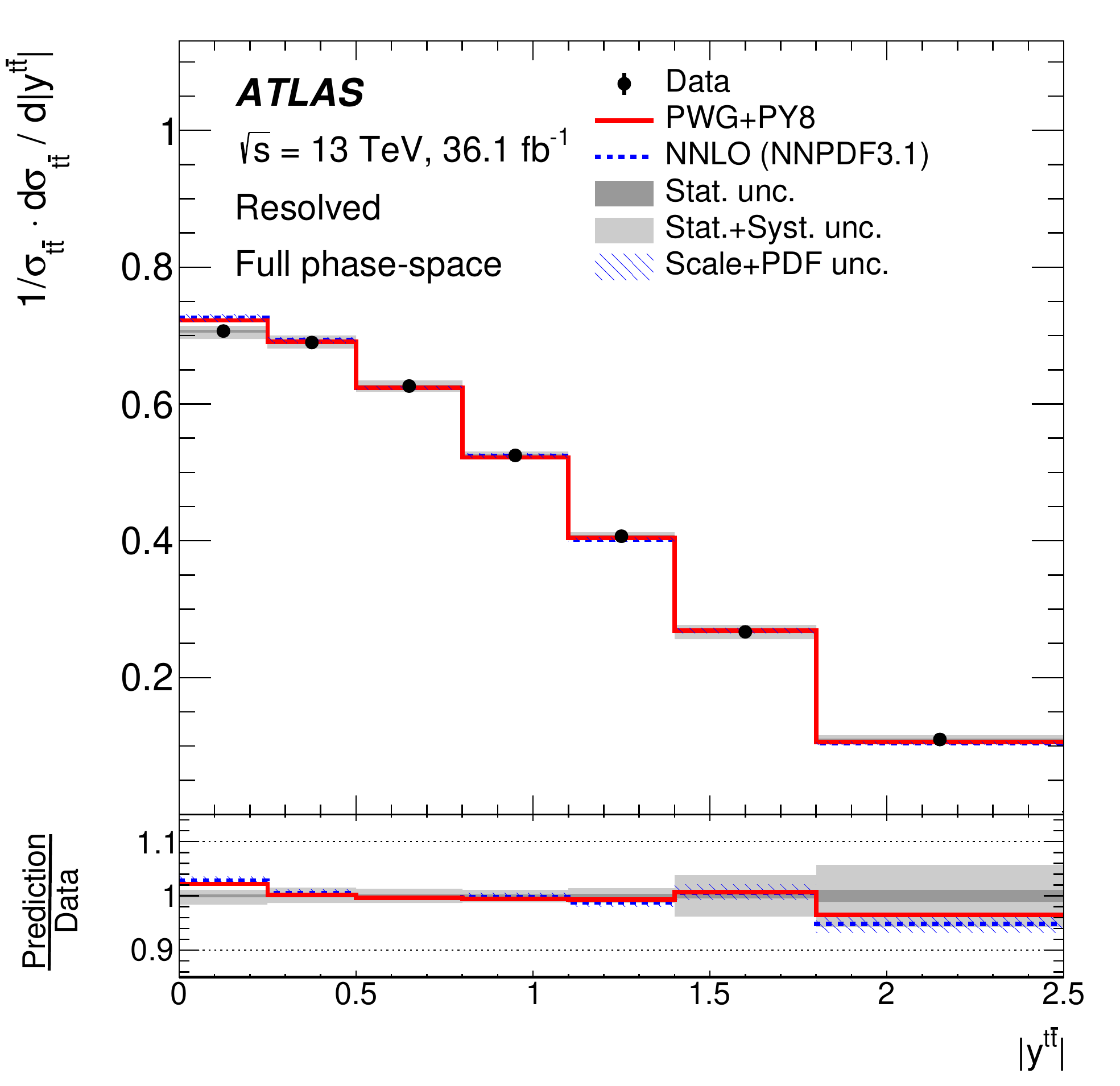}
\label{fig:results:NNLO:resolved:ttbar_y:rel}}
\caption{Parton-level normalised differential cross-sections as a function of~\subref{fig:results:NNLO:resolved:ttbar_m:rel} the mass,
~\subref{fig:results:NNLO:resolved:ttbar_pt:rel} \pt{} and~\subref{fig:results:NNLO:resolved:ttbar_y:rel} absolute value of the rapidity of the \ttb{} system  in the resolved topology, compared with the NNLO predictions obtained
using the NNPDF3.1 NNLO PDF set and the  predictions obtained with the \Powheg+\PythiaEight{} MC generator.  The hatched band represents the total uncertainty in the NNLO prediction. The solid bands represent the statistical and total uncertainty in the data. Data points are placed at the centre of each bin. The lower panel shows the ratios of the predictions to data.}
\label{fig:results:NNLO:rel:parton:resolved:1D:ttbar}
\end{figure*}

\begin{figure*}[t]
\centering
\subfigure[]{\includegraphics[width=0.45\textwidth]{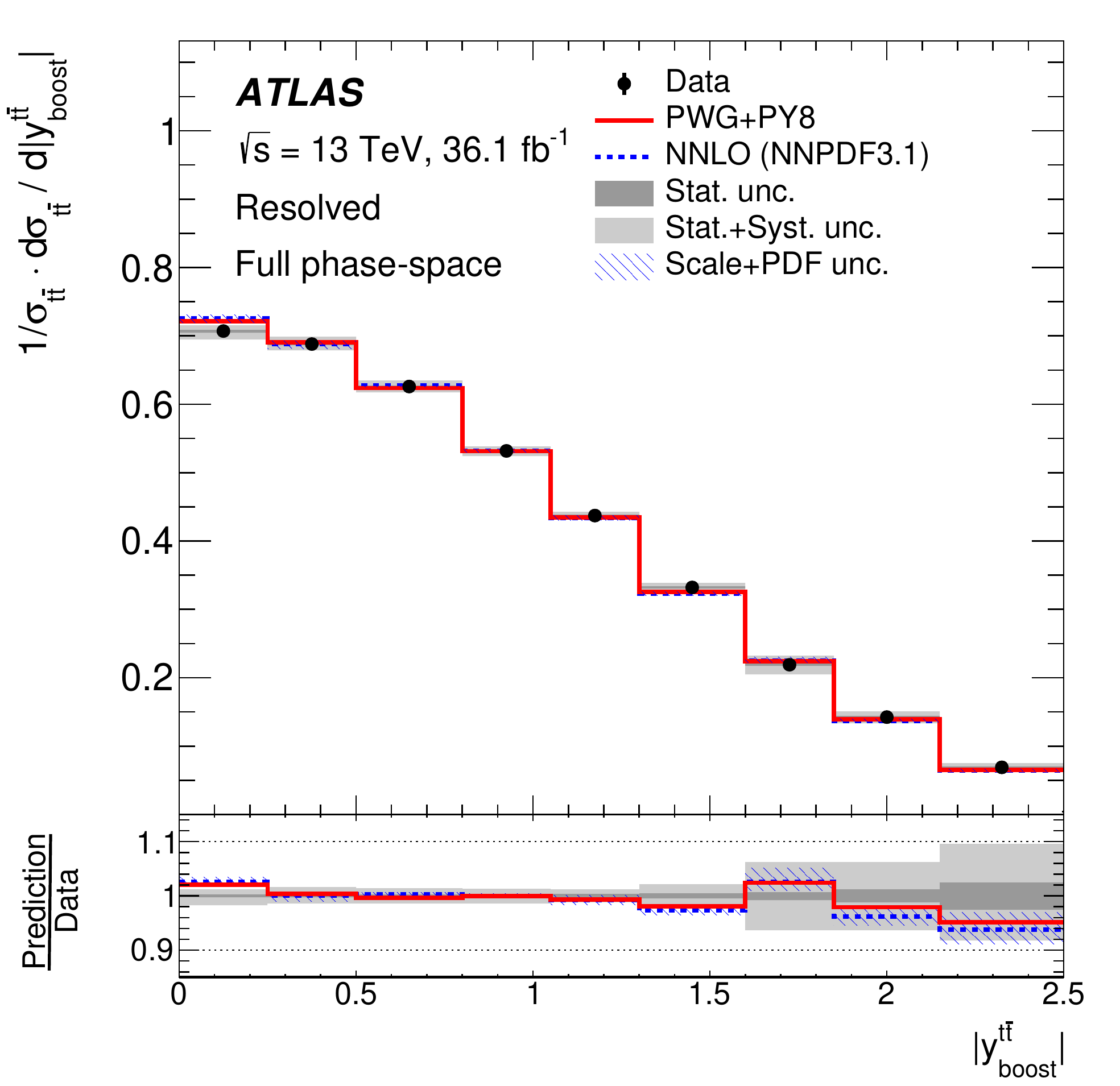}
\label{fig:results:NNLO:resolved:y_boost:rel}}
\subfigure[]{\includegraphics[width=0.45\textwidth]{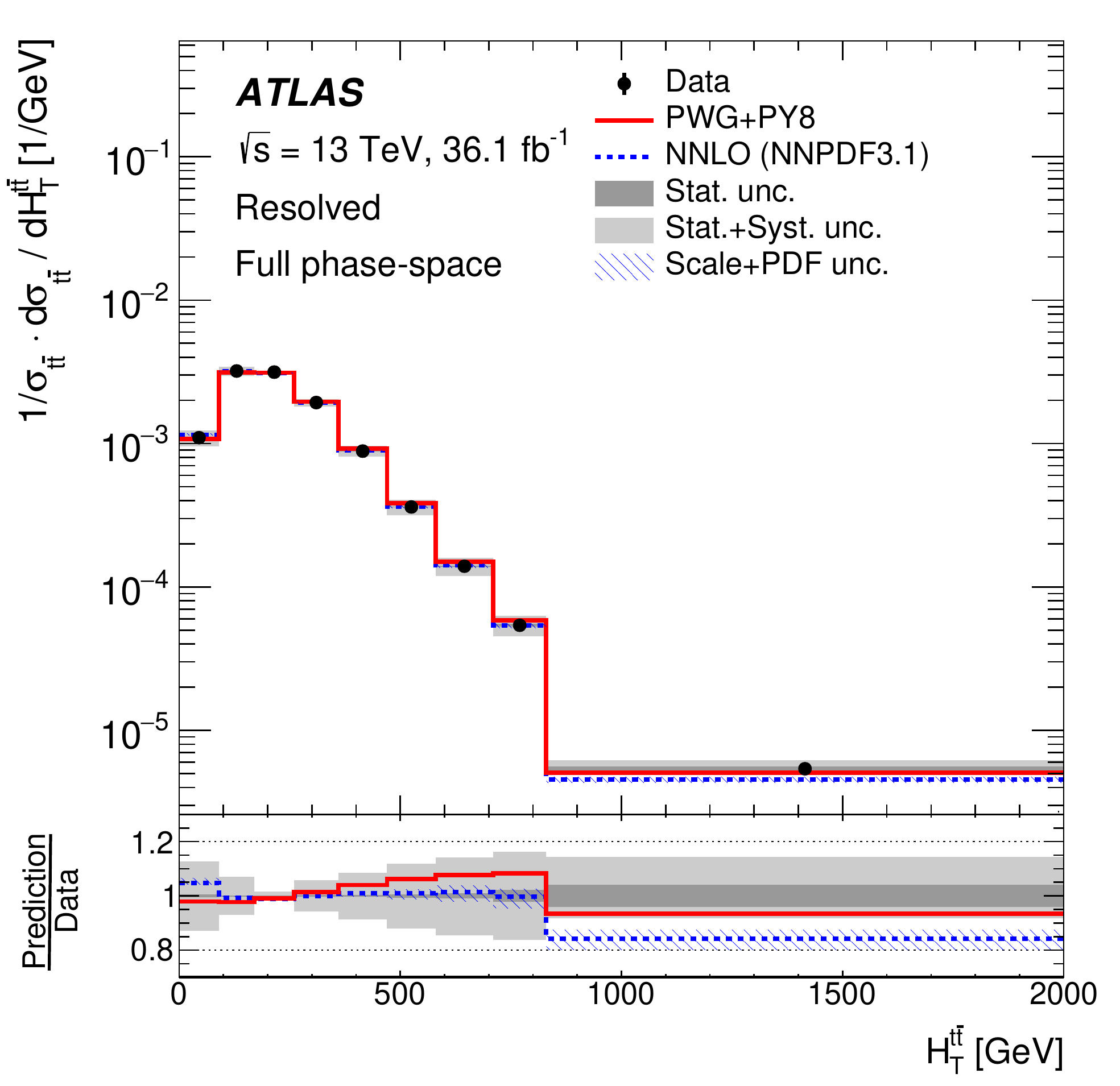}
\label{fig:results:NNLO:resolved:ht:rel}}\\
\subfigure[]{\includegraphics[width=0.45\textwidth]{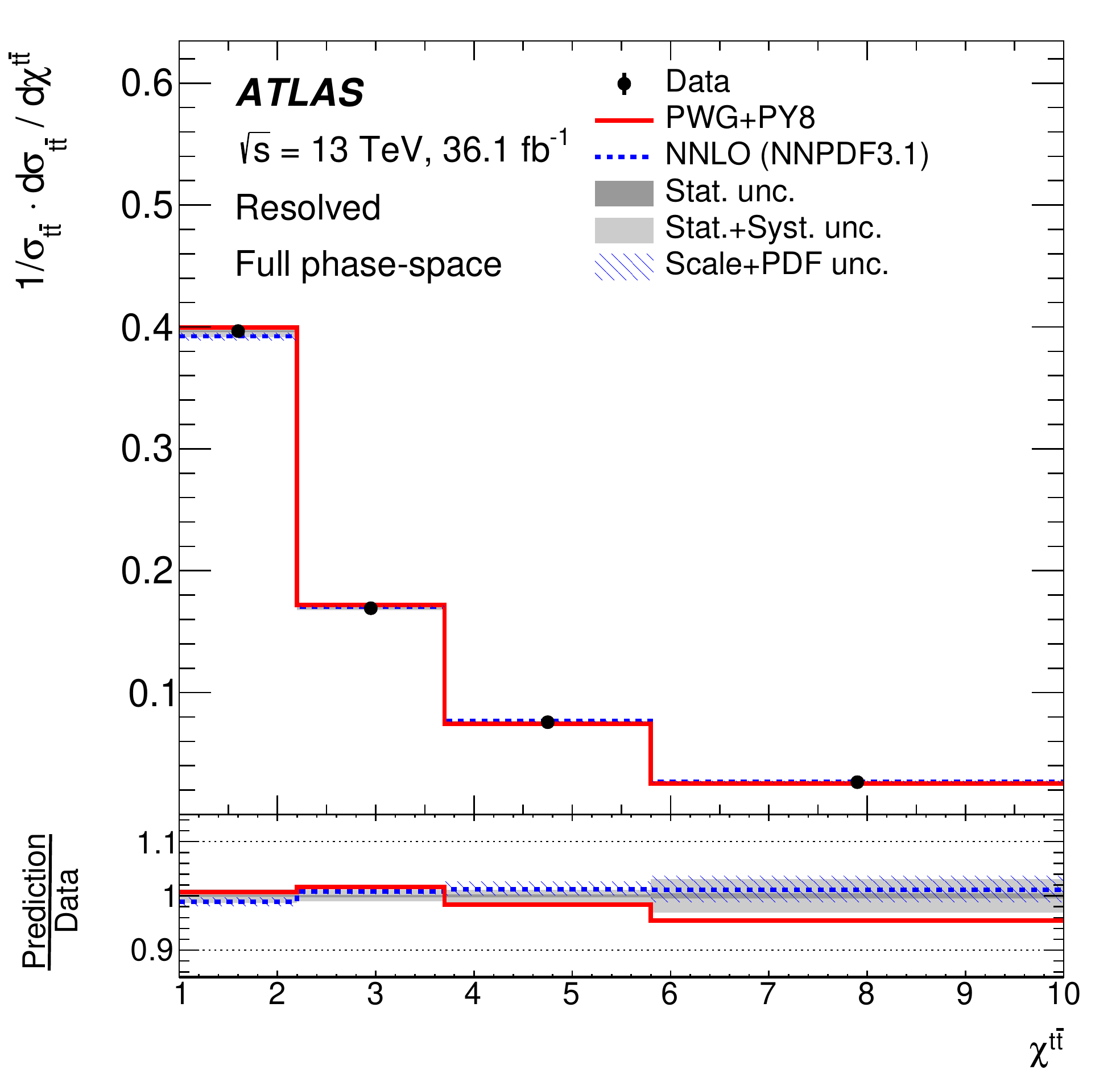}
\label{fig:results:NNLO:resolved:chitt:rel}}
\caption{Parton-level normalised differential cross-sections as a function of~\subref{fig:results:NNLO:resolved:y_boost:rel} $|\yboost|$,
~\subref{fig:results:NNLO:resolved:ht:rel} \Httbar{} and~\subref{fig:results:NNLO:resolved:chitt:rel} \chitt{} in the resolved topology, compared with the NNLO predictions obtained
using the NNPDF3.1 NNLO PDF set and the  predictions obtained with the \Powheg+\PythiaEight{} MC generator.  The hatched band represents the total uncertainty in the NNLO prediction. The solid bands represent the statistical and total uncertainty in the data. Data points are placed at the centre of each bin. The lower panel shows the ratios of the predictions to data.}
\label{fig:results:NNLO:rel:parton:resolved:1D:additional_variables}
\end{figure*}

\begin{figure*}[t]
\centering
\subfigure[]{\includegraphics[width=0.38\textwidth]{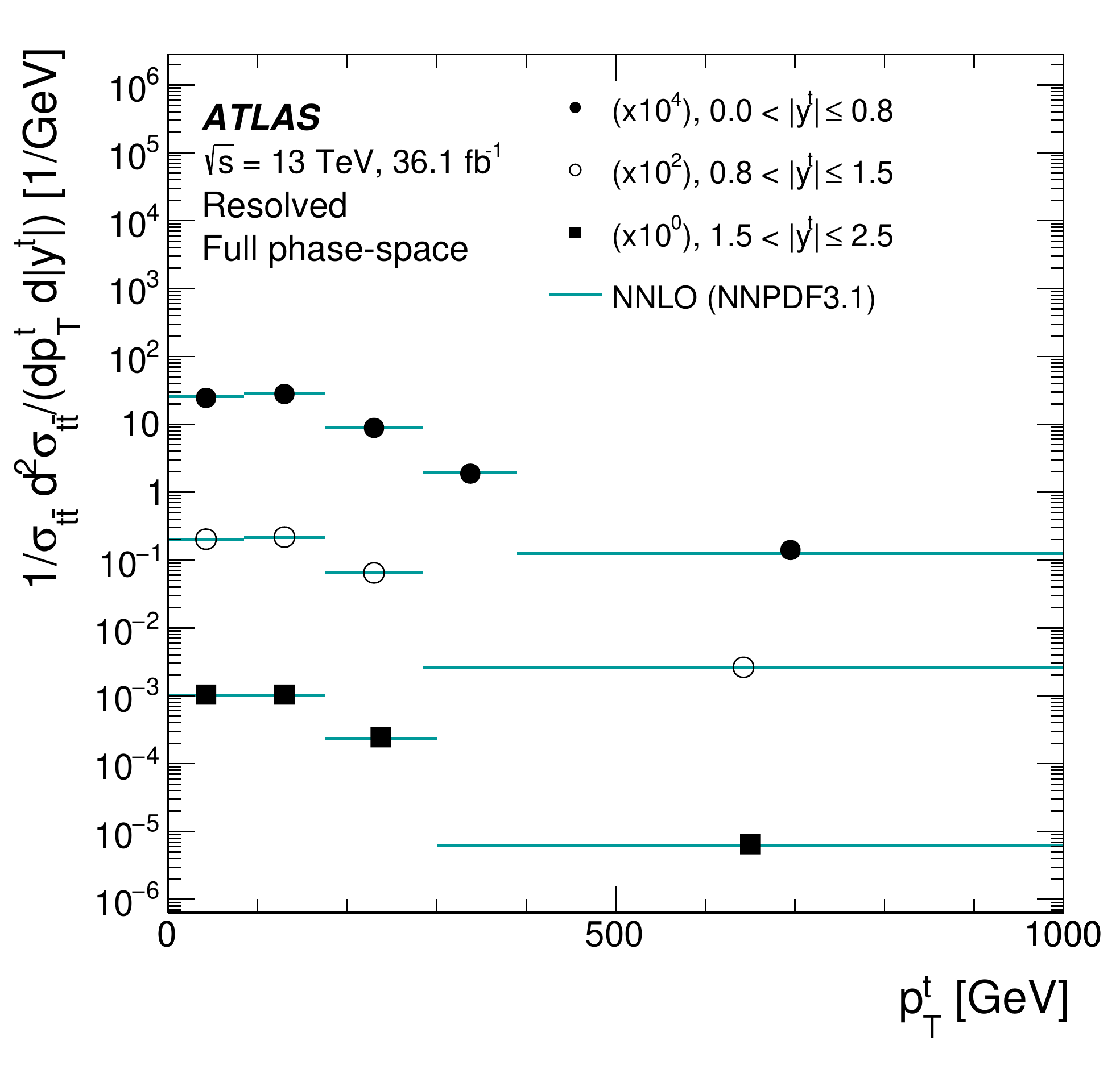}
\label{fig:results:NNLO:resolved:top_pt:top_y:rel}}
\subfigure[]{\includegraphics[width=0.58\textwidth]{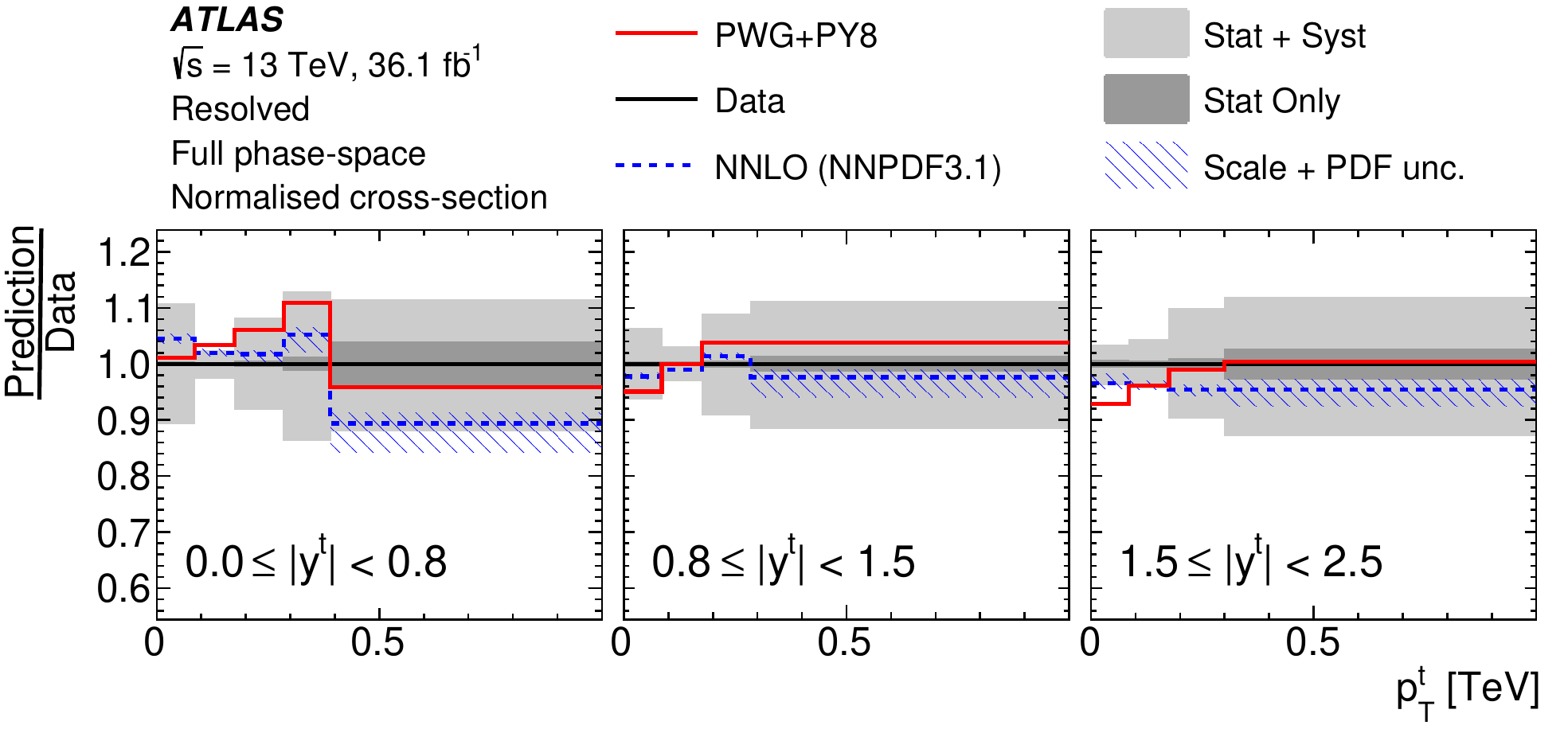}
\label{fig:results:NNLO:resolved:top_pt:top_y:rel:ratios}}
\caption{\small{\subref{fig:results:NNLO:resolved:top_pt:top_y:rel} Parton-level normalised differential cross-section as a function
of \ptt{} in bins of \absyt{} in the resolved topology compared with the NNLO prediction obtained
using the NNPDF3.1 NNLO PDF set.  Data points are placed at the centre of each bin.
\subref{fig:results:NNLO:resolved:top_pt:top_y:rel:ratios} The ratio of the measured cross-section to the NNLO prediction  and the  prediction obtained with the \Powheg+\PythiaEight{} MC generator.  The hatched band represents the total uncertainty in the NNLO prediction. The solid bands represent the statistical and total uncertainty in the data. }}
\label{fig:results:NNLO:rel:parton:resolved:2D:top_abs_y:top_pt}
\end{figure*}

\begin{figure*}[t]
\centering
\subfigure[]{\includegraphics[width=0.38\textwidth]{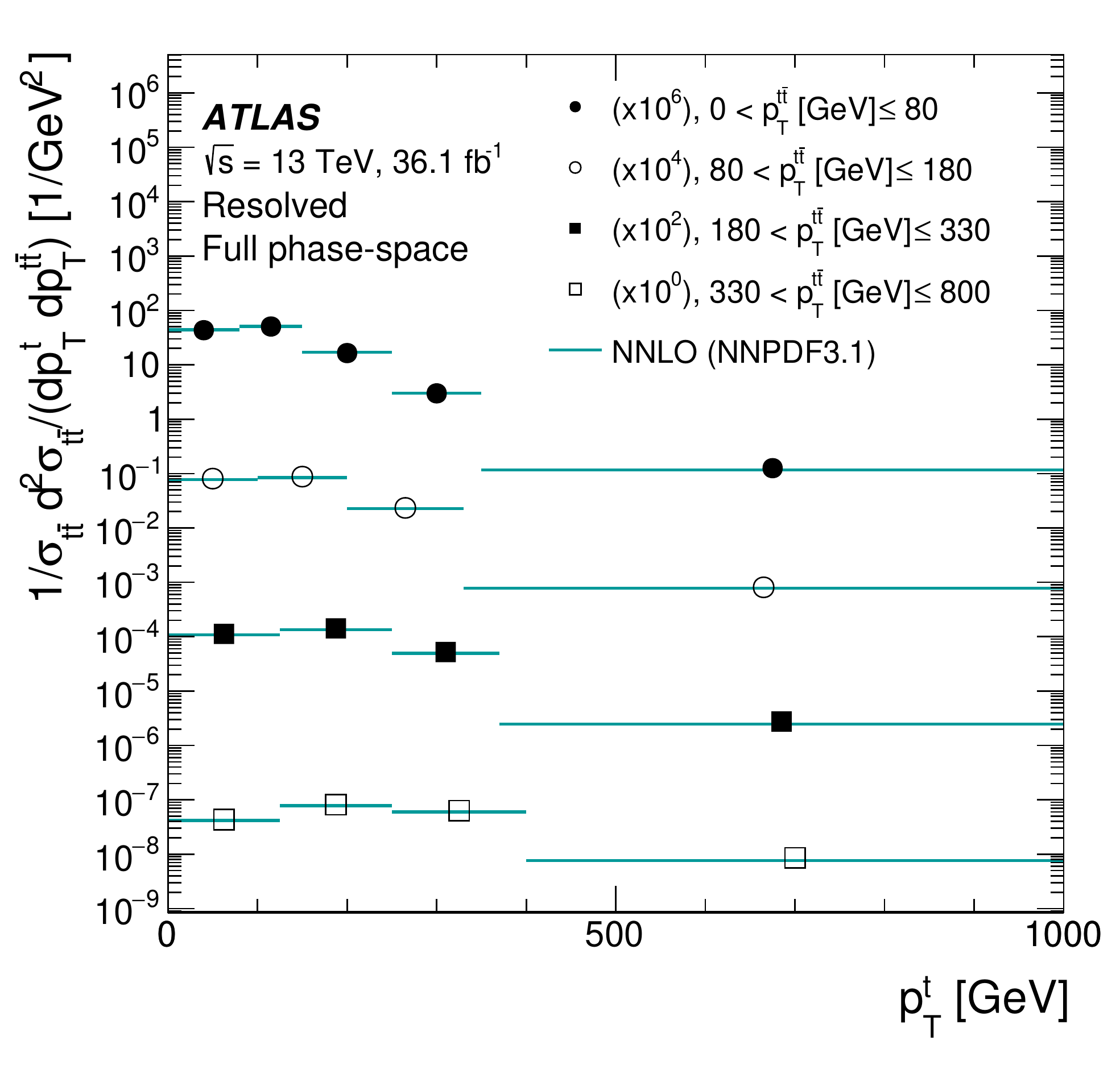}
\label{fig:results:NNLO:resolved:top_pt:ttbar_pt:rel}}
\subfigure[]{\includegraphics[width=0.58\textwidth]{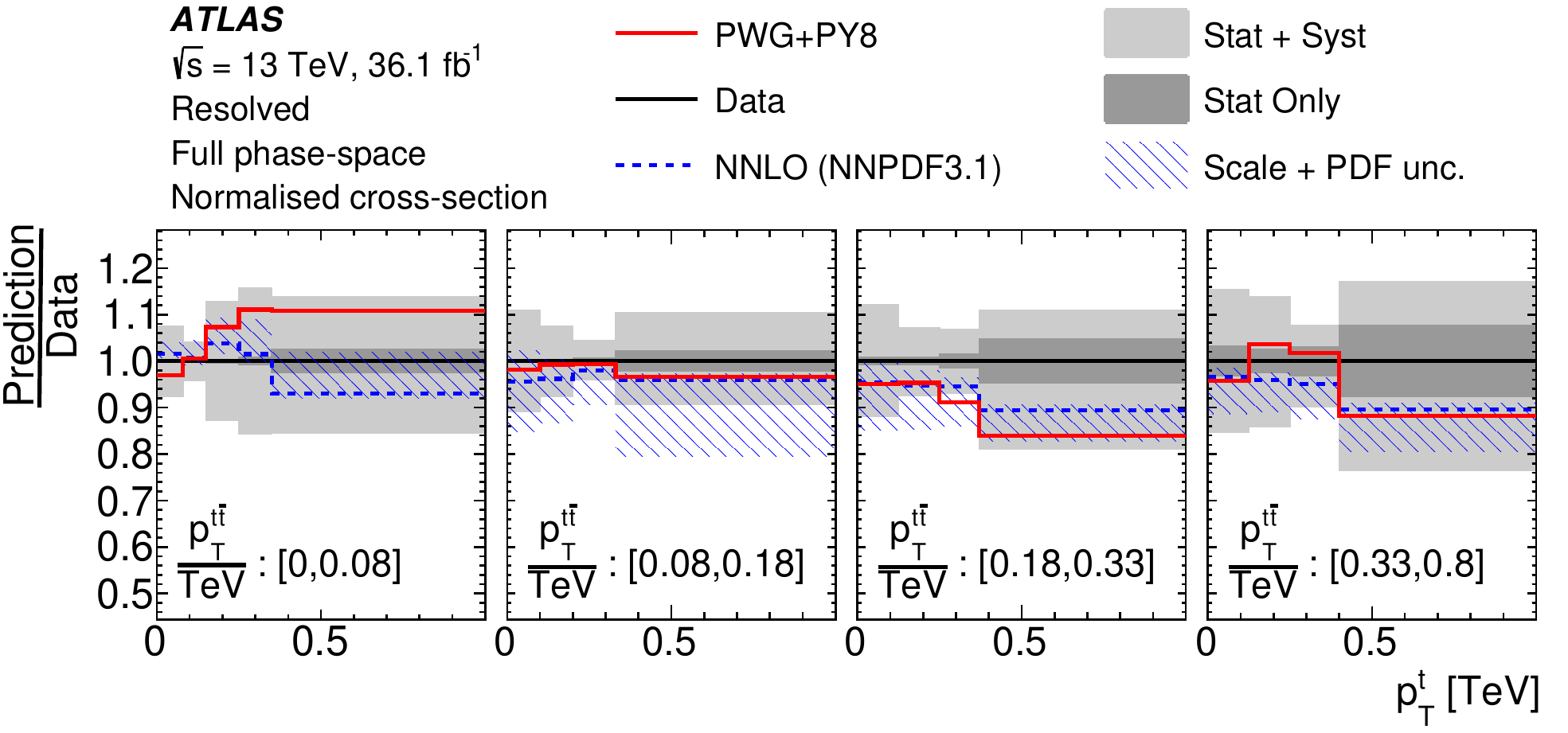}
\label{fig:results:NNLO:resolved:top_pt:ttbar_pt:rel:ratios}}
\caption{\small{\subref{fig:results:NNLO:resolved:top_pt:ttbar_pt:rel} Parton-level normalised differential cross-section as a function
of \ptt{} in bins of \pttt{} in the resolved topology compared with the NNLO prediction obtained
using the NNPDF3.1 NNLO PDF set.  Data points are placed at the centre of each bin.
\subref{fig:results:NNLO:resolved:top_pt:ttbar_pt:rel:ratios} The ratio of the measured cross-section to the NNLO prediction  and the  prediction obtained with the \Powheg+\PythiaEight{} MC generator.  The hatched band represents the total uncertainty in the NNLO prediction. The solid bands represent the statistical and total uncertainty in the data. }}
\label{fig:results:NNLO:rel:parton:resolved:2D:ttbar_pt:top_pt}
\end{figure*}

\begin{figure*}[t]
\centering
\subfigure[]{\includegraphics[width=0.38\textwidth]{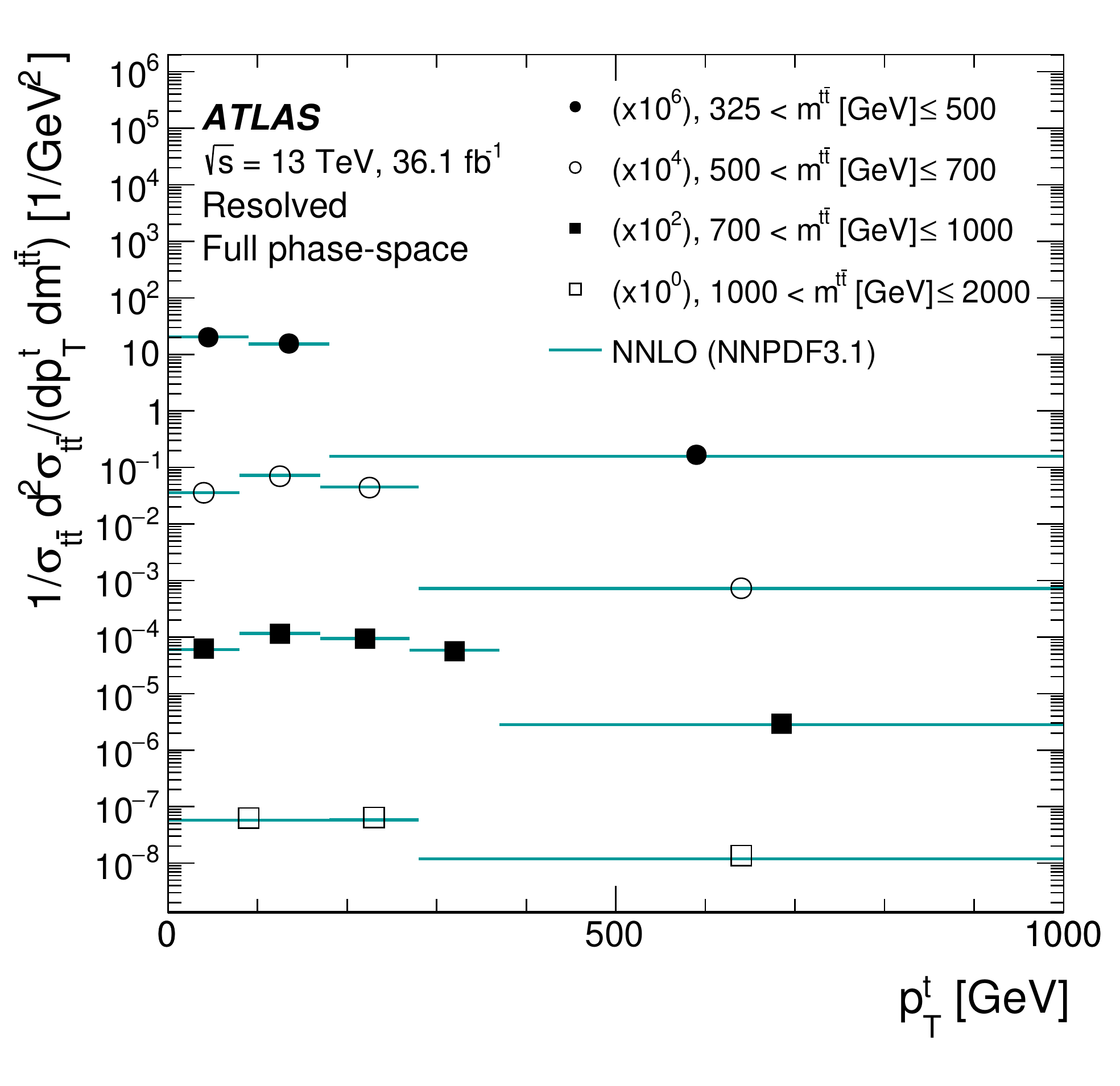}
\label{fig:results:NNLO:resolved:top_pt:ttbar_m:rel}}
\subfigure[]{\includegraphics[width=0.58\textwidth]{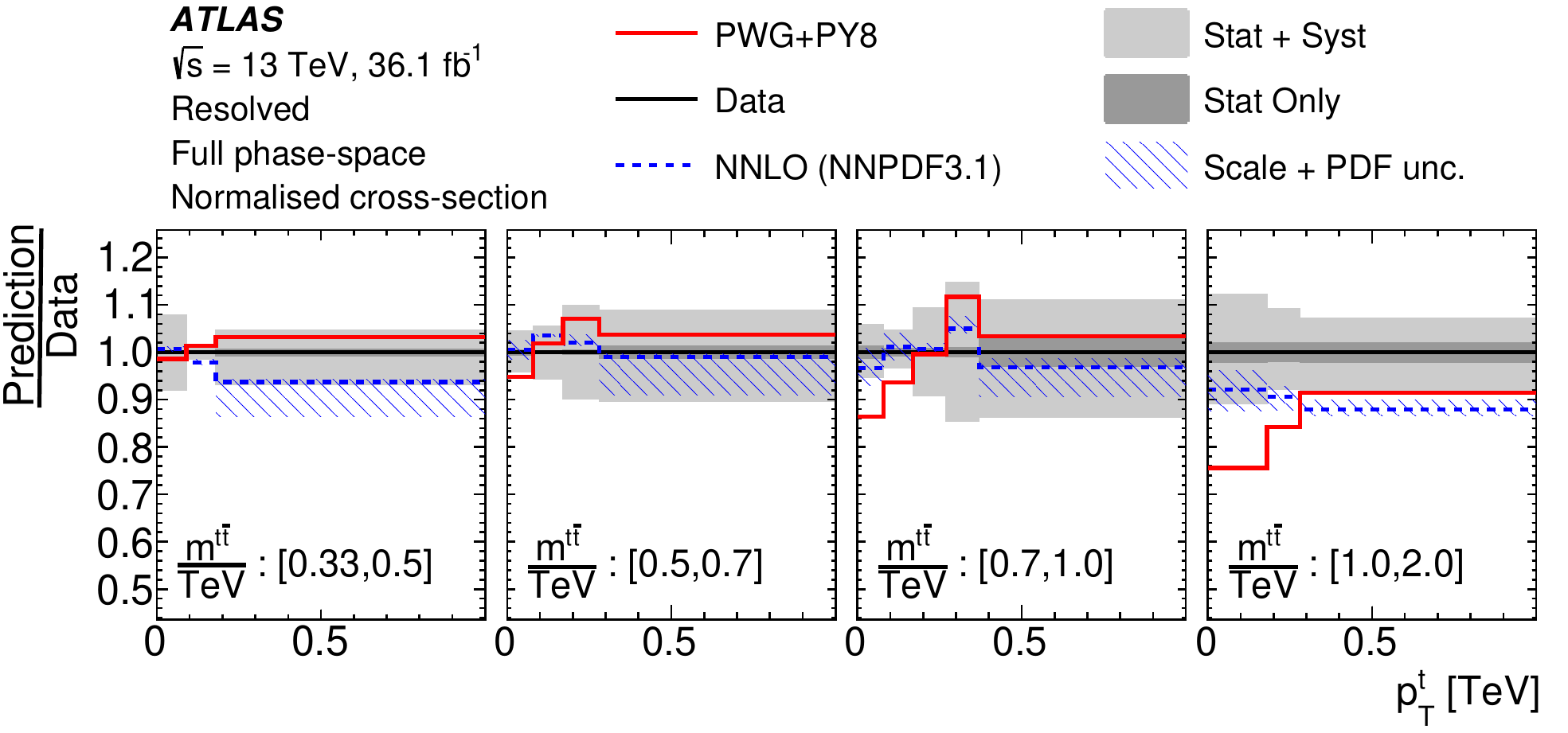}
\label{fig:results:NNLO:resolved:top_pt:ttbar_m:rel:ratios}}
\caption{\small{\subref{fig:results:NNLO:resolved:top_pt:ttbar_m:rel} Parton-level normalised differential cross-section as a function of \ptt{} in bins of \mtt{} in the resolved topology compared with the NNLO prediction obtained
using the NNPDF3.1 NNLO PDF set.  Data points are placed at the centre of each bin.
\subref{fig:results:NNLO:resolved:top_pt:ttbar_m:rel} The ratio of the measured cross-section to the NNLO prediction  and the  prediction obtained with the \Powheg+\PythiaEight{} MC generator.  The hatched band represents the total uncertainty in the NNLO prediction. The solid bands represent the statistical and total uncertainty in the data.}}
\label{fig:results:NNLO:rel:parton:resolved:2D:ttbar_m:top_pt}
\end{figure*}

\begin{figure*}[t]
\centering
\subfigure[]{\includegraphics[width=0.38\textwidth]{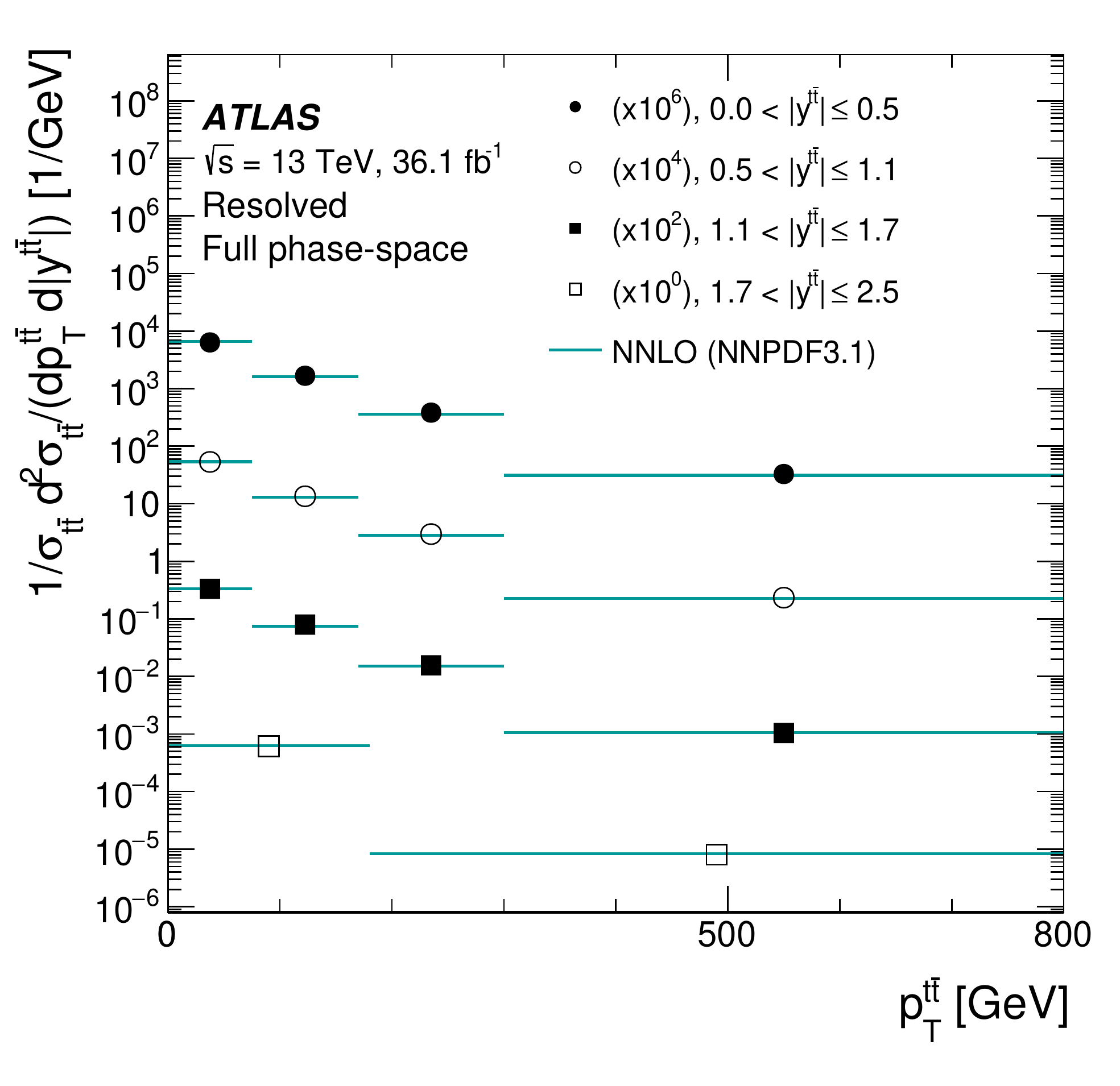}
\label{fig:results:NNLO:resolved:ttbar_pt:ttbar_y:rel}}
\subfigure[]{\includegraphics[width=0.58\textwidth]{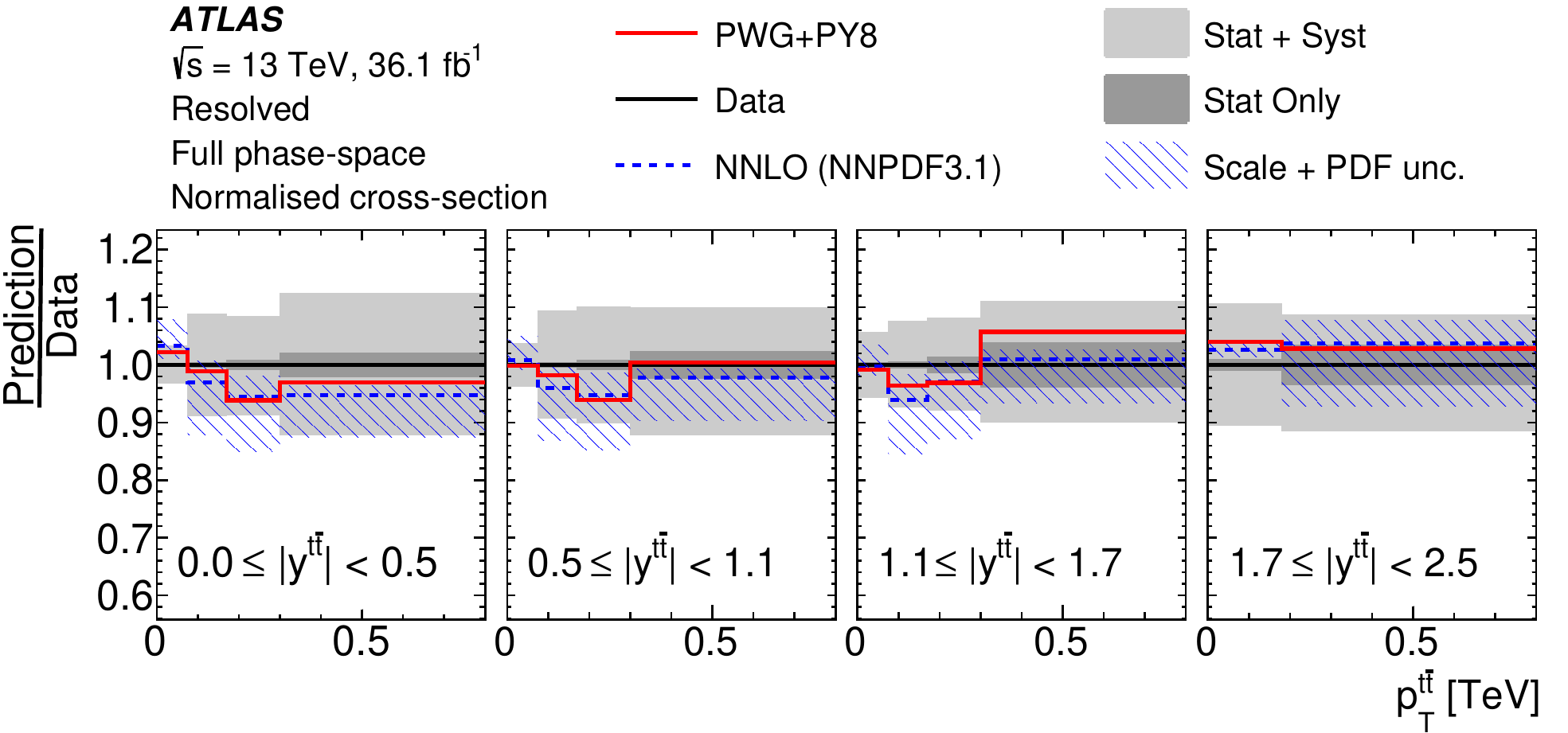}
\label{fig:results:NNLO:resolved:ttbar_pt:ttbar_y:rel:ratios}}
\caption{\small{\subref{fig:results:NNLO:resolved:ttbar_pt:ttbar_y:rel} Parton-level normalised differential cross-section as a function
of \pttt{} in bins of \absyttbar{} in the resolved topology compared with the NNLO prediction obtained
using the NNPDF3.1 NNLO PDF set.  Data points are placed at the centre of each bin.
\subref{fig:results:NNLO:resolved:ttbar_pt:ttbar_y:rel:ratios} The ratio of the measured cross-section to the NNLO prediction  and the  prediction obtained with the \Powheg+\PythiaEight{} MC generator.  The hatched band represents the total uncertainty in the NNLO prediction. The solid bands represent the statistical and total uncertainty in the data. }}
\label{fig:results:NNLO:rel:parton:resolved:2D:ttbar_abs_y:ttbar_pt}
\end{figure*}

\begin{figure*}[t]
\centering
\subfigure[]{\includegraphics[width=0.38\textwidth]{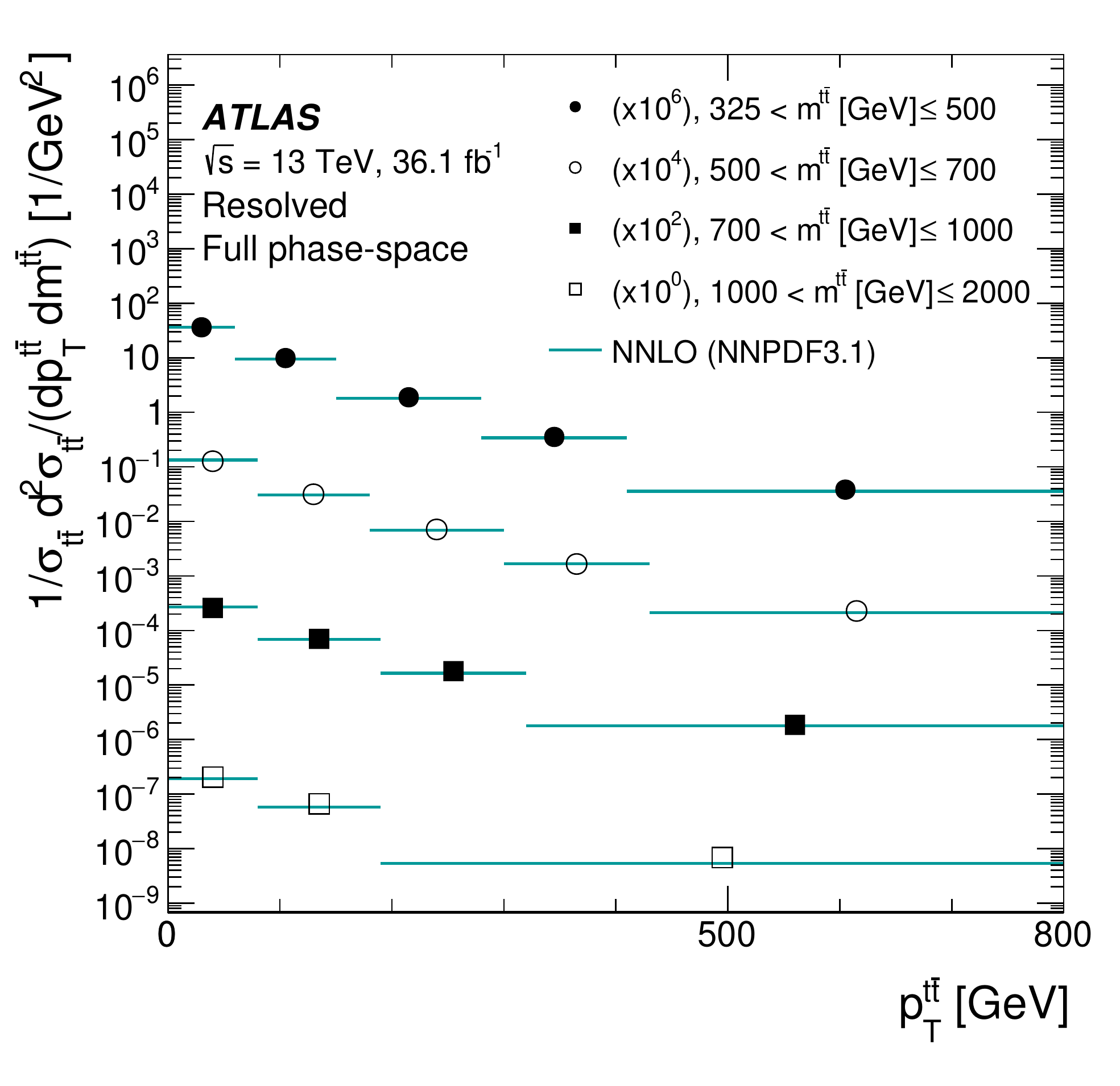}
\label{fig:results:NNLO:resolved:ttbar_pt:ttbar_m:rel}}
\subfigure[]{\includegraphics[width=0.58\textwidth]{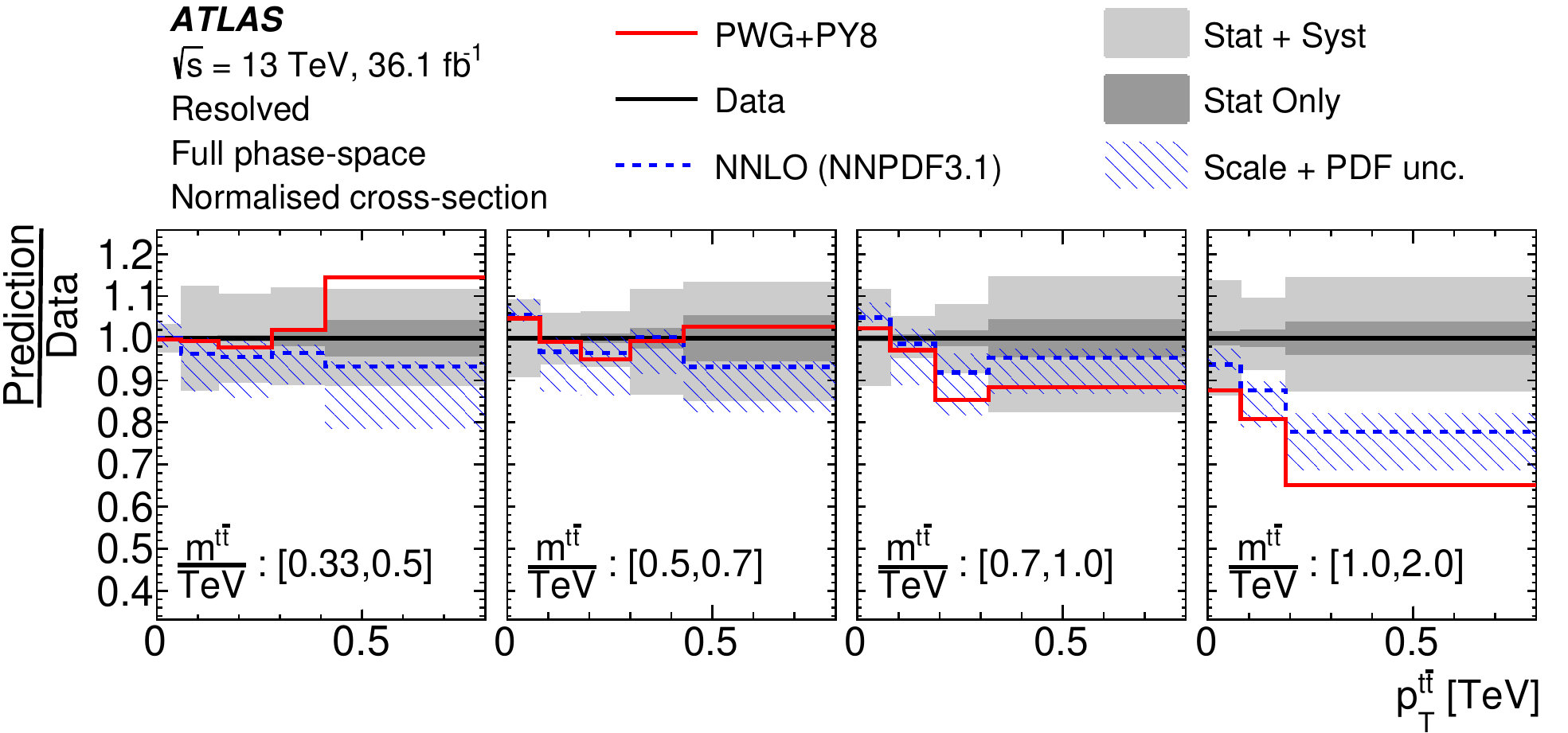}
\label{fig:results:NNLO:resolved:ttbar_pt:ttbar_m:rel:ratios}}
\caption{\small{\subref{fig:results:NNLO:resolved:ttbar_pt:ttbar_m:rel} Parton-level normalised differential cross-section as a function
of \pttt{} in bins of \mtt{} in the resolved topology compared with  the NNLO prediction obtained
using the NNPDF3.1 NNLO PDF set.  Data points are placed at the centre of each bin.
\subref{fig:results:NNLO:resolved:ttbar_pt:ttbar_m:rel:ratios} The ratio of the measured cross-section to the NNLO prediction  and the  prediction obtained with the \Powheg+\PythiaEight{} MC generator.  The hatched band represents the  total uncertainty in the NNLO prediction. The solid bands represent the statistical and total uncertainty in the data. }}
\label{fig:results:NNLO:rel:parton:resolved:2D:ttbar_m:ttbar_pt}
\end{figure*}

\begin{figure*}[t]
\centering
\subfigure[]{\includegraphics[width=0.38\textwidth]{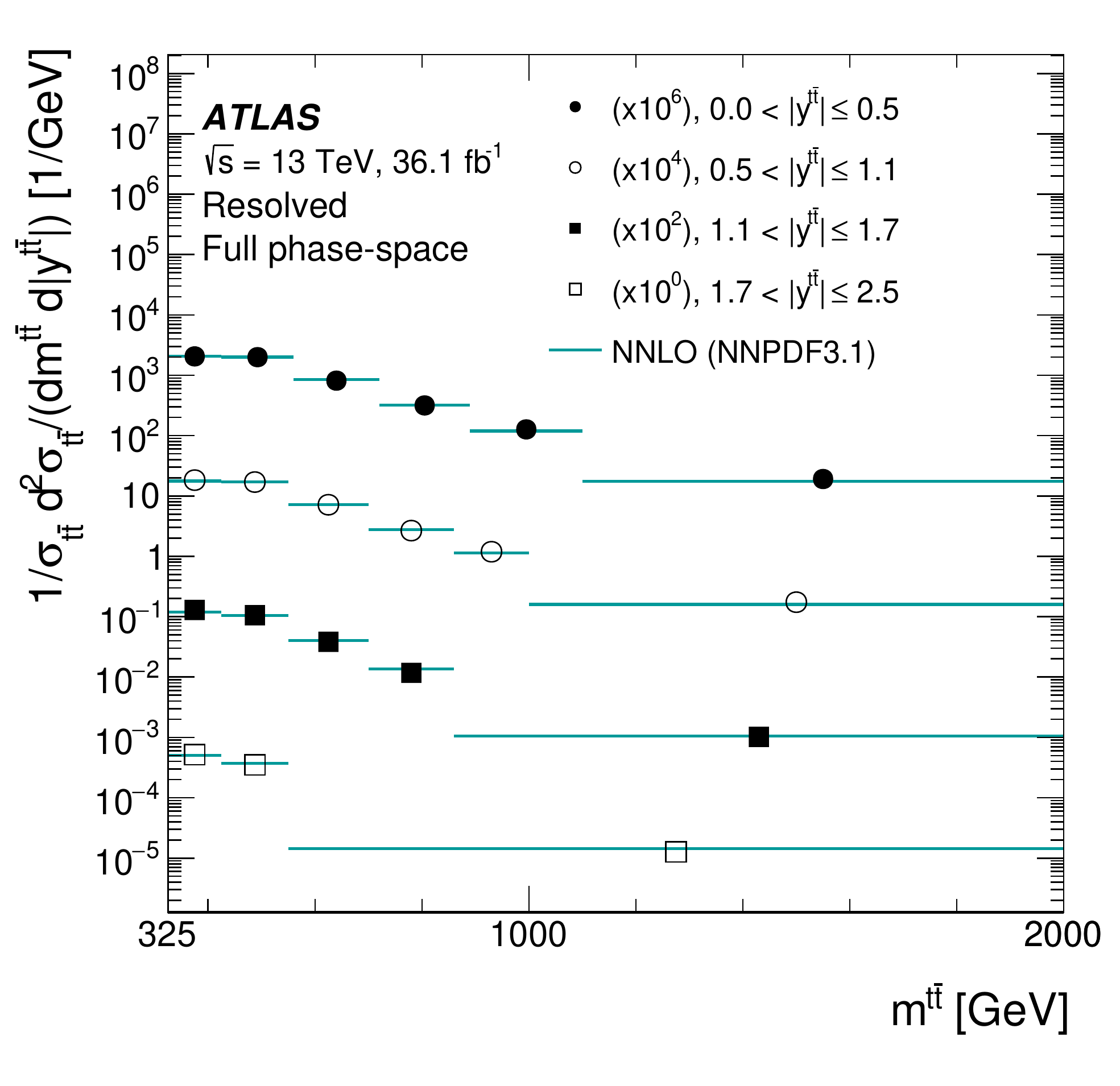}
\label{fig:results:NNLO:resolved:ttbar_m:ttbar_y:rel}}
\subfigure[]{\includegraphics[width=0.58\textwidth]{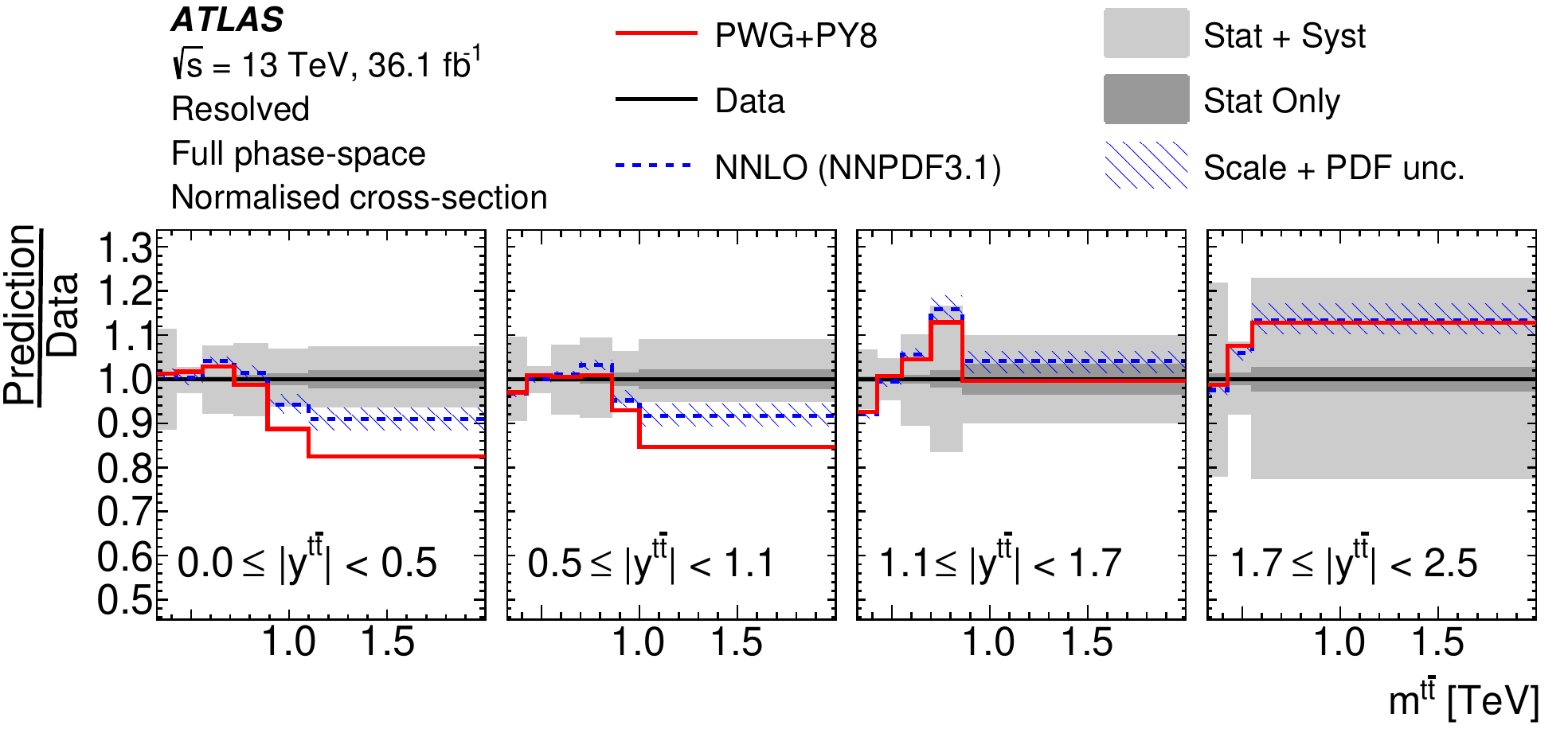}
\label{fig:results:NNLO:resolved:ttbar_m:ttbar_y:rel:ratio}}
\caption{\small{\subref{fig:results:NNLO:resolved:ttbar_m:ttbar_y:rel} Parton-level normalised differential cross-section as a function of \mtt{} in bins of \absyttbar{} in the resolved topology compared with  the NNLO prediction obtained
using the NNPDF3.1 NNLO PDF set.  Data points are placed at the centre of each bin. \subref{fig:results:NNLO:resolved:ttbar_m:ttbar_y:rel:ratio} The ratio of the measured cross-section to the NNLO prediction  and the  prediction obtained with the \Powheg+\PythiaEight{} MC generator.  The hatched band represents the total uncertainty in the NNLO prediction. The solid bands represent the statistical and total uncertainty in the data. }}
\label{fig:results:NNLO:rel:parton:resolved:2D:ttbar_abs_y:ttbar_m}
\end{figure*}

The single-differential normalised cross-sections are measured
as a function of the transverse momentum and absolute value of the  rapidity  of the top quark and as a function of the mass,
transverse momentum and absolute value of the rapidity of the \ttb{} system and of the additional variables $\left|\yboost\right|$, \Httbar{} and \chitt{}.
The results are shown in Figures~\ref{fig:results:NNLO:rel:parton:resolved:1D:top}--\ref{fig:results:NNLO:rel:parton:resolved:1D:additional_variables}.  The quantitative comparisons among the parton-level results and MC predictions, obtained with a $\chi^2$ test statistic, are shown in Tables~\ref{tab:chisquare:relative:1D:allpred:resolved:parton} and~\ref{tab:chisquare:absolute:1D:allpred:resolved:parton}, for normalised and absolute single-differential distributions, respectively.
The double-differential cross-sections, presented in Figures~\ref{fig:results:NNLO:rel:parton:resolved:2D:top_abs_y:top_pt}--\ref{fig:results:NNLO:rel:parton:resolved:2D:ttbar_abs_y:ttbar_m}, are measured as a function of \ptt{} in bins of \mtt{}, \pttt{}
and \absyt{}, as a function  of  \pttt{} in bins of \mtt{} and \absyttbar{} and finally as a function of \mtt{} in bins of  \absyttbar{}. The quantitative comparisons among the parton-level results and MC predictions, obtained with a $\chi^2$ test statistic, are shown in Tables~\ref{tab:chisquare:relative:2D:allpred:resolved:parton} and~\ref{tab:chisquare:absolute:2D:allpred:resolved:parton}, for normalised and absolute single-differential distributions, respectively.
 
The measured  differential cross-sections are compared with the fixed-order NNLO pQCD predictions and with
the \Powheg+\PythiaEight{} NLO+PS parton-level predictions.  In the case of the top-quark \pt{} and rapidity, NNLO predictions are available for the distributions of the top/anti-top average, which are calculated not on an event-by-event basis but by averaging the results of the histograms of the distributions of the top and anti-top quark~\cite{Czakon:2017wor}. For these variables, the measured differential cross-sections are taken as a function of the hadronic top quark's kinematics.

The NNLO pQCD predictions are obtained, for the optimised binning of this analysis, using the NNLO NNPDF3.1 PDF set~\cite{NNPDF3.1} with the renormalisation ($\mu_{\rm R}$) and factorisation ($\mu_{\rm F}$) scales both set to $H_{\rm T}/4$ (with $H_{\rm T}$ equal to the sum of the transverse masses of the top and anti-top quark) for all the measured differential cross-sections with the exception of the  differential cross-section as a function of $\ptt$, for which both scales were set to $m_{\mathrm{T}}/2$~\cite{Czakon:2016dgf}.\footnote{$m_{\mathrm{T}}=\sqrt{m_t^2+p_{\mathrm{T},t}^2}$.}  The top-quark pole mass is set to $172.5$ \GeV{}.
The theoretical uncertainty in the central NNLO predictions is obtained by summing in quadrature the uncertainty
due to the higher-order terms, estimated from the envelope of the predictions obtained by independently increasing and decreasing
$\mu_{\rm R}$ and $\mu_{\rm F}$ by a factor of two relative to the central scale choice, and the uncertainty due to the PDFs obtained according to the prescription of the NNPDF Collaboration.  The quantitative comparisons among the parton-level results and the NNLO pQCD predictions, obtained with a $\chi^2$ test statistic, are shown in Tables~\ref{tab:chisquare:relative:1D:NNLO:resolved:parton} and~\ref{tab:chisquare:absolute:1D:NNLO:resolved:parton} and Tables~\ref{tab:chisquare:relative:2D:NNLO:resolved:parton} and~\ref{tab:chisquare:absolute:2D:NNLO:resolved:parton}, for single- and double-differential distributions, respectively.
 
For the single-differential cross-sections the NNLO and NLO+PS predictions give a good and comparable description of the data, with the exception of \mtt{} that is poorly described by several NLO+PS predictions. Regarding the measured double-differential cross-sections, tensions are observed for several variables with respect to the NLO+PS predictions while a better description is observed when comparing the measurements with the NNLO calculations.
In the double-differential cross-sections as a function of \ptt{} in bins of \mtt{}, shown in Figure~\ref{fig:results:NNLO:rel:parton:resolved:2D:ttbar_m:top_pt}, the NNLO and NLO+PS central predictions show a contrasting behaviour, with the \Powheg+\PythiaEight{} predictions giving a better description of the data in the low \mtt{} region while the NNLO predictions better model the measurements in the high \mtt{} region.
 
The absolute differential cross-sections as a function of \ptt, \yt{}, \pttt{}, \absyttbar{} and \mtt{} are also measured using a  coarser binning,\footnote{The binning used for this comparison is tested and fully validated against the stability of the unfolding procedure.} used in a recent measurement from the CMS Collaboration~\cite{CMS-TOP-17-014}, to test the impact of including EW corrections in the NNLO pQCD predictions. These EW corrections~\cite{Czakon:2017wor} include the NLO EW effects of ${\cal O}(\alphas^2 \alpha)$, all subleading NLO (${\cal O}(\alphas \alpha^2)$ and ${\cal O}( \alpha^3)$) terms as well as the LO (${\cal O}(\alphas \alpha)$ and ${\cal O}( \alpha^2)$) contributions in the QCD and EW coupling constants. For these predictions, the mass of the top quark is set to 173.3~\GeV.
 
\begin{figure*}[t]
\centering
\subfigure[]{\includegraphics[width=0.45\textwidth]{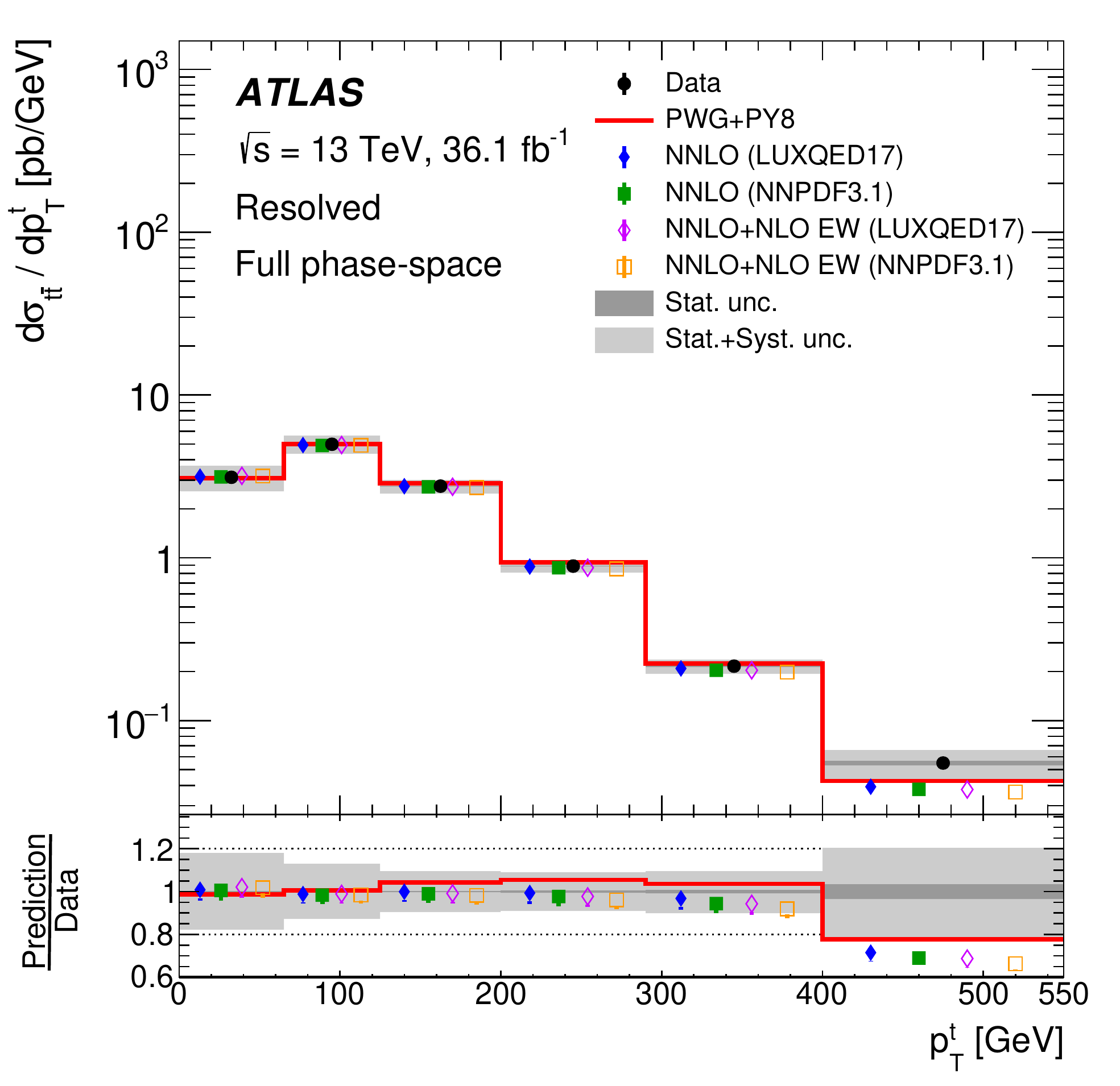}\label{fig:results:NNLO_EW:resolved:top_pt:abs}}
\subfigure[]{\includegraphics[width=0.45\textwidth]{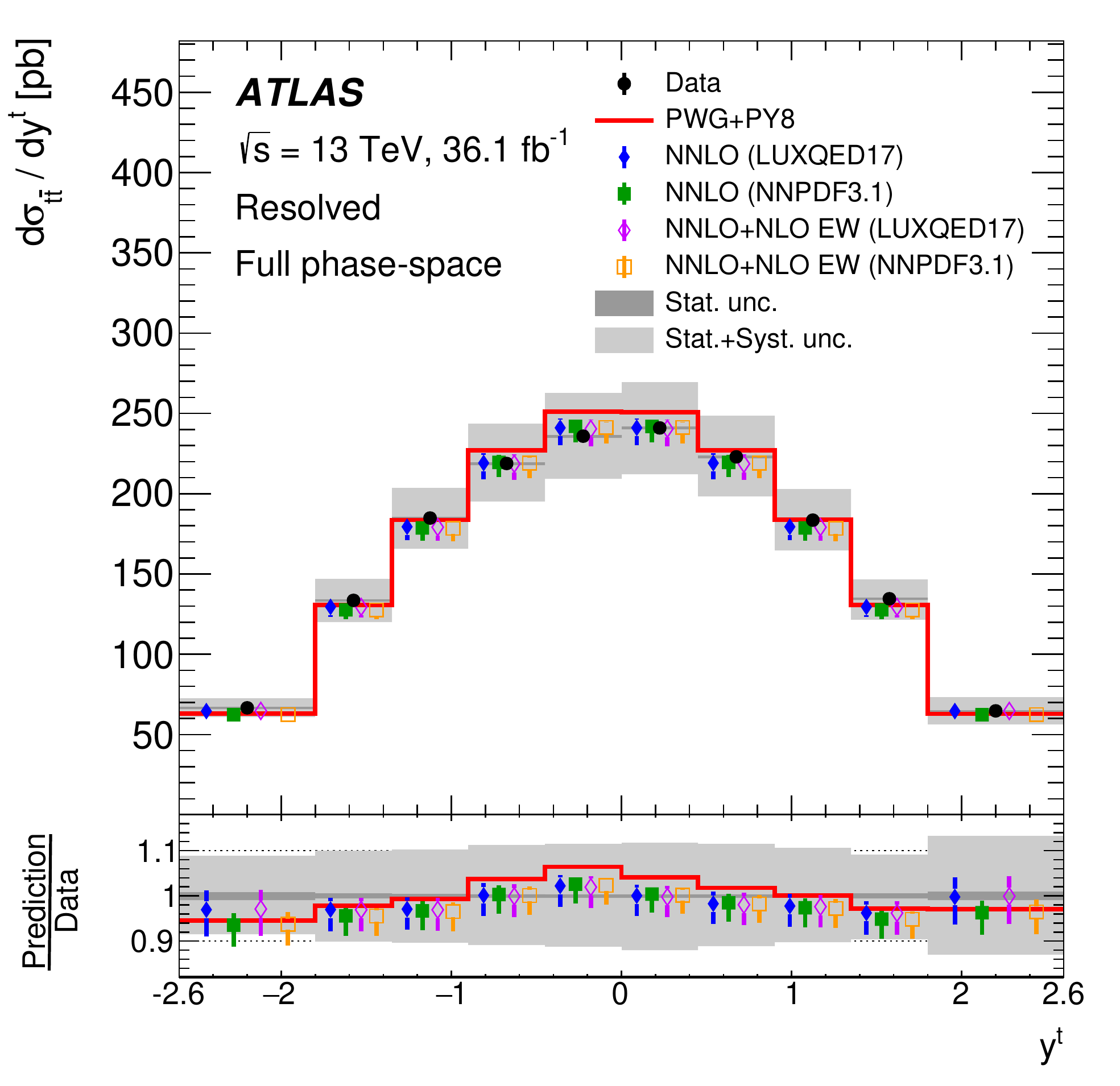}\label{fig:results:NNLO_EW:resolved:top_y:abs}}
 
\caption{\small{Parton-level absolute differential cross-sections as a function of~\subref{fig:results:NNLO_EW:resolved:top_pt:abs} \ptt{}
and~\subref{fig:results:NNLO_EW:resolved:top_y:abs} $y^t$ in the resolved topology. The results are compared with NNLO QCD and NNLO QCD+NLO EW theoretical calculations using the NNPDF3.1 and LUXQED17 PDF sets.  The vertical bars on each marker represents the total uncertainty in the prediction. The solid line is the nominal NLO \Powheg+\PythiaEight{} prediction. The bands represent the statistical and total uncertainty in the data.  Data points are placed at the centre of each bin. The lower panel shows the ratios of the predictions to data. The binning adopted in these distributions is the same used in a recent measurement from the CMS collaboration~\cite{CMS-TOP-17-014}. }}\label{fig:results:NNLO_EW:abs:parton:resolved:top}
\end{figure*}

\begin{figure*}[t]
\centering
\subfigure[]{\includegraphics[width=0.45\textwidth]{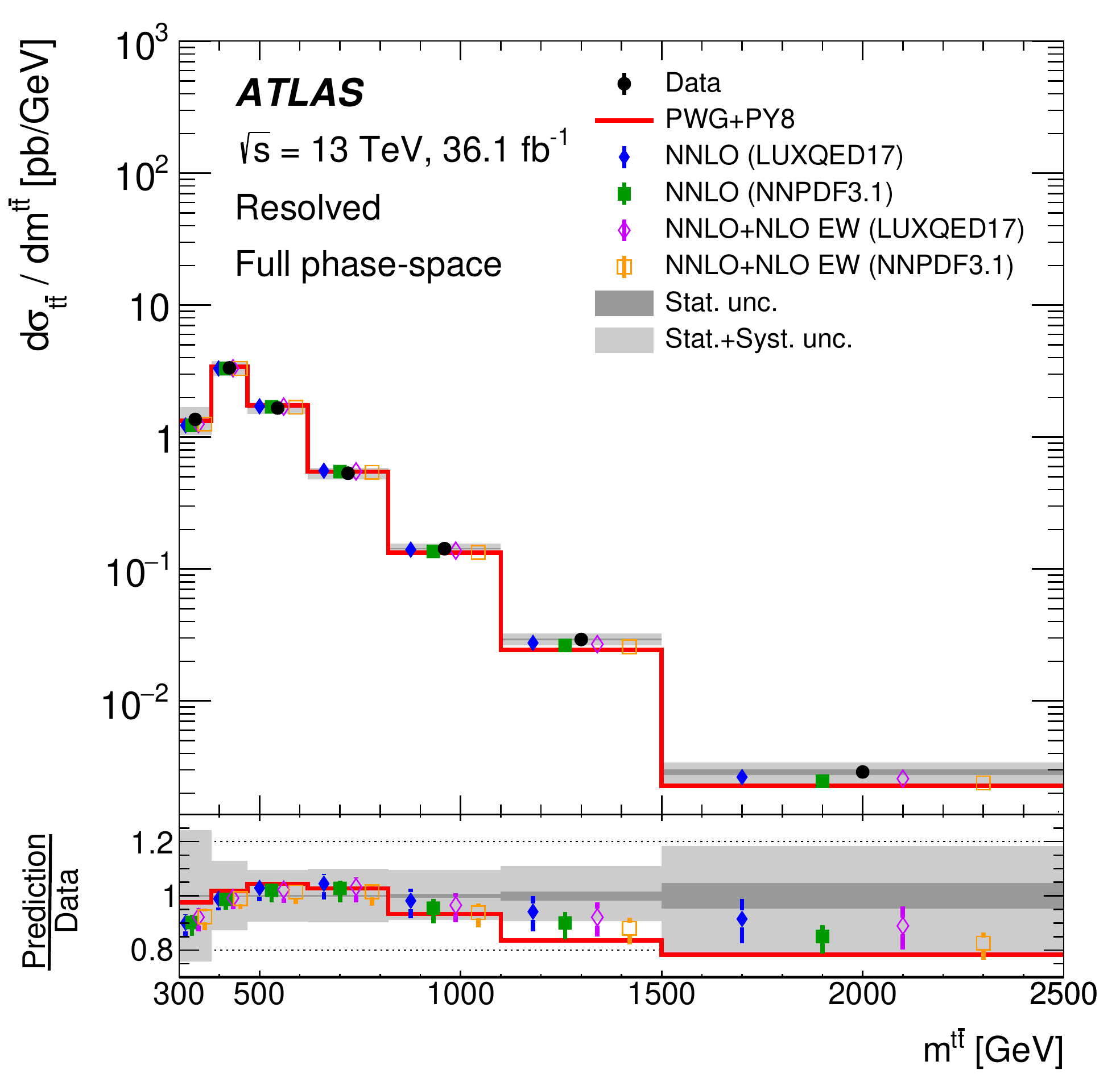}\label{fig:results:NNLO_EW:resolved:ttbar_m:abs}}
\subfigure[]{\includegraphics[width=0.45\textwidth]{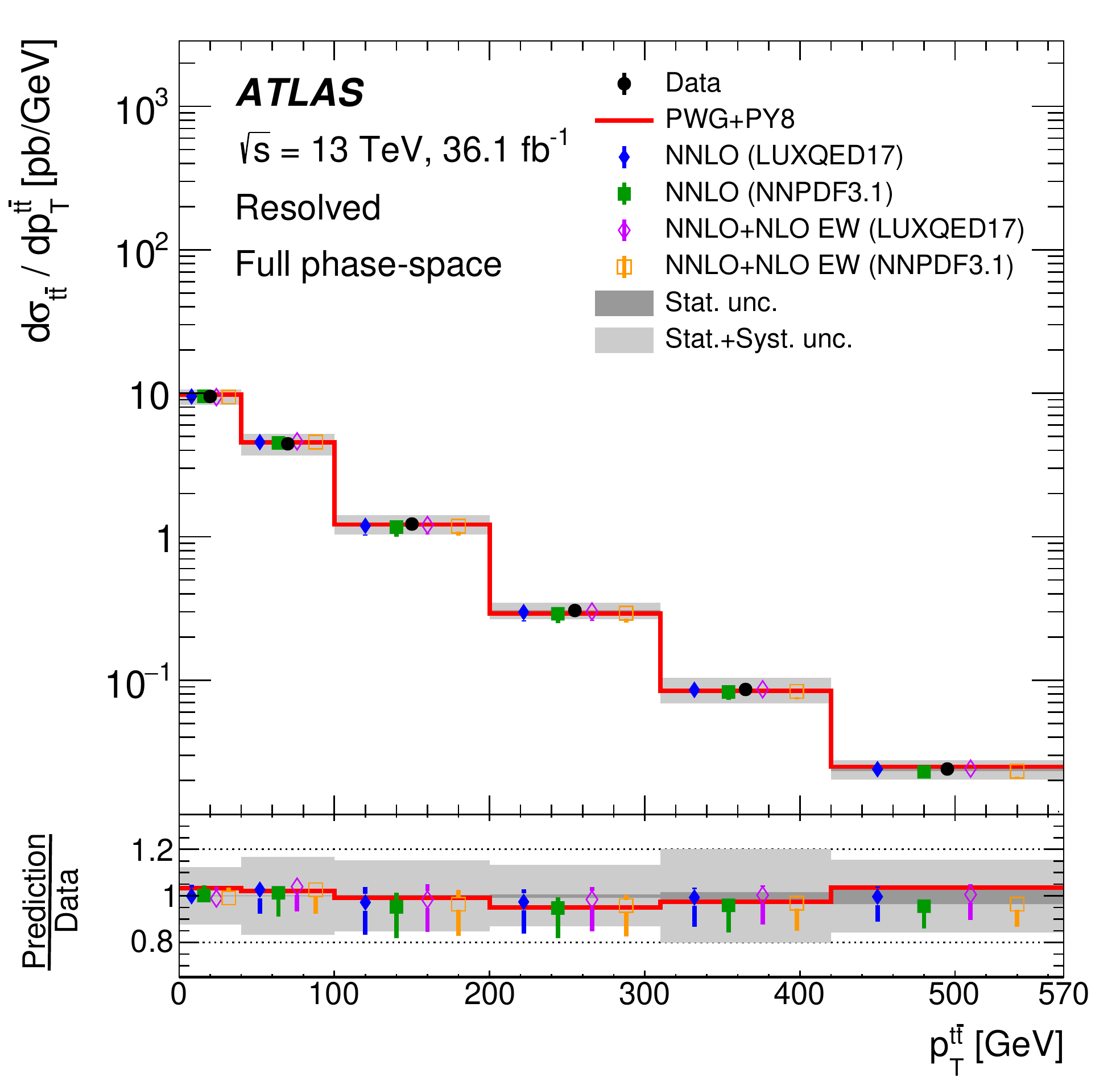}\label{fig:results:NNLO_EW:resolved:ttbar_pt:abs}}
\subfigure[]{\includegraphics[width=0.45\textwidth]{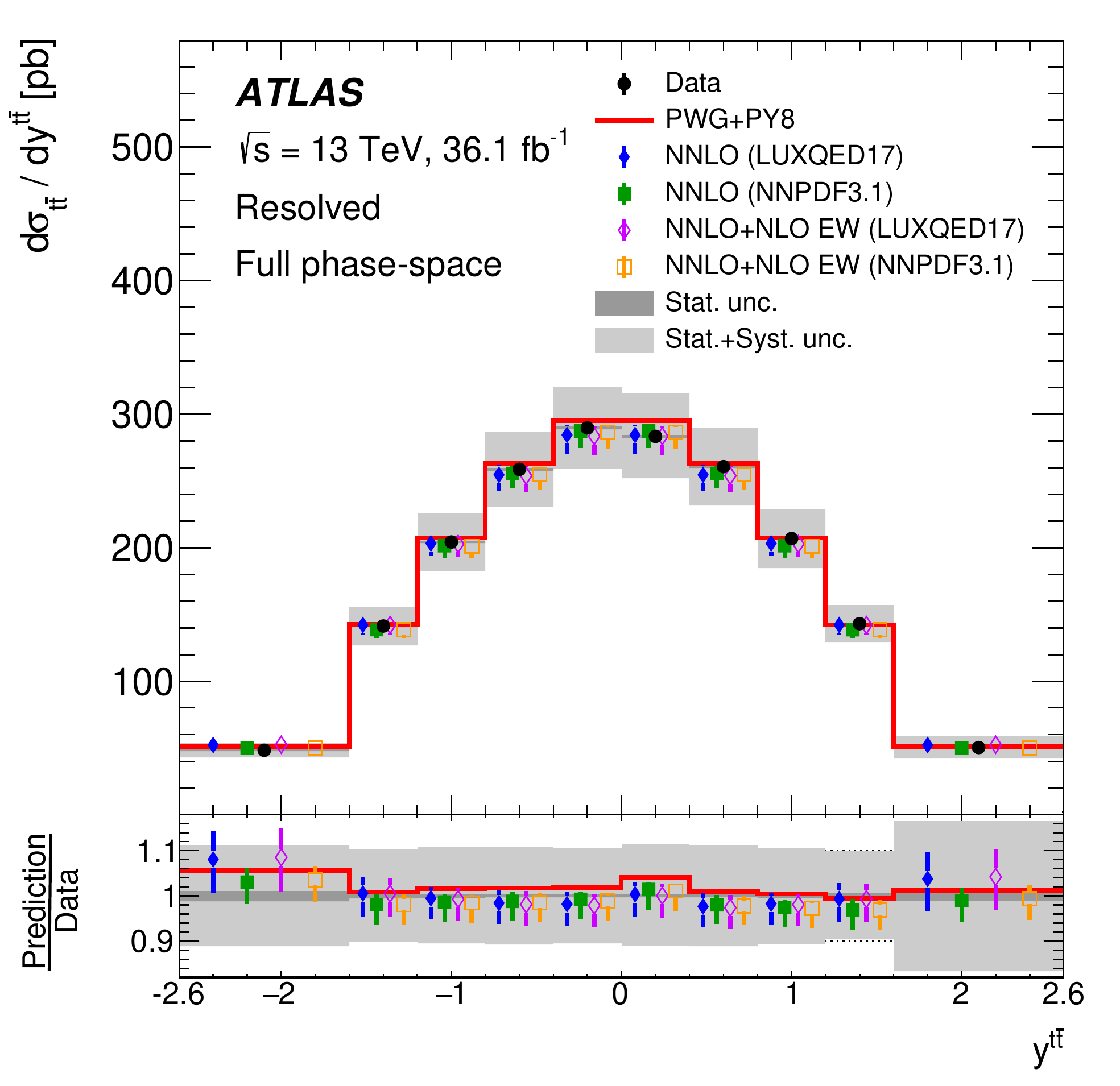}\label{fig:results:NNLO_EW:resolved:ttbar_y:abs}}
 
\caption{\small{Parton-level absolute differential cross-sections as a function of~\subref{fig:results:NNLO_EW:resolved:ttbar_m:abs}  \mtt, \subref{fig:results:NNLO_EW:resolved:ttbar_pt:abs}~\pttt{} and~\subref{fig:results:NNLO_EW:resolved:ttbar_y:abs}~\ytt{} in the resolved topology. The results are compared with NNLO QCD and NNLO QCD+NLO EW theoretical calculations using the NNPDF3.1 and LUXQED17 PDF sets. The vertical bars on each marker represents the total uncertainty in the prediction. The solid line is the nominal NLO \Powheg+\PythiaEight{} prediction. The bands represent the statistical and total uncertainty in the data.  Data points are placed at the centre of each bin. The lower panel shows the ratios of the predictions to data. The binning adopted in these distributions is the same used in a recent measurement from the CMS collaboration~\cite{CMS-TOP-17-014}.}}\label{fig:results:NNLO_EW:abs:parton:resolved:ttbar}
\end{figure*}

These additional measurements are shown in Figures~\ref{fig:results:NNLO_EW:abs:parton:resolved:top} and~\ref{fig:results:NNLO_EW:abs:parton:resolved:ttbar} and are compared with theoretical predictions obtained, with and without EW corrections, with two different PDF sets: the NNLO NNPDF3.1 PDF set and the LUXQED17 PDF set~\cite{Manohar:2017eqh}, the latter includes in addition to the standard partonic structure of the proton its photon component. The still rather limited range covered by the transverse momenta of top and anti-top quarks does not yet allow quantitative tests of the impact of the EW corrections as well as  the contribution of  the PDF of the photon in the proton to the production of top-quark pairs.

\FloatBarrier
 
\begin{table}[t]
\footnotesize
\centering\noindent\makebox[\textwidth]{
\renewcommand*{\arraystretch}{1.2}\begin{tabular}{|c | r @{/} l r  | r @{/} l r  | r @{/} l r  | r @{/} l r  | r @{/} l r |}
\hline
Observable
& \multicolumn{3}{c|}{\textsc{Pwg+Py8}}& \multicolumn{3}{c|}{\textsc{Pwg+Py8} Rad.~Up}& \multicolumn{3}{c|}{\textsc{Pwg+Py8} Rad.~Down}& \multicolumn{3}{c|}{\textsc{Pwg+H7}}& \multicolumn{3}{c|}{\textsc{Sherpa} 2.2.1}\\
& \multicolumn{2}{c}{$\chi^{2}$/NDF} &  ~$p$-value& \multicolumn{2}{c}{$\chi^{2}$/NDF} &  ~$p$-value& \multicolumn{2}{c}{$\chi^{2}$/NDF} &  ~$p$-value& \multicolumn{2}{c}{$\chi^{2}$/NDF} &  ~$p$-value& \multicolumn{2}{c}{$\chi^{2}$/NDF} &  ~$p$-value\\
\hline
\hline
$H_{\mathrm{T}}^{t\bar{t}}$ &{\ } 3.8 & 8 & 0.88 & {\ } 2.9 & 8 & 0.94 & {\ } 4.0 & 8 & 0.86 & {\ } 2.1 & 8 & 0.98 & {\ } 10.1 & 8 & 0.26\\
$|y_{\mathrm{boost}}^{t\bar{t}}|$ &{\ } 4.9 & 8 & 0.77 & {\ } 5.3 & 8 & 0.73 & {\ } 5.1 & 8 & 0.74 & {\ } 4.8 & 8 & 0.78 & {\ } 5.6 & 8 & 0.70\\
$     \chi^{t\bar{t}}$ &{\ } 9.7 & 3 & 0.02 & {\ } 4.2 & 3 & 0.24 & {\ } 20.9 & 3 & $<$0.01 & {\ } 5.8 & 3 & 0.12 & {\ } 19.1 & 3 & $<$0.01\\
$             |y^{t}|$ &{\ } 9.4 & 4 & 0.05 & {\ } 8.8 & 4 & 0.07 & {\ } 10.3 & 4 & 0.03 & {\ } 8.4 & 4 & 0.08 & {\ } 9.8 & 4 & 0.04\\
$  p_{\mathrm{T}}^{t}$ &{\ } 6.4 & 7 & 0.49 & {\ } 5.8 & 7 & 0.56 & {\ } 6.8 & 7 & 0.45 & {\ } 4.7 & 7 & 0.69 & {\ } 7.6 & 7 & 0.37\\
$      |y^{t\bar{t}}|$ &{\ } 4.1 & 6 & 0.67 & {\ } 4.5 & 6 & 0.61 & {\ } 4.3 & 6 & 0.63 & {\ } 4.1 & 6 & 0.66 & {\ } 4.4 & 6 & 0.62\\
$        m^{t\bar{t}}$ &{\ } 32.1 & 8 & $<$0.01 & {\ } 26.7 & 8 & $<$0.01 & {\ } 37.6 & 8 & $<$0.01 & {\ } 29.6 & 8 & $<$0.01 & {\ } 17.1 & 8 & 0.03\\
$p_{\mathrm{T}}^{t\bar{t}}$ &{\ } 7.8 & 8 & 0.45 & {\ } 41.7 & 8 & $<$0.01 & {\ } 25.0 & 8 & $<$0.01 & {\ } 11.9 & 8 & 0.15 & {\ } 22.1 & 8 & $<$0.01\\
\hline
\end{tabular}}
\caption{Comparison of the measured parton-level normalised single-differential cross-sections in the resolved topology with the predictions from several MC generators. For each prediction a $\chi^2$ and a $p$-value are calculated using the covariance matrix of the measured spectrum. The NDF is equal to the number of bins in the distribution minus one.}
\label{tab:chisquare:relative:1D:allpred:resolved:parton}
\end{table}
 
\begin{table}[t]
\footnotesize
\centering\noindent\makebox[\textwidth]{
\renewcommand*{\arraystretch}{1.2}\begin{tabular}{|c | r @{/} l r  | r @{/} l r  | r @{/} l r  | r @{/} l r  | r @{/} l r |}
\hline
Observable
& \multicolumn{3}{c|}{\textsc{Pwg+Py8}}& \multicolumn{3}{c|}{\textsc{Pwg+Py8} Rad.~Up}& \multicolumn{3}{c|}{\textsc{Pwg+Py8} Rad.~Down}& \multicolumn{3}{c|}{\textsc{Pwg+H7}}& \multicolumn{3}{c|}{\textsc{Sherpa} 2.2.1}\\
& \multicolumn{2}{c}{$\chi^{2}$/NDF} &  ~$p$-value& \multicolumn{2}{c}{$\chi^{2}$/NDF} &  ~$p$-value& \multicolumn{2}{c}{$\chi^{2}$/NDF} &  ~$p$-value& \multicolumn{2}{c}{$\chi^{2}$/NDF} &  ~$p$-value& \multicolumn{2}{c}{$\chi^{2}$/NDF} &  ~$p$-value\\
\hline
\hline
$H_{\mathrm{T}}^{t\bar{t}}$ &{\ } 9.9 & 9 & 0.36 & {\ } 10.1 & 9 & 0.34 & {\ } 9.9 & 9 & 0.36 & {\ } 6.7 & 9 & 0.67 & {\ } 19.6 & 9 & 0.02\\
$|y_{\mathrm{boost}}^{t\bar{t}}|$ &{\ } 5.9 & 9 & 0.75 & {\ } 6.4 & 9 & 0.70 & {\ } 6.2 & 9 & 0.72 & {\ } 5.8 & 9 & 0.76 & {\ } 6.4 & 9 & 0.70\\
$     \chi^{t\bar{t}}$ &{\ } 10.7 & 4 & 0.03 & {\ } 4.5 & 4 & 0.34 & {\ } 23.6 & 4 & $<$0.01 & {\ } 6.3 & 4 & 0.18 & {\ } 22.1 & 4 & $<$0.01\\
$             |y^{t}|$ &{\ } 10.8 & 5 & 0.06 & {\ } 10.0 & 5 & 0.08 & {\ } 12.2 & 5 & 0.03 & {\ } 9.5 & 5 & 0.09 & {\ } 10.9 & 5 & 0.05\\
$  p_{\mathrm{T}}^{t}$ &{\ } 9.9 & 8 & 0.27 & {\ } 8.8 & 8 & 0.36 & {\ } 10.8 & 8 & 0.21 & {\ } 8.2 & 8 & 0.42 & {\ } 11.9 & 8 & 0.15\\
$      |y^{t\bar{t}}|$ &{\ } 5.0 & 7 & 0.66 & {\ } 5.5 & 7 & 0.60 & {\ } 5.2 & 7 & 0.63 & {\ } 4.9 & 7 & 0.67 & {\ } 5.2 & 7 & 0.63\\
$        m^{t\bar{t}}$ &{\ } 29.1 & 9 & $<$0.01 & {\ } 22.9 & 9 & $<$0.01 & {\ } 36.8 & 9 & $<$0.01 & {\ } 25.6 & 9 & $<$0.01 & {\ } 15.4 & 9 & 0.08\\
$p_{\mathrm{T}}^{t\bar{t}}$ &{\ } 8.6 & 9 & 0.47 & {\ } 42.4 & 9 & $<$0.01 & {\ } 24.3 & 9 & $<$0.01 & {\ } 14.1 & 9 & 0.12 & {\ } 20.6 & 9 & 0.01\\
\hline
\end{tabular}}
\caption{ Comparison of the measured parton-level absolute single-differential cross-sections in the resolved topology with the predictions from several MC generators. For each prediction a $\chi^2$ and a $p$-value are calculated using the covariance matrix of the measured spectrum. The NDF is equal to the number of bins in the distribution. 
}
\label{tab:chisquare:absolute:1D:allpred:resolved:parton}
\end{table}
 
\FloatBarrier
\begin{table}[t]
\centering
\footnotesize
\centering\noindent\makebox[\textwidth]{
\renewcommand*{\arraystretch}{1.2}\begin{tabular}{|c | r @{/} l r  | r @{/} l r |}
\hline
Observable
& \multicolumn{3}{c|}{NNPDF31 NNLO}& \multicolumn{3}{c|}{\textsc{Pwg+Py8}}\\
& \multicolumn{2}{c}{$\chi^{2}$/NDF} &  ~$p$-value& \multicolumn{2}{c}{$\chi^{2}$/NDF} &  ~$p$-value\\
\hline
\hline
$H_{\mathrm{T}}^{t\bar{t}}$ &{\ } 5.0 & 8 & 0.76 & {\ } 3.8 & 8 & 0.88\\
$|y_{\mathrm{boost}}^{t\bar{t}}|$ &{\ } 8.6 & 8 & 0.38 & {\ } 4.9 & 8 & 0.77\\
$     \chi^{t\bar{t}}$ &{\ } 2.4 & 3 & 0.50 & {\ } 9.7 & 3 & 0.02\\
$|y^{t,\mathrm{had}}|$ &{\ } 8.2 & 4 & 0.09 & {\ } 9.4 & 4 & 0.05\\
$  p_{\mathrm{T}}^{t}$ &{\ } 6.3 & 7 & 0.51 & {\ } 6.4 & 7 & 0.49\\
$      |y^{t\bar{t}}|$ &{\ } 6.1 & 6 & 0.41 & {\ } 4.1 & 6 & 0.67\\
$        m^{t\bar{t}}$ &{\ } 17.2 & 8 & 0.03 & {\ } 32.1 & 8 & $<$0.01\\
$p_{\mathrm{T}}^{t\bar{t}}$ &{\ } 3.7 & 8 & 0.88 & {\ } 7.8 & 8 & 0.45\\
\hline
\end{tabular}}
\caption{Comparison of the measured parton-level normalised single-differential in the resolved topology cross-sections with the  NNLO predictions and the nominal \Powheg+\PythiaEight{} predictions. For each prediction a $\chi^2$ and a $p$-value are calculated using the covariance matrix of the measured spectrum. The NDF is equal to the number of bins in the distribution minus one.
}
\label{tab:chisquare:relative:1D:NNLO:resolved:parton}
\end{table}
 
\begin{table}[t]
\centering
\footnotesize
\centering\noindent\makebox[\textwidth]{
\renewcommand*{\arraystretch}{1.2}\begin{tabular}{|c | r @{/} l r  | r @{/} l r |}
\hline
Observable
& \multicolumn{3}{c|}{NNPDF31 NNLO}& \multicolumn{3}{c|}{\textsc{Pwg+Py8}}\\
& \multicolumn{2}{c}{$\chi^{2}$/NDF} &  ~$p$-value& \multicolumn{2}{c}{$\chi^{2}$/NDF} &  ~$p$-value\\
\hline
\hline
$H_{\mathrm{T}}^{t\bar{t}}$ &{\ } 10.4 & 9 & 0.32 & {\ } 9.9 & 9 & 0.36\\
$|y_{\mathrm{boost}}^{t\bar{t}}|$ &{\ } 10.9 & 9 & 0.28 & {\ } 5.9 & 9 & 0.75\\
$     \chi^{t\bar{t}}$ &{\ } 2.6 & 4 & 0.63 & {\ } 10.7 & 4 & 0.03\\
$|y^{t,\mathrm{had}}|$ &{\ } 9.5 & 5 & 0.09 & {\ } 10.8 & 5 & 0.06\\
$  p_{\mathrm{T}}^{t}$ &{\ } 7.8 & 8 & 0.45 & {\ } 9.9 & 8 & 0.27\\
$      |y^{t\bar{t}}|$ &{\ } 7.2 & 7 & 0.41 & {\ } 5.0 & 7 & 0.66\\
$        m^{t\bar{t}}$ &{\ } 14.0 & 9 & 0.12 & {\ } 29.1 & 9 & $<$0.01\\
$p_{\mathrm{T}}^{t\bar{t}}$ &{\ } 4.9 & 9 & 0.84 & {\ } 8.6 & 9 & 0.47\\
\hline
\end{tabular}}
\caption{Comparison of the measured parton-level absolute single-differential in the resolved topology cross-sections with the  NNLO predictions and the nominal \Powheg+\PythiaEight{} predictions. For each prediction a $\chi^2$ and a $p$-value are calculated using the covariance matrix of the measured spectrum. The NDF is equal to the number of bins in the distribution.}
\label{tab:chisquare:absolute:1D:NNLO:resolved:parton}
\end{table}

\FloatBarrier
 
\FloatBarrier
 
\begin{table}[t]
\footnotesize
\centering\noindent\makebox[\textwidth]{
\renewcommand*{\arraystretch}{1.2}\begin{tabular}{|c | r @{/} l r  | r @{/} l r  | r @{/} l r  | r @{/} l r  | r @{/} l r |}
\hline
Observable
& \multicolumn{3}{c|}{\textsc{Pwg+Py8}}& \multicolumn{3}{c|}{\textsc{Pwg+Py8} Rad.~Up}& \multicolumn{3}{c|}{\textsc{Pwg+Py8} Rad.~Down}& \multicolumn{3}{c|}{\textsc{Pwg+H7}}& \multicolumn{3}{c|}{\textsc{Sherpa} 2.2.1}\\
& \multicolumn{2}{c}{$\chi^{2}$/NDF} &  ~$p$-value& \multicolumn{2}{c}{$\chi^{2}$/NDF} &  ~$p$-value& \multicolumn{2}{c}{$\chi^{2}$/NDF} &  ~$p$-value& \multicolumn{2}{c}{$\chi^{2}$/NDF} &  ~$p$-value& \multicolumn{2}{c}{$\chi^{2}$/NDF} &  ~$p$-value\\
\hline
\hline
$|y^{t}|\textrm{ vs }p_{\mathrm{T}}^{t}$ &{\ } 30.9 & 12 & $<$0.01 & {\ } 30.2 & 12 & $<$0.01 & {\ } 34.7 & 12 & $<$0.01 & {\ } 22.9 & 12 & 0.03 & {\ } 44.3 & 12 & $<$0.01\\
$|y^{t\bar{t}}|\textrm{ vs }m^{t\bar{t}}$ &{\ } 51.8 & 19 & $<$0.01 & {\ } 47.0 & 19 & $<$0.01 & {\ } 56.6 & 19 & $<$0.01 & {\ } 49.4 & 19 & $<$0.01 & {\ } 41.4 & 19 & $<$0.01\\
$|y^{t\bar{t}}|\textrm{ vs }p_{\mathrm{T}}^{t\bar{t}}$ &{\ } 17.6 & 13 & 0.17 & {\ } 61.8 & 13 & $<$0.01 & {\ } 32.4 & 13 & $<$0.01 & {\ } 28.3 & 13 & $<$0.01 & {\ } 39.5 & 13 & $<$0.01\\
$m^{t\bar{t}}\textrm{ vs }p_{\mathrm{T}}^{t}$ &{\ } 64.6 & 14 & $<$0.01 & {\ } 118.0 & 14 & $<$0.01 & {\ } 129.0 & 14 & $<$0.01 & {\ } 60.9 & 14 & $<$0.01 & {\ } 63.4 & 14 & $<$0.01\\
$m^{t\bar{t}}\textrm{ vs }p_{\mathrm{T}}^{t\bar{t}}$ &{\ } 62.6 & 16 & $<$0.01 & {\ } 163.0 & 16 & $<$0.01 & {\ } 82.1 & 16 & $<$0.01 & {\ } 66.4 & 16 & $<$0.01 & {\ } 118.0 & 16 & $<$0.01\\
$p_{\mathrm{T}}^{t\bar{t}}\textrm{ vs }p_{\mathrm{T}}^{t}$ &{\ } 37.4 & 16 & $<$0.01 & {\ } 87.1 & 16 & $<$0.01 & {\ } 95.0 & 16 & $<$0.01 & {\ } 50.7 & 16 & $<$0.01 & {\ } 47.2 & 16 & $<$0.01\\
\hline
\end{tabular}}
\caption{Comparison of the measured parton-level normalised double-differential cross-sections in the resolved topology with the predictions from several MC generators. For each prediction a $\chi^2$ and a $p$-value are calculated using the covariance matrix of the measured spectrum. The NDF is equal to the number of bins in the distribution minus one.
}
\label{tab:chisquare:relative:2D:allpred:resolved:parton}
\end{table}
 
\begin{table}[t]
\footnotesize
\centering\noindent\makebox[\textwidth]{
\renewcommand*{\arraystretch}{1.2}\begin{tabular}{|c | r @{/} l r  | r @{/} l r  | r @{/} l r  | r @{/} l r  | r @{/} l r |}
\hline
Observable
& \multicolumn{3}{c|}{\textsc{Pwg+Py8}}& \multicolumn{3}{c|}{\textsc{Pwg+Py8} Rad.~Up}& \multicolumn{3}{c|}{\textsc{Pwg+Py8} Rad.~Down}& \multicolumn{3}{c|}{\textsc{Pwg+H7}}& \multicolumn{3}{c|}{\textsc{Sherpa} 2.2.1}\\
& \multicolumn{2}{c}{$\chi^{2}$/NDF} &  ~$p$-value& \multicolumn{2}{c}{$\chi^{2}$/NDF} &  ~$p$-value& \multicolumn{2}{c}{$\chi^{2}$/NDF} &  ~$p$-value& \multicolumn{2}{c}{$\chi^{2}$/NDF} &  ~$p$-value& \multicolumn{2}{c}{$\chi^{2}$/NDF} &  ~$p$-value\\
\hline
\hline
$|y^{t}|\textrm{ vs }p_{\mathrm{T}}^{t}$ &{\ } 33.2 & 13 & $<$0.01 & {\ } 32.4 & 13 & $<$0.01 & {\ } 37.3 & 13 & $<$0.01 & {\ } 24.5 & 13 & 0.03 & {\ } 48.5 & 13 & $<$0.01\\
$|y^{t\bar{t}}|\textrm{ vs }m^{t\bar{t}}$ &{\ } 55.6 & 20 & $<$0.01 & {\ } 50.4 & 20 & $<$0.01 & {\ } 61.3 & 20 & $<$0.01 & {\ } 52.9 & 20 & $<$0.01 & {\ } 44.6 & 20 & $<$0.01\\
$|y^{t\bar{t}}|\textrm{ vs }p_{\mathrm{T}}^{t\bar{t}}$ &{\ } 18.8 & 14 & 0.17 & {\ } 67.1 & 14 & $<$0.01 & {\ } 35.1 & 14 & $<$0.01 & {\ } 30.2 & 14 & $<$0.01 & {\ } 42.9 & 14 & $<$0.01\\
$m^{t\bar{t}}\textrm{ vs }p_{\mathrm{T}}^{t}$ &{\ } 70.5 & 15 & $<$0.01 & {\ } 126.0 & 15 & $<$0.01 & {\ } 138.0 & 15 & $<$0.01 & {\ } 65.5 & 15 & $<$0.01 & {\ } 73.3 & 15 & $<$0.01\\
$m^{t\bar{t}}\textrm{ vs }p_{\mathrm{T}}^{t\bar{t}}$ &{\ } 69.8 & 17 & $<$0.01 & {\ } 174.0 & 17 & $<$0.01 & {\ } 89.5 & 17 & $<$0.01 & {\ } 75.5 & 17 & $<$0.01 & {\ } 128.0 & 17 & $<$0.01\\
$p_{\mathrm{T}}^{t\bar{t}}\textrm{ vs }p_{\mathrm{T}}^{t}$ &{\ } 44.2 & 17 & $<$0.01 & {\ } 92.7 & 17 & $<$0.01 & {\ } 112.0 & 17 & $<$0.01 & {\ } 57.6 & 17 & $<$0.01 & {\ } 51.4 & 17 & $<$0.01\\
\hline
\end{tabular}}
\caption{ Comparison of the measured parton-level absolute double-differential cross-sections in the resolved topology with the predictions from several MC generators. For each prediction a $\chi^2$ and a $p$-value are calculated using the covariance matrix of the measured spectrum. The NDF is equal to the number of bins in the distribution.}
\label{tab:chisquare:absolute:2D:allpred:resolved:parton}
\end{table}
 
\begin{table}[t]
\centering
\footnotesize
\centering\noindent\makebox[\textwidth]{
\renewcommand*{\arraystretch}{1.2}\begin{tabular}{|c | r @{/} l r  | r @{/} l r |}
\hline
Observable
& \multicolumn{3}{c|}{NNPDF31 NNLO}& \multicolumn{3}{c|}{\textsc{Pwg+Py8}}\\
& \multicolumn{2}{c}{$\chi^{2}$/NDF} &  ~$p$-value& \multicolumn{2}{c}{$\chi^{2}$/NDF} &  ~$p$-value\\
\hline
\hline
$|y^{t}|\textrm{ vs }p_{\mathrm{T}}^{t}$ &{\ } 25.4 & 12 & 0.01 & {\ } 30.9 & 12 & $<$0.01\\
$|y^{t\bar{t}}|\textrm{ vs }m^{t\bar{t}}$ &{\ } 39.9 & 19 & $<$0.01 & {\ } 51.8 & 19 & $<$0.01\\
$|y^{t\bar{t}}|\textrm{ vs }p_{\mathrm{T}}^{t\bar{t}}$ &{\ } 15.9 & 13 & 0.26 & {\ } 17.6 & 13 & 0.17\\
$m^{t\bar{t}}\textrm{ vs }p_{\mathrm{T}}^{t}$ &{\ } 55.7 & 14 & $<$0.01 & {\ } 64.4 & 14 & $<$0.01\\
$m^{t\bar{t}}\textrm{ vs }p_{\mathrm{T}}^{t\bar{t}}$ &{\ } 40.6 & 16 & $<$0.01 & {\ } 62.6 & 16 & $<$0.01\\
$p_{\mathrm{T}}^{t\bar{t}}\textrm{ vs }p_{\mathrm{T}}^{t,\mathrm{had}}$ &{\ } 22.2 & 16 & 0.14 & {\ } 37.4 & 16 & $<$0.01\\
\hline
\end{tabular}}
 
\caption{Comparison of the measured parton-level normalised double-differential cross-sections in the resolved topology with the  NNLO predictions and the nominal \Powheg+\PythiaEight{} predictions. For each prediction a $\chi^2$ and a $p$-value are calculated using the covariance matrix of the measured spectrum. The NDF is equal to the number of bins in the distribution minus one.}
\label{tab:chisquare:relative:2D:NNLO:resolved:parton}
\end{table}
\begin{table}[t]
\centering
\footnotesize
\centering\noindent\makebox[\textwidth]{
\renewcommand*{\arraystretch}{1.2}\begin{tabular}{|c | r @{/} l r  | r @{/} l r |}
\hline
Observable
& \multicolumn{3}{c|}{NNPDF31 NNLO}& \multicolumn{3}{c|}{\textsc{Pwg+Py8}}\\
& \multicolumn{2}{c}{$\chi^{2}$/NDF} &  ~$p$-value& \multicolumn{2}{c}{$\chi^{2}$/NDF} &  ~$p$-value\\
\hline
\hline
$|y^{t}|\textrm{ vs }p_{\mathrm{T}}^{t}$ &{\ } 26.8 & 13 & 0.01 & {\ } 33.2 & 13 & $<$0.01\\
$|y^{t\bar{t}}|\textrm{ vs }m^{t\bar{t}}$ &{\ } 43.7 & 20 & $<$0.01 & {\ } 55.6 & 20 & $<$0.01\\
$|y^{t\bar{t}}|\textrm{ vs }p_{\mathrm{T}}^{t\bar{t}}$ &{\ } 17.1 & 14 & 0.17 & {\ } 18.8 & 14 & 0.17\\
$m^{t\bar{t}}\textrm{ vs }p_{\mathrm{T}}^{t}$ &{\ } 60.7 & 15 & $<$0.01 & {\ } 70.5 & 15 & $<$0.01\\
$m^{t\bar{t}}\textrm{ vs }p_{\mathrm{T}}^{t\bar{t}}$ &{\ } 47.4 & 17 & $<$0.01 & {\ } 69.8 & 17 & $<$0.01\\
$p_{\mathrm{T}}^{t\bar{t}}\textrm{ vs }p_{\mathrm{T}}^{t,\mathrm{had}}$ &{\ } 25.6 & 17 & 0.08 & {\ } 44.2 & 17 & $<$0.01\\
\hline
\end{tabular}}
\caption{Comparison of the measured parton-level absolute double-differential cross-sections in the resolved topology with the  NNLO predictions and the nominal \Powheg+\PythiaEight{} predictions. For each prediction a $\chi^2$ and a $p$-value are calculated using the covariance matrix of the measured spectrum. The NDF is equal to the number of bins in the distribution.
}
\label{tab:chisquare:absolute:2D:NNLO:resolved:parton}
\end{table}

\FloatBarrier

\FloatBarrier
\subsubsection{Boosted topology}
 
\label{sec:results:full:boosted}
\begin{figure*}[t]
\centering
\subfigure[]{\includegraphics[width=0.45\textwidth]{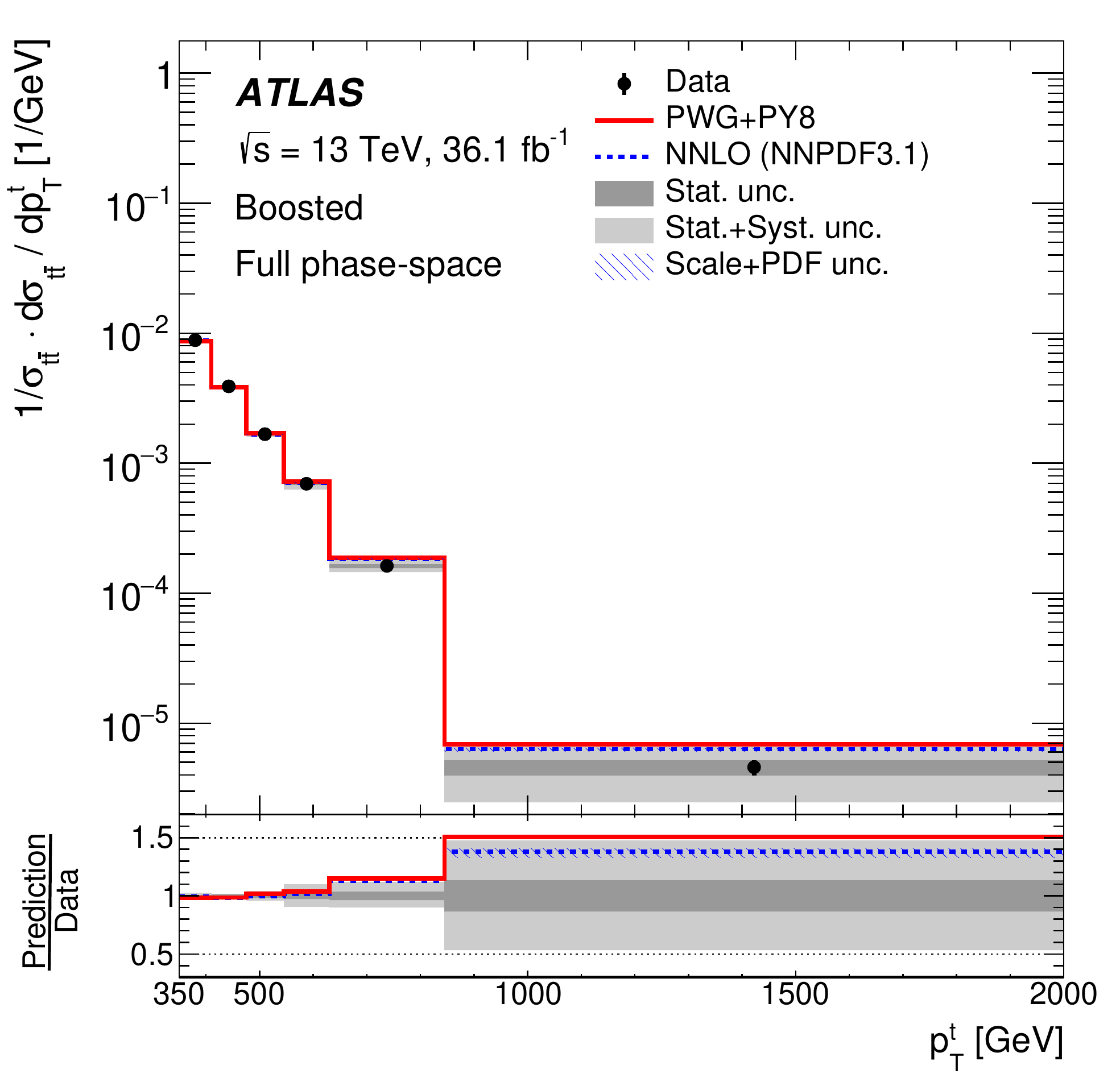}
\label{fig:results:NNLO:boosted:top_pt:rel}}
\subfigure[]{\includegraphics[width=0.45\textwidth]{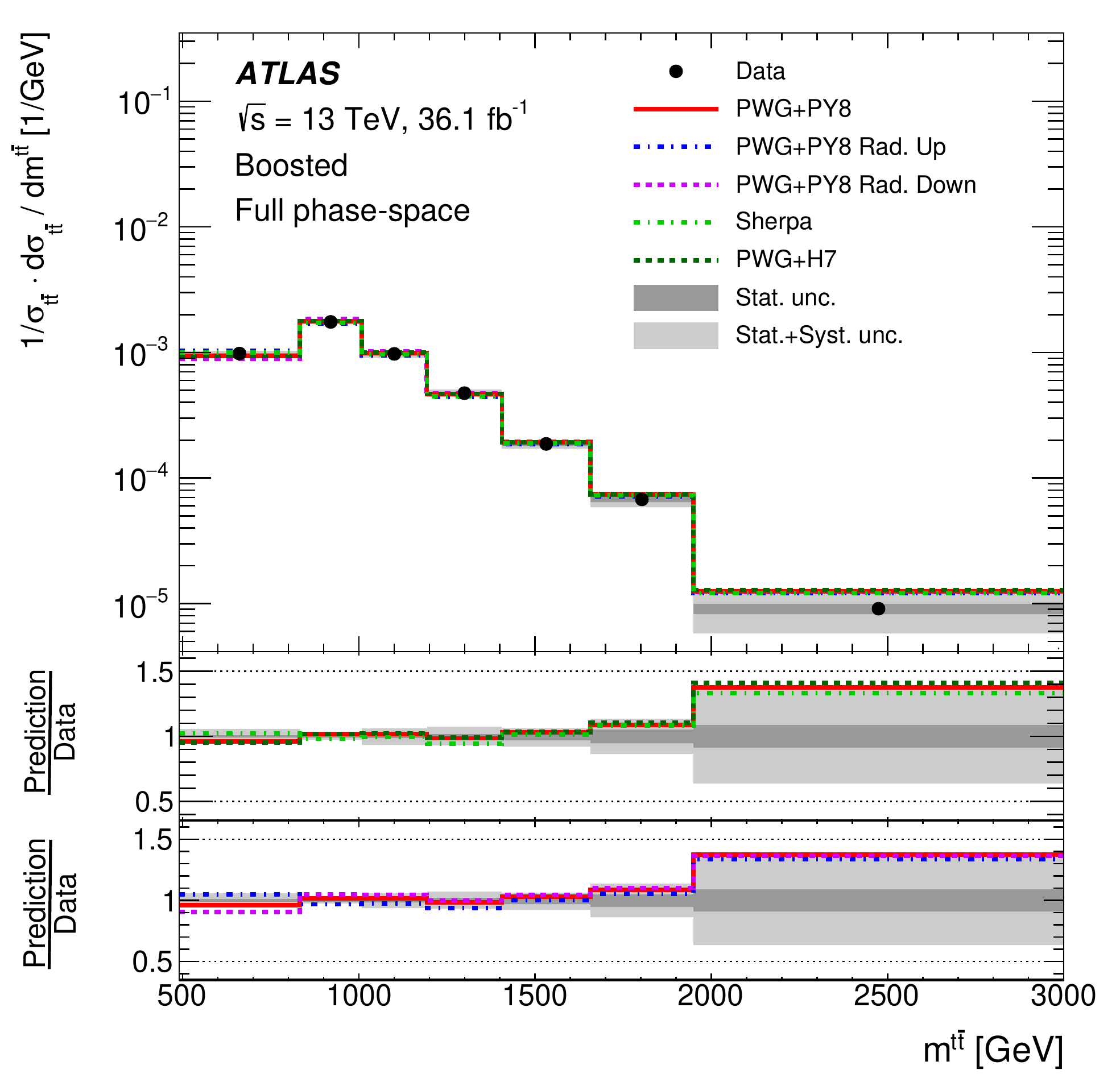}
\label{fig:results:parton:boosted:ttbar_m:rel}}
\caption{\small{\subref{fig:results:NNLO:boosted:top_pt:rel} Parton-level normalised differential cross-section as a function of \ptt{} in the boosted topology, compared with the NNLO predictions obtained
using the NNPDF3.1 NNLO PDF set and the  predictions obtained with the \Powheg+\PythiaEight{} MC generator. The hatched band represents the total uncertainty in the NNLO prediction. \subref{fig:results:parton:boosted:ttbar_m:rel} Parton-level normalised differential cross-section as a function of \mtt{} in the boosted topology, compared with predictions obtained  with different  MC generators. The bands represent the statistical and total uncertainty in the data.  Data points are placed at the centre of each bin. The lower panel shows the ratios of the predictions to data.}}
\label{fig:results:parton:rel:boosted:NNLO:1D}
\end{figure*}

\begin{figure*}[t]
\centering
\subfigure[]{\includegraphics[width=0.38\textwidth]{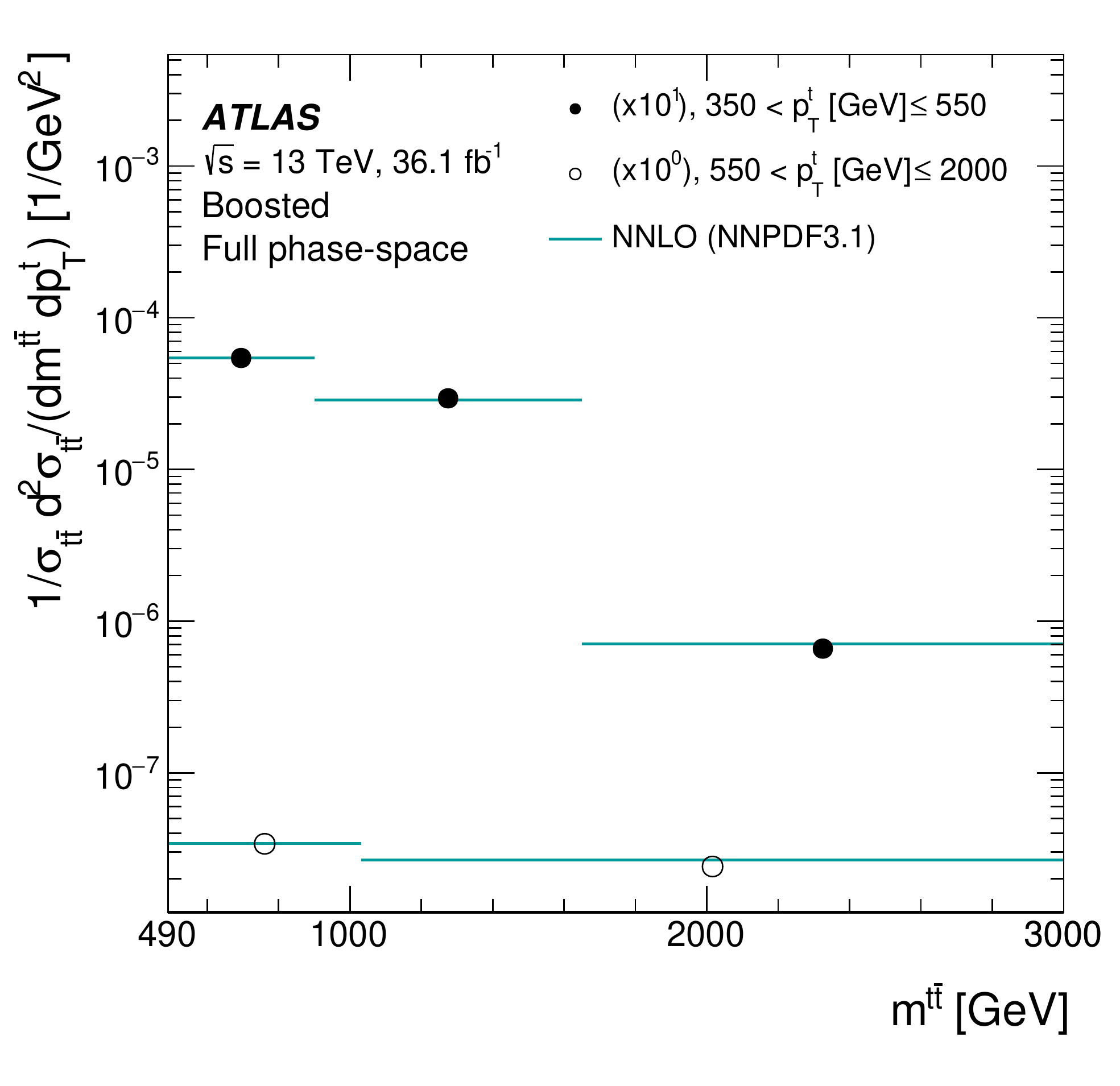}
\label{fig:results:NNLO:boosted:ttbar_m:top_pt:rel}}
\subfigure[]{\includegraphics[width=0.58\textwidth]{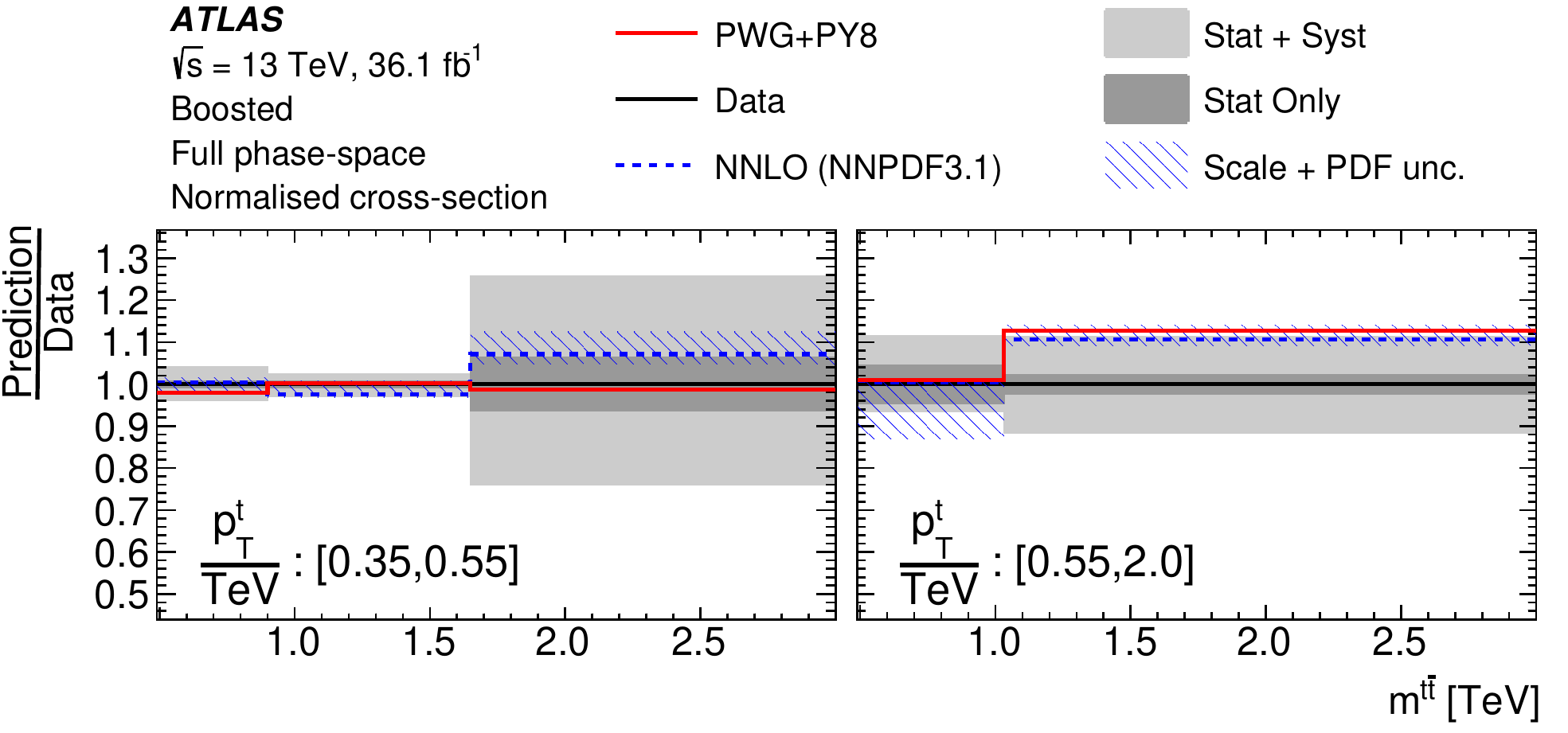}
\label{fig:results:NNLO:boosted:ttbar_m:top_pt:rel:ratio}}
\caption{\small{\subref{fig:results:NNLO:boosted:ttbar_m:top_pt:rel} Parton-level normalised differential cross-section as a function
of \mtt{} in bins of \ptt{} in the boosted topology compared with the NNLO prediction obtained
using the NNPDF3.1 NNLO PDF set.  Data points are placed at the centre of each bin.
\subref{fig:results:NNLO:boosted:ttbar_m:top_pt:rel:ratio} The ratio of the measured cross-section to the NNLO prediction  and the  prediction obtained with the \Powheg+\PythiaEight{} MC generator.  The hatched band represents the total uncertainty in the NNLO prediction. The bands represent the statistical and total uncertainty in the data. }}
 
\label{fig:results:parton:rel:boosted:NNLO:2D}
\end{figure*}

In the boosted topology, the parton-level normalised differential cross-sections are extracted in a region of the phase-space where the top quark is produced with $\pt > 350$~\GeV. The single-differential cross-sections are measured
as a function of the transverse momentum of the top quark and of the invariant mass of the \ttbar system.
The results are shown in Figure~\ref{fig:results:parton:rel:boosted:NNLO:1D}. The parton-level double-differential cross-sections, presented in Figure~\ref{fig:results:parton:rel:boosted:NNLO:2D}, are measured as a function of \mtt{} in bins of \ptt{}.
 
\begin{figure*}[t]
\centering
 
\includegraphics[width=0.58\textwidth]{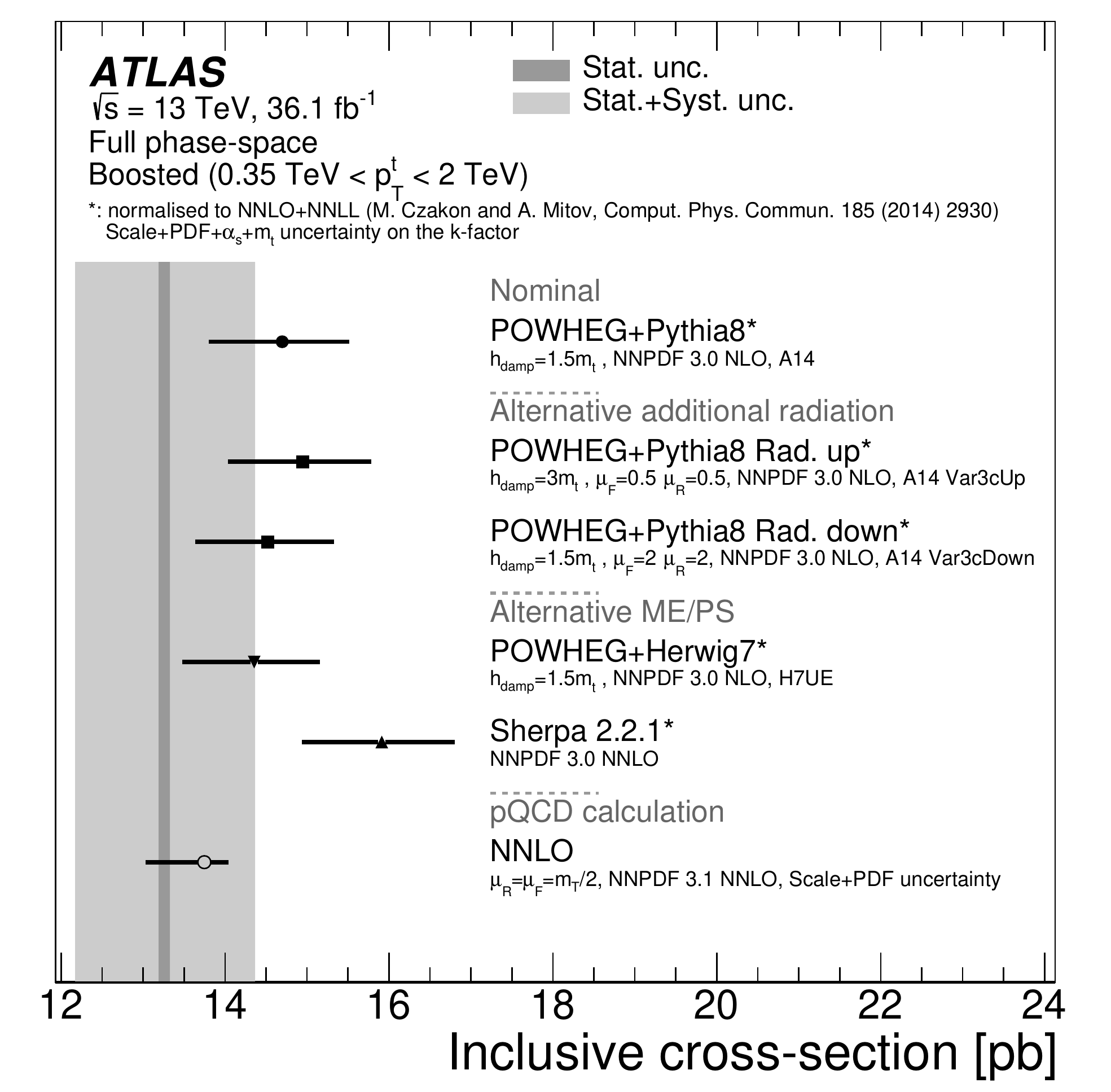}
\caption{Comparison of the measured inclusive parton-level cross-section in the boosted topology with the predictions from several MC generators and the NNLO prediction obtained using the NNPDF3.1 NNLO PDF set. The uncertainties associated to the NNLO prediction have been calculated starting from the scale and PDF uncertainties associated to the NNLO prediction of the differential cross-section as a function of \ptt{}. The uncertainty on the cross-section predicted by each NLO MC generator only includes the uncertainty (due to PDFs, $m_{t}$ and $\alpha_{s}$) affecting the $k$-factor used in the normalisation. The bands represent the statistical and total uncertainty in the data. }
\label{fig:results_parton:boosted:totalXs}
 
\end{figure*}

The inclusive parton-level cross-section measured in the boosted topology is shown in Figure~\ref{fig:results_parton:boosted:totalXs}, where it is compared with the MC predictions previously described and the NNLO calculation. The total cross-section predicted by each NLO MC generator is normalised to the NNLO+NNLL prediction as quoted in Ref.~\cite{Czakon:2011xx} and the corresponding uncertainty only includes the uncertainty affecting the $k$-factor used  in the normalisation. Since the parton-level definition in the boosted topology doesn't cover the full phase space, the inclusive cross-section predicted is different for each generator and differs from the normalisation value described in Section~\ref{sec:samples}. The prediction given by the NNLO calculation shows better agreement with the measured inclusive cross-section, while several NLO predictions overestimate data.
 
The measured single- and double-differential cross-sections are compared with the fixed-order NNLO pQCD predictions,
obtained using the same parameter settings already described for the resolved topology,  and with
the \Powheg+\PythiaEight{} NLO+PS parton-level predictions. A trend is observed in the agreement between the predictions and the measured single-differential cross-sections in the high \ptt{} and \mtt{} regions, where both the NLO+PS and NNLO (when available) predictions lie at the edge of the uncertainty band. Both the predictions, however, give a good description of the double-differential cross-section as a function of \mtt{} in bins of \ptt{}.
 
Tables~\ref{tab:chisquare:relative:allpred:boosted:parton} and~\ref{tab:chisquare:absolute:allpred:boosted:parton} and Tables~\ref{tab:chisquare:relative:NNLO:boosted:parton} and~\ref{tab:chisquare:absolute:NNLO:boosted:parton} show the quantitative comparisons among the parton-level results and the Monte Carlo and NNLO predictions. The normalised and absolute single- and double-differential cross-sections are shown.

\begin{figure*}[t]
\centering
\subfigure[]{\includegraphics[width=0.45\textwidth]{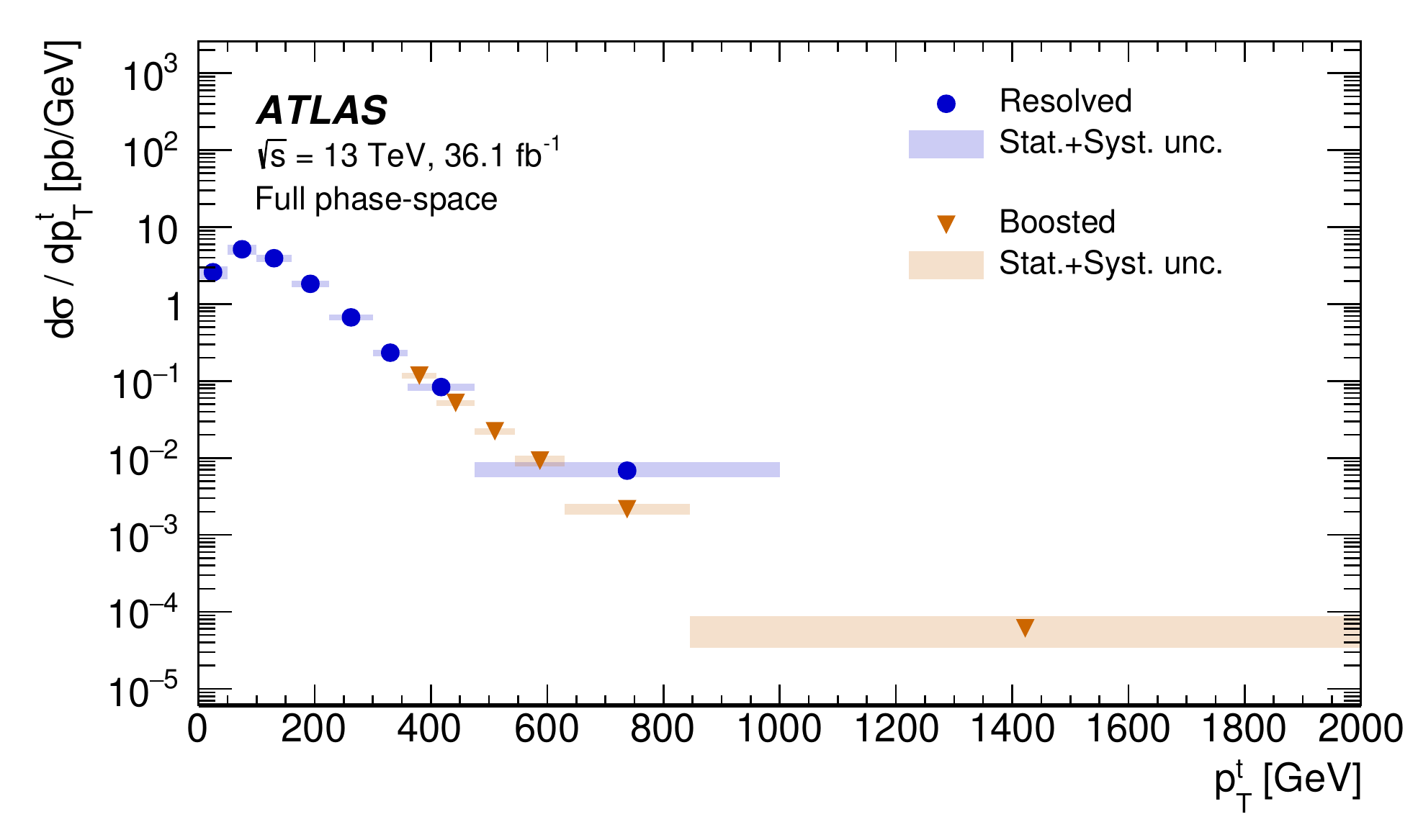}
\label{subfig:resolved_boosted:parton:comparison}}
\subfigure[]{\includegraphics[width=0.45\textwidth]{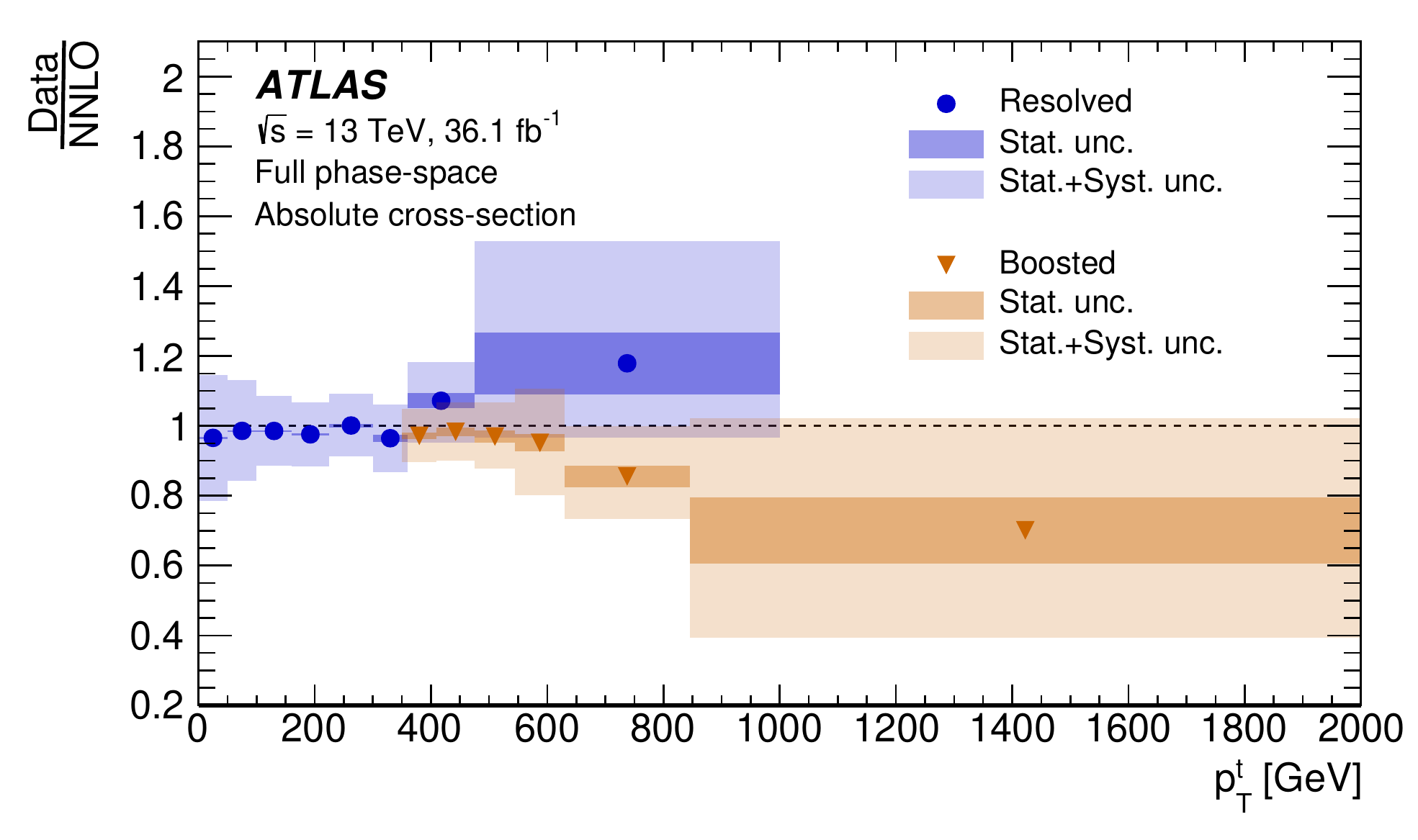}
\label{subfig:resolved_boosted:parton:ratio}}
\caption{\small{\subref{subfig:resolved_boosted:parton:comparison} Comparison between the measured full phase-space normalised differential cross-sections in the resolved and boosted topologies as a function of the transverse momentum of the top quark. \subref{subfig:resolved_boosted:parton:ratio} The ratios of the measured full phase-space absolute differential cross-sections to the NNLO predictions in the resolved and boosted topologies as a function of the transverse momentum of the top quark. The bands indicate the statistical and total uncertainties of the data in each bin. }}
\label{fig:resolved_boosted:parton}
\end{figure*}

Unlike the particle-level measurements, at parton level the definition of the top-quark observables is identical between the resolved and boosted topologies. This allows a direct comparison to be made between the measured  differential cross-sections as a function of the $\pt$ of the top quark in the two topologies, shown in Figure~\ref{fig:resolved_boosted:parton}. The two measurements are consistent in the overlap region.

\FloatBarrier
\begin{table}[t]
\footnotesize
\centering\noindent\makebox[\textwidth]{
\renewcommand*{\arraystretch}{1.2}\begin{tabular}{|c | r @{/} l r  | r @{/} l r  | r @{/} l r  | r @{/} l r  | r @{/} l r |}
\hline
Observable
& \multicolumn{3}{c|}{\textsc{Pwg+Py8}}& \multicolumn{3}{c|}{\textsc{Pwg+Py8} Rad.~Up}& \multicolumn{3}{c|}{\textsc{Pwg+Py8} Rad.~Down}& \multicolumn{3}{c|}{\textsc{Pwg+H7}}& \multicolumn{3}{c|}{\textsc{Sherpa} 2.2.1}\\
& \multicolumn{2}{c}{$\chi^{2}$/NDF} &  ~$p$-value& \multicolumn{2}{c}{$\chi^{2}$/NDF} &  ~$p$-value& \multicolumn{2}{c}{$\chi^{2}$/NDF} &  ~$p$-value& \multicolumn{2}{c}{$\chi^{2}$/NDF} &  ~$p$-value& \multicolumn{2}{c}{$\chi^{2}$/NDF} &  ~$p$-value\\
\hline
\hline
$m^{t\bar{t}}\textrm{ vs }p_{\mathrm{T}}^{t}$ &{\ } 0.5 & 4 & 0.97 & {\ } 11.6 & 4 & 0.02 & {\ } 4.9 & 4 & 0.30 & {\ } 0.7 & 4 & 0.95 & {\ } 9.0 & 4 & 0.06\\
$p_{\mathrm{T}}^{t}$ &{\ } 4.9 & 5 & 0.43 & {\ } 6.9 & 5 & 0.23 & {\ } 5.0 & 5 & 0.41 & {\ } 4.6 & 5 & 0.46 & {\ } 10.4 & 5 & 0.07\\
$m^{t\bar{t}}$ &{\ } 4.3 & 6 & 0.64 & {\ } 7.5 & 6 & 0.28 & {\ } 19.2 & 6 & $<$0.01 & {\ } 5.4 & 6 & 0.49 & {\ } 5.0 & 6 & 0.55\\
\hline
\end{tabular}}
\caption{Comparison of the measured parton-level normalised differential cross-sections in the boosted topology with the predictions from several MC generators. For each prediction a $\chi^2$ and a $p$-value are calculated using the covariance matrix of the measured spectrum. The NDF is equal to the number of bins in the distribution minus one.}
\label{tab:chisquare:relative:allpred:boosted:parton}
\end{table}
 
\begin{table}[t]
\footnotesize
\centering\noindent\makebox[\textwidth]{
\renewcommand*{\arraystretch}{1.2}\begin{tabular}{|c | r @{/} l r  | r @{/} l r  | r @{/} l r  | r @{/} l r  | r @{/} l r |}
\hline
Observable
& \multicolumn{3}{c|}{\textsc{Pwg+Py8}}& \multicolumn{3}{c|}{\textsc{Pwg+Py8} Rad.~Up}& \multicolumn{3}{c|}{\textsc{Pwg+Py8} Rad.~Down}& \multicolumn{3}{c|}{\textsc{Pwg+H7}}& \multicolumn{3}{c|}{\textsc{Sherpa} 2.2.1}\\
& \multicolumn{2}{c}{$\chi^{2}$/NDF} &  ~$p$-value& \multicolumn{2}{c}{$\chi^{2}$/NDF} &  ~$p$-value& \multicolumn{2}{c}{$\chi^{2}$/NDF} &  ~$p$-value& \multicolumn{2}{c}{$\chi^{2}$/NDF} &  ~$p$-value& \multicolumn{2}{c}{$\chi^{2}$/NDF} &  ~$p$-value\\
\hline
\hline
$m^{t\bar{t}}\textrm{ vs }p_{\mathrm{T}}^{t}$ &{\ } 6.2 & 5 & 0.29 & {\ } 29.6 & 5 & $<$0.01 & {\ } 18.7 & 5 & $<$0.01 & {\ } 3.9 & 5 & 0.56 & {\ } 41.5 & 5 & $<$0.01\\
$p_{\mathrm{T}}^{t}$ &{\ } 4.7 & 6 & 0.58 & {\ } 6.2 & 6 & 0.41 & {\ } 5.8 & 6 & 0.45 & {\ } 4.1 & 6 & 0.67 & {\ } 9.7 & 6 & 0.14\\
$m^{t\bar{t}}$ &{\ } 5.9 & 7 & 0.55 & {\ } 18.8 & 7 & $<$0.01 & {\ } 18.5 & 7 & $<$0.01 & {\ } 6.0 & 7 & 0.54 & {\ } 23.8 & 7 & $<$0.01\\
\hline
\end{tabular}}
\caption{ Comparison of the measured parton-level absolute differential cross-sections in the boosted topology with the predictions from several MC generators. For each prediction a $\chi^2$ and a $p$-value are calculated using the covariance matrix of the measured spectrum. The NDF is equal to the number of bins in the distribution.}
\label{tab:chisquare:absolute:allpred:boosted:parton}
\end{table}
\FloatBarrier
\begin{table}[t]
\centering
\footnotesize
\centering\noindent\makebox[\textwidth]{
\renewcommand*{\arraystretch}{1.2}\begin{tabular}{|c | r @{/} l r  | r @{/} l r |}
\hline
Observable
& \multicolumn{3}{c|}{NNPDF3.1 NNLO}& \multicolumn{3}{c|}{PWG+PY8}\\
& \multicolumn{2}{c}{$\chi^{2}$/NDF} &  ~$p$-value& \multicolumn{2}{c}{$\chi^{2}$/NDF} &  ~$p$-value\\
\hline
\hline
$m^{t\bar{t}}\textrm{ vs }\ptt$ &{\ } 6.2 & 4 & 0.18 & {\ } 0.5 & 4 & 0.97\\
$\ptt$ &{\ } 4.8 & 5 & 0.44 & {\ } 4.9 & 5 & 0.43\\
\hline
\end{tabular}}
\caption{Comparison of the measured parton-level normalised differential cross-sections in the boosted topology with the  NNLO predictions and the nominal \Powheg+\PythiaEight{} predictions. For each prediction a $\chi^2$ and a $p$-value are calculated using the covariance matrix of the measured spectrum. The NDF is equal to the number of bins in the distribution minus one.}
\label{tab:chisquare:relative:NNLO:boosted:parton}
\end{table}
\begin{table}[t]
\centering
\footnotesize
\centering\noindent\makebox[\textwidth]{
\renewcommand*{\arraystretch}{1.2}\begin{tabular}{|c | r @{/} l r  | r @{/} l r |}
\hline
Observable
& \multicolumn{3}{c|}{NNPDF3.1 NNLO}& \multicolumn{3}{c|}{PWG+PY8}\\
& \multicolumn{2}{c}{$\chi^{2}$/NDF} &  ~$p$-value& \multicolumn{2}{c}{$\chi^{2}$/NDF} &  ~$p$-value\\
\hline
\hline
$m^{t\bar{t}}\textrm{ vs }\ptt$ &{\ }  6.3 & 5 & 0.28 & {\ } 6.2 & 5 & 0.29 \\
\ptt{} &{\ } 4.3 & 6 & 0.64 & {\ } 4.7 & 6 & 0.58 \\
\hline
\end{tabular}}
\caption{Comparison of the measured parton-level absolute differential cross-sections in the boosted topology with the  NNLO predictions and the nominal \Powheg+\PythiaEight{} predictions. For each prediction a $\chi^2$ and a $p$-value are calculated using the covariance matrix of the measured spectrum. The NDF is equal to the number of bins in the distribution.}
\label{tab:chisquare:absolute:NNLO:boosted:parton}
\end{table}

\FloatBarrier
\FloatBarrier

\section{Conclusion}
\label{sec:conclusions}
Single- and double-differential cross-sections for the production of top-quark pairs are measured in the \ljets{} channel at particle and parton level, in the resolved and boosted topologies,
using data from $pp$ collisions at $\sqrt{s}=13$~\TeV\ collected  in 2015 and 2016 by the ATLAS detector
at the CERN Large Hadron Collider and corresponding to an integrated luminosity of $36.1\,\mathrm{fb}^{-1}$. The differential cross-sections  are presented as a function of the main kinematic variables of the \ttb{} system, jet multiplicities and observables sensitive to extra QCD radiation and PDFs.
 
The particle-level measurements are compared with NLO+PS MC predictions as implemented
in state-of-the-art MC generators. At the particle level, the predictions agree with the single-differential measurements over a wide kinematic region for both the resolved and boosted topologies, although poorer modelling is observed in specific regions of the probed phase-space. In the boosted topology, which is focused in the region where the hadronic top quark is produced with high \pt{}, a disagreement between the measured inclusive cross-section and several predictions is observed. Overall, the NLO+PS MC generators show poorer modelling of the double-differential distributions and no combination that includes $\pttt$ can be described by the generators in the resolved topology. Overall, the \Powheg+\PythiaEight{} and, in the boosted topology, \Powheg+\herwig{}7 are the two generators able to give a good prediction of the largest fraction of the probed variables.
The measurements show high sensitivity to the different aspects of the predictions of the MC generators and are hence relevant for the tuning of the MC generators and will contribute to improving the description of the \ttbar{} final state and to reducing the systematic uncertainties related to top-quark modelling.
 
The measured parton-level differential cross-sections are compared with state-of-the-art fixed-order NNLO QCD predictions and a general improvement relative to the NLO+PS MC generators is found in the level of agreement of the single- and double-differential cross-sections in both the resolved and boosted regimes. The comparison of double-differential distributions with NNLO predictions provides a very stringent test of the SM description of \ttbar{} production. The comparison with the NNLO pQCD predictions including EW corrections, due to the still rather limited range probed for the measured transverse momenta of the top and anti-top quarks, does not yet allow the impact of the EW corrections in the production of top-quark pairs to be quantified.
 
The measured differential cross-sections at the parton level will be able to be used in detailed phenomenological studies and in particular to improve the determination of the gluon density in the proton and of the top-quark pole mass.

\FloatBarrier

\section*{Acknowledgements}
 
 
We thank CERN for the very successful operation of the LHC, as well as the
support staff from our institutions without whom ATLAS could not be
operated efficiently.
 
We acknowledge the support of ANPCyT, Argentina; YerPhI, Armenia; ARC, Australia; BMWFW and FWF, Austria; ANAS, Azerbaijan; SSTC, Belarus; CNPq and FAPESP, Brazil; NSERC, NRC and CFI, Canada; CERN; CONICYT, Chile; CAS, MOST and NSFC, China; COLCIENCIAS, Colombia; MSMT CR, MPO CR and VSC CR, Czech Republic; DNRF and DNSRC, Denmark; IN2P3-CNRS, CEA-DRF/IRFU, France; SRNSFG, Georgia; BMBF, HGF, and MPG, Germany; GSRT, Greece; RGC, Hong Kong SAR, China; ISF and Benoziyo Center, Israel; INFN, Italy; MEXT and JSPS, Japan; CNRST, Morocco; NWO, Netherlands; RCN, Norway; MNiSW and NCN, Poland; FCT, Portugal; MNE/IFA, Romania; MES of Russia and NRC KI, Russian Federation; JINR; MESTD, Serbia; MSSR, Slovakia; ARRS and MIZ\v{S}, Slovenia; DST/NRF, South Africa; MINECO, Spain; SRC and Wallenberg Foundation, Sweden; SERI, SNSF and Cantons of Bern and Geneva, Switzerland; MOST, Taiwan; TAEK, Turkey; STFC, United Kingdom; DOE and NSF, United States of America. In addition, individual groups and members have received support from BCKDF, CANARIE, CRC and Compute Canada, Canada; COST, ERC, ERDF, Horizon 2020, and Marie Sk{\l}odowska-Curie Actions, European Union; Investissements d' Avenir Labex and Idex, ANR, France; DFG and AvH Foundation, Germany; Herakleitos, Thales and Aristeia programmes co-financed by EU-ESF and the Greek NSRF, Greece; BSF-NSF and GIF, Israel; CERCA Programme Generalitat de Catalunya, Spain; The Royal Society and Leverhulme Trust, United Kingdom.
 
The crucial computing support from all WLCG partners is acknowledged gratefully, in particular from CERN, the ATLAS Tier-1 facilities at TRIUMF (Canada), NDGF (Denmark, Norway, Sweden), CC-IN2P3 (France), KIT/GridKA (Germany), INFN-CNAF (Italy), NL-T1 (Netherlands), PIC (Spain), ASGC (Taiwan), RAL (UK) and BNL (USA), the Tier-2 facilities worldwide and large non-WLCG resource providers. Major contributors of computing resources are listed in Ref.~\cite{ATL-GEN-PUB-2016-002}.
 
 
\FloatBarrier

\printbibliography

\clearpage
 
\begin{flushleft}
{\Large The ATLAS Collaboration}

\bigskip

G.~Aad$^\textrm{\scriptsize 101}$,    
B.~Abbott$^\textrm{\scriptsize 128}$,    
D.C.~Abbott$^\textrm{\scriptsize 102}$,    
A.~Abed~Abud$^\textrm{\scriptsize 70a,70b}$,    
K.~Abeling$^\textrm{\scriptsize 53}$,    
D.K.~Abhayasinghe$^\textrm{\scriptsize 93}$,    
S.H.~Abidi$^\textrm{\scriptsize 167}$,    
O.S.~AbouZeid$^\textrm{\scriptsize 40}$,    
N.L.~Abraham$^\textrm{\scriptsize 156}$,    
H.~Abramowicz$^\textrm{\scriptsize 161}$
% ...
\bigskip
\\

% ...
$^{40}$Niels Bohr Institute, University of Copenhagen, Copenhagen; Denmark.\\
% ...
$^{53}$II. Physikalisches Institut, Georg-August-Universit\"{a}t G\"ottingen, G\"ottingen; Germany.\\
% ...
$^{70}$$^{(a)}$INFN Sezione di Pavia;$^{(b)}$Dipartimento di Fisica, Universit\`a di Pavia, Pavia; Italy.\\
% ...
$^{93}$Department of Physics, Royal Holloway University of London, Egham; United Kingdom.\\
% ...
$^{101}$CPPM, Aix-Marseille Universit\'e, CNRS/IN2P3, Marseille; France.\\
$^{102}$Department of Physics, University of Massachusetts, Amherst MA; United States of America.\\
% ...
$^{128}$Homer L. Dodge Department of Physics and Astronomy, University of Oklahoma, Norman OK; United States of America.\\
% ...
$^{156}$Department of Physics and Astronomy, University of Sussex, Brighton; United Kingdom.\\
% ...
$^{161}$Raymond and Beverly Sackler School of Physics and Astronomy, Tel Aviv University, Tel Aviv; Israel.\\
% ...
$^{167}$Department of Physics, University of Toronto, Toronto ON; Canada.\\
% ...
$^{a}$ Also at Borough of Manhattan Community College, City University of New York, New York NY; United States of America.\\
$^{b}$ Also at CERN, Geneva; Switzerland.\\
% ...

\end{flushleft}

 
\end{document}